%% file: iosart2x.tex
\documentclass[jnr]{iosart2x}

\usepackage{hyperref}
\usepackage{appendix}
\usepackage{booktabs}
\usepackage[utf8]{inputenc}
\usepackage{graphicx} 
\usepackage{times} 		% Add graphics capabilities
\usepackage{amsmath}  		% Better maths support
\usepackage{natbib}	% bibliography style
\usepackage{eurosym}
\usepackage{xcolor}
\usepackage{color, colortbl}
\usepackage{longtable}
\usepackage{wrapfig}
\usepackage{textcomp}
\usepackage[font=small,labelfont=bf]{caption}
\usepackage{pagecolor,lipsum}% http://ctan.org/pkg/{pagecolor,lipsum}
\usepackage{fancyvrb}
\usepackage{rotating}
\usepackage{subcaption}
\usepackage{caption}
\usepackage{amsmath}
\usepackage{array}

\usepackage[noabbrev, capitalize]{cleveref}

\usepackage{siunitx}
\DeclareSIUnit\angstrom{\text {Å}}%Added by Nicola
\usepackage{booktabs}%Added by Nicola
\usepackage{placeins}%Added by Nicola
\usepackage[version=4]{mhchem}%Added by Nicola

\usepackage[version=4]{mhchem}
\usepackage{float}
\usepackage{multirow} %added by Ben
\usepackage[labelformat=simple]{subcaption}
\DeclareCaptionLabelFormat{andtable}{#1~#2  \&  \tablename~\thetable} %Added by Nicola

\usepackage{makecell}

\usepackage{array}
\usepackage{makecell}
\usepackage{subfloat}

\usepackage{booktabs}

\usepackage[version=4]{mhchem}
\usepackage{appendix}
\usepackage{listings}
\usepackage{bm}
\newcommand*\diff{\mathop{}\!\mathrm{d}}        %% differential symbol, \diff
\newcommand*\Diff[1]{\mathop{}\!\mathrm{d^#1}}  %% high order differential symbol, \Diff3
\usepackage[frozencache=true]{minted}

\usepackage{enumitem}
\setitemize{noitemsep,topsep=0pt,parsep=0pt,partopsep=0pt,leftmargin=*}
\usepackage{amssymb}

\usepackage{enumitem}
\usepackage{fancyhdr}
\newenvironment{itemize*}%
  {\begin{itemize}%
    \setlength{\itemsep}{0pt}%
    \setlength{\parskip}{0pt}}%
  {\end{itemize}}

\usepackage{enumitem}

\newcommand{\babar}{\mbox{\slshape B\kern-0.1em{\smaller A}\kern-0.1em B\kern-0.1em{\smaller A\kern-0.2em R} }}

\newcolumntype{g}{>{\columncolor[gray]{.8}}c}
\newcolumntype{w}{>{\columncolor{white}}c}

\usepackage{dcolumn}

\newcolumntype{d}[1]{D{.}{.}{#1}}

\firstpage{1}
\lastpage{5}
\volume{1}
\pubyear{2021}

\usepackage{alphalph}
\renewcommand{\theaddress}{\alphalph{\value{address}}}

\begin{document}
\makeatletter
\let\put@numberlines@box\relax
\makeatother
\begin{frontmatter} % The preamble begins here.

%
%\pretitle{Pretitle}
\title{HighNESS Conceptual Design Report: Volume I} %title of paper
\runtitle{HighNESS Conceptual Design Report: Volume I}

\begin{aug}
\author[1,2]{V.~Santoro}
\author[3]{O.~Abou~El~Kheir}
\author[3]{D.~Acharya}
\author[4]{M.~Akhyani}
\author[5]{K.H.~Andersen}
\author[6,7]{J.~Barrow}
\author[1]{P.~Bentley}
\author[3]{M.~Bernasconi}
\author[1]{M.~Bertelsen}
\author[8]{Y.~Be{\ss}ler}
\author[1]{A.~Bianchi}
\author[9]{G.~Brooijmans}
\author[5]{L.~Broussard}
\author[1]{T.~Brys}
\author[10]{M.~Busi}
\author[3]{D.~Campi}
\author[11]{A.~Chambon}
\author[8]{J.~Chen} 
\author[12]{V.~Czamler}
\author[1]{P.~Deen}
\author[1]{D.D.~DiJulio}
\author[13,14]{E.~Dian}
\author[14]{L.~Draskovits}
\author[15]{K.~Dunne}
\author[8]{M.~El Barbari}
\author[1]{M.J.~Ferreira,}   
\author[16]{P.~Fierlinger}
\author[17]{V.T.~Fr\"ost }
\author[1,3]{B.T.~Folsom}
\author[1]{U.~Friman-Gayer}
\author[1]{A.~Gaye}
\author[3]{G.~Gorini}
\author[17]{A.~Gustafsson} 
\author[8]{T.~Gutberlet}
\author[8]{C.~Happe}
\author[28,29,30]{X. Han}
\author[1]{M.~Hartl}
\author[1]{M.~Holl}
\author[1]{A.~Jackson}
\author[18]{E.~Kemp}
\author[19]{Y.~Kamyshkov}
\author[1]{T. Kittelmann}
\author[11]{E.B.~Klinkby}
\author[20]{R.~Kolevatov}
\author[3]{S.I.~Laporte}
\author[11]{B.~Lauritzen}
\author[15]{W.~Lejon}
\author[1]{R.~Linander}
\author[1]{M.~Lindroos}
\author[14]{M. Marko} 
\author[1]{J.I.~M\'arquez~Dami\'an}
\author[5]{T. C.~McClanahan}
\author[15,2]{B.~Meirose}
\author[13]{F.~Mezei}
\author[1]{K.~Michel}
\author[15]{D.~Milstead}
\author[1]{G.~Muhrer}
\author[21]{A.~Nepomuceno}
\author[12]{V.~Neshvizhevsky}
\author[22]{T.~Nilsson}
\author[1]{U.~Od\'en}
\author[23]{T.~Plivelic}
\author[1]{K.~Ramic}
\author[1,2]{B.~Rataj}
\author[5]{I.~Remec}
\author[11]{N.~Rizzi}
\author[19]{J.~Rogers}
\author[8]{E.~Rosenthal}
\author[14]{L.~Rosta}
\author[8]{U.~R\"ucker}
\author[10]{S.~Samothrakitis}
\author[24]{A.~Schreyer}
\author[1]{J.R.~Selknaes}
\author[13]{H.~Shuai}
\author[15]{S.~Silverstein}
\author[25]{W.M.~Snow}
\author[10]{M.~Strobl}
\author[8]{M.~Strothmann}
\author[1]{A.~Takibayev}
\author[12]{R.~Wagner}
\author[1,11]{P.~Willendrup}
\author[1]{S.~Xu}
\author[15]{S.C.~Yiu}  
\author[17]{L.~Yngwe} 
\author[26]{A.R.~Young}
\author[27]{M.~Wolke}  %Temporarily remove one author to get it to compile
\author[8]{P.~Zakalek}
\author[5]{L.~Zavorka}
\author[1]{L.~Zanini}
\author[12]{O.~Zimmer}

% Affiliations can be added in the order they should appear. For breaks in addresses, use either \\ or \tabularnewline
\address[1]{European Spallation Source ERIC, Lund, Sweden}
\address[2]{Lund University, Lund, Sweden}
\address[3]{University of Milano-Bicocca, Milano, Italy}
\address[4]{\'Ecole Polytechnique F\'ed\'erale de Lausanne (EPFL), Lausanne, Switzerland}
\address[5]{Oak Ridge National Laboratory, Oak Ridge, USA}
\address[6]{Massachusetts Institute of Technology (MIT), Cambridge, USA}
\address[7]{Tel Aviv University, Tel Aviv, Israel}
\address[8]{Forschungszentrum J\"ulich GmbH, J\"ulich, Germany }
\address[9]{Department of Physics, Columbia University, New York, USA}
\address[10]{Paul Scherrer Institut (PSI), Villigen, Switzerland}
\address[11]{DTU Physics, Technical University of Denmark, Lyngby, Denmark}
\address[12]{Institut Laue-Langevin ILL, Grenoble, France}
\address[13]{Mirrotron Ltd., Budapest, Hungary}
\address[14]{Centre for Energy Research, Budapest, Hungary}
\address[15]{Stockholm University, Stockholm, Sweden}
\address[16]{Technical University of Munich, Garching, Germany}
\address[17]{Sweco AB, Malm\"o, Sweden}
\address[28]{Institute of High Energy Physics, Chinese Academy of Science, Beijing, 100049, China}
\address[29]{University of Chinese Academy of Science, Beijing, 100049, China}
\address[30]{Spallation neutron source science center, Dongguan, 523803, Guangdong, China}
\address[18]{State University of Campinas, Campinas, Brazil}
\address[19]{University of Tennessee, Knoxville, USA}
\address[20]{ESS consultant, Oslo, Norway}
\address[21]{Departamento de Ci\^encias da Natureza, Universidade Federal Fluminense, Niter\'oi, Brazil}
\address[22]{Institutionen f{\"o}r Fysik, Chalmers Tekniska H\"{o}gskola, Sweden}
\address[23]{MAX IV Synchrotron, Lund University, Lund, Sweden}
\address[24]{Helmholtz-Zentrum hereon GmbH, Geesthacht, Germany}
\address[25]{Department of Physics, Indiana University, Bloomington, USA}
\address[26]{Department of Physics, North Carolina State University, Raleigh, USA}
\address[27]{Department of Physics and Astronomy, Uppsala University, Uppsala, Sweden}
\end{aug}

\newpage

\begin{abstract}
The European Spallation Source, currently under construction in Lund, Sweden, is a multidisciplinary international laboratory. Once completed to full specifications, it will operate the world's most powerful pulsed neutron source. Supported by a 3 million Euro Research and Innovation Action within the EU Horizon 2020 program, a design study (HighNESS) has been completed to develop a second neutron source located below the spallation target. Compared to the first source, designed for high cold and thermal brightness, the new source has been optimized to deliver higher intensity, and a shift to longer wavelengths in the spectral regions of cold (CN, 2--20\,\AA), very cold (VCN, 10--120\,\AA), and ultracold (UCN, ${>}\,{500}$\,\AA) neutrons. The second source comprises a large liquid deuterium moderator designed to produce CN and support secondary VCN and UCN sources. Various options have been explored in the proposed designs, aiming for world-leading performance in neutronics. These designs will enable the development of several new instrument concepts and facilitate the implementation of a high-sensitivity neutron-antineutron oscillation experiment (NNBAR). This document serves as the Conceptual Design Report for the HighNESS project, representing its final deliverable.
\end{abstract}
\end{frontmatter}

\newpage
%\tableofcontents 
\newpage
\section*{List of Acronyms}
\begin{table}[!htbp]
{\small
\begin{tabular}{lr}
 \hline
{\bf Acronym/term} & {\bf Meaning}  \\
\toprule
BNV & Baryon number violation \\
BZ & Brillouin zone \\
BNC & Budapest Neutron Center \\
CAD & Computer Aided Design \\ 
COMSOL & A finite element analysis and simulation software package \\ 
CDR & Conceptual design report \\ 
CEF &Current ENDF Format  \\
CMTF & Cold Moderator Test Facility  \\
CN & Cold neutron \\
CNS & Cold neutron source \\
COSY & COoler SYnchrotron \\
DF-DND & Deagglomerated F-DND \\
DFPT & Density Functional Perturbation Theory \\
DFT & Density Functional Theory\\
DFTB & Density Functional Tight Binding\\
DOS & Density of States\\
DND & Detonation Nanodiamond\\
EDM & Electric dipole moment \\
EGO & Efficient global optimization \\
ESS & European Spallation Source \\
F-DND &Fluorinated DND \\
FOM & Figure of merit \\ 
FRM-II & Forschungsreaktor München II \\
FZJ &  Forschungszentrum Jülich\\
GEANT4 & A Monte Carlo simulation program for GEometry ANd Tracking \\
HDPE &  High-density polyethylene \\
HiCANS & High Current Accelerator-driven Neutron Source \\
%HighNESS & HIGH intensity Neutron source at the ESS \\
HighNESS & High intensity Neutron Source at the European Spallation Source \\
IKP & Institute of Nuclear Physics \\
ILL & Institut Laue Langevin \\
JCNS & Jülich Centre for Neutron Science  \\
JULIC & JUelich Light Ion Cyclotron  \\
LANSCE & Los Alamos Neutron Science Center \\
 LANL & Los Alamos National Laboratory \\
 LBP & Large beamport \\
 LD$_2$ & Liquid deuterium \\
 LH$_2$ & Liquid hydrogen \\
\bottomrule
\end{tabular}}
\end{table}
\newpage
\section*{List of Acronyms (cont.)}
\begin{table}[!htbp]
{\small
\begin{tabular}{lr}
 \hline
{\bf Acronym/term} & {\bf Meaning}  \\
\toprule
 MCB & Moderator cooling block \\
 MCNP & Monte Carlo N Particle \\
 MCPL & Monte Carlo Particle Lists \\ 
 ML & Machine Learning \\
 MSD & Mean squared displacement \\
 MEF & Mixed Elastic Format \\
 MP & Monoplanar reflector \\
 NMO & Nested Mirror Optics \\
 NNBAR & An experiment to search for neutrons converting to anti-neutrons at the ESS \\
 ND & Nanodiamond\\
 NW & North-west\\
 PBE & Perdew-Burke-Ernzerhof\\
PHITS & Particle and Heavy Ion Transport code System \\
 PMT & Photo-Multiplier Tube \\
PSI & Paul Scherrer Institut \\ 
R\&D & Research and development \\
SiPM & Silicon photomultiplier \\ 
SANS &Small Angle Neutron Scattering\\
SD$_2$ & Solid deuterium \\ 
SNR  & Signal-to-Noise ratio \\ 
SW & South-west \\ 
TEM & Transmission electron microscopy \\
TMR & Target-Moderator-Reflector \\
SNS & Spallation Neutron Source \\
TSL & Thermal Scattering Library\\
 UCN & Ultracold neutron \\ 
VCN & Very cold neutron \\
VDOS & Vibrational density of states\\
WP & Work package \\
\bottomrule
\end{tabular}}
\end{table}

%%%%%%% INTRO %%%%%
%\begin{center}
%\LARGE{\textbf{HighNESS Conceptual Design Report}}\\
%\end{center}

\input{Introduction}

\FloatBarrier \newpage
%%%%%% Luca is the reviewers and Ignacio David and Markus
%
%%%%% COLD SOURCE %%%%
%\begin{center}
%\LARGE{\textbf{HighNESS Conceptual Design Report (Cold Source)}}\\
%\end{center}
\input{coldsource}

\FloatBarrier \newpage
%%%%%%%Ask Franz Gallmeier ????
%
%%%%%%% VCN SOURCE %%%%%
%\begin{center}
%\LARGE{\textbf{HighNESS Conceptual Design Report (Very Cold Neutron) }}\\
%\end{center}
\input{vcnsource}
\FloatBarrier \newpage
%%% Reviewers  Egor 
%
%%%%% UCN SOURCE %%%%%
%\begin{center}
%\LARGE{\textbf{HighNESS Conceptual Design Report (Ultra Cold Neutron) }}\\
%\end{center}
\input{ucnsource}
\FloatBarrier \newpage
%%%%%Reviewers Oliver Zimmer  Skyler and integration parts  
%
%%%%%% NEUTRON SCATTERING %%%%%%
%
%\begin{center}
%\LARGE{\textbf{HighNESS Conceptual Design Report (Neutron Scattering)}}\\
%\end{center}
\input{neutronscattering}

\FloatBarrier \newpage
%%%%Reviewers Ken Andersen or Andrew Jackson 
%
%%%%%% NNBAR %%%%%
%\begin{center}
%\LARGE{\textbf{HighNESS Conceptual Design Report (NNBAR) }}\\
%\end{center}
% \input{NNBAR-sections/nnbar-cdr}
% \FloatBarrier \newpage
%%%% reviewers Yuri 
%
%%%%% TSL %%%%%%
%\begin{center}
%\LARGE{\textbf{HighNESS Conceptual Design Report (Thermal Scattering Libraries) }}\\
%\end{center}
\input{tsl}
\FloatBarrier \newpage
%%%%Reviewes Rolando Richard or Oliver % Nicola 
%
%%%%% ADVANCED REFLECTORS %%%%%
%\begin{center}
%\LARGE{\textbf{HighNESS Conceptual Design Report (Advanced Reflector) }}\\
%\end{center}

\input{adreflectors}
\FloatBarrier \newpage
%%%%Reviewers Doug and Ignacio 
%
%%%%% CONCLUSIONS %%%%%
%\begin{center}
%\LARGE{\textbf{HighNESS Conceptual Design Report (Conclusions)}}\\
%\end{center}
\input{conclusions}
\FloatBarrier \newpage
%
%%%% APPENDICES %%%%%

\begin{appendices}

\end{appendices}

%\bibliographystyle{plain}

%\AtNextBibliography{}
%\newpage
%{\footnotesize \printbibliography}
\bibliographystyle{plain} % We choose the "plain" reference style
\bibliography{biblio} % Entries are in the refs.bib file

\end{document}

%% file: Introduction.tex
\newpage
\section{Introduction and Scientific Motivations }

\label{sec:Introduction}

\subsection{The European Spallation Source }
\label{essdes}

The European Spallation Source (ESS) is a cutting-edge scientific research facility currently under construction in Lund, Sweden. When completed to full specifications, it will be the world's most powerful accelerator-based source of neutrons for scientific research~\cite{Garoby_2017}. This unparalleled capability will open new scientific avenues across various disciplines, including materials science, life sciences, energy research, environmental technology, and fundamental physics.

Currently, the ESS is actively constructing 15 instruments, which represent a subset of the envisioned 22-instrument suite essential for fulfilling the facility's scientific mission, as outlined in the ESS statutes. Notably, the ESS mandate encompasses a fundamental physics program and the absence of a dedicated beamline for fundamental physics has been identified as a critical gap~\cite{ess-gap}.

The remarkable neutron flux produced by ESS is due to its housing the world's most powerful linear accelerator and the high beam power directed on target. At full design capability, the proton beam operates at a current of 62.5\,mA and is accelerated to 2\,GeV, employing a 14 Hz pulse structure, with each pulse lasting 2.86 ms. Consequently, this configuration yields an average power of 5\,MW and a peak power of 125\,MW.

\begin{figure}[h!]
	\centering
	\includegraphics[width=10cm]{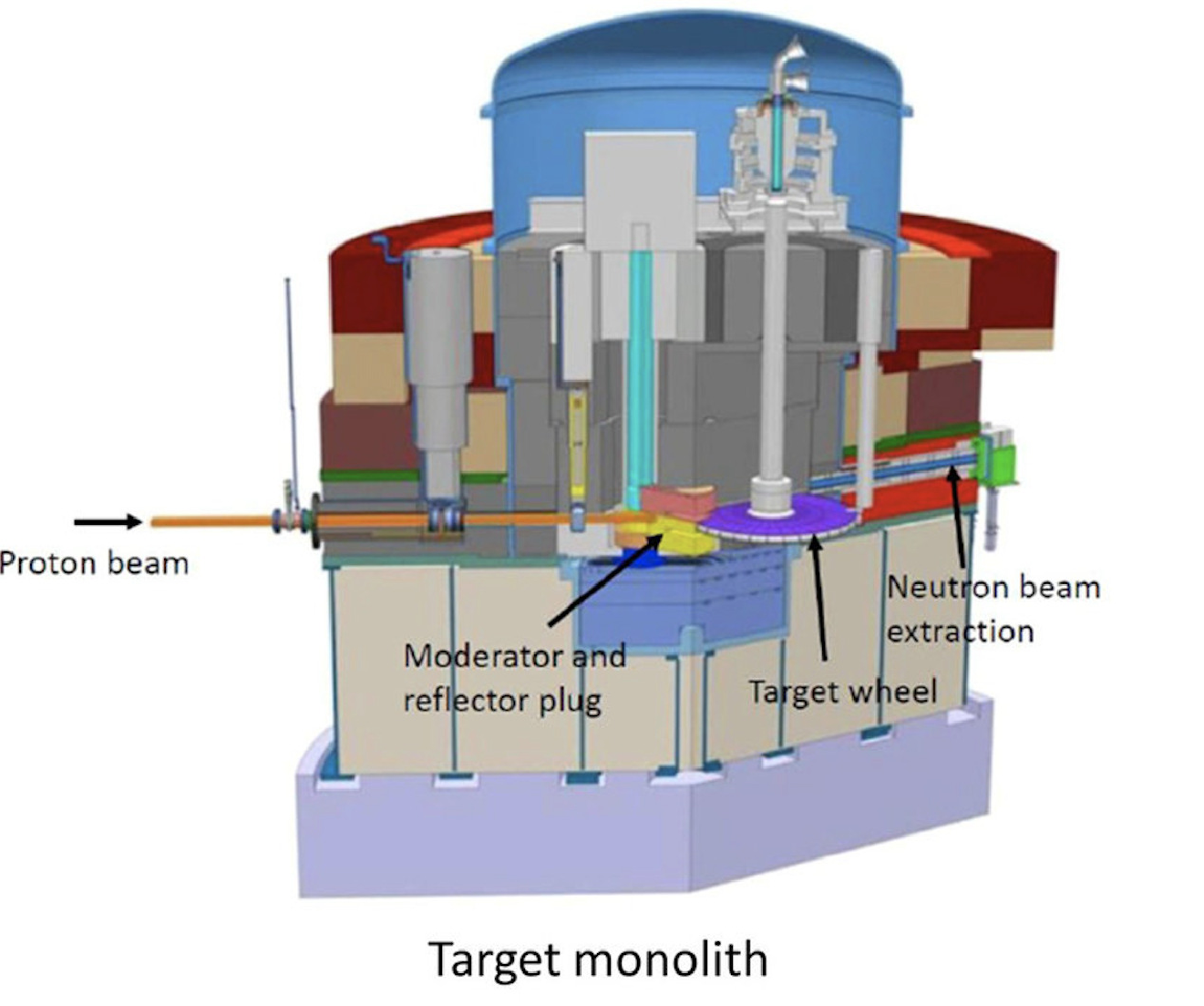}
 \caption{The ESS target monolith with key components indicated: moderator and reflector plug, spallation target, and one of the 42 neutron beam ports.}
	\label{monolith}
\end{figure}

\begin{figure}[h!]
	\centering
	\includegraphics[width=.8\textwidth]{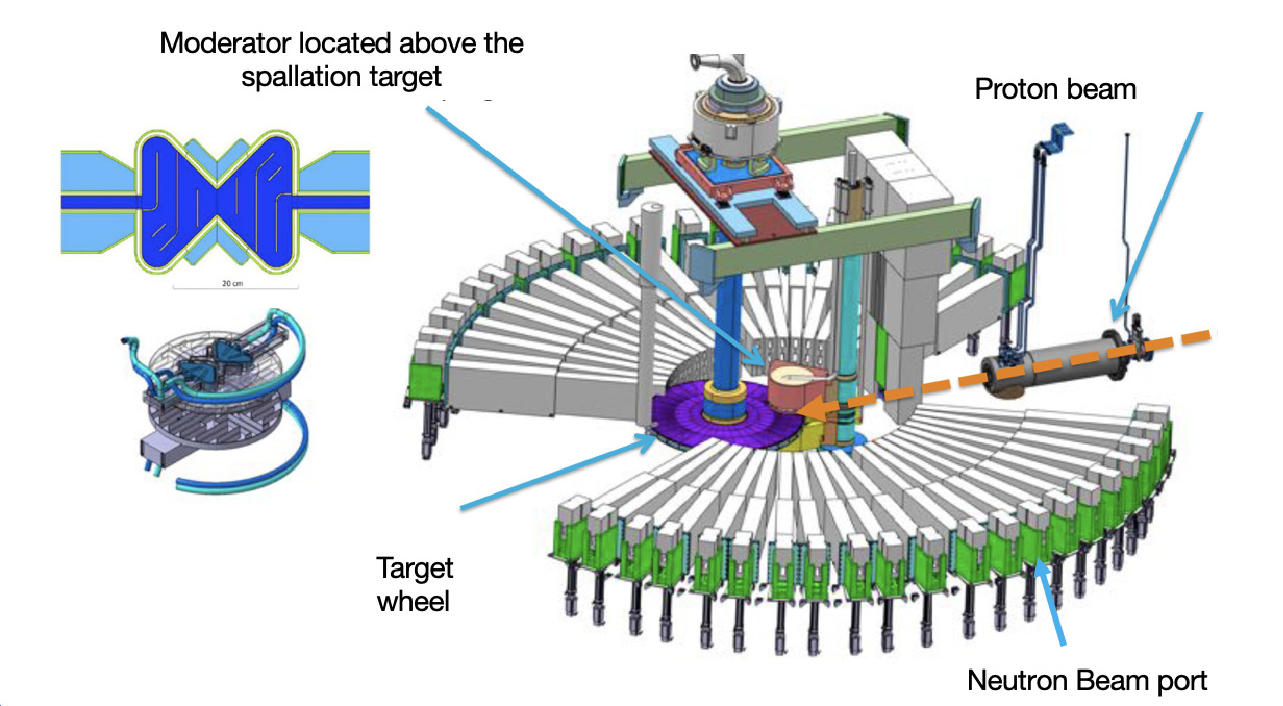}
 \caption{The ESS target-moderator system. The picture depicts the proton beam, the spallation target, the structure housing the moderator, and the neutron beam ports.}
	%\centerfloat	
	%form the layer
 
	\label{targetmoderator}
\end{figure}

Once the proton beam reaches its final energy, it impacts a rotating tungsten target (see \cref{monolith}), inducing spallation and primarily generating evaporation neutrons with energies around 2\,MeV. The high-energy spallation neutrons are decelerated within the neutron moderators situated inside the moderator-reflector plug, as described in~\cref{uppermoderator} and shown in~\cref{targetmoderator}.

%This configuration yields an average power of 5\,MW\footnote{It's worth noting that ESS is currently committed to delivering 2\,MW as accelerator power, with a planned upgrade to 5\,MW.} 

\begin{figure}
\begin{center}
\includegraphics[width=.86\textwidth]{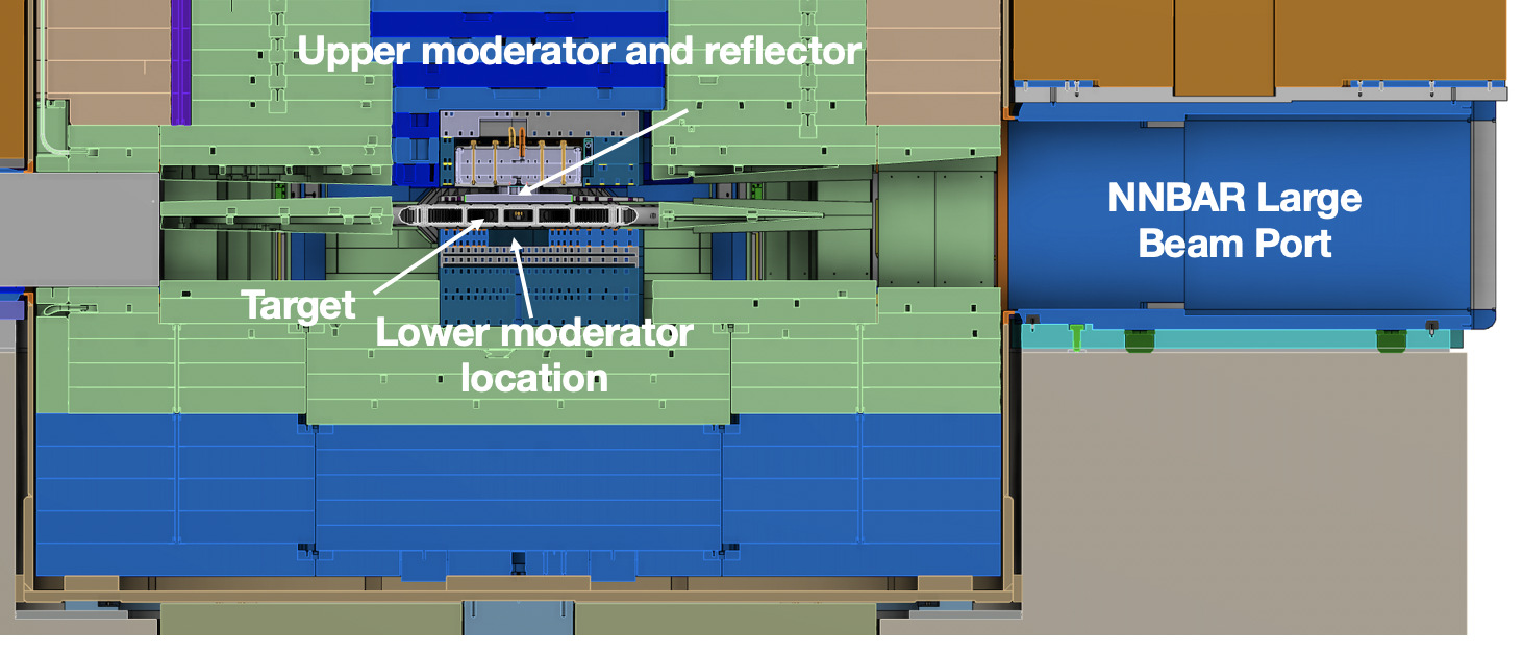}
\caption{Cross-sectional view of the ESS target/moderator area and inner shielding. The figure displays the location of the ESS moderator above the spallation target, referred to as the ``upper moderator", and the moderator below the spallation target, referred to as the ``lower moderator". The NNBAR experiment (see \cref{sec:essupgrade}) will utilize both moderators.} %The location is initially occupied by steel shielding in the first version of the moderator-reflector plug. }
\label{targetarea}
\end{center}
\end{figure}

%\begin{figure}
%\begin{center}
%\includegraphics[width=.76\textwidth]{int_fig/TM1.eps}
%\caption{ The ESS target monolith with key components indicated: moderator and reflector plug, spallation target, and the 42 neutron beamports.

%}
%\label{monolith}
%\end{center}
%\end{figure}

The target area is encircled by a cylindrical steel structure with an 11-meter diameter. This structure, consisting of 6000 tons of steel shielding, is known as the 'monolith'. The monolith is designed to absorb most of the beam power and is capable of withstanding the heat generated during operations. Additionally, the monolith includes a cooling system that utilizes pressurized helium gas. This cooling system effectively reduces the peak temperature by 150°C between successive beam shots. % 
This structure also houses the neutron beam ports, which are essential for extracting thermal and cold neutrons from the moderators. ESS has positioned its beam port system around the moderators, allowing for neutron extraction both above and below the target as shown in \cref{targetarea}. 

\begin{figure}
\begin{center}
\includegraphics[width=.76\textwidth]{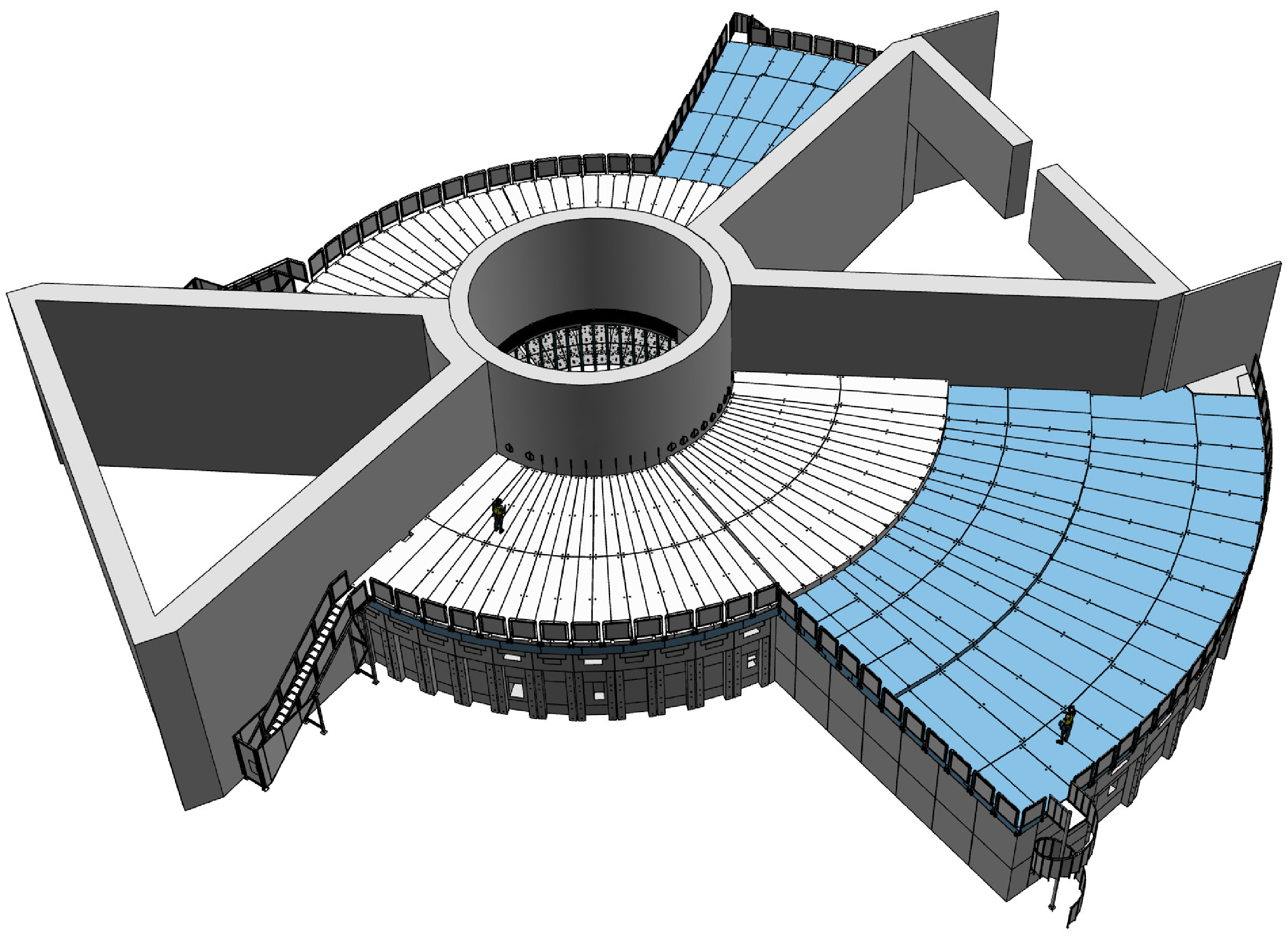}
\caption{CAD drawing of the ESS bunker, depending on the sector, the bunker has different lengths.}
\label{bunkersk}
\end{center}
\end{figure}

Beyond the ESS monolith, the beamlines reside in the bunker, as referenced in~\cite{bunkerpaper}. The bunker serves as a crucial shielding area, enveloping the ESS monolith and shielding the instrument area from the significant ionizing radiation generated during operation. The shielding structure of the bunker consists of 3.5-meter-thick walls (see \cref{bunkersk}) constructed from heavy magnetite concrete, complemented by a roof of variable thickness, also composed of heavy concrete.

\begin{figure}
\begin{center}
\includegraphics[width=.76\textwidth]{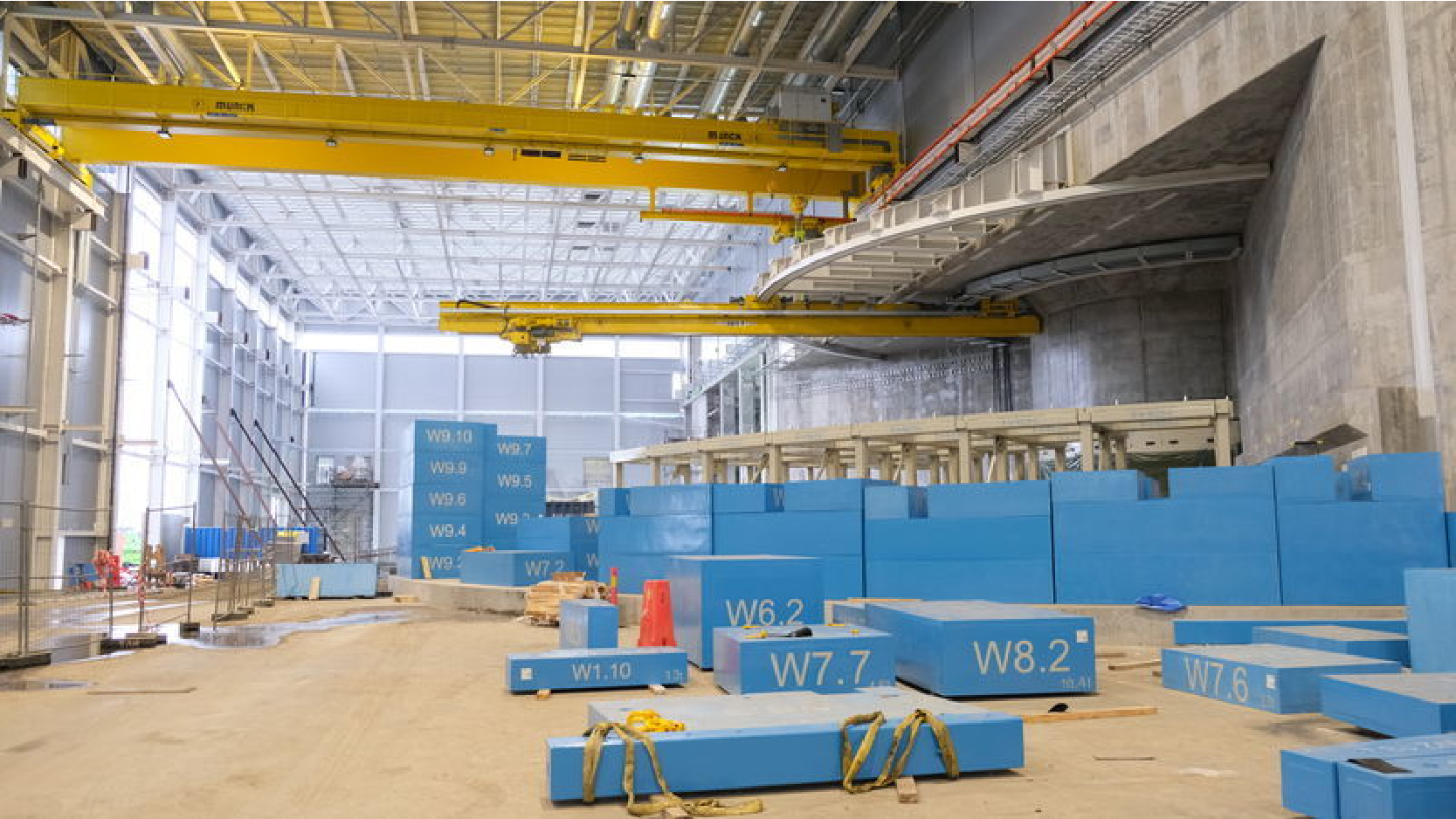}
\caption{Overview of the complete bunker area in the ESS D03 instrument hall.}
\label{bunkerpicture}
\end{center}
\end{figure}

\begin{figure}
\begin{center}
\includegraphics[width=.96\textwidth]{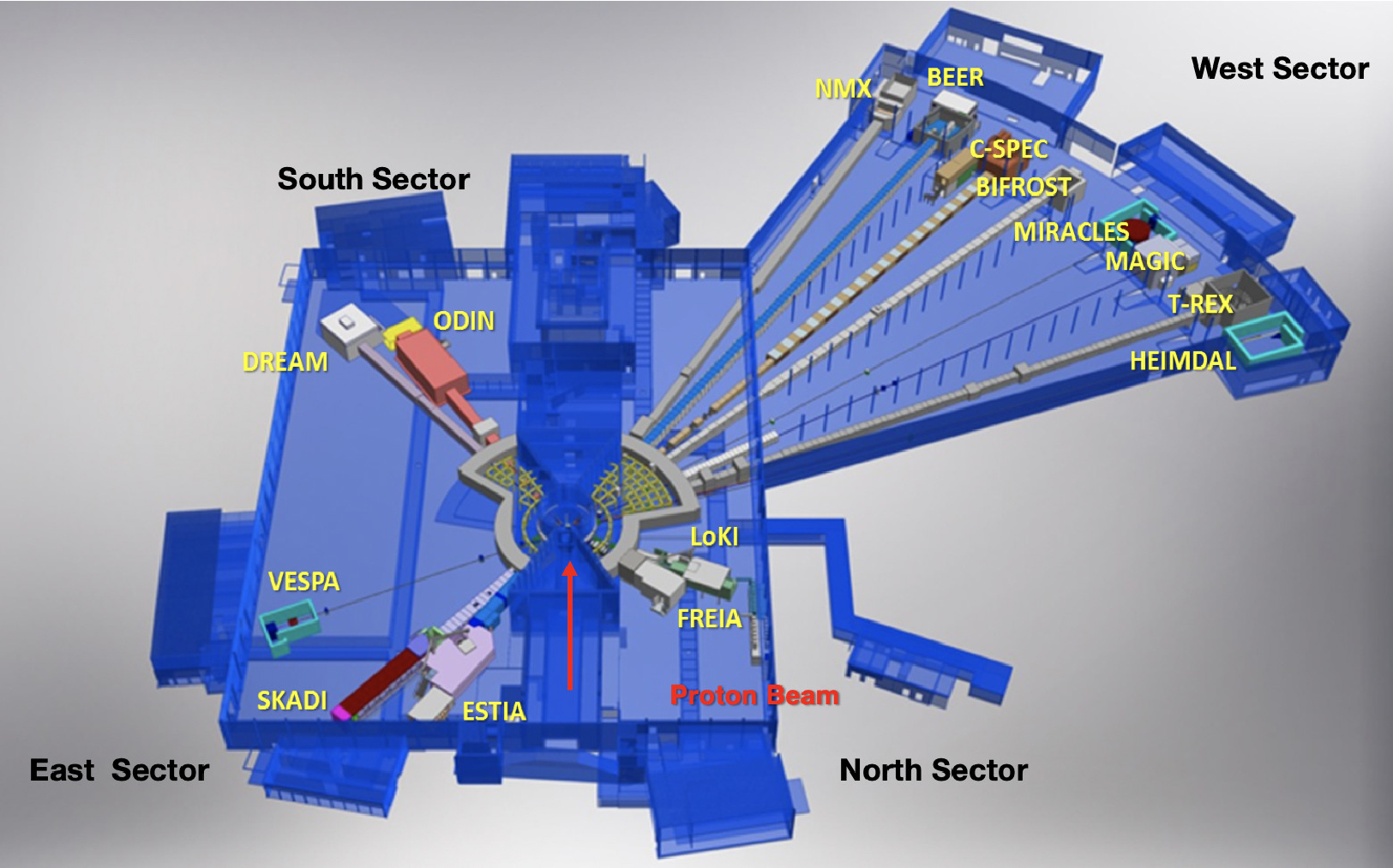}
\caption{Overview of the 15 instruments currently under construction at the ESS and the instrument halls.}
\label{essinstruments}
\end{center}
\end{figure}

\begin{figure}
\begin{center}
\includegraphics[width=.96\textwidth]{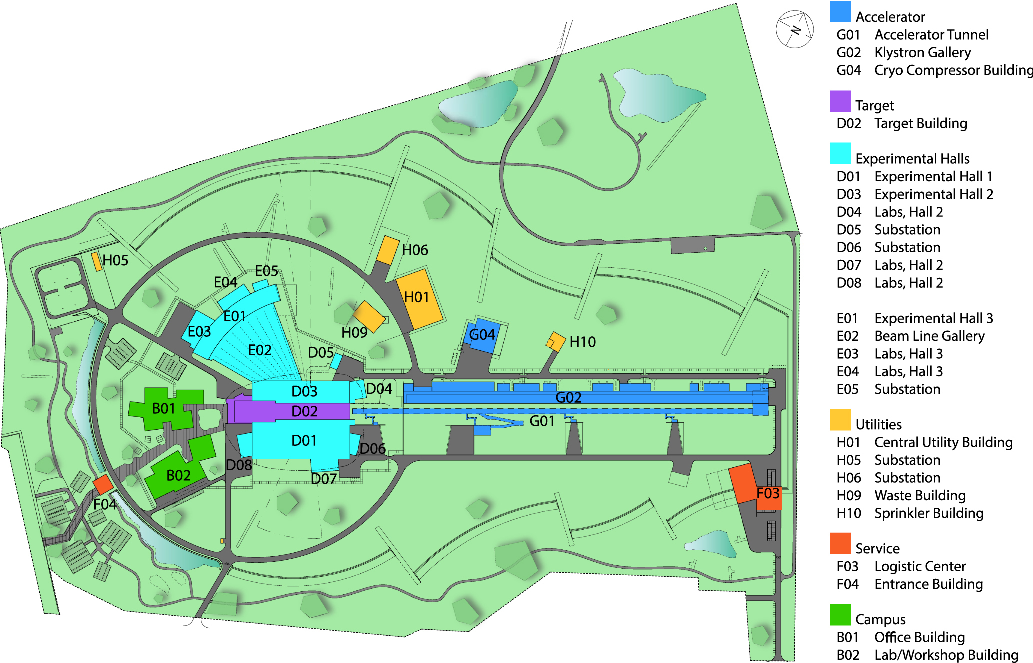}
\caption{ESS layout and building description.}
\label{esslayout}
\end{center}
\end{figure}

\cref{bunkerpicture} shows the completed bunker in one of the instrument hall. After the bunker, the neutron beamlines reach the instrument halls, as shown in \cref{essinstruments}.  \cref{esslayout} shows the complete layout of the facility. 
The instrument hall areas are located in the D01, D03, and E01 buildings.

%The exceptional neutron flux of the ESS can be attributed \textcolor{red}{(change: is due)} to its possession of the world's most powerful accelerator and \textcolor{red}{consequently} the highest beam power on target.
%The proton beam at full design specifications will have  a current of 62.5 mA, proton are  accelerated to 2 GeV and follows a pulse structure of 14 Hz, with each pulse lasting 2.86 ms. Consequently, this configuration yields an average power of 5\,MW and a peak power of 125\,MW. Once the proton beam  reaches its final energy, it impacts a rotating tungsten target (see \cref{targetmonolith}), inducing spallation and generating primarily evaporation neutrons with energies around 2 MeV.
%The user program is scheduled to begin in 2027-2028 when a suite of 15 neutron scattering instruments will be installed \textcolor{red}{(change: after the installation of the first 15 instruments will be completed)}. The maximum accelerator power and proton beam energy will be 2\,MW and 800 MeV, respectively. The possibility of operating the accelerator at 2 GeV beam energy with 5\,MW power would then be part of an ESS upgrade project.
%\textcolor{red}{comment: need to rephrase it better because it seems in contradiction with the 5\,MW stated 2 paragraphs above}

\subsubsection{The ESS current moderator system }
\label{uppermoderator}

Initially, the spallation source will be equipped with a single compact, low-dimensional moderator specifically designed to produce high-brightness neutron beams for condensed matter experiments. This design is optimized for small samples and offers flexibility for parametric studies. In ~\cref{targetmoderator}, the left side showcases the high-brightness moderator, while the right side illustrates the current configuration of the ESS target-moderator-reflector system. 

\begin{figure}[htbp!] % replace 't' with 'b' to force it to be on the bottom
  \centering
  \includegraphics[width=0.89\columnwidth]{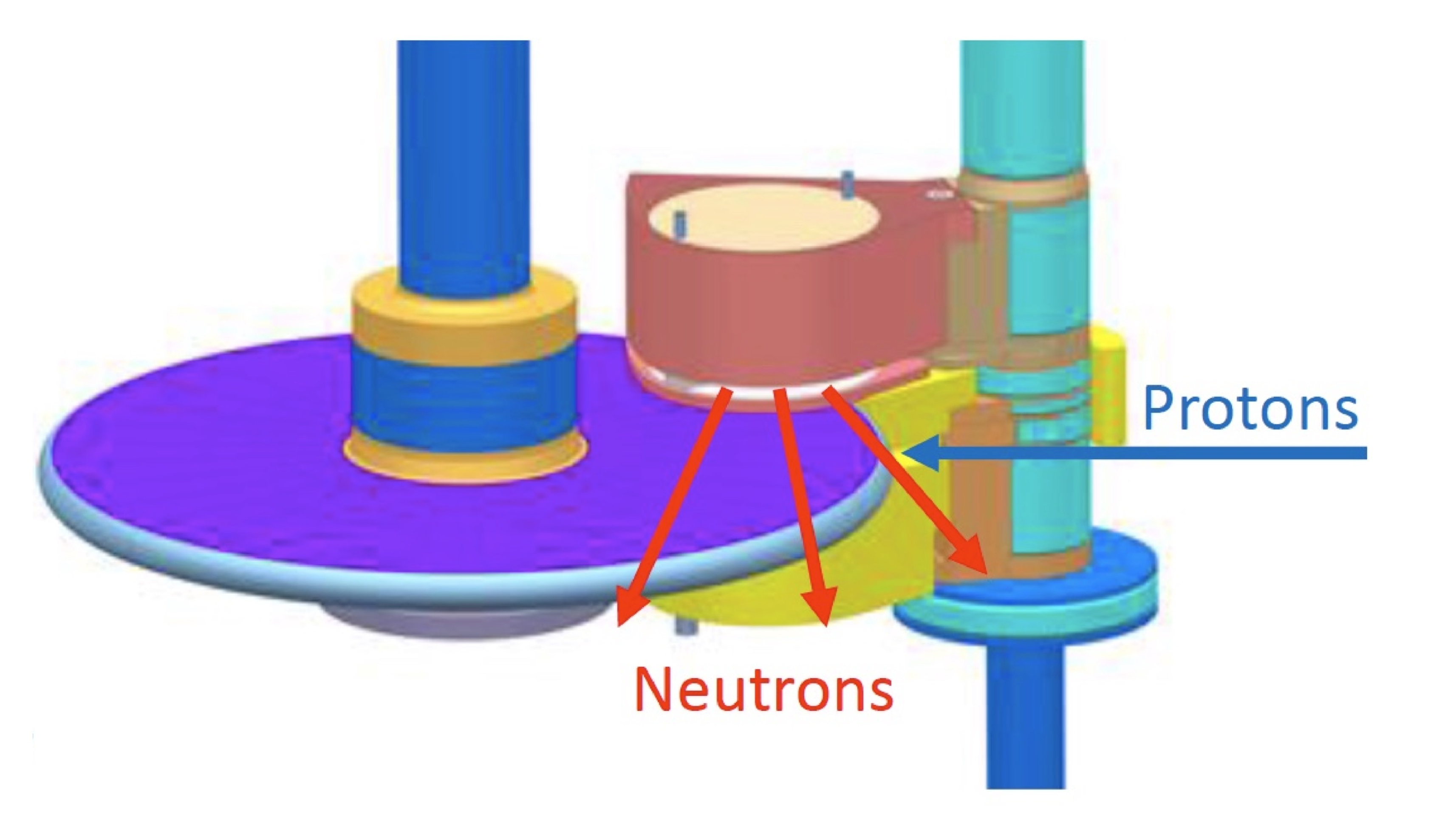}
  \caption{Schematic view of the ESS target-moderator-reflector system. The proton beam 
impinges on a rotating target consisting of tungsten (purple target in the figure). A cylindrical steel structure (dark red) placed above the target contains the moderator and reflector from which neutrons are extracted to the beam lines (red arrows). A similar container (yellow) is placed below the target, and is the intended location for the placement of the high-intensity moderator in the HighNESS project. These two structures and the shaft (light blue) form the so-called twister (cf. Section \ref{sec:fac}).}
  \label{fig:scheme}
\end{figure}

The core neutron production at ESS happens in the upper and lower moderator, two moderator-reflector systems positioned both above and below the spallation target, as shown at the center of~\cref{fig:scheme}. A cylindrical steel structure (shown in dark red) located above the target houses the moderator and reflector, from which neutrons are extracted for the beamlines. A similar container (depicted in yellow) is positioned below the target.

~\cref{targetmoderator} shows the openings in the shielding, which are the beam extraction channels and are present above and below the spallation target. These channels allow for the extraction of neutrons from either or both moderators. The space shown in \cref{targetarea} below the spallation target is presently occupied by a steel plug and currently un-utilized. This space has the potential to accommodate an additional moderator-reflector system. The design of such a system constitutes one of the primary objectives of the HighNESS project.

%\begin{figure}[htbp!] % replace 't' with 'b' to force it to be on the bottom
 % \centering
 % \includegraphics[width=0.99\columnwidth]{int_fig/nnbar.eps}
 % \caption{View of the geometry of the monolith, courtesy of R. Holmberg. This is a cross-sectional view perpendicular to the proton beam. See explanation in the text.}
 % \label{fig:nnbar}
%\end{figure}

\subsubsection{The ESS upgrade area }
\label{sec:essupgrade}

\begin{figure}[h!]
	\centering
	\includegraphics[width=.8\textwidth]{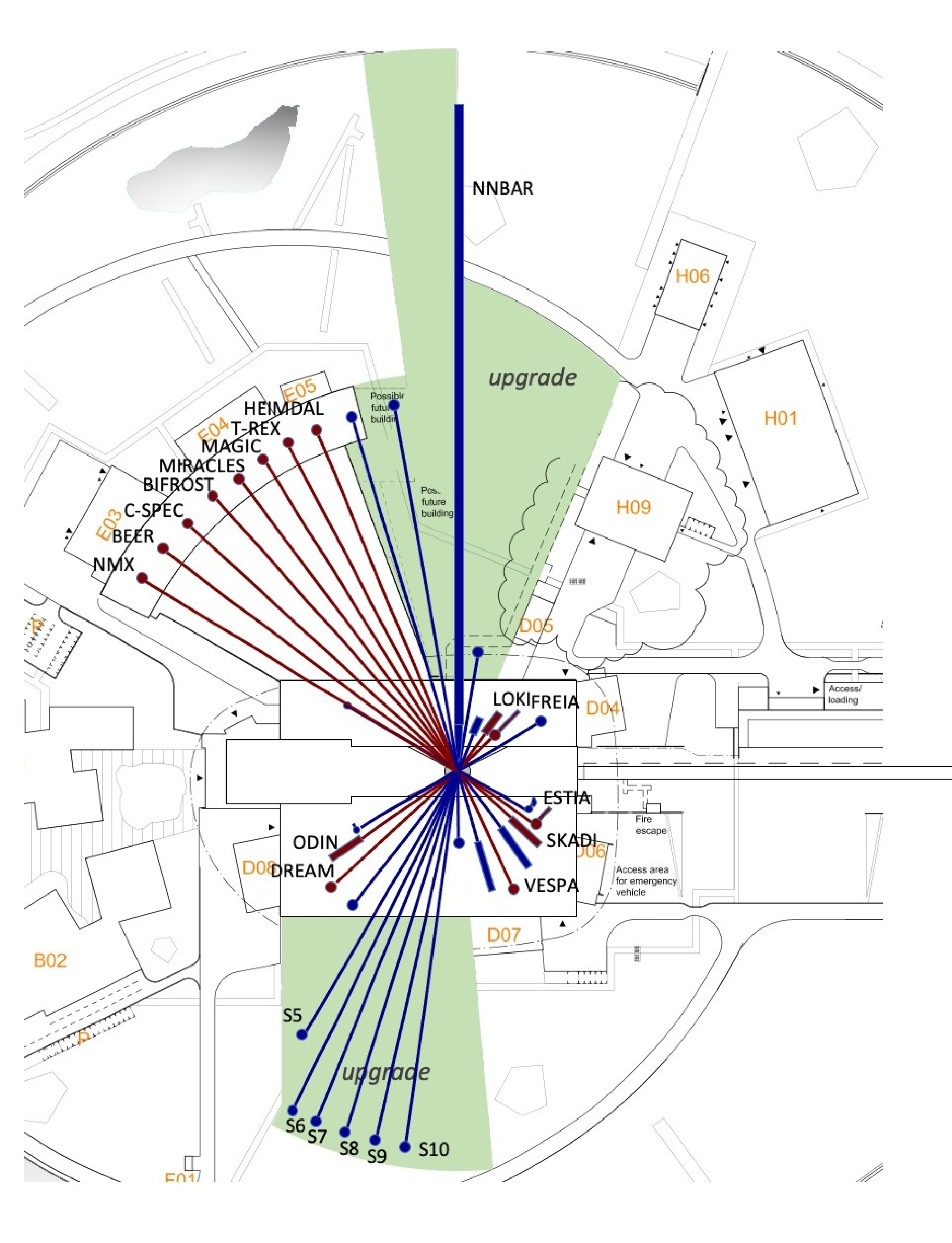}
 \caption{The ESS facility. The location of the instruments of the initial suite is shown. In green the upgrade areas, where the HighNESS instruments and experiments could be placed, are highlighted. }
	%\centerfloat	
	%form the layer
	\label{upgradearea}
\end{figure}

In addition to utilizing the space beneath the spallation target for the placement of a second moderator system, the HighNESS project will take advantage of the availability of extra beam ports.~\cref{upgradearea} shows the available sectors of experimental zones, the distribution of the 15 instruments of the initial suite, and the additional beam ports and areas for additional future instruments that can be fed by a second moderator system. These areas, highlighted in green in~\cref{upgradearea}, serve as the primary locations for designing instruments and experiments that will utilize the moderators developed in the HighNESS project.

\begin{figure}[h!]
	\centering
	\includegraphics[width=.8\textwidth]{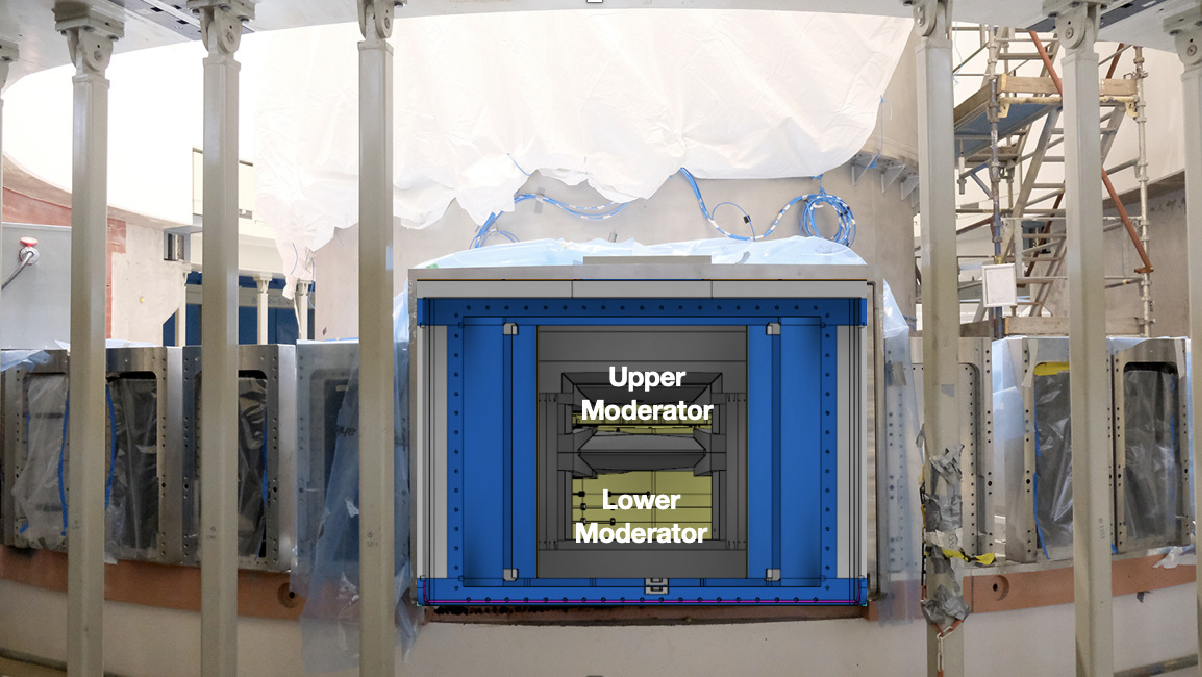}
 \caption{Photograph of the monolith beam ports around the ESS Large Beam Port. An overlaid drawing shows the visible surface of the upper and lower moderators, with a shielding block in between.}
	%\centerfloat	
	%form the layer
 
	\label{esslargebeamport}
\end{figure}

\begin{figure}[htbp!]   
\begin{subfigure}{0.95\textwidth}
  \centering
  \includegraphics[width=\textwidth]{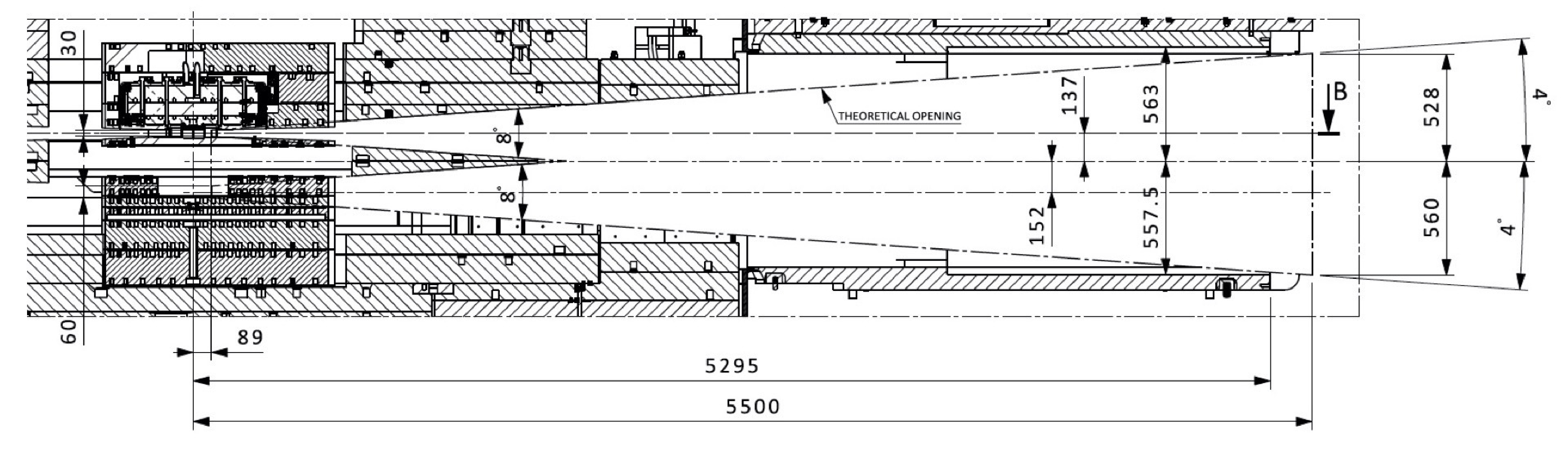}
  \caption{}
  \label{fig:nnbar2}
\end{subfigure}

\begin{subfigure}{0.95\textwidth}
  \centering
  \includegraphics[width=\textwidth]{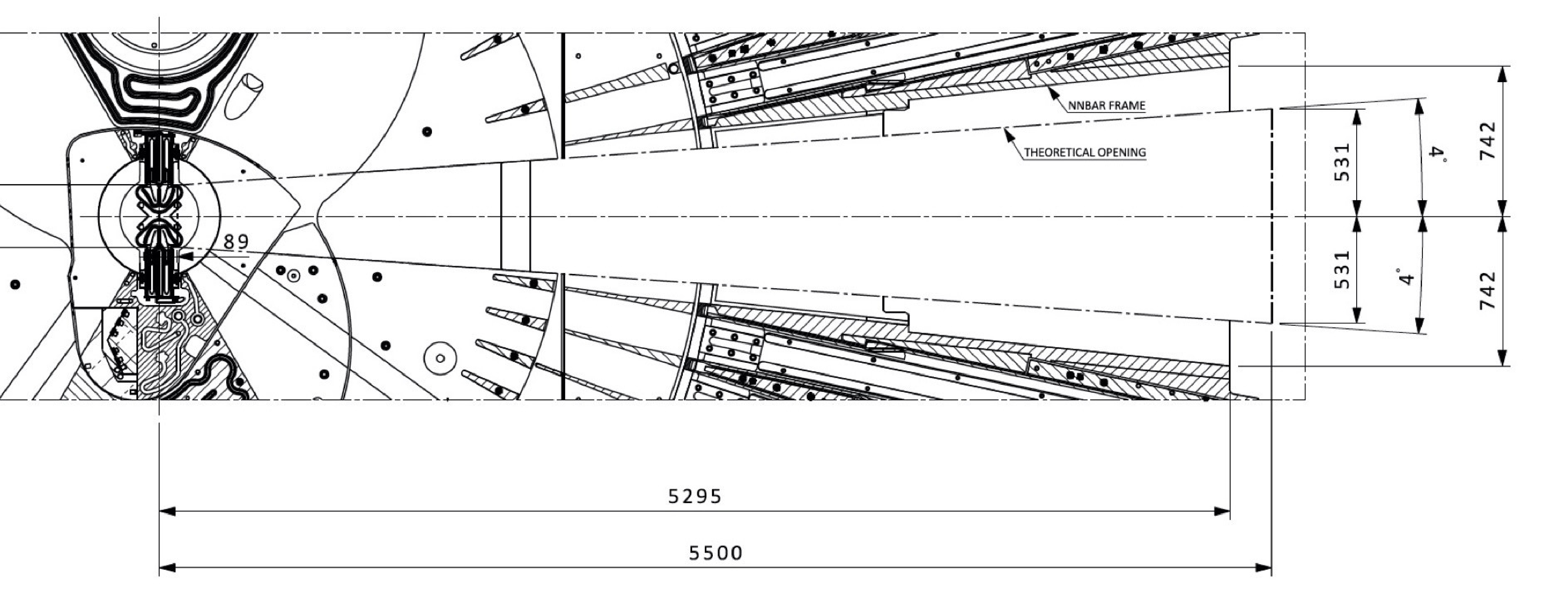}
  \caption{}
  \label{fig:nnbar3}
\end{subfigure}
\caption{(a) Side view of opening in correspondence of NNBAR. (b) Top view of the large beam-port opening for NNBAR.}
\label{fig:nnbar2and3}
\end{figure}

ESS is also equipped with a special beam port located in the monolith for neutron extraction. This beam port, often referred to as the Large Beam Port (LBP), due to its size, is equivalent in size to three standard ESS beam ports and is illustrated in \cref{esslargebeamport,fig:nnbar2and3}. This infrastructure has been installed in anticipation of the neutron-antineutron oscillation experiment  NNBAR~\cite{Addazi_2021,Backman_2022}, to allow the experiment to achieve its design goals. 
%(see HighNESS Conceptual Design Report Volume II).
At the time of this writing, no other existing or planned neutron facility will have a beam port of similar dimensions, making ESS the best possible facility worldwide for the NNBAR experiment. The LBP also serves a broader purpose beyond the NNBAR experiment, since it can be used for various potential designs of Ultracold Neutron (UCN) sources, as detailed in~\cref{sec:UCN_intro}. 

\begin{figure}
\begin{center}
\includegraphics[width=.66\textwidth]{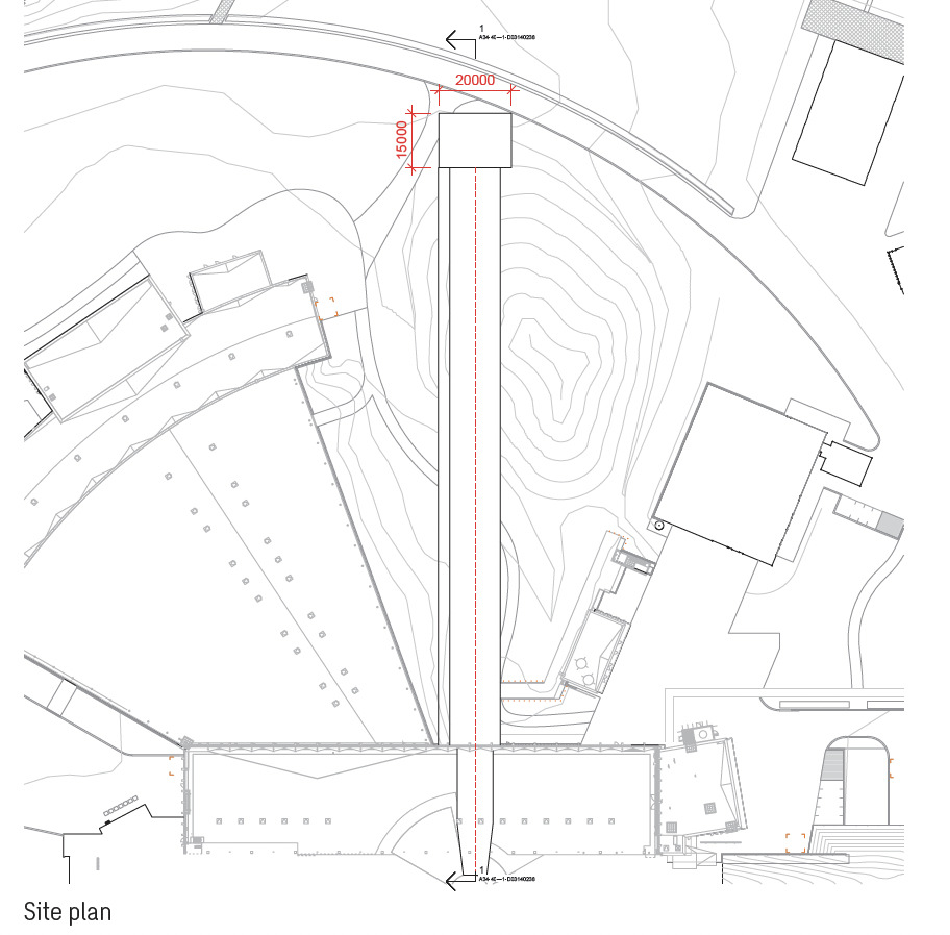}
\caption{Overview of the NNBAR beamline extension from the current ESS building. }
\label{nnbarex1}
\end{center}
\end{figure}

\begin{figure}
\begin{center}
\includegraphics[width=.86\textwidth]{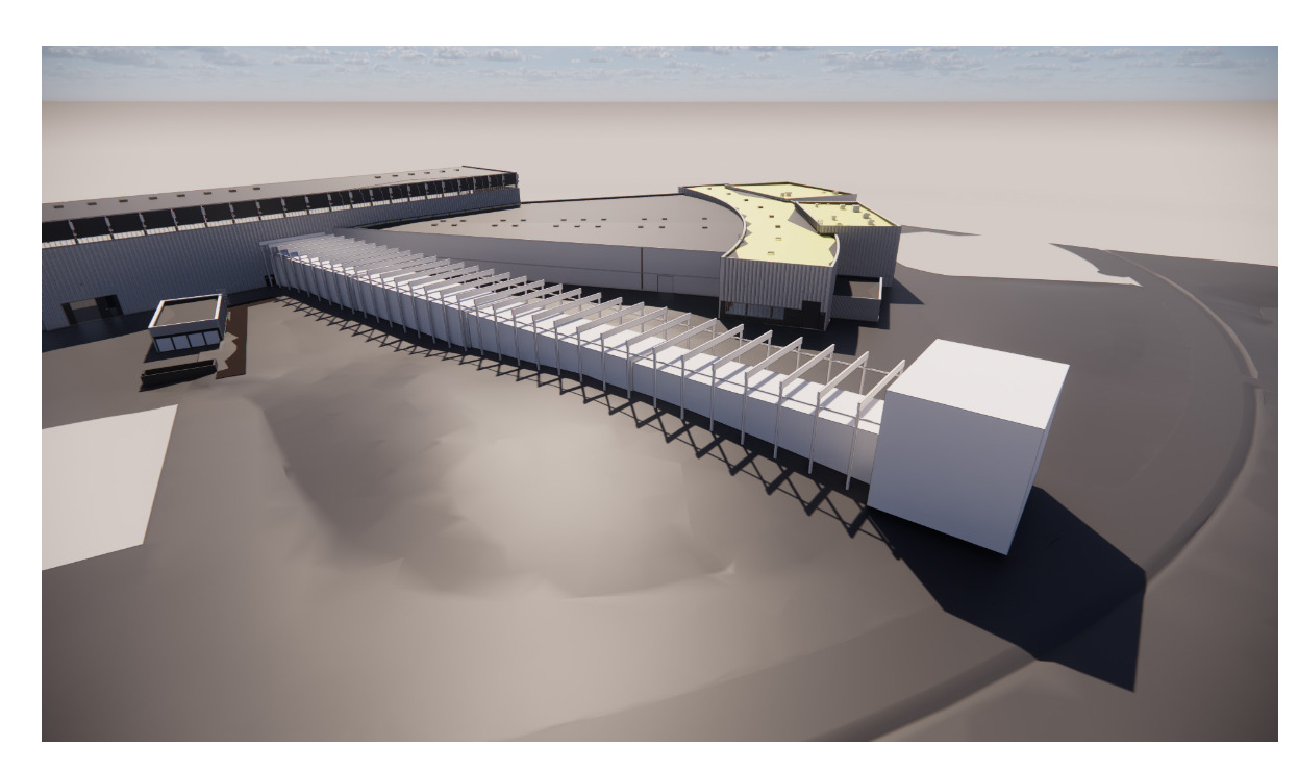}
\caption{The NNBAR beamline and the detector hall located 200m from the source.  }
\label{nnbarex2}
\end{center}
\end{figure}

\cref{nnbarex1,nnbarex2} depict how the NNBAR experiment will require an extension of the instrument hall. The current building perimeter extends approximately 21 meters away from the moderator. This area is available for future upgrades and will be used by the NNBAR experiment described in HighNESS Conceptual Design Report Volume II.

%as can be seen from \cref{essfromthetop}. 

\subsubsection{ESS timeline and beam power projections}
\label{timeline}

Owing to delays from the covid-19 pandemic and other technical challenges, the ESS in 2021 conducted a rebaseline exercise to revise its construction and commissioning schedule \footnote{\url{https://europeanspallationsource.se/article/2021/12/10/ess-revises-project-plan-and-budget}}
. The revised baseline plan introduces a two-year delay with respect to the previous schedule and will enable the ESS to start full operations and be open for scientific users working with up to fifteen instruments in late 2027. The maximum accelerator power and proton beam energies will be 2~MW and 800~MeV, respectively. The possibility of operating the accelerator at its nominal design power of 5\,MW is taken as part of an ESS upgrade project.

%\subsection{HighNESS scientific motivations}

%As described in the previous section, the flexible design of ESS, with only one moderator located above the target, leaves a great opportunity to implement a second source with complementary characteristics going well beyond the initial goals of the facility development. This new infrastructure has been designed with the purpose of delivering the highest possible total intensity (as opposed to brightness which was the main design criterion for the first source) of cold neutrons with wavelengths above 4 Å (see \cref{sec:coldsource}), including neutrons in the long-wavelength part of the spectrum, referred to as Very Cold Neutrons (VCN) and Ultracold Neutrons (UCN). 
%With these new performances ambitious projects in fundamental physics like the search for neutron to antineutron oscillations (the NNBAR program ) become possible, as well as complementary scattering techniques requiring a more intense source with a different spectral range than the first source. 
%With these new source capabilities, various condensed matter instrument concepts have been considered, as detailed in \cref{sec:neutronscattering}.
%Furthermore, the presence of a high-intensity neutron source is of paramount importance for the NNBAR experiment and for the development of UCN and VCN sources.\\
%that offer exceptional potential for particle physics research \cite{ABELE20231}.\\

\subsection{HighNESS objectives}

The HighNESS project, \cite{santoro2020development,HighnessCurrentStatus}, is an EU-funded 3-year project, that commenced in October 2020. The primary objective of the project is to develop a second cold neutron source at ESS that complements the first source located above the spallation target. For the initial instrument suite, the emphasis was on creating a source capable of providing a high brightness of thermal and cold neutrons. In contrast, the new sources, designed as part of the HighNESS project, focused on two distinct aspects:
\vspace{0.2cm}
\begin{itemize}
    \item[-] Increase in source intensity: this entails boosting the total number of neutrons emitted from the source. Achieving a more intense source necessitates larger moderators and emission surfaces to enhance the count rate for instruments or experiments that require high flux.
     \item[-] Shift toward colder neutrons: this shift has manifested in the development of a very cold neutron (VCN) source and an ultracold neutron (UCN) source, in addition to the second cold neutron source. To create a highly intense VCN source as part of the project, we have invested significant effort in extensively characterizing promising candidate materials. This comprehensive characterization is essential for fully developing the capabilities required to design a source utilizing these materials. The project has placed specific emphasis on distinct materials: solid deuterium and deuterated clathrate hydrates, suitable for use as dedicated VCN moderators. While nanodiamonds, graphite intercalation compounds, magnesium hydride, and clathrate hydrates (see \cref{sec:dch}) can serve as advanced reflectors to enhance the VCN flux.
     
\end{itemize} 
\vspace{0.2cm}

The second source was designed as a liquid-deuterium (LD$_{2}$) moderator, based on proven technology. One of the main results of this work is a complete engineering design of the LD$_{2}$ moderator, taking into account the actual layout of the ESS target station. In addition to this, the project has explored several designs for VCN and UCN sources. These design possibilities are described in~\cref{sec:vcn}, and~\cref{sec:UCN_intro} respectively. Furthermore, the project has successfully developed a set of neutron scattering instruments (see~\cref{sec:neutronscattering}) and conducted a conceptual design study for the NNBAR experiment. 
%(see HighNESS Conceptual Design Report Volume II).

As mentioned previously, the first 15 instruments for the user program of ESS will view the top moderator. The start of operation of ESS will be at low power, which will be increased stepwise to finally reach the time average power of 2\,MW. It is planned to reach 2\,MW power around 2028. Beyond 2030, a second moderator system, like the one proposed herein, could become available. Around that date the facility would ideally be equipped with two separate neutron sources with the following features:

\vspace{0.2cm}
\begin{itemize}
\item[-]	A high-brightness bi-spectral (thermal--cold) moderator, based on water and liquid para-hydrogen, placed above the spallation target, able to serve all the 42 beam ports of the ESS grid with brightness for thermal and cold neutrons higher than any other facility worldwide.
\item[-]	A high-intensity LD$_{2}$ moderator, placed below the target station, capable of directly serving instruments demanding a cold-neutron flux of unprecedented intensity or acting as a primary source for secondary VCN and UCN sources for experiments requiring sub-cold neutrons. In particular, the ESS would provide the first high-intensity VCN source in the world.
\end{itemize}
\vspace{0.2cm}

Thus, in this scenario of separate moderators optimized for different neutron characteristics, ESS would offer,
with respect to the single-moderator case,
a versatile neutron source of outstanding performance, spanning a larger neutron wavelength range, and delivering high brightness and intensity to the instruments according to their requirements. This will enable a plethora of multi-disciplinary activities which fit the original plan for ESS but offer many more possibilities beyond, for which there is strong topical scientific demand.  

At the time of the writing, it is challenging to predict whether this moderator will be installed before ESS reaches the 5\,MW design power. As a consequence, the majority of the findings presented in this work have been developed with consideration of both the 2\,MW and 5\,MW options.

\subsection{The HighNESS project in the neutron landscape}

Some examples of instruments and experiments HighNESS aims to make possible are already given in this introduction, but more are outlined in~\cref{sec:vcn}, \cref{sec:neutronscattering}, and in the HighNESS Conceptual Design Report Volume II that describes the NNBAR experiment. There is currently no alternative source with comparable parameters to ESS to enable the experiments mentioned.
There are two other high-power, MW-class, pulsed spallation sources in the world, J-PARC in Japan and SNS in the USA. Both facilities have upgrade plans for the implementation of second target stations, and in both cases, the focus is on using low-dimensional high-brightness moderators, similar to the approach of the ESS first source.

These upgrades aim to achieve peak brightness superior to that of the ESS first source, with a potential increase of up to a factor of 5. However, the time-average brightness is expected to remain approximately 5 times lower than that of the first ESS source when operating at \SI{5}{MW}, and for the majority of applications targeted by the HighNESS project, a high time-average intensity is essential. Furthermore, it is worth noting that the upgrades at SNS and J-PARC primarily focus on cold neutron ranges.
Neither of these facilities have plans to install a VCN source near the spallation target. As a result, ESS, with its intense source and focus on delivering longer wavelengths, could provide intensities in the long-wavelength regime that are more than one order of magnitude higher.
 The installation of a long-wavelength facility at the ESS is uniquely favorable due to the high proton power and long-pulse time structure.
 %In other words, no existing facilities can be upgraded in this way.  
 
 Finally, it is worth highlighting that no other facility houses a feature akin to ESS's Large Beam Port, which opens exceptional opportunities not only for the NNBAR experiment but also for the development of world-leading UCN sources.

\subsection{HighNESS project structure}

\label{sec:highwp}

\begin{figure}[ht]
    \centering
    \includegraphics[height=8cm]{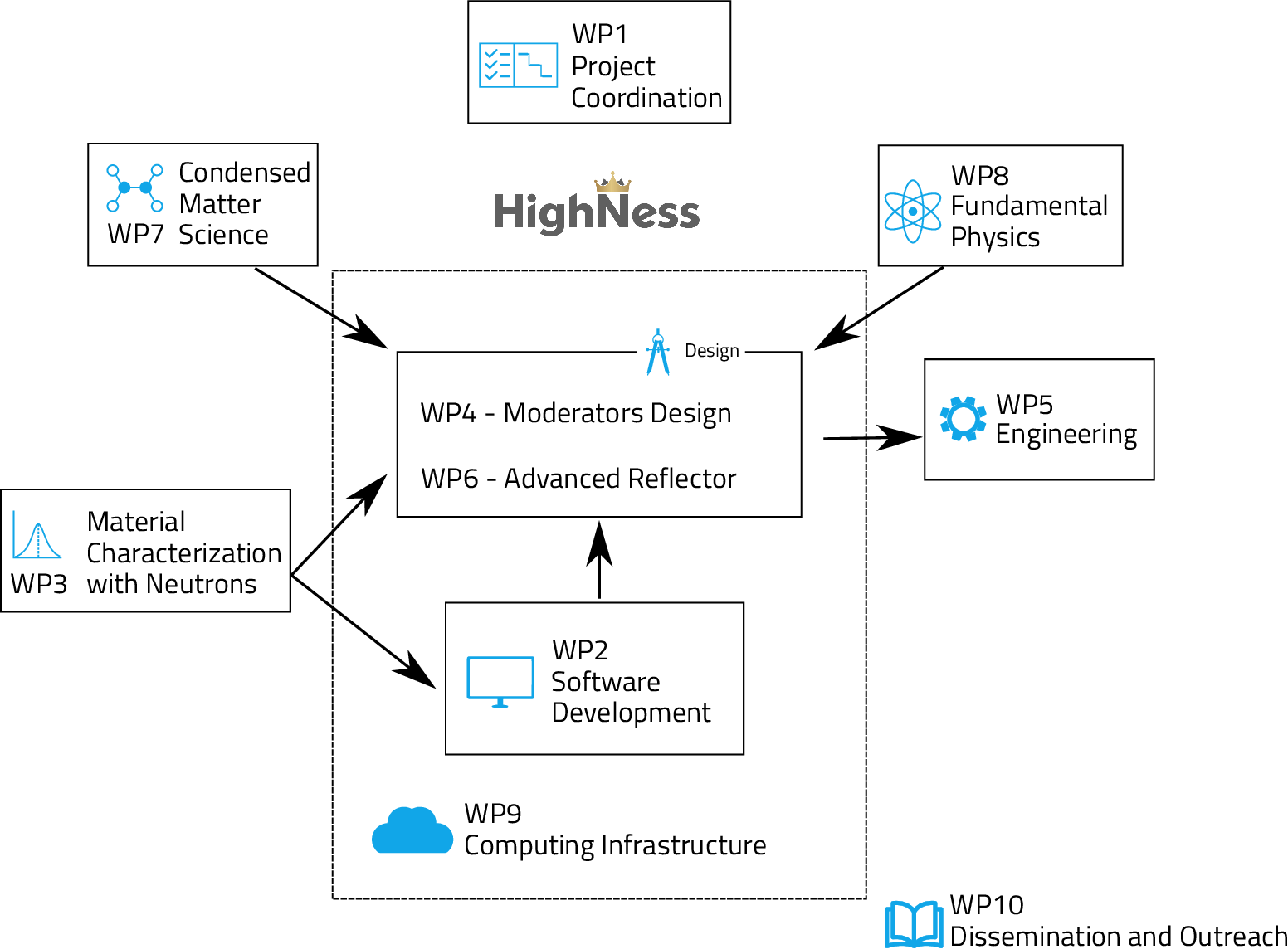}
    % JIMD: I added this figure because the work package structure was missing
    \caption{HighNESS project Work Package structure}.  
    \label{highnesswps}
\end{figure}

The HighNESS project is organized into ten distinct Work Packages (WPs), as illustrated in \cref{highnesswps}. WP1, titled ``Project Coordination", serves as the central hub for coordinating all project activities. WP2, known as ``Software Development", focuses on creating the necessary computational tools required for the analysis and design of high-intensity moderators.

The computational tools developed in WP2 rely on experimental measurements conducted in WP3, titled ``Material Characterization with Neutrons". This work package provides material-property data, essential for generating thermal-neutron scattering kernels and validating the required models.

WP4 and WP6, designated as ``Moderator Design" and ``Advanced Reflectors", respectively, serve as the foundation for neutron design efforts, specifically the development of cold sources (CN), very cold neutron sources, and ultracold neutron sources. These two work packages require inputs and specifications from the scientific-oriented WPs, WP7 (``Condensed Matter Science") and WP8 (``Fundamental Physics"). The inputs from these scientific WPs ensure that the source designs deliver the necessary neutron fluxes and spectra, enabling groundbreaking scientific experiments at ESS that are beyond the capabilities offered by the upper moderator. WP5 (``Engineering") is responsible for handling all the engineering aspects related to the different sources. 

The objective of WP9 (``Computing Infrastructure'') is to make software and data developments carried out by WP2 and WP6 accessible to the public through the creation of cloud computing resources. Subsequently, WP10 manages the distribution of all scientific content generated as part of this project.

%% file: coldsource.tex
%\usepackage{placeins}
\section{Design of the cold source}
\label{sec:coldsource}

The design of the \ce{LD_2} moderator is one of the main objectives of the HighNESS project. This moderator should provide a significantly higher intensity than the low-dimensional moderator located above the spallation target, thus enabling new types of experiments in condensed matter research and fundamental physics.

The moderator design has been carried out using the MCNP6~\cite{mcnp6} code. The work has been performed incrementally, considering initial, simple figures of merit (FOMs), and later more elaborated ones to account for the specific needs of instruments. Likewise, the work started considering a single opening (channel for neutron extraction), and then moved to a configuration closer to the one needed for ESS, with openings for both fundamental physics and condensed matter experiments. Engineering details were progressively added in the design process, following feedback from the engineering team.
%The results from the study with a single opening are detailed in~\cite{D4.2} and will not be described in this chapter, where the focus will be on the design with two openings.

\subsection{General description of the ESS moderator area}
\label{sec:fac}

\begin{figure}[htbp!] % replace 't' with 'b' to force it to be on the bottom
  \centering
  \includegraphics[width=0.8\columnwidth]{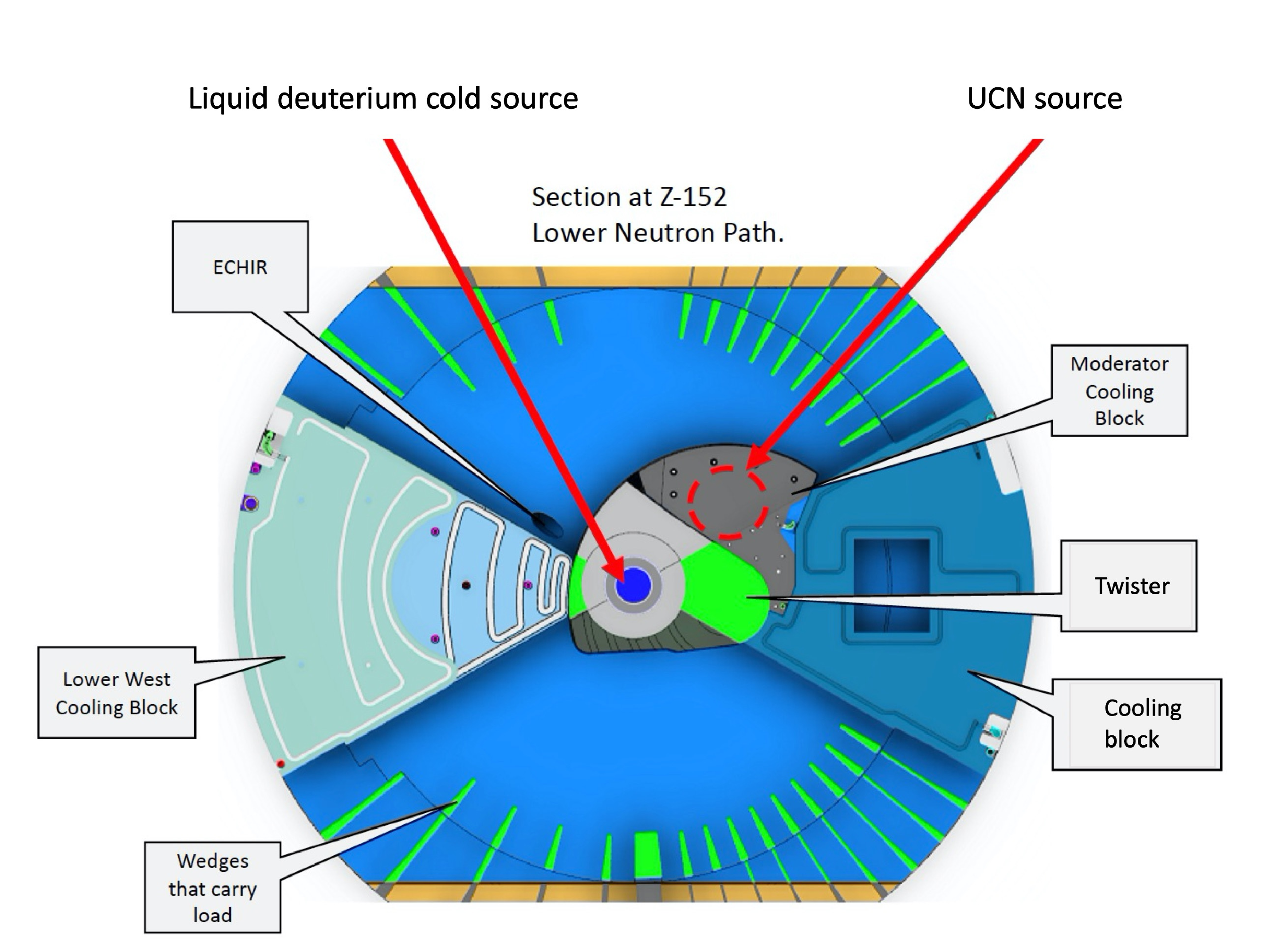}
  \caption{Horizontal cut in the geometry of the target monolith in correspondence to the center of the lower moderator, courtesy of R. Holmberg. The grey circle inside the twister represents the maximum available space for a Be reflector (maximum diameter 72 cm). The blue circle inside represents the location of the lower moderator. One of several candidate UCN source locations is indicated in the pciture by the red dotted circle. For more details see \cref{sec:UCN_intro}. }
  \label{fig:XYshield}
\end{figure}

\cref{fig:XYshield} shows a horizontal cut in the geometry of the monolith, highlighting the available space below the target for the neutron extraction, consisting of two areas, each with an angle of 120$^\circ$.  
%Overall, they cover two areas of 120$^\circ$ each.
These openings allow for the diffusion of slow neutrons towards beamlines starting outside the monolith, and arranged in four sectors which are conventionally labelled with their cardinal directions (cf. \cref{essinstruments}).

The lower moderator system will be installed inside the so-called twister,
a steel structure containing the upper and lower moderator-reflector assemblies. Its name is derived from the fact that it is rotated around its shaft during the periodic exchange procedures. 
The coordinate system used to describe the geometry is the so-called TCS (Target Coordinate System) which has the origin inside the spallation target, with the vertical axis (X,Y=0) passing through the centre of the upper moderator. We have the following constraints for the liquid deuterium moderator and reflector: 40\,cm in height, and 72\,cm in diameter. This corresponds to between -7.8\,cm and -47.8\,cm along the Z-TCS axis. The twister starts at -61 cm with respect to Z-TCS, hence we have 13.2\,cm between the maximum size of the liquid deuterium moderator and the inner shielding. However, some of this space, depending on the engineering design, must be reserved for the structural support of the lower moderator-reflector plug.

\begin{figure}[htbp!] % replace 't' with 'b' to force it to be on the bottom
  \centering
  \includegraphics[width=0.85\columnwidth]{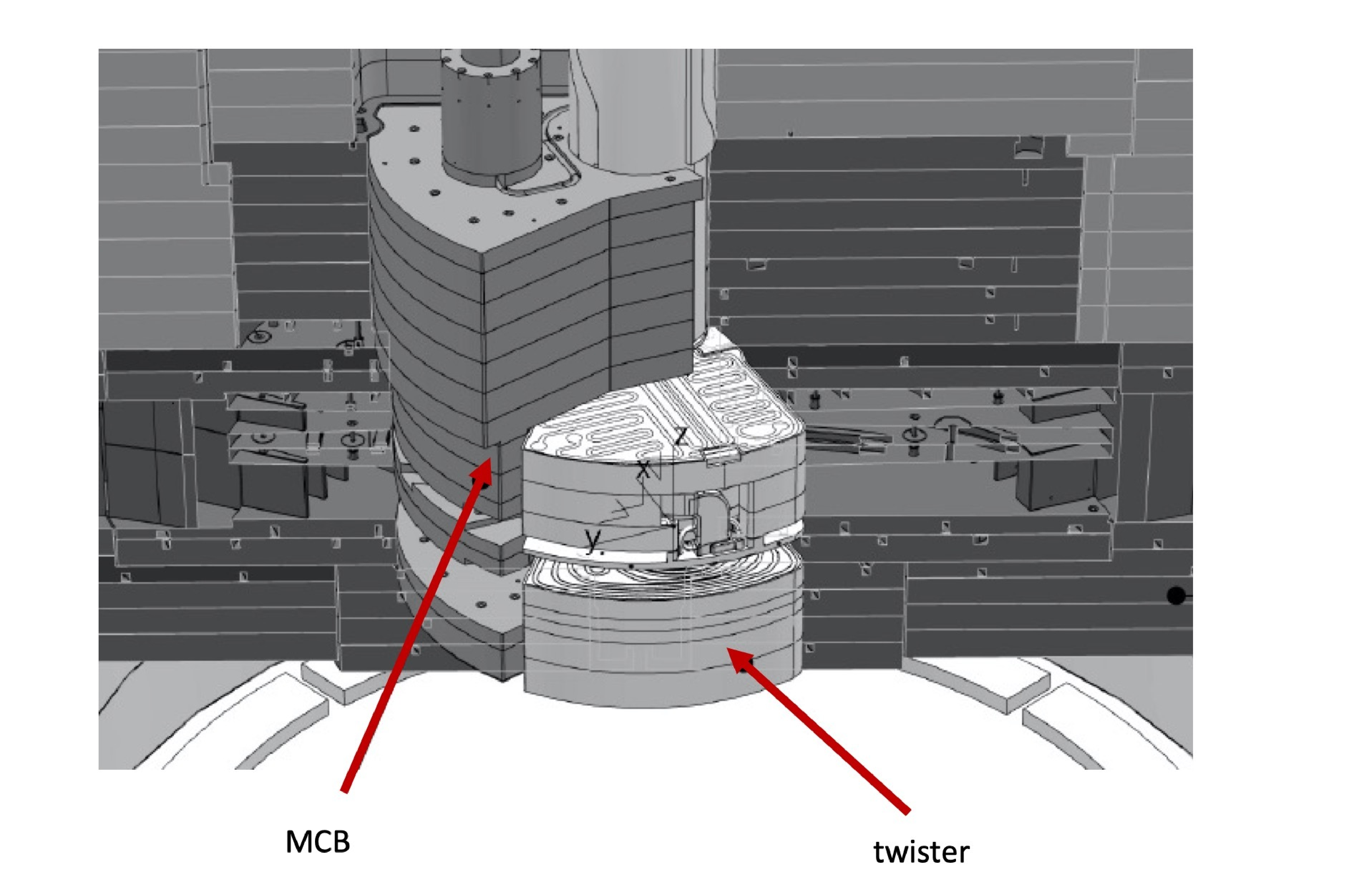}
  \caption{Twister and moderator cooling block (MCB). 
  %The twister consists of two steel structures containing the upper and lower moderator-reflectors assemblies, and its name is derived from the fact that it rotates around its shaft during the periodic exchange procedures. 
  Drawing courtesy of C. Jones, ESS. See explanation in the text.}
  \label{fig:MCB}
\end{figure}

\begin{figure}[htbp!] % replace 't' with 'b' to force it to be on the bottom
  \centering
  \includegraphics[width=0.75\columnwidth]{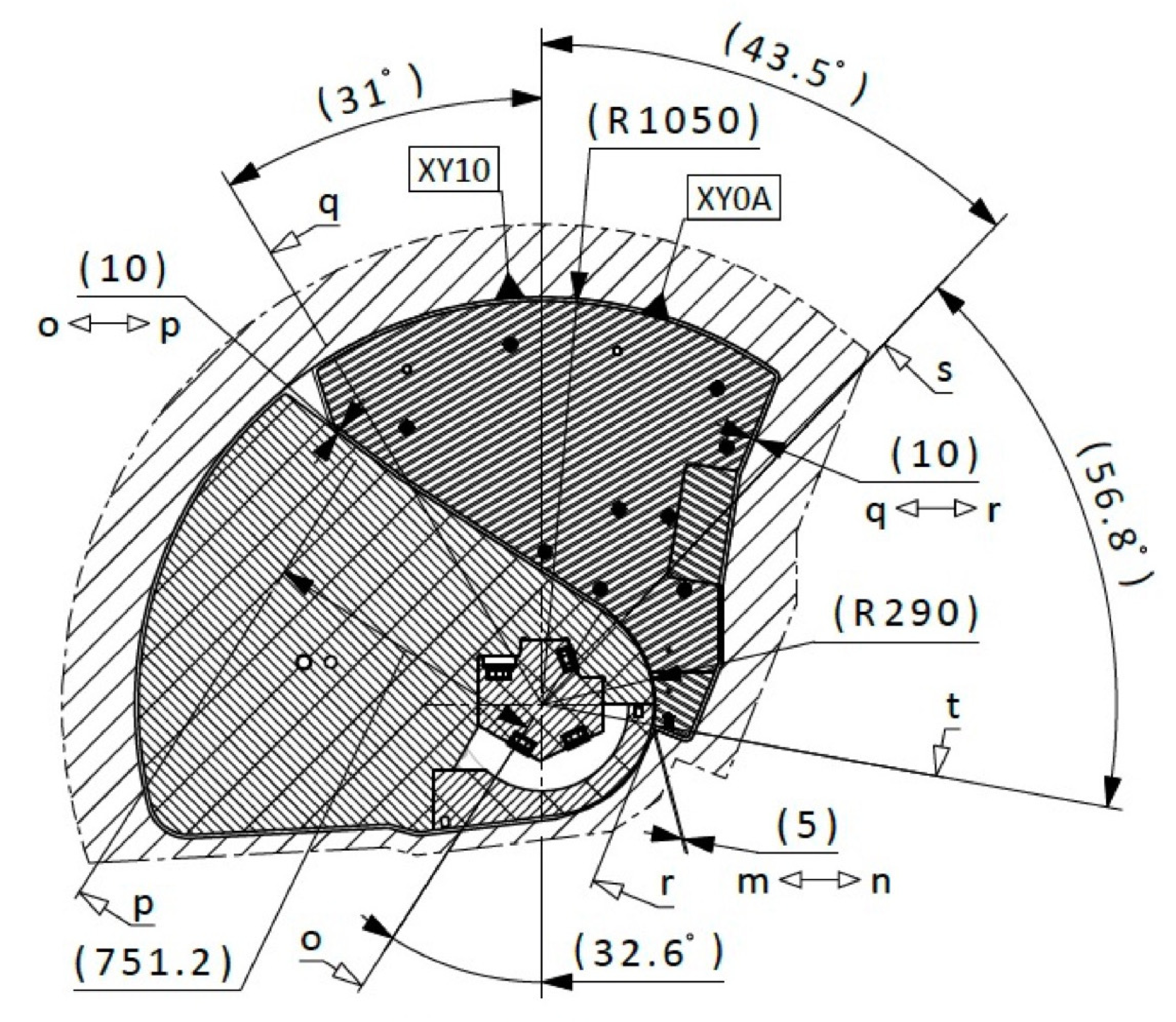}
  \caption{Twister (left block) and moderator cooling block (MCB, right block), top view. The twister is the location for both the upper high-brightness bispectral moderator and the lower liquid deuterium moderator. The MCB is a potential location for a UCN in-pile source.}
  \label{fig:MCB3}
\end{figure}

In addition to the space inside the twister structure, which contains the moderator and reflector below the spallation target, another potential location for a moderator system was identified during the early stages of the HighNESS project: the Moderator Cooling Block (MCB). The MCB, shown in \cref{fig:MCB}, is essentially a removable and replaceable block of shielding material that is positioned adjacent to the twister and is water cooled. Its location has been considered for the placement of a secondary source, such as a UCN source (see Section ~\cref{sec:UCN_intro}). The placement of a source in that location would require a modification of the MCB. This possibility has been discussed with the ESS Target engineering team and is considered as feasible. It has therefore been explored, along with a variety of other UCN source candidate locations, as part of the HighNESS project (see ~\cref{sec:UCN_intro}). \cref{fig:MCB} and \cref{fig:MCB3} show the twister and the MCB next to each other. Such a source embedded in the MCB would be best suited to beamlines placed in the North sector.

\subsection{Figures of merit}

It is important to define the wavelength ranges for the three groups of neutrons of interest in HighNESS (\cref{tab:def}). Cold Neutrons (CN) are defined in the range from about 2\,{\AA} to about 20\,\AA, which is the typical range for the instruments using the cold neutrons from the upper moderator~\cite{andersen}. VCNs, as defined in~\cite{carpenterloong}, are neutrons in the range between about 10\,\AA~and 120\,\AA. It is worth noting that the lower range, 10--40\,\AA, is of particular interest for condensed matter research, while the upper range, 40--120\,\AA, is more applicable to fundamental physics experiments. 
%\textcolor{red}{we need to refer to the sections where this is discussed, and check that it is consistent with this statement}
Finally, UCNs are defined as neutrons with wavelengths above 500\,\AA. The present section reports on the design and optimization of the lower moderator in terms of cold neutron intensity; the neutronic design was performed using the Monte Carlo code MCNP version 6.2 \cite{ref_MCNP6}. The design of the VCN and UCN sources are discussed in~\cref{sec:vcn} and~\cref{sec:UCN_intro}, respectively. Finally, in ~\cref{sourceintegretion} options for the integration of the different sources, aiming at providing the best possible solutions to deliver CNs, VCNs and UCNs to the users, are discussed.

 \begin{table}[h!]
\caption{The wavelength range that defines cold, very cold, and ultracold neutrons.}
\label{tab:def}
\centering
\begin{tabular}{ c  c }
\toprule
%\texttt{iel} & Description \\ 
\midrule
cold & 2-20 {\AA} \\
very cold & $\sim$ 10--120 {\AA} \\
ultracold & $>$ 500 {\AA} \\
\bottomrule
\end{tabular}
\end{table}

The whole process from the neutron moderation to the neutron transport to an experimental area needs to be taken into account when designing a neutron source. This is because the sensitivity of an experiment is proportional to the count rate of neutrons in a given wavelength and divergence range at the location of experiment.
% Blahoslav: Needs to be re-phrased
Consequently, the design of the source must be done considering the initial cold neutron flux calculations at the emission surface of the source provided by MCNP as well as the subsequent neutron transport using ray-tracing Monte Carlo simulations using the McStas code~\cite{mcstas}.

The two processes are done at different stages. In the case of the HighNESS project, MCNP simulations are done by the neutronic team (Work Package 4 of the HighNESS project, see \cref{sec:highwp}), while the McStas ones are mainly performed by instrument designers for both condensed matter instruments and fundamental physics experiments. In order to link the two parts, MCNP must provide a source file to be used in McStas. This can be done in several ways, such as by fitting neutron spectra from the moderator, or by writing individual neutron trajectories in a text file (MCPL format \cite{MCPL}), which is then read by McStas. The latter is the preferred option within HighNESS.

There is however an additional factor that complicates the work of the source designers in HighNESS, which is the use of advanced reflectors that are intended to increase the flux from the source to the instruments. Various options have been investigated, such as the use of nanodiamond layers at the exit of the moderator, which increase the flux towards the neutron guides via quasi-specular reflection. Other materials have also been considered. To investigate these options, thermal scattering libraries were developed and included in the MCNP simulations, as well as in other particle transport tools (see \cref{sec:adref}).  

To optimize the moderator design it is necessary to define a metric, or figure of merit (FOM). Several FOMs are needed to accommodate the needs of the various instruments. \cref{tab:fom} provides the summary of the FOMs used for the design of the moderators in WP4.

\subsubsection{Figures of Merit for condensed matter }
\label{sec:fom7}
For the neutron scattering instrument designs developed by HighNESS, the number of neutrons reaching the sample within the desired wavelength range and divergence is essential for the instrument optimization. 

Choosing a FOM is not always as straightforward as, for instance, in the case of NNBAR discussed below. The reasons are as follows: first, there are three classes of instruments under study: SANS, imaging, and spin-echo. Second, for a given class of instruments, different solutions for the optics can be considered, which might work best for competing features of a moderator.

In general, however, a few guidelines have been established that can be used to define the way of working and a useful FOM. Since the neutron divergence at the sample can only be studied with McStas, and is not a parameter controlled with MCNP, it has been  omitted from the FOM. The FOM to be used in designing the source should rather concentrate on the intensity and spectrum:

\vspace{0.25cm}
\begin{itemize}
\item[-] Different emission surfaces from the moderator are of interest \textit{a priori} for the relevant instruments. The reference emission surfaces are: 5$\times $5 cm$^2$, 10$\times $10 cm$^2$, 15$\times $15 cm$^2$, 20$\times $20 cm$^2$.
\item[-] For each of these surfaces, an optimization of the moderator can be done, using the integrated intensity in a given range of wavelengths as FOM.
\item[-] For some instruments, in particular spin-echo -- and potentially future SANS instruments -- the instrument design team (WP7) is also interested in a colder spectrum, i.e., an increase in the spectrum for wavelengths greater than 10\,\AA. For this purpose, the combined efforts of WP4 and the advanced reflector team, WP6, were needed (i.e., using advanced reflectors to enhance the coldest part of the spectrum from the cold moderator). It is likely that the greatest increase will be achieved using a smaller viewing surface than the reference ones under study by WP7. This could be obtained with a smaller extraction surface, such as 3$\times $3~cm$^2$, surrounded by a layer of advanced reflectors such as nanodiamonds. This option has also been investigated, and in this case smaller openings will also be considered, in addition to the larger openings mentioned above. %The outcome of this research of a colder spectrum beamline could require the extraction of two separate neutron beams from the moderator from the scattering experiments opening or "WP7 side", which should be possible thanks to the large dimensions of the lower moderator.
\end{itemize}

\subsubsection{Figures of Merit for fundamental physics }
\label{sec:fom8}

\begin{figure}[htbp!] % replace 't' with 'b' to force it to be on the bottom
		\centering
		\includegraphics[width=1.1\columnwidth]{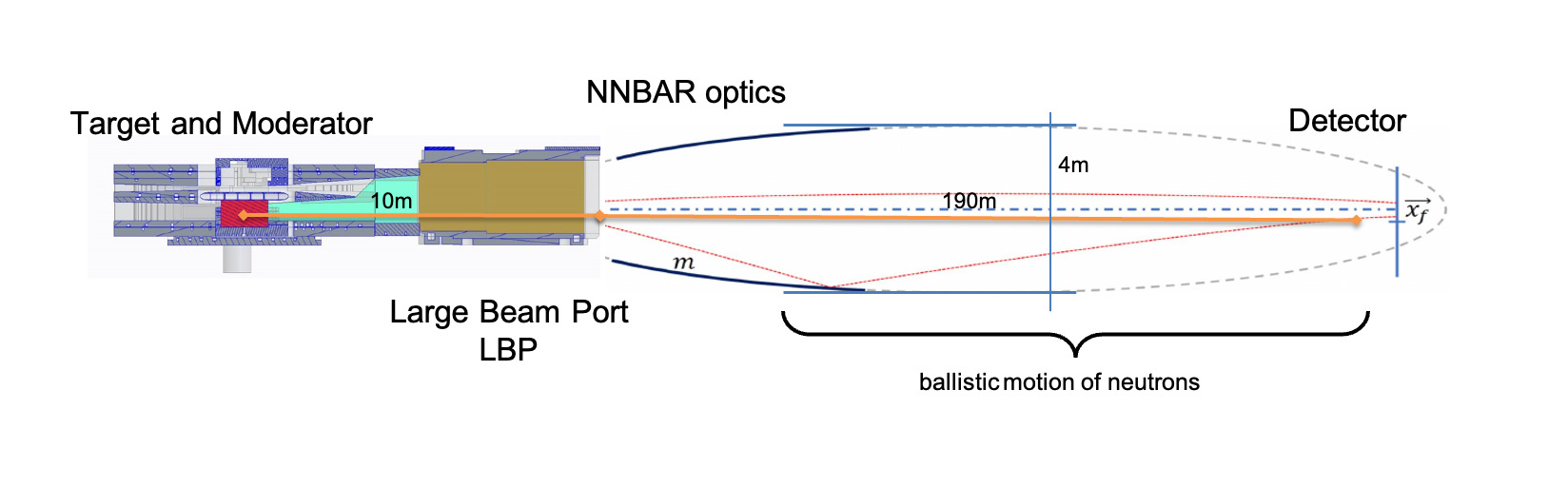}
		\caption{Schematics of the NNBAR Experiment (not to scale).}
		\label{fig:NNBAR_schematic}
\end{figure}

The NNBAR experiment will search for baryon number violation (BNV) via neutron ($n$) -- antineutron ($\bar{n}$) oscillation. The full experiment and its scientific motivation are described in the HighNESS Conceptual Design Report Volume II. In \cref{fig:NNBAR_schematic} a schematic overview of the planned NNBAR experiment ~\cite{Addazi_2021} is depicted.
Neutrons that are generated in the target are moderated and traverse the LBP. A system of elliptically shaped neutron guides is placed in the region after the LBP's exit to focus the neutrons in the direction of the detector located downstream. After having passed the optics, neutrons fly free to the detector region at the end of the instrument. The moderator-to-detector distance is 200\,m.  

The probability of an oscillation is proportional to the (uninterrupted) flight time squared $t^2$ of the neutrons. The FOM of the NNBAR experiment is then defined as $\sum_{i} N_i t_i^2$ with $N_i$ being the number of neutrons with a specific flight time $t_i$. Thus, slower neutrons are favored.
%and a trade-off between intensity and wavelength will be established.

\begin{figure}[htbp!] % replace 't' with 'b' to force it to be on the bottom
	\centering
	\includegraphics[width=0.96\columnwidth]{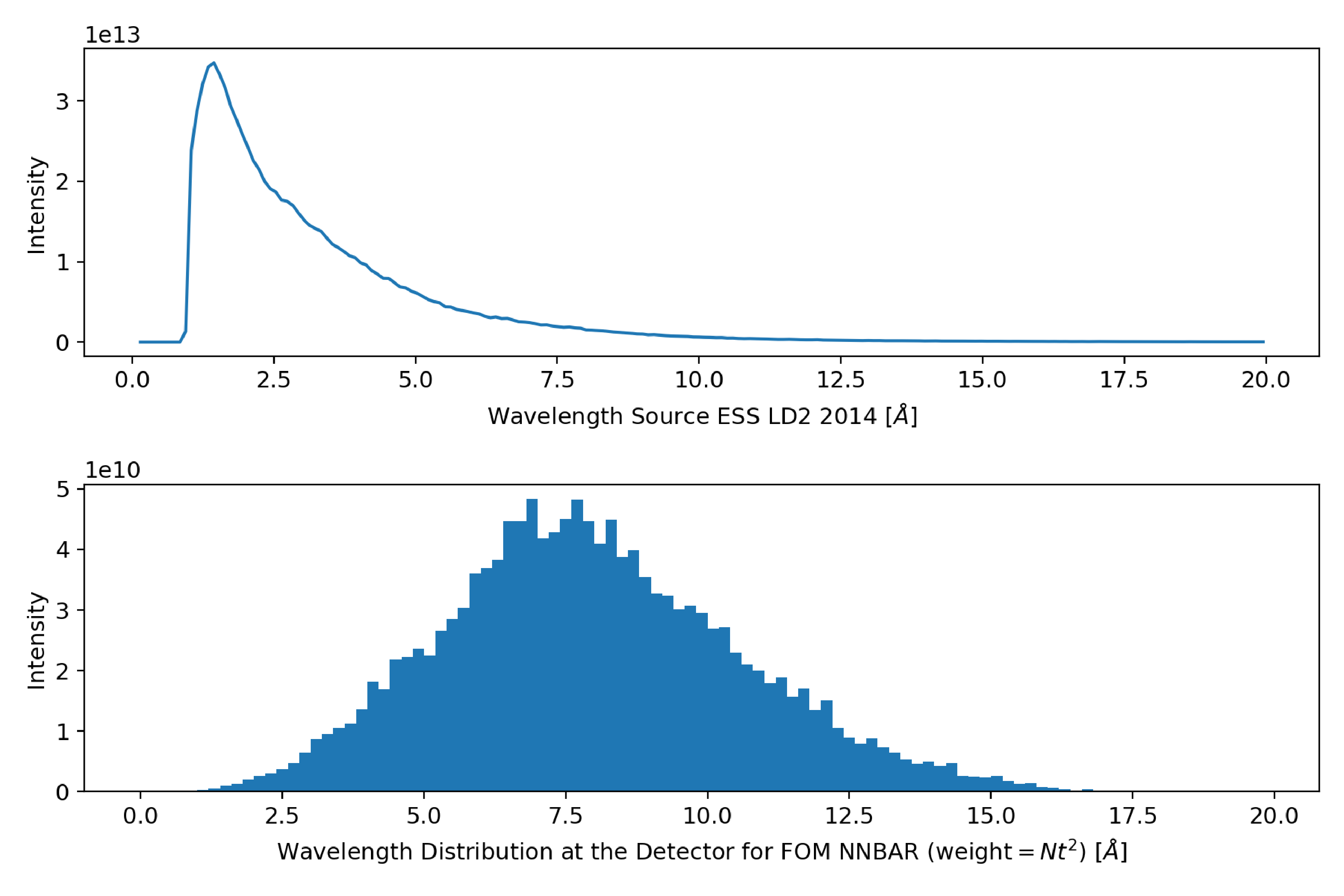}
	\caption{Wavelength spectrum of the \ce{LD_2} source (upper plot) and the distribution of wavelength contribution to FOM for NNBAR (lower plot).}
	\label{fig:Wavelength_Contribution_to_FOM_NNBAR}
\end{figure}

In the lower plot of \cref{fig:Wavelength_Contribution_to_FOM_NNBAR}, the distribution of wavelengths and how they contribute to the FOM is shown for a representative simulation result. The upper plot displays the wavelength distribution in the \ce{LD_2} source for comparison.

%To define a FOM for the moderator that is independent of the optic it 
A possible FOM for the moderator optimization can therefore be the wavelength intensity distribution that covers the solid angle defined by the exit of the LBP in the range above 5\,{\AA}  (or alternatively, 2.5~\AA) at the detector weighted with $ \lambda^2$. We note that there will be a natural cutoff at $\lambda$ of about 15\,\AA,  due to the effect of gravity on neutrons with longer wavelengths. For example, for neutrons of wavelength 12.5\,{\AA} the vertical drop due to gravity would already be 2\,m. Thus, neutrons with wavelengths that are too long will not be able to reach the detector, which has a radius of 1\,m. Moderator designs that perform better under this criteria should lead to a higher FOM for NNBAR. 

The FOM used for the moderator design for NNBAR is therefore the intensity of neutrons, weighted with $\lambda^2$, for two ranges of neutrons: above 2.5\,{\AA} and above 5\,\AA, with an upper cut of 15\,\AA.

\subsubsection{Figure of merit for VCN}

%A full study has been performed using advanced reflectors such as nanodiamonds to increase the reflection of the colder part of the spectrum from the \ce{LD_2} moderator. Additionally, at a conceptual level, we have studied the VCN source using a dedicated moderator/converter, such as the deuterated clathrates hydrates and \ce{SD_2} studied by the software (WP2) and materials characterization teams (WP2 and WP3, respectively). In the first case we aim at obtaining a gain in reflection of neutrons above 10\,\AA, but there is not an increase in VCN production, only an improved extraction. In the second case, a dedicated VCN moderator or converter will deliver a colder spectrum. These results are reported in detail in~\cref{sec:intro_VCN}.
 
%Some possible uses of VCN in fundamental physics experiments are listed in \cite{zeilinger1986_TestingQuantum,zeilinger1989_NewVeryColdNeutron,gahler1991_WaveOptical}. For both design options that will be explored within WP4, we will therefore consider as figure of merit the intensity of neutrons integrated above 40\,{\AA} for fundamental physics applications. 

The emphasis on colder neutron spectra is one of the key aspects of HighNESS, and consequently a great effort has been devoted to the design of a source delivering high intensity of neutrons above 10 \AA.

Concerning the FOM, for fundamental physics applications using VCNs, we are mostly interested in neutrons in the range 40-120\,\AA.

For condensed matter applications, the range of 10-40\,{\AA} is in general more useful (see~\cref{sec:neutronscattering}), as neutron scattering experiments are interested in extending existing techniques such as SANS and spin-echo to the use of neutrons of longer wavelengths than traditionally used. 

This distinction is however somewhat arbitrary and in the design described in 
~\cref{sec:vcn} the intensity in both ranges was calculated by looking at two wavelength bins, 10-40\,{\AA} and above 40\,{\AA}.

\FloatBarrier
\subsubsection{Figure of merit for UCN}
%we need to define what are the quantities to be calculated for UCN (e.g. neutrons per unit volume)
A further goal of the HighNESS project is to design  a source of ultracold neutrons. This UCN source can be located in-pile or in-beam. 
The design work on the different concepts is described in detail in ~\cref{sec:UCN_intro}. 
%In this section we discuss the FOMs used in the design.
Two materials have been considered in HighNESS as UCN converters: superfluid helium (He-II) and solid deuterium 
(\ce{SD_2}). In both materials, the UCN production happens via low-energy excitations induced by cold neutrons, primarily 8.9 {\AA} for He-II, and  neutrons in the range 
from 2.5--7\,{\AA} for \ce{SD_2}. 

There are several quantities that can be used to characterize the performance of a UCN source, as discussed in
~\cref{sec:production}.
In this work we focused on the calculation of the production rate density ${P_\text{UCN}}$  [cm$^\text{-3}$ s$^\text{-1}$], which is the rate of production of UCN inside the source, per unit volume of the source. 
The methods to calculate ${P_\text{UCN}}$ in He-II and \ce{SD_2} are described in 
~\cref{sec:MethodsForCalculatingUCN}.
Knowledge of the volume of the UCN converter and of the UCN storage lifetime, which is dependent on the material used, but can also depend on the source location for a different material, gives the total UCN number in the source.

Calculating the production of UCN in the source is the first step in the study; the performance of the source depends on several factors that determine losses of UCNs either in the source or in the extraction. This makes it difficult to make reliable predictions and underlines the importance of benchmarking simulations to experiments.

\begin{table}[h!]
\caption{List of FOMs. $I_P$ is the proton beam intensity (for 2.5 mA it is $1.56 \times 10^{16}$ protons/s). $\Delta \Omega$ is the solid angle defined by the emission surface and the distance at which the  tally is calculated. The integral in $dA$ is over the emission surface of the moderator; the integral in $d \Omega$ is over 4$\pi$ sr and the integral in $dt$ is over all times; collimators are used in the calculations to limit integration to the desired $\Delta \Omega$. For VCN, two different wavelength ranges are defined for fundamental physics and condensed matter, respectively, even though the separation is somewhat arbitrary. In practice, in the design described in ~\cref{sec:vcn}, the intensity in both ranges was considered.  
For UCN sources, the methods to calculate the FOM, ${P_\text{UCN}}$, are described in 
~\cref{sec:MethodsForCalculatingUCN}.}
\label{tab:fom}
\centering
\begin{tabular}{ c l }
\toprule
%\texttt{iel} & Description \\ 
%\midrule
% _{t=0}^{\infty}
COLD SOURCE\\ 
\midrule
NNBAR & $\frac {I_P}{ \Delta \Omega} \int dA \int d \Omega \int dt  \int_{\lambda=2.5 \AA}^{ 15 \AA} d\lambda \phi(\lambda,\Omega,\boldsymbol{r},t) \lambda^2  $\\
 & $\frac {I_P}{ \Delta \Omega} \int dA \int d \Omega \int dt  \int_{\lambda=5 \AA}^{ 15 \AA} d\lambda \phi(\lambda,\Omega,\boldsymbol{r},t) \lambda^2  $\\
 condensed matter & 
 $\frac {I_P}{ \Delta \Omega} \int dA \int d \Omega \int dt  \int_{\lambda=2.5 \AA}^{ 40 \AA} d\lambda \phi(\lambda,\Omega,\boldsymbol{r},t)   $\\
% $ I_P \int dA  \int_{t=0}^{\infty} dt  \int_{\lambda=2.5 \AA}^{\infty} d\lambda \phi(\lambda,\Omega,\boldsymbol{r},t)  $  \\
\midrule
\midrule
VERY COLD SOURCE\\ 
\midrule
fundamental physics  & 
$\frac {I_P}{ \Delta \Omega} \int dA \int d \Omega \int dt  \int_{\lambda=40 \AA}^{ 120 \AA} d\lambda \phi(\lambda,\Omega,\boldsymbol{r},t)   $ \\
condensed matter  &
$\frac {I_P}{ \Delta \Omega} \int dA \int d \Omega \int dt  \int_{\lambda=10 \AA}^{ 40 \AA} d\lambda \phi(\lambda,\Omega,\boldsymbol{r},t)   $ \\
%fundamental physics  & $ I_P \int dA  \int_{t=0}^{\infty} dt  \int_{\lambda=40 \AA}^{\infty} d\lambda \phi(\lambda,\Omega,\boldsymbol{r},t)  $ \\
%condensed matter  & $ I_P \int dA  \int_{t=0}^{\infty} dt  \int_{\lambda=10 \AA}^{\infty} d\lambda \phi(\lambda,\Omega,\boldsymbol{r},t)  $ \\
%\midrule
%\midrule
%ULTRA COLD SOURCE \\ 
% He-II & Flux of 9\,{\AA} neutrons inside the helium converter \\
% \ce{SD_2} & Flux of 2.5--7\,{\AA} neutrons inside the \ce{SD_2} volume \\
\bottomrule
\end{tabular}
\end{table}

\subsection{Neutronic design}

The purpose of the cold moderator is to provide cold neutrons to beamlines for
fundamental physics and neutron scattering instruments. The "NNBAR opening" is intended for fundamental physics experiments in general and the NNBAR experiment in particular, while the opening for one or more neutron scattering instruments is referred to as the "WP7 opening".

\subsubsection{Iterative design process}

The design of the cold moderator  followed several iterations, which are summarized in the following sections. The design steps performed were:
\vspace{0.25cm}
\begin{itemize}
\item[-] Preliminary neutronic studies using a cylindrical moderator shape and one opening for beam extraction (\cref{sec:1open}).
\item[-] First neutronic optimization (\cref{sec:1iter}). In this step, a box shape for the moderator was selected, and the design was performed with two openings. The FOM used for the optimization was the intensity of cold neutrons above 4 \AA.
\item[-] Second neutronic optimization (\cref{sec:2iter}). The optimization from the previous step was repeated using the FOM from \cref{tab:fom}.
\item[-] Engineering study based on the model from second neutronic optimization resulting in new engineering requirement for adding aluminum into LD$_2$.
% , see \cref{ch:1}
\item[-] Third neutronic optimization, see \cref{sec:3iter}, motivated by updated constraint on adding aluminum into LD$_2$ and the effort to minimize heatload by reducing the volume of the moderator. 
%\item[-] Engineering study based on the model from third neutronic optimization, see \cref{ch:1}.
\item[-] Engineering design based on the model from the third neutronic optimization, see \cref{ch:1}.
\item[-] Evaluation of the impact of the engineering design on neutronic performance, \cref{sec:engeffect}.
\item[-] Additional neutronic model with separated LD$_2$ box and Be filter, U-shaped reentrant hole and rounded walls of the LD$_2$ box based on new engineering studies which are in a preliminary stage, see \cref{sec:roundedshape}.

\end{itemize}

\subsubsection{Summary of the results with a single opening cylindrical moderator}
\label{sec:1open}

The initial neutronic study of the \ce{LD_2} moderator was meant to investigate its  performance with respect to the variation of a large set of parameters. For this purpose, a simple   geometry consisting of a cylindrical  moderator with a single opening for neutron extraction was used.

Here we summarize the most important findings, which provide important indications for the design of the liquid deuterium moderator with two openings.

\vspace{0.25cm}

\begin{itemize}
    \item[-] \textbf{Ortho-para deuterium fraction}. A 100\% orthodeuterium composition gives 7 \% higher cold neutron intensity than the natural composition (67 \% ortho - 33\% para). It may therefore be beneficial to use a catalyzer to increase the fraction of orthodeuterium.
    \item[-] \textbf{Temperature effect}. A comparison of intensities was performed with scattering kernels at different temperatures, ranging from 19\,K to 23\,K, for a composition of (67\% ortho -- 33\% para). The results indicate that the lower the temperature, the higher the intensity, as expected. A temperature of 20\,K gives an intensity 7 \% higher than at 23\,K. It is therefore recommended to run at a temperature closer to 20\,K.
    \item[-] \textbf{Top premoderator thickness}. The top premoderator (i.e. the water layer between moderator and target) was found to have an ideal thickness of about 3\,cm.
    \item[-] \textbf{Side and bottom premoderator thickness}. The side and bottom premoderator (i.e. the water layer between moderator and reflector) has an ideal thickness of about 1\,cm. There is a large (20-30\%) difference between having a water layer, and having no layer at all. This effect is explained by Be reflector becoming semi-transparent to colder neutrons delivered by the \ce{LD_2} moderator.
    \item[-] \textbf{Size of emission window}. For a given, fixed moderator size, the intensity increases linearly with the height of the emission window opening, as expected.
    \item[-] \textbf{MgH$_2$ as side and bottom premoderator}. In the configuration considered, there is no gain in using a MgH$_2$ side and bottom layer instead of the water layer.   
    \item[-] \textbf{Effect of HD}. With the tests performed, little (if any) gain in intensity was observed using a small amount of hydrogen in the liquid deuterium.
    \item[-] \textbf{Intensity vs moderator size for a fixed 24$\times$24 cm$^2$  opening}. The general trend observed is that the intensity is increasing with the moderator size.
    \item[-] \textbf{Reentrant hole}. A preliminary study of the effect of a reentrant hole on the intensity of a 10$\times$10 cm$^2$ emission window indicated a significant increase, at the level of 20-30 \% in the intensity. A box-shaped reentrant hole was found a preferable option over the wedge-shaped one after assessing the neutronic performance of both options.
    \item[-] \textbf{Be filter/reflector}. A first study of the effect of a Be filter/reflector on the intensity was performed. The study was also done with two openings. The first results, later investigated in more detail, indicate very promising gains, on the order of 20-30 \% especially for the NNBAR experiment.
    
\end{itemize}

\subsubsection{Design optimization}
The design of the cold source is extremely important, as two objectives must be achieved: first, to provide high-intensity cold beams to the experiments; second, to serve as primary source for a secondary VCN or UCN source. Therefore, the  design study of the cold source was performed throughout the whole duration of the HighNESS project. % The results of the first iteration of optimization are reported in Deliverable 4.2~\cite{D4.2}.
% This first optimization led to a box-shaped moderator using of LD$_2$; such moderator would provide the best neutronic performance using the NNBAR and WP7 FOMs. This model of moderator with Be filter and 24 x 40 cm$^2$ opening on the NNBAR side and reentrant hole and 15 x 15 cm$^2$ opening on the WP7 side was further tested and developed during the second iteration of optimisation while keeping it inside the ”standard” MR plug. 

\subsubsection{First iteration}
\label{sec:1iter}

For the first baseline model, a box-shaped geometry was selected, 
which was found to have better neutronic performance than the cylindrical-shaped moderator. Two box-shaped geometries were initially investigated, giving a similar neutronic  performance. The geometry with the lower   heat load was selected in the continuation of the project.

For the selection of the dimensions, among the three optimization criteria chosen for input into the optimization code Dakota~\cite{Dakota_6.18} (sum of intensities, sum of brightness, and NNBAR intensity), the sum of brightness for the two openings was selected. The reason for this choice is that, for a fixed dimension of the emission window (as was the case in all these optimizations), the sum of brightness gives equal weights to NNBAR and WP7, and is the de facto equivalent to optimizing for intensity for each of the two emission windows. 
However, among the many results provided by Dakota, with the criteria of maximization of the sum of brightness, in this first iteration a preference was given to 
the ones giving more intensity to NNBAR, while still having a high enough intensity to WP7.

%The model that has been submitted to WP5 for initial engineering design is the box shape 1 moderator from \cref{sec:box_shape_moderator1}.

The Be filter was introduced into the model adjacent to the NNBAR and WP7 emission surfaces as an efficient reflector of neutrons with wavelengths below about 4\,Å, leading to a considerable gain in cold neutron intensity (see \cref{Section_Effect_BeFilter_REH}). 

%The configuration has been chosen among the designs with the highest sum of brightness (de facto equivalent to optimizing for intensity when the size of the window is fixed) that would also satisfy the condition of giving high intensity output on the NNBAR side. For this configuration, the NNBAR intensity for neutrons above $>$ \SI{4}{\angstrom} is \num{6.89e+15} and for WP7 intensity for the same energy range is \num{2.50e+15}, with heat load of 56.4\,kW. 

\begin{figure}[bt!]
	\begin{center}
		\includegraphics[width=0.45\textwidth]{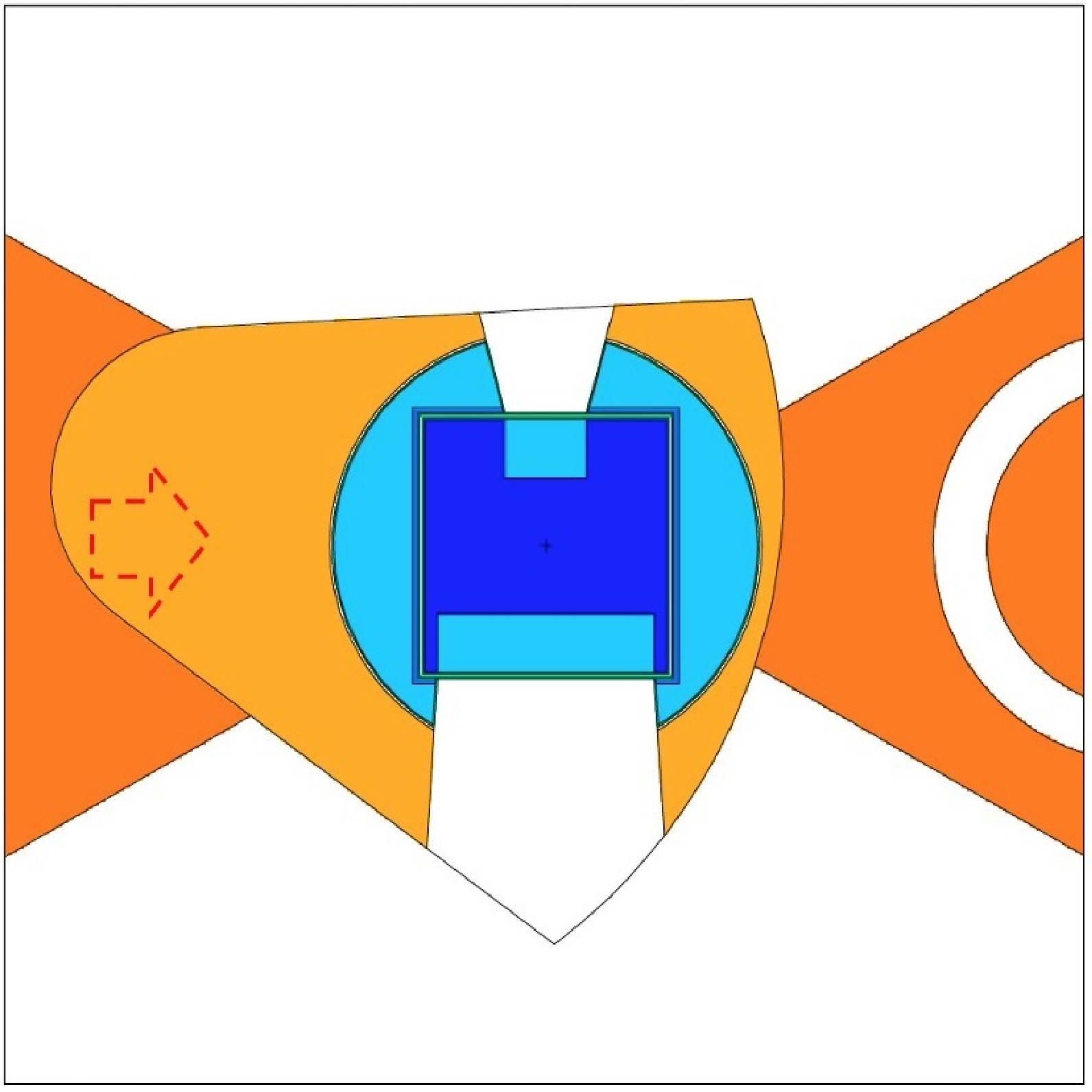}
		\includegraphics[width=0.45\textwidth]{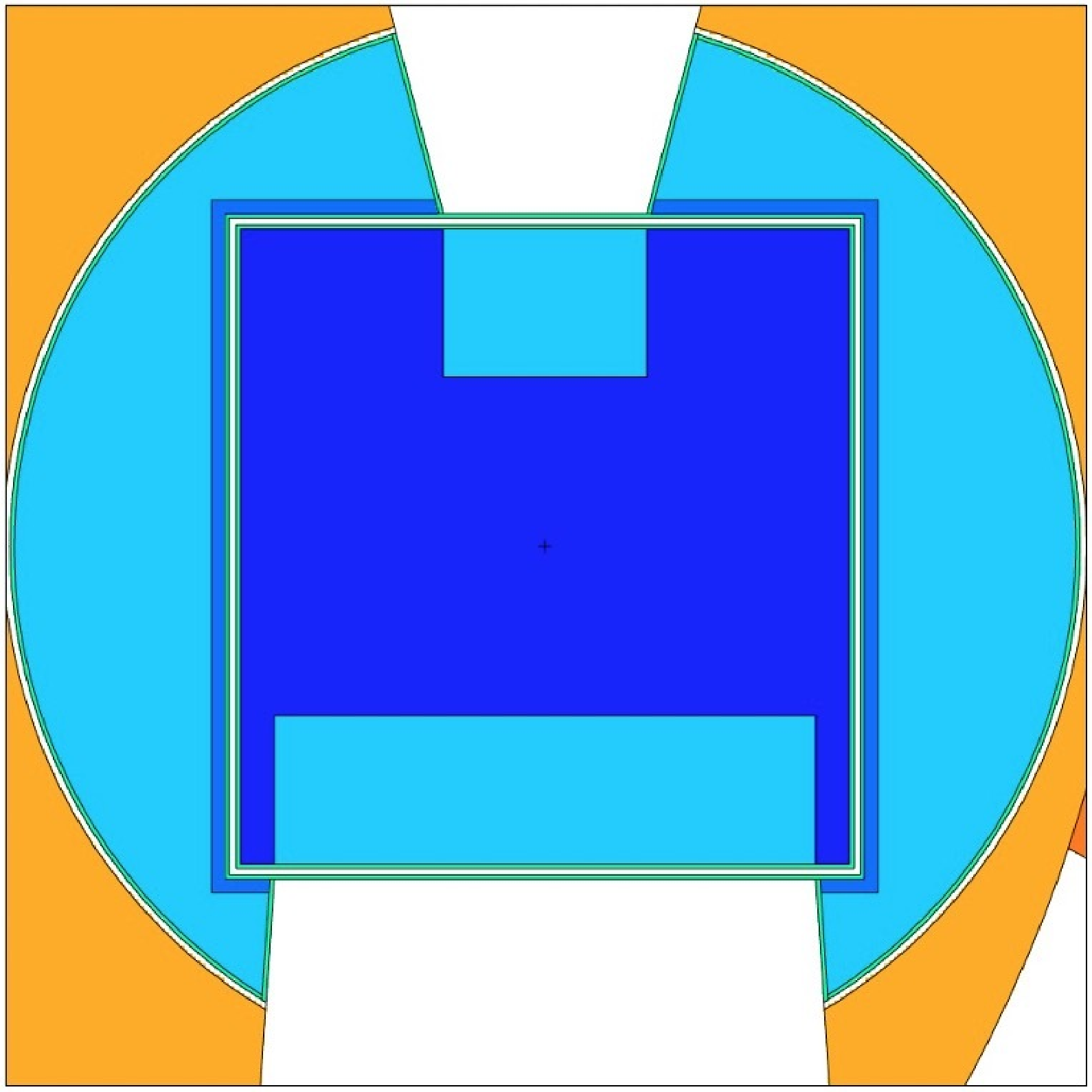}
		\includegraphics[width=0.45\textwidth]{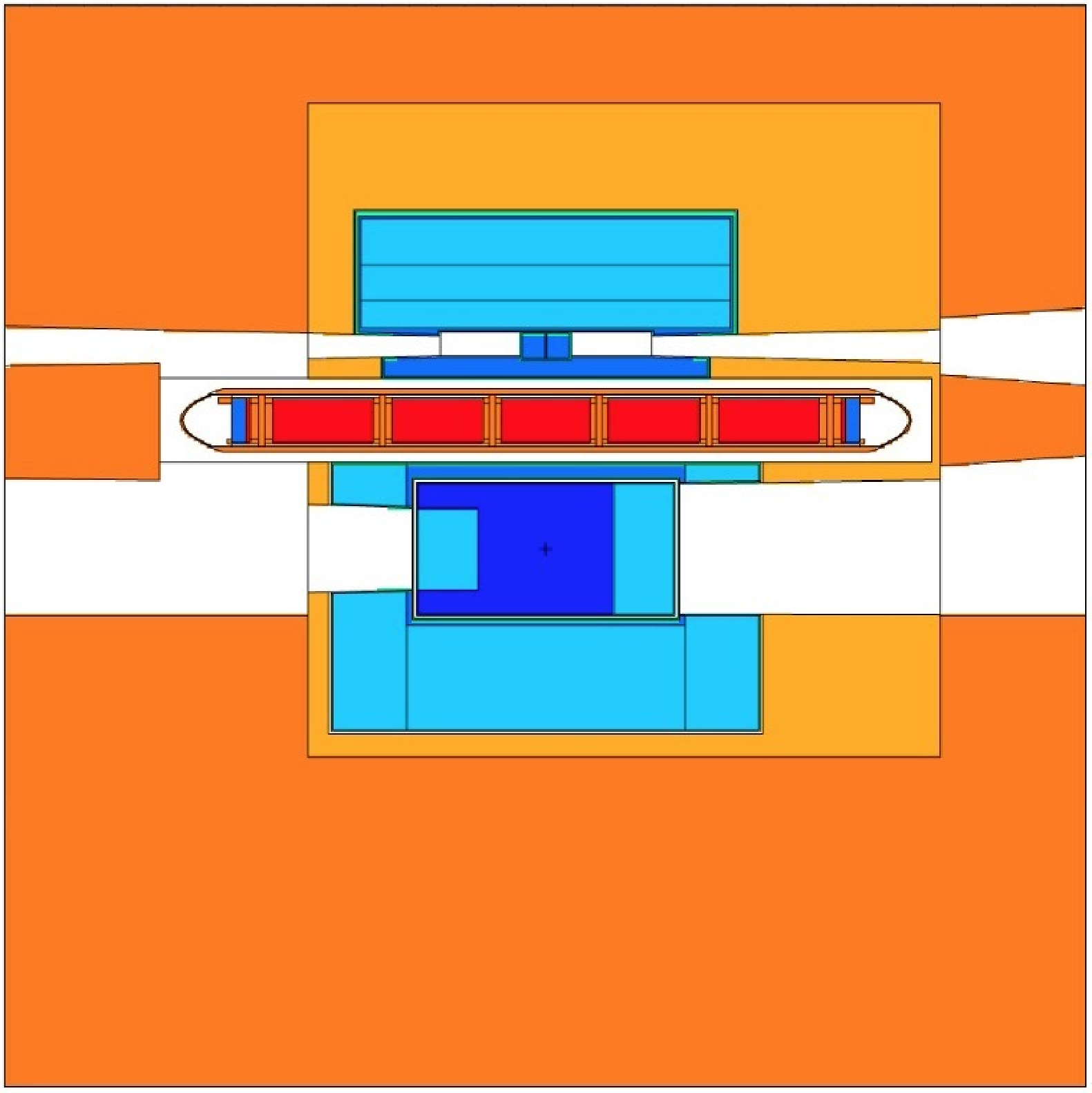}
		\includegraphics[width=0.45\textwidth]{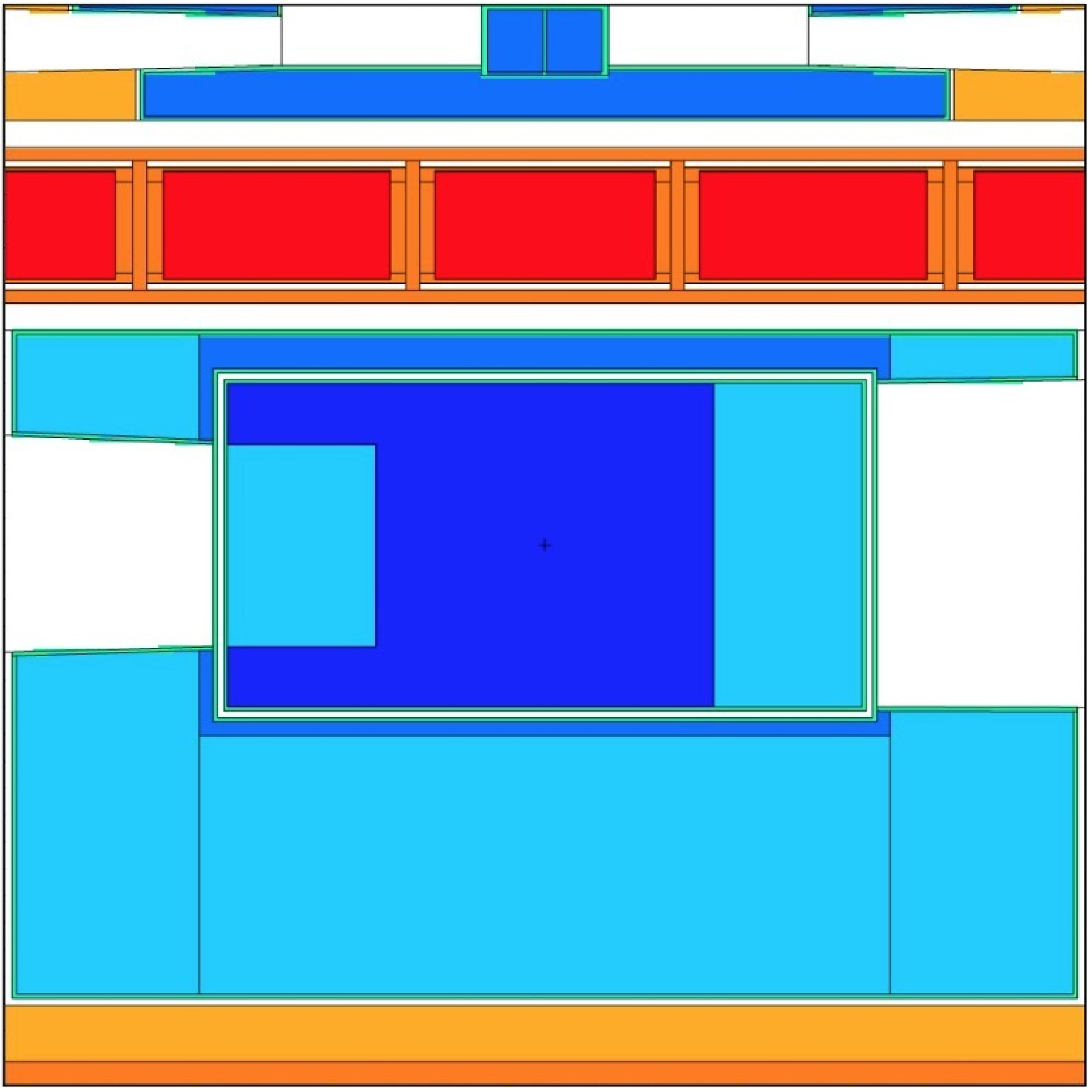}
		\includegraphics[width=0.45\textwidth]{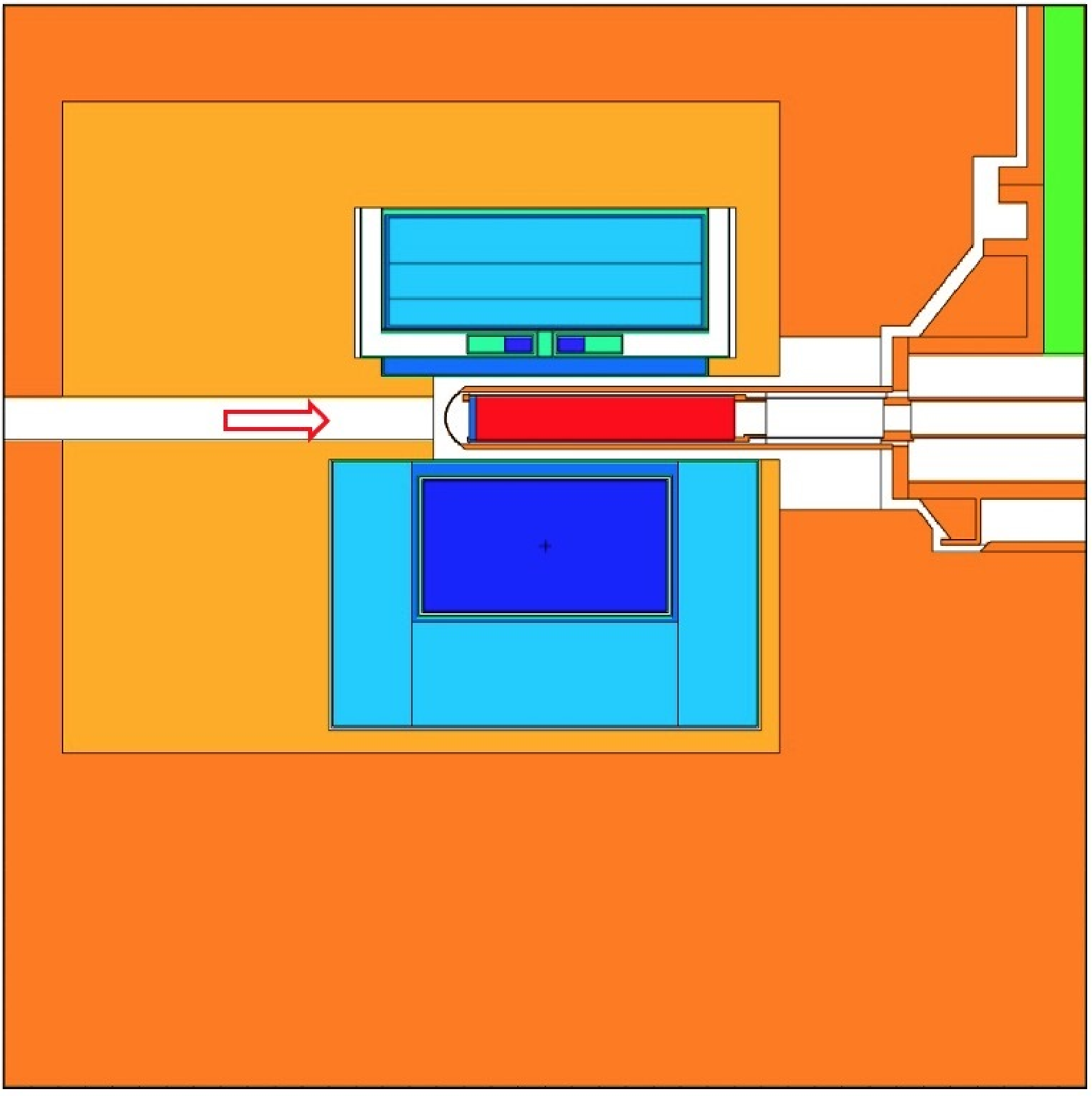}
		\includegraphics[width=0.45\textwidth]{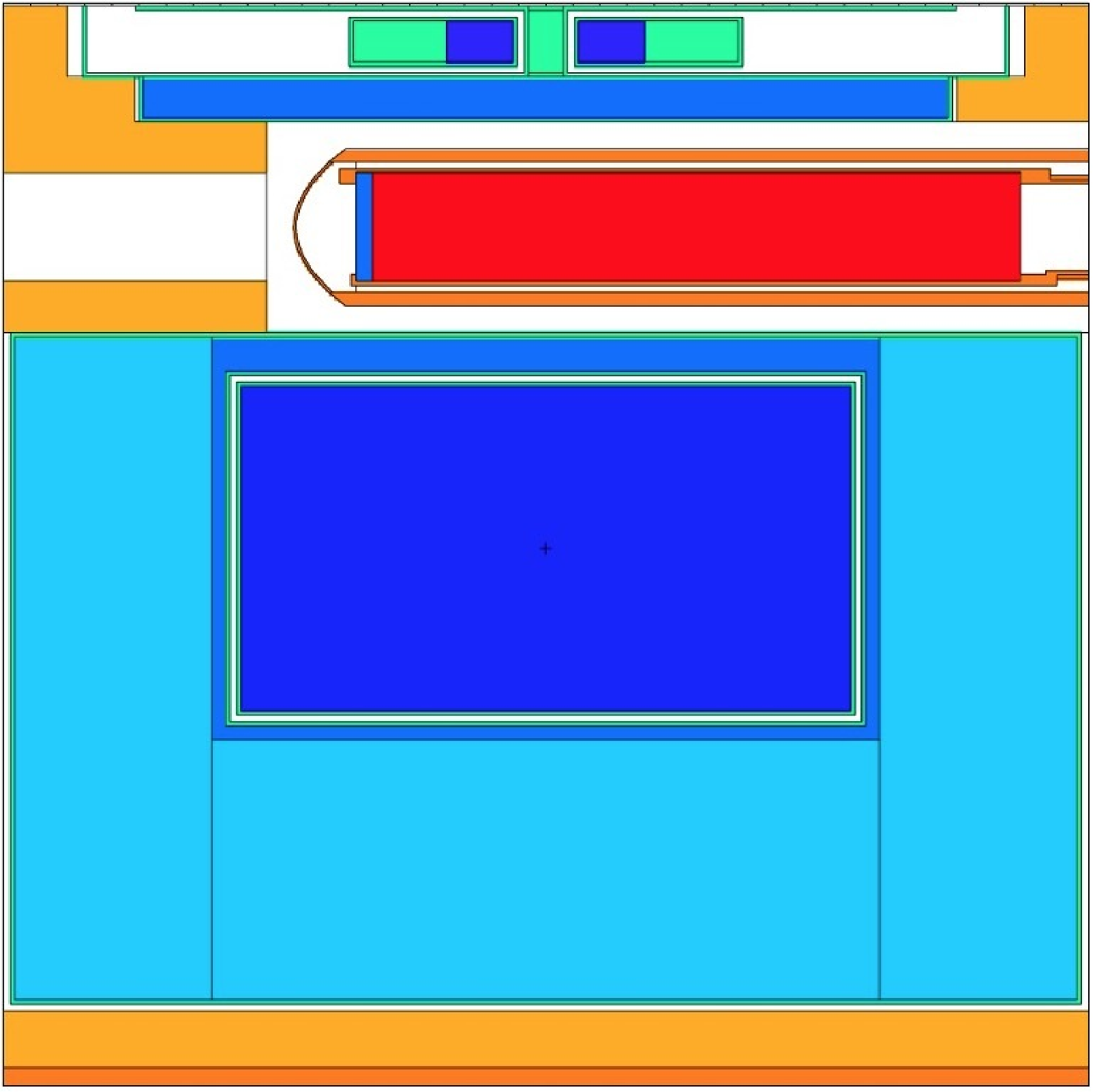}
		\caption{Graphical representation of the baseline design for the \ce{LD_2} moderator submitted to WP5 for engineering studies.
		The color codes are the following: orange: steel (twister frame, inner shielding, etc); dark blue: liquid ortho-deuterium; blue: light water; light blue: beryllium; green: aluminum. Note that cold Be filters and ambient Be reflector are shown using the same color; the same note applies to Al. The direction of the incoming proton beam is indicated in the upper and lower left figures (for the upper left figure, the dashed arrow means that the proton beam is at a different vertical level).
		}
		\label{fig:ld2_baseline_model}
	\end{center}
\end{figure}

Below is the description of the "first iteration" model of the \ce{LD_2} moderator. This model is illustrated in \cref{fig:ld2_baseline_model}.

%All dimensions indicated are tentative and subject to future change both by engineering and by neutronics.

\vspace{0.25cm}
 
\textbf{A}.
The \ce{LD_2} moderator is represented by a rectangular vessel made of aluminum. Its inner dimensions are:

\begin{itemize}[leftmargin=1cm]
    \item[-] 45\,cm long in the proton beam direction ("width"),
    \item[-] 47\,cm long in the direction transverse to proton beam direction ("length"),
    \item[-]  24\,cm long in the vertical direction ("height").
\end{itemize} 

\vspace{0.25cm}

In neutronics calculations, the thickness of the vessel Al walls is set to 0.3\,cm for the sake of consistency and comparison with work on the upper moderator. The vessel itself is filled with liquid ortho-deuterium at 20\,K and the cold Be filters are immersed in deuterium.

\vspace{0.25cm}
 
\textbf{B}.
There are two openings, both perpendicular to the incoming proton beam, but in opposite directions:

\begin{itemize}[leftmargin=1cm]
\item[-] NNBAR opening serving the North-West (NW) sector (see \cref{essinstruments});
\item[-] WP7 opening serving the defined South-East (SE) sector.
\end{itemize}

\vspace{0.25cm}

The NNBAR opening has an emission surface of 40\,cm width by 24\,cm height, while the WP7 opening has an emission surface of 15\,cm width by 15\,cm height. The geometric centers of both emission surfaces and of the \ce{LD_2} moderator are situated at the same level, -23.7\,cm  below the center of the spallation target.
 
The emission surfaces are covered from the inside by rectangular cold Be filters. The filters are 110\,mm thick each, while their widths and heights correspond to those of their respective emission surfaces. Cold Be filters are in direct contact with deuterium, however, it is important to note that the bulk temperature of cold Be filters is not required to be 20\,K: as long as their temperature is kept below about 80\,K, they will provide the expected performance increase.

\vspace{0.25cm}

\textbf{C}.
The \ce{LD_2} moderator is placed inside a standard moderator-reflector plug, thus, surrounded by the following layers:

 %\begin{itemize}
 %\item[-] 5 mm vacuum gap,
 %\item[-] 3 mm Al walls,
 %\item[-] 10 mm H2O layer below and on the sides,
 %\item[-] 25 mm H2O layer above (between the moderator and the target),
 %\item[-] Be reflector below and on the side, of as much size as possible,
 %\item[-] 3 mm Al walls.
 %\end{itemize}
\begin{itemize}[leftmargin=1cm]
  \item[-] 0.3\,cm thick Al vessel around \ce{LD_2} and Be filter at the same temperature as \ce{LD_2};
  \item[-] 0.5\,cm vacuum gap between the Al vessel and premoderator (engineering constraint);
  \item[-] 0.3\,cm thick Al walls at ambient temperature surrounding the H$_2$O premoderator and Be reflector; 
  \item[-] 1.0\,cm thick layer of H$_2$O at ambient temperature below and on the sides to serve as a premoderator;
  \item[-] 2.5 cm H$_2$O layer at ambient temperature above (between the moderator and the target);
  \item[-] Be at ambient temperature below and on the side to serve as reflector, with as much volume as possible.
\end{itemize}

\vspace{0.25cm}
 
%\textbf{D}.
%Heat load on cryogenic parts of the MR plug:
 
% \begin{itemize}[leftmargin=1cm]
%     \item[-] Deuterium: 29.4 kW,
%    \item[-] Al walls: 7.2 kW,
%    \item[-] Be filter for nnbar opening: 15.8 kW,
%    \item[-] Be filter for material opening: 4.2 kW,
%    \item[-] Total: 56.6 kW.
% \end{itemize}

The summary of the expected performances, integrated over various wavelength ranges, is shown in \cref{tab:performance_table_iteration1}.

\begingroup
    \setlength{\tabcolsep}{10pt} 
    \begin{table}[bt!]
    \centering
%    \begin{threeparttable}
      \caption{Neutronic performance and characteristics of the model from the first iteration of the design of the cold moderator. Brightness and intensity are integrated over different wavelength ranges (NNBAR and WP7 openings) and compared with the upper moderator. The values for the upper moderator are from Ref. \cite{zanini_design_2019}. To calculate the intensity we considered the following size of the emission windows: NNBAR -- 960 cm$^2$ (40 cm $\times$ 24 cm); WP7 -- 225 cm$^2$ (15 cm $\times$ 15 cm); upper moderator -- 42 cm$^2$ (sum of two openings of the butterfly moderator of 7 cm $\times$ 3 cm for the beamline perpendicular to the proton beam, i.e., in the NNBAR direction, cf. also \cite{zanini_design_2019}). All values are for 5 MW average beam power. The heatload on Be filters accounts for both the contribution at NNBAR and WP7 openings. The moderator volume is the sum of the volumes of the LD$_2$ box and Be filters.}
    \begin{tabular}{lccc}
    \toprule   & \multicolumn{3}{c}{BRIGTHNESS [\si{n/cm\squared/s/sr}] }\\
    \cmidrule{2-4}
       & $>$ \SI{2}{\angstrom} & $>$ \SI{4}{\angstrom} & $>$ \SI{10}{\angstrom} \\
    \midrule
    NNBAR & \num{9.11e+12} & \num{7.04e+12} & \num{5.51e+11}  \\
    WP7 & \num{1.45e+13} & \num{1.08e+13} & \num{8.13e+11} \\
    upper moderator & \num{5.3e13} & \num{1.7e13} & \num{9.9e11}  \\
    \midrule
             & \multicolumn{3}{c}{INTENSITY [\si{n/s/sr}] }\\
    \cmidrule{2-4}
       & $>$ \SI{2}{\angstrom} & $>$ \SI{4}{\angstrom} & $>$ \SI{10}{\angstrom} \\
    \midrule
    NNBAR & \num{8.74e+15} & \num{6.76e+15} & \num{5.29e+14}  \\
    WP7 & \num{3.26e+15} & \num{2.42e+15} & \num{1.83e+14} \\
    upper moderator & \num{2.2e15} & \num{7.0e14} & \num{4.2e13}  \\
    \midrule
    NNBAR FOM & \num{3.00e+17} [\si{n/s/sr} $\times$ $\lambda^2$] &  & \\
    WP7 FOM & \num{2.96e+15} [\si{n/s/sr}] & & \\
    Heatload on LD$_2$ moderator & \num{29.3} [kW] & & \\
    Heatload on Be filter & \num{20.0} [kW] & & \\
    Heatload on Al vessel & \num{7.1} [kW] & & \\
    Total heatload & \num{56.4} [kW] & & \\
    Moderator volume  & \num{50.8} [liters] & & \\
    \bottomrule
    \end{tabular}
  \label{tab:performance_table_iteration1}
%  \end{threeparttable}
\end{table}
\endgroup

\subsubsection{Second iteration}
\label{sec:2iter}

In the first iteration, all the optimizations and calculations were done using the intensity for neutrons above \SI{4}{\angstrom} as FOM. This was a FOM meant to generally target cold neutrons in the early optimization steps. The second round of optimization was performed using the more elaborate FOMs, cf. \cref{tab:fom}. The optimum size of the \ce{LD_2} moderator was found to be similar (45 cm width, 48.5 cm length, 24 cm height). A reentrant hole (REH) of 12.5 cm depth was introduced at the emission surface of the WP7 opening replacing the Be filter, providing a considerable gain in cold neutron intensity and brightness (see \cref{Section_Effect_BeFilter_REH})). The NNBAR opening is covered by an 11-cm thick Be filter. 
%The optimization is described in detail in \cite{D4.2}.
 
% The resulting geometry is shown in \cref{fig:ld2_baseline_modelb}.

% \begin{figure}[bt!]
%	\begin{center}
%		\includegraphics[width=0.45\textwidth]{cn_fig/alann1.eps}
%		\includegraphics[width=0.45\textwidth]{cn_fig/alann2.eps}
%		\includegraphics[width=0.45\textwidth]{cn_fig/alann3.eps}
%		\includegraphics[width=0.45\textwidth]{cn_fig/alann4.eps}
%		\includegraphics[width=0.45\textwidth]{cn_fig/alann5.eps}
%		\includegraphics[width=0.45\textwidth]{cn_fig/alann6.eps}
%		\caption{Graphical representation of  design for \ce{LD_2} moderator after the second iteration of optimization. Compared to the model shown in \cref{fig:ld2_baseline_model}, note the removal of the Be filter on the WP7 side.
%		The color codes are the following: orange: steel (twister frame, inner shielding, etc); dark blue: liquid ortho-deuterium; blue: light water; light blue: beryllium; green: aluminum. Note that cold Be filters and ambient Be reflector are  shown using the same color; the same note applies to Al.
%		}
%		\label{fig:ld2_baseline_modelb}
%	\end{center}
% \end{figure}

The design of the moderator, optimized according to the procedure outlined above, gives the intensity of the source at the NNBAR side, integrated above \SI{4}{\angstrom}, of \SI{6.7e15}{n/s/sr}. 
This is the intensity of the emission window of 24$\times$40 cm$^2$, for 5\,MW of beam power (i.e. 2\,GeV for 2.5\,mA average proton current). The intensity at the WP7 side is lower, because of the smaller size of the opening, although it has a greater brightness.
For a beam power of 2\,MW (800\,MeV for 2.5\,mA average proton beam current) the intensity is simply a factor of 2.5 lower, because as shown later in \cref{fig:spectra_MOGA}, the difference in emitted spectra from the moderator between 800\,MeV and 2\,GeV proton beam is negligible. Therefore, at 2\,MW the intensity of the moderator is \SI{2.7e15}{n/s/sr}. 

To have an idea of what these numbers mean in terms of performance of the experiments intended to use this moderator, we can compare with the original calculations done for the NNBAR experiment in \cite{klinkby2014voluminous}: at 5\,MW beam power, the intensity was predicted to be \SI{2.9e15}{n/s/sr}. That calculation was the basis for the original estimate of the performance increase of NNBAR compared to the ILL measurement~\cite{Baldo-Ceolin:1994hzw}. It would therefore mean that the NNBAR performance estimate of this second iteration of the moderator at 2 MW is practically equivalent to the previous estimate for 5 MW. We note however that a large part of this gain is due to the increase in the size of the emission window, which almost doubled; as neutrons are emitted from a more extended source,  their transport towards the NNBAR detector is expected to be more challenging.

A summary of the expected performances for the moderator designed in this iteration, integrated in different wavelength ranges, is shown in \cref{tab:performance_iteration2}. This model was sent to WP5 team for engineering design.

\begingroup
    \setlength{\tabcolsep}{10pt} 
    \begin{table}[bt!]
    \centering
%    \begin{threeparttable}
      \caption{Neutronic performance and characteristics of the model from second iteration of cold moderator. Brightness and intensity integrated over different wavelength ranges (NNBAR and WP7 openings) and compared with the upper moderator. The values for the upper moderator are from Ref \cite{zanini_design_2019}. To calculate the intensity we considered the following size of emission windows: NNBAR -- 960 cm$^2$; WP7 -- 225 cm$^2$; upper moderator -- 42 cm$^2$. The moderator volume is the sum of volume of the LD$_2$ box and Be filter.}
    \begin{tabular}{lccc}
    \toprule   & \multicolumn{3}{c}{BRIGTHNESS [\si{n/cm\squared/s/sr}] }\\
    \cmidrule{2-4}
       & $>$ \SI{2}{\angstrom} & $>$ \SI{4}{\angstrom} & $>$ \SI{10}{\angstrom} \\
    \midrule
    NNBAR & \num{8.87e+12} & \num{6.95e+12} & \num{5.77e+11}  \\
    WP7 & \num{1.79e+13} & \num{1.07e+13} & \num{9.16e+11} \\
    upper moderator & \num{5.3e13} & \num{1.7e13} & \num{9.9e11}  \\
    \midrule
             & \multicolumn{3}{c}{INTENSITY [\si{n/s/sr}] }\\
    \cmidrule{2-4}
       & $>$ \SI{2}{\angstrom} & $>$ \SI{4}{\angstrom} & $>$ \SI{10}{\angstrom} \\
    \midrule
    NNBAR & \num{8.52e+15} & \num{6.67e+15} & \num{5.54e+14}  \\
    WP7 & \num{4.03e+15} & \num{2.42e+15} & \num{2.06e+14} \\
    upper moderator & \num{2.2e15} & \num{7.0e14} & \num{4.2e13}  \\
    \midrule
    NNBAR FOM & \num{3.03e+17} [\si{n/s/sr} $\times$ $\lambda^2$] &  & \\
    WP7 FOM & \num{3.68e+15} [\si{n/s/sr}] & & \\
    Heatload on LD$_2$ moderator & \num{29.8} [kW] & & \\
    Heatload on Be filter & \num{15.1} [kW] & & \\
    Heatload on Al vessel & \num{7.2} [kW] & & \\
    Total heatload & \num{52.1} [kW] & & \\
    Moderator volume  & \num{49.6} [liters] & & \\
%    Heatload & \num{0} [kW] & & \\
    \bottomrule
    \end{tabular}
  \label{tab:performance_iteration2}
%  \end{threeparttable}
\end{table}
\endgroup

\subsubsection{Third iteration}
\label{sec:3iter}

The third iteration of neutronic optimization of the moderator was performed using Dakota with the model from the second iteration updated by the engineering constraint of requiring a 2.5\% volume of aluminium in LD$_2$ to account for flow guides and support structures. Moreover, the third neutronic optimization was also motivated by an attempt to reduce the heat load in the moderator while preserving or even improving its performance. The temperature of LD$_2$ in the moderator vessel was increased from 20 K to 22 K, following indications from the engineering team about the expected temperature of the liquid deuterium in the moderator.  %The results of third iteration of optimisation are presented in this report in more detail. 

%\textcolor{red}{following part not clear, please go through my comments in red}
%The following outer layers around LD$_2$ and Be were designed based on the engineering constraints and results in  \cite{D4.2} as part of the first and second iteration of optimisation and were used for the third iteration of optimisation:
%\begin{itemize}
%  \item[-] 3 mm thick Al vessel around \ce{LD_2} and Be filter (engineering constraint) - the innermost layer
%  \textcolor{red}{I don't think this is an engineering constraint, actually the thickness by engineering was 8 mm!}
%  \item[-] 5 mm vacuum gap (engineering constraint)
%  \item[-] 3 mm Al walls (engineering constraint) \textcolor{red}{which Al walls, external, after the vacuum gap? - ANSW: These bullters should follow a logic from inermost to outermost layer}
%  \item[-] 10 mm H$_2$O layer below and on the sides, as a result of the neutronic optimization;
%  \item[-] 25 mm H2O layer above (between the moderator and the target), as a result of the neutronic optimization;
%  \item[-] Be reflector below and on the side, of as much size as possible
%  \item[-] 3 mm Al walls - outermost layer
% \end{itemize}

For the purpose of the third iteration of optimization, a parametric model of the moderator was used with the following variable parameters:

\vspace{0.25cm}

\begin{itemize}[leftmargin=1cm]
  \item[-] Moderator length; 
  \item[-] Moderator width;
  \item[-] Be filter thickness;
  \item[-] Reentrant hole depth.
\end{itemize}

\vspace{0.25cm}

\begin{figure}[bt!]
	\begin{center}
		\includegraphics[width=0.45\textwidth]{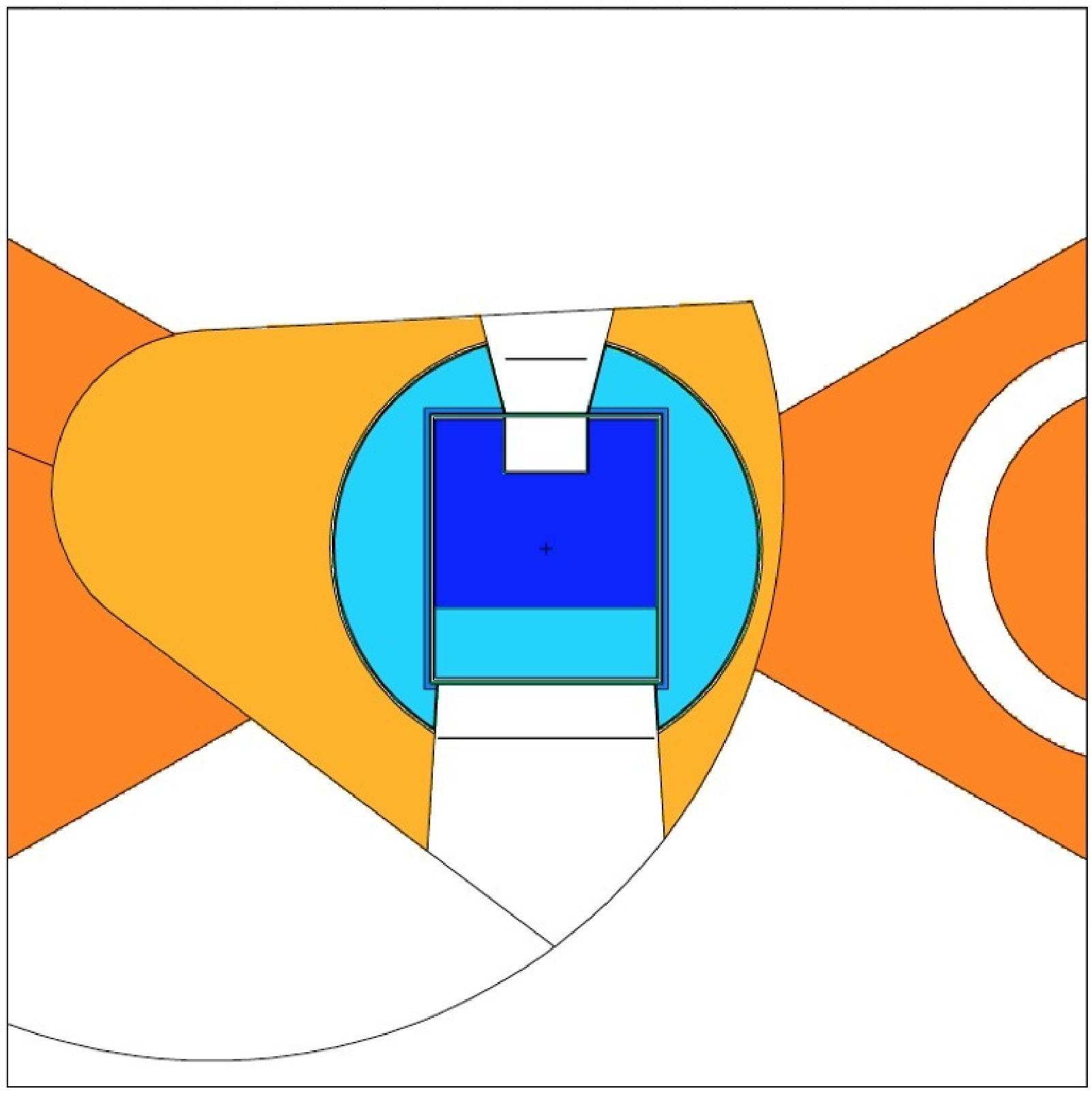}
		\includegraphics[width=0.45\textwidth]{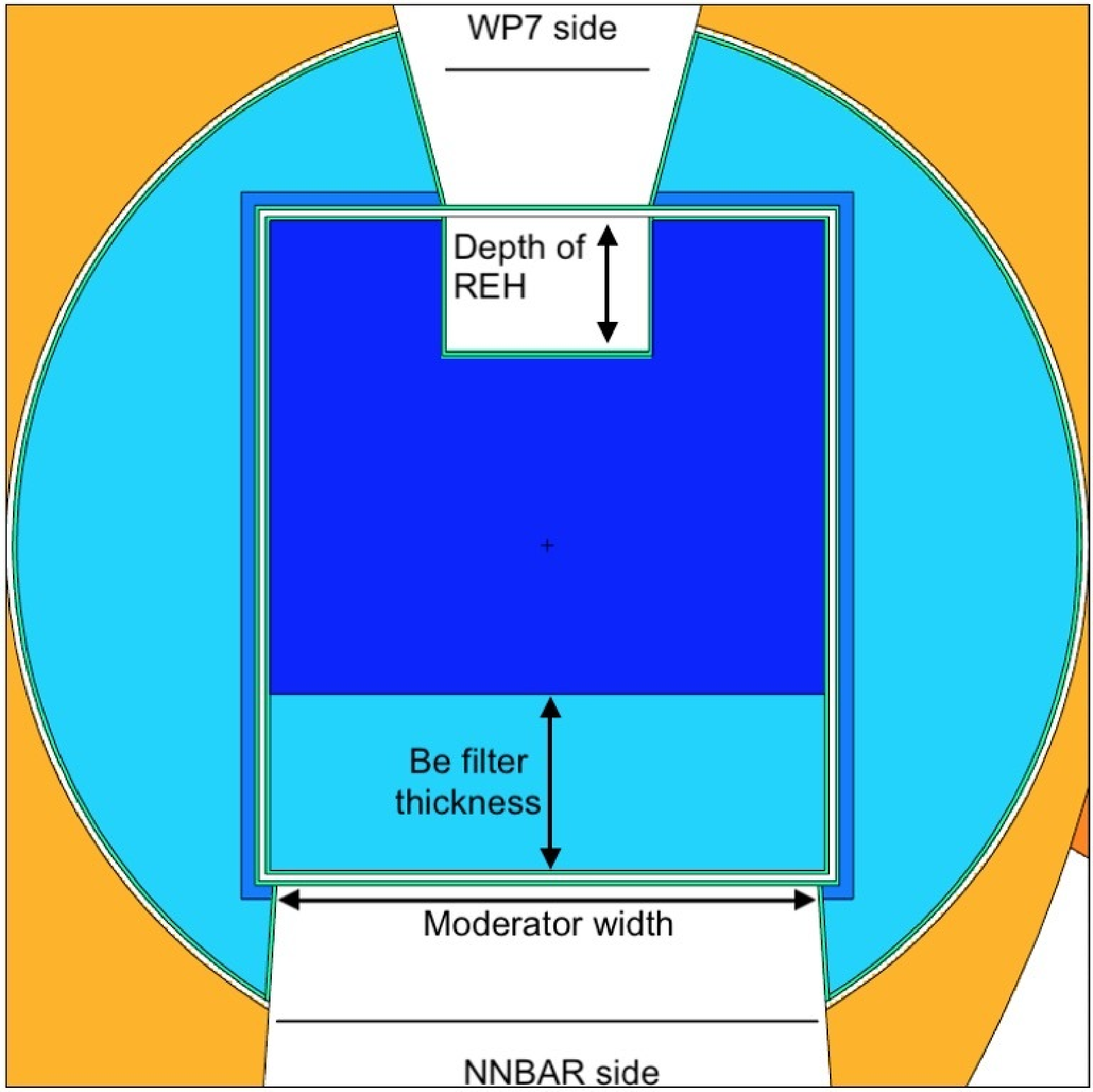}
		\includegraphics[width=0.45\textwidth]{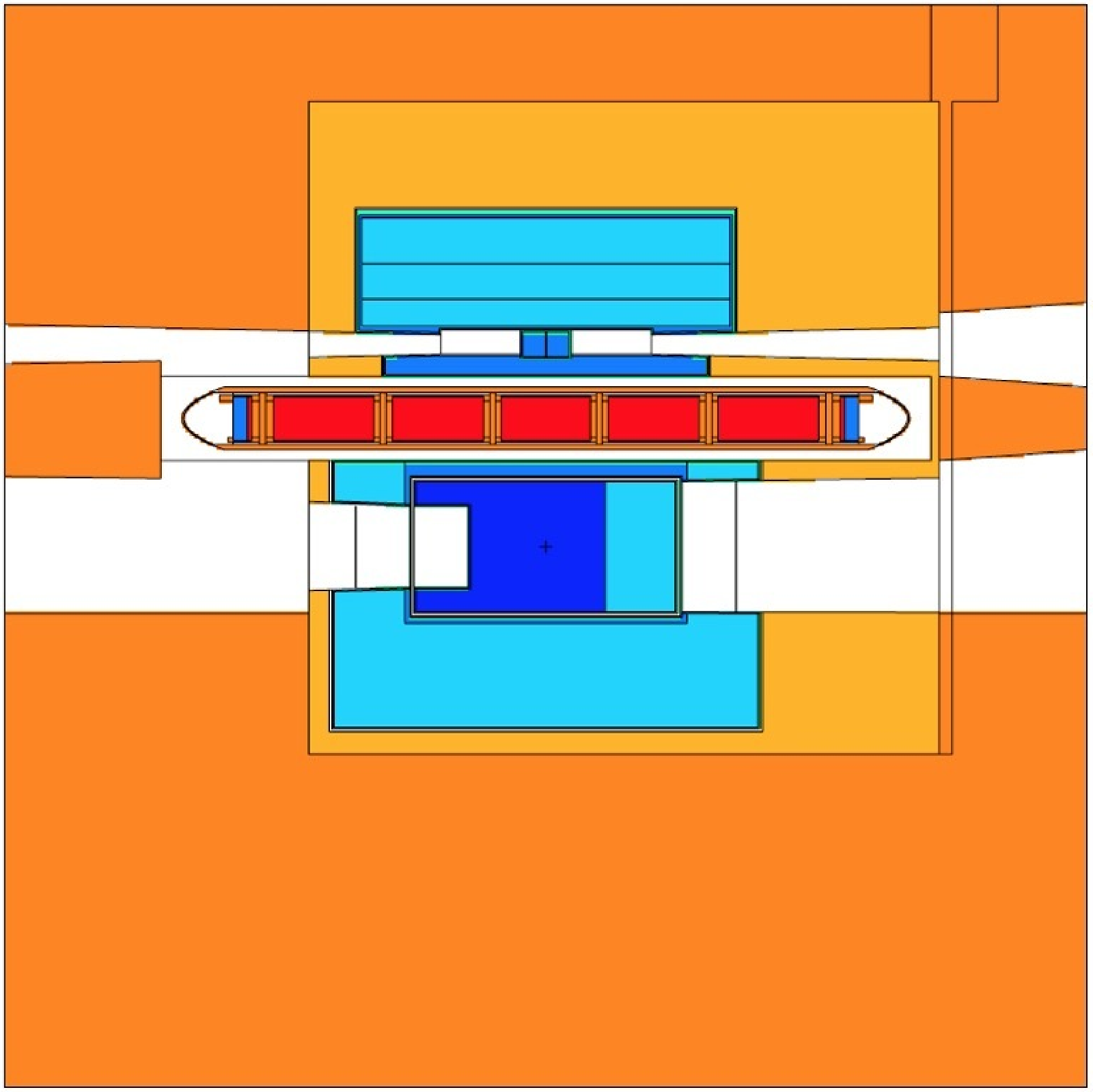}
		\includegraphics[width=0.45\textwidth]{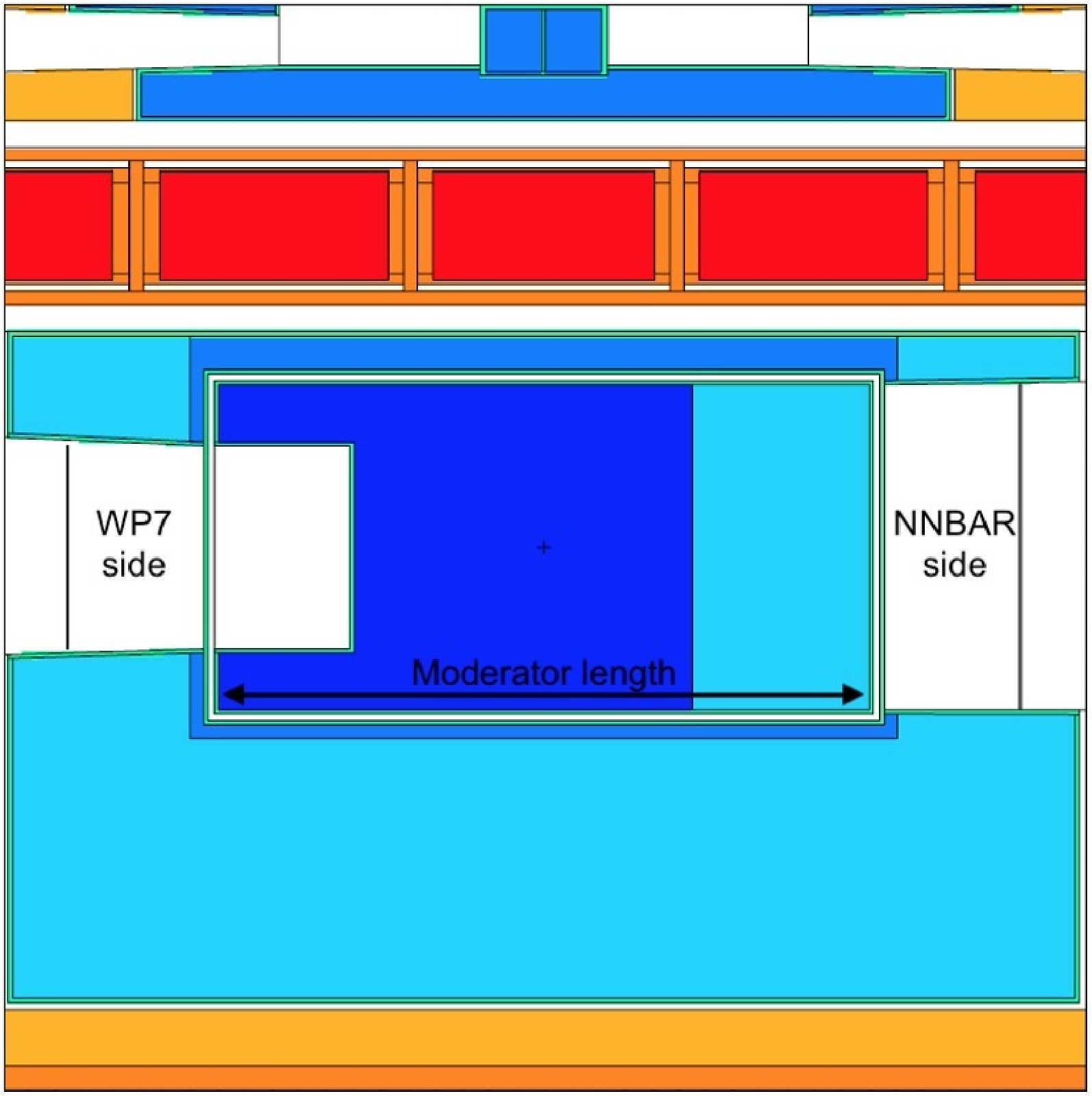}
		\includegraphics[width=0.45\textwidth]{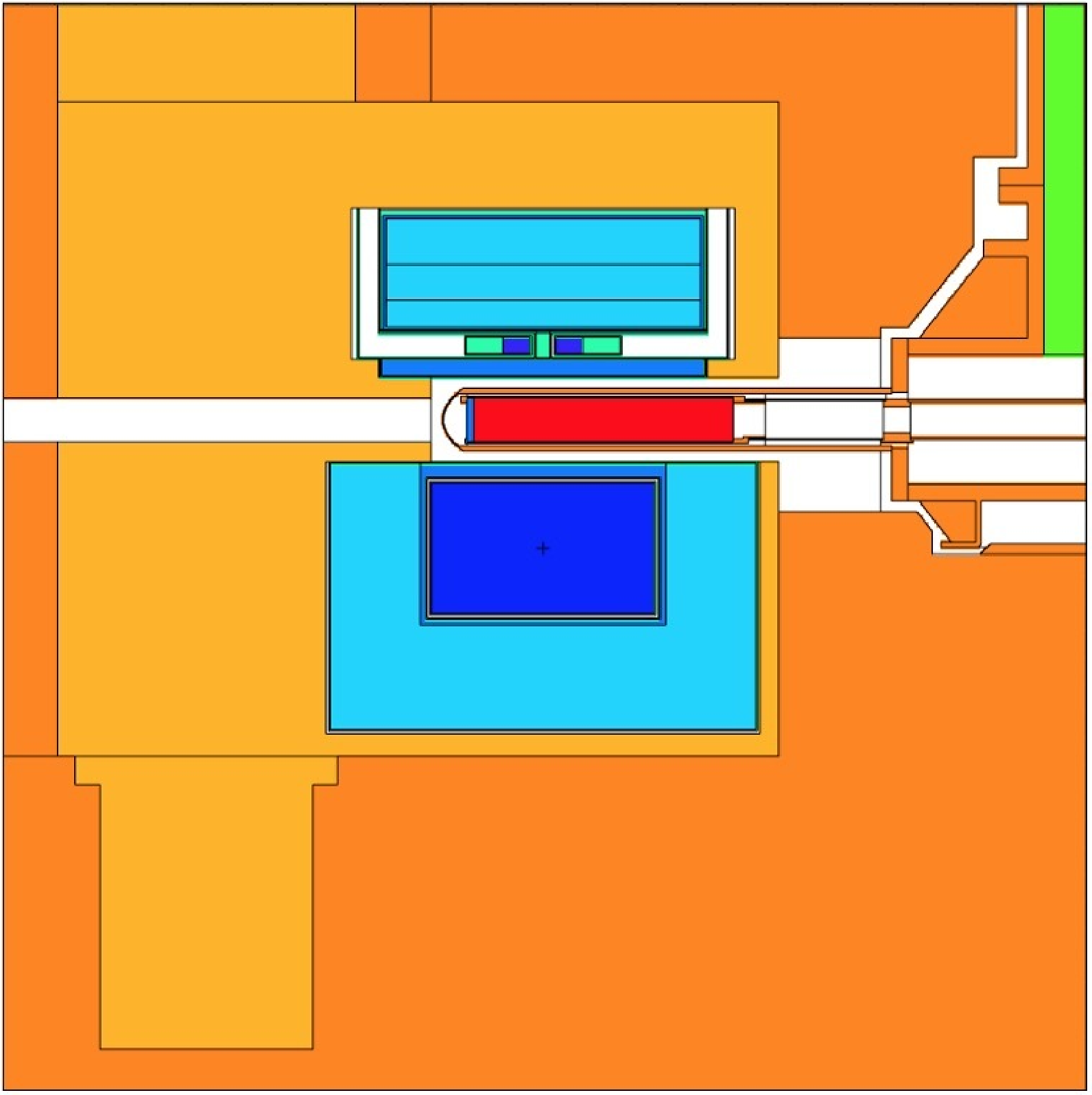}
		\includegraphics[width=0.45\textwidth]{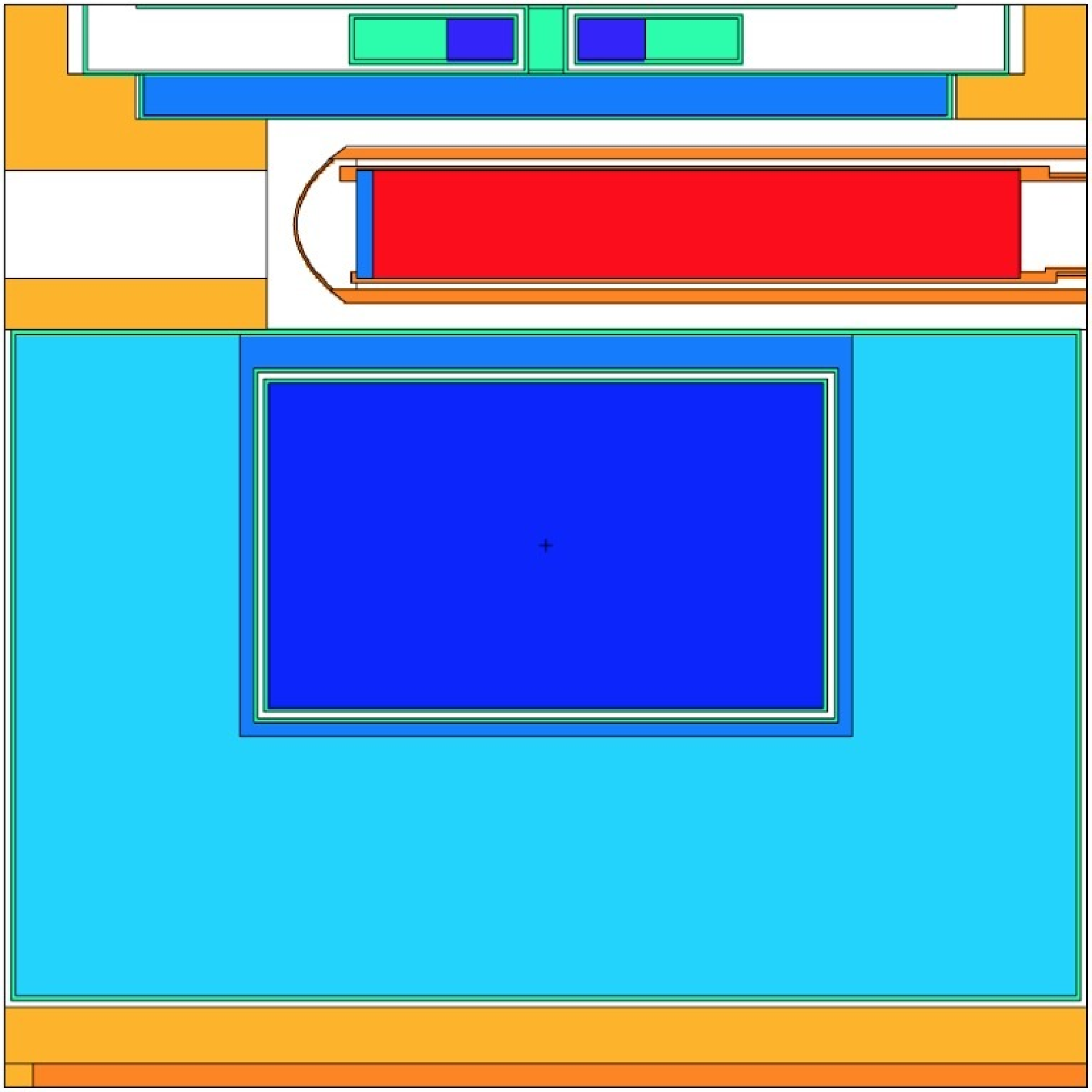}
		\caption{Graphical representation of the design for the \ce{LD_2} moderator after the third iteration. 
		The color codes are the following: orange: steel (twister frame, inner shielding, etc); dark blue: liquid ortho-deuterium; blue: light water; light blue: beryllium; green: aluminum. Note that cold Be filters and ambient Be reflector are  shown using the same color; the same note applies to Al.
		}
		\label{fig:ld2_baseline_model_Iteration3}
	\end{center}
\end{figure}

\begin{figure}[hbt!]
\begin{center}
\includegraphics[width=0.75\textwidth]{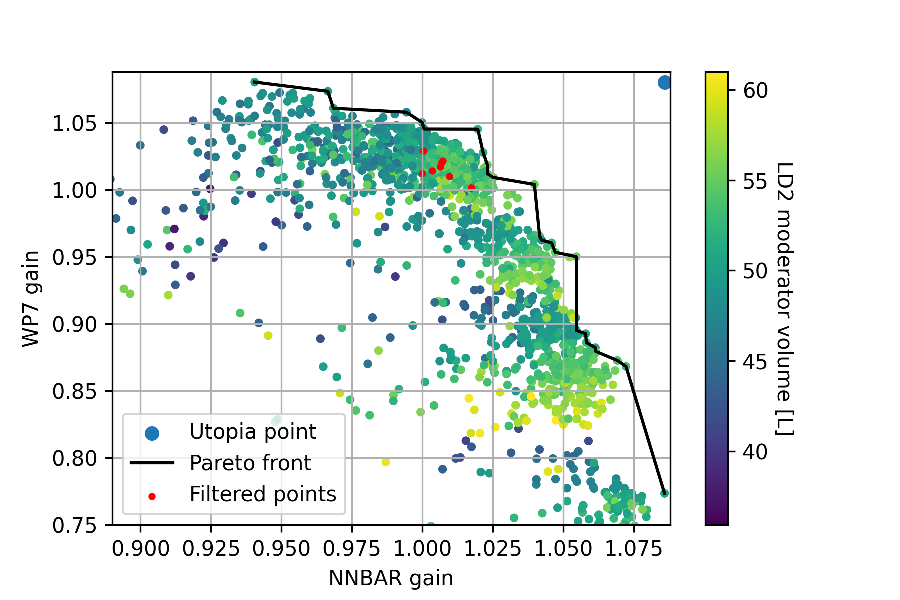}
\caption{Third iteration of optimization: each point represents the gain in FoMs for NNBAR and WP7 with a given set of parameters with respect to the baseline model obtained in the second iteration. The Pareto front was fitted through points closest to Utopia point, an idealized point with highest possible NNBAR and WP7 gains. The aim was to select a model that lies as close as possible to the Utopia point. The moderator volume was calculated as the sum of volume of LD$_2$ box and Be filter. 
The red circles depict models after applying the limits on moderator parameters. }
\label{fig:LD2_Optimisation_12_Blahoslav.eps}
\end{center}
\end{figure}

\begin{figure}[tb!]      
    \begin{subfigure}[b]{0.48\textwidth}
        \centering
        \includegraphics[width=\textwidth]{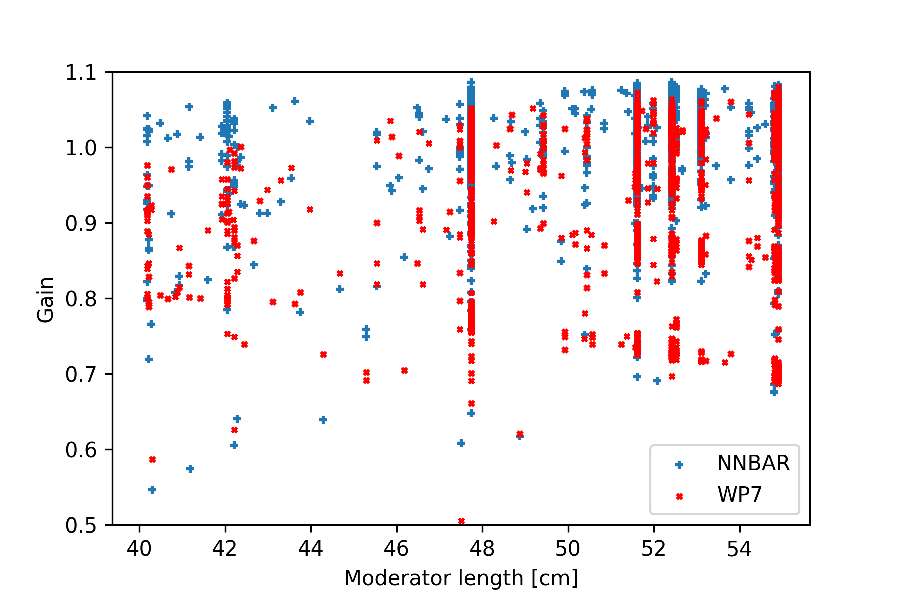}
        \subcaption{}
        \label{fig:LD2_Optimisation_1_Blahoslav.eps}
    \end{subfigure}
    \hfill
    \begin{subfigure}[b]{0.48\textwidth}
        \centering        
        \includegraphics[width=\textwidth]{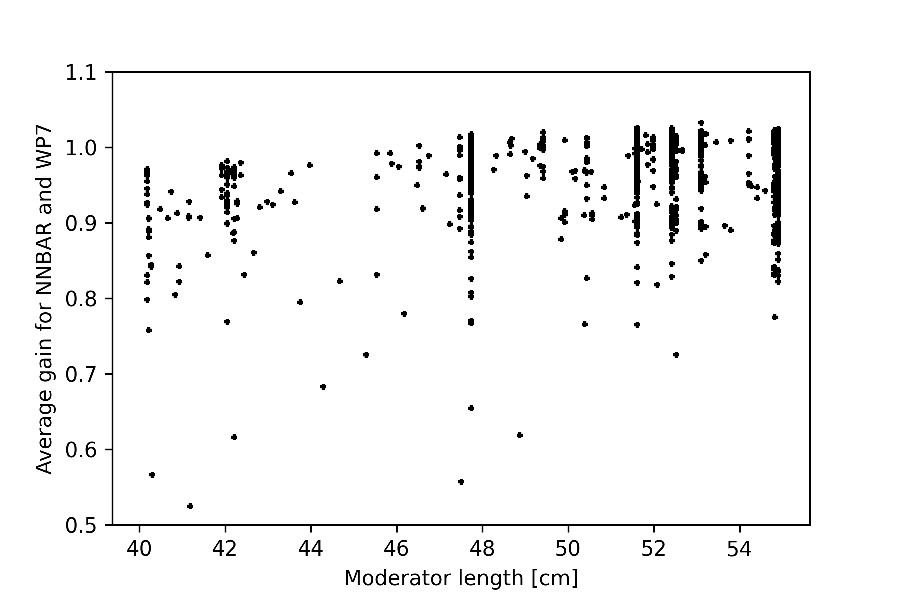}
        \subcaption{}
        \label{fig:LD2_Optimisation_101_Blahoslav.eps}
    \end{subfigure}
    \caption{Performance of models with respect to moderator length. (a) Gain in NNBAR and WP7 FOMs with respect to the baseline model. (b) Average gain in NNBAR and WP7 FOM with respect to the baseline model.}
        \label{fig:LD2_Optimisation_1}
    \end{figure}

\begin{figure}[tb!]      
    \begin{subfigure}[b]{0.48\textwidth}
        \centering
        \includegraphics[width=\textwidth]{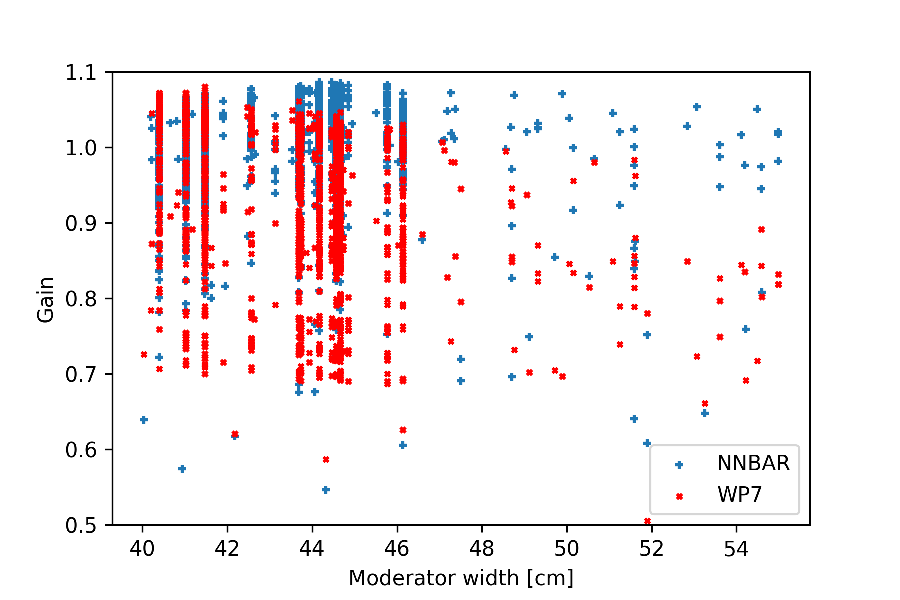}
        \subcaption{}
        \label{fig:LD2_Optimisation_3_Blahoslav.eps}
    \end{subfigure}
    \hfill
    \begin{subfigure}[b]{0.48\textwidth}
        \centering        
        \includegraphics[width=\textwidth]{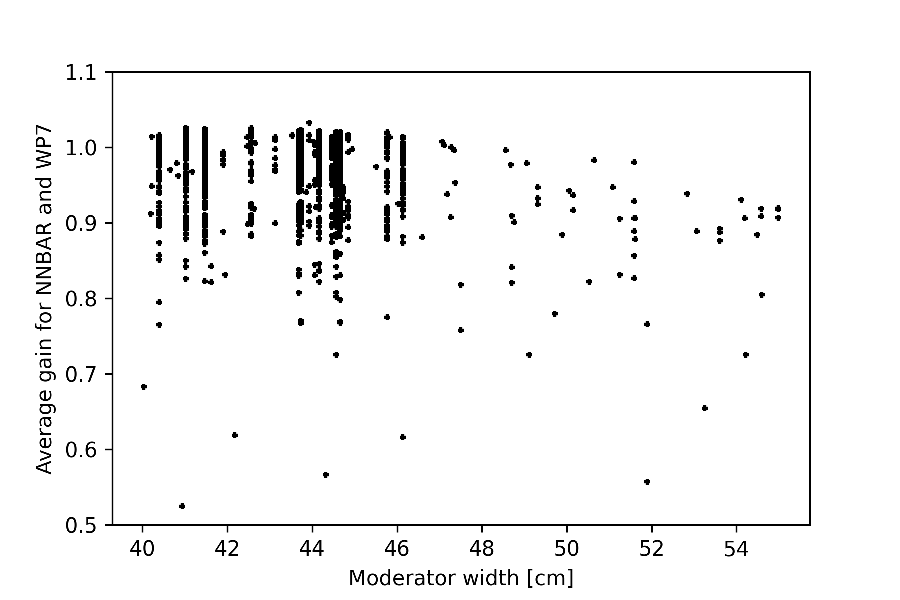}
        \subcaption{}
        \label{fig:LD2_Optimisation_301_Blahoslav.eps}
    \end{subfigure}
    \caption{Performance of models with respect to moderator width. (a) Gain in NNBAR and WP7 FOM with respect to the baseline model (b) Average gain in NNBAR and WP7 FOM with respect to the baseline model.}
        \label{fig:LD2_Optimisation_3}
    \end{figure}

The height of the moderator was fixed at 24\,cm. The listed parameters (see \cref{fig:ld2_baseline_model_Iteration3}) were optimised with respect to NNBAR and WP7 FOMs. A multi-objective genetic algorithm (MOGA) \cite{murata1995moga} was implemented resulting in a set of solutions over a wide range of NNBAR and WP7 FOMs. These values were compared to the baseline model from the second optimization iteration; however the baseline model was modified adding 2.5 vol\% Al in the liquid deuterium, and increasing the temperature of LD$_2$ from 20 K to 22 K. In \cref{fig:LD2_Optimisation_12_Blahoslav.eps} the gains made over the second iteration FoMs for WP7 are shown as a function of gains for NNBAR. A Pareto front was fitted through points closest to Utopia point, an idealized point with highest possible NNBAR and WP7 gains. In the subsequent analysis of possible solutions, these gains in NNBAR and WP7 FOMs were analyzed with regards to moderator length (see \cref{fig:LD2_Optimisation_1}), moderator width (see \cref{fig:LD2_Optimisation_3}), Be filter thickness (see \cref{fig:LD2_Optimisation_4}) and the depth of the reentrant hole (see \cref{fig:LD2_Optimisation_5}). 
The following limits on variable parameters used in the optimisation to maximize NNBAR and WP7 FOMs were defined:  

\vspace{0.25cm}

\begin{itemize}[leftmargin=1cm]
  \item[-] 45 cm $\leq$ moderator length $\leq$ 50 cm;
  \item[-] 40 cm $\leq$ moderator width $\leq$ 45 cm;
  \item[-] 11 cm $\leq$ Be filter thickness $\leq$ 15 cm;
  \item[-] 9 cm $\leq$ depth of reentrant hole $\leq$ 15 cm;
  \item[-] Gain wrt to the model from second iteration ("baseline") $\geq$ 1.
\end{itemize}

\vspace{0.25cm}

A smaller subset of solutions was filtered out based on these limits (represented as red circles in \cref{fig:LD2_Optimisation_12_Blahoslav.eps}). The goal was to select a solution that is close to the Utopia point, and which also minimizes the moderator volume in order to reduce heatload. In the end, this combination of parameters was selected for the \ce{LD_2} moderator:

\vspace{0.25cm}

\begin{itemize}[leftmargin=1cm]
  \item[-] Moderator length = 48 cm;
  \item[-] Moderator width = 41 cm;
  \item[-] Be filter thickness = 13 cm;
  \item[-] Reentrant hole depth = 10 cm.
\end{itemize}

\vspace{0.25cm}

The volume of LD$_2$ in the moderator is 32.2 l. For reference, we can also compare the intensity of the lower cold moderator from this neutronic optimization with the model of the upper moderator. The upper moderator has a higher brightness; however, because of the smaller size, its intensity is lower. This can be seen in \cref{fig:LD2_Optimisation_third_iteration_spectra.eps}(a) and (b) which show the time-averaged brightness and intensity respectively for the thermal and cold neutrons measured at the NNBAR and WP7 openings including the time-averaged brightness and intensity calculated for the upper moderator. For the upper moderator, the size of the emission window depends on the angle of extraction: at the NNBAR location, an emission window with a surface of about 42 cm$^2$ would be available\footnote{In the direction of neutron extraction perpendicular to the incoming proton beam, neutrons can be extracted from both the cold wings of the butterfly moderator. For each wing the height is 3 cm, and the width is of about 7 cm, cf. Figure 32 in \cite{zanini_design_2019}, thus giving a total emission surface of about 42 cm$^2$.}, and this give an intensity of the upper moderator above \SI{4}{\angstrom} of \SI{7e14}{n/s/sr} (i.e. 10 times lower than for the NNBAR opening of the lower moderator). 

After completion of the third iteration of optimisation, the model of the cold moderator was delivered to the engineering team, WP5, where a thorough engineering study was conducted. 

A summary of the expected performances, integrated in different wavelength ranges, is shown in \cref{tab:performance_iteration3}. There is a decrease in NNBAR and WP7 FOMs in comparison to the model from the second iteration of neutronic optimisation (see \cref{tab:performance_iteration2}) due to the increase in temperature of LD$_2$ from 20 K to 22 K and the presence of 2.5 vol\% Al in LD$_2$.

\begingroup
    \setlength{\tabcolsep}{10pt} 
    \begin{table}[bt!]
    \centering
%    \begin{threeparttable}
      \caption{Neutronic performance and characteristics of the model from third iteration of cold moderator. Brightness and intensity integrated over different wavelength ranges (NNBAR and WP7 openings) and compared with the upper moderator. The values for the upper moderator are from \cite{zanini_design_2019}. To calculate the intensity we considered the following size of emission windows: NNBAR -- 960 cm$^2$; WP7 -- 225 cm$^2$; upper moderator -- 42 cm$^2$. The moderator volume is the sum of volume of the LD$_2$ box and Be filter.}
    \begin{tabular}{lccc}
    \toprule   & \multicolumn{3}{c}{BRIGTHNESS [\si{n/cm\squared/s/sr}] }\\
    \cmidrule{2-4}
       & $>$ \SI{2}{\angstrom} & $>$ \SI{4}{\angstrom} & $>$ \SI{10}{\angstrom} \\
    \midrule
    NNBAR & \num{7.89e+12} & \num{6.07e+12} & \num{4.46e+11}  \\
    WP7 & \num{1.60e+13} & \num{8.76e+12} & \num{6.89e+11} \\
    upper moderator & \num{5.3e13} & \num{1.7e13} & \num{9.9e11}  \\
    \midrule
             & \multicolumn{3}{c}{INTENSITY [\si{n/s/sr}] }\\
    \cmidrule{2-4}
       & $>$ \SI{2}{\angstrom} & $>$ \SI{4}{\angstrom} & $>$ \SI{10}{\angstrom} \\
    \midrule
    NNBAR & \num{7.57e+15} & \num{5.83e+15} & \num{4.28e+14}  \\
    WP7 & \num{3.60e+15} & \num{1.97e+15} & \num{1.55e+14} \\
    upper moderator & \num{2.2e15} & \num{7.0e14} & \num{4.2e13}  \\
    \midrule
    NNBAR FOM & \num{2.51e+17} [\si{n/s/sr} $\times$ $\lambda^2$] &  & \\
    WP7 FOM & \num{3.22e+15} [\si{n/s/sr}] & & \\
    Heatload on LD$_2$ moderator & \num{28.5} [kW] & & \\
    Heatload on Be filter & \num{19.5} [kW] & & \\
    Heatload on Al vessel & \num{7.3} [kW] & & \\
    Total heatload & \num{55.3} [kW] & & \\
    Moderator volume  & \num{45.0} [liters] & & \\
    \bottomrule
    \end{tabular}
  \label{tab:performance_iteration3}
%  \end{threeparttable}
\end{table}
\endgroup

\subsubsection{Effect of Be filter and reentrant hole}
\label{Section_Effect_BeFilter_REH}
\begin{figure}[tb!]      
    \begin{subfigure}[b]{0.48\textwidth}
        \centering
        \includegraphics[width=\textwidth]{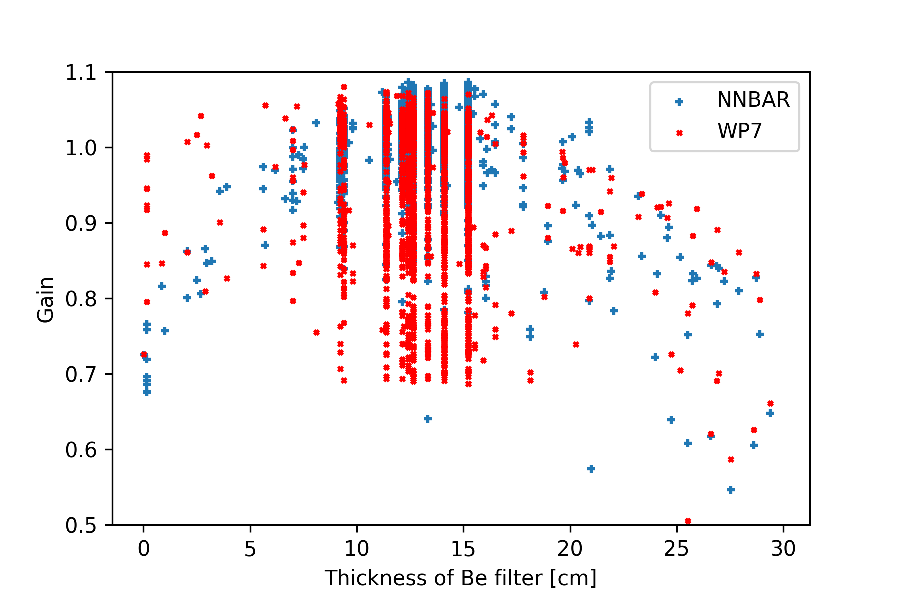}
        \subcaption{}
        \label{fig:LD2_Optimisation_4_Blahoslav.eps}
    \end{subfigure}
    \hfill
    \begin{subfigure}[b]{0.48\textwidth}
        \centering        
        \includegraphics[width=\textwidth]{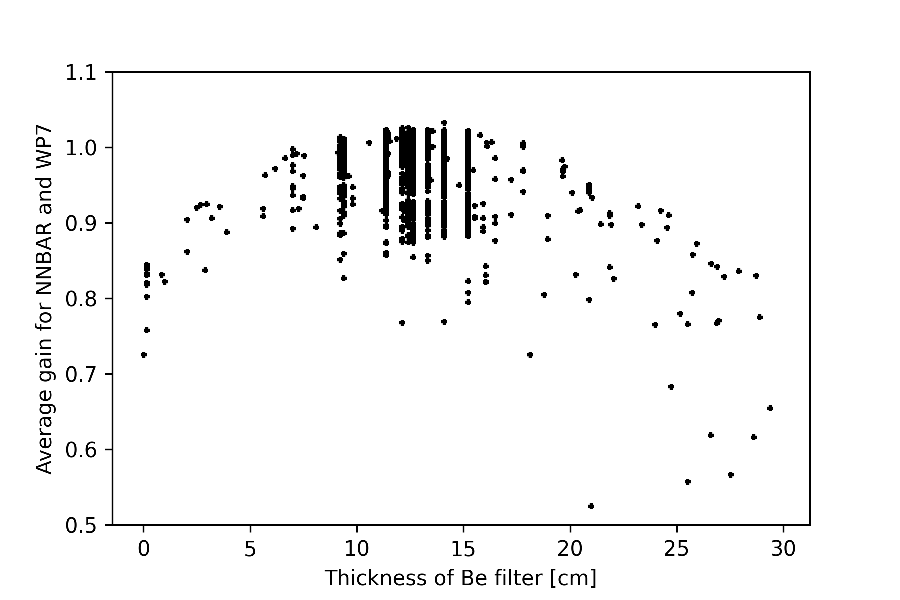}
        \subcaption{}
        \label{fig:LD2_Optimisation_401_Blahoslav.eps}
    \end{subfigure}
    \caption{Performance of models with respect to Be filter thickness. (a) Gain in NNBAR and WP7 FOM with respect to the baseline model (b) Average gain in NNBAR and WP7 FOM with respect to the baseline model.}
        \label{fig:LD2_Optimisation_4}
    \end{figure}

\begin{figure}[tb!]      
    \begin{subfigure}[b]{0.48\textwidth}
        \centering
        \includegraphics[width=\textwidth]{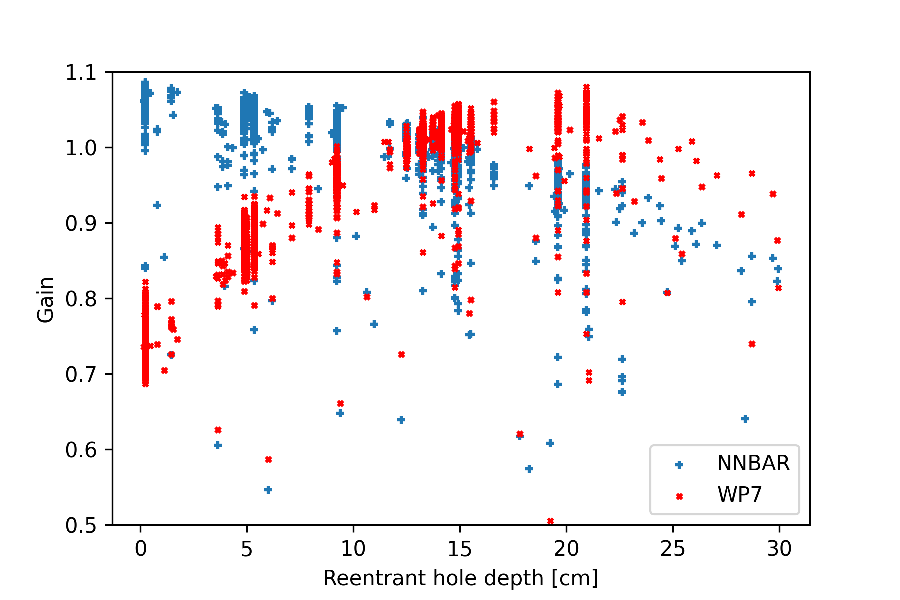}
        \subcaption{}
        \label{fig:LD2_Optimisation_5_Blahoslav.eps}
    \end{subfigure}
    \hfill
    \begin{subfigure}[b]{0.48\textwidth}
        \centering        
        \includegraphics[width=\textwidth]{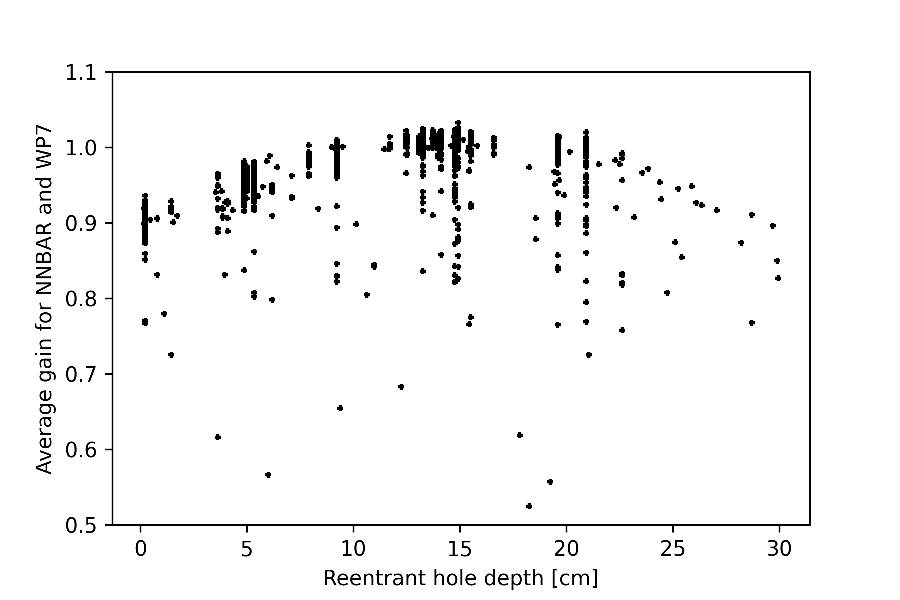}
        \subcaption{}
        \label{fig:LD2_Optimisation_501_Blahoslav.eps}
    \end{subfigure}
    \caption{Performance of models with respect to the reentrant-hole depth. (a) Gain in NNBAR and WP7 FOM with respect to the baseline model (b) Average gain in NNBAR and WP7 FOM with respect to the baseline model.}
        \label{fig:LD2_Optimisation_5}
    \end{figure}

\begin{figure}[tb!]
  \centering
  \begin{tabular}{@{}c@{}}
    \includegraphics[width=.9\linewidth]{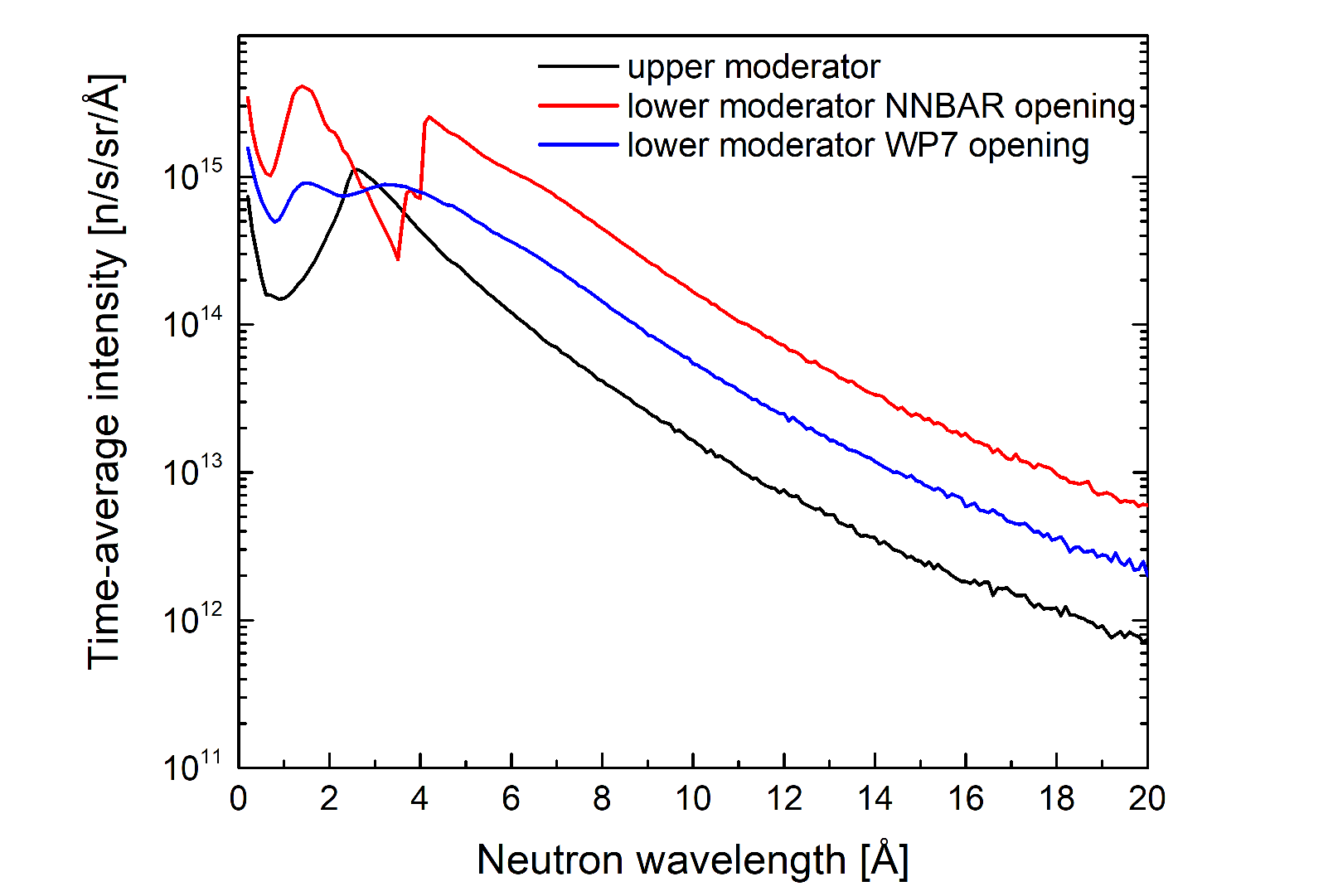} \\[\abovecaptionskip]
    (a)
  \end{tabular}

  \vspace{\floatsep}

  \begin{tabular}{@{}c@{}}
    \includegraphics[width=.9\linewidth]{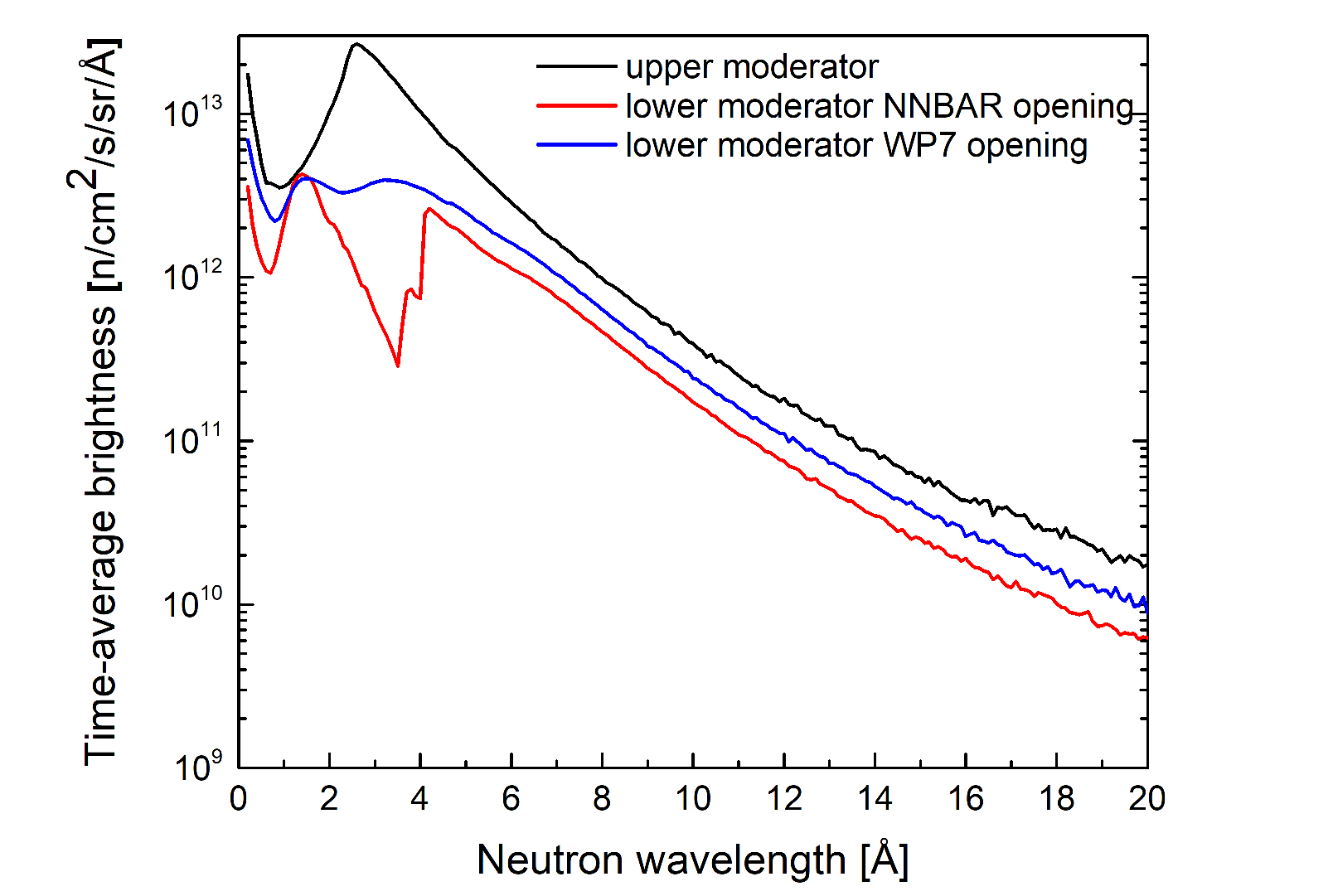} \\[\abovecaptionskip]
    (b)
  \end{tabular}

  \caption{Comparison of neutron spectra for the lower and upper moderators for the third iteration of optimisation. (a) Time-averaged intensity. (b) Time-averaged brightness.}\label{fig:LD2_Optimisation_third_iteration_spectra.eps}
\end{figure}

A quantification of the effect of the Be filter and reentrant hole on the neutronic performance was done in the third iteration of optimisation. \cref{fig:LD2_Optimisation_4_Blahoslav.eps} clearly shows that the NNBAR side benefits from the presence of the Be filter. The gain in the NNBAR FOM observed when comparing the case with no Be filter (Be filter thickness = 0\,cm) and the cases with Be filters having thicknesses between about 10 and 15\,cm is about 30\%. Similarly, the WP7 side benefits from the presence of a reentrant hole (see \cref{fig:LD2_Optimisation_5_Blahoslav.eps}). In this case, the WP7 FOM increases by about 20\% for reentrant holes with depths between about 10 and 15\,cm. 

\subsubsection{Shape of the reentrant hole}

\begin{figure}[tb!]      
    \begin{subfigure}[b]{0.48\textwidth}
        \centering
        \includegraphics[width=\textwidth]{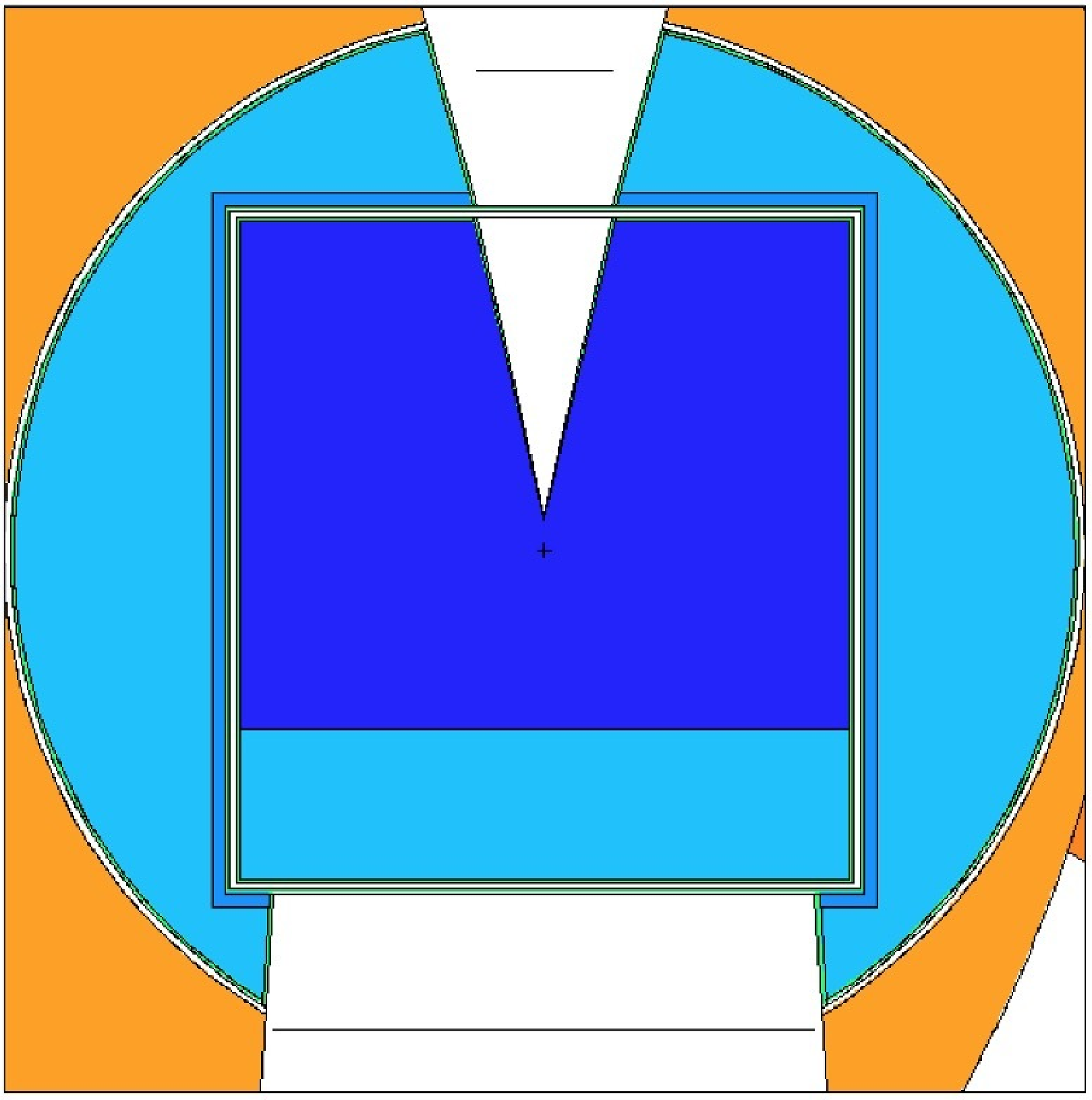}
        \subcaption{}
        \label{fig:cn_13.jpg}
    \end{subfigure}
    \hfill
    \begin{subfigure}[b]{0.48\textwidth}
        \centering        
        \includegraphics[width=\textwidth]{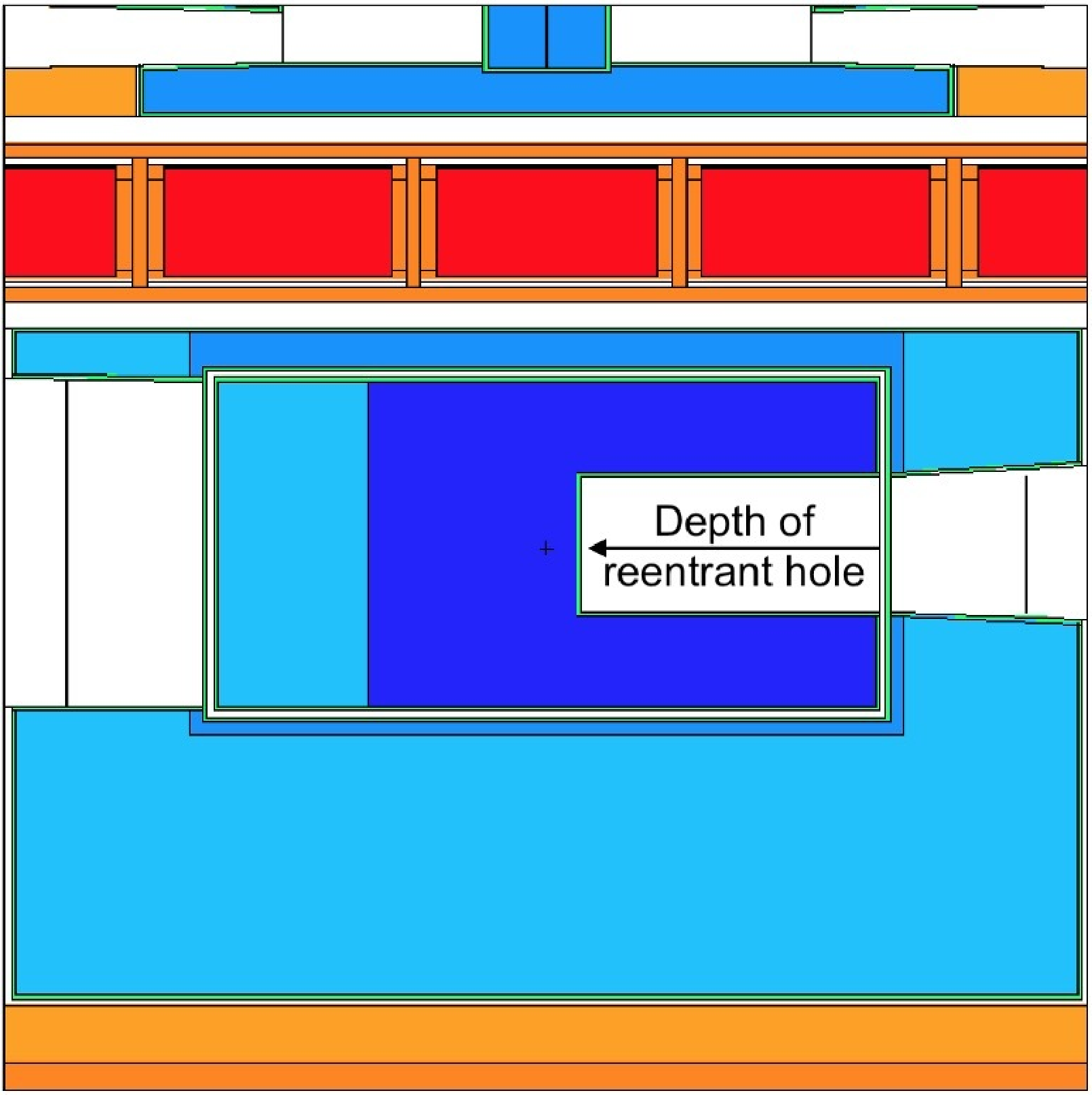}
        \subcaption{}
        \label{fig:LD2_Optimisation_1301_Blahoslav.eps}
    \end{subfigure}
    \caption{Model of the wedge-shaped reentrant hole. (a) Horizontal view. (b) Vertical view.}
        \label{fig:LD2_Optimisation_13_Blahoslav.eps}
    \end{figure}

\begin{figure}[hbt!]
\begin{center}
\includegraphics[width=0.75\textwidth] {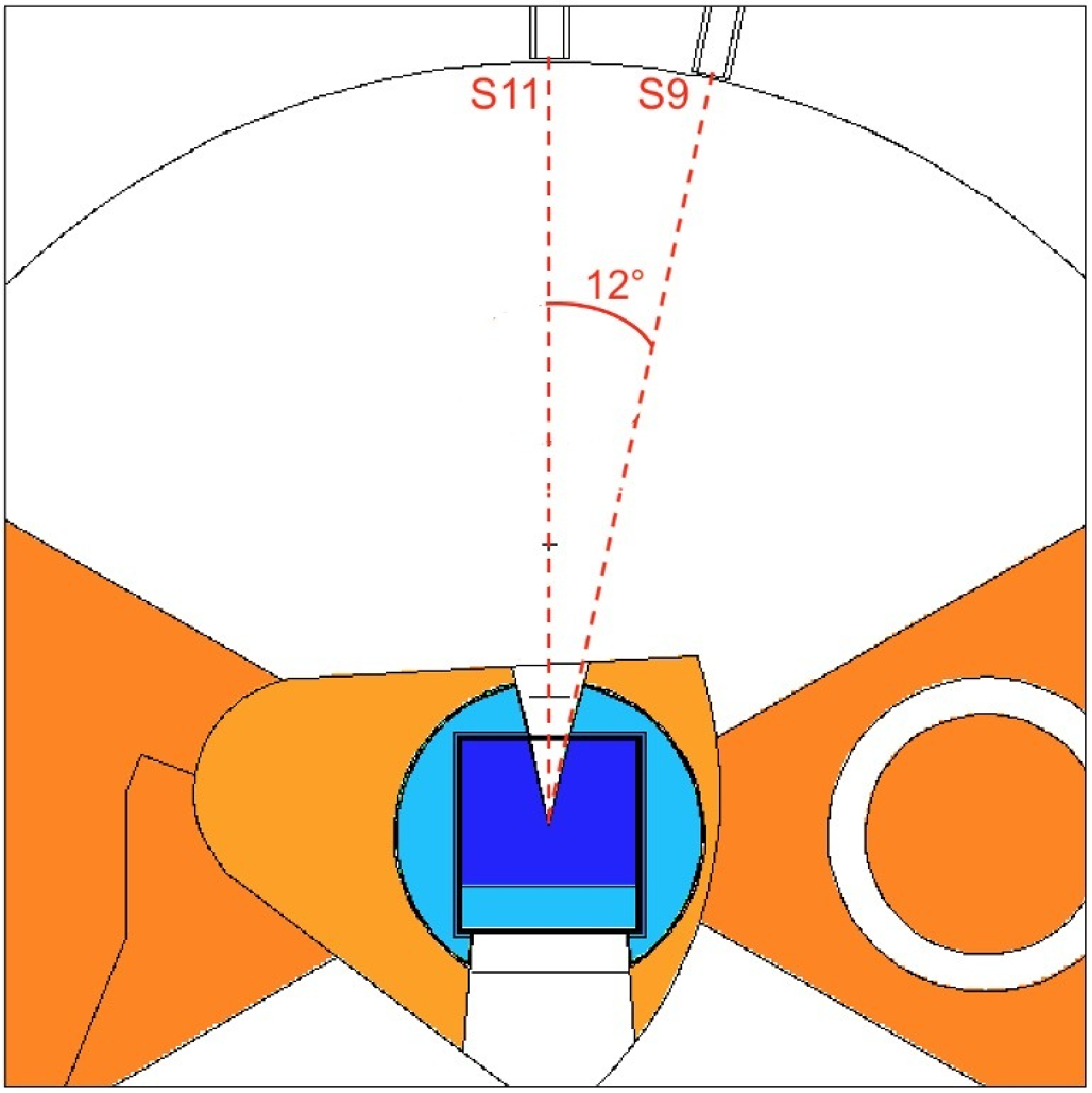}
\caption{Horizontal view of the wedge-shaped reentrant hole, the on-axis S11 beamport, and the off-axis S9 beamport.}
\label{fig:LD2_Optimisation_1302_Blahoslav.eps}
\end{center}
\end{figure}

\begin{figure}[hbt!]
\begin{center}
\includegraphics[width=0.75\textwidth]{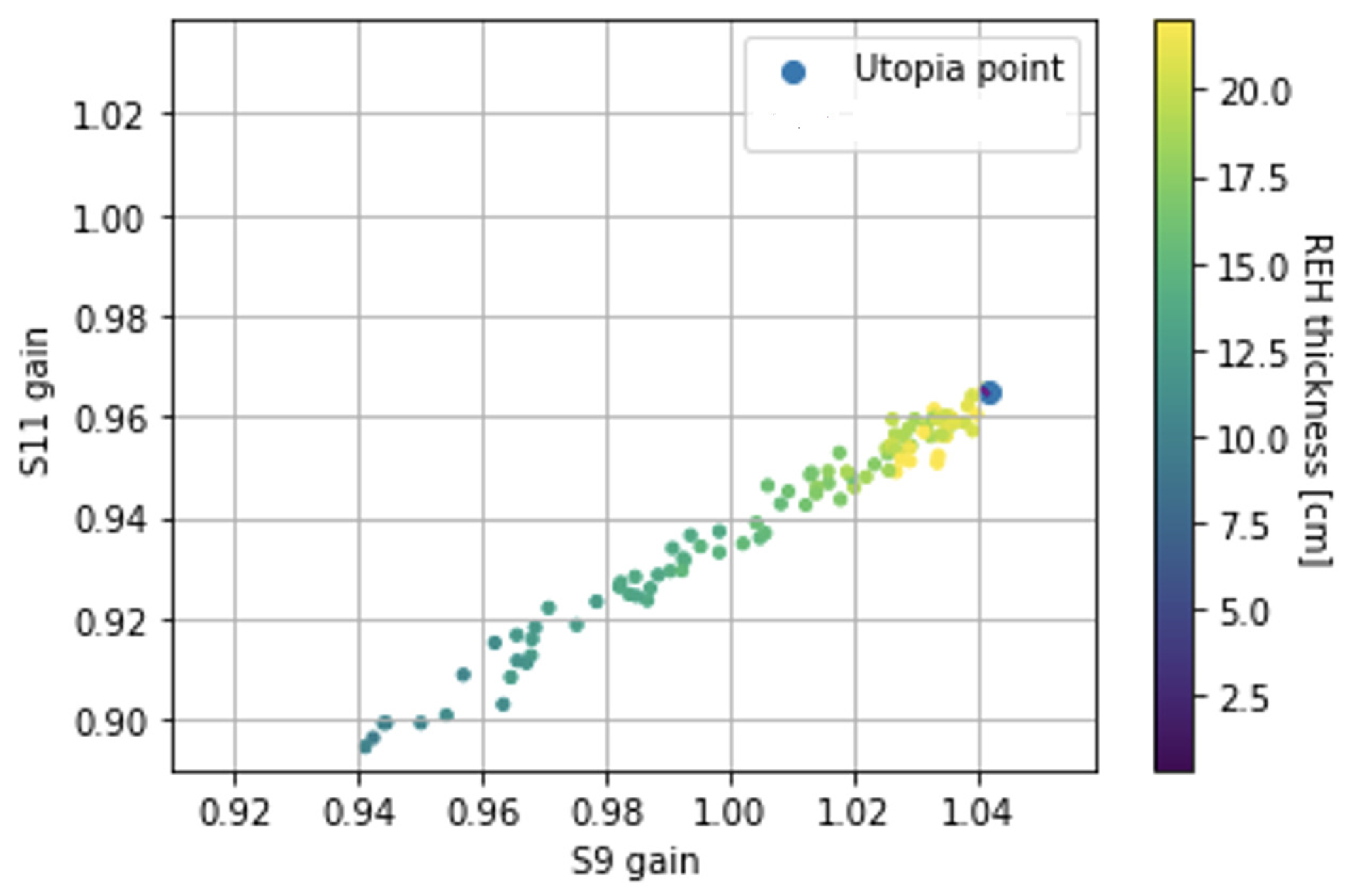}
\caption{Performance of a wedge-shaped reentrant hole at different depths. }
\label{fig:LD2_Optimisation_16_Blahoslav.eps}
\end{center}
\end{figure}

The development of the cold moderator continued with a study of the impact of different shapes of reentrant holes. More specifically, the initially rectangular shaped reentrant hole was replaced by, for example, a wedge-shaped reentrant hole with variable depth (see \cref{fig:LD2_Optimisation_13_Blahoslav.eps}). Subsequently, the neutronic performance on the WP7 side was estimated with this shape of reentrant hole for the beamport S11 ("S" stands for South Sector, and "11" refers to the number of the beamport; it is the beamport perpendicular to both the incoming proton beam and the emission surface of the moderator at the WP7 side) and the beamport S9 with an offset of 12$^{\circ}$ with respect to S11 (see \cref{fig:LD2_Optimisation_1302_Blahoslav.eps} 
for a depiction of the beamport positions). The calculated values of the WP7 FOM at various reentrant hole depths are shown in \cref{fig:LD2_Optimisation_16_Blahoslav.eps}. The maximal gain with respect to the rectangular reentrant hole at the S9 beamport was of about 5\%, but with a penalty to the WP7 FOM for the S11 beamport of about 3\%. Other shapes of reentrant holes were investigated, but with a consistent result that the gain for the off-axis beamport S9 was only marginal with a small loss for the on-axis S11 beamport at the same time. Therefore, the shape of the reentrant hole was kept rectangular in this round of optimization.

% \begin{figure}[tb!]      
%    \begin{subfigure}[b]{0.7\textwidth}
%        \centering
%        \includegraphics[width=\textwidth]{cn_fig/LD2Moderator_Sep2022_time-averaged_intensity_Blahoslav.eps}
%        \subcaption{}
%        \label{fig:LD2Moderator_Sep2022_time-averaged_intensity_Blahoslav.eps}
%    \end{subfigure}
%    \hfill
%    \begin{subfigure}[b]{0.7\textwidth}
%        \centering        
%        \includegraphics[width=\textwidth]{cn_fig/LD2Moderator_Sep2022_Time-averaged_brightness_Blahoslav.eps}
%        \subcaption{}
%        \label{fig:LD2Moderator_Sep2022_Time-averaged_brightness_Blahoslav.eps}
%    \end{subfigure}
%    \caption{Comparison of neutron spectra for the lower and upper moderators for the third iteration of optimisation. (a) Time-averaged intensity. (b) Time-averaged brightness.}
%        \label{fig:LD2_Optimisation_second_iteration_spectra.eps}
% \end{figure}

\FloatBarrier
\subsubsection{Moderator characterization}

%\textcolor{red}{we have to say that this is for the second optimization (nicola) and that the difference with the third one (blahoslav) is minimal)}

The model of the cold moderator from the second neutronic iteration was used to perform a comprehensive moderator characterization. 
Given the similarity of the designs from the different iterations, 
the results presented in this section also apply to the design from 
 from the third iteration.  

\begin{figure}[bt!]      
    \centering
    \includegraphics[width=0.94\textwidth]{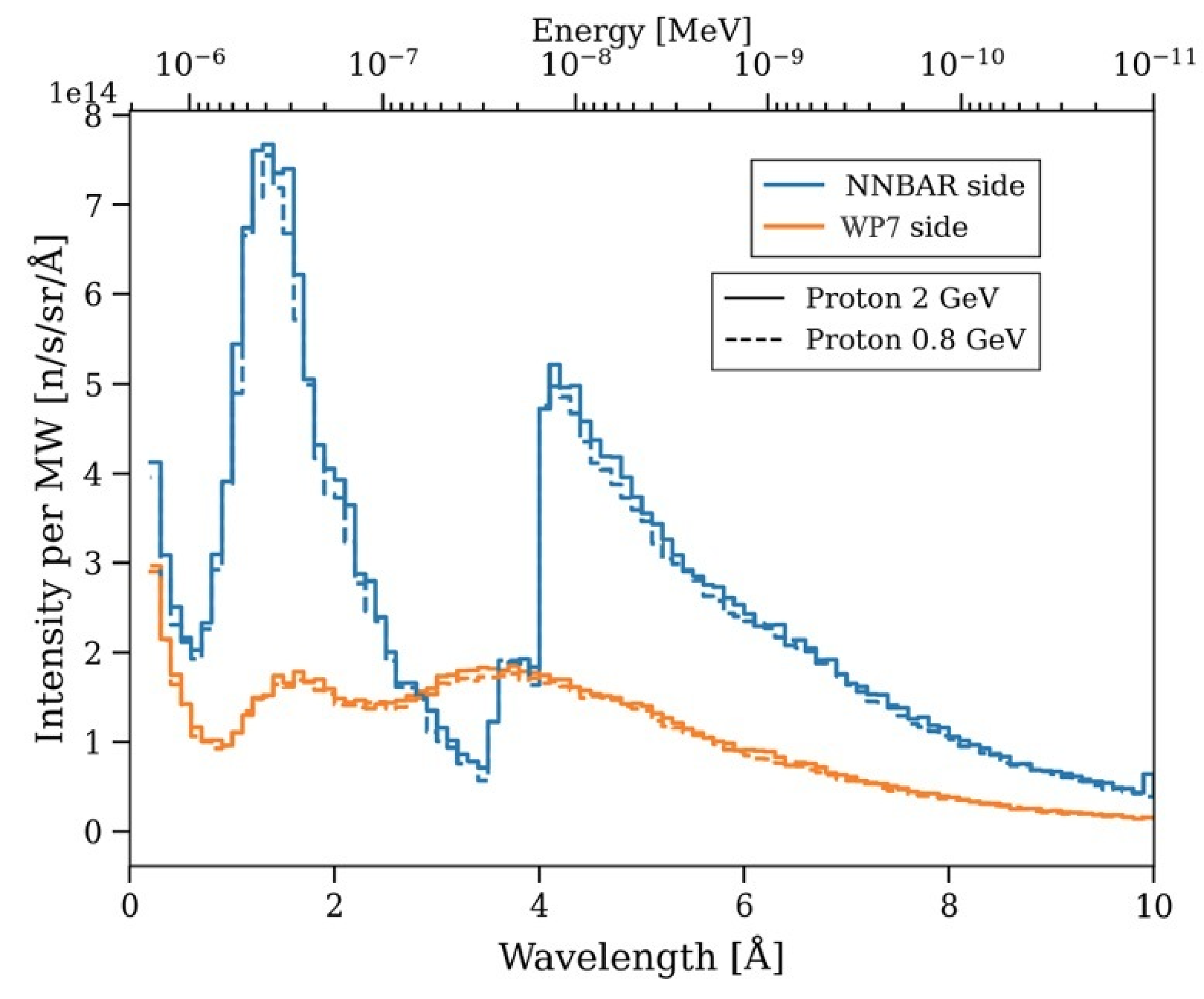}
	\caption{Spectra comparison between the NNBAR and the neutron scattering instrument (WP7) sides of the \ce{LD_2} using MOGA optimization for protons impinging on the target at \SI{2}{GeV} and \SI{800}{MeV}, per unit power.}
	\label{fig:spectra_MOGA}
\end{figure}

Intensity spectra from the two moderator openings are shown in 
\cref{fig:spectra_MOGA}. Spectra are shown for two proton energies of 2 GeV and 800 MeV, which for an average current of 2.5 mA correspond to operation at average power of 5 MW and 2 MW, respectively. For both openings, intensity spectra per unit power are almost equivalent, indicating that the results presented in most of this work, which are for 5 MW average power, can be directly scaled to the 2 MW average power (the initial operating power of ESS) dividing by a factor of 2.5.

%The energy distribution binned in wavelength for the MOGA optimization is shown in \cref{fig:spectra_MOGA} (this was the best-performing optimization approach for a variety of algorithms, for a comparison of these methods see \cite{D4.2}.) The baseline case has protons impinging on the target at \SI{2}{GeV}, and the additional case shows protons at \SI{800}{MeV}. Both cases are normalized per unit of the accelerator power in \si{MW}. The fact that spectra match for the both openings suggests that lowering the power of the accelerator will not affect the energy distribution besides the obvious drop in absolute intensity.

The main feature of the spectral brightness distribution on the NNBAR side is the very sharp cut-off at \SI{4}{\angstrom} due to the Be filter. The enhanced peak between
\SI{1}{\angstrom} and \SI{2}{\angstrom}
%\SIrange{1}{2}{\angstrom} 
is a feature introduced by the filter. A smoother spectral distribution is observed on the neutron scattering instruments (WP7) opening with the thermal and cold peaks clearly visible. 

\begin{figure}[bpht!]      
    \begin{subfigure}[c]{\textwidth}
        \centering
        \includegraphics[height=0.31\textheight,keepaspectratio]{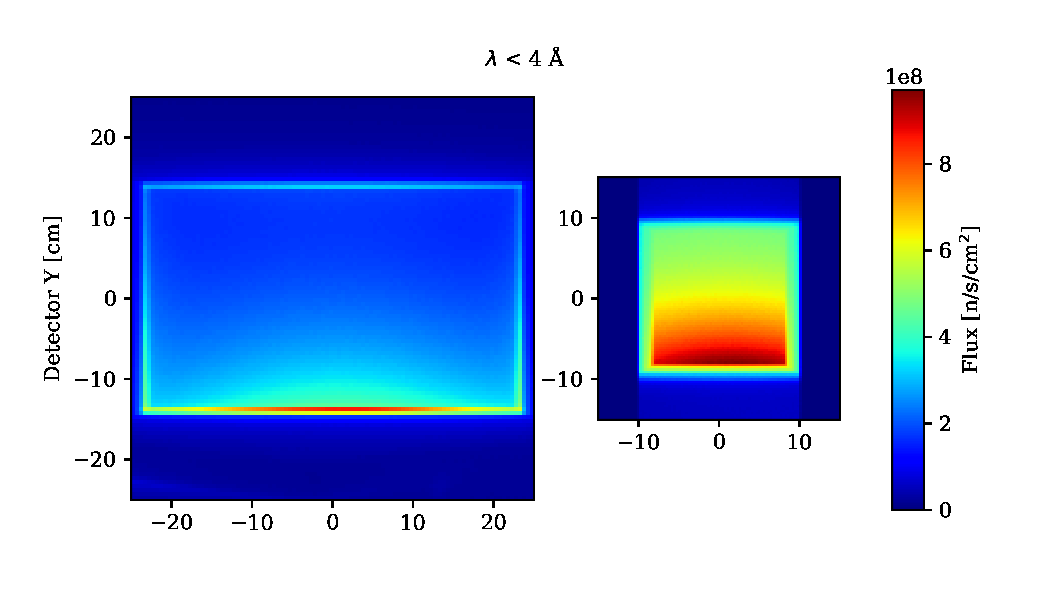}
    \end{subfigure}
    \begin{subfigure}[c]{\textwidth}
        \centering
        \includegraphics[height=0.31\textheight,keepaspectratio]{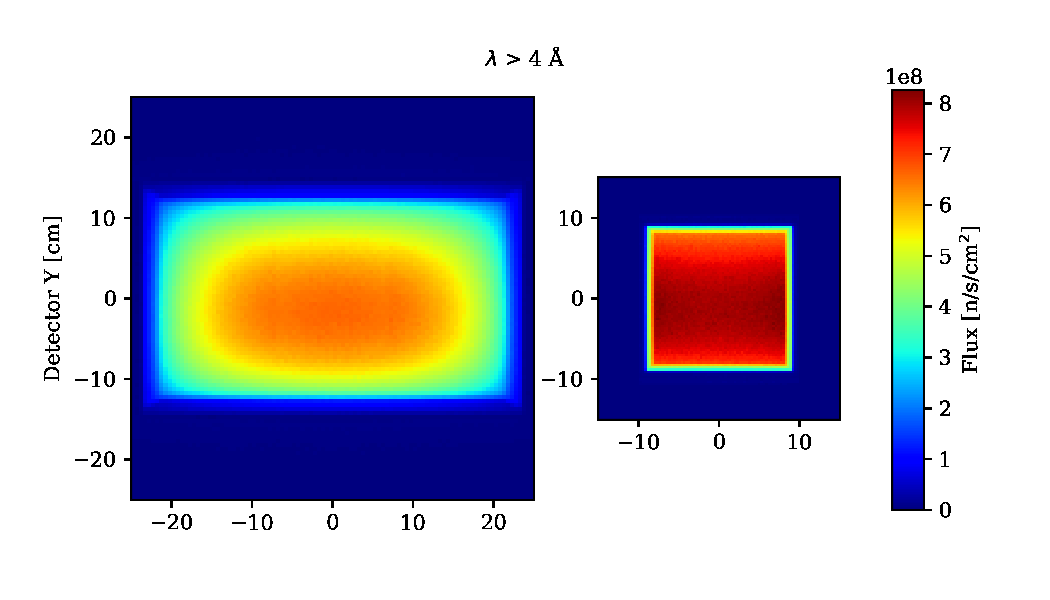}
    \end{subfigure}
    \begin{subfigure}[c]{\textwidth}
        \centering
        \includegraphics[height=0.31\textheight,keepaspectratio]{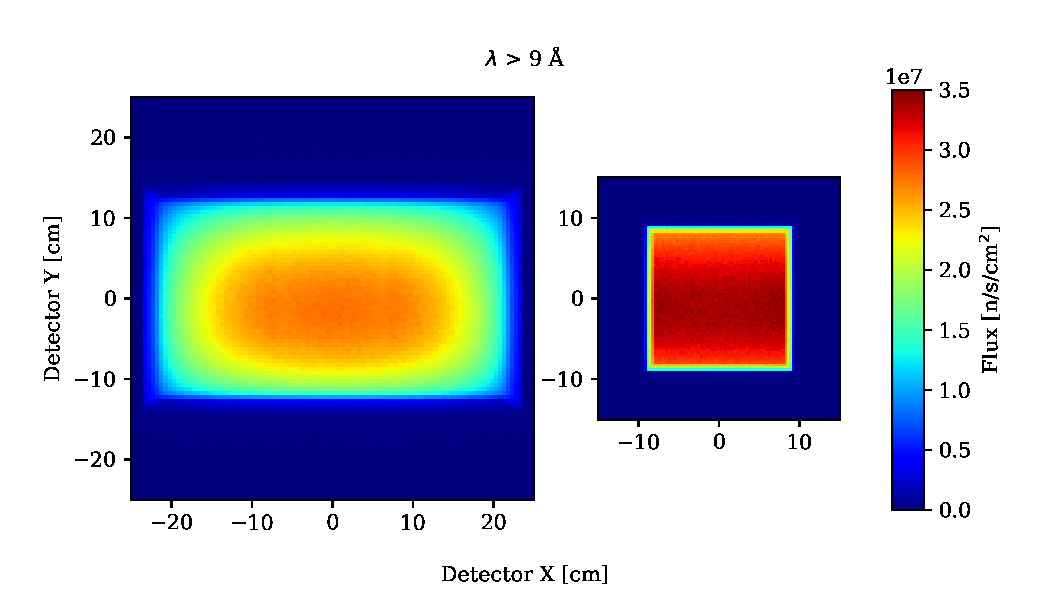}
    \end{subfigure}
	\caption{Pinhole images of the \qtyproduct{24x40}{cm} NNBAR opening (left) and the \qtyproduct{15x15}{cm} WP7 opening (right) with filters for neutrons at different energies. The pinholes are placed \SI{2}{m} away from the center of the moderator along the central axis, while the detector arrays are \SI{2}{m} away from the pinholes on the same axis and with the dimensions \qtyproduct{50x50}{cm} and \qtyproduct{30x30}{cm} for NNBAR and WP7, respectively.  The pictures have the same scale and appear inverted along the vertical direction due to the camera obscura effect.}
 \label{fig:Pinhole}
\end{figure}

The  spatial distribution of neutrons coming out of the surface of the moderator was studied with a pinhole-camera tally integrated in MCNP. The pinholes are placed \SI{2}{m} away from the center of the moderator along the central axis for both openings, while the detector arrays are \SI{2}{m} away from the pinholes on the same axis and have dimensions of \qtyproduct{50x50}{cm} and \qtyproduct{30x30}{cm} for the NNBAR and WP7 sides, respectively. The results are shown in \cref{fig:Pinhole} (NNBAR on the left and WP7 on the right) for three wavelength ranges: $\lambda < \SI{4}{\angstrom}$, $\lambda > \SI{4}{\angstrom}$, and $\lambda > \SI{9}{\angstrom}$.

The distribution of the neutrons is quite different between the two openings. The bigger NNBAR opening has a non-uniform profile along both axes, with a central hot-spot slightly off-centered toward the top of the moderator (the pinhole images appear inverted in the vertical direction). The smaller opening for the neutron scattering instruments is a more uniform source. Although the integrated intensity is higher on the NNBAR side, the effective contribution of the neutrons from the corners is quite limited due to current optics design~\cite{WAGNER2023168235}. In any case, the advantage of having such a large opening in the moderator has been established, but further improvement can be made to increase the intensity farther from the center.

\begin{figure}[tbhp!]
	\centering
	\includegraphics[width=0.7\textwidth]{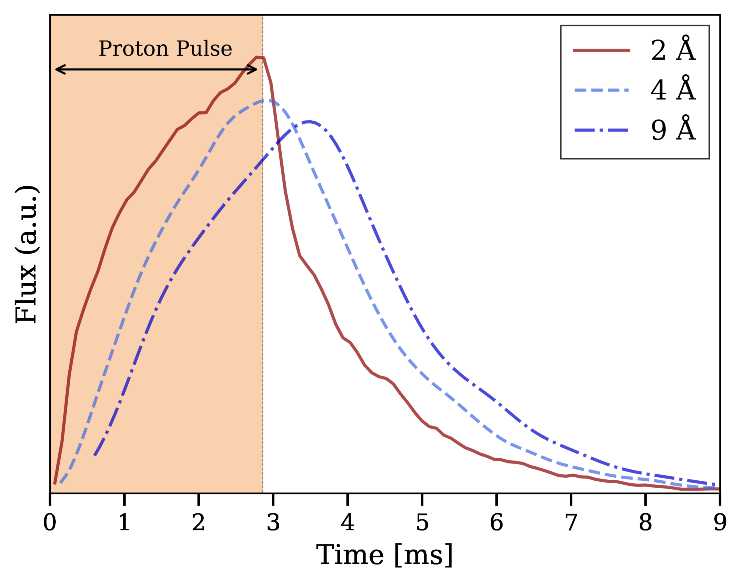}
	\caption{Pulse spectra of neutrons from the moderator at 2, 4, and \SI{9}{\angstrom} wavelength. The proton beam impinging on the target lasts \SI{2.86}{ms}, indicated by the shaded region on the left. 
 %The neutrons are binned in elapsed time in milliseconds and the respective 
 Histograms are normalized to probability density functions (a.u.) for comparison. The smoothed curves are obtained with Gaussian kernel density estimation.}
	\label{fig:pulse-shape}
\end{figure}

Finally, in \cref{fig:pulse-shape} the pulse shape of the neutrons coming out of the moderator at \SI{2}{\angstrom}, \SI{4}{\angstrom}, and \SI{9}{\angstrom} is presented. The curves were obtained by recording the neutrons leaving the moderator by crossing the surface on the NNBAR side through a SSW card \cite{ref_MCNP6} . The neutrons were then filtered in energy using the Monte Carlo Particle Lists (MCPL) interface \cite{MCPL} and binned in elapsed time in \si{ms}. The histograms were normalized in order to be comparable and smoothed with Gaussian kernel density estimation. The pulse width is significantly broader than the proton pulse (\SI{2.86}{ms}). This is due to the slow \ce{LD_2} thermalization time and will likely impact neutron scattering experiments. In particular, the use of choppers in conventional neutron scattering experiments will certainly lead to a reduction in peak fluxes. On the other hand, for NNBAR there is no foreseen impact since the quantity of interest is the time-averaged flux.

\subsubsection{Effects of engineering design on neutronic performance}
\label{sec:engeffect}

%\textcolor{red}{add more text (Luca)}

An engineering design was made based on the optimized neutronic model (see \cref{ch:1}). In this section, the impact of  various engineering solutions on the neutronic performance is analyzed. 
%This step of the process is necessary because for the selection of the best available options in terms of neutronics, while satisfying the engineering constraints.
Two features of the engineering design were found to particularly affect the neutronic performance: first, the large thickness (0.8 cm) of the Al vessel required for the engineering design of a vessel with flat walls; second, the relatively complex design of the Al flow channels, due to the presence of a reentrant hole on the neutron scattering side.

Additionally, on recommendation from the engineering team, the beryllium was placed outside the LD$_2$ vessel. This was done to solve two issues: 1) there is a significant reduction of the heat load which has to be handled by the LD$_2$ flow inside the moderator when the beryllium is not inside; 2) there was an issue with film boiling between beryllium blocks in the thermomechanical simulations when the blocks are inside the LD$_2$.

To simplify the flow structure and to reduce the amount of Al inside the vessel, a change in the shape of the reentrant hole was implemented: the original square-shaped REH was replaced by a U-shaped one of the same width (15 cm) but with the height of the moderator vessel (24 cm).
 
\begin{figure}[hbt!]
\begin{center}
\includegraphics[width=0.99\textwidth]{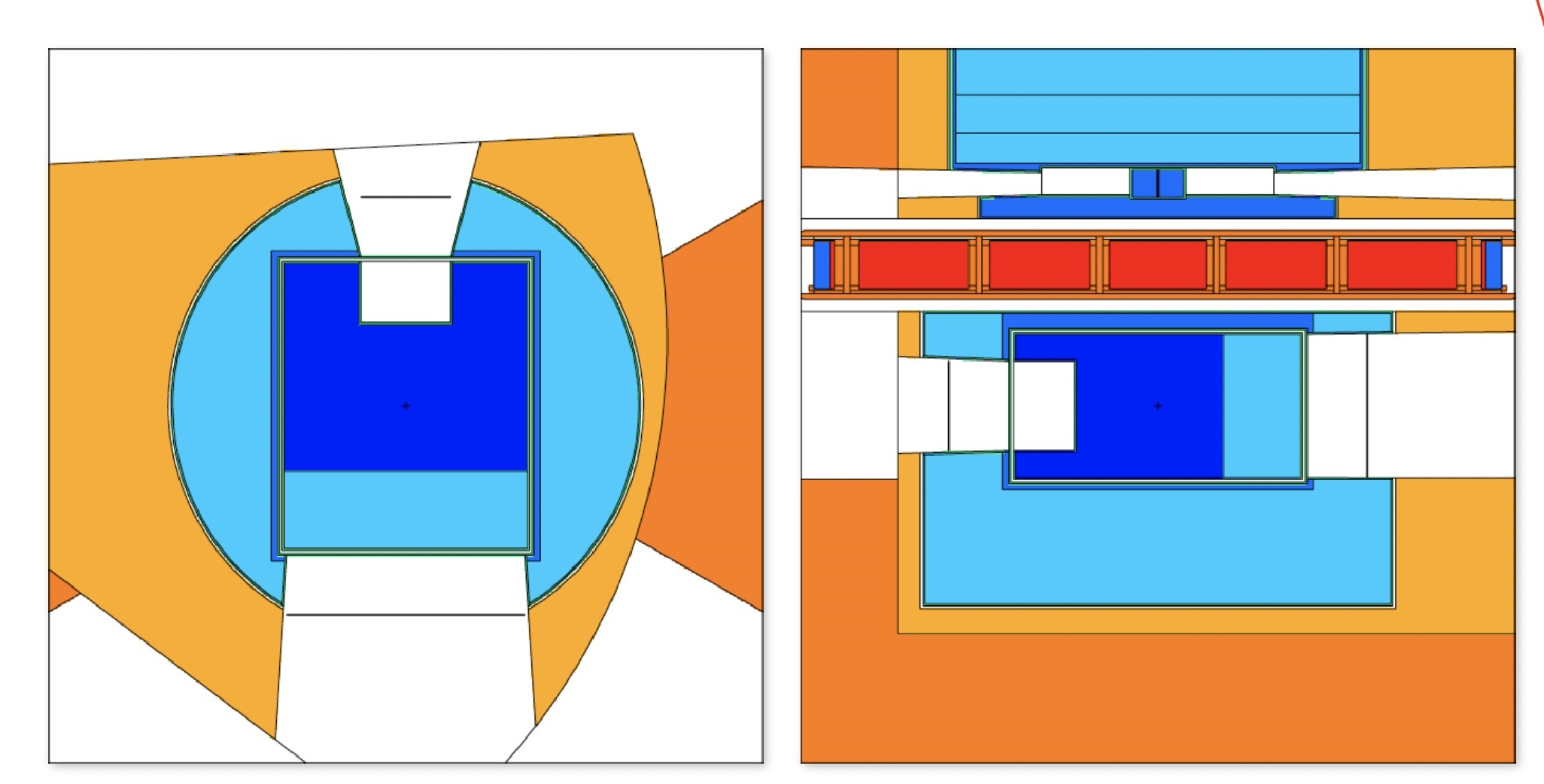}
\caption{Original optimized design for the \ce{LD_2} source.}
\label{fig:alan1}
\end{center}
\end{figure}

\begin{figure}[hbt!]
\begin{center}
\includegraphics[width=0.99\textwidth]{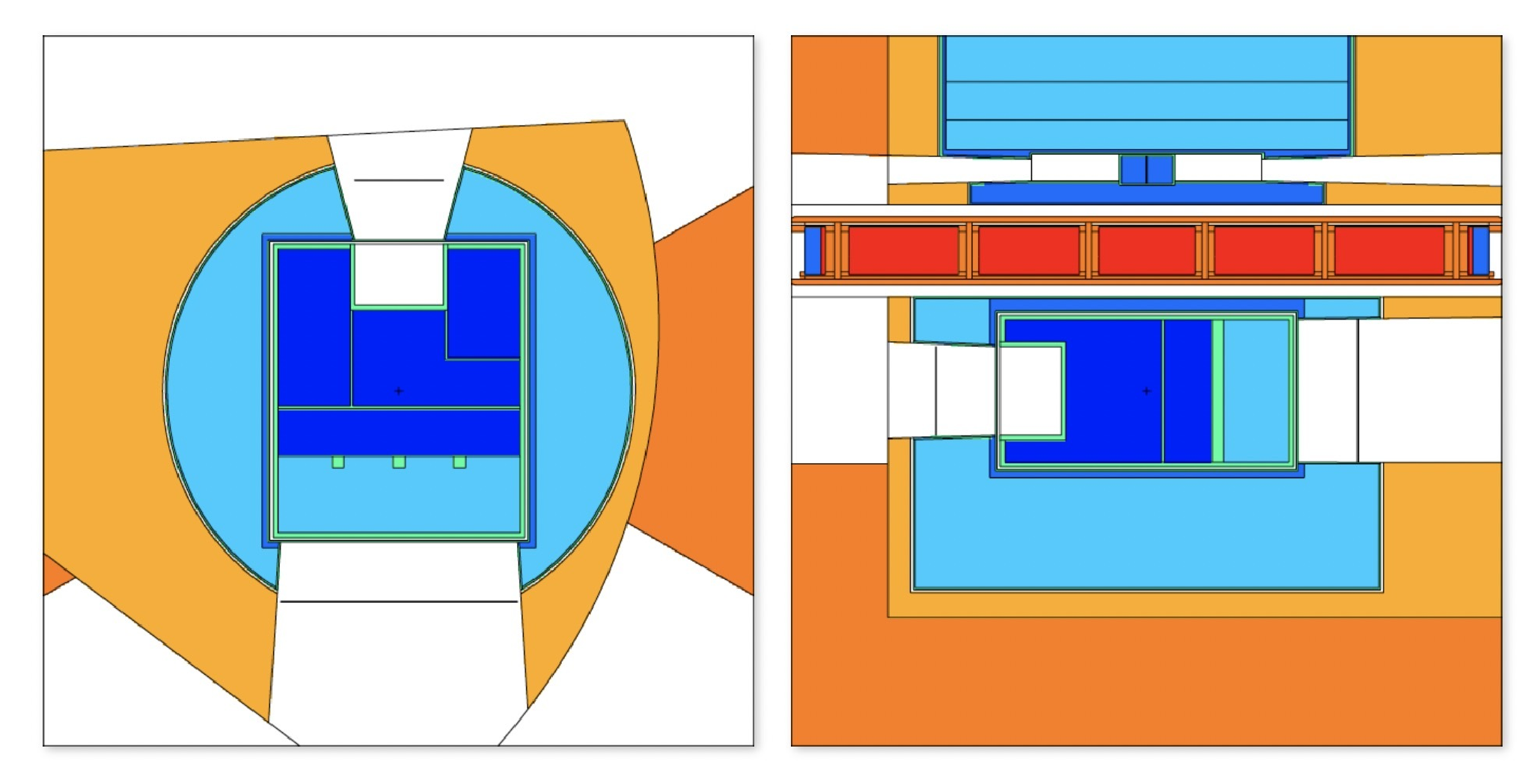}
\caption{As in \cref{fig:alan1}, but  Al flow guides modeled explicitly, while the wall thickness increased from 0.3\,cm to 0.8\,cm.}
\label{fig:alan3}
\end{center}
\end{figure}

\begin{figure}[hbt!]
\begin{center}
\includegraphics[width=0.99\textwidth]{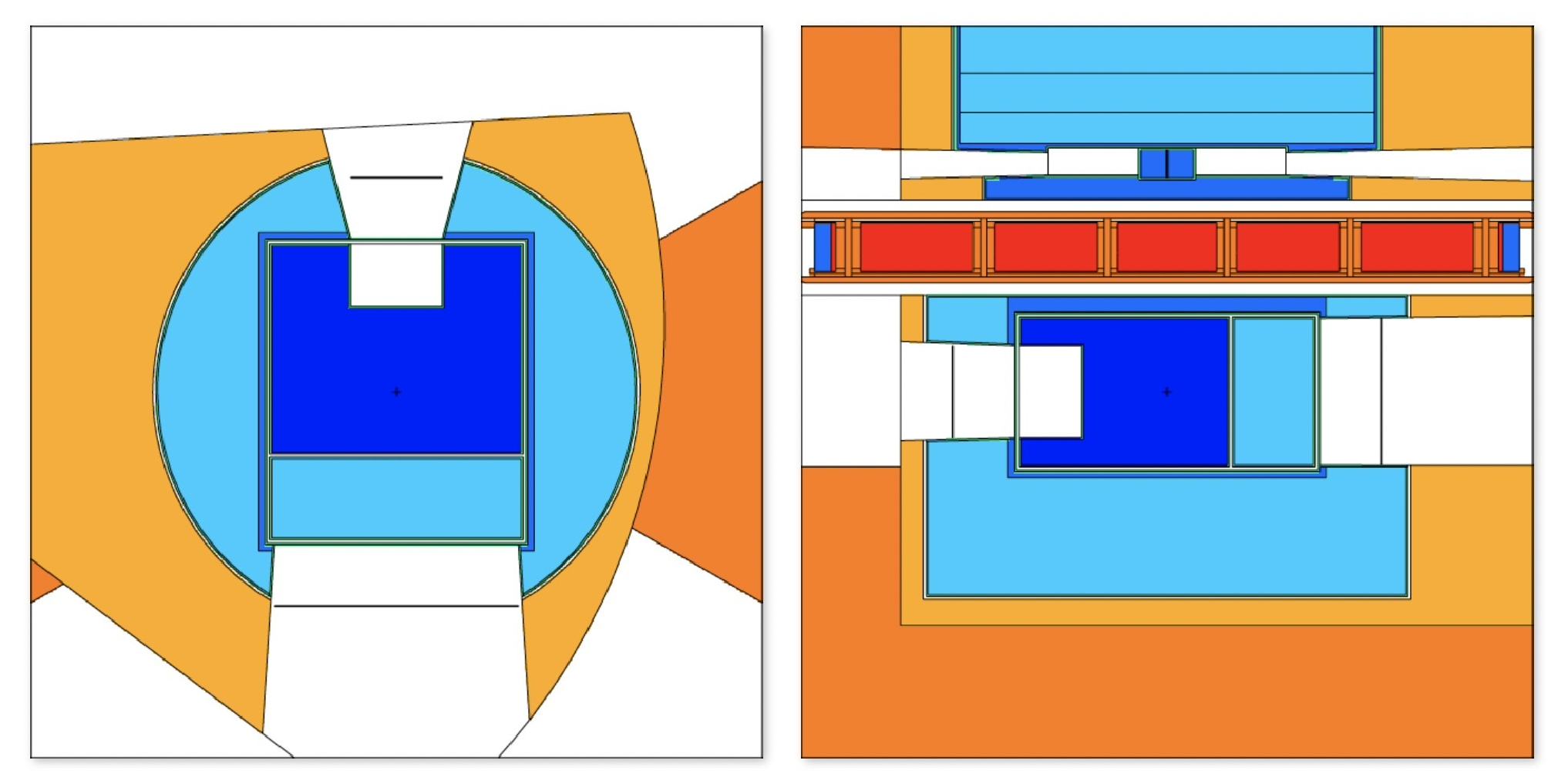}
\caption{As in \cref{fig:alan1}, but Be filter separated from \ce{LD_2} moderator.}
\label{fig:alan4}
\end{center}
\end{figure}

\begin{figure}[hbt!]
\begin{center}
\includegraphics[width=0.99\textwidth]{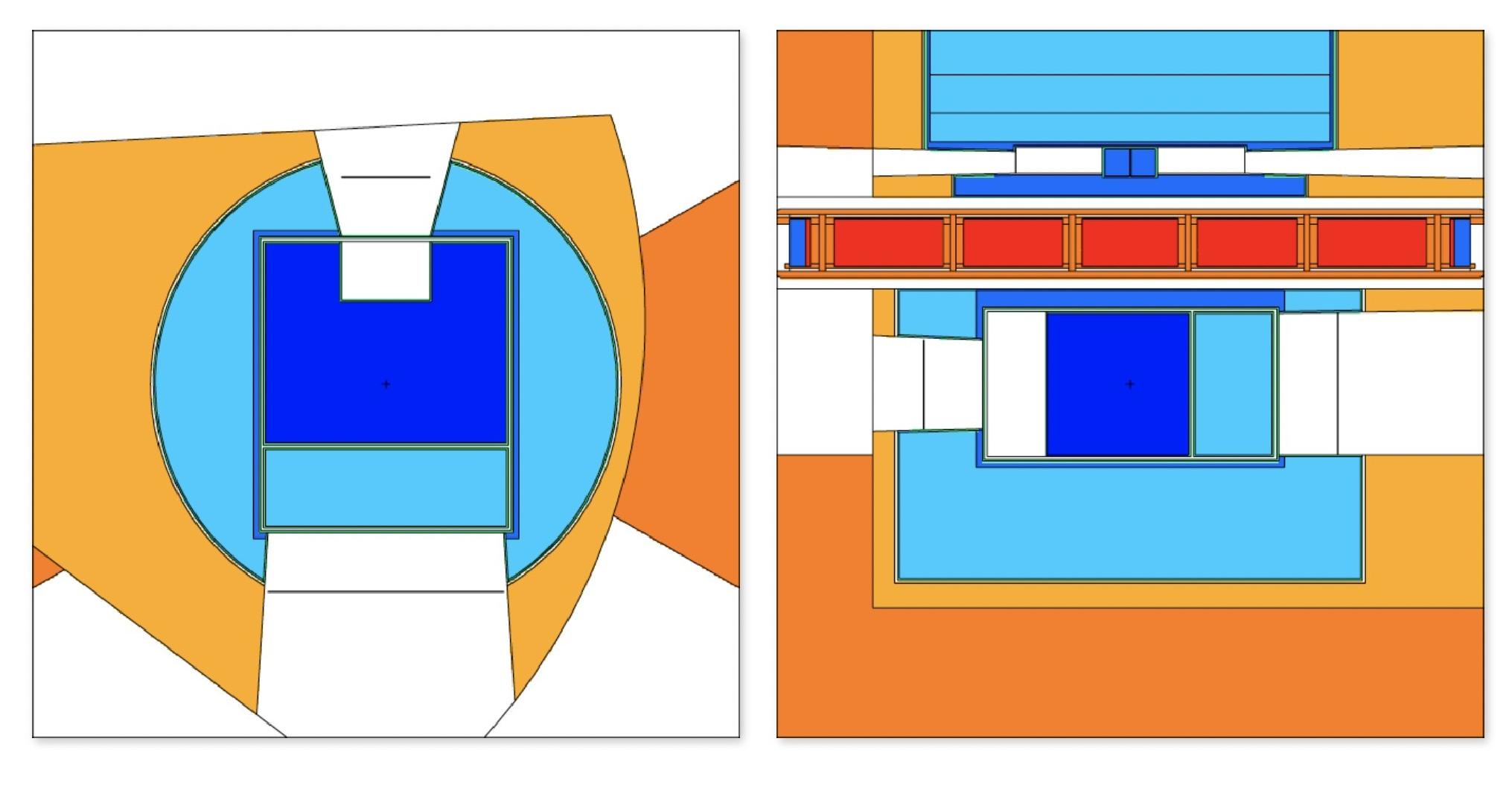}
\caption{As in \cref{fig:alan4}, but with an alternative reentrant hole shape and a few parameters slightly changed to streamline the model.}
\label{fig:alan6}
\end{center}
\end{figure}

\begin{figure}[hbt!]
\begin{center}
\includegraphics[width=0.99\textwidth,trim={1.5cm 1.5cm 1.cm 1.5cm},clip]{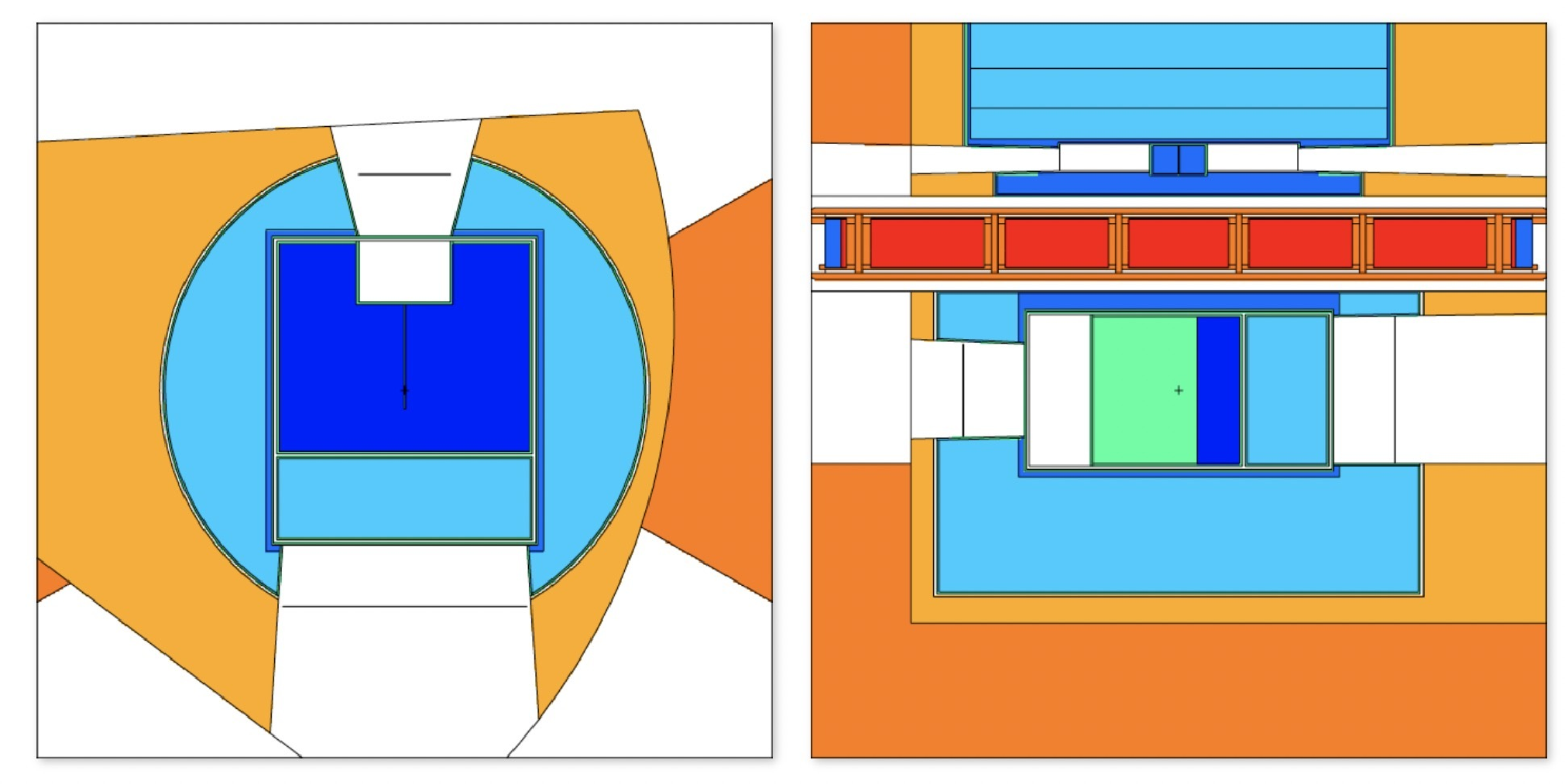}
\caption{As in \cref{fig:alan6}, but  Al flow guides modeled explicitly, while the wall thickness increased from 0.3\,cm to 0.4\,cm.}
\label{fig:alan8}
\end{center}
\end{figure}

\begin{figure}[hbt!]
\begin{center}
\includegraphics[width=0.99\textwidth,trim={1.5cm 1.5cm 1.cm 1.5cm},clip]{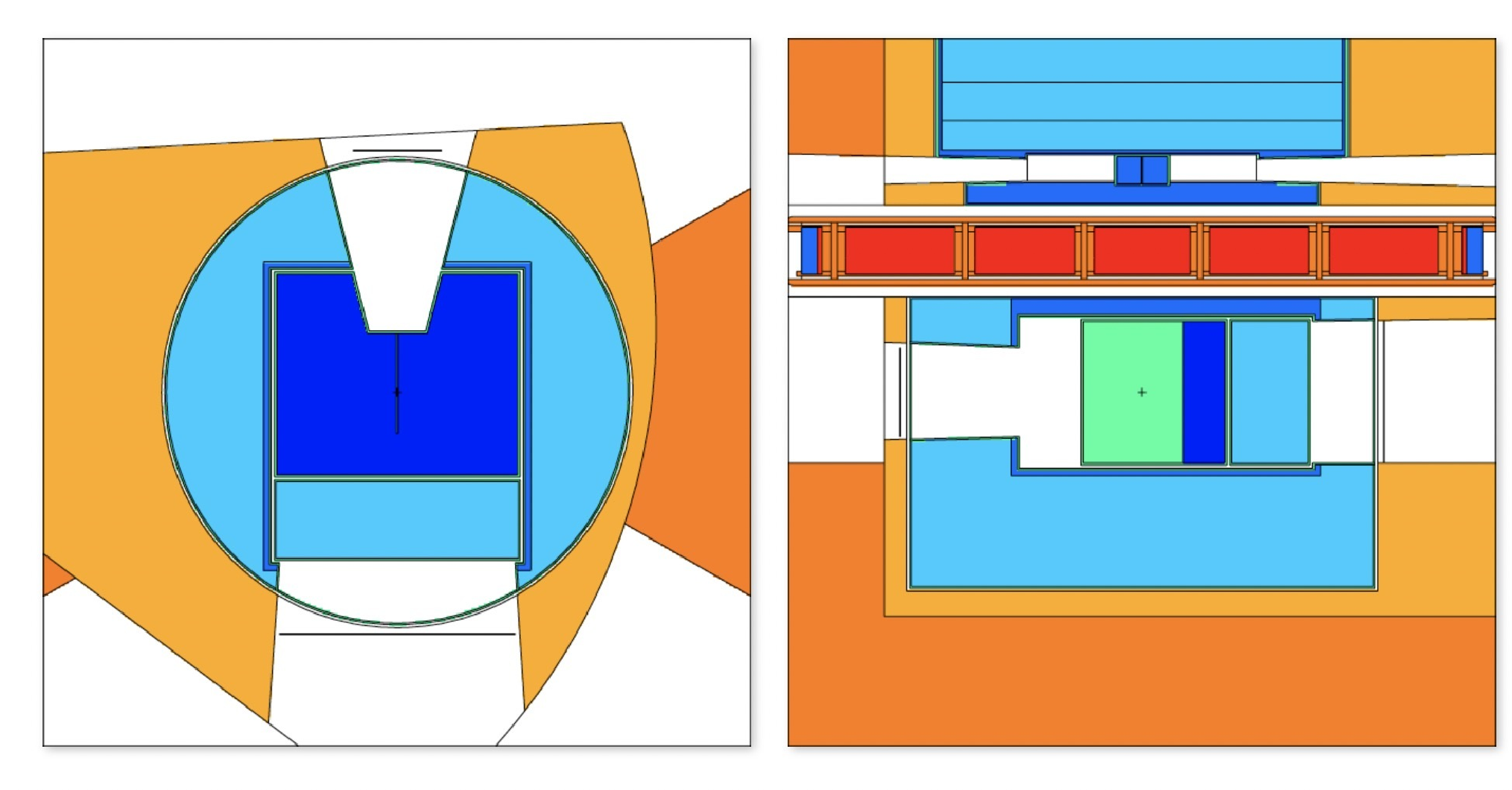}
\caption{As in \cref{fig:alan8}, but rectangular REH is tapered and the geometry is slightly changed to increase the overall performance.}
\label{fig:alan9}
\end{center}
\end{figure}

To quantify the effects of these design changes, a set of 
tests was performed with the following modifications of the third iteration model, starting from the original optimized model (\cref{fig:alan1}):

\vspace{0.25cm}
\begin{itemize}
    \item \ce{LD_2} with Al flow guides modeled explicitly and wall thickness increased from 0.3\,cm to 0.8\,cm (\cref{fig:alan3}).
    \item Be filter separated from \ce{LD_2} moderator (\cref{fig:alan4}).
    \item U-shaped reentrant hole with a few parameters slightly changed to streamline the model (\cref{fig:alan6}).
    \item \ce{LD_2} with Al flow guides modeled explicitly and wall thickness increased from 0.3\,cm to 0.4\,cm (\cref{fig:alan8}).
    \item Tapered rectangular REH (\cref{fig:alan9}).
\end{itemize}
\vspace{0.25cm}

The conclusions from this analysis are that a thickness of 0.8 cm of Al gives too much neutronic penalty and should be reduced. On the other hand, moving the Be filter outside the Al vessel (and envisaging a separate cooling) and using a simplified U-shaped reentrant hole do not give a large penalty. 

%A summary of the geometrical configurations analyzed and their impact on the neutronic performance is given in \cref{tab:def} and illustrated in \cref{fig:alan1,fig:alan9}. This includes the reshaping of the reentrant hole (REH) to optimize extraction and the viewing surface from all beam port openings, the addition of support structures necessary for cooling, and increasing wall thickness for the aluminum enclosure. 
%\textcolor{red}{need to write more text and explain all the figures}

 \begin{table}[h!]
\caption{Impact of engineering refinements on neutronic performance, normalized to the optimized \ce{LD_2} baseline.}
\label{tab:impact}
\centering
\begin{tabular}{ c  c  c }
\toprule
%\texttt{iel} & Description \\ 
\midrule
 & NNBAR & WP7 \\
\cref{fig:alan1} & 100 & 100 \\
%\cref{fig:alan2} & 82 & 80 \\
\cref{fig:alan3} & 75 & 77\\
\cref{fig:alan4} & 92 & 98 \\
%\cref{fig:alan5} & 91 & 93 \\
\cref{fig:alan6} & 91 & 93 \\
%\cref{fig:alan7} & 88-91 & 89-88 \\
\cref{fig:alan8} & 94 & 93 \\
\cref{fig:alan9} & 94 & 94 \\
\bottomrule
\end{tabular}
\end{table}

% \begin{figure}[hbt!]
% \begin{center}
% \includegraphics[width=0.99\textwidth,trim={1.5cm 1.5cm 1.cm 1.5cm},clip]{cn_fig/alan2.eps}
% \caption{As in \cref{fig:alan1}, but the wall thickness increased from 3 mm to 8 mm.}
% \label{fig:alan2}
% \end{center}
% \end{figure}

% \begin{figure}[hbt!]
% \begin{center}
% \includegraphics[width=0.99\textwidth,trim={1.5cm 1.5cm 1.cm 1.5cm},clip]{cn_fig/alan5.eps}
% \caption{As in \cref{fig:alan4}, but original REH is remodeled to a U-shaped one.}
% \label{fig:alan5}
% \end{center}
% \end{figure}

% \begin{figure}[hbt!]
% \begin{center}
% \includegraphics[width=0.99\textwidth,trim={1.5cm 1.5cm 1.cm 1.5cm},clip]{cn_fig/alan7.eps}
% \caption{As in \cref{fig:alan6}, but the wall thickness increased from 3 mm to 4 mm.}
% \label{fig:alan7}
% \end{center}
% \end{figure}

%The engineering design of the box-shaped moderator from the first neutronic iterations required an 8\,mm-thick Al vessel around the \ce{LD_2} preserve the mechanical stability of moderator. 
% Thus, thicker Al walls were implemented in the model, together with 
 %This led to a drop of about 25\% in the NNBAR FOM (see \cref{fig:alan3} and \cref{tab:impact}). 

%The solution to reduce the thickness of Al vessel to 4 mm proposed by WP5 was to round the walls of moderator. The WP5 team also provided a detailed CAD model of the \ce{LD_2} moderator with Al vessel. It was additionally required to separate \ce{LD_2} and Be vessels with a gap of thickness of 5\,mm. An MCNP model was designed according to this CAD model (see \cref{fig:LD2_Moderator_Optimisation_6.eps} and \cref{fig:LD2_Moderator_Optimisation_7.eps}). This last step of the design is described in \cref{sec:roundedshape}.

\subsubsection{Rounded shape model} 
\label{sec:roundedshape}

%Additional neutronic model with separated LD$_2$ box and Be filter, U-shaped reentrant hole and rounded walls of the LD$_2$ box based on new engineering studies which are in preliminary stage

As a final step of the design of the liquid deuterium moderator, we investigated a model
with separated LD$_2$ box and Be filter, U-shaped reentrant hole and rounded walls of the LD$_2$ box based on new preliminary engineering studies.
%updated following a preliminary engineering study based on the model from third neutronic optimisation of cold moderator. 
As shown in the previous section, a 0.8-cm thick Al vessel and additional Al flow guides would be required for a vessel with flat walls (see model on \cref{fig:alan3}) leading to a significant drop in NNBAR FOM of about 25 \% (see \cref{tab:impact}). To reduce the negative impact on the neutronic performance of the cold moderator, rounded walls were proposed for the moderator. Moreover, as stated in the previous section, the Be filter and LD$_2$ box were separated and the shape of reentrant hole was simplified to an U-shape. This allowed reducing the thickness of the Al vessel to 0.4 cm.

\begin{figure}[tb!]      
    \begin{subfigure}[b]{0.48\textwidth}
        \centering
        \includegraphics[width=\textwidth]{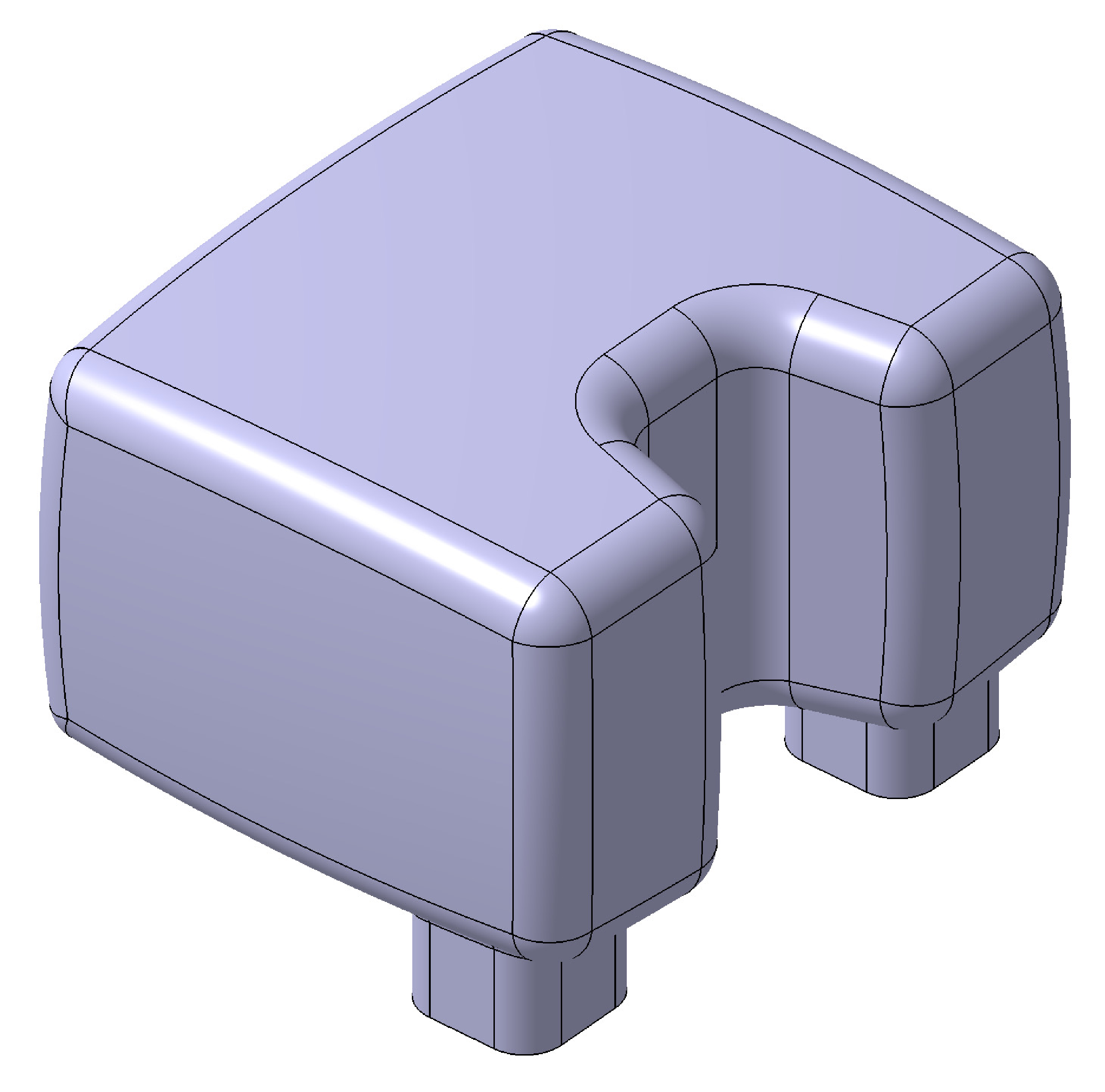}
        \subcaption{}
        \label{fig:LD2-1}
    \end{subfigure}
    \hfill
    \begin{subfigure}[b]{0.48\textwidth}
        \centering        
        \includegraphics[width=\textwidth]{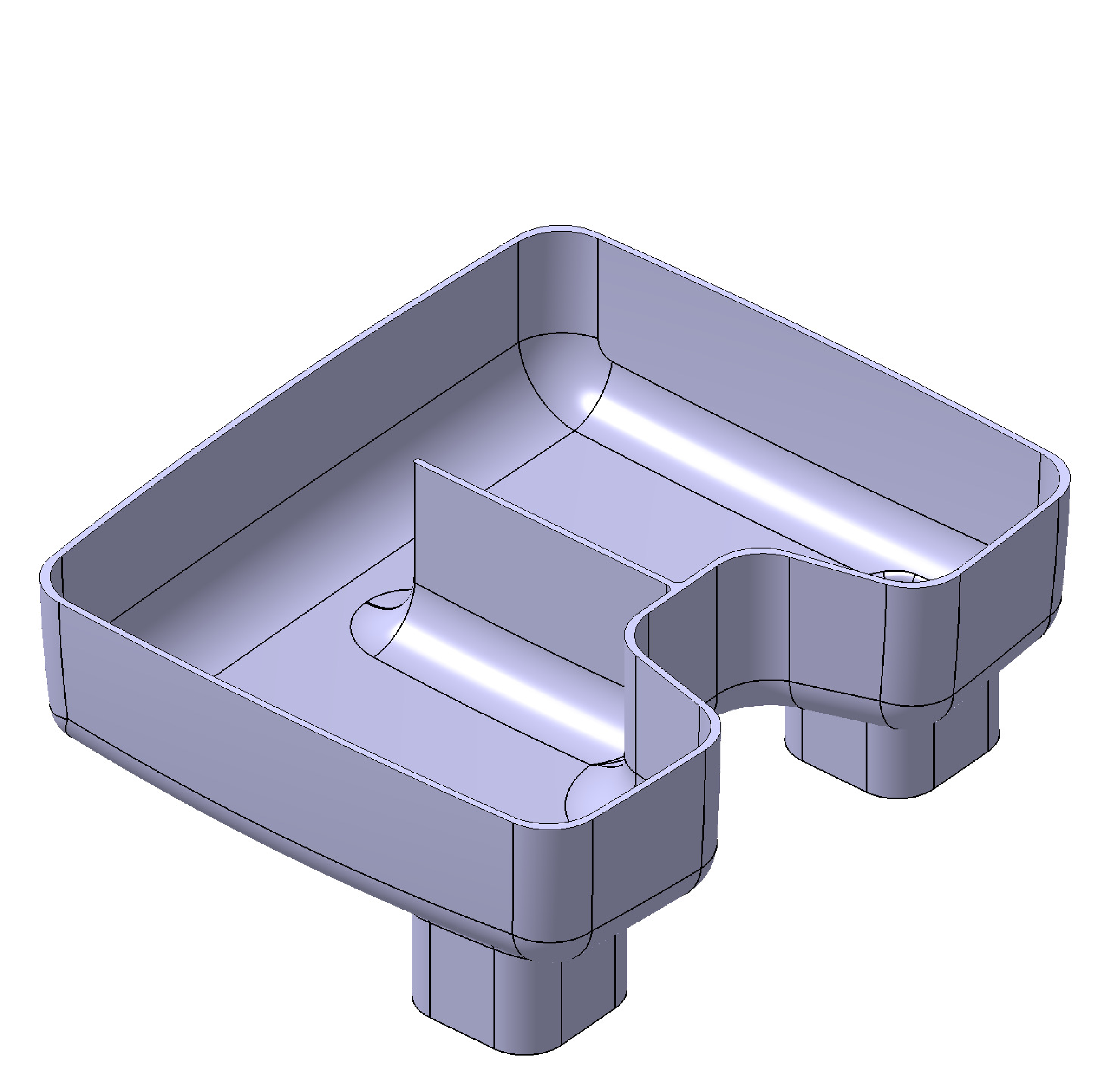}
        \subcaption{}
        \label{fig:rounded}
    \end{subfigure}
    \caption{Design of \ce{LD_2} moderator with rounded shape, developed by the engineering team. (a) Full model. (b) Horizontal cut.}
        \label{fig:engineering_1}
    \end{figure}

\begin{figure}[bt!]
	\begin{center}
		\includegraphics[width=0.45\textwidth]{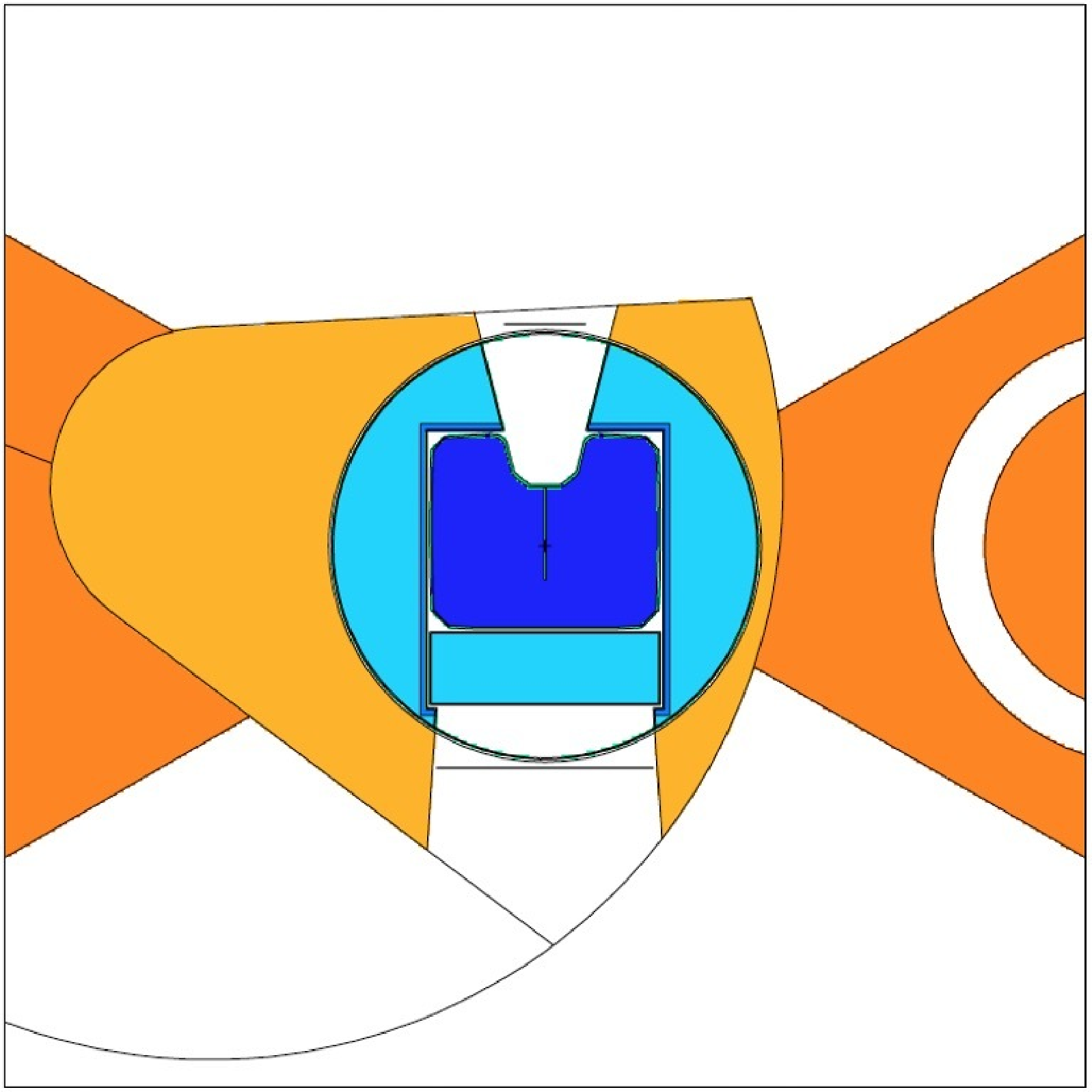}
		\includegraphics[width=0.45\textwidth]{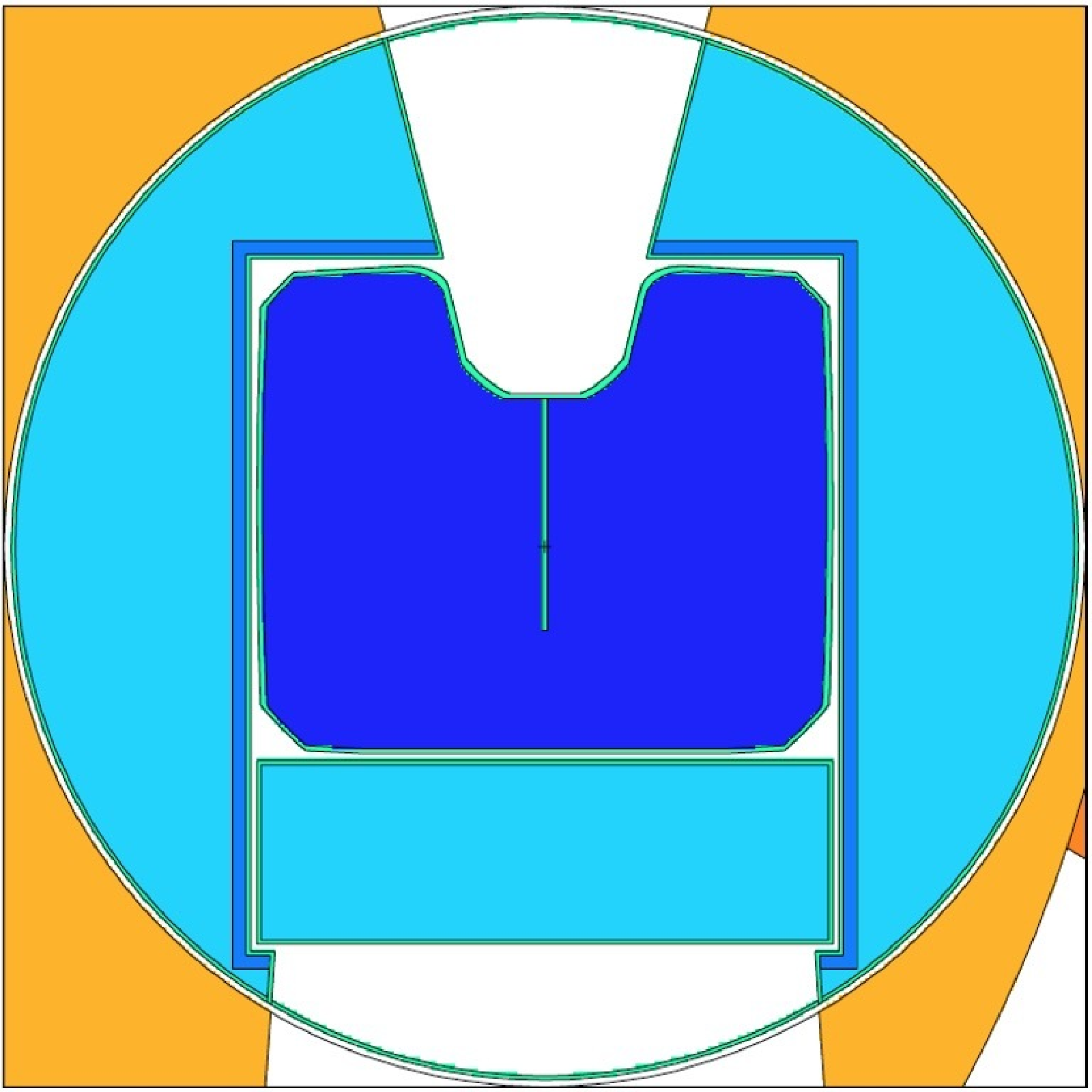}
		\includegraphics[width=0.45\textwidth]{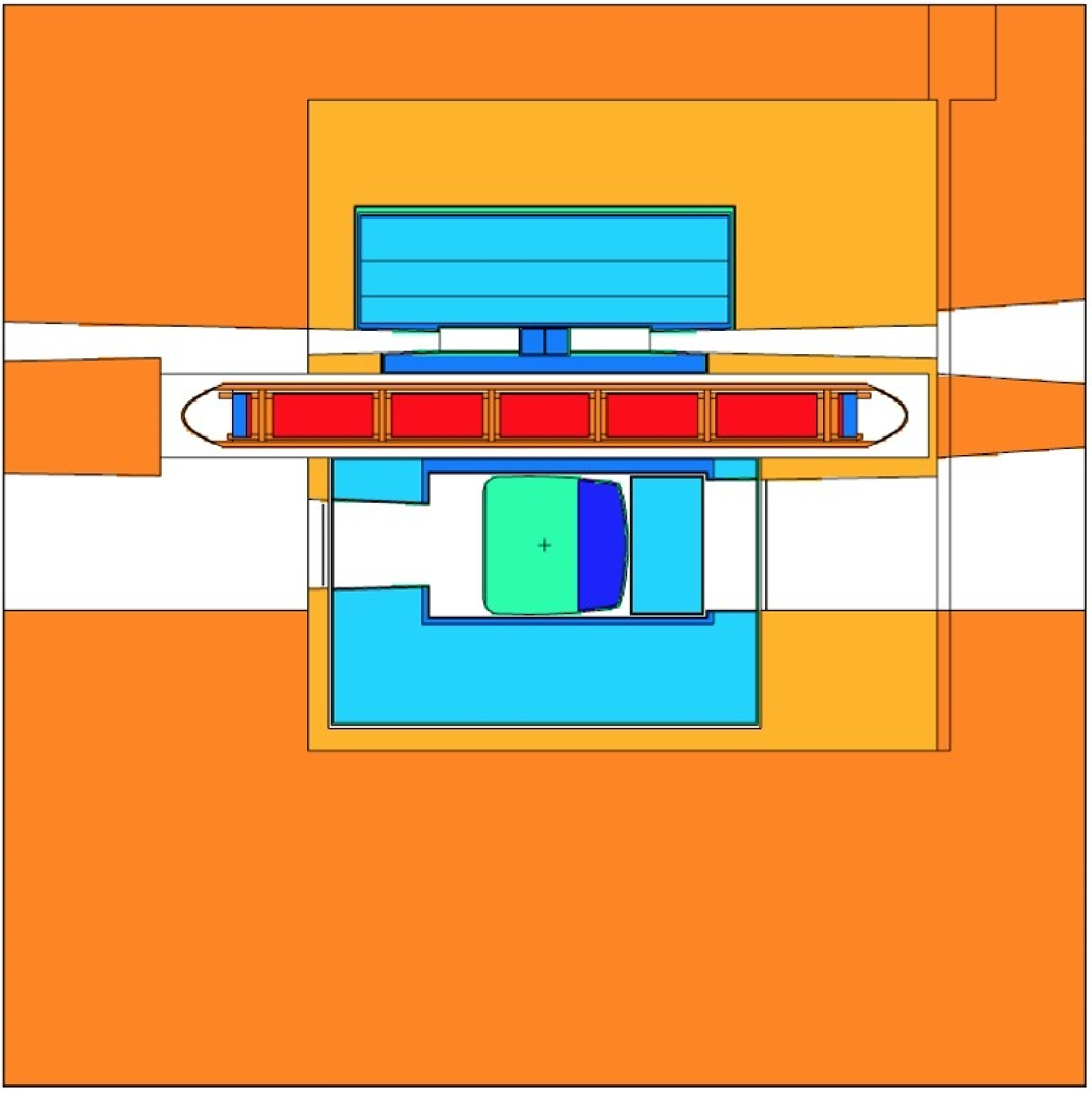}
		\includegraphics[width=0.45\textwidth]{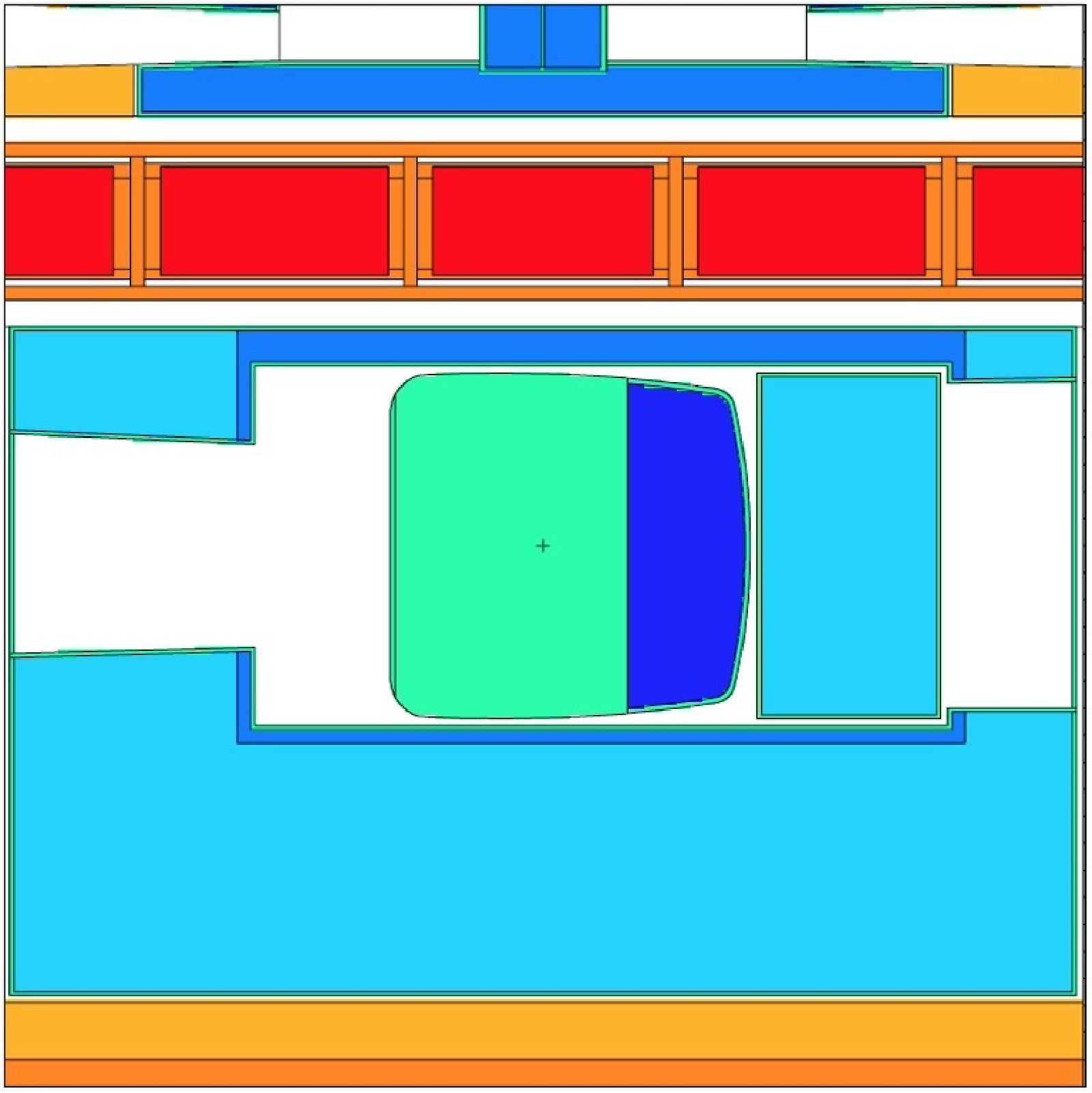}
		\includegraphics[width=0.45\textwidth]{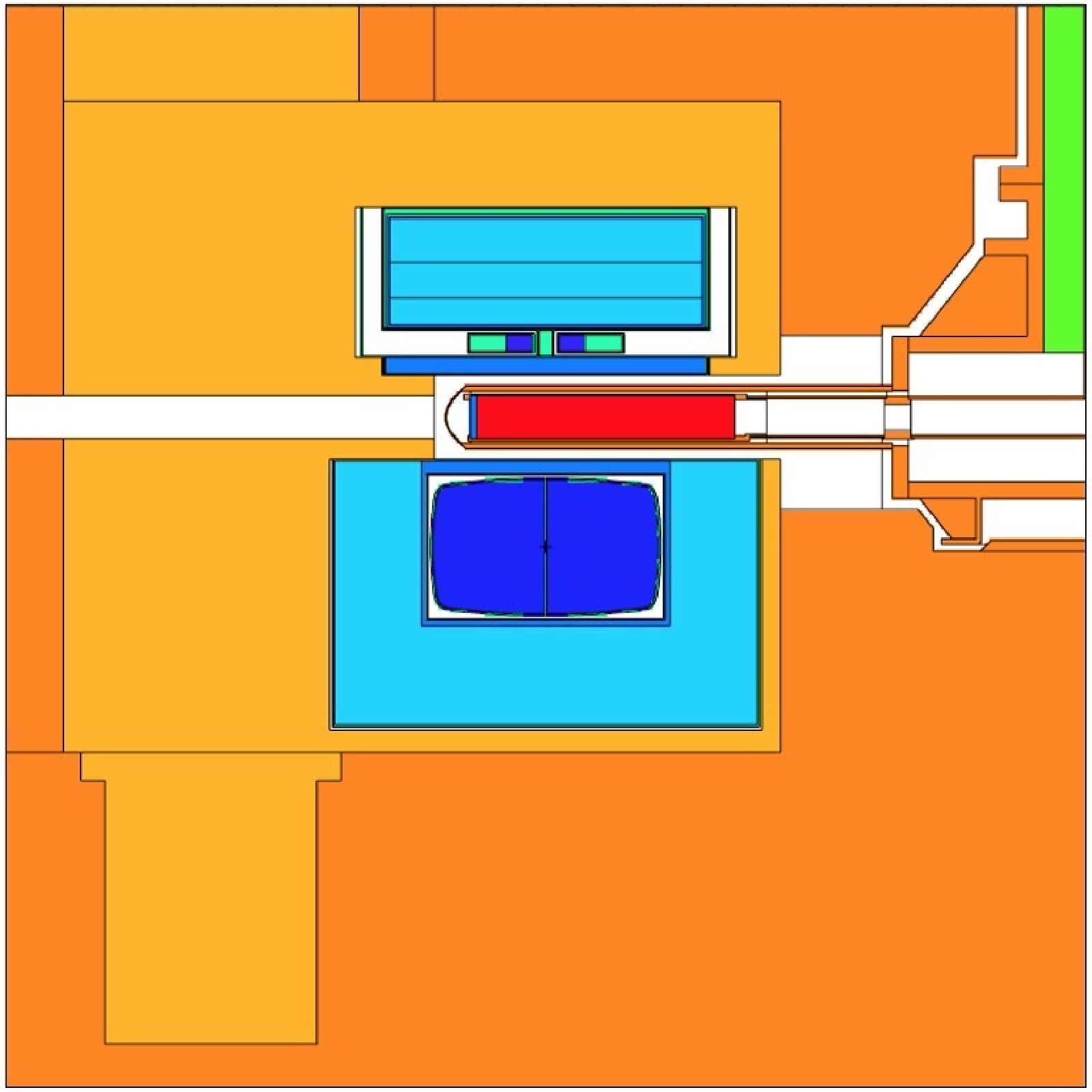}
		\includegraphics[width=0.45\textwidth]{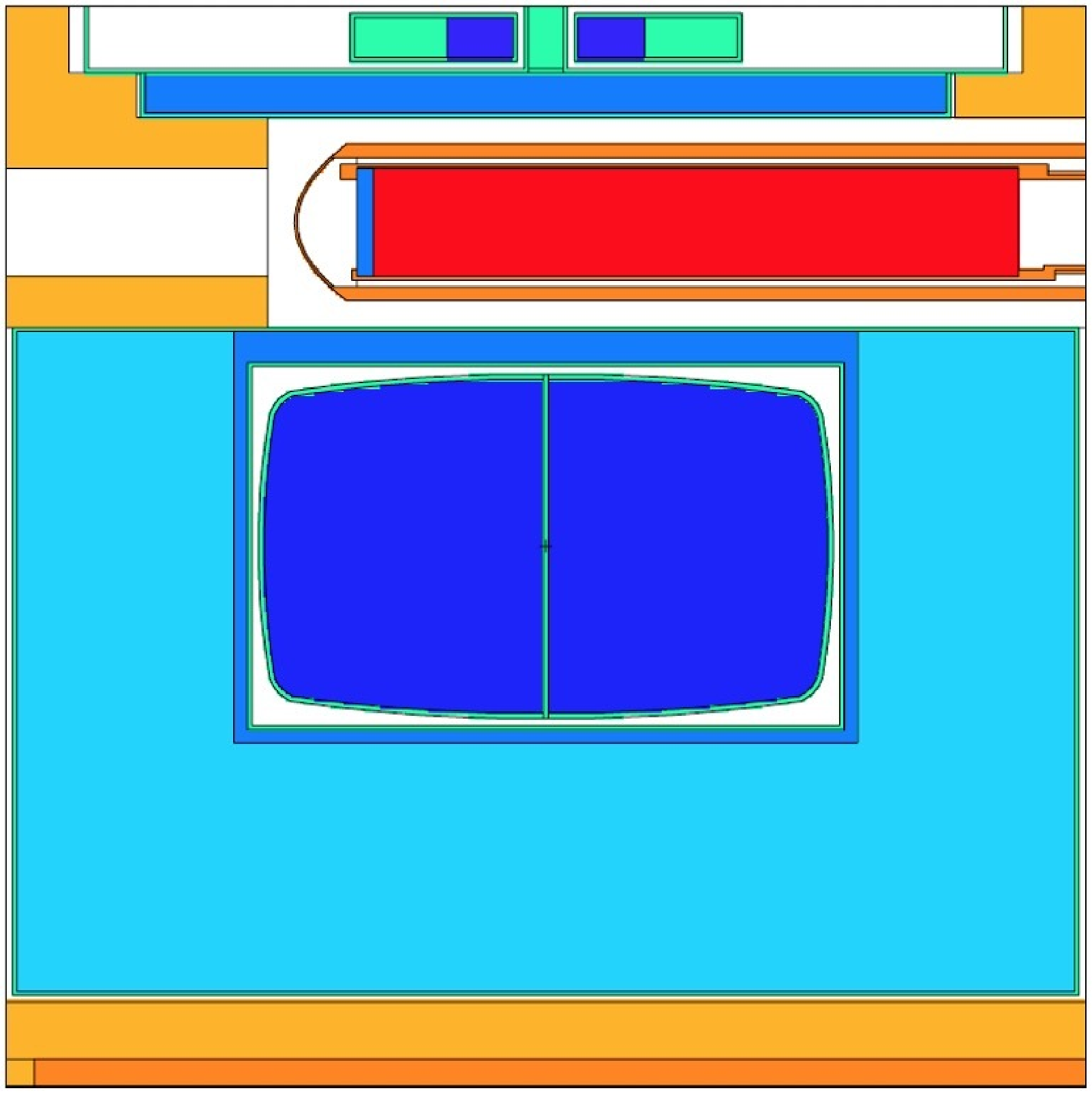}
		\caption{Graphical representation of the final design for the \ce{LD_2} moderator with an U-shaped reentrant hole, separated Be filter and \ce{LD_2} and rounded walls. The color codes are the following: orange: steel (twister frame, inner shielding, etc); dark blue: liquid ortho-deuterium; blue: light water; light blue: beryllium; green: aluminum. Note that cold Be filters and ambient Be reflector are  shown using the same color; the same note applies to Al.
		}
		\label{fig:ld2_baseline_model_Iteration4}
	\end{center}
\end{figure}

\begin{figure}[tb!]      
%    \begin{subfigure}[b]{0.5\textwidth}
    \begin{subfigure}[b]{0.46\textwidth}
        \centering
        \includegraphics[width=\textwidth]{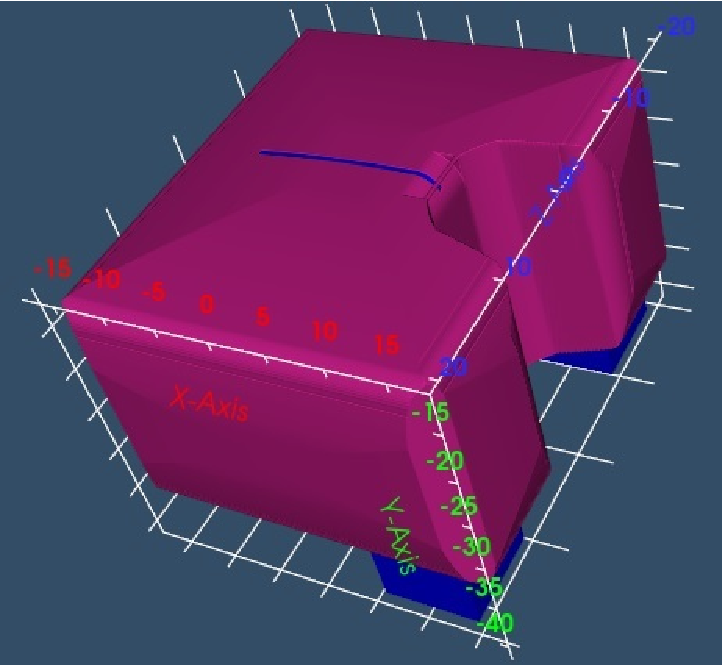}
        \subcaption{}
        \label{fig:LD2_Moderator_Optimisation_6.eps}
    \end{subfigure}
    \hfill
%    \begin{subfigure}[b]{0.524\textwidth}
    \begin{subfigure}[b]{0.48\textwidth}
        \centering        
        \includegraphics[width=\textwidth]{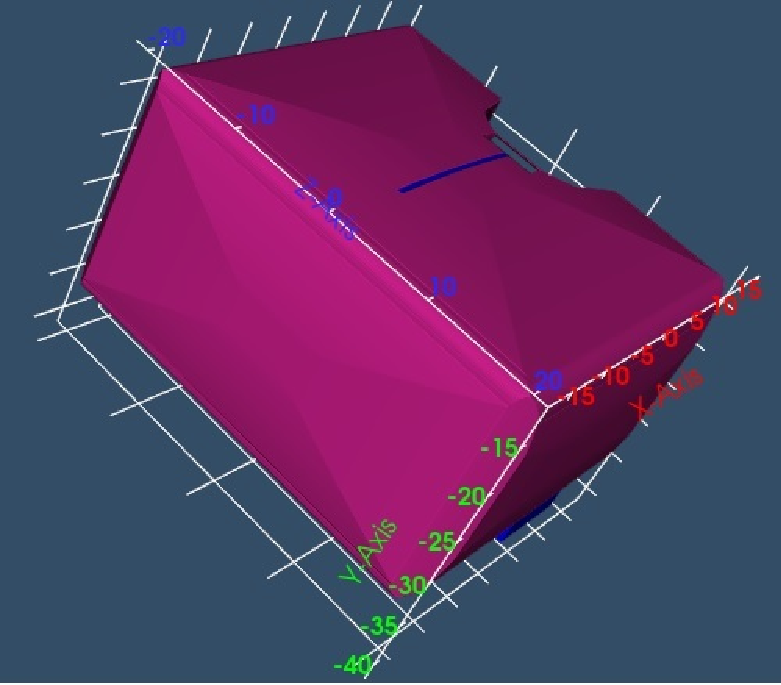}
        \subcaption{}
        \label{fig:LD2_Moderator_Optimisation_7.eps}
    \end{subfigure}
    \caption{MCNP model of the cold moderator with rounded shapes according to the technical drawing from the engineering team. This view shows the outer surface of cold moderator covered by the Al vessel.}
    \label{fig:mcnpround}
    \end{figure}

The WP5 team provided a detailed CAD model of the \ce{LD_2} moderator vessel (\cref{fig:LD2-1,fig:rounded}). An MCNP model designed according to this CAD model is shown in \cref{fig:ld2_baseline_model_Iteration4,fig:mcnpround}.

% \begin{figure}[hbt!]
% \begin{center}
% \includegraphics[width=0.48\textwidth]{cn_WP5/LD2-1.eps}
% \caption{Design of \ce{LD_2} moderator with rounded shape, developed by the engineering team.}
% \label{fig:LD2-1}
% \end{center}
% \end{figure}

% \begin{figure}[hbt!]
% \begin{center}
% \includegraphics[width=0.48\textwidth]{cn_WP5/LD2-2.eps}
% \caption{\ce{LD_2} moderator with rounded shape shown with horizontal cut, developed by the engineering team.}
% \label{fig:rounded}
% \end{center}
% \end{figure}

We note that there was no time available in the project to perform a full engineering study of this option. Nevertheless, we report these results in this document since it should be further investigated in the future. The summary of the expected performances, integrated in different wavelength ranges, is shown in \cref{tab:performance_iteration4}.

\begingroup
    \setlength{\tabcolsep}{10pt} 
    \begin{table}[bt!]
    \centering
%    \begin{threeparttable}
      \caption{Neutronic performance and characteristics of the most recent model of the cold moderator based on the preliminary engineering study. Brightness and intensity integrated over different wavelength ranges (NNBAR and WP7 openings) and compared with the upper moderator. The values for the upper moderator are from Ref \cite{zanini_design_2019}. To calculate the intensity we considered the following size of emission windows: NNBAR -- 960 cm$^2$; WP7 -- 225 cm$^2$; upper moderator -- 42 cm$^2$.}
    \begin{tabular}{lccc}
    \toprule   & \multicolumn{3}{c}{BRIGTHNESS [\si{n/cm\squared/s/sr}] }\\
    \cmidrule{2-4}
       & $>$ \SI{2}{\angstrom} & $>$ \SI{4}{\angstrom} & $>$ \SI{10}{\angstrom} \\
    \midrule
    NNBAR & \num{7.05e+12} & \num{5.40e+12} & \num{3.92e+11}  \\
    WP7 & \num{1.47e+13} & \num{7.77e+12} & \num{6.22e+11} \\
    upper moderator & \num{5.3e13} & \num{1.7e13} & \num{9.9e11}  \\
    \midrule
             & \multicolumn{3}{c}{INTENSITY [\si{n/s/sr}] }\\
    \cmidrule{2-4}
       & $>$ \SI{2}{\angstrom} & $>$ \SI{4}{\angstrom} & $>$ \SI{10}{\angstrom} \\
    \midrule
    NNBAR & \num{6.77e+15} & \num{5.18e+15} & \num{3.76e+14}  \\
    WP7 & \num{3.30e+15} & \num{1.75e+15} & \num{1.40e+14} \\
    upper moderator & \num{2.2e15} & \num{7.0e14} & \num{4.2e13}  \\
    \midrule
    NNBAR FOM & \num{2.24e+17} [\si{n/s/sr} $\times$ $\lambda^2$] &  & \\
    WP7 FOM & \num{2.91e+15} [\si{n/s/sr}] & & \\
    Heatload on LD$_2$ moderator & \num{25.8} [kW] & & \\
    Heatload on Be filter & \num{16.3} [kW] & & \\
    Heatload on Al vessel & \num{10.5} [kW] & & \\
    Total heatload & \num{52.6} [kW] & & \\
    \bottomrule
    \end{tabular}
  \label{tab:performance_iteration4}
%  \end{threeparttable}
\end{table}
\endgroup

% \begin{figure}[hbt!]
% \begin{center}
% \includegraphics[width=0.9\textwidth]{cn_fig/rounded.eps}
% \caption{Technical drawing for a \ce{LD_2} moderator with rounded shape}
% \label{fig:rounded_detail}
% \end{center}
% \end{figure}

\subsection{Summary of the cold moderator design}

We have successfully designed a cold neutron source with a neutron intensity above 4 \AA~exceeding the one from the upper moderator by a factor of 10. This accomplishment aligns with one of the primary objectives of HighNESS, which aimed to create a complementary source capable of supporting various experiments where neutron intensity holds greater significance than brightness. This has been achieved using the proven technology of liquid deuterium moderators which are the workhorses at high power research reactors and continuous sources like SINQ \footnote{\url{https://www.psi.ch/en/sinq}}.

The design of the cold moderator has been refined throughout the whole duration of the project, with several iterations with the engineering team. Details on the engineering design are provided in the next section. Although an engineering design of the rounded-shape model has not been performed due to time constraints, there are strong indications that this design could be the best one for the cold source and should be further investigated. 
%In the following sections, the design of VCN and UCN sources are described. A key aspect is the integration between the different sources. As it will be shown, such integration can be performed in different ways, which in some cases require a modification of the cold source, or even a replacement with a VCN source. 

\FloatBarrier \newpage
\section{Engineering design of the liquid deuterium moderator} \label{ch:1}
The first moderator and reflector system (so called twister I) will have two liquid parahydrogen moderators above the target wheel only. However, there is available space below the target wheel (lower moderator plug), which has been reserved for future moderator upgrades that could be used for a LD$_\text{2}$ volume moderator.
In the HighNESS project, Work Package 5 (WP5) was in charge of   the engineering implementation of new cold moderator concepts.  %Deliverable 5.1). 
This includes the definition of fluid parameters; cooling process concept design; structural materials choice; detailed CAD design of the advanced moderator concepts for the lower moderator plug including manufacturability verification; weldability analysis; structural-mechanics and fluid-dynamics simulations; and integrability into the ESS facility \cite{HighnessCurrentStatus}.

From an engineering point of view, the cooling and integration into the existing facility of such a large cryogenic pressure vessel in the high-power spallation source ESS is critical and must be analyzed in combination with the mechanical design~\cite{ESSproposal}.

%\subsection{Engineering description of the cold source (Yannick and WP5 team)}  
%Within the framework of work package 5 (WP5), deliverable 5a, the engineering implementation of new cold moderator concepts is analyzed. Thereby, in particular, the definition of fluid parameters, cooling process concept design, structural materials choice, detail CAD design of the advanced moderator concepts for the lower moderator plug including manufacturability verification, weldability analysis, structural-mechanics and fluid-dynamics simulations and integrability into the ESS facility will be verified. [1]
%The first moderator and reflector system (so called Twister I) will have two liquid para hydrogen moderators above the target wheel only. However, there is available space below the target wheel (lower moderator plug), which has been reserved for future moderator upgrades that could be used for a LD$_\text{2}$ volume moderator. 
%From the engineering point of view the cooling and integration into the existing facility of such a large cryogenic pressure vessel in the high-power spallation source ESS is critical and must be analyzed in combination with the mechanical design. [2]

\subsection{Introduction} \label{ch:1-1}

\begin{figure}[hbt!]
\begin{center}
\includegraphics[width=0.7\textwidth]{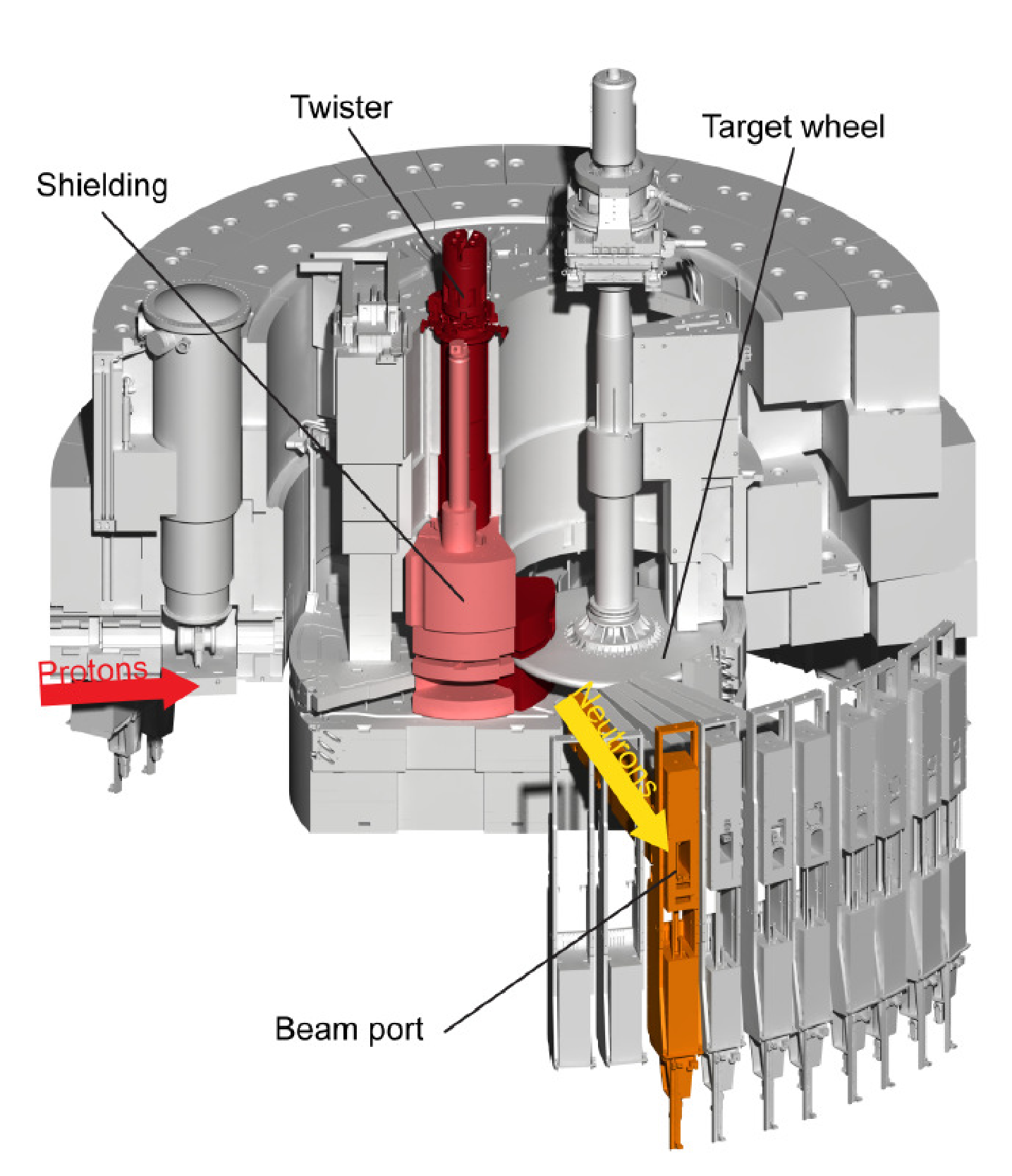}
\caption{ESS Target Monolith, isometric cutaway view.}
\label{fig:cn_1-1}
\end{center}
\end{figure}

%\subsection{Moderator Engineering - Introduction} \label{ch:1-1}
The first generation ESS moderator system consists of two liquid parahydrogen low-dimensional moderators, located above the tungsten target wheel. As mentioned previously, the moderator support structure, the twister, also allows for the use of the space below the target wheel for future moderator upgrades. \cref{fig:cn_1-1} shows the isometric view of the ESS Target Monolith. The colored components are subjects of the planned upgrades of the ESS Target station~\cite{santoro2020development}.

\FloatBarrier

\begin{figure}[hbt!]
\begin{center}
\includegraphics[width=0.75\textwidth]{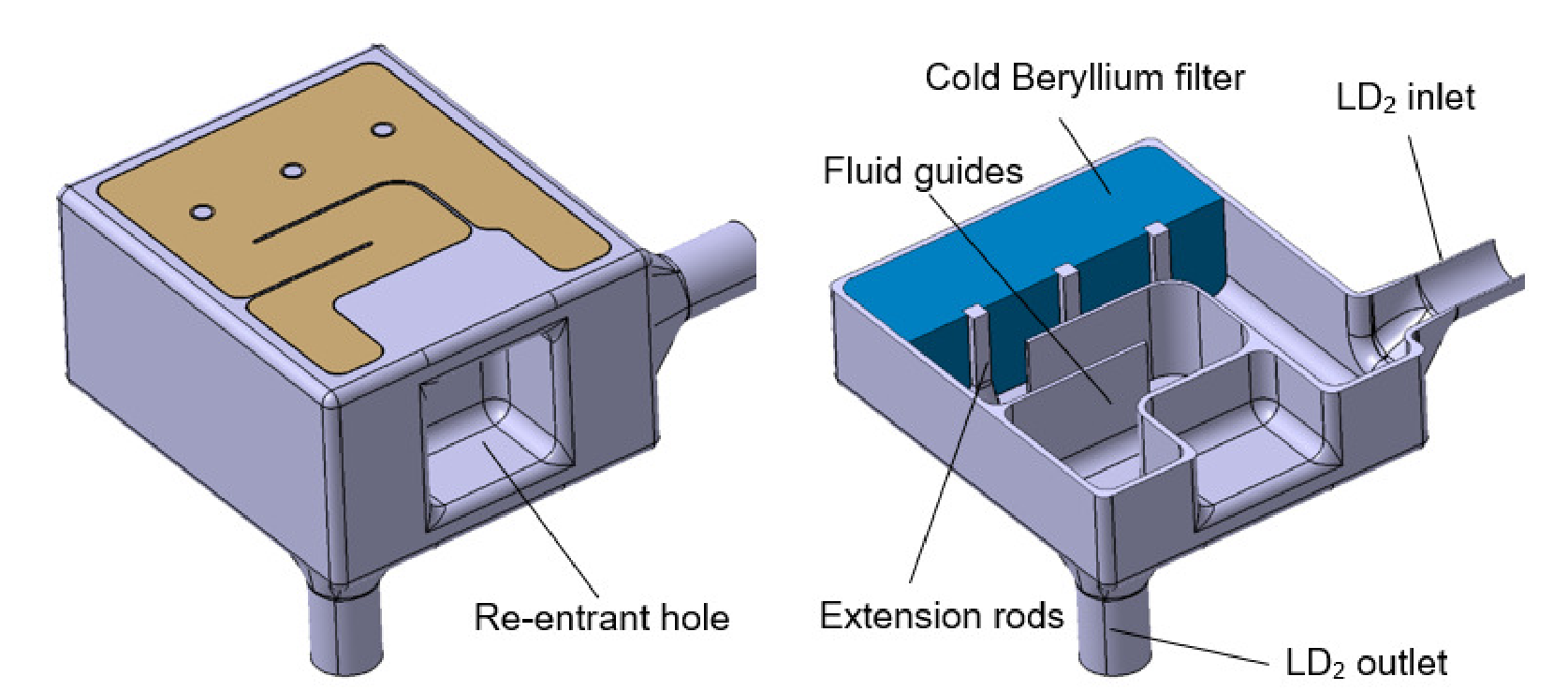}
\caption{Pre-design of the LD$_\text{2}$ moderator. Engineering design based on the third neutronic iteration (Fig. \ref{fig:ld2_baseline_model_Iteration3}).}
\label{fig:cn_1-2}
\end{center}
\end{figure}

The major engineering challenge here is to handle the enormous heat load into the LD$_\text{2}$ moderator of around \SI{60}{kW} resulting from the spallation process of a \SI{5}{MW} accelerator-driven neutron source. Following the several iterations discussed in the previous section, the following preliminary engineering design of the LD$_\text{2}$ moderator was developed. The moderator vessel consists of high-strength Aluminum alloy EN AW-6061 T6, which allows local stresses up to \SI{87}{MPa} and will be filled with approximately \SI{30}{L} of liquid deuterium. The cold moderator is surrounded by a vacuum jacket followed by a light water premoderator and a warm beryllium reflector. In addition, one cold beryllium filter ($\leq$ \SI{80}{K}) is installed inside the cold moderator vessel on the NNBAR side serving the large beam port  (see \cref{fig:cn_1-2} blue block)~\cite{HighnessCurrentStatus}.

\FloatBarrier

\begin{figure}[hbt!]
\begin{center}
\includegraphics[width=0.75\textwidth]{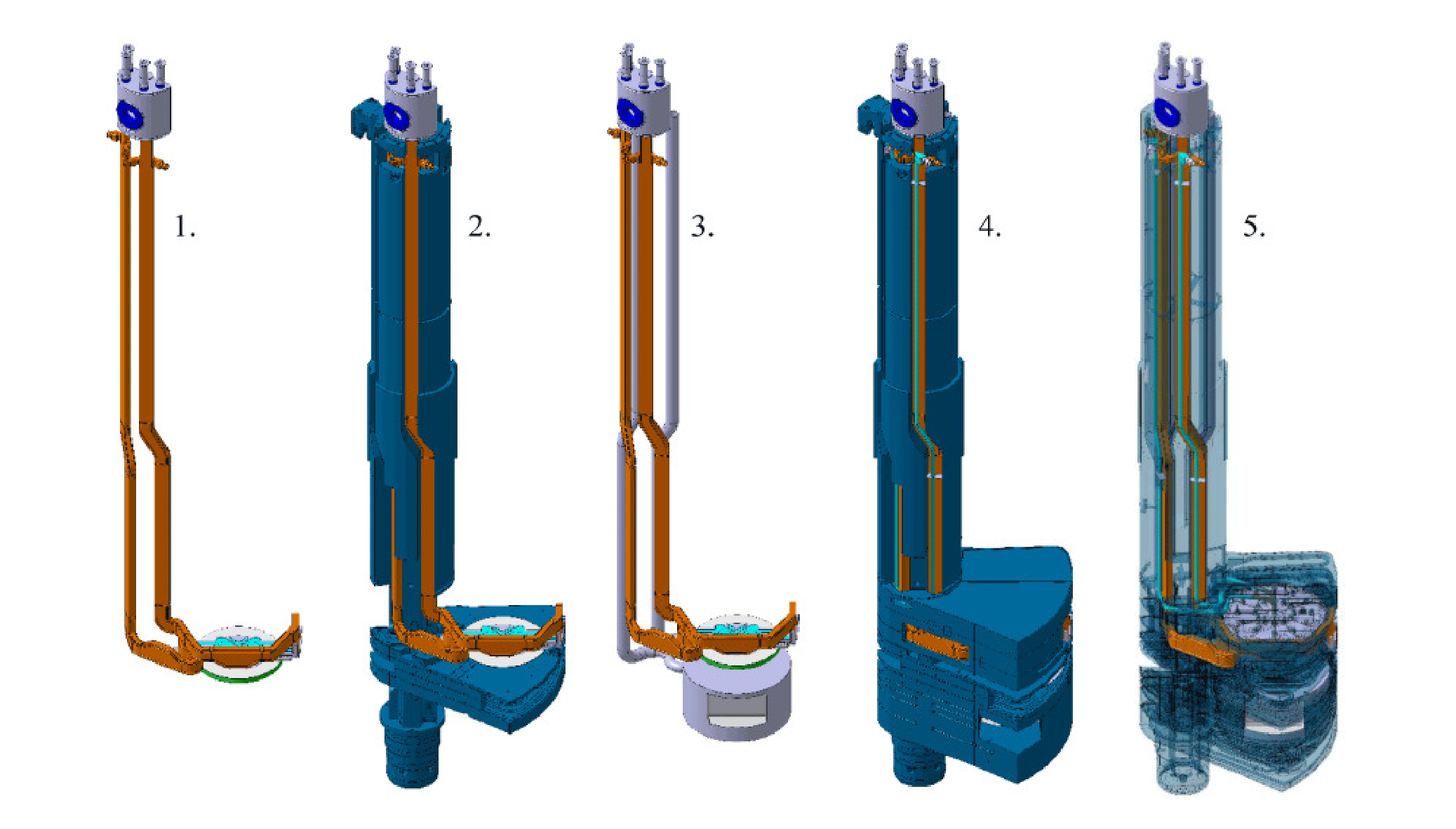}
\caption{Integration of LD$_\text{2}$ moderator: 1. Upper moderator plug; 2. upper moderator plug in the moderator support structure; 3. moderator plug with LD$_\text{2}$ moderator; 4. and 5. moderator plug in the moderator support structure near the outer reflector~\cite{HighnessCurrentStatus}.}
\label{fig:cn_1-3}
\end{center}
\end{figure}

The fluid guides ensure that flow separation, dead areas, and swirls do not occur. The extension rods ensure additional mechanical stability since the vessel walls are flat and need to be as thin as possible to minimize neutron losses. The moderator will be milled and partially Electrical Discharge Machining (EDM) machined from a solid block of aluminum EN AW-6061 T6, and the cover will be finally welded to the main body with “low-distortion” electron beam welding. In addition to the usual structural and fluid mechanics issues, the integrability must also be checked, since the components must be installed into an existing source. Originally, the twister was designed for low-dimensional hydrogen moderators only. The integrability is critical, since the LD$_\text{2}$ volume moderator has a substantially larger volume and thus dissipates significantly more heat. Therefore, larger supply and dissipation cross-sections are required. All supply lines must be routed through the twister shaft. However, the diameter of the shaft cannot be increased because it is surrounded by non-replaceable shielding elements. \cref{fig:cn_1-3} shows the twister with the upper moderator on the left-hand side and on the right-hand side the integrated volume LD$_\text{2}$ moderator in the lower moderator plug. \cite{HighnessCurrentStatus}

\subsection{Definition of fluid parameters and cooling process concept design} \label{ch:1-2}
\begin{figure}[hbt!]
\begin{center}
\includegraphics[width=0.9\textwidth]{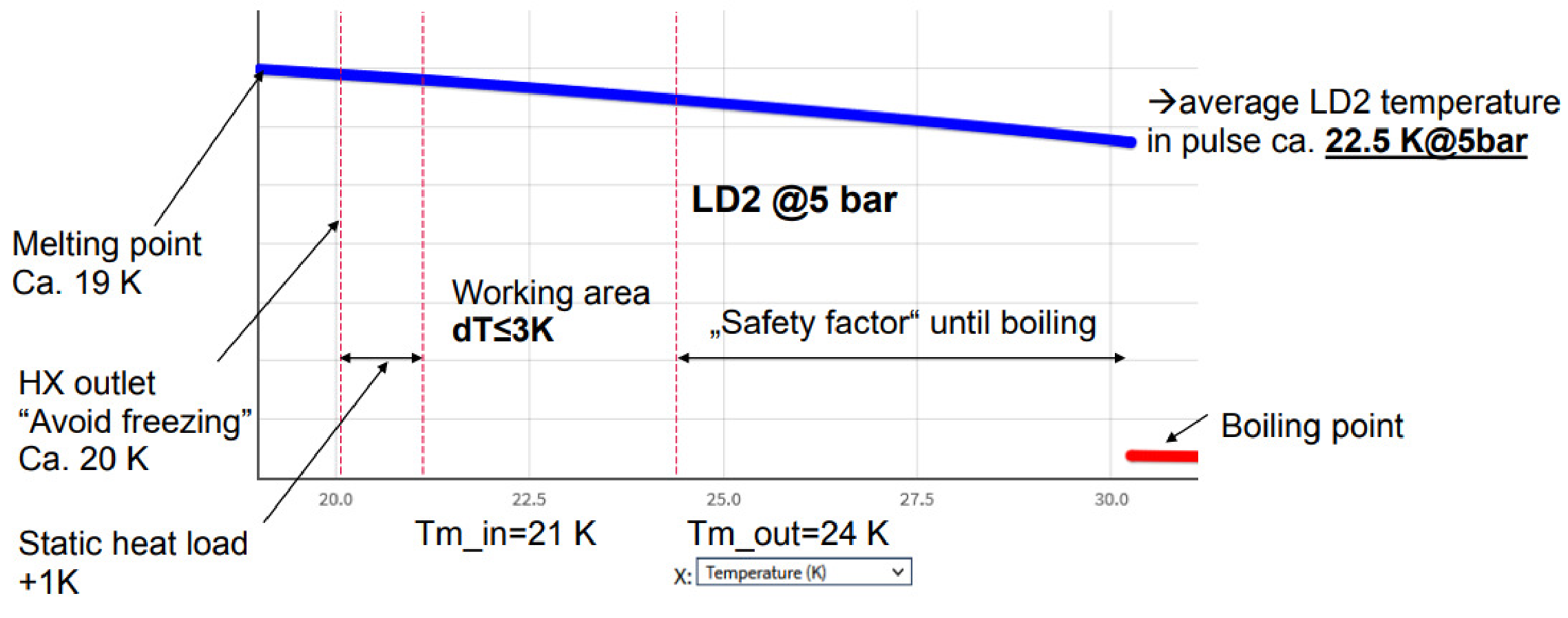}
\caption{Definition of fluid parameters for the moderator \ce{LD_2} volume.}
\label{fig:cn_1-4}
\end{center}
\end{figure}

The fluid parameters of liquid deuterium are defined in the following section, since these are required for the following designs and calculations of the new cold moderator system. \cref{fig:cn_1-4} shows a simplified representation of \ce{LD_2} (blue line) at the chosen operation pressure of \SI{5}{bar}. 
%In \cref{fig:cn_1-4} 
It can be seen that the deuterium solidifies at a temperature of {$\approx$}\SI{19}{K} and vaporizes at {$\approx$}\SI{30}{K}, which defines the temperature range in the liquid phase for the moderator system. In order to avoid freezing during operation, the minimum temperature at the outlet of the heat exchanger is set to \SI{20}{K}. In addition, it is assumed that additional heat load from the circulation pumps and insulation losses will increase the deuterium temperature by \SI{1}{K} before it arrives at the moderator inlet. The average temperature increases in the moderator due to particle heating of \SI{59.8}{kW} will be up to \SI{3}{K}, which means that the average outlet temperature will be around \SI{24}{K}. As a result, there is still a contingency factor for local temperature peaks of approximately \SI{6}{K} before the deuterium evaporates.

The mass flow can now be estimated by using the chosen operating pressure, the inlet- and outlet- temperature, and the heat load. A mass flow of at least \SI{3400}{\g\per\second} liquid deuterium is needed to remove the enormous particle heat and to keep the average temperature increase below $\leq$\,\SI{3}{K}. Assuming a flow velocity of up to $\leq$\,\SI{5}{\m\per\second}, an inlet/outlet pipe diameter of \SI{70}{mm} would be required, which has to fit into the existing twister structure.

\FloatBarrier
%The critical design parameters for the moderator are as follows:

%\vspace{0.2cm}

%\begin{itemize}[leftmargin=1cm]
%\item[-] Liquid ortho-Deuterium volume: \quad V$_{Mo} \approx$ \SI{30}{L}
%\item[-] Heat load into oLD$_\text{2}$-moderator: \quad $\Sigma$Q$_{Mo}$ = \SI{59.8}{kW} (@ \SI{5}{MW})
%\item[-] Total heat load: \qquad $\Sigma$Q$_{To} \approx$ \SI{70}{kW} (\SI{60}{kW} + \SI{10}{kW})
%\item[-] Average pressure (Moderator) \quad p = \SI{5}{bar}
%\item[-] Average density \quad $\rho$ = \SI{173.8}{\kilogram\per\cubic\m}
%\item[-] Flow velocity: \quad w $\leq$ \SI{5}{\m\per\second}
%\item[-] Average mass flow \quad \.{m} $\geq$ \SI{3400}{\gram\per\second}
%\item[-] Average temperature increase (mo.): \quad dT $\leq$ \SI{3}{K}
%\item[-] Average temperature \quad T$_{Av}$ = \SI{22.5}{K}
%\item[-] Average inlet temperature \quad T$_{AVI}$ = \SI{21}{K}
%\item[-] Average outlet temperature \quad T$_{AVO}$ = \SI{24}{K}

%\end{itemize}

%\vspace{0.3cm}

\begin{table}[h!]
\caption{Critical design parameters.}
\label{tab:design_parameters}
\centering
\begin{tabular}{ l l }
%\toprule
%Item & Cost [M€] & Comments \\ 
\midrule
Liquid ortho-Deuterium volume & V$_{Mo} \approx$ \SI{30}{L} \\
Heat load into oLD$_\text{2}$-moderator & $\Sigma$Q$_{Mo}$ = \SI{59.8}{kW} (@ \SI{5}{MW}) \\
Total heat load & $\Sigma$Q$_{To} \approx$ \SI{70}{kW} (\SI{60}{kW} + \SI{10}{kW}) \\
Average pressure (Moderator) & p = \SI{5}{bar}\\
Average density & $\rho$ = \SI{173.8}{\kilogram\per\cubic\m} \\
Flow velocity & w $\leq$ \SI{5}{\m\per\second} \\
Average mass flow & \.{m} $\geq$ \SI{3400}{\gram\per\second} \\
Average temperature increase (mo.) & dT $\leq$ \SI{3}{K} \\
Average temperature & T$_{Av}$ = \SI{22.5}{K} \\	
Average inlet temperature & T$_{AVI}$ = \SI{21}{K} \\
Average outlet temperature & T$_{AVO}$ = \SI{24}{K} \\
\end{tabular}
\end{table}

\begin{figure}[hbt!]
\begin{center}
\includegraphics[width=0.90\textwidth]{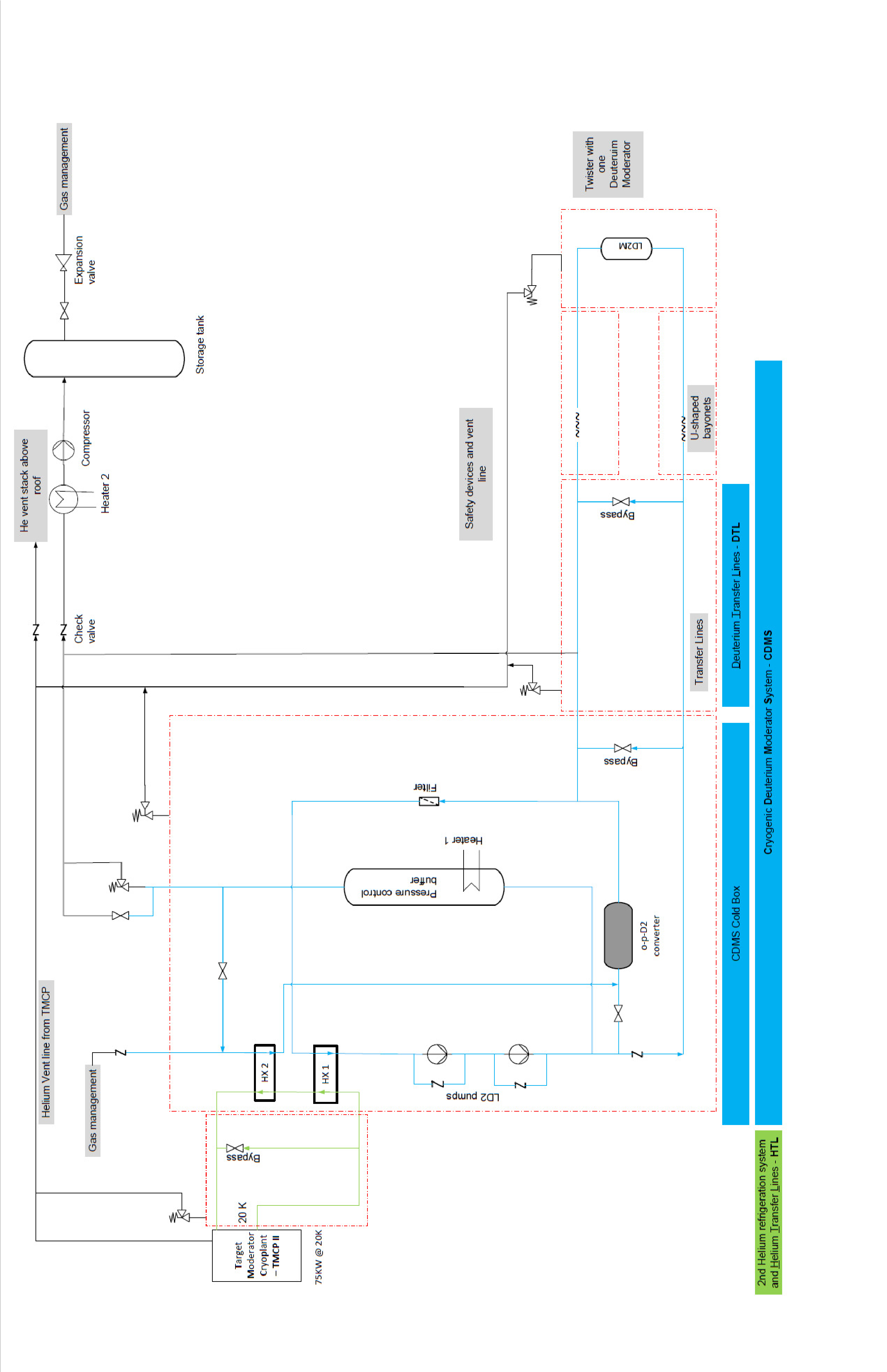}
\caption{Piping and instrumentation diagram for the \ce{LD_2} moderator system.}
\label{fig:cn_1-5}
\end{center}
\end{figure}

With the defined fluid parameters, a first piping and instrumentation diagram (P\& ID, see \cref{fig:cn_1-5}) was created, and a working CAD design of the lower moderator plug was completed for use in structural mechanical simulations and fluid dynamic simulations. The P\& ID in \cref{fig:cn_1-5} is a simplified system flow diagram of the cryogenic deuterium moderator system (CDMS), consisting of the helium refrigeration system (TMCP II) with \SI{75}{kW} cryo-power at \SI{20}{K}, the helium transfer lines (HTL), the deuterium liquefaction cryostat (CDMS Cold Box), the deuterium transfer lines (DTL), and finally the twister with the lower moderator plug.

The helium refrigeration system is responsible for providing the required cryo-power to the deuterium cryostat to liquify the gaseous deuterium and to remove the heat load of the moderator system during operation. For this, gaseous helium is pumped in a closed loop at a minimum temperature of 20\,K via the helium transfer lines to the deuterium cryostat heat exchanger and back to the helium refrigeration system. The main components of the deuterium cryostat itself are the heat exchanger, the circulation pumps, the ortho--para converter and the pressure control buffer. The circulation pumps are installed in the deuterium cryostat to supply the moderator with the liquefied deuterium in a closed circuit. In addition, the para-deuterium content is converted into almost 100\% ortho deuterium by using a catalyst in the cryostat, since this has better moderator properties. An active heated pressure-control buffer is also installed in the cryostat to compensate pressure changes caused by the pulsed proton beam and beam trips.

The cryostat provides the needed liquid deuterium to the moderator via the deuterium transfer lines, where the deuterium is heated up by particle interactions and returned to the heat exchanger of the cryostat to cool the deuterium down again. In addition, a storage tank for the deuterium will be necessary since deuterium produces tritium under irradiation. Therefore, the deuterium cannot be vented like hydrogen, for example, when maintenance work or similar actions have to take place. For this reason, in case of maintenance work, the deuterium will be heated and then fed into the storage tank using a heater and compressor. Then, completing maintenance, the deuterium would be reused.

This is also advantageous, because of the high cost of deuterium. The size of the storage tank is determined by the total inventory of the circuit, cryostat and moderator, which must be kept as small as possible for safety and costs reasons. Therefore, it is very important to place the cryostat as close as possible to the moderator to reduce the total amount of deuterium. Since selecting a final location of the components was not within the framework of the HighNESS project, this will be necessary in the final design phase.

\FloatBarrier

\subsection{Costs estimation of LD$_\text{2}$ infrastructure}
The rough cost estimation shown in \cref{tab:costs} is based on the cost of the existing ESS hydrogen system and is scaled up according to the heat load of the deuterium moderator system (ca. \SI{75}{kW}) compared to the hydrogen moderator system (ca. \SI{30}{kW}). In addition, a storage tank system is considered because of the issue mentioned above that deuterium can not be vented in maintenance cases due to tritium production. In addition, significantly higher costs for deuterium itself are considered as well. Since the location of the needed new building for the helium refrigeration system and the location of the deuterium cryostate inside the Target Station building are unknown, the distance to each other is estimated in order to be able to estimate the costs for the respective transfer lines.

 \begin{table}[h!]
\caption{Cost estimation of the LD$_\text{2}$ moderator system.}
\label{tab:costs}
\centering
\begin{tabular}{ l r c }
\toprule
Item & Cost [M€] & Comments \\ 
\midrule
Helium refrigeration system & \SI{35}{} & Ca. \SI{75}{kW} @ \SI{20}{K} \\
Helium transfer lines & \SI{1.93}{} & Assumed: 2x100m \\
D$_{2}$ cryostat & \SI{14}{} &  \\
LD$_\text{2}$ transfer lines & \SI{2.75}{} & Assumed: 2x40m \\
LD$_\text{2}$ moderator plug & \SI{5.5}{} & Complete new twister \\
D$_{2}$ storage system & \SI{2.5}{} &  \\
New building for He-refrigaration system \\ Room preparation for LD$_\text{2}$ cryostat, ATEX, ... & \SI{5}{} & No room selected so far \\
\bottomrule
Total: & \SI{66.68}{} & \\
\end{tabular}
\end{table}

The real costs can deviate significantly upwards or downwards from the estimated ones if the locations (distance) of components change and if the heat load is lower or higher, for example.

\subsection{Selection of materials and overall design} \label{ch:1-2-1}
For the newly designed moderator, the moderating medium of \ce{LD_2} is kept at \SI{20}{K}. The structural material for the moderator vessel is required to be tolerant to radiation damage, as transparent for neutrons as possible, and suitable for cryogenic temperatures. The chosen structural material shall also be sufficiently ductile for the fast load changes caused by the pulsed proton beam and sufficient in terms of strength to withstand the internal pressure of the LD$_\text{2}$.

Aluminum is nearly transparent for cold and thermal neutrons and is therefore typically selected as a moderator vessel material. However, the strength of pure aluminum is much too low to keep the stress from inner pressure within the allowable range, especially for large volumes as in the case of the LD$_\text{2}$ moderator. Therefore, heat treatable aluminum alloys (6000 series) or non-heat treatable (5000 series) are the preferred choice~\cite{ESSproposal}.

There is no ideal aluminum alloy, with each choice having advantages and drawbacks. Many facilities use EN AW-6061 T6, for example at SNS, J-PARC, and for the liquid hydrogen moderator of ESS. In addition, sufficient data of irradiated samples are available to evaluate the radiation-related lifetime of the moderator vessel~\cite{ESSproposal}.
Because of its well-known radiation properties, the high strength values and wide application fields in the frame of spallation sources, it was ultimately decided to use EN AW-6061 T6 for the lower moderator.

\subsection{Design overview of the lower moderator plug} \label{ch:1-3}
\begin{figure}[hbt!]
\begin{center}
\includegraphics[width=0.9\textwidth]{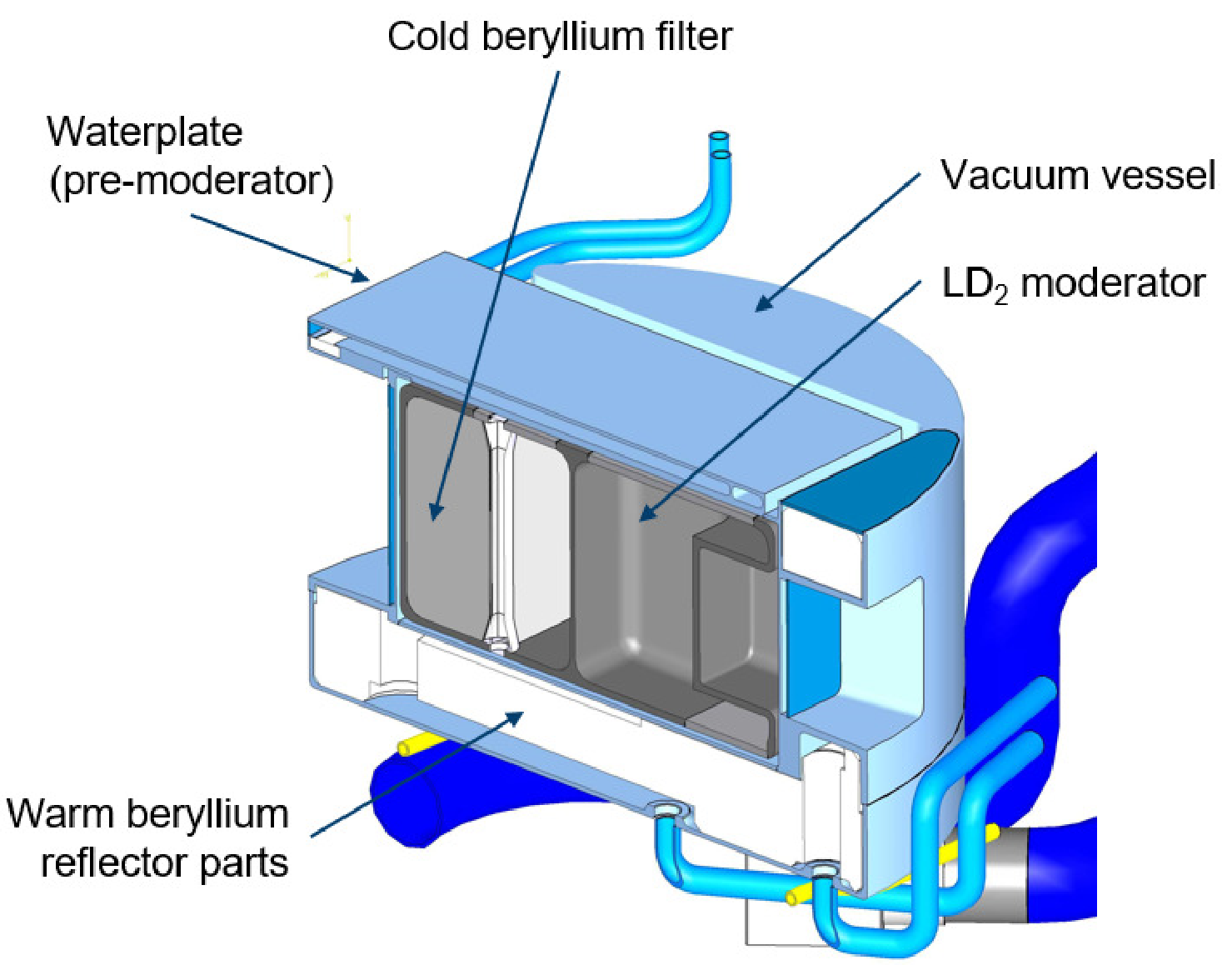}
\caption{3D cross section illustration of the lower moderator plug.}
\label{fig:cn_1-6}
\end{center}
\end{figure}

The new moderator is located in the frames of the twister just below the target wheel, as can be seen is \cref{fig:cn_1-3}. All details about the integration of the LD$_\text{2}$ moderator assembly into the already existing ESS twister can be found in \cref{ch:1-8}.
The lower moderator plug consists of the inner LD$_\text{2}$ moderator and a cold beryllium filter, that are surrounded by an insulation vacuum, a warm beryllium reflector, and a water premoderator on the top side, facing the target wheel. On each side of the outer vacuum vessel there are actively helium-cooled neutron windows (part of the vacuum vessel) where the cold neutrons exit the moderator (see \cref{fig:cn_1-6}).

\begin{figure}[hbt!]
\begin{center}
\includegraphics[width=0.9\textwidth]{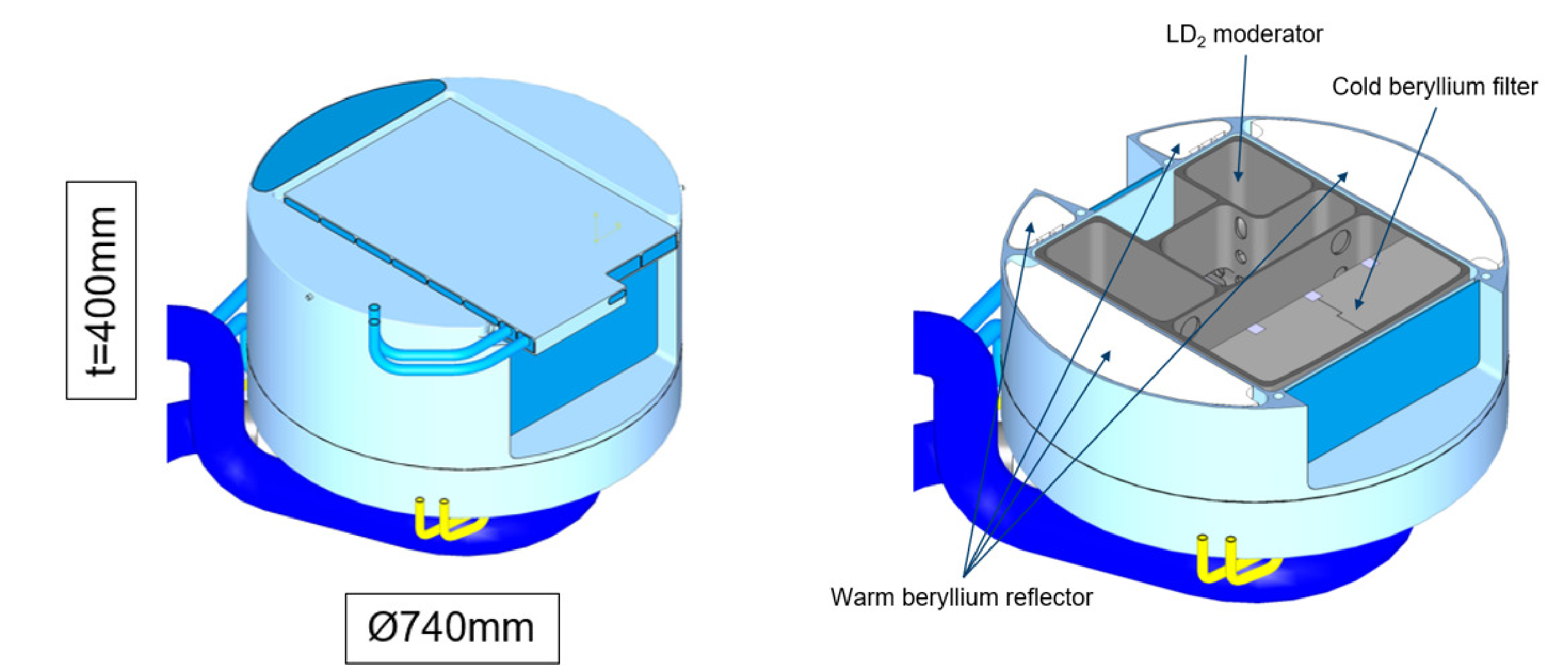}
\caption{Dimensions and layout of the lower moderator plug.}
\label{fig:cn_1-7}
\end{center}
\end{figure}

\cref{fig:cn_1-7} shows the overall box dimensions of the lower moderator plug without the media supply pipes and a horizontal cross-section of the assembly with the parts inside the plug. In principle, the LD$_\text{2}$ moderator sits within the surrounding vacuum vessel with a \SI{5}{mm} insulation vacuum gap to the adjacent aluminum walls. A layer of water and the warm beryllium reflector segments surround the moderator, with the reflector segments also placed inside a closed aluminum vessel.

\FloatBarrier

\begin{figure}[hbt!]
\begin{center}
\includegraphics[width=0.9\textwidth]{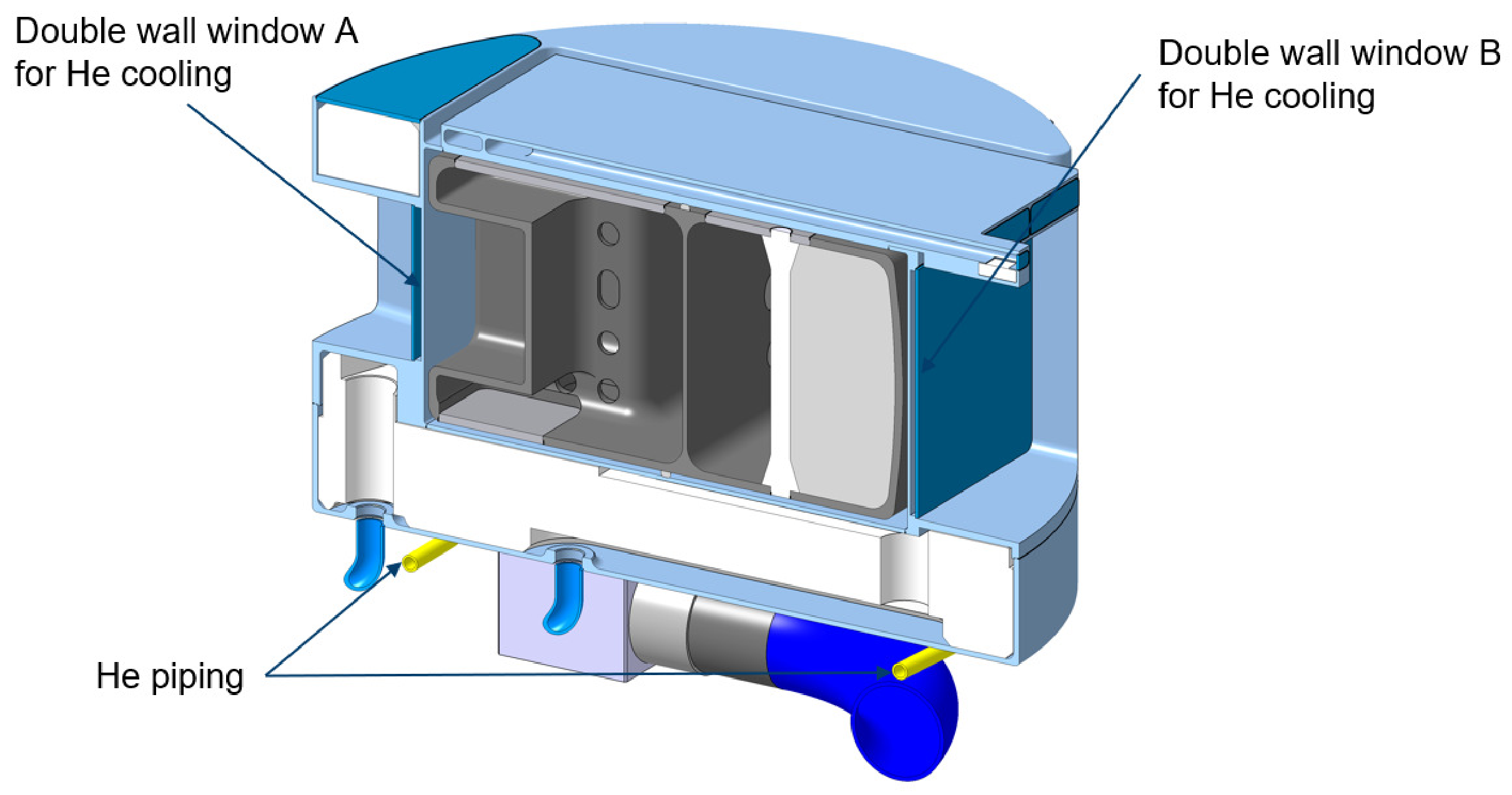}
\caption{3D cross-section of the neutron windows.}
\label{fig:cn_1-8}
\end{center}
\end{figure}

The lower moderator plug has two openings on the front and the back side, one smaller neutron window A (WP7 side) and a larger neutron window B (NNBAR side). Both neutron windows are actively cooled by a transversal helium flow and are therefore designed with a thin double-wall design, as shown in \cref{fig:cn_1-8}.

\subsection{Assembly and welding of the moderator} \label{ch:1-4}
\begin{figure}[hbt!]
\begin{center}
\includegraphics[width=0.9\textwidth]{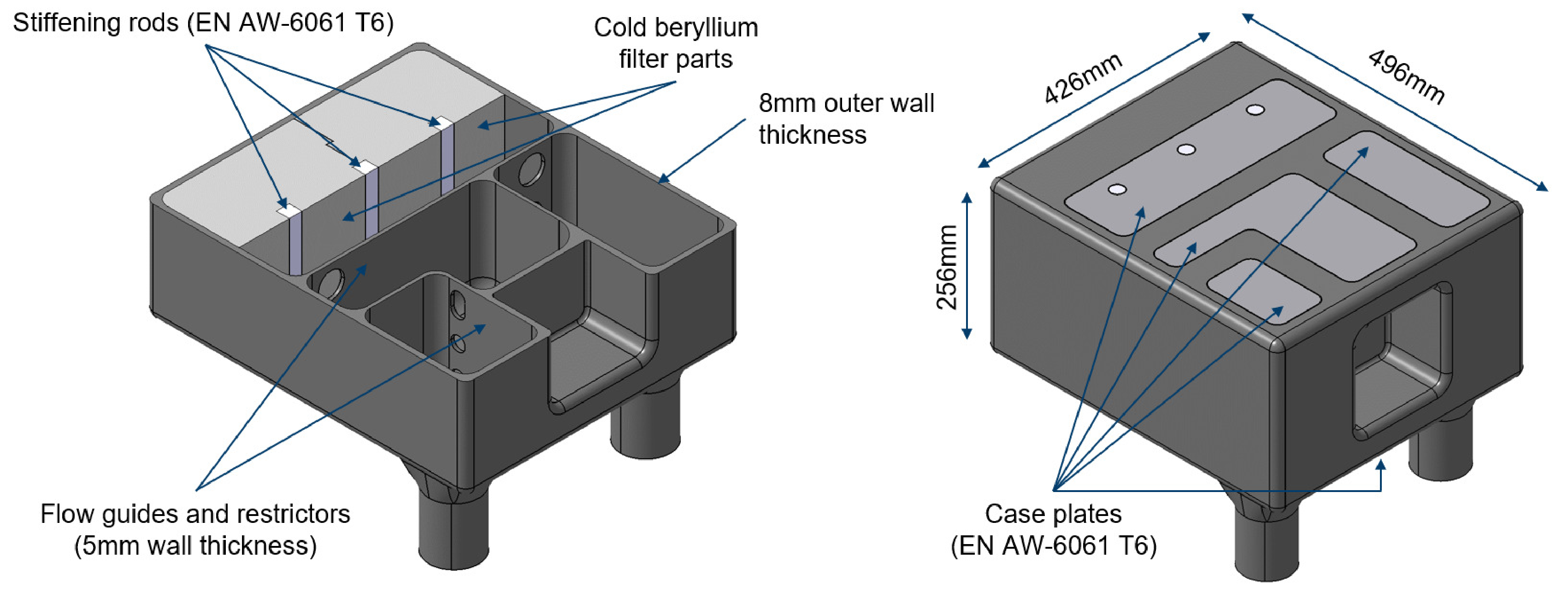}
\caption{Detailed view of the LD$_\text{2}$ moderator structure within the moderator plug.}
\label{fig:cn_1-9}
\end{center}
\end{figure}

A cold beryllium filter is embdedded within the \ce{LD_2} moderator on the NNBAR side, adjacent window B. This filter element is divided into two pieces for installation purposes. Three stiffening rods are also needed to support the structural integrity of the vessel. The flow guides and restrictors ensure a defined flow of the liquid deuterium from the inlet to the outlet. The whole moderator is milled and eroded from a single piece of EN AW-6061 T6 aluminum. The openings on the top and the bottom side are necessary for inserting the cold beryllium filter parts and for manufacturing reasons. The box dimensions of the moderator vessel with a wall thickness of \SI{8}{mm} are shown in \cref{fig:cn_1-9}. 

\FloatBarrier

\begin{figure}[hbt!]
\begin{center}
\includegraphics[width=0.9\textwidth]{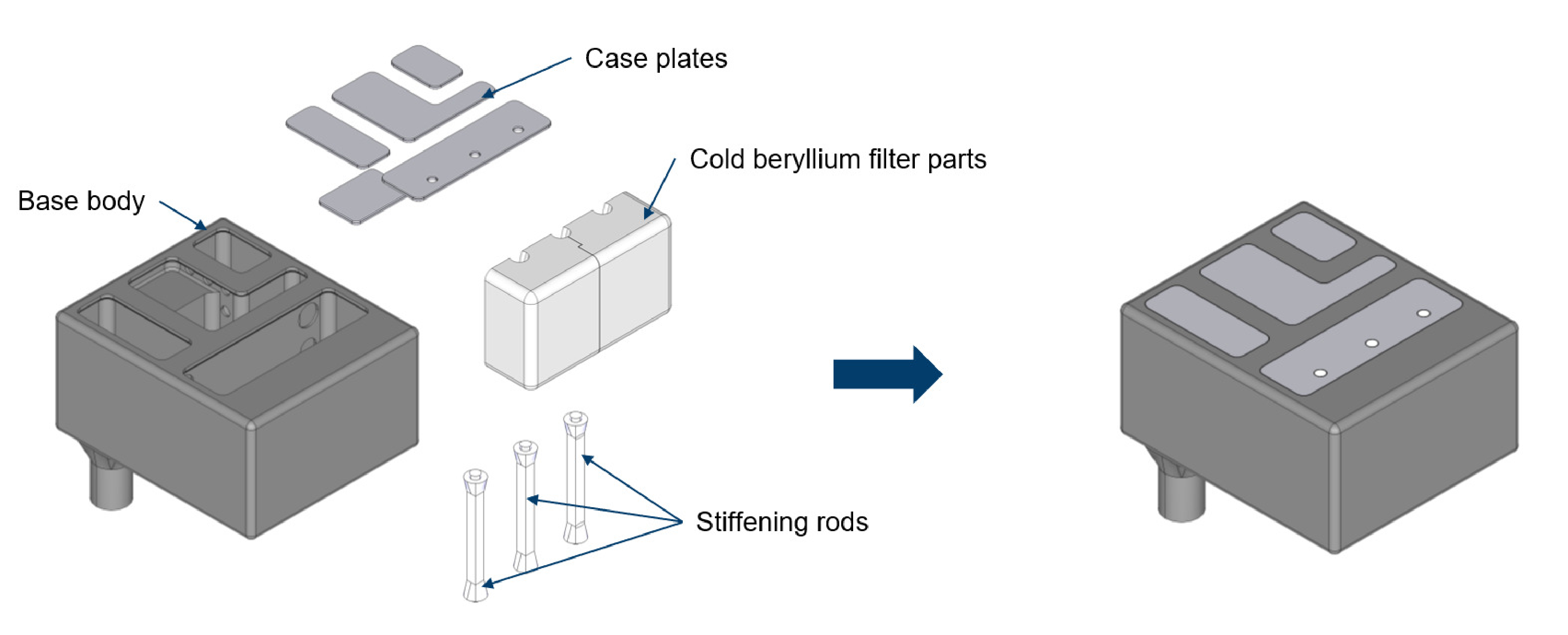}
\caption{Assembly of the LD$_\text{2}$ moderator vessel – parts overview.}
\label{fig:cn_1-10}
\end{center}
\end{figure}

\begin{figure}[hbt!]
\begin{center}
\includegraphics[width=0.9\textwidth]{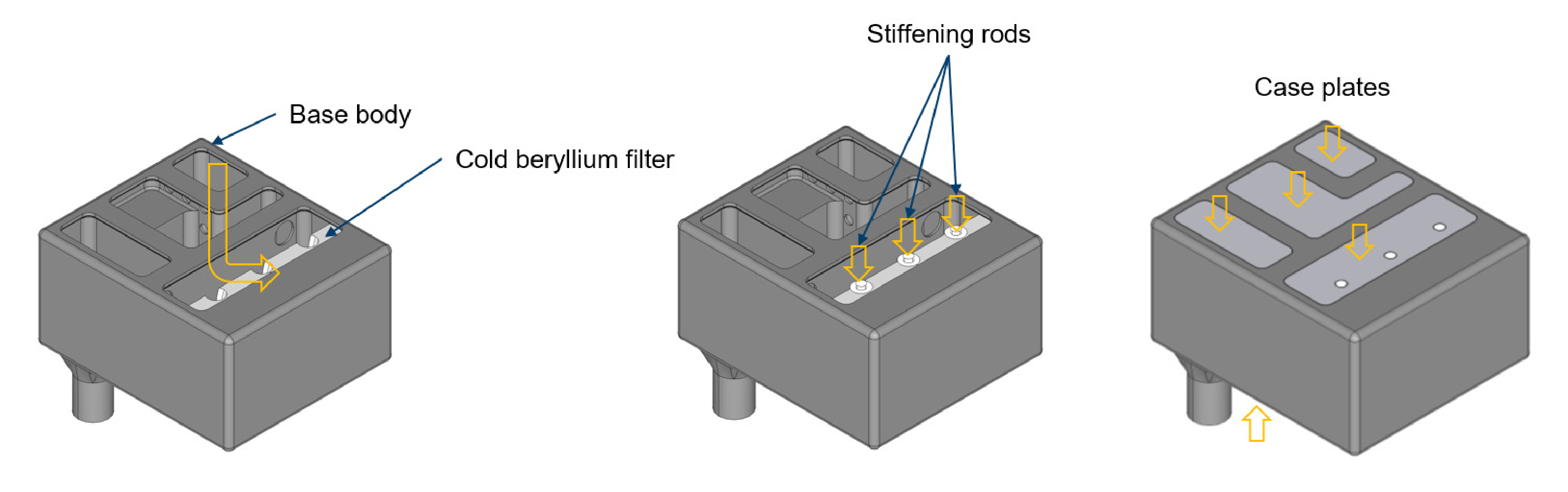}
\caption{Assembly of the LD$_\text{2}$ moderator vessel – assembly sequence.}
\label{fig:cn_1-11}
\end{center}
\end{figure}

The assembly sequence of the LD$_\text{2}$ moderator is shown in \cref{fig:cn_1-10} and \cref{fig:cn_1-11}. First, the parts of the cold beryllium filter are inserted through the top cut-out into the base body of the moderator. Then the 3 stiffening rods are put into the base body and into their grooves in the cold beryllium filter parts, ensuring the correct positioning. Afterwards, the moderator vessel is closed with the case plates.

\begin{figure}[hbt!]
\begin{center}
\includegraphics[width=0.9\textwidth]{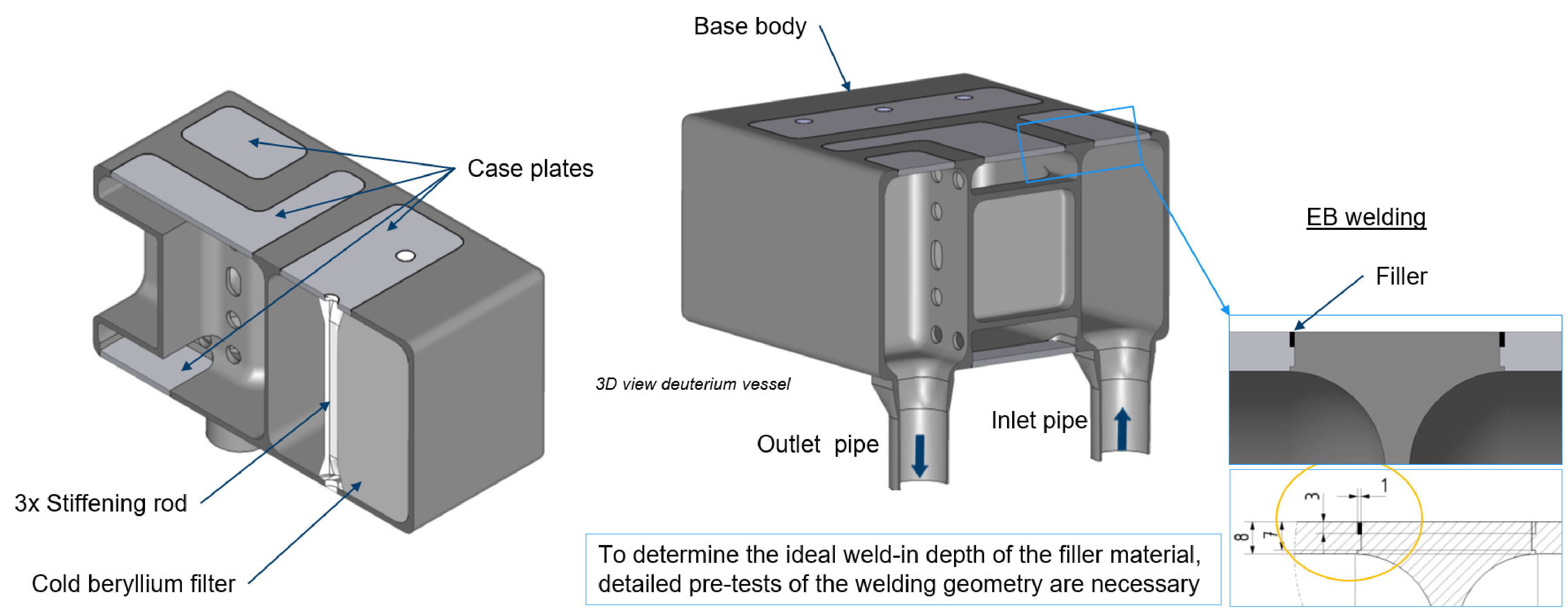}
\caption{EB-welding of the LD$_\text{2}$ moderator vessel.}
\label{fig:cn_1-12}
\end{center}
\end{figure}

One challenge when manufacturing the inner LD$_\text{2}$ moderator vessel is the welding process of the case plates, because of the high structural loads that result from the interior operating pressure of the liquid deuterium of \SI{5}{bar} and the unusual box shape for a pressurized vessel.
\cref{fig:cn_1-12} shows a close-up of the welding area of one of the case plates. To determine the ideal weld-in depth of the filler material, that is needed to avoid heat cracks during the welding process of the EN AW-6061 T6 aluminum alloy, detailed pre-tests of the welding geometry will be necessary in the future before building of such a vessel can be realized.

\subsection{Structural mechanical calculations} \label{ch.1-5}
\begin{figure}[hbt!]
\begin{center}
\includegraphics[width=0.9\textwidth]{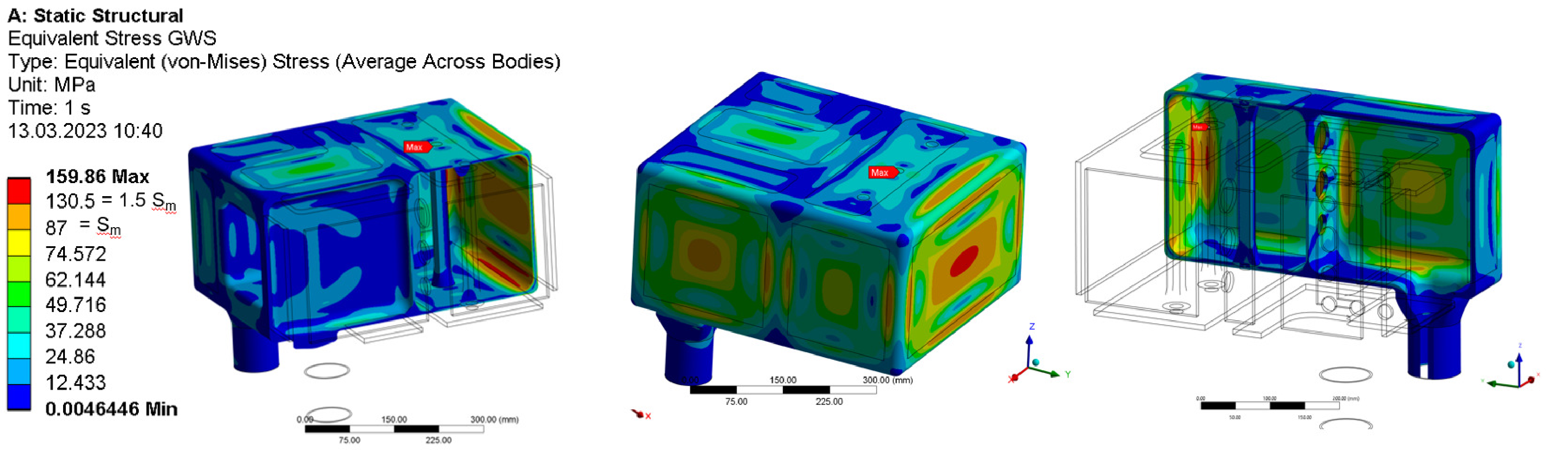}
\caption{Equivalent primary stress at design pressure.}
\label{fig:cn_1-13}
\end{center}
\end{figure}

Structural strength simulations with a design pressure of \SI{7}{bar} (operation pressure \SI{5}{bar}) following the design rules of RCC-MRx 2012 code were performed to analyze the static behavior of the LD$_\text{2}$ moderator, which basically behaves as a pressure vessel. \cref{fig:cn_1-13} illustrate that the large flat wall on the front-facing side -- where the cold beryllium filter is placed (not included in the simulation) -- the stress is particularly high due to the unusual box shape. In this area, the stress exceeds the limit of $1.5 \cdot S_{m}$ for EN AW-6061 T6, with $S_{m}$ = \SI{87}{MPa} and $1.5 \cdot S_{m}$ = \SI{130.5}{MPa}.

To further optimize the design and meeting the strength criteria regarding the allowable stress, optimizations on this large wall were performed. At first, the wall was shaped cylindrically, which reduced the bending stresses compared to a flat wall. Nevertheless, the sum of the local membrane and bending stress $\overline{P_{L}+P_{b}}$ on the fillet at the boundary of the rear wall still slightly exceeded the limit of $1.5 \cdot S_{m}$. 

\begin{figure}[hbt!]
\begin{center}
\includegraphics[width=0.8\textwidth]{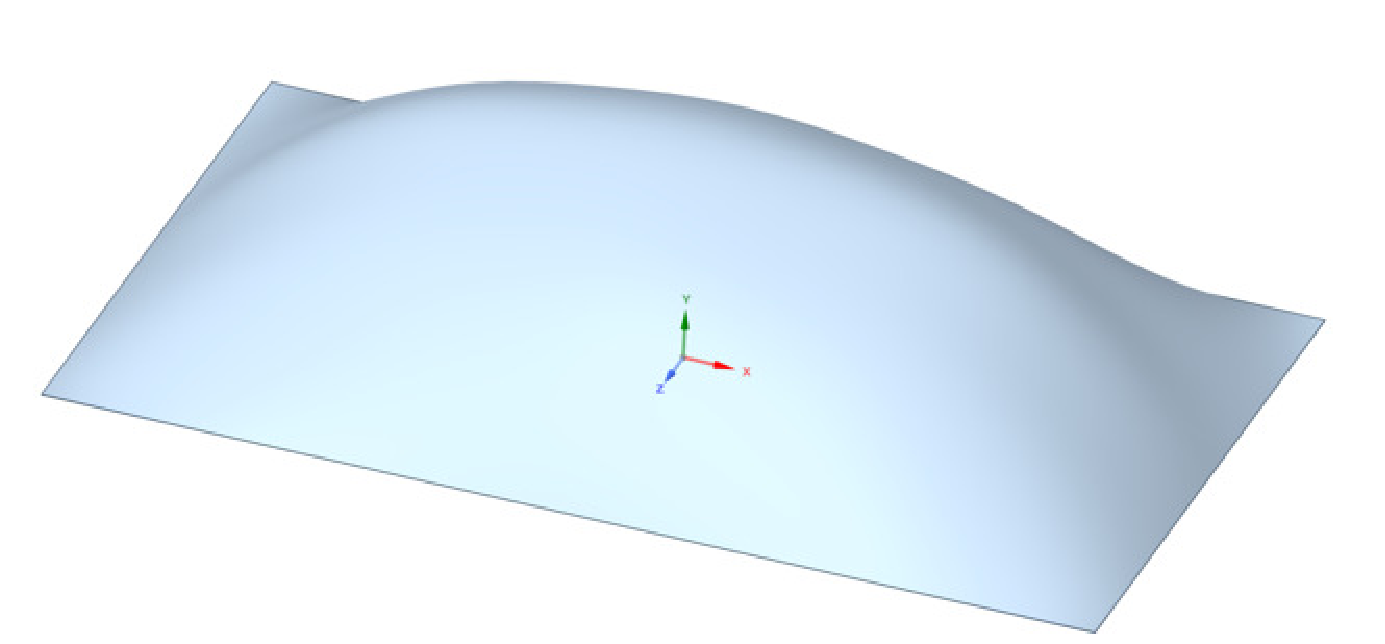}
\caption{Shape of membrane under pressure, magnified 10{$\times$} for visualization purpose}
\label{fig:cn_1-14}
\end{center}
\end{figure}

Finally, the large and originally flat surface was transformed into a perfectly plastic membrane, which is deformed by pressure to create a shape capable of bearing the load with reduced bending stress. This derived shape is used as a contour for the geometry of the rear wall. This free-formed, or ``membrane-shaped'' wall under pressure is shown in \cref{fig:cn_1-14}, shown with a 10{$\times$} magnification of the deformation for visualization. To limit the extent of the free-forming effect, the membrane shape was selected with a maximum deformation of \SI{10}{mm}.

\begin{figure}[hbt!]
\begin{center}
\includegraphics[width=0.9\textwidth]{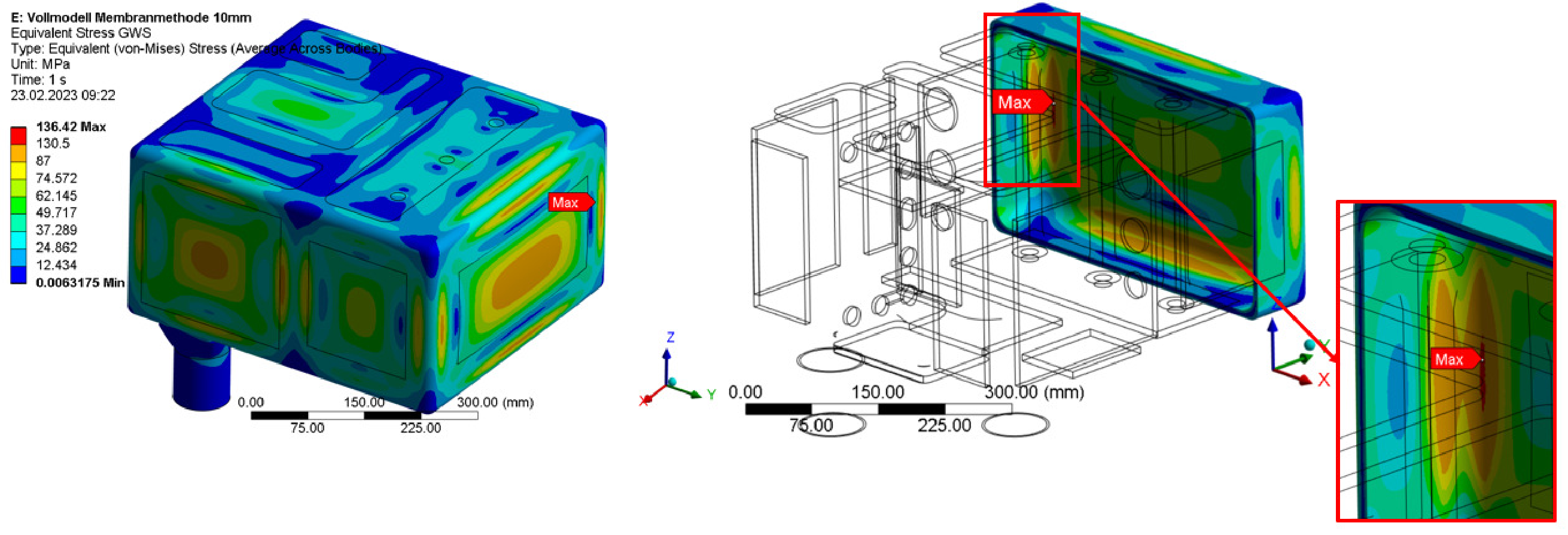}
\caption{Updated equivalent primary stress at design pressure.}
\label{fig:cn_1-15}
\end{center}
\end{figure}

\begin{figure}[hbt!]
\begin{center}
\includegraphics[width=0.9\textwidth]{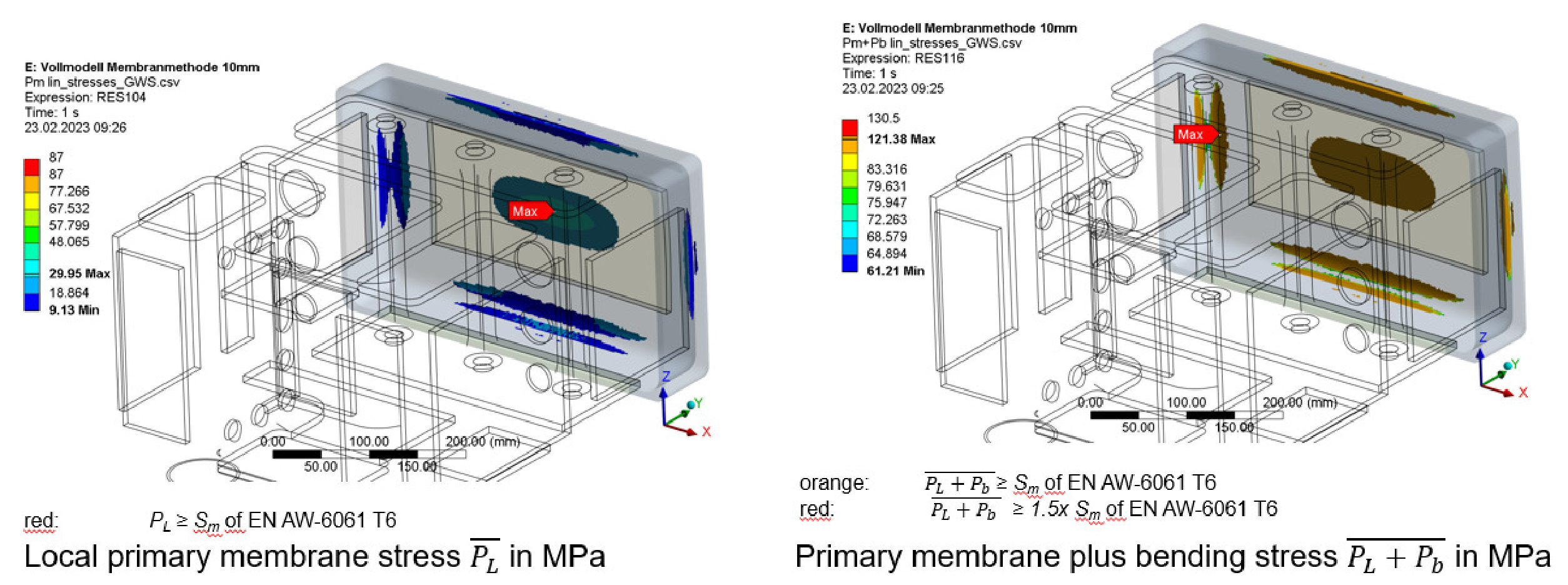}
\caption{Results for membrane stress on the NNBAR-side outer wall.}
\label{fig:cn_1-16}
\end{center}
\end{figure}

\cref{fig:cn_1-15} shows the updated equivalent primary stresses at a design pressure of \SI{7}{bar}. Here, it is indicated that the stress only exceeds the maximum of \SI{130.5}{MPa} by \SI{0.92}{MPa} at a single spot in a rounded corner. However, the linearized stresses at surfaces with $\sigma_{eqv}\geq S_{m}$ shown in \cref{fig:cn_1-16} demonstrate that the local primary membrane stress (left) and the primary membrane plus bending stress (right) completely satisfy the allowed strength criteria of $\overline{P_{m}}\leq\overline{P_{L}}\leq S_{m}(\Theta_{m})$ = \SI{87}{MPa} and $\overline{P_{L}+P_{b}}\leq 1.5\cdot S_{m}(\Theta_{m})$ = \SI{130.5}{MPa}.

\FloatBarrier

In summary, a free-shaped rear wall significantly reduces the sum of local membrane and bending stress compared to a flat rear wall or a cylindrical rear wall with identical offset of \SI{10}{mm}. The sum of local membrane and bending stress $\overline{P_{L}+P_{b}}$ on the fillet at the boundary of the rear wall is below the limit of $1.5 \cdot S_{m}$. With this design modification the strength criteria are met.

\subsection{Analysis of weld seams} \label{ch:1-5-1}
In principle, all moderator vessels are welded. Depending on the number of welds, the distance to each other and the welding process itself (TIG, Laser, Electron-Beam, etc.) the entire vessel can lose the additional strength gained by the selected tempered material. The TIG welding process requires a pre-heating of the vessel, which will often damage the previous heat treatments that have been made to achieve the high material strength, such as T6. This can be avoided to a large degree by using electron beam welding (with no pre-heating necessary) and a carefully planned welding design. 

Nevertheless, reduced strength must be expected in the area around a weld. Thus, by putting the welds in a low stress area where possible, it is still possible to benefit from the T6 conditions. On the other hand, electron beam (EB) welding requires a very high manufacturing tolerance.

Another issue with using EN AW-6061 T6 is the need of a welding filler to avoid heat crack formation. These heat cracks occur during the rapid solidification of the melt (thanks to the high thermal conductivity of aluminum) and can be mitigated by the alloy composition of the EN AW-6061 T6, particularly owing to the silicon and magnesium content. Therefore, a good choice of filler would have, for example, a relatively high silicon content, which is offered by EN AW-4047A with 12\% silicon. Increasing the content of magnesium is not helpful in this case since it evaporates when the aluminum melt is produced.

\begin{figure}[hbt!]
\begin{center}
\includegraphics[width=0.7\textwidth]{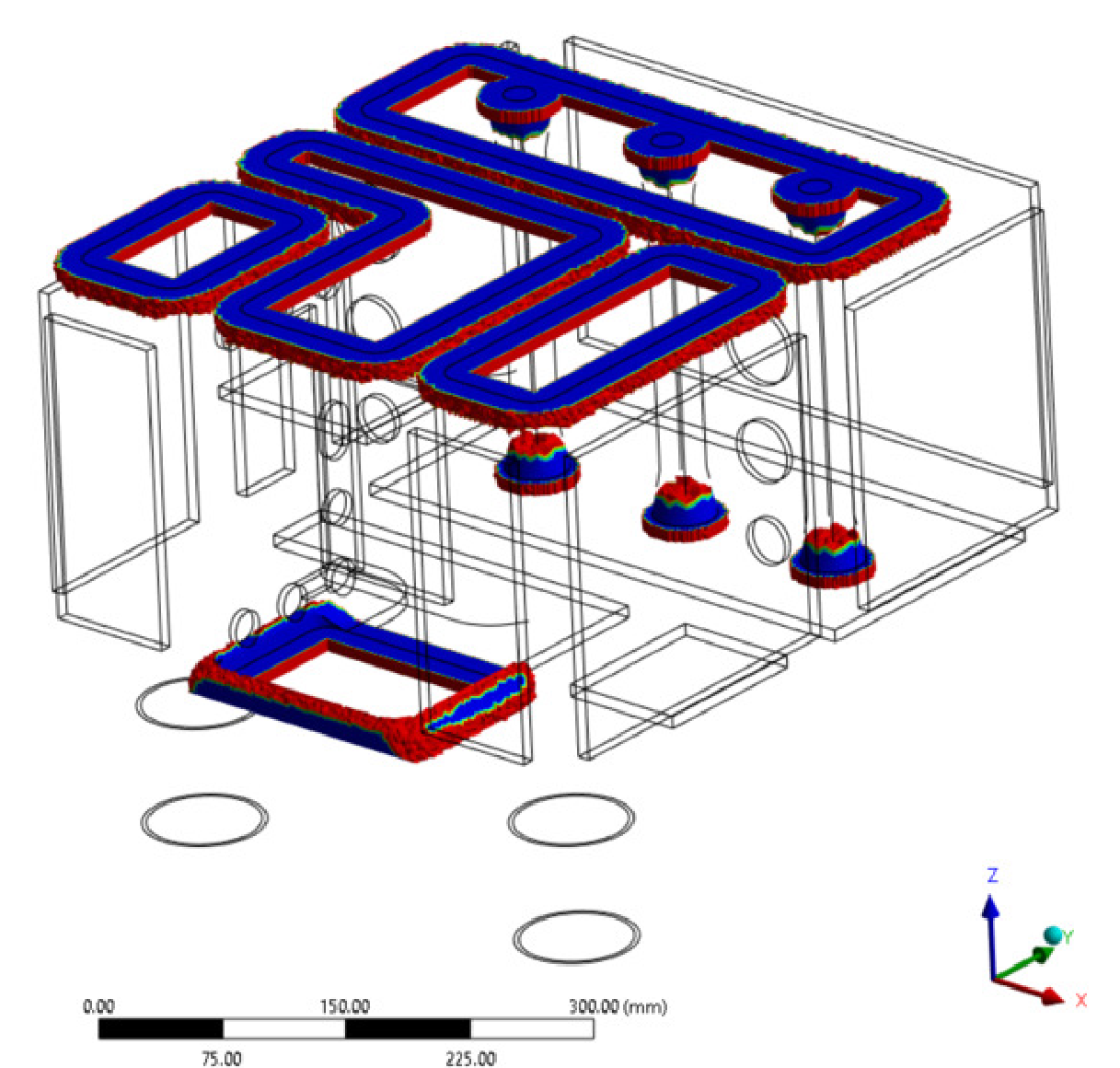}
\caption{Heat affected zones of the LD$_\text{2}$ moderator.}
\label{fig:cn_1-17}
\end{center}
\end{figure}

The filler material needs to be set precisely into the welding area. Also, the weld-in depth is a very sensitive value. For EB welding, it is ideal that there are no gaps between the welded parts and the filler material, because even small gaps $>$\,\SI{0.1}{mm} will result in irregularities in the weld seam and a faulty welding process. This is very challenging for the manufacturing process of the welded parts in terms of accuracy and the manufacturing strategy. The loss of the T6 condition of the aluminum alloy and its influence on the strength of the weld seams and the heat affected zone (HAZ) around them was investigated by simulation.
For this purpose, all material within \SI{15}{mm} of weld seams is assumed to be heat affected and therefore loses the T6 condition, with a significant corresponding loss in strength values which must be considered. This was approximated by conservatively upscaling the observations from previously tested components with lower wall thicknesses. An overview of the heat affected zones of the LD$_\text{2}$ moderator is given in \cref{fig:cn_1-17}.

\FloatBarrier

\begin{figure}[hbt!]
\begin{center}
\includegraphics[width=0.9\textwidth]{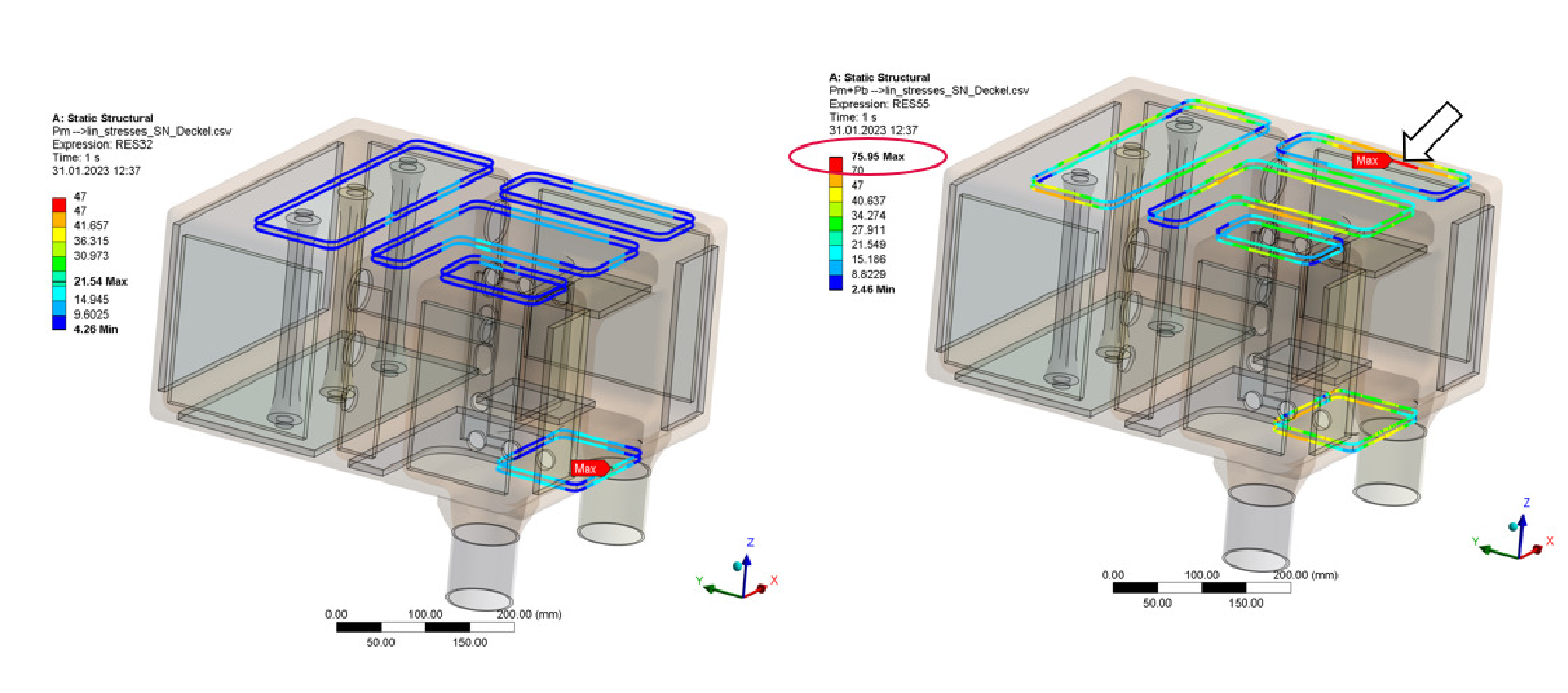}
\caption{Linearized stresses at the weld seams of cover plates.}
\label{fig:cn_1-18}
\end{center}
\end{figure}

The structural strength simulations with a static design pressure of \SI{7}{bar} following the design rules of RCC-MRx 2012 code showed that the linearized stresses at the weld seams of the cover plates did satisfy the strength criteria of $\overline{P_{m}}\leq \overline{P_{L}}\leq S_{m}(\Theta_{m})$ = \SI{47}{MPa} at all locations, but the criteria of $\overline{P_{L}+P_{b}}\leq 1.5 \cdot S_{m}(\Theta_{m})$ = \SI{70.5}{MPa} was not satisfied for one weld of one cover plate. These results are visualized in \cref{fig:cn_1-18}.

\begin{figure}[hbt!]
\begin{center}
\includegraphics[width=0.9\textwidth]{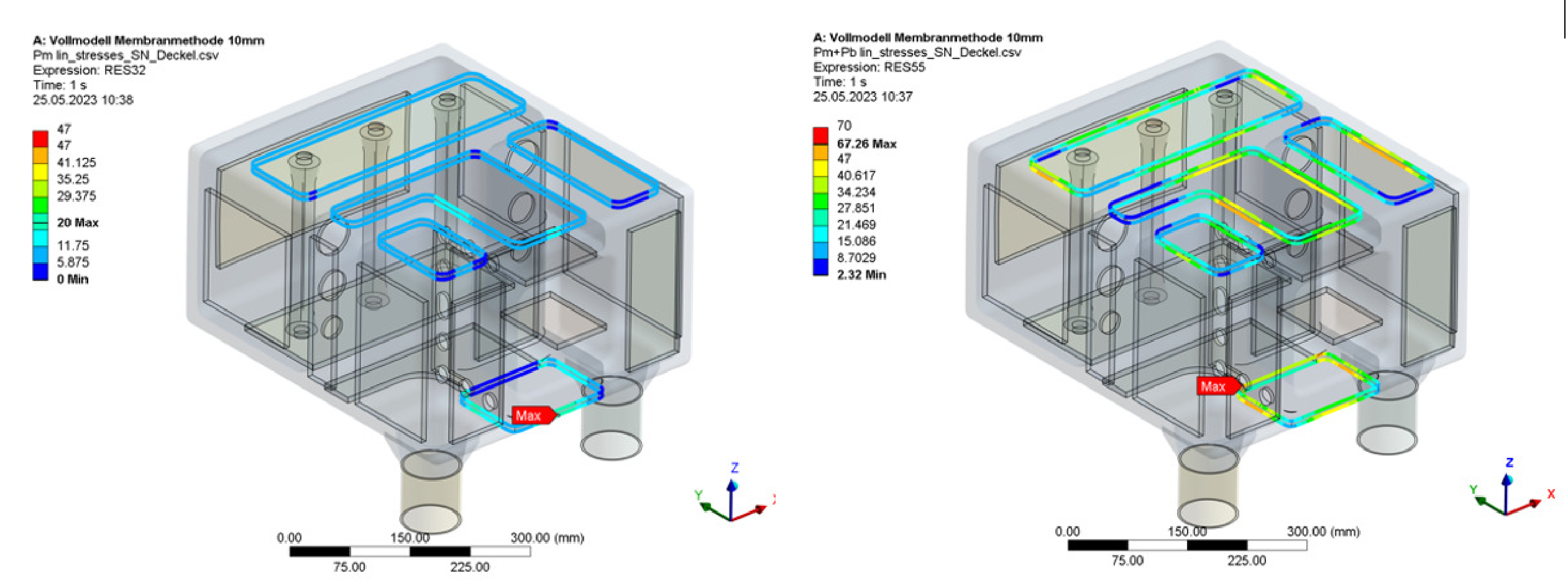}
\caption{Updated linearized stresses at weld seams of cover plates.}
\label{fig:cn_1-19}
\end{center}
\end{figure}

A further simulation was performed with the same boundary conditions, but an updated design model. In this case the position of the weld seam of the former, unsatisfactory cover plate was moved 6\,mm toward a less stressed region of the vessel. \cref{fig:cn_1-19} shows that the criteria of $\overline{P_{m}}\leq\overline{P_{L}}\leq S_{m} (\Theta_{m})$ = \SI{47}{MPa} as well as the criteria of $\overline{P_{L}+P_{b}}\leq 1.5 \cdot S_{m} (\Theta_{m})$ = \SI{70.5}{MPa} are satisfied for the welds of all cover plates.

\begin{figure}[hbt!]
\begin{center}
\includegraphics[width=0.9\textwidth]{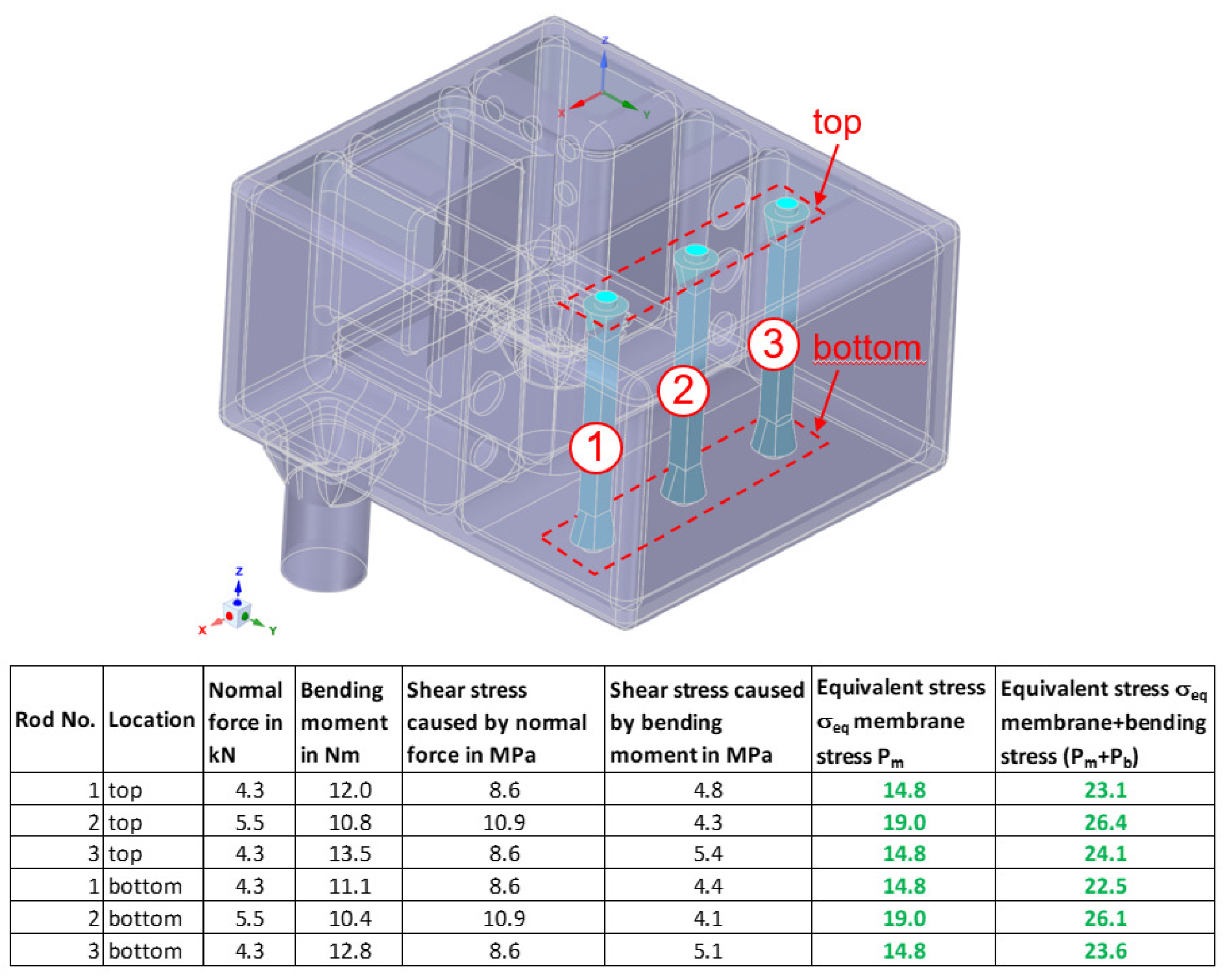}
\caption{Nominal stresses on the weld seams of the stiffening rods.}
\label{fig:cn_1-20}
\end{center}
\end{figure}

The weld seams at the three stiffening rods were also subject of the structural strength simulation. The results overview in \cref{fig:cn_1-20} shows that the nominal stresses at the weld seams of the stiffening rods all satisfy the strength criteria of $\overline{P_{m}}\leq S_{m}$ = \SI{27.5}{MPa} and $\overline{P_{L}+P_{b}}\leq 1.5\cdot S_{m}$ = \SI{41.3}{MPa}.

In conclusion, it can be stated that all requirements from a structural mechanical point of view are met, considering the RCC-MRx code. 

\FloatBarrier
\subsection{Thermohydraulic simulations} \label{ch:1-6}
The fluid parameters and cooling concepts have been verified by full-scale fluid dynamic simulations. Here, the fluid modeled is the subcooled moderating media \ce{LD_2} with an average inlet temperature of approximately \SI{21}{K}. The heat input by neutrons is pulsed, with a \SI{14}{Hz} pulse frequency and \SI{2.86}{ms} pulse length based on the ESS proton beam parameters. The allowed temperature increase of deuterium by neutron heat is limited in order to avoid boiling and particularly film boiling. Therefore, the goal for the fluid dynamical simulation is that the fluid temperature does not exceed the boiling temperature at any time.

This goal could not be achieved for the \SI{5}{MW} load case with the given design and requirements. Further details on this issue and remedial measures are discussed in \cref{ch:1-9}. The simulations were thus carried out using the design shown at reduced beam power of \SI{2}{MW}, since ESS will run at this power in the initial operations phase. A corresponding total heat deposition of \SI{28.4}{kW} was therefore considered in the simulations, using distribution functions derived for the design of the liquid hydrogen moderator of ESS.

\begin{figure}[hbt!]
\begin{center}
\includegraphics[width=0.9\textwidth]{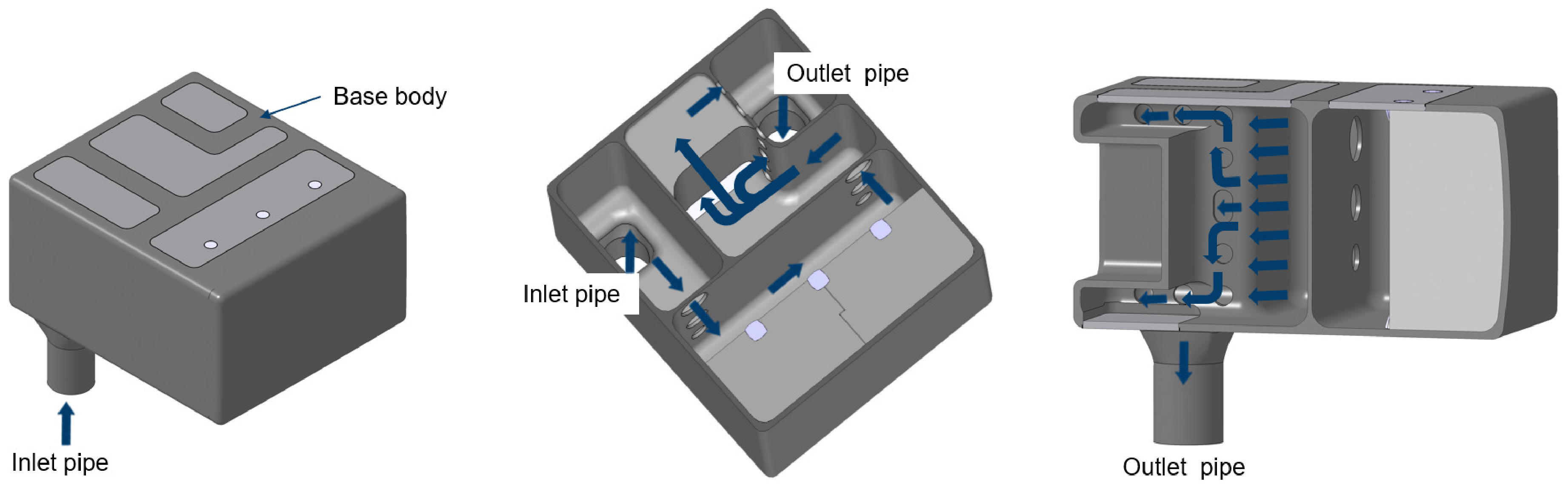}
\caption{Flow pattern of LD$_\text{2}$ through the moderator vessel, below and above the reentrant hole on window A.}
\label{fig:cn_1-21}
\end{center}
\end{figure}

The reentrant hole on the WP7 side (window A) of the moderator, opposite to the cold beryllium filter requires special consideration, since it is fully surrounded by LD$_\text{2}$ on the interior faces of the moderator. To achieve this, a complex flow pattern of the liquid deuterium through the moderator vessel is needed. The deuterium is guided from the inlet to the outlet via flow distributors and restrictors to ensure a homogeneous flow inside the vessel. In the area of the reentrant hole the flow is split, so that the deuterium can flow above and below it (see \cref{fig:cn_1-21}).

\begin{figure}[hbt!]
\begin{center}
\includegraphics[width=0.6\textwidth]{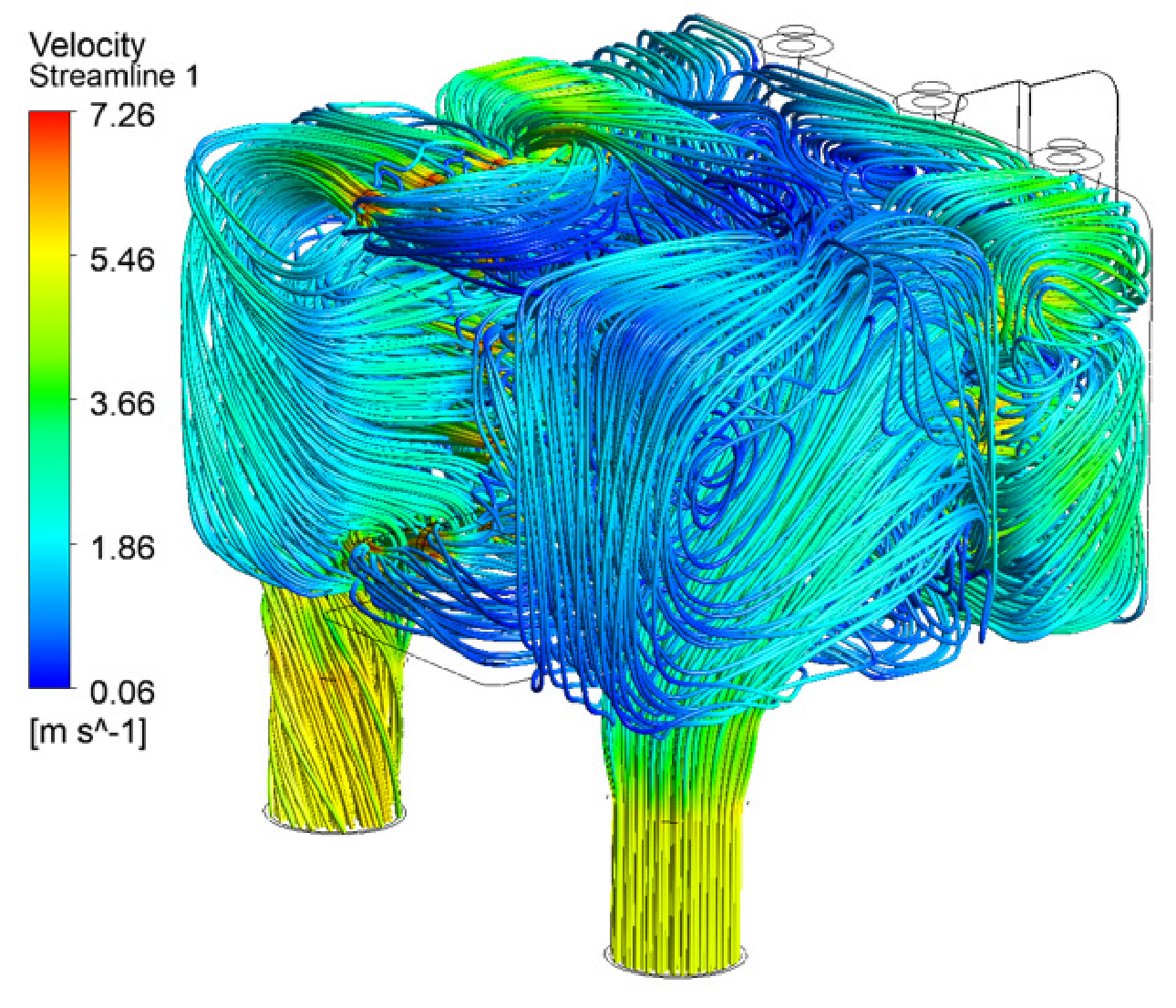}
\caption{Velocity streamlines of LD$_\text{2}$ inside the moderator vessel}
\label{fig:cn_1-22}
\end{center}
\end{figure}

\cref{fig:cn_1-22} shows the velocity streamlines of the liquid deuterium inside the moderator vessel that result from the CFX analysis~\cite{ansys_cfx}. Time-averaged results confirm a pressure drop of \SI{0.15}{bar} in the moderator and a maximum fluid velocity of \SI{7.26}{\m\per\s} at an inlet velocity of \SI{5.16}{\m\per\s}.

\FloatBarrier

\begin{figure}[hbt!]
\begin{center}
\includegraphics[width=0.9\textwidth]{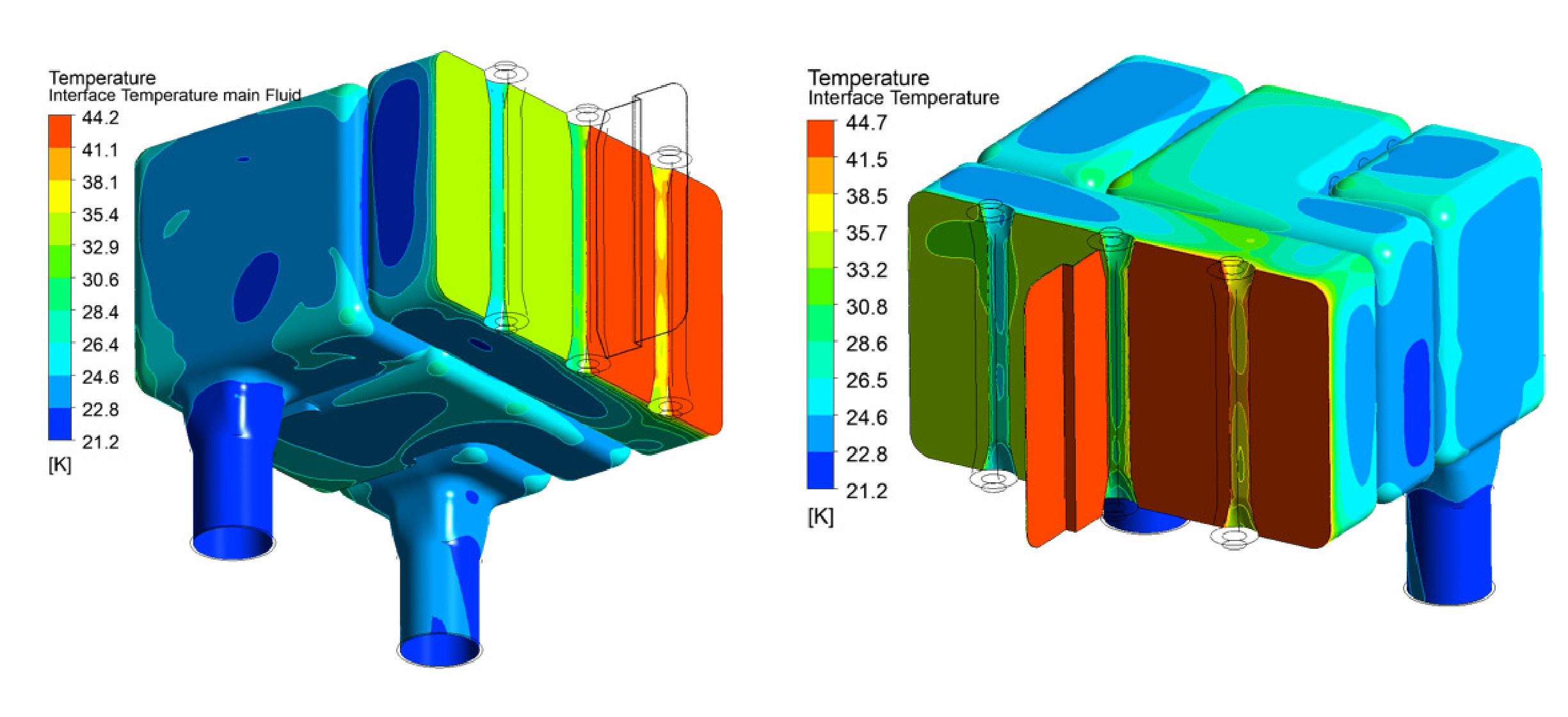}
\caption{Time-averaged fluid temperatures for the \ce{LD_2} moderator.}
\label{fig:cn_1-23}
\end{center}
\end{figure}

The temperature difference from the inlet to the outlet is \SI{1.4}{K} with an outlet temperature of \SI{22.4}{K} (time averaged). The boiling temperature of deuterium at a pressure of \SI{5}{bar} is \SI{30.6}{K}. This temperature is reached locally at the inner interface layer between the deuterium and the cold beryllium filter and in the small gap between the cold beryllium filter parts, as shown in \cref{fig:cn_1-23}, but the main flow is still sufficiently subcooled. Therefore, there is no risk of film boiling.

In summary, it can be said that the fluid mechanical requirements for the \SI{2}{MW} load case are met, if local boiling at the beryllium filter -- fluid inner interface is allowed. So, there are no concerns from an operational safety point of view.

\subsection{Design solution of the vacuum vessel} \label{ch:1-7}

\begin{figure}[hbt!]
\begin{center}
\includegraphics[width=0.9\textwidth]{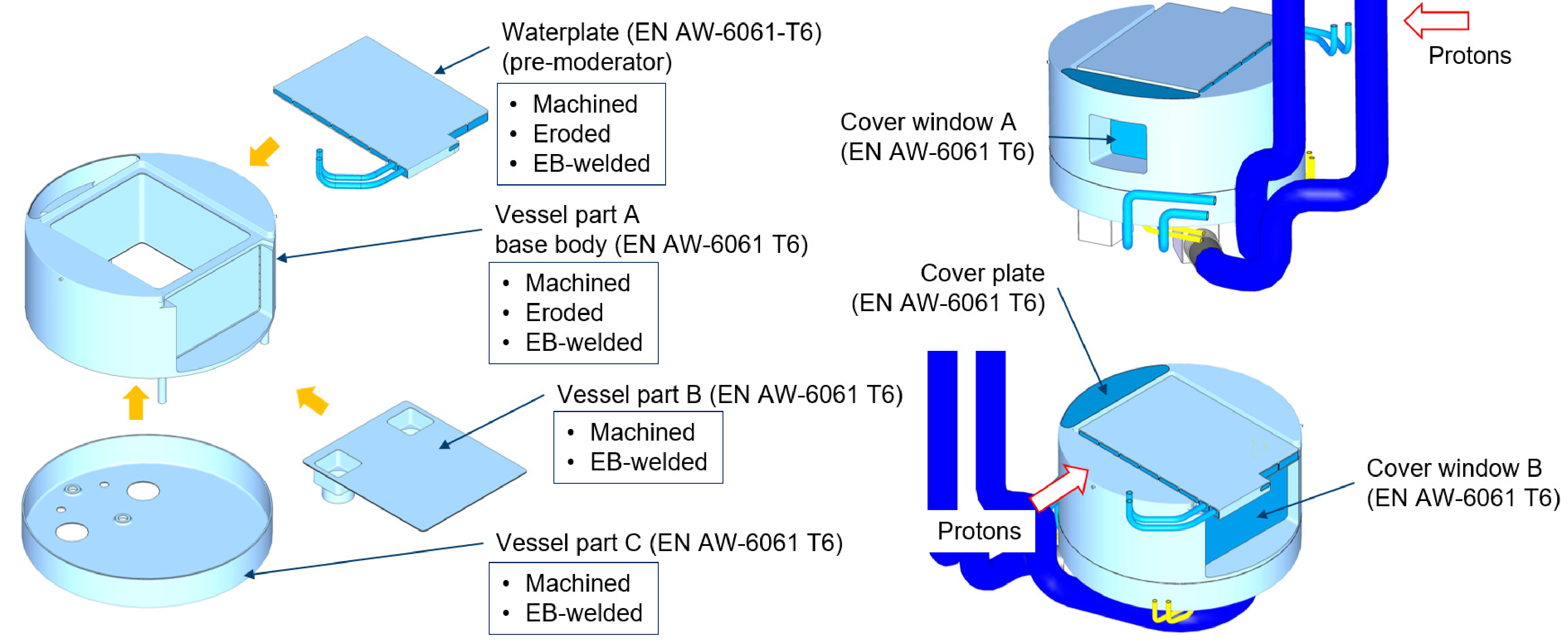}
\caption{Assembly of outer vessel for the lower moderator.}
\label{fig:cn_1-24}
\end{center}
\end{figure}

\FloatBarrier

The outer vessel shown in \cref{fig:cn_1-24} consists of multiple parts that are combined inside a vessel made of EN AW-6061 T6 aluminum alloy. This vessel and the involved reflector parts also enable the water cooling of the beryllium (Be) reflector. Between the Be-reflector and the inner LD$_\text{2}$ moderator there is an insulation vacuum gap of \SI{5}{mm} to minimize the heat transfer between the cold and warm parts.

\begin{figure}[hbt!]
\begin{center}
\includegraphics[width=0.9\textwidth]{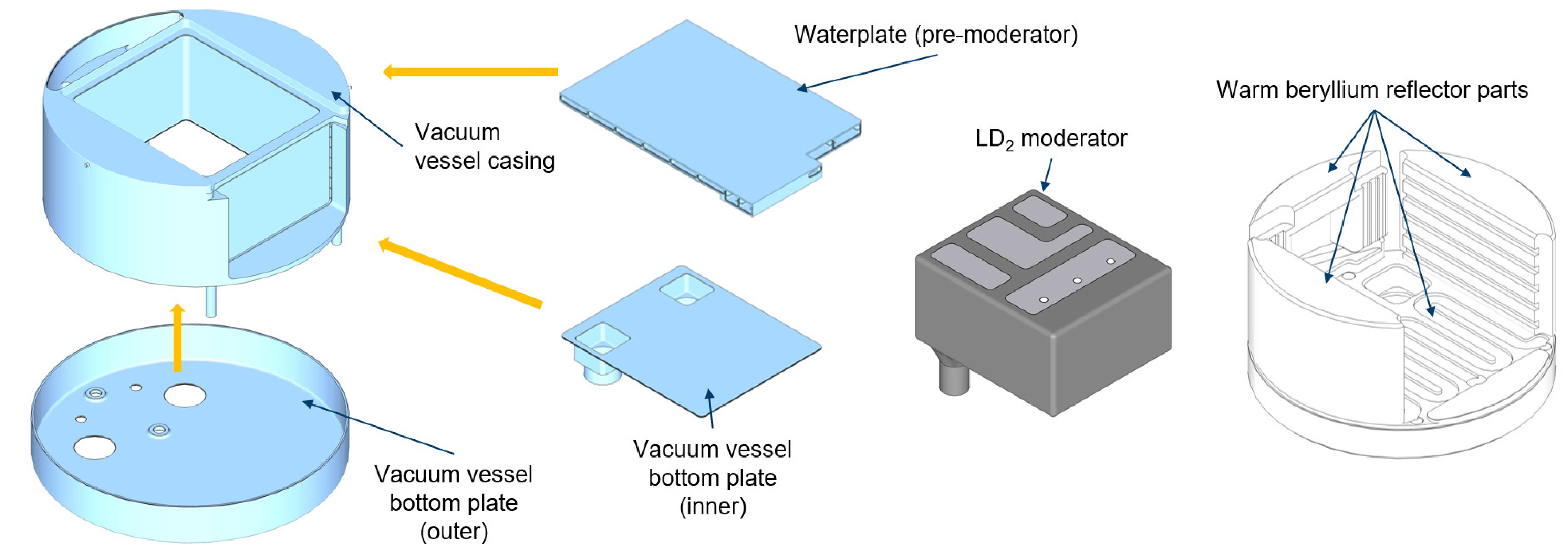}
\caption{Assembly of the lower moderator plug -- parts overview.}
\label{fig:cn_1-25}
\end{center}
\end{figure}

\cref{fig:cn_1-25} shows the different parts of the assembly, that consists of the center LD$_\text{2}$ moderator, the warm beryllium reflector parts, and the outer aluminum vessel.

\begin{figure}[hbt!]
\begin{center}
\includegraphics[width=0.9\textwidth]{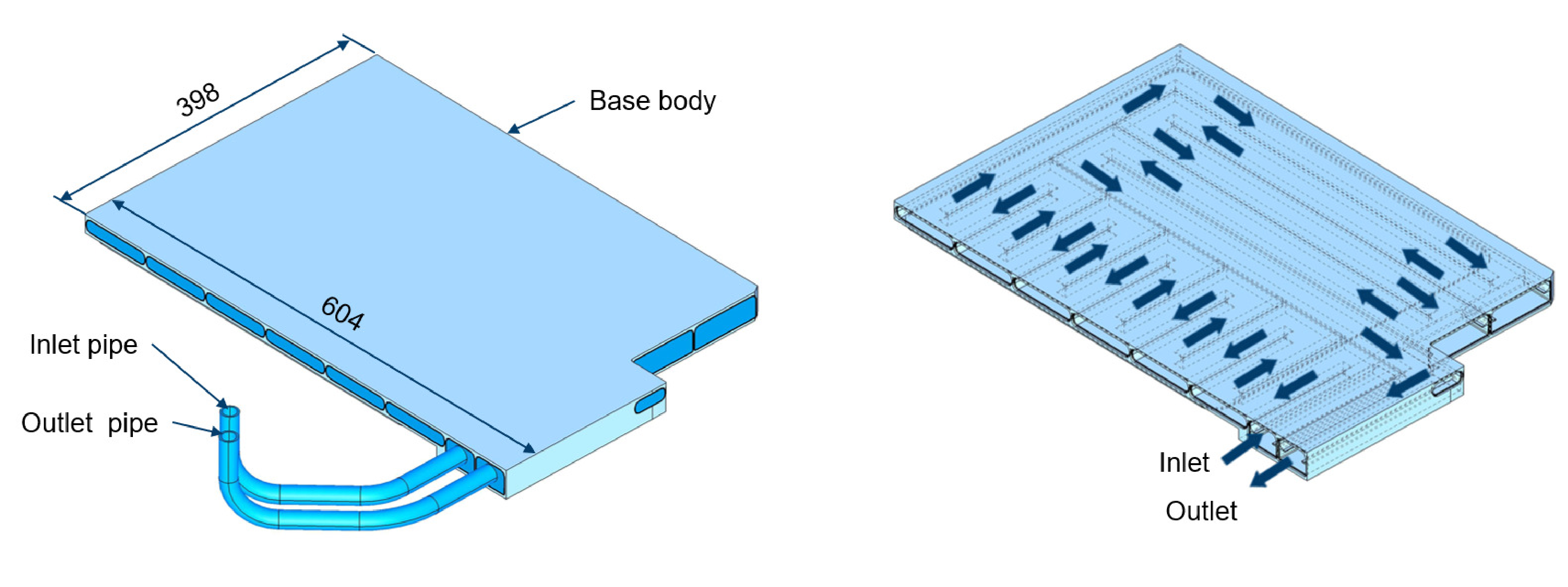}
\caption{Flow pattern of water through the premoderator.}
\label{fig:cn_1-26}
\end{center}
\end{figure}

The water disc on top of the moderator plug assembly works as a premoderator for the fast neutrons released by the tungsten target. \cref{fig:cn_1-26} shows the flow pattern of water through the premoderator, that is located on top of the lower moderator plug. 

\FloatBarrier

\begin{figure}[hbt!]
\begin{center}
\includegraphics[width=0.9\textwidth]{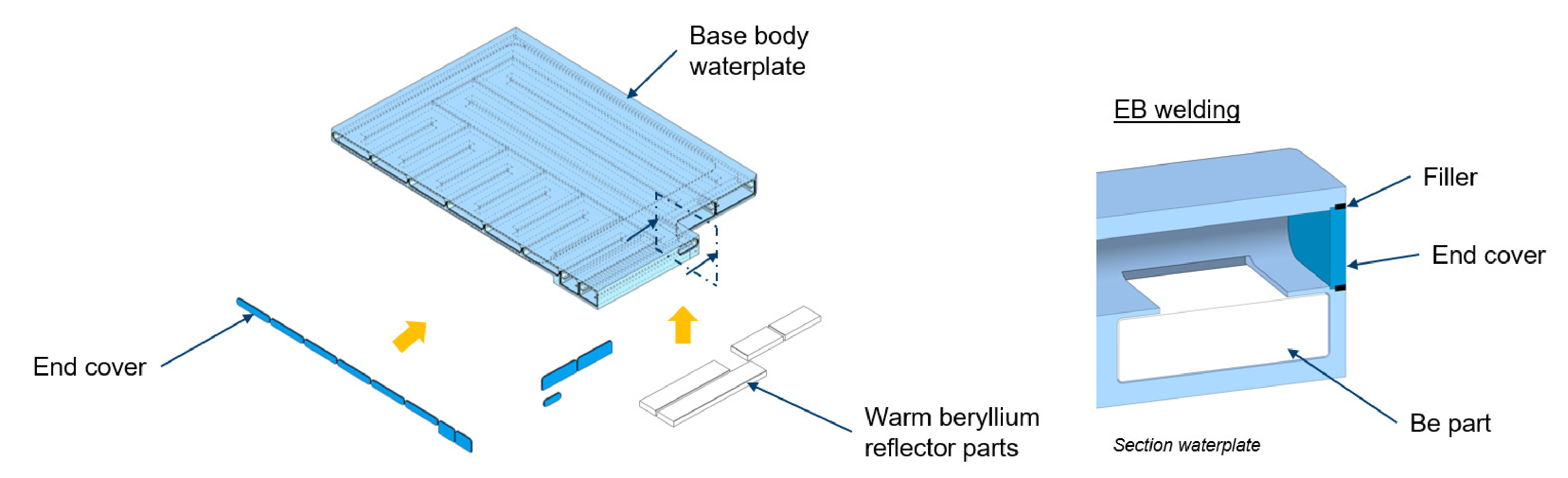}
\caption{Assembly and EB-welding detail of the premoderator.}
\label{fig:cn_1-27}
\end{center}
\end{figure}

It is assembled as shown in \cref{fig:cn_1-27}. Above the bigger neutron window B, also warm beryllium reflector parts are inserted and the milled and eroded water channels inside the water disc are closed via electron-beam welding.

\begin{figure}[hbt!]
\begin{center}
\includegraphics[width=0.9\textwidth]{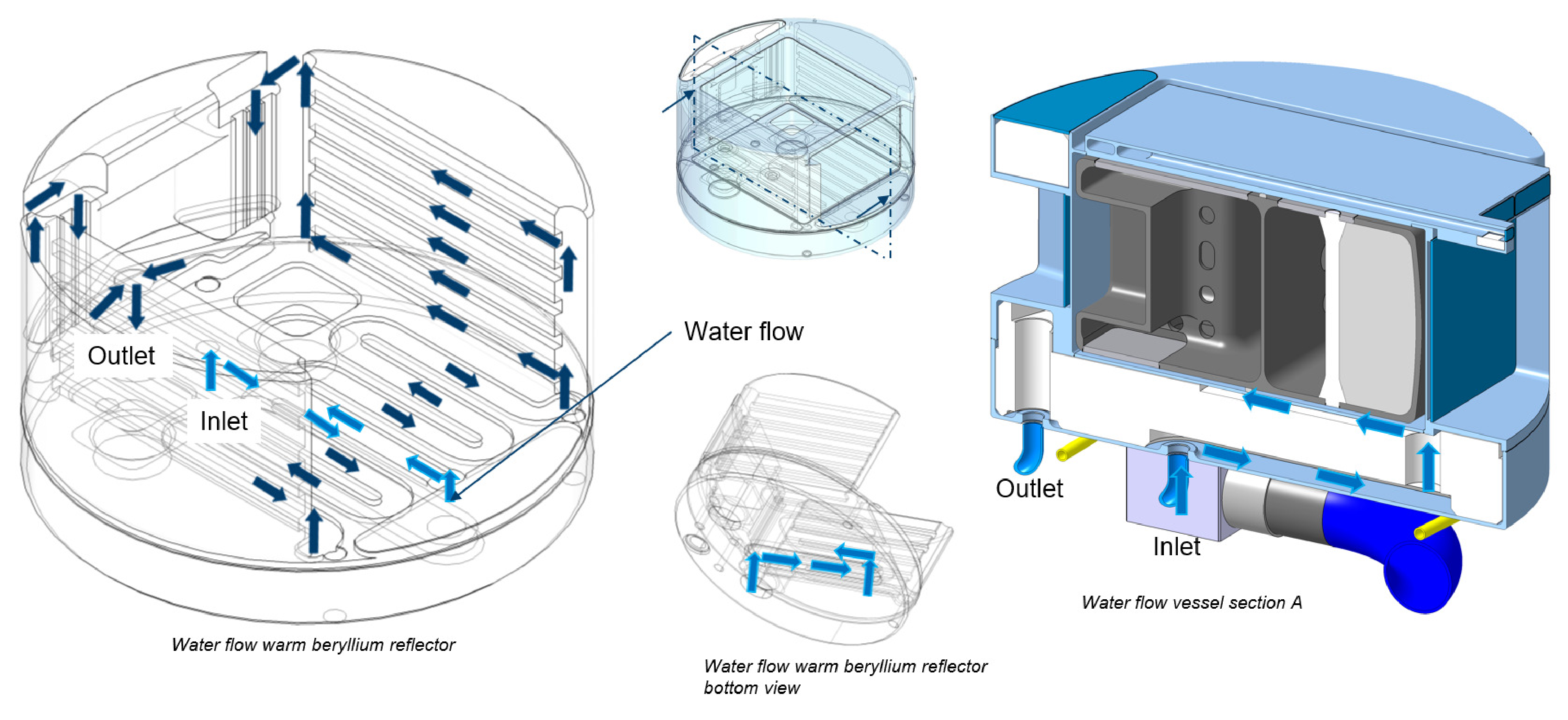}
\caption{Flow pattern of water through the warm beryllium reflector parts, section A.}
\label{fig:cn_1-28}
\end{center}
\end{figure}

\begin{figure}[hbt!]
\begin{center}
\includegraphics[width=0.9\textwidth]{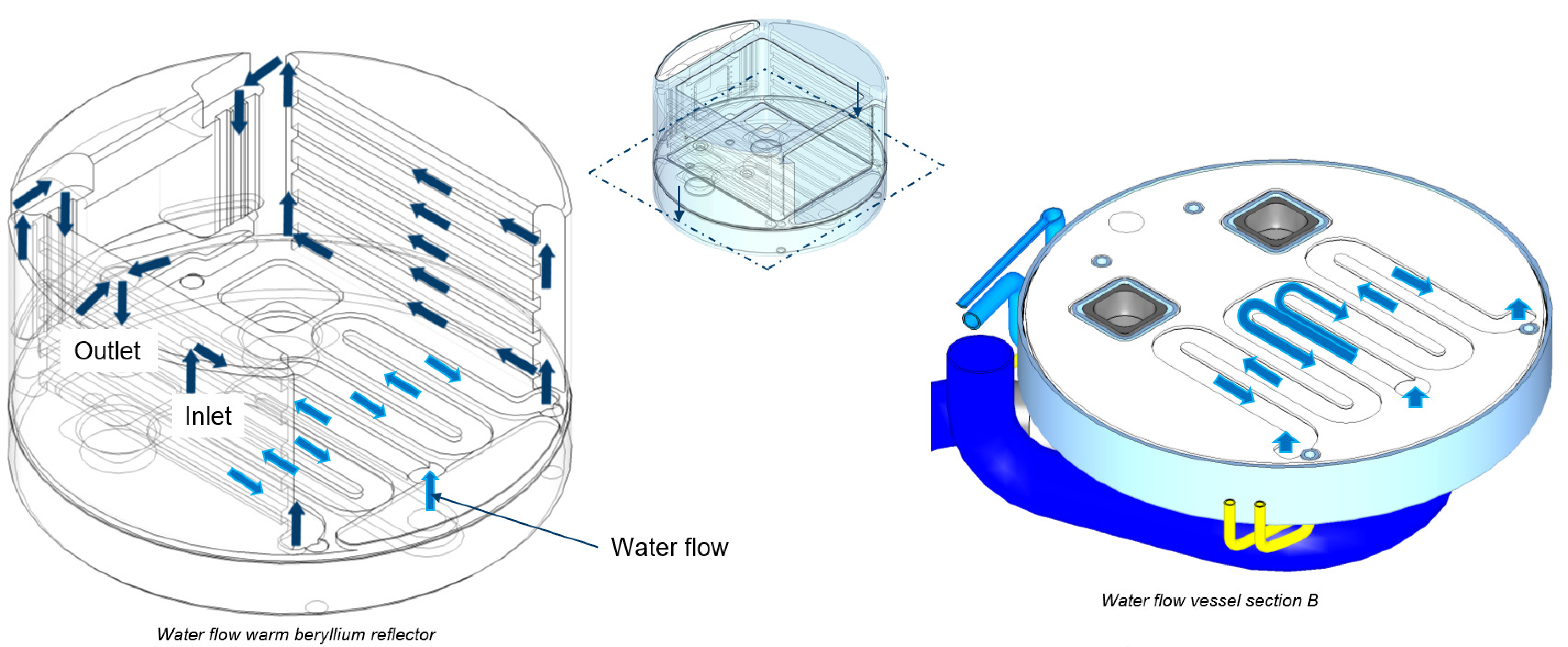}
\caption{Flow pattern of water through the warm beryllium reflector parts, section B.}
\label{fig:cn_1-29}
\end{center}
\end{figure}

\begin{figure}[hbt!]
\begin{center}
\includegraphics[width=0.9\textwidth]{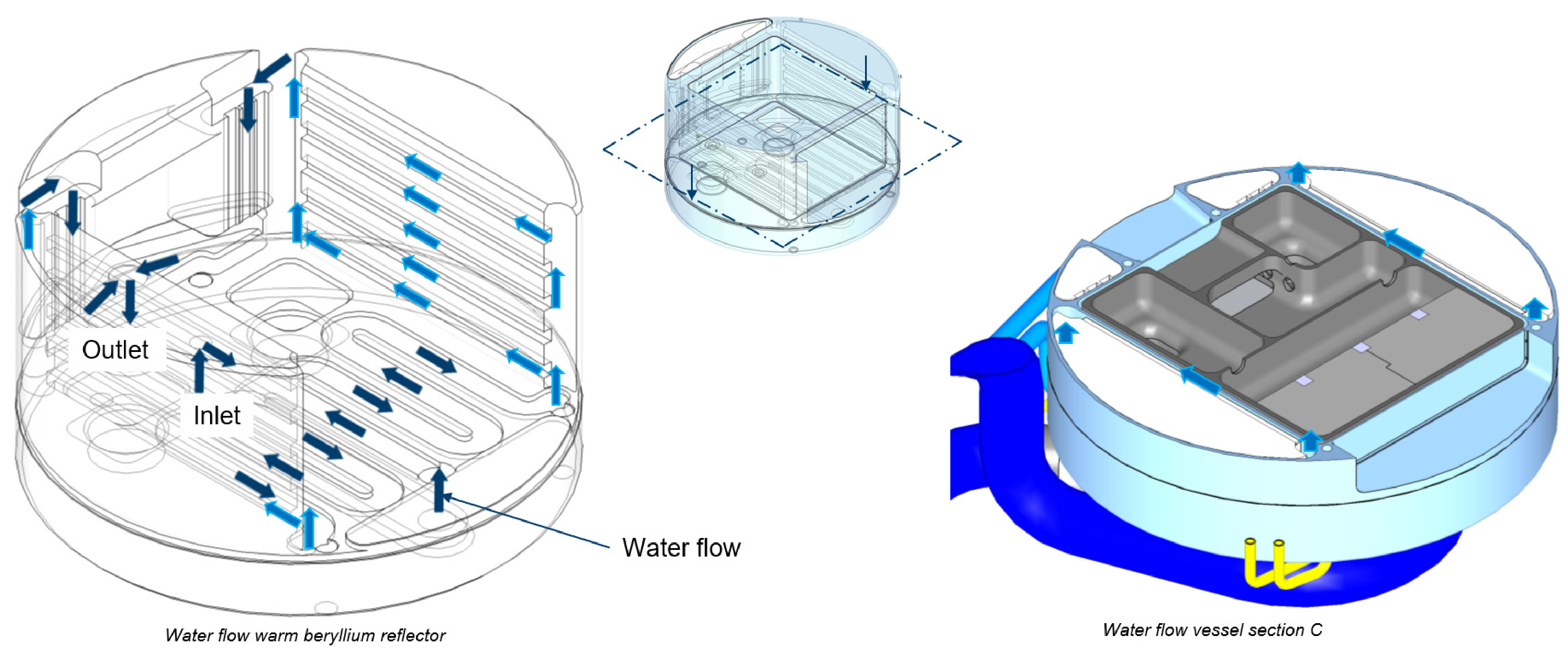}
\caption{Flow pattern of water through the warm beryllium reflector parts, section C.}
\label{fig:cn_1-30}
\end{center}
\end{figure}

\begin{figure}[hbt!]
\begin{center}
\includegraphics[width=0.9\textwidth]{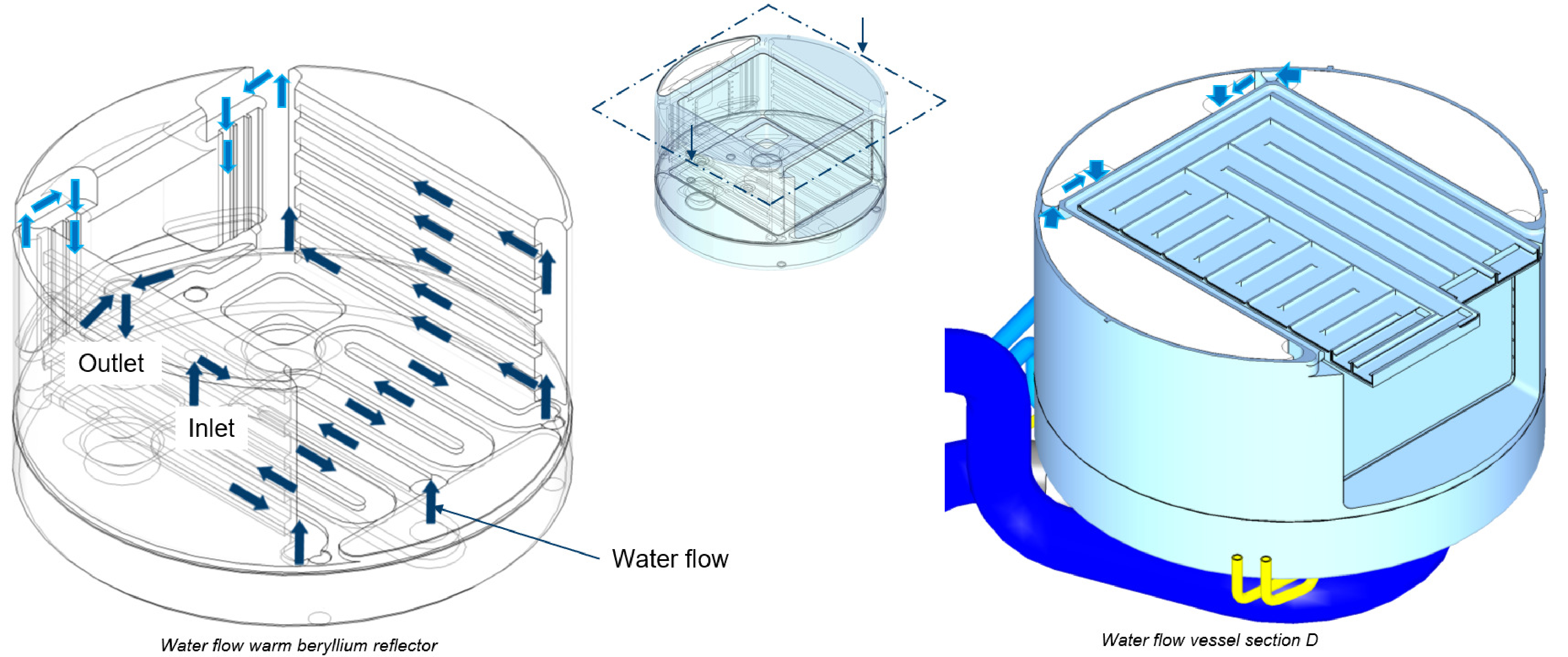}
\caption{Flow pattern of water through the warm beryllium reflector parts, section D.}
\label{fig:cn_1-31}
\end{center}
\end{figure}

\begin{figure}[hbt!]
\begin{center}
\includegraphics[width=0.9\textwidth]{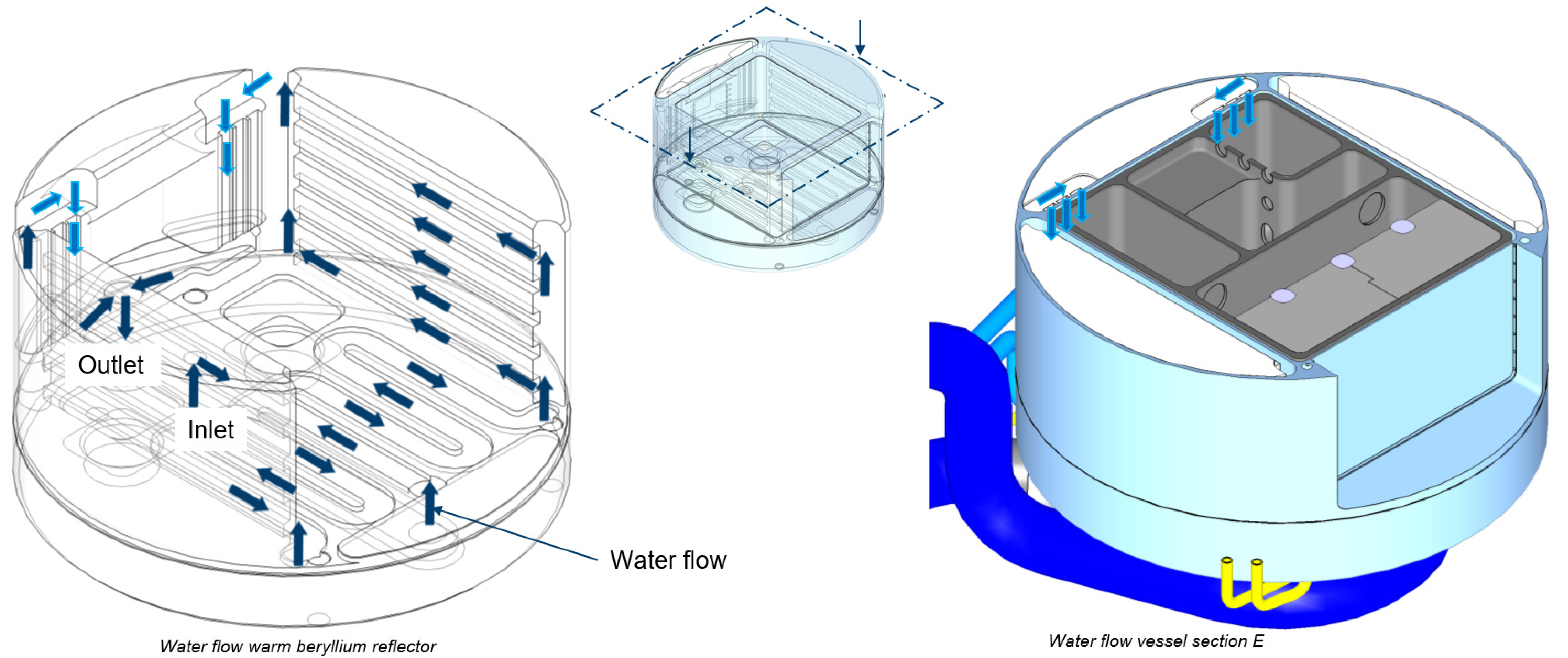}
\caption{Flow pattern of water through the warm beryllium reflector parts, section E.}
\label{fig:cn_1-32}
\end{center}
\end{figure}

\begin{figure}[hbt!]
\begin{center}
\includegraphics[width=0.9\textwidth]{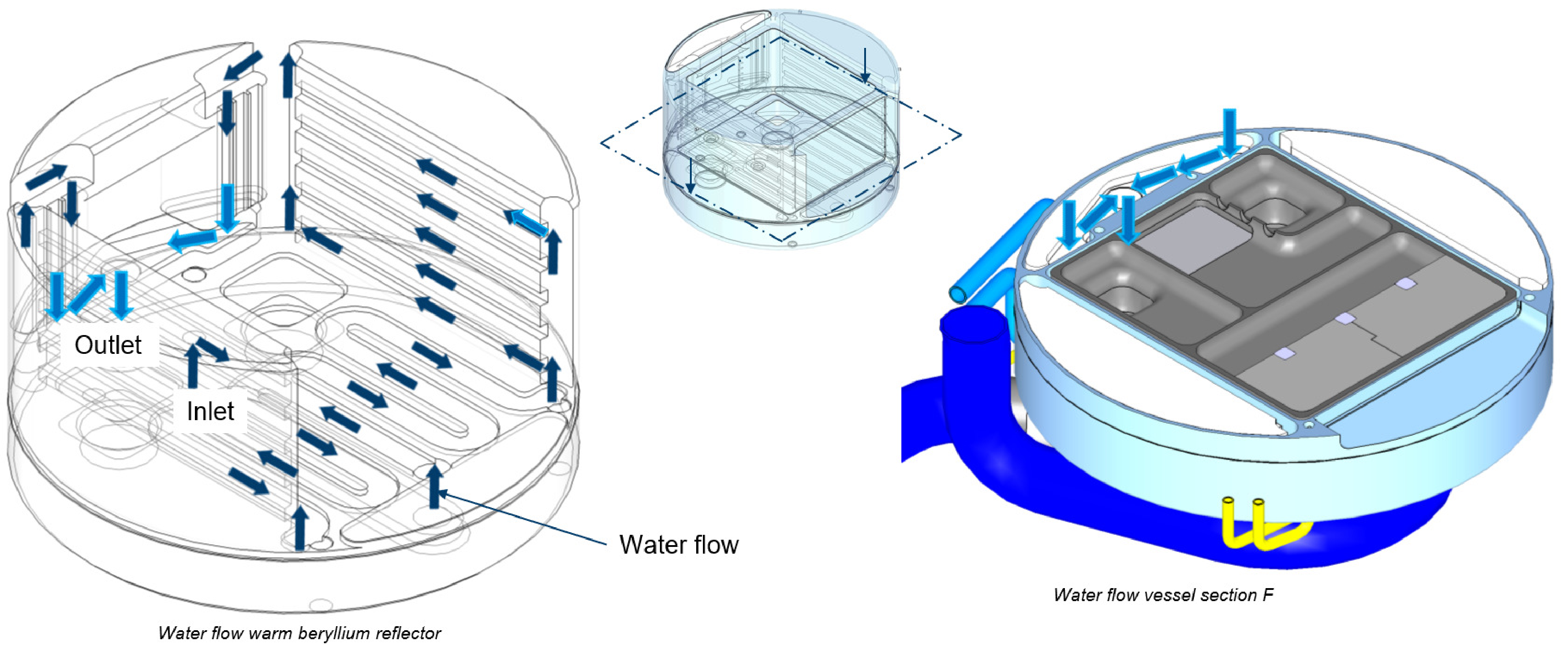}
\caption{Flow pattern of water through the warm beryllium reflector parts}
\label{fig:cn_1-33}
\end{center}
\end{figure}

\begin{figure}[hbt!]
\begin{center}
\includegraphics[width=0.9\textwidth]{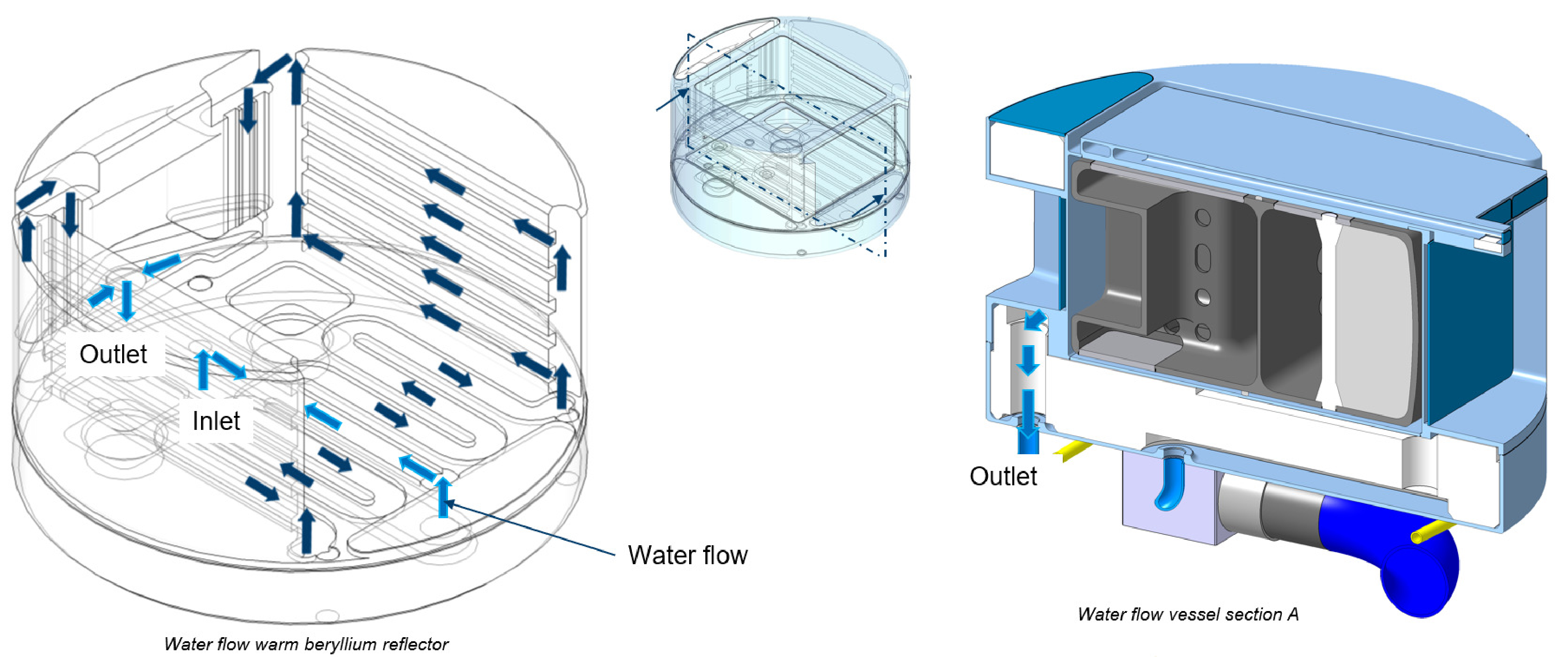}
\caption{Flow pattern of water through the warm beryllium reflector parts}
\label{fig:cn_1-34}
\end{center}
\end{figure}

\cref{fig:cn_1-28} to \cref{fig:cn_1-34} show the flow pattern of water through the warm beryllium reflector parts inside the outer vessel, shown in progressive vertical cross-sections. At first the water enters through the inlet on the bottom side of the outer vessel and is afterwards distributed to the bottom part of the warm beryllium reflector.

\FloatBarrier

In this bottom part, the flow is separated into two parallel flows that go upwards along each side of the NNBAR opening (\cref{fig:cn_1-29}).

\FloatBarrier

While flowing toward the top, the water is also distributed alongside the side pieces of the warm beryllium reflector, as shown in \cref{fig:cn_1-30}.

\FloatBarrier

When the smaller neutron window A opening is reached in \cref{fig:cn_1-31} and \cref{fig:cn_1-32}, the water flows downwards again.

\FloatBarrier

\FloatBarrier

Next, the water flow is guided below the smaller neutron window A opening, through the warm beryllium filter parts underneath it (\cref{fig:cn_1-33}).

\FloatBarrier

Finally, the flow is guided to the outlet pipe at the bottom side of the lower moderator plug, as shown in \cref{fig:cn_1-34}.

\FloatBarrier

\begin{figure}[hbt!]
\begin{center}
\includegraphics[width=0.9\textwidth]{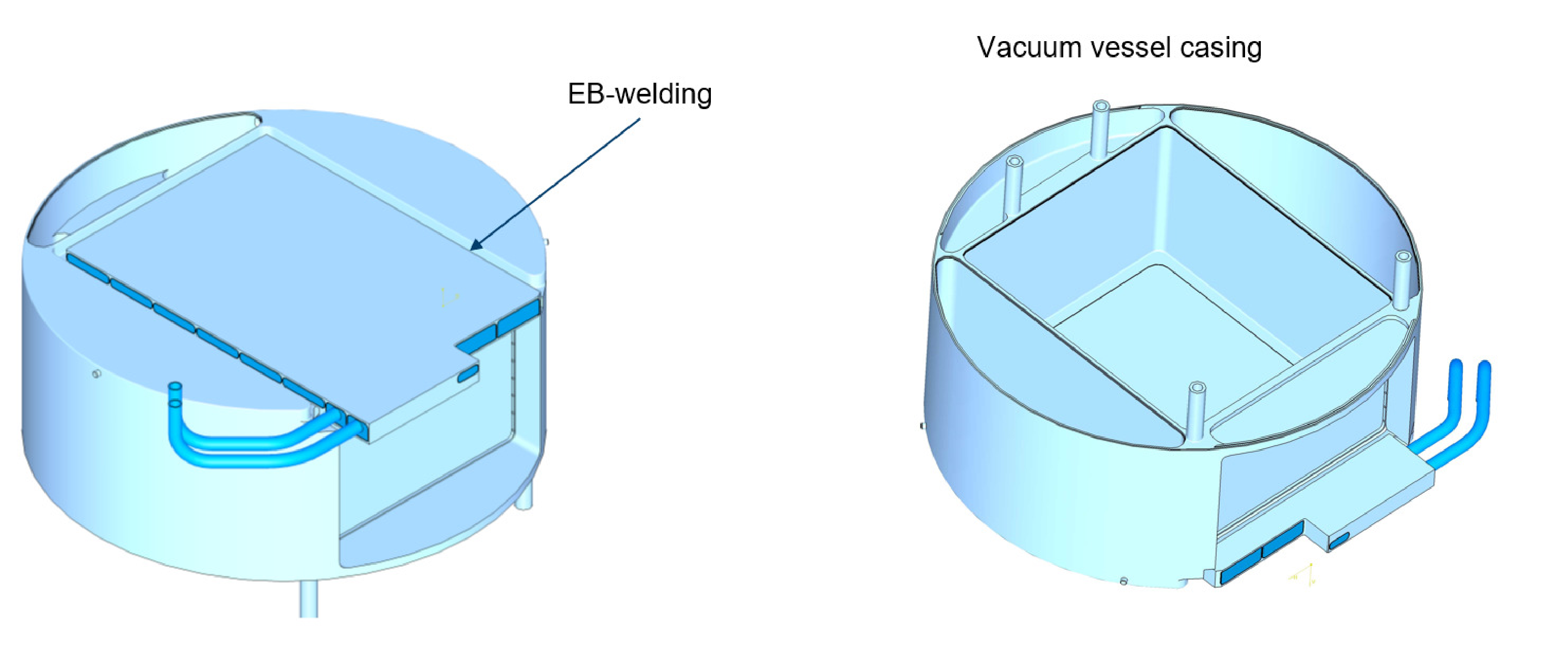}
\caption{Assembly and EB-welding of the vacuum vessel (premoderator).}
\label{fig:cn_1-35}
\end{center}
\end{figure}

\begin{figure}[hbt!]
\begin{center}
\includegraphics[width=0.9\textwidth]{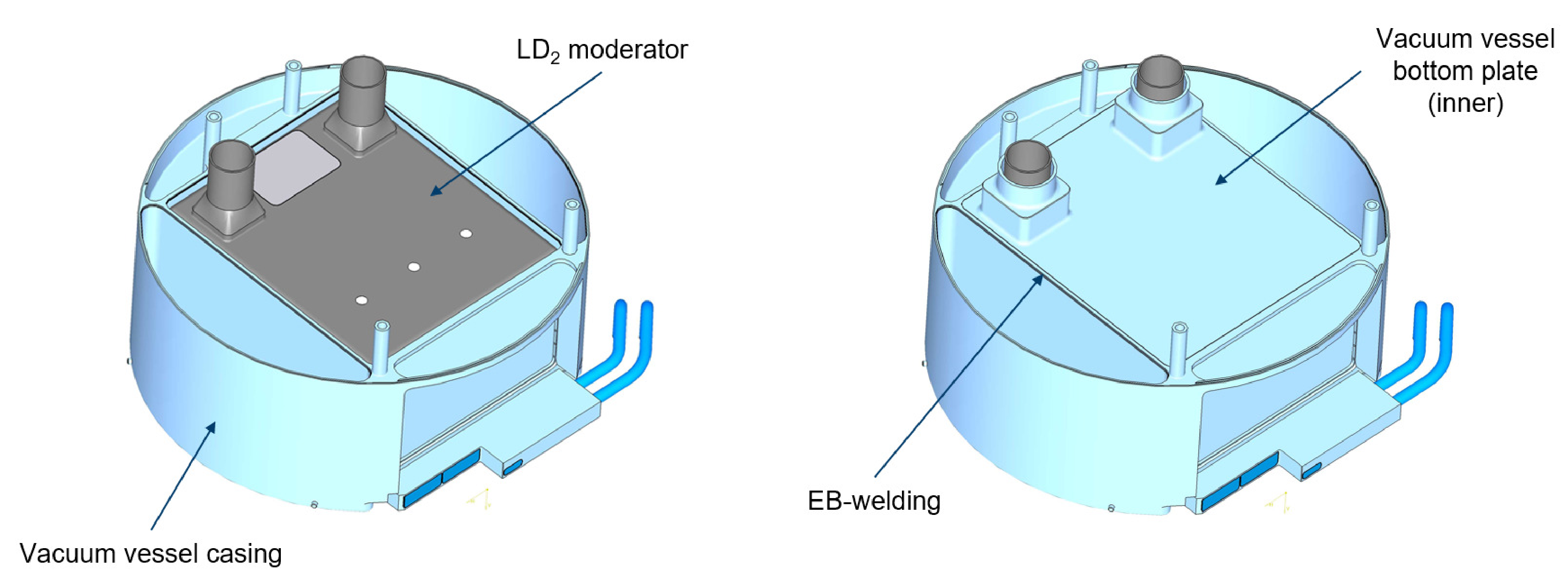}
\caption{Assembly and EB-welding of the vacuum vessel (inner bottom plate).}
\label{fig:cn_1-36}
\end{center}
\end{figure}

\begin{figure}[hbt!]
\begin{center}
\includegraphics[width=0.9\textwidth]{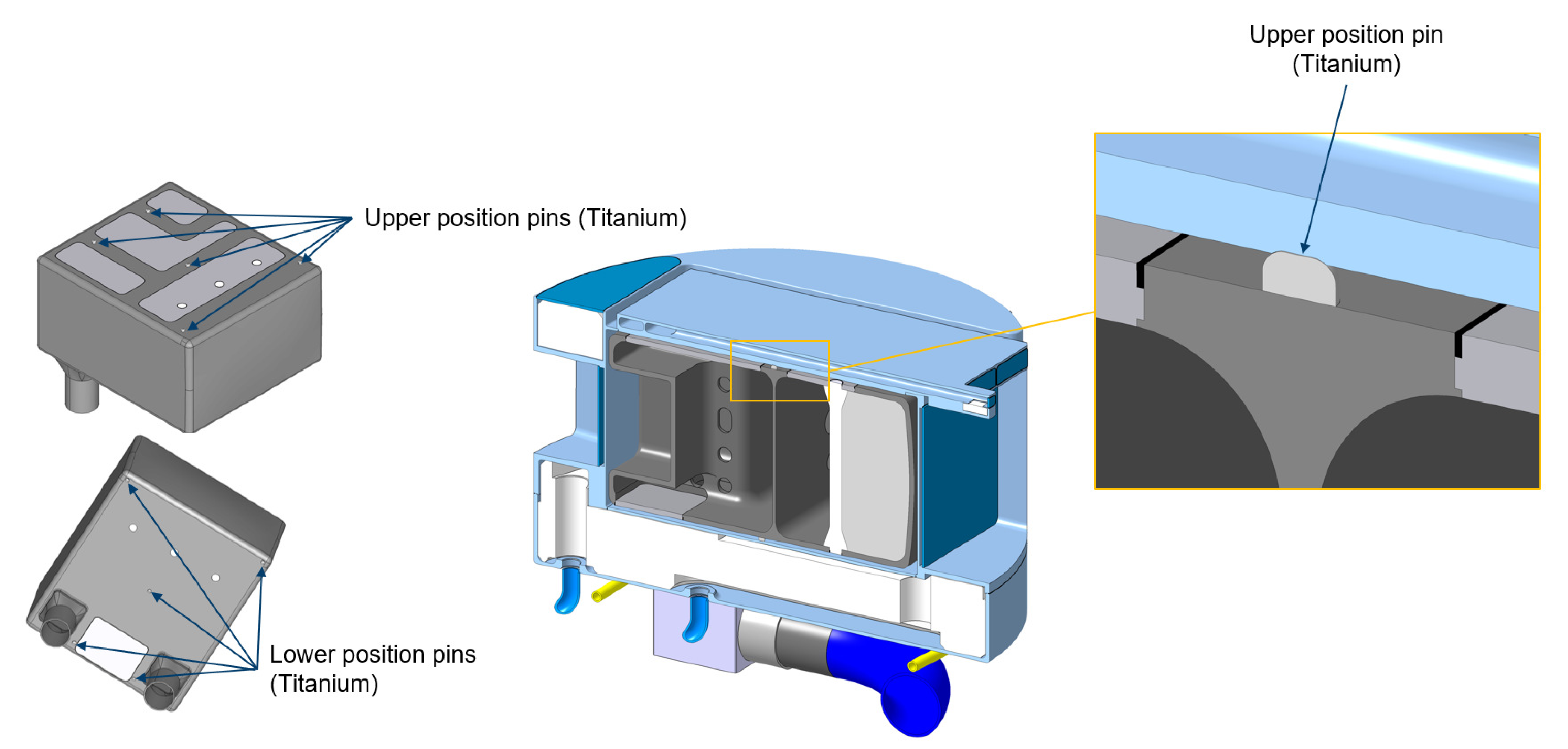}
\caption{Positioning of the LD$_\text{2}$ moderator inside the vacuum vessel.}
\label{fig:cn_1-37}
\end{center}
\end{figure}

\begin{figure}[hbt!]
\begin{center}
\includegraphics[width=0.9\textwidth]{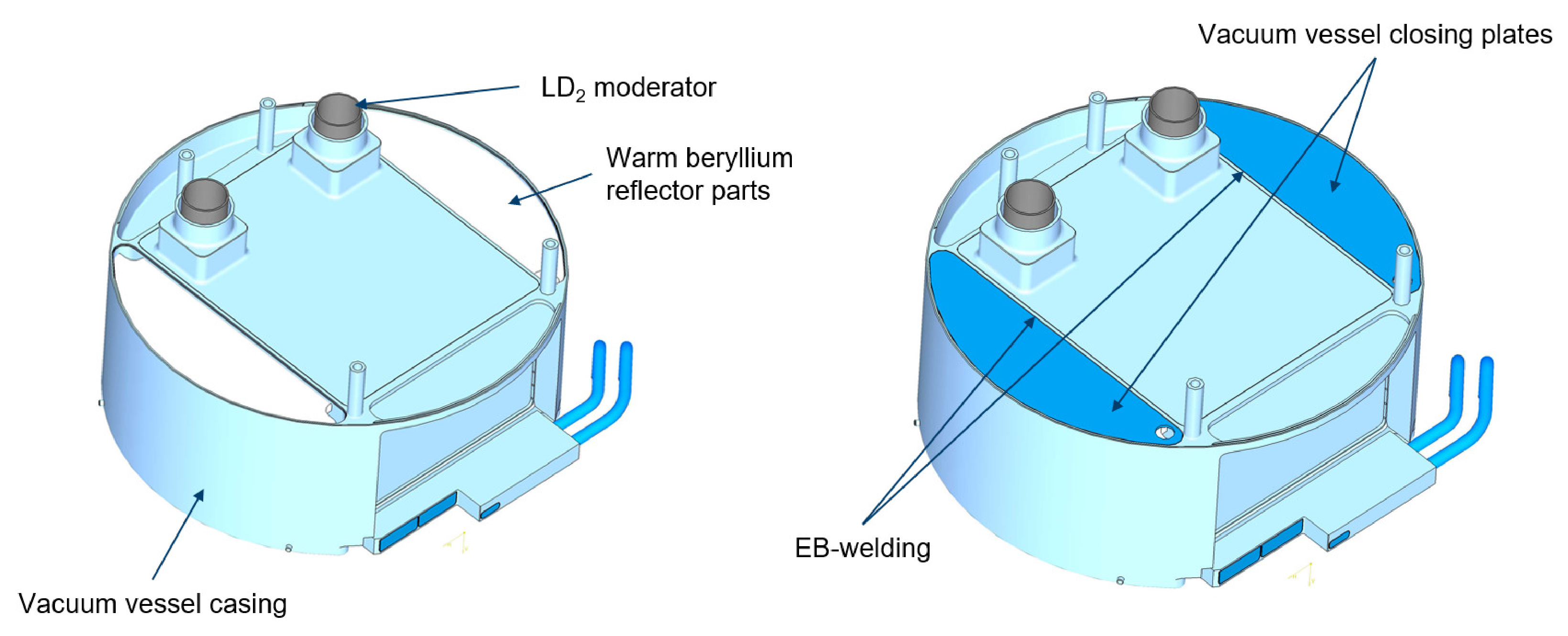}
\caption{Assembly and EB-welding of the vacuum vessel (Be reflector side segments).}
\label{fig:cn_1-38}
\end{center}
\end{figure}

\begin{figure}[hbt!]
\begin{center}
\includegraphics[width=0.9\textwidth]{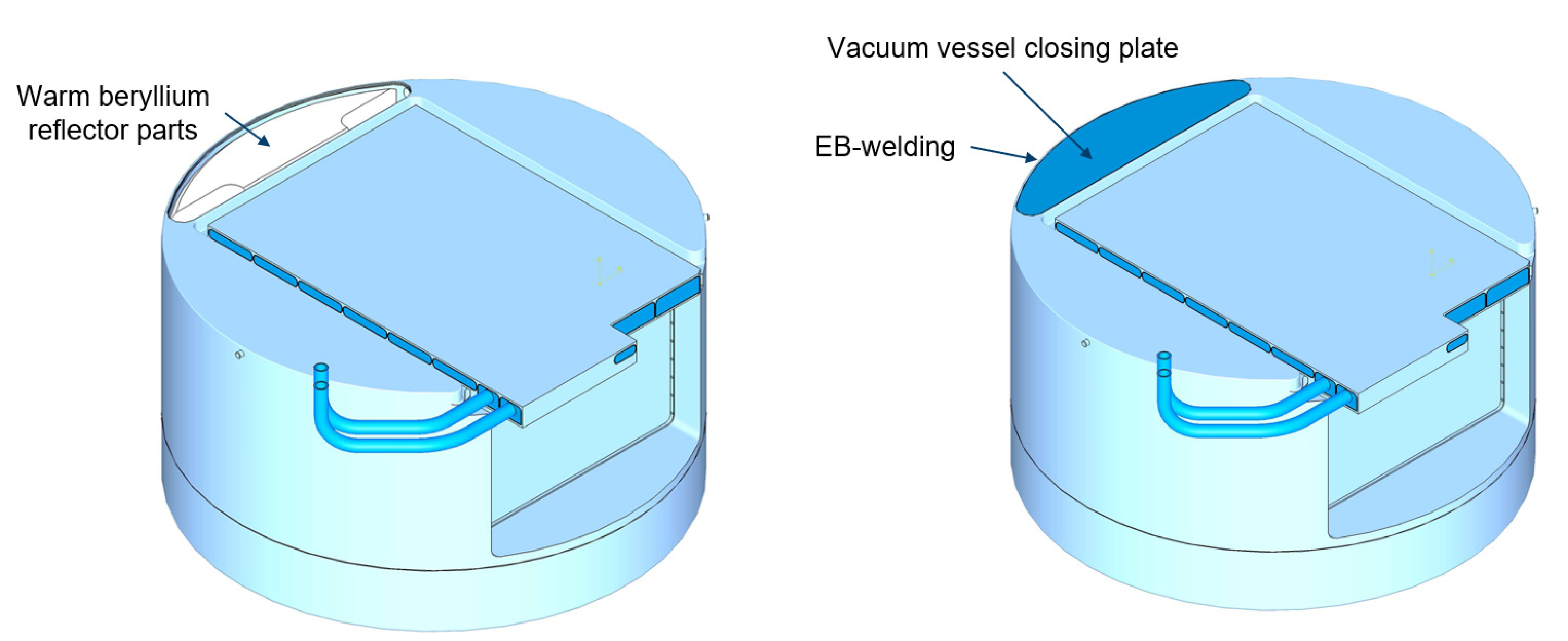}
\caption{Assembly and EB-welding of the vacuum vessel (Be reflector segment above WP7 window A).}
\label{fig:cn_1-39}
\end{center}
\end{figure}

\begin{figure}[hbt!]
\begin{center}
\includegraphics[width=0.9\textwidth]{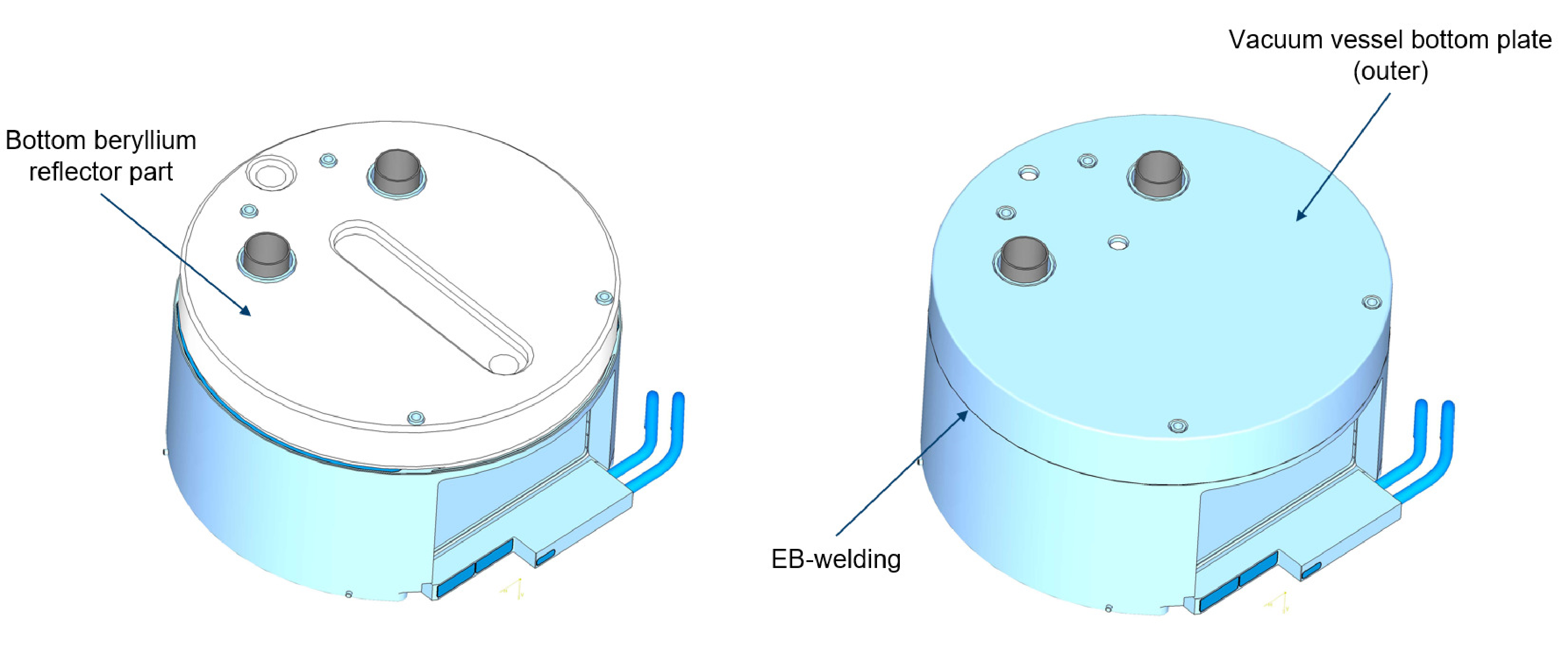}
\caption{Assembly and EB-welding of the vacuum vessel (large Be reflector below the moderator).}
\label{fig:cn_1-40}
\end{center}
\end{figure}

\begin{figure}[hbt!]
\begin{center}
\includegraphics[width=0.9\textwidth]{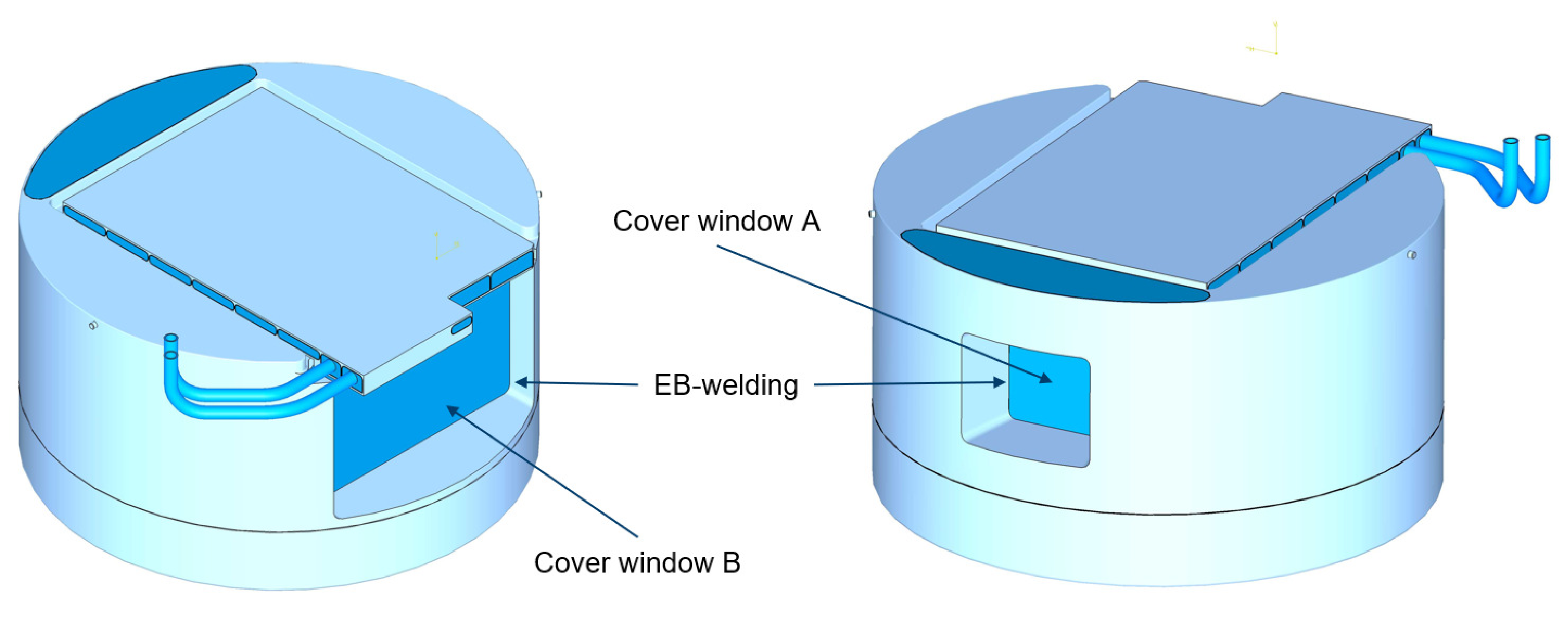}
\caption{Assembly and EB-welding of the vacuum vessel (cover plates for beam windows).}
\label{fig:cn_1-41}
\end{center}
\end{figure}

The assembly sequence of the outer vessel is then shown in \cref{fig:cn_1-35} to \cref{fig:cn_1-41}. At first, the water disc premoderator is welded to the top of the outer vessel via electron-beam welding (\cref{fig:cn_1-35}).

\FloatBarrier

The vessel is then flipped upside down and the LD$_\text{2}$ moderator in inserted. After this is complete, the inner bottom plate of the vacuum vessel is welded on via electron-beam welding (\cref{fig:cn_1-36}).

\FloatBarrier

\cref{fig:cn_1-37} shows how the moderator is positioned inside the insulation vacuum. It will rest on titanium pins on the top and the bottom of the moderator vessel. Titanium as a material is chosen for the positioning pins, because of its relatively low thermal conductivity -- to minimize the heat transfer between the moderator and the vacuum vessel -- and for its radiation resistance.

\FloatBarrier

When the insulation vacuum around the moderator is successfully closed, the side segments of the warm beryllium reflector are inserted and the outer vessel is closed with the vacuum vessel closing discs via electron-beam welding (\cref{fig:cn_1-38}).

\FloatBarrier

Then the warm beryllium reflector parts above the WP7 window A (\cref{fig:cn_1-39}) and the large bottom reflector part below the moderator (\cref{fig:cn_1-40}) can be mounted and afterwards covered and sealed via electron-beam welding.

\FloatBarrier

\FloatBarrier

In a final step, the cover plates for the actively cooled neutron windows A \& B can be welded on via electron-beam welding, as shown in \cref{fig:cn_1-41}.

\FloatBarrier

At this point, the  pre-assembly is ready for integration into the twister, which will be described in the following section.

\subsection{Integrability into the ESS twister} \label{ch:1-8}
\begin{figure}[hbt!]
\begin{center}
\includegraphics[width=0.9\textwidth]{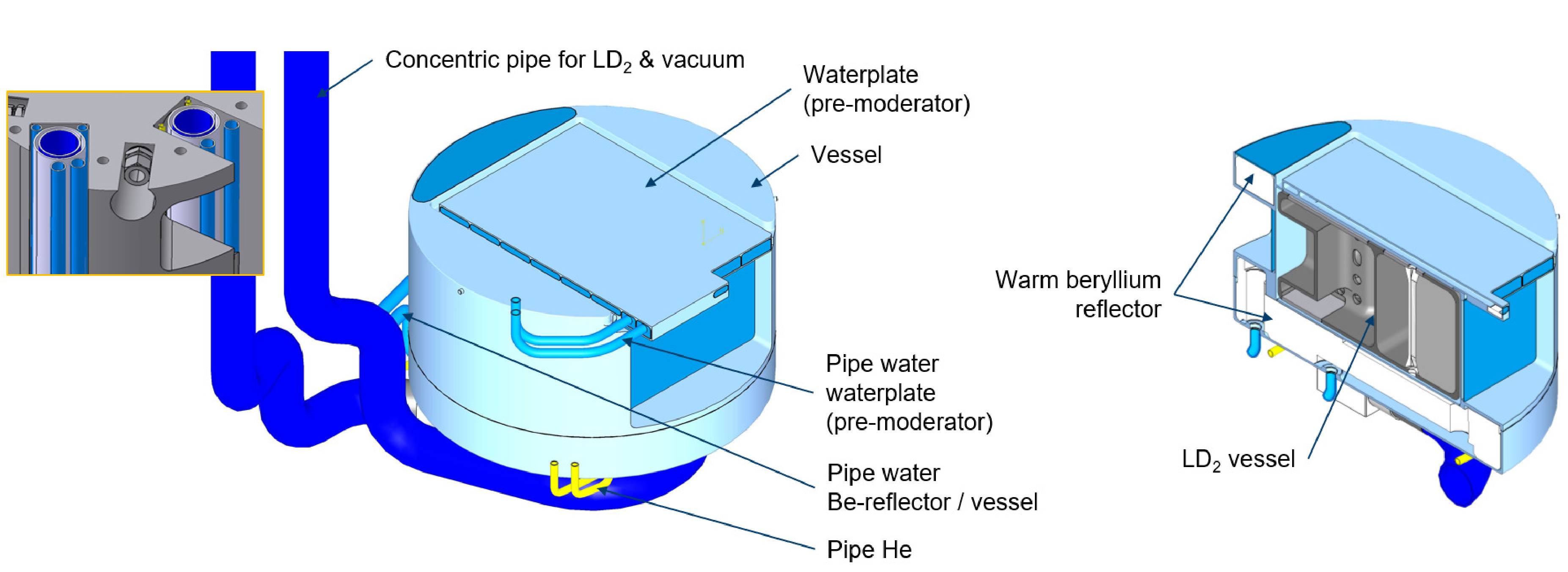}
\caption{Media interfaces on the lower moderator plug.}
\label{fig:cn_1-42}
\end{center}
\end{figure}

\begin{figure}[hbt!]
\begin{center}
\includegraphics[width=0.9\textwidth]{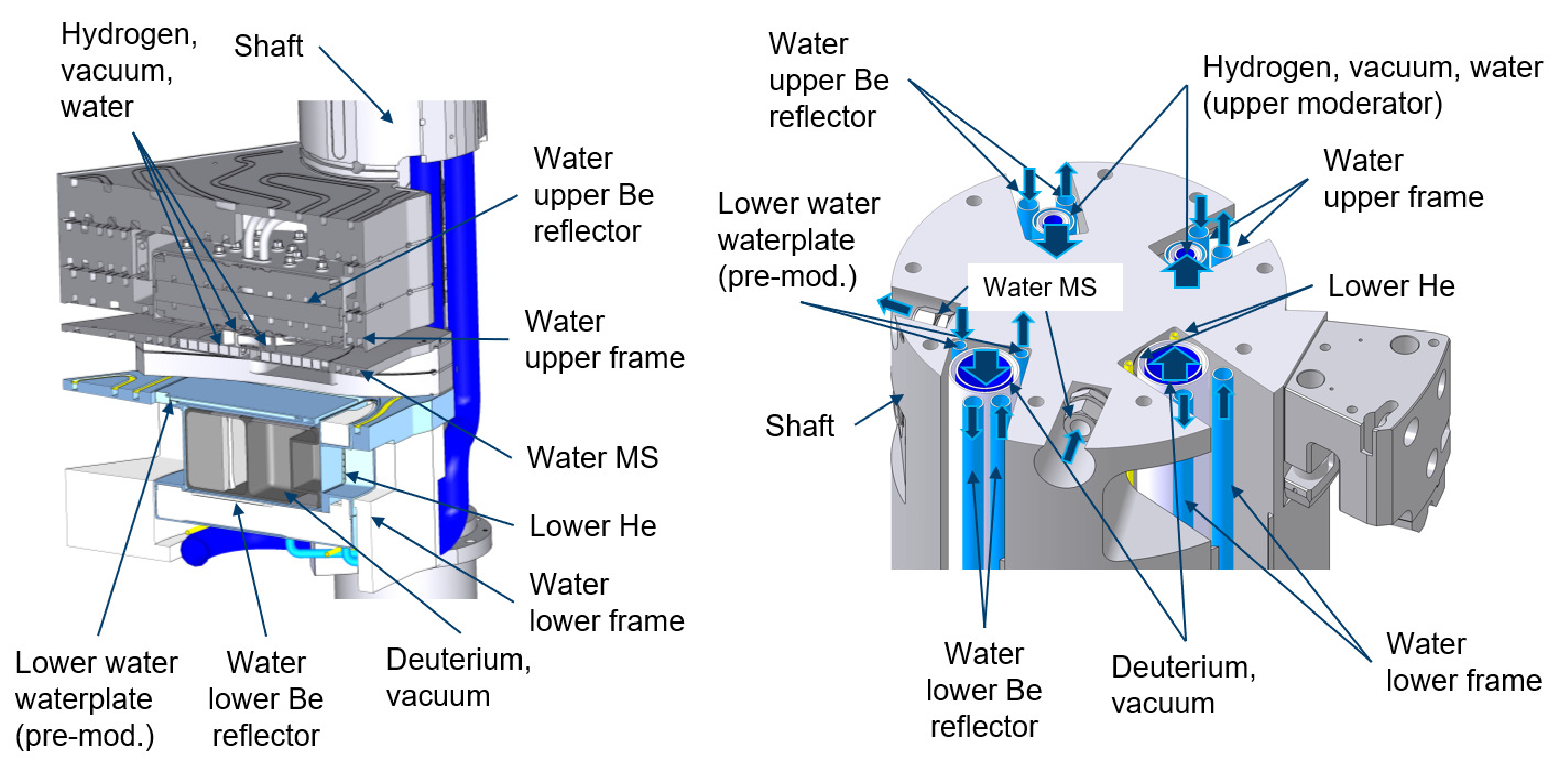}
\caption{Twister integration and interface description.}
\label{fig:cn_1-43}
\end{center}
\end{figure}

The newly designed moderator and reflector plug must fit into the existing twister frame below the target wheel. Also, all media supply for the moderators and the reflectors needs to be guided through the existing design of the twister shaft, since the surrounding shielding elements are not exchangeable and therefore a larger twister shaft diameter is not possible. In addition to the supply media for the first generation LH$_\text{2}$ moderator and its beryllium reflector above the target wheel, the lower moderator plug needs an additional supply lines for liquid deuterium, water for the premoderator, helium for the cooling of the neutron windows, water for the warm beryllium reflector, and vacuum for insulation. The designated location of the different media and the supply-pipe routing through the existing twister shaft are shown in \cref{fig:cn_1-42,fig:cn_1-43}.

\FloatBarrier

\FloatBarrier

\begin{figure}[hbt!]
\begin{center}
\includegraphics[width=0.9\textwidth]{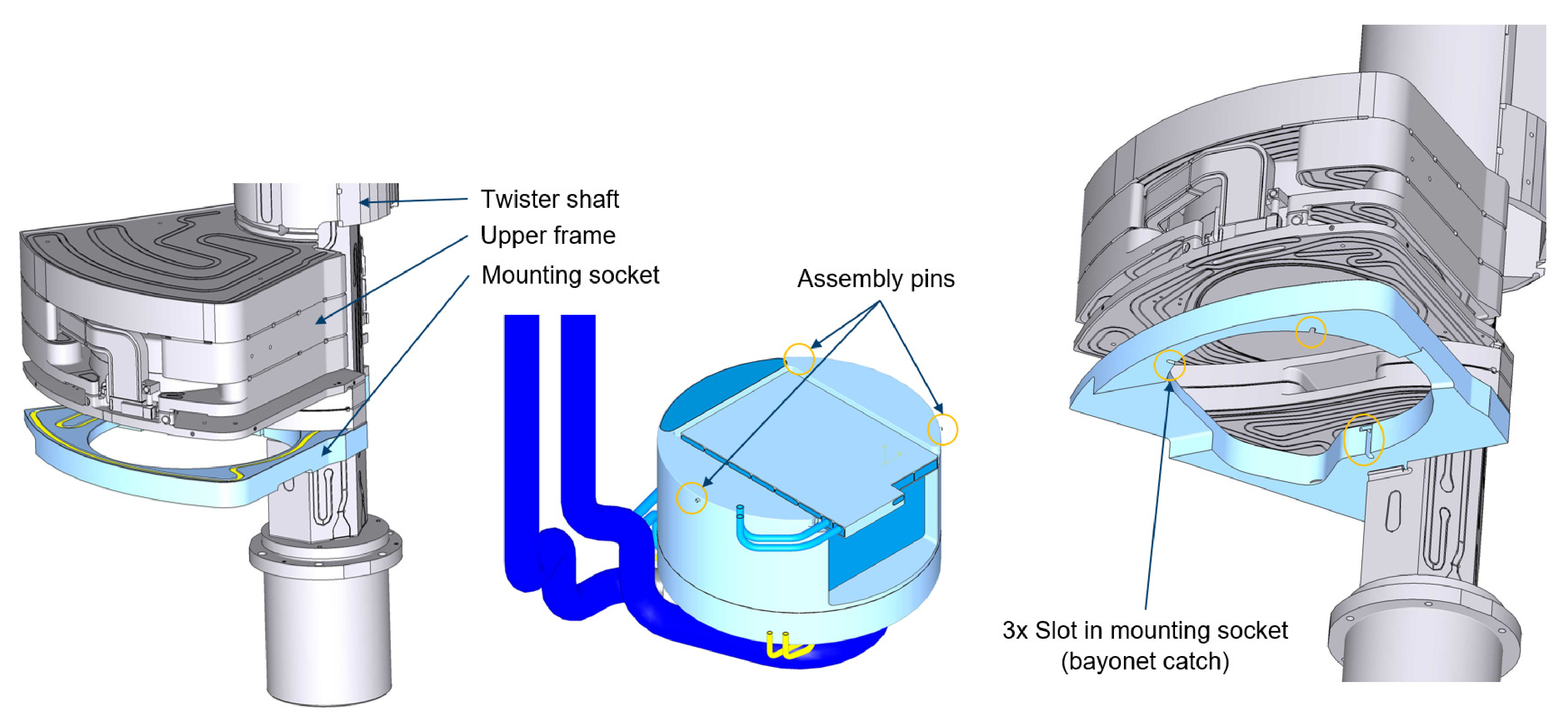}
\caption{Assembly of the lower moderator plug to the twister (mounting details).}
\label{fig:cn_1-44}
\end{center}
\end{figure}

\begin{figure}[hbt!]
\begin{center}
\includegraphics[width=0.9\textwidth]{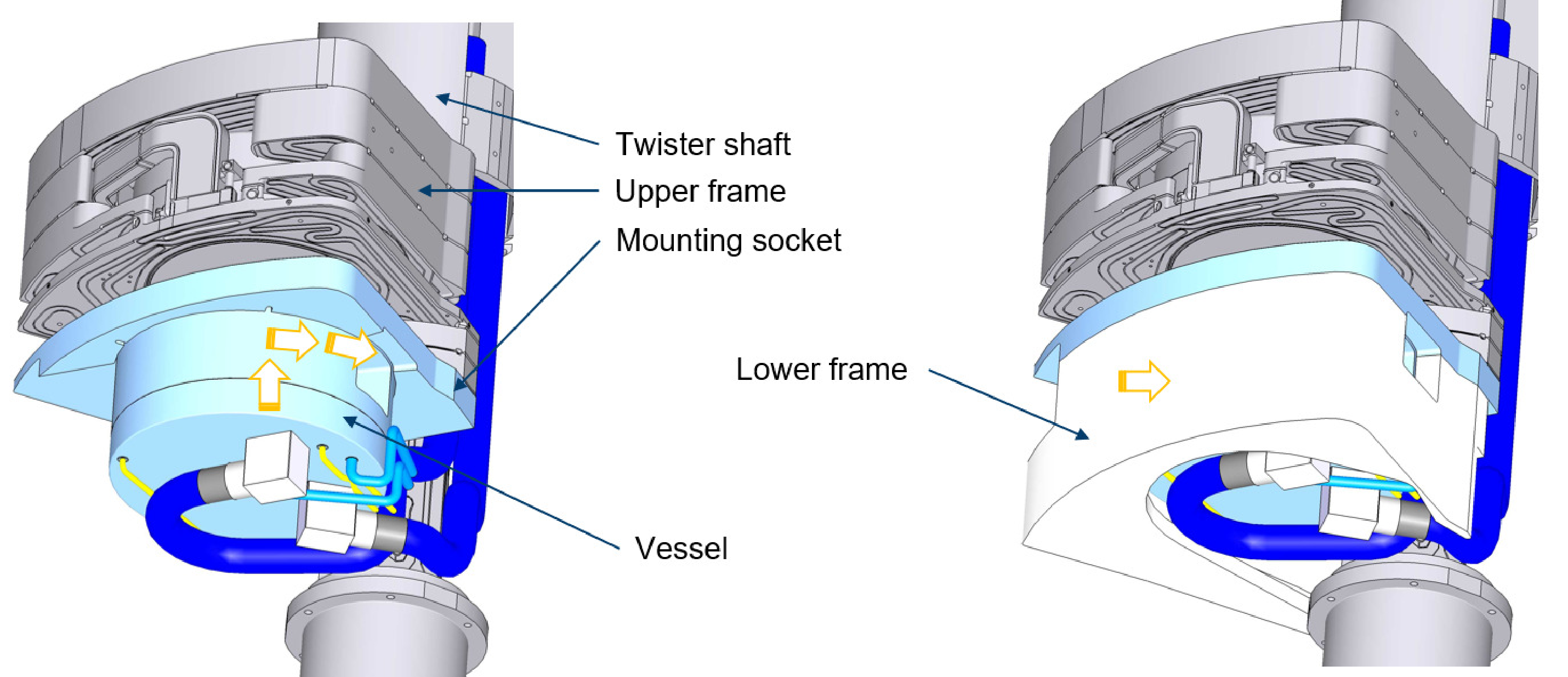}
\caption{Assembly of the lower moderator plug to the twister (insertion and turning).}
\label{fig:cn_1-45}
\end{center}
\end{figure}

The attachment and integration of the lower moderator plug into the twister, underneath the target wheel, is illustrated in \cref{fig:cn_1-44,fig:cn_1-45}. The outer vessel has three assembly pins radially installed on the outer cylindrical surface. The corresponding part in the twister is the lower mounting socket, where three slots with a bayonet-catch shape are implemented.

\FloatBarrier

With this mechanism, the moderator plug can be inserted into the mounting socket from below and then is aligned to its final position by turning. After the plug is fixed in the mounting socket, the lower frame can be assembled to the twister shaft and welded to the mounting socket, to complete the stainless-steel shielding in this area.

In summary, it can be stated that the integration of the new LD$_\text{2}$ moderator system into the existing twister is a very complex process, but feasible.

\subsection{Summary and outlook: engineering implementation of deuterium moderator} \label{ch:1-9}
In general, it can be concluded that a \ce{LD_2} moderator system for ESS appears to be feasible from an engineering point of view. It was demonstrated that a volume moderator, including all necessary process lines, can be integrated into the existing twister structure under the given restrictions. Furthermore, the detailed design shows that the moderator system can withstand all mechanical loads and that the manufacturing and welding, although very complex, is feasible.

The conceptual design for the cooling process describes all the additional infrastructure needed to realize such a moderator upgrade. A key point is to reduce the deuterium inventory as much as possible. On one hand, this can be accomplished by choosing a location for the cryostat as close as possible to the moderator and, on the other hand, by reducing the heat load.

The reduction of the heat load is also decisive from a thermo-mechanical point of view in order to be able to use the moderator at full beam power of \SI{5}{MW}. Without further optimization, the proton beam power is currently limited to \SI{2}{MW}. However, the first optimizations have already shown that the separation of the cold beryllium filter already leads to significant improvements. In contrast to the demonstrated box-shaped moderator (worst case), slightly shaped vessel walls would allow for a reduction in the wall thickness, which means in turn that less heat is generated and the neutronic performance of the moderator is improved as well. However, the following recommendations should be examined more closely in a follow-up project, before a final decision for the moderator upgrade will be made:

\subsubsection*{Recommended further engineering investigations} %\label{ch:1-9-1}

\begin{itemize}
    \item[-] Building a full-scale prototype is recommended, because of the stringent requirements on the manufacturing process and specifically because of the welding process. Electron-beam welding of the aluminum alloy EN AW-6061 T6 can be critical, because of the weld-in depth of the needed filler material. Therefore, thorough welding tests should be performed to determine the final design.
    \item[-] Thermo-hydraulic analysis needs further investigations, especially for \SI{5}{MW} beam power.
    \item[-] Separation of the beryllium filter from the LD$_\text{2}$ moderator, to reduce the heat load.
    \item[-] Reducing aluminum content by changing the shape of the moderator to reduce the heat load and to improve the neutronic performance.
    \item[-] Better understanding of local nuclear boiling on the wall with subcooled main flow.
    \item[-] Thermo-hydraulic analysis, considering boiling on the wall (by experiment).
    \item[-] Possibly relaxing ESS requirements regarding local boiling and measurement of critical heat flux.
    \item[-] Analysis of the risk of cavitation due to wall boiling in-pulse (by experiment).
\end{itemize}

%% file: vcnsource.tex
\section{Design of a Very Cold Neutron Source}
\label{sec:vcn}

\subsection{VCN: a long-wished-for source}

An intense VCN source has been a desire of the neutron community for at least 20 years. A dedicated workshop~\cite{osti_1248367} was held in 2005 to discuss ideas for sources and applications. While several ideas on possible materials for VCN were raised (cf. \cref{sec:SD2}), it was clear at that time that the lack of knowledge of the properties of the candidate materials, and, in particular, the thermal scattering libraries, constituted a major problem for realistic design studies. It is worth recalling that at this workshop a program to study VCN sources was outlined,
including topics such as: 1) experiments on selected moderator types; 2) development of scattering kernels for candidate materials; 3) benchmark measurements of VCN production; 4) investigations of thermal and superthermal production of VCN. The HighNESS project has taken on most of these tasks, thus offering for the first time the concrete possibility of designing realistic VCN sources.

\subsection{VCN neutrons: a brief introduction}

Very cold neutrons cover a wide spectral range within the long-wavelength tail of cold neutron sources\footnote{The standard cold moderator materials for both spallation and fission sources are liquid or solid deuterium ($\mathrm{S_2}$), liquid hydrogen ($\mathrm{H_2}$) and liquid or solid hydrocarbons (e.g. methane($\mathrm{CH_4}$)).}, with energies ranging from below 1 meV (\SI{9}{\angstrom}) to few hundreds of neV ($>$ several  \SI{100}{\angstrom}), the domain of UCN. 
A VCN source offers new possibilities in neutron scattering applications \cite{mezeiworkshop2022} and fundamental physics research. Extended neutron wavelengths ($\lambda$) inherently enhance the capabilities of diverse instrument categories.
%, making it a pivotal endeavor.
This impact can be observed in the $\lambda$ dependencies of the gains in instrumental resolution at fixed geometry and in intensity at fixed resolution. For example, 
gains in resolutions  proportional to
$\lambda^{-1}$, $\lambda^{-3}$  and $\lambda^{-3}$ are expected for reflectometers, time-of-flight instruments and neutron spin echo (NSE), respectively, while gains in intensity at fixed resolution are expected to be proportional to  $\lambda^{2}$, $\lambda^{2}$  and 
$\lambda^{3}$ for these three classes of instruments \cite{osti_1248367}.

%$\lambda^{-1}$ and $\lambda^{2}$ are relevant for reflectometers, $\lambda^{-3}$ and $\lambda^{2}$ for Time-of-Flight instruments, and $\lambda^{-3}$ and $\lambda^{3}$ for neutron spin echo (NSE) \cite{osti_1248367}.

Particle physics experiments can benefit from enhanced VCN fluxes, including the search for neutron-antineutron oscillations (described in HighNESS Conceptual Design Report Volume II), beam experiments looking for  a non-zero neutron electric dipole moment and search for novel fundamental forces (discussed in \cref{sec:fundamental_physics_VCN}). 

%The design of a VCN source is one of the main objectives of the HighNESS project. This stems from the fact that having access to intense fluxes of VCNs can be a game-changer in several neutron scattering applications \cite{mezeiworkshop2022}, as well as in fundamental physics research involving neutrons . Longer neutron wavelengths ($\lambda$) have a positive impact on the performance of various classes of instruments. This impact can be observed in the $\lambda$ dependencies of instrumental resolution at fixed geometry and intensity at fixed resolution. For example, $\lambda^{-1}$ and $\lambda^{2}$ are relevant for reflectometers, $\lambda^{-3}$ and $\lambda^{2}$ for Time-of-Flight (ToF) instruments, and $\lambda^{-3}$ and $\lambda^{3}$ for neutron spin echo (NSE)\\.
%%%add reference to WP7 \label{prismasec}

%There are also particle physics experiments that can benefit from higher VCN fluxes. These include the search for neutron-antineutron oscillations, which is discussed in Section [Section Number], where the discovery potential is proportional to $\lambda^{2}$. Additionally, there are projects related to in-beam searches for a non-vanishing neutron electric dipole moment and experiments aimed at investigating new fundamental forces. 
% Their main advantage in fundamental physics experiments are their very low energies. This causes them to exhibit long interaction times with the experimental apparatus. At the same time, unlike UCN, they can still be considered as beams, making beam optics and related methods applicable.

In this section, we first provide an overview of the scientific case for VCNs and later describe the different design concepts that have been developed in the course of the HighNESS project.

\subsection{Fundamental physics with VCN}\label{sec:fundamental_physics_VCN}
The primary advantage of VCNs in fundamental physics experiments is their extremely low energy, which results in long interaction times with the experimental apparatus. Unlike UCNs, VCNs can still be considered as beams, allowing for the application of beam optics and related methods. Consequently, VCNs enhance the sensitivity of most fundamental physics experiments that utilize beams of slow neutrons or exploit the wave-optical properties of these beams.

\subsubsection{Beam Experiments}
Recently suggested particle physics experiments using beams of slow neutrons include:\footnote{Note that this list is not exhaustive.}
\begin{enumerate}
    \item[-] In-beam searches for a permanent neutron electric dipole moment (nEDM) \cite{PiegsaFlorianM2013Ncfa,piegsa_nEDM_2018,ABELE20231}, and searches for new fundamental forces \cite{piegsa_limits_2012}. These kinds of experiments profit linearly from the interaction time $t$ of the neutron in the apparatus. Therefore, their figure of merit can be considered proportional to $\lambda$. 
    
    \item[-] Measuring the neutron lifetime $\tau_n$ by means of the \textit{beam method}. Typically these experiments contain a trapping region of some length $L$ that intercepts a neutron beam. The neutron decay is observed by detecting decay protons within the volume of this region. In an experiment similar to \cite{PhysRevC.71.055502}, the mean number of neutrons in the decay region at any time is given by, 
    \begin{equation} \label{eq_beam_lifetime}
    N_n = L A \int I(v) \frac{1}{v} dv \;,
    \end{equation}
    where $A$ is the cross-sectional area of the trap and $I(v)$ is the velocity-dependent fluence rate \cite{PhysRevC.71.055502}, which is independent of the position in the trap for a perfectly collimated beam without attenuation.  
    By increasing the VCN flux through such an experiment, the neutron density and hence the observed neutron decay rate increases, which in turn improves the statistical accuracy of the result. Similarly to the nEDM experiments, there is an expected $\mathrm{FOM} \propto \lambda$. It should be noted that these experiments still suffer from a systematic uncertainty, associated with the determination of the absolute neutron flux. Further progress towards the reduction of this uncertainty, as pointed out in \cite{PhysRevLett.111.222501}, could increase the precision of VCN beam lifetime experiments significantly and  shed light on the neutron lifetime puzzle \cite{Paul_2009}.
    
    \item[-] Experiments searching for neutron anti-neutron oscillations \cite{Addazi_2021, Backman_2022}. Here it is not the interaction time, but quasi-free flight time of VCN that allow to significantly increase the sensitivity of the experiment. Under these conditions the probability for a transition can be expressed by: 
    \begin{equation}
        P_{n \rightarrow \Bar{n}} = \left( \frac{t}{\tau_{n \rightarrow \Bar{n}}}\right)^2 \; ,
    \end{equation}
    with the free oscillation time $\tau_{n \rightarrow \Bar{n}}$ \cite{Addazi_2021}. This results in a $\mathrm{FOM} = N_n t^2$, with $N_n$ being the total number of neutrons observed.
    Due to the proportionality of the ($\mathrm{FOM}$) to $\lambda^2$, VCN exhibit specific potential for neutron oscillation experiments.
    \end{enumerate}
%% This needs more care%%    
One should note, that, under the assumption of equivalent phase-space density, in most experiments the long interaction time or the longer free flight time of VCN is in competition with their lower flux. It is therefore crucial, in order to truly increase the sensitivity of the experiments described above, to increase the phase space density of VCN in the source. This has been discussed in \cite{ABELE20231} and \cite{nesvizhevsky_why_2022}.

\subsubsection{Wave‐Optical Experiments}

Another class of experiments that could benefit from intense  VCN beams exploits the wave-optical properties of the neutron, in particular diffraction and interferometry. The potential of long wavelength neutrons for wave optics experiments was already pointed out in \cite{EDER1989171}. 
A  list of fundamental physics experiments utilizing neutron interferometry techniques is given in \cite{RauchH.Helmut2015Ni:l}. Some of these experiments can benefit from lower energy neutrons. \\ %While certain proposed experiments may have become less relevant due to advancements in other fields, wave-optical experiments involving VCN remain an exciting avenue for fundamental physics research.\\
A  VCN source could be used for the following studies:
\begin{enumerate}    \item[-] Further investigation of gravitationally induced quantum interference as measured in the COW\footnote{Acronym for R. Colella, A. W. Overhauser, and S. A. Werner.} experiment \cite{COW}.  This effect was already measured with VCN \cite{VANDERZOUW2000568}, but the experiment could be significantly improved with higher fluxes  and modern VCN optics.\footnote{It should be noted that VCN might not be competitive with the accuracy of atom interferometers, as pointed out in \cite[p. 263f.]{RauchH.Helmut2015Ni:l}.} With regard to the latter, especially the field of holographic gratings, as described in \cite{klepp_2011}, made significant progress.
    \item[-] Improvement of the current limit on the neutron charge \cite{PhysRevD.37.3107} with a Talbot-Lau interferometer \cite{PhysRevC.98.045503}. 
    \item[-] Measurement of the gravitational constant G using an active gravitational mass other than the Earth itself, in a \textit{Cavendish experiment}, as described in \cite{KleinA.G.1984WOWC}.
\end{enumerate}

\subsection{VCNs in condensed matter research}
\label{sec:vcncase}

In condensed matter research with neutrons, the momentum and energy transfer of scattering events is measured to probe the structure and dynamics of a sample material. Very cold neutrons hold significant potential for advancing this field of study. While low-energy neutrons provide the highest sensitivity to low momentum and energy transfers (and are therefore well suited for maximizing resolution) their usage is hindered by the low available flux in the VCN regime. The flux of VCNs in the tail of a Maxwellian spectrum typically available from conventional cold neutron moderators exhibits a wavelength dependence of $\lambda^{-5}$.

High-resolution techniques that are expected to gain using longer wavelength neutrons are 
 small-angle neutron scattering (SANS), time-of-flight spectroscopy, and neutron spin-echo (NSE)~\cite{mezei_very_2023}.
 Moreover, a real-space technique
 like neutron imaging might benefit in some cases from improved fluxes in the VCN range.

In small angle scattering, the resolution of small angles enables studying large-scale structures on the nanometer scale, i.e. from a few to a few thousand nanometers. Because of flux limitations it is usually not possible  to probe  larger structures up to the micrometer scale using a conventional pinhole SANS instrumentation. Other instrumentation exploring even smaller momentum transfers, and thus giving access to the low-micrometer scale, are typically limited to one dimension and require separate measurements. It is therefore desirable to increase the resolution of conventional SANS instruments.

A simple analysis~\cite{osti_1248367} shows that a $\lambda^{-5}$ slope of the long wavelength tail of the neutron spectrum of a cold moderator marks the break-even for gaining either efficiency or resolution at longer wavelengths with SANS. The momentum transfer is $q=2\pi/\lambda\sin\theta$, and therefore larger wavelengths enable lower collimation and larger sample sizes in both dimensions across the beam to achieve the same angular resolution. Together with the increase of $d\lambda$ for a constant $d\lambda/\lambda$, the gain is proportional to $\lambda^5$. Hence, for a long-wavelength tail of the neutron spectrum that decays at a rate lower than $\lambda^{-5}$ 
there will be a performance gain,
either in data rate or resolution, or a balanced gain in both.

%The large size of the candidate moderator is essential in order to benefit from increased sample sizes and divergence. These applications carry significant potential to elucidate complex and hierarchical structures and their transformations on an extended range of length scales, likely drawing an increasing interest in general science and technology fields, and in particular, with advanced soft-matter research and its applications.   

It is however important to point out some downsides that can be expected with VCNs, in particular the increased probability of absorption, and the effect of gravity.
%However, these simplified considerations do not yet take into account some additional aspects, in particular the downsides, of utilizing very long wavelength neutrons -- namely their increased absorption
%and scattering  
%probability and the significant impact of gravity on their trajectories. 
Increased absorption increases the requirements on all materials used in the neutron beam, in particular neutron beam windows of sources, guides, and sample cells and environments. The increased absorption probability along with an increased scattering probability will favor the use of thin samples to avoid absorption losses and multiple scattering. Effects from gravity can in principle be countered by instrument design but might increase the complexity and lead to additional absorption issues. 
%Thus, gains that translate to increased performance and advancing scientific capabilities require significant gains in flux at long wavelengths and a careful balance in the choice of the utilized neutron energies.

Considering experimental methods that probe energy transfers and thus dynamics, the same break-even principle is found for time-of-flight spectroscopy probing inelastic scattering. 
Simple estimates take into account that the energy change $dE$ is proportional to $dt/\lambda^3$, where $dt$ is the time resolution. The estimated gains, taking into account a lower repetition rate, and thus scaling with $\lambda^5$ again, imply a wavelength-independent performance at constant resolution for a long wavelength tail of a Maxwellian spectrum decaying with $\lambda^{-5}$ \cite{osti_1248367,mezei_very_2023}, keeping in mind the corresponding relaxation of instrument requirements for monochromatization -- initial pulse length -- and neutron pulse length at the sample. However, in time-of-flight spectroscopy, the momentum transfer resolution would increase, and relaxing collimation and increased sample size could result, corresponding to the previous considerations for SANS, with gains proportional to $\lambda^4$. Only the required momentum-transfer range limits such gains, where for example, for quasi-elastic measurements the wavelength must be limited to $\lambda < 4\pi/k_{max}$, where $k_{max}$ denotes the maximum momentum transfer.

In NSE, the use of longer wavelengths allows improving the energy resolution based on the relation $dE\propto dt/\lambda^3$, however without being able to profit from relaxing the requirements on incoming wavelength resolution, divergence, or increased beam size, like in the cases discussed above. Results from high-resolution NSE instruments such as IN15 at ILL indicate that the current technology of magnetic field configurations does not allow for further gains above about \SI{25}{\angstrom}~\cite{mezei_very_2023}. 

For neutron imaging, different considerations need to be taken into account. Currently, thanks to progress in detector technology, neutron imaging can reach resolutions of a few micrometers. However, in the range of single micrometers, the technique meets two limitations. The first one is related to contrast, since only a few materials would provide sufficient image contrast on a thickness scale of micrometers. A colder spectrum would be beneficial, since absorption increases with longer wavelengths,  enabling the resolution of smaller structures. The second limitation is neutron flux. 
Current resolutions and the utilization of the entire white beam spectrum already indicate a limitation in flux, suggesting that a decay in flux close to $\lambda^{-5}$ prevents the effective use of very cold neutrons for high-resolution imaging. An  VCN source with a decay in intensity lower than 
$\lambda^{-5}$ would then be beneficial.

 Novel lens systems for neutrons promise gains of the order of \num{e4} \cite{HUSSEY2021164813}, and long wavelengths are advantageous for lenses due to the refractive index increasing with $\lambda^2$ , the focal length increasing with $\lambda^{-2}$, and the critical angle proportional to $\lambda$. 
The concern about increased absorption makes reflective focusing optics, like Wolter optics \cite{MILDNER2011S7}, a very attractive choice, and, by using shorter focal lengths, the impact of gravity can be minimized. However, it is important to take into account and address any potential aberrations that may arise and require correction~\cite{kubec_achromatic_2022}.
The potential of true neutron microscopy to unlock unexplored applications and fields of study is particularly promising in soft matter and biology, where deuteration can provide unique opportunities for neutron imaging. Moreover, efficient neutron lenses designed for long wavelengths can also have a significant impact, especially in techniques such as SANS.

%\textcolor{red}{suggest to add a paragraph on reflectometry (Ott paper) }
Finally, in \cite{ottworkshop2022} the possible advantages of using VCNs in reflectometry techniques are discussed for several configurations. 
Gains are expected for a shift of the neutron spectrum towards colder energies, provided that the brightness is preserved. The use of VCNs for reflectometry is also considered promising for compact sources; new compact sources could be designed aiming at optimizing the VCN flux, having more freedom in the advanced VCN moderator and reflector materials used in the design, with respect to existing facilities.
%This would clearly be an advantage in comparison to operating facilities, where constraints to design upgrades of new sources are present.

%one could exploit the advantage designed for optimal use of VCNs.

\subsection{VCN design options }
\label{sec:intro_VCN}
In the original proposal of the HighNESS project, two approaches were considered for the design of a VCN source. One is based on a dedicated VCN source using a suitable material that would provide a high flux of 
VCNs, while another one, similarly to the PF2 beamline 
\footnote{\url{https://www.ill.eu/users/instruments/instruments-list/pf2/description/instrument-layout}}
at the ILL, extracts VCNs directly
from  the cold source. At the dedicated workshop in 2022 \cite{santoro_workshop}, an additional concept was suggested, that can be considered a merger of  the two approaches \cite{Valery_why_VCN_2022}.

\subsubsection{Dedicated VCN moderator}
In previous studies, a set of candidate materials for which thermal scattering libraries were available was analyzed as a potential VCN source.
Results indicated that solid orthodeuterium and solid methane should have a higher VCN production performance than LD$_2$~\cite{gallmeier}. 
As discussed below, 
%However, as discussed below, 
solid methane  is likely not suitable in a high-power spallation facility,
%On the other hand, 
while solid deuterium is  well-established as an in-pile converter medium for UCN sources 
%(see \cref{sec:MethodsForCalculatingUCN} and \cref{sec:sd2_intwister}), 
and is therefore a suitable candidate material for an in-depth investigation in HighNESS. Another class of candidate materials are fully deuterated clathrate hydrates; in these inclusion compounds, guest molecules occupy cages formed by a rigid network of water molecules. Various local modes can be excited in incoherent inelastic scattering events, which provide a path for cascaded neutron cooling that is not kinematically restricted by a dispersion relation~\cite{zimmer}. A strong candidate material investigated within HighNESS is the fully deuterated tetrahydrofuran (THF-d) clathrate hydrate,
%(17D2O.C4D8O), 
which has a broad band of low-energy modes, as experimentally demonstrated for its undeuterated version~\cite{V1}.

For any selected moderator material the highest VCN fluxes are to be expected at source locations near the spallation target. Their use is justified only if they outperform the main source in the wavelength range of VCNs. This is very difficult to achieve, as shown in Ref.~\cite{gallmeier}; 
    the highest possible incident flux is required, meaning that
   the best location for a VCN source would be below the spallation target, where the cold source is located and would need to be replaced. Additionally, this solution would have to cope with the challenge of cooling the material to the needed temperature in a region with  a high radiation field and high heat loads.
 While for SD$_2$ a temperature of 5 K seems optimal~\cite{R8}, the THF-d clathrate hydrate would best be operated below 2 K. At such low temperatures, the local modes are predominantly populated in the ground state, which suppresses up-scattering. From a practical point of view, He-II can be used for effective cooling of the weakly thermal-conducting clathrate hydrate. 

Designs with dedicated VCN moderators are reported in \cref{sec:full_SD2} for SD$_2$ and in \cref{sec:dch} for deuterated clathrate hydrates.

\subsubsection{Use of advanced reflectors}

In the original proposal  of the HighNESS project 
the use of advanced reflectors for a VCN source was
intended in combination with the liquid deuterium moderator; this option is discussed in \cref{sec:NDextraction}. However, in the course of the project we have developed designs where
advanced reflectors, specifically
nanodiamonds, have also been applied in combination with SD$_2$, either  in the full   SD$_2$ design, as detailed in \cref{sec:full_SD2}, or in the hybrid
LD$_2$/SD$_2$
design.
%,  described in \cref{sec:combined}.

%\subsubsection{Hybrid design}
The hybrid design, proposed by Nesvizhevsky, supplements the baseline \ce{LD_2} moderator with an adjacent or embedded block of \ce{SD_2}. The original rationale for such an option was to improve the performance for NNBAR. However, potential gains for general VCN performance may also be feasible. As such, this option has been investigated in several configurations, for both the NNBAR and WP7 beam openings, and is discussed in-depth in \cref{sec:combined}.

\subsection{Solid \texorpdfstring{ortho-\ce{D_2}}{ortho-D2}}
\label{sec:SD2}
%\textcolor{red}{This is taken from Nicola's thesis. Do we need to change the wording?}
In the first workshop dedicated to applications of a Very Cold Neutron Source \cite{osti_1248367}, the designs  proposed  for a VCN  moderator already made use of solid materials like Be, graphite, \ce{D_2} and \ce{D_2O}  in the configuration of a bed of small pellets between 5 and \SI{10}{K} with liquid helium at \SI{2}{K} flowing  through  the gaps as a coolant.
%It is well known that deuterium is one of the best neutron moderator materials thanks to its small nuclear mass and absorption cross section.
During the workshop, it was acknowledged that deuterium-based materials are expected to be generally better than graphite and beryllium, whose total cross sections have no incoherent contribution and a steep fall for energies  below  the  lowest  Bragg  edge  (\SI{5}{meV}  in  Be  and \SI{2}{meV}  in graphite).  
However, the higher heat capacities of beryllium and graphite left open the possibility that they might be favored when optimizing the source for the neutron flux weighted on the heating during the accelerator pulse.  At the end of the workshop, there was a consensus that the lack of neutron scattering kernels and measurements was hindering the design of a VCN source based on calculations.

As mentioned above, a more recent study by Gallmeier et al. \cite{gallmeier} looked for a VCN source option for the second target station at SNS. The set of materials investigated was chosen based on the thermal scattering libraries available. In particular, para-\ce{H_2}, water ice, beryllium \cite{Bernnat_proceedings,bernnat2002scattering}, ortho-\ce{D_2} \cite{granada2009neutron}, \ce{CH_4} \cite{SHIN2010382}, and neon, all solid at cryogenic temperatures of 4-\SI{6}{K}, were compared. The figure of merit in that study was the time-integrated pulse brightness over a liquid para-\ce{H_2} moderator at \SI{20}{K}. The authors indicated that solid ortho-deuterium (\ce{SD_2}) and solid methane are the most promising material for delivering a higher VCN production than liquid hydrogen, with the former able to outperform a conventional cold source by a factor 2, at the expense of a pulse five time longer. While solid \ce{CH_4} is generally considered a good candidate material for a VCN source due its high hydrogen content, it is likely not suitable in a high-power spallation facility like ESS. Irradiation of methane leads to the formation of radiolysis-induced radicals \ce{-(CH_2)_n-}, followed by polymerization into carbon-based deposits and release of \ce{H_2}, both of which have been observed to accumulate in the moderator vessel \cite{Barnert-Wiemer_2002,damage}.
On the other hand, solid ortho-deuterium, well-established as an in-pile converter medium for UCN sources \cite{serebrov_studies_2000,lavelle2010ultracold,atchison2009investigation,thomasBrys,korobkina_growing_2022,morris_measurements_2002}, has no physical impediment to the VCN production in high radiation fields, despite the many challenges related to keeping such a source at low temperature.

\begin{figure}[tb!]
    \centering
    \includegraphics[width=0.8\textwidth]{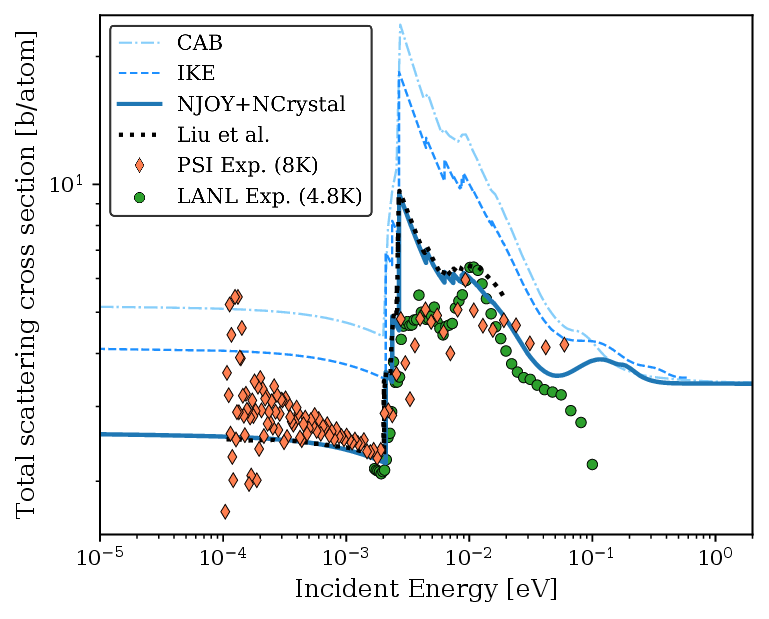}
    \caption{Total scattering cross section per atom for \ce{SD_2} as a function of incident energy. Here, NJOY+NCrystal is the new library, CAB is the evaluation generated by \emph{Centro Atomico Bariloche} \cite{granada2009neutron,granada2011neutron}, and IKE is the evaluation from \emph{Institut für Kernenergetik und Energiesysteme} \cite{bernnat2002scattering}. Also the more recent model developed in \cite{liu2010coherent} is shown for comparison. Models are also validated with measurements from SINQ/PSI \cite{atchison2009investigation} and LANL/LANSCE \cite{lavelle2010ultracold}. Only every fourth data point is shown for clarity.}
    \label{fig:xs_sd2}
    \end{figure}

In this work, a new scattering kernel was used to model the interactions of slow neutrons with \ce{SD_2}. The library, developed by the Spallation Physics Group at ESS, follows the physics of the thermal scattering model from Granada \cite{granada2009neutron,granada2011neutron}, and uses the new mixed-elastic scattering format to correctly include both the coherent and incoherent components of elastic scattering \cite{ramic2022njoy}. The cross section and the benchmark against experimental data \cite{atchison2009investigation,lavelle2010ultracold} can be seen in \cref{fig:xs_sd2}. The related files are available in the ESS Gitlab repository \cite{tsl_repo}.
\subsection{Dedicated \texorpdfstring{\ce{SD_2}}{SD2} source}
\label{sec:full_SD2}
\begin{figure}[hb!]      
    \begin{subfigure}[b]{0.48\textwidth}
        \centering
        \includegraphics[width=\textwidth]{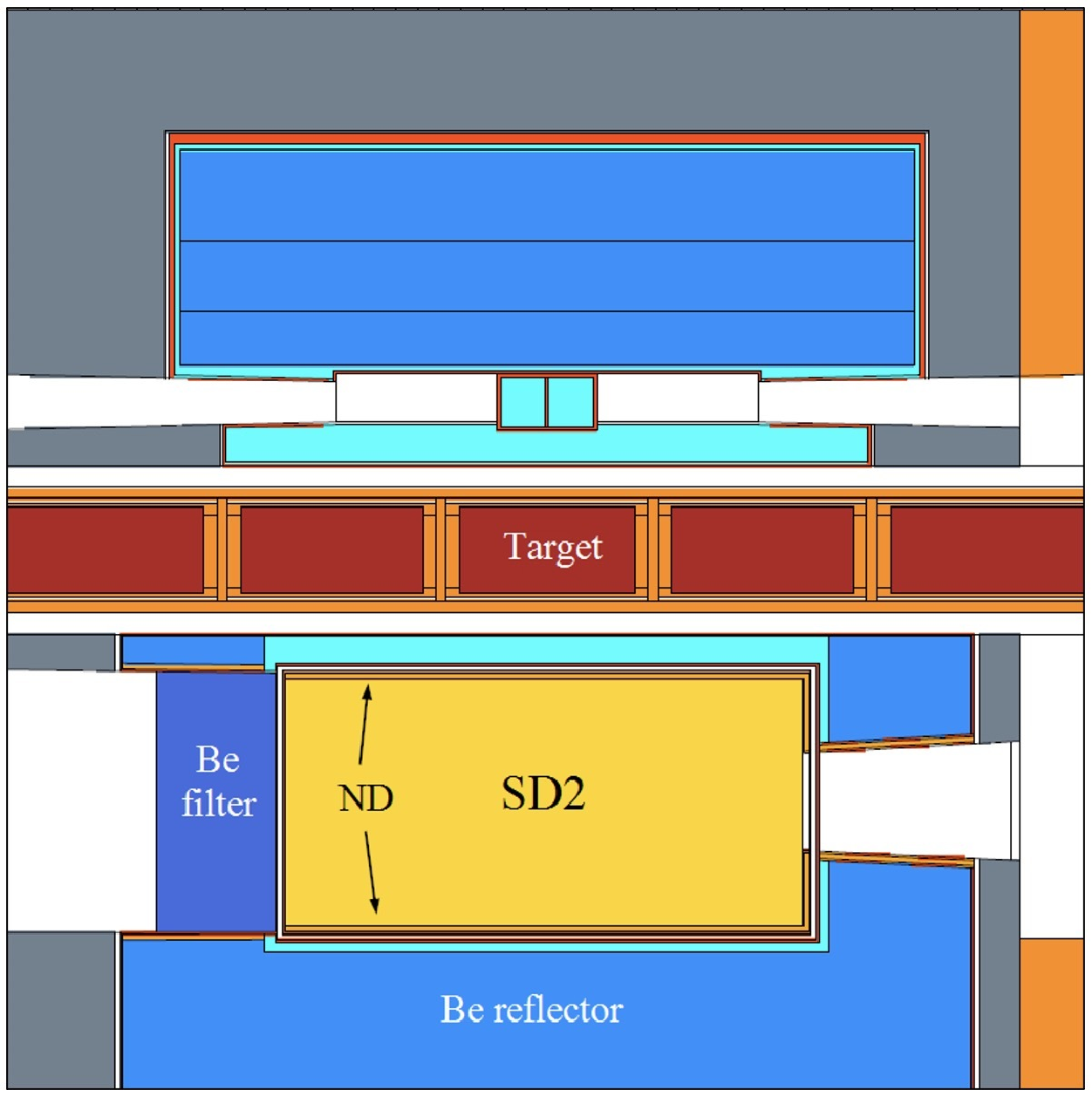}
        \subcaption{}
        \label{fig:full_SD2_XY}
    \end{subfigure}
    \hfill
    \begin{subfigure}[b]{0.48\textwidth}
        \centering        
        \includegraphics[width=\textwidth]{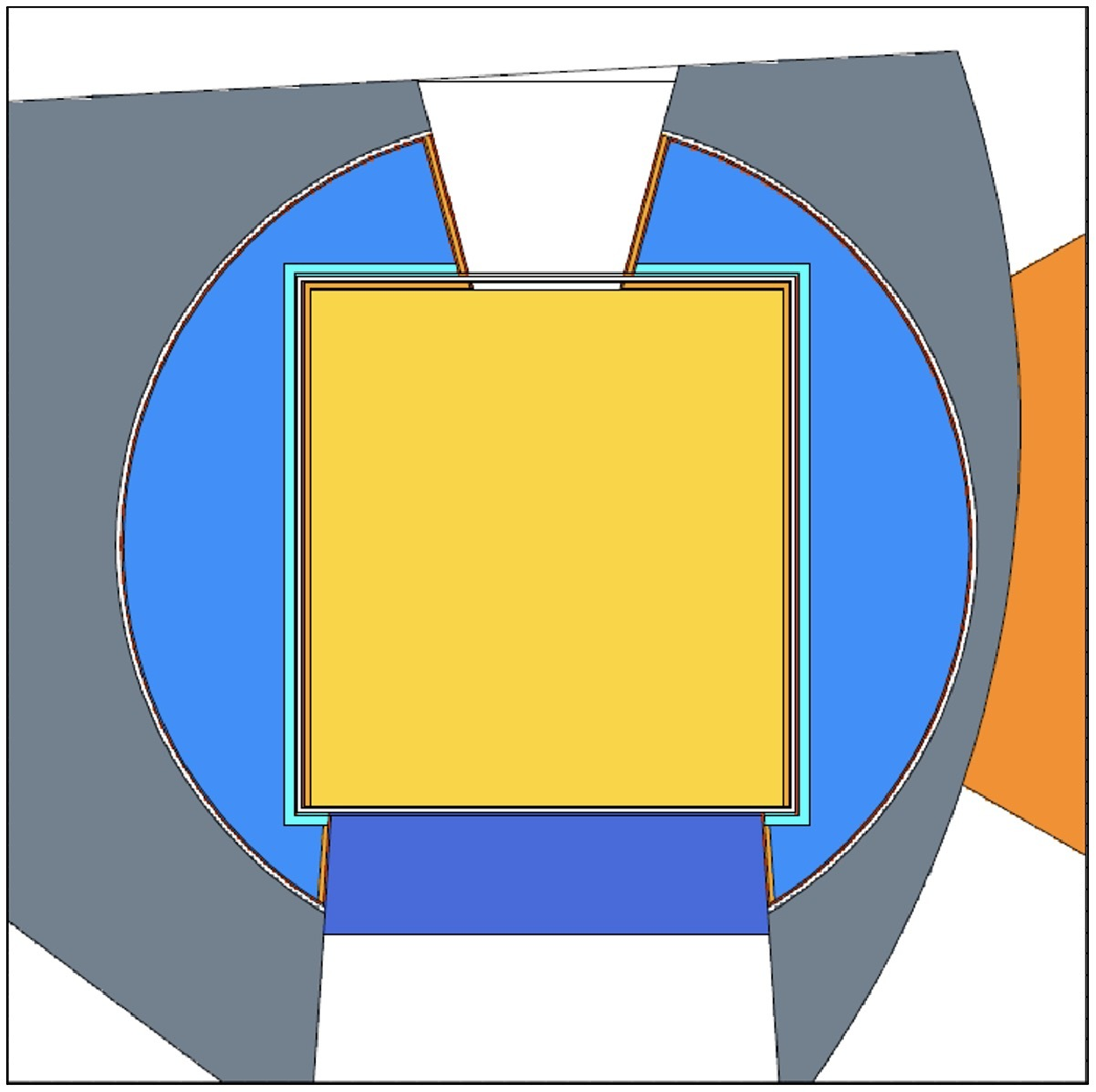}
        \subcaption{}
        \label{fig:full_SD2_XZ}
    \end{subfigure}
    \caption[MCNP model of the full \ce{SD_2} ]{MCNP model of the full \ce{SD_2} moderator (a) vertical cut, perpendicular to the proton beam direction. (b) cut parallel to the target plane with the proton beam impinging from the left.}
        \label{fig:fullSD2_geom}
    \end{figure}

Neutronic simulations for this analysis used \textsc{MCNP~6.2}~\cite{osti_1419730}. Our first tests involved a moderator which had a similar total volume to the baseline \ce{LD_2} source, where the total \ce{LD_2} moderator volume was partially displaced by \ce{SD_2}. Different configurations were tested, varying the relative amounts of \ce{SD_2} and \ce{LD_2}. The highest intensity in the VCN region was obtained for \ce{SD_2} completely replacing the \ce{LD_2}. Therefore, the focus in this section is on the properties and performance of a full \ce{SD_2} moderator. Additionally, we have investigated the effect of surrounding the \ce{SD_2} with a layer of nanodiamonds (NDs), see \cref{sec:nanodiamonds_finite_spectra}. 
The ND layer, separated from the \ce{SD_2} by a \SI{0.5}{mm} Al sheet, has a thickness of \SI{5}{mm} and a density of \SI{0.6}{g/cm^3}. \cref{fig:fullSD2_geom} presents the MCNP modelling of the geometry of the \ce{SD_2} moderator.
    
\indent 
In comparison to the \ce{LD_2} baseline, the \ce{SD_2} moderator vessel has thinner Al walls (\SI{2}{mm}), which is possible because of the lower pressure exerted by the solid block, and due to the simpler geometry, as no reentrant hole is considered in that design.
The design benefits from further improvement by the addition of a \SI{20}{K} beryllium filter on the left-hand side, facing the NNBAR experiment. Unlike the design of the \ce{LD_2} moderator, the Be filter would be added outside the vessel, thus avoiding the need to remove moderator material. The addition of that filter  leads to an increase of the performance on the NNBAR side in the range from 4 \AA~to 10 \AA~of about 40\% in comparison to the case in which no filter is considered, while sacrificing 70\% of the intensity in the range from 2.5 \AA~to 4 \AA~
and 20\% above 40 \AA. 
The addition of such a filter on the opposite side would also be beneficial for the scattering instruments' opening as this would lead to a flux increase of 5\% in the CN and VCN ranges.

\indent \cref{tab:sd2_vs_ld2} summarizes the gain factors for the best performing \ce{SD_2} model (\cref{fig:fullSD2_geom}) over the baseline \ce{LD_2} design for the different wavelength ranges of interest. 
\begin{table}[tbp!]
\centering
\caption{Gain factors of the \ce{SD_2} moderator over the \ce{LD_2} baseline design for the  NNBAR opening and the  general-purpose neutron scattering instruments opening (WP7). No internal cooling structures are considered (pure ortho-\ce{D_2} is used for both cases). \vspace{2pt}}\label{tab:sd2_vs_ld2}
    \begin{tabular}{c c c c c}
\toprule
$\lambda$  & \SIrange{2.5}{4}{\angstrom} &\SIrange{4}{10}{\angstrom} &\SIrange{10}{40}{\angstrom}& $>\SI{40}{\angstrom}$ \\
\midrule
\textbf{WP7}  &0.7 &1.1 &2.3 &27 \\
\textbf{NNBAR} &0.6 &1.3 &2.1 &14 \\
\bottomrule
\end{tabular}
\vspace{0.75cm} %only needed if the table must sit directly above a figure.
\end{table}

\begin{figure}[tb!]
    \centering
    \includegraphics[width=0.7\columnwidth]{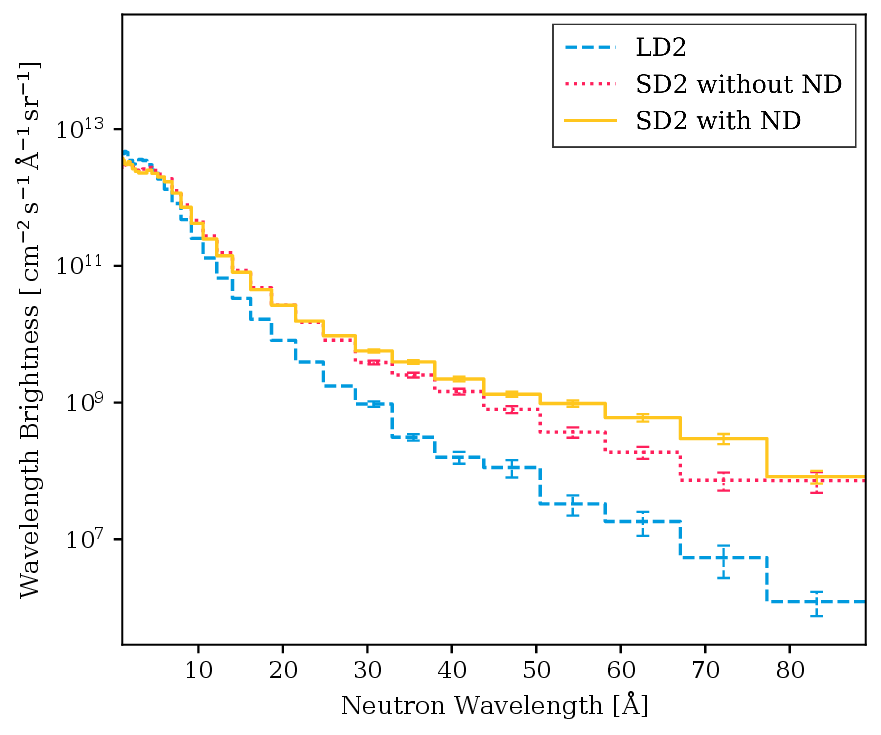}
    \caption{Brightness spectra comparison for VCNs emitted by lower-moderator designs using \ce{LD_2}, \ce{SD_2}, and \ce{SD_2} + ND reflection layer. Tallies are taken for neutrons traveling $\pm \SI{2}{\degree}$ from the normal of the emission surface; the recording surface is placed at the twister exit in the direction of the beam port for neutron scattering instruments.}
    \label{fig:sD2_spectrum_nd_nond}
\end{figure}

\cref{fig:sD2_spectrum_nd_nond} presents the comparison of the spectral brightness, calculated for the cases with and without ND, with the  \ce{LD_2} design. To obtain the absolute brightness this procedure was followed:
\begin{enumerate}
    \item the raw counts of neutrons, per incident proton on the spallation target, are recorded at the exit of the twister ($\approx \SI{50}{cm}$ from the emission surface);
    \item the counts within a cone with $\SI{2}{\degree}$ half opening angle to the surface normal are divided by the emission surface, by the solid angle defined by such cone and by the bin width in wavelength;
    \item  finally, the result is scaled by \SI{1.56e16}{proton/s}, corresponding to the average proton current on the ESS target at \SI{5}{MW}.
\end{enumerate}
The brightness is given by:
\begin{equation}
    B \left[\si{n/s/cm^2/sr/\angstrom}\right] = \frac{C\left[\si{n/nps}\right] P\left[\si{nps/s}\right]}{S\left[\si{cm^2}\right]\Omega\left[\si{sr}\right]\Delta\lambda\left[\si{\angstrom}\right]}
    \label{eq:brightness}
\end{equation}
For benchmark purpose, the brightness of the \ce{LD_2} moderator calculated with this method is compared with the one obtained using the conventional method described in \cite{zanini_design_2019,Backman_2022} based on point detector tallies. This comparison is possible only for the \ce{LD_2} case due to a technical limitation in the use of a point detector tally (F5), on which the second method is based, when neutrons are transmitted through a ND layer. The results showed a difference of \numrange[range-phrase = --]{10}{20}\% across the whole energy spectrum, a  value deeemed reasonable  as a systematic approximation of brightness.

\begin{figure}[tb!]
    \centering
    \includegraphics[width=0.7\columnwidth]{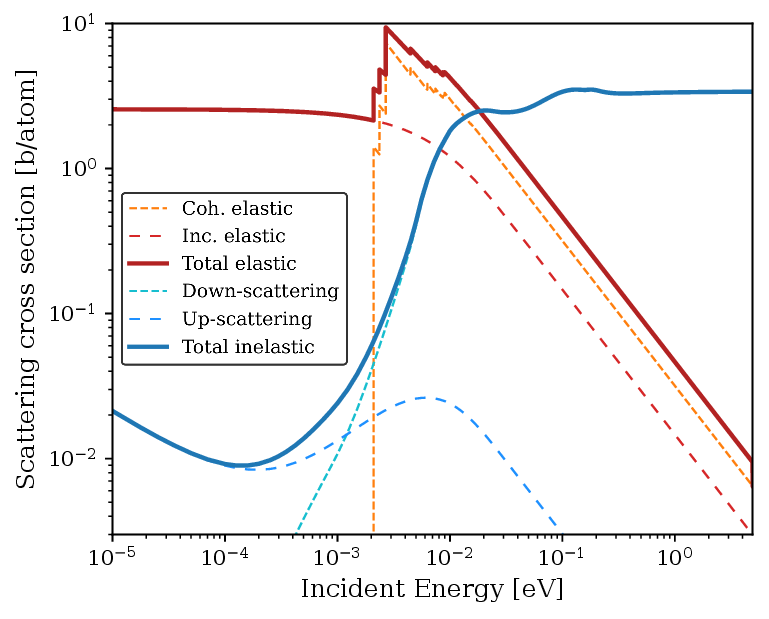}
    \caption{Thermal scattering cross section per atom for ortho-\ce{SD_2} used in this study. The elastic, both coherent and incoherent, and inelastic contributions are shown. The fall of the upscattering cross section at low energy makes \ce{SD_2} an excellent moderator for the VCN range. }
    \label{fig:xs_sod}
\end{figure}

\begin{figure}[tb!]
    \centering
    \includegraphics[width=0.7\columnwidth]{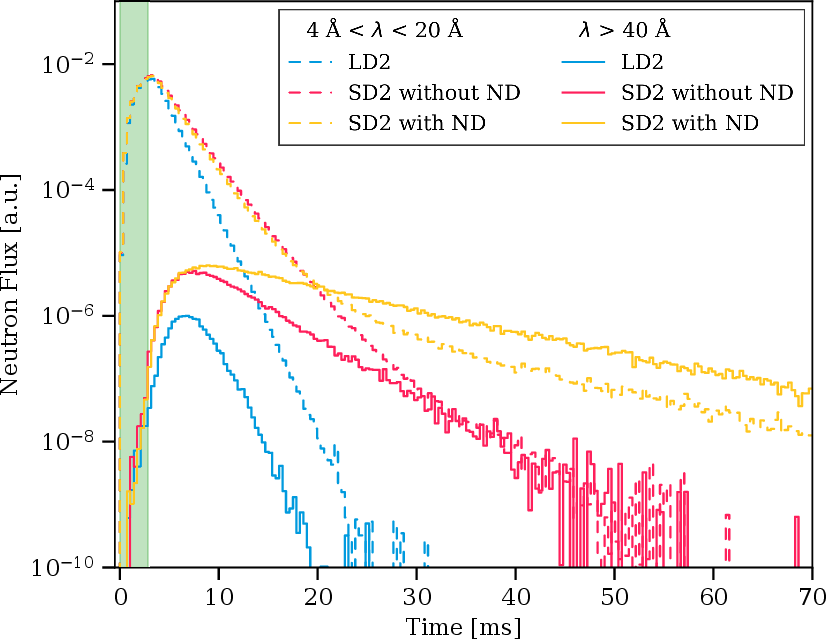}
    \caption{Temporal response of the moderator  for CNs in the range $\num{4} < \lambda < \SI{20}{\angstrom}$ (dashed lines) and VCN $\lambda > \SI{20}{\angstrom}$ (solid lines) for the \ce{LD_2} baseline and the \ce{SD_2} moderator with and without NDs. 
    %Counts are normalized to the maximum values for the $\num{4} < \lambda < \SI{20}{\angstrom}$ spectra.
    The recording surface is placed at the twister exit in the direction of the beam port for neutron scattering instruments.}
     \label{fig:time_response_sd2_vs_lD2}
\end{figure}

Two factors contribute to the improved performance of the \ce{SD_2} moderator with respect to \ce{LD_2}. The first factor is the high VCN yield of the ortho-\ce{D_2} crystal arising from the very small value of the upscattering cross-section for energies lower than \SI{20}{meV} ($\lambda >$ \SI{2}{\angstrom}) in comparison with \ce{LD_2}, as seen in \cref{fig:xs_sod}. This results in a longer mean free path for upscattering and thus in a significant deviation from a Maxwellian thermalization spectrum at long wavelengths (an estimated $\propto \lambda^{-3.5}$, compared with the Maxwellian $\lambda^{-5}$).
%, an effect that can be exploited for designing new neutron scattering instruments.
The second factor yielding  an additional gain in the VCN range is the addition of ND. Thanks to their high albedo for VCNs, NDs limit losses and trap neutrons in the moderator material. \cref{fig:time_response_sd2_vs_lD2}, presenting the different moderators' time response at the twister exit for CNs and VCNs,  shows that NDs are effective in increasing the neutron travel path. The neutron count values recorded are normalized to the peak value in the cold energy range. The VCN counts, already much higher in \ce{SD_2} than in \ce{LD_2}, start to drop after \SI{7.5}{ms}. The addition of reflective layer increases the number of reflections of VCNs, thus the outgoing flux increases. In the cold energy range, NDs are transparent and therefore no significant difference in the time response is observed.

 As another interesting result, it was found that the addition of a thin ND layer to the \ce{LD_2} baseline did not produce the same relative gains described for \ce{SD_2}. On the NNBAR side, a significant 40\% increase was calculated for $\lambda>\SI{40}{\angstrom}$, but at the expense of lower  wavelength  neutrons, which are the most relevant for the current NNBAR design. However, no gain was found on the WP7 side, while the same losses at lower wavelengths were observed. These findings suggest that in \ce{LD2} the good VCN reflective properties of ND are most likely not sufficient to counterbalance the upscattering from the medium.

For neutrons in the $\lambda >\SI{40}{\angstrom}$ range, it was challenging to obtain good statistics and MCNP simulations were very time consuming in comparison with shorter wavelengths, especially when modelling the full geometry of the ESS target monolith and shielding. In order to improve the statistics, the energy-splitting variance reduction technique was introduced. Despite the use of biasing, the relative error for this energy range remains at roughly 10\% -- even performing the simulations on the ESS computing cluster using 400+ cores.

It should be noted that these basic simulations are not taking into account the following parameters that could have a negative effect on the gains: 

\begin{itemize}
\item[-] Internal structure for cooling;
\item[-] Para-\ce{D_2} and H impurities in ortho-\ce{D_2};
\item[-] Cracks and imperfections in the \ce{D_2} crystal.
\end{itemize}

Before giving more details about the effects of those three parameters in the next sections, it should be mentioned that a full optimization of the moderator's dimensional parameters (size, shape, position, etc.) has not yet been performed. This is expected to have a positive impact the final VCN yield. 
For example, substantial gains are observed when adding NDs as reflector in the extraction. Similarly, reducing the thickness of the \SI{0.5}{mm} Al sheet located between the \ce{D_2} crystal and the ND layer and/or using a different low-absorbing material like beryllium or magnesium can also contribute to better performance \cite{chernyavsky_enhanced_2022}. A full optimization which accounts for such engineering considerations requires a wider effort and is outside the scope of this preliminary study. Nevertheless, possible detrimental factors can be assessed and, when possible, their impact is estimated at the best of the author's knowledge.

\subsubsection{Internal Cooling Structure}
\label{ics}
The most critical aspect to cover to assess the feasibility of a large-volume solid deuterium moderator is the cooling of the system made by the crystal and its aluminum vessel, which is on the scale of tens of kilowatts.  When the \ce{SD_2} temperature is exceeding \SI{5}{K}, the VCN flux is predictibly decreasing with respect to its optimum; however, it was estimated that \ce{SD_2} still outperforms the \ce{LD_2} baseline by roughly a factor of 5 at wavelengths greater than \SI{10}{\angstrom} at \SI{15}{K}, near its melting point. Nevertheless, considering greater thermal stresses, there is a greater likelihood for defect formation, and the thermal conductivity of the \ce{SD_2} itself drops between 5 and \SI{12}{K}~\cite{anghel_solid_2018} from approximately 30 to \SI{1}{W\,m^{-1}\,K^{-1}}, indicating the imperative need of keeping as much of the \ce{SD_2} bulk as possible at a temperature below roughly \SI{10}{K}.

Preliminary steady-state thermal simulations done by the HighNESS engineering team in FZJ showed that, for pure solid ortho-\ce{D_2} at \SI{2}{MW} beam power, it could be feasible to keep the moderator at the foreseen operating temperature even within the ESS environment, by implementing an embedded cooling structure. The simulations assumed a cooling structure made of two main components. First, a metallic (aluminum or beryllium) or graphite foam inlay to increase the thermal conductivity of the solid block; and second, an additional system of inner cooling pipe (or pipes) with flowing liquid helium that would extract heat from both the outer surfaces and the very deep core of the solid.

Considering both neutronic and thermal conductivity aspects, three possible material options for the inlay can be considered: aluminum, beryllium, or graphite~\cite{zhang_2022_foam_thermal_cond,Khalid_2020,GALLEGO20031461,KLETT201943,inagaki_graphite_foams_2014}. Those foams are known to have very high effective thermal conductivities~\cite{MOEINISEDEH2013117,YE2015127,CHUNHUI20181049,klett_role_2004}, performing comparably with their bulk counterparts at cryogenic temperatures~\cite{woodcraft_predicting_2005,weng_cryogenic_2021,baudouy_low_2014,woodcraft_thermal_2009,Uher1980ThermalCO}. More complex materials presenting exceptional thermal conductivity, such as diamond/copper composites \cite{CHEN2019249} or aluminum/beryllium alloys, may also be worth studying but are not considered further in the context of this work.

The steady-state thermal simulation tool of Ansys\textsuperscript{\textregistered}\, Workbench 19.2 was used for a preliminary assessment.
Material data for deuterium with variable ortho-para concentrations, helium, aluminum (99.994\% and 6061-T6) and beryllium, taken from the MPDB Material Property Database v7.91, were considered in the simulation as a function of the temperature \cite{MPDB} except for the deuterium data \cite{osti_4244810}.
The combined thermal conductivity $\lambda_{eff}$ of solid deuterium with the foam made of pure aluminum (99.994\%), aluminum alloy 6061-T6 or beryllium was calculated as follows:
\begin{eqnarray}
\lambda_{eff}=f_{A}(\Phi\lambda_{1}+(1-\Phi)\lambda_{2})+\frac{1-f_{A}}{\frac{\Phi}{\lambda_{1}}+\frac{1-\Phi}{\lambda_{2}}} ,
\label{eq:thermalconductivity}
\end{eqnarray}
where $f_{A}=0.33$ is the Bhattacharya correlation factor, $\Phi=0.85-0.95$ the porosity of the foam, $\lambda_{1}(T)$ the thermal conductivity of solid deuterium and $\lambda_{2}(T)$ the thermal conductivity of the foam.
\begin{table}[tb!]
	\centering
    	\caption{Heat conductivity $\lambda_{eff} [\si{Wm^{-1}K^{-1}}]$ in different porosities $\Phi$ for pure aluminum (99.994\%).}	\label{tab:thermalconductivity}	
	\begin{tabular}{ l c c c}
		\toprule
%   $f_{A}$ & 0.33 & 0.33 & 0.33 \\
	$\Phi$ & 0.85 & 0.93 & 0.95 \\
	\midrule
	$\lambda_{eff}\ (\SI{4}{K})$ & 85.97 & 41.02 & 29.78  \\
	$\lambda_{eff}\ (\SI{10}{K})$ & 200.24 & 101.90 & 77.35  \\
	$\lambda_{eff}\ (\SI{20}{K})$ & 213.12 & 99.59 & 71.2 \\
	\bottomrule
	\end{tabular}
\end{table}

A uniformly distributed volumetric heat source induced by the spallation reaction was considered over the moderator volume. Only steady-state simulations have been performed (no pulsed heat). Two cases where studied, one corresponding to full-load operation (\SI{5}{MW} proton beam power Q = \SI{60}{kW}) and one to partial-load operation (\SI{2}{MW} proton beam power Q = \SI{24}{kW}).
The heat is removed by a surface cooling only. Helium is the cooling medium, having an average temperature of \SI{3.75}{K}. The heat transfer coefficient of the He-wall is considered to be infinite. The thermal contact of the foam to the aluminum vessel wall is also assumed as optimal.
All of the approximations listed need to be checked performing detailed simulations. 

\begin{figure}[tbp!]
    \centering    \includegraphics[width=0.7\columnwidth]{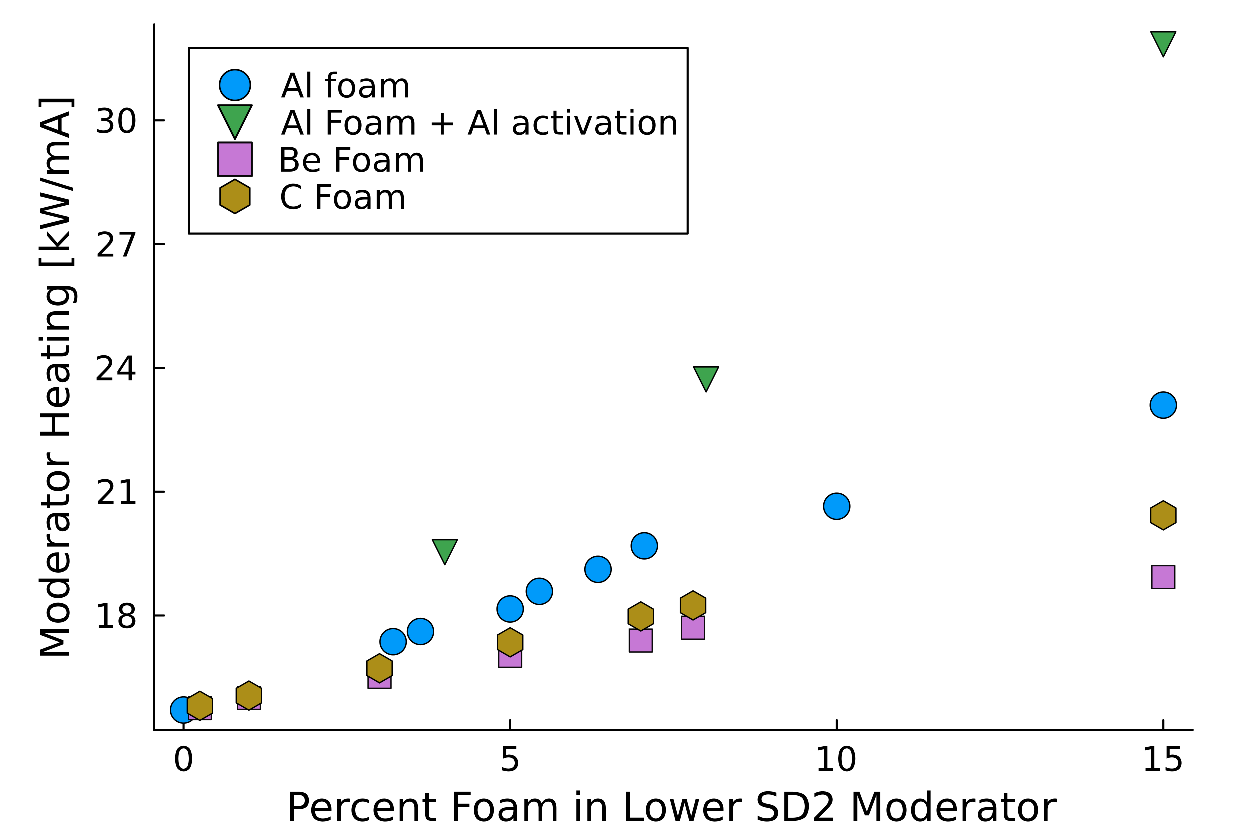}
    \caption{Lower-moderator cell heat deposition for metallic (Al and Be) and graphite (C) foam configurations, divided by the proton current in \si{mA} impinging on the target (for a \SI{5}{MW} beam). The green triangles show results accounting for additional heating due to \ce{^{27}Al} activation.}
    \label{fig:test_heat_foams}
\end{figure}

Neutronics simulations aimed at investigating the effect of metallic foam on the VCN production are summarized in \cref{tab:sd2_pls_alm}. The system of crystal and foam was first approximated in MCNP by a uniform mixture of 85\% volume fraction of \ce{SD_2} and 15\% foam inlay. The gains are reported as count ratios at \SI{2}{m} (Beam Port Tally) over the baseline \ce{LD_2} moderator with 4\% aluminum added, which approximates the \ce{LD_2} cooling-flow channels, allowing for fairer comparison. The possibility of reducing the amount of foam material by having an inlay with different densities between the core region and the outer corners, where it is less important, was also explored. In these calculations, the boundary between these two regions is a sphere of \SI{10}{cm} radius, and the volume percentage of the outer foam is reduced from 15\% to 7\%. This last configuration has a significant advantage over the uniform case in terms of neutron yield, while keeping the required cooling power almost unchanged. We also studied the increase of heat deposition in the moderator cell as a function of the foam volume percentage (\cref{fig:test_heat_foams}). 

According to the results presented, graphite and beryllium outperform aluminum regarding neutronic, self-heating, and thermal conductivity. In terms of performance, beryllium is a promising candidate but manufacturing  a beryllium foam would be challenging due to the higher melting point and non-standard safety measures in handling this material. Graphite then appears as an appealing candidate material in general terms. 

However, at roughly \SI{3200}{W m^{-1} K^{-1}}~\cite{Marquardt2002} near the melting point of \ce{SD_2}, the thermal conductivity of high-quality bulk beryllium at the relevant temperature range of 5--20\,K is particularly high with an effective thermal conductivity value of roughly \SI{200}{W m^{-1} K^{-1}} for a porosity of approximately 85\%;  the thermal conductivity of graphite foam is likely to be lower by at least a factor of two ~\cite{ZHAO2022544,klett_role_2004}. Depending on the purity of the material, it should also be mentioned that the thermal conductivity for aluminum and beryllium can vary by orders of magnitude~\cite{woodcraft_predicting_2005,touloukian_thermal_conductivity_1970}, for example, the thermal conductivity of ultra-high-purity aluminum is \SI{35,000}{W m^{-1} K^{-1}} at \SI{10}{K}~\cite{tanaka2013refining}.

A foam with open pores (I) or a 3D printed grid structure (II) would be two interesting options worth investigating regarding the manufacturability of the inlay~\cite{parveez_2022_openporefoamsreview,HUTTER2011_3d_sls_foams}. In the case of option (I), a melting of the optimized alloy can be frothed through the addition of a foaming agent. The density and porosity can be tuned homogeneously for the inlay in this process. The second option (II) would consist of a grid structure manufactured using the selective-laser-sintering (SLS) 3D printing technique; a thin layer of metal powder is locally sintered by a laser beam. A defined three-dimensional geometry can be created by repeating the operation layer-by-layer. The advantage of this method in comparison with option (I) is the capability to adapt the geometry to the need of an optimal thermal conduction while minimizing the reduction of \ce{SD_2} volume. This is overcoming the limitations due to the homogeneous properties of option (I). 
\begin{table}[tb]
\centering
%\begin{threeparttable}
\caption{Intensity gain factors in the \ce{SD_2} lower moderator at different volume percentages of aluminum (Al), beryllium (Be), and graphite (C) approximating embedded foam densities. Counts are measured at the Beam Port Tally and gains are reported as ratios over the baseline with 4\% aluminum added. In the non-uniform cases (second rows), the boundary is a higher-density 10-cm-radius sphere encompassing the \ce{SD_2} core; only such non-uniform trials are reported for graphite.}\label{tab:sd2_pls_alm}
\begin{tabular}{ l c c c c c c}
\toprule
$\lambda$ range & Al \% & \SIrange{2.5}{4}{\angstrom} &\SIrange{4}{10}{\angstrom} &\SIrange{10}{40}{\angstrom}& $>\SI{40}{\angstrom}$ \\
\midrule
\multirow{2}{*}{\textbf{WP7}} &15\% &0.57&0.68 &0.95  &2.5 \\
&15/7 \%& 0.68& 0.94 & 1.52  & 5.73 \\
\multirow{2}{*}{\textbf{NNBAR}} &15\% &0.48&0.71 &0.95 &1.9  \\
&15/7 \% & 0.56& 0.96 & 1.50 & 4.15  \\
\bottomrule
\end{tabular}

\begin{tabular}{ l c c c c c c} 
\\ \toprule
$\lambda$ range & Be \% & \SIrange{2.5}{4}{\angstrom} &\SIrange{4}{10}{\angstrom} &\SIrange{10}{40}{\angstrom}& $>\SI{40}{\angstrom}$ \\
\midrule
\multirow{2}{*}{\textbf{WP7}} &15\% &0.64&1.10 &2.12 &10.0 \\
&15/7 \%& 0.77 & 1.27 & 2.47 & 18.3 \\
\multirow{2}{*}{\textbf{NNBAR}} &15\% &0.56&1.10 &1.93 &10.1 \\
&15/7 \% & 0.63& 1.26 & 2.33  & 13.8 \\
\bottomrule
\end{tabular}
    \begin{tabular}{ l c c c c c c} 
\\ \toprule
$\lambda$ range & C \% & \SIrange{2.5}{4}{\angstrom} &\SIrange{4}{10}{\angstrom} &\SIrange{10}{40}{\angstrom}& $>\SI{40}{\angstrom}$ \\
\midrule
\multirow{1}{*}{\textbf{WP7}} &15/7 \% &0.79& 1.28  & 2.57 & 22.0  \\
\multirow{1}{*}{\textbf{NNBAR}} &15/7 \% &0.64& 1.25 & 2.42 & 16.1  \\
\bottomrule
\end{tabular}
%\end{threeparttable}
\end{table}
It should be noted that this heatsink inlay is not intended to act as a structural support, its only goal is to improve the thermal conductivity. When incorporating such an inlay into the moderator design, a critical aspect is its thermal coupling with the surrounding moderator vessel walls and the liquid helium channel. This coupling could be realized by a defined press fit or a soldering process with a matched plumb alloy. Inlay-type (II) would allow for the printing of a thin-wall structure in the geometry to enlarge the connecting interface surface. This would increase the thermal coupling compared with the press fit and soldering option. Further experimental investigations will be necessary to investigate more in details this point. 

\subsubsection{Hydrogen and Para-\ce{D_2} impurities}
Thanks to its small nuclear mass and absorption cross section, deuterium is one of the best neutron moderator materials. The importance of having high-purity ortho-\ce{D_2} crystals for UCN production through the use of \ce{SD_2} have been highlighted in the literature~\cite{liu_ultracold_2000,morris_measurements_2002}. Hydrogen and para-deuterium are the elements usually considered as impurities due to their respectively higher absorption and upscattering cross sections \cite{liu_ultracold_2000}. However, in view of the similarities between the two spin isomers in the VCN energy range and the achievable ortho-\ce{D_2} purity, the impact of these impurities on the VCN yield are expected to be small, as discussed in the following.

\begin{figure}[tbp!]
    \centering
    \includegraphics[width=0.8\columnwidth]{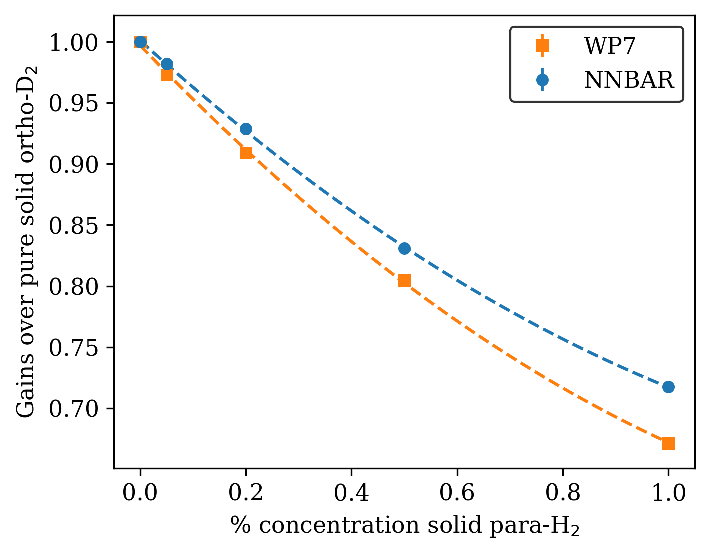}
    \caption{Effects of introducing small vol\% of solid para-\ce{H_2} in the ortho-\ce{D_2} crystal for neutrons with $\lambda > \SI{40}{\angstrom}$. }
    \label{fig:sensitivity_solid-pH2}
\end{figure}

The effect of hydrogen impurities was simulated by adding increasing percentages of solid para-\ce{H_2} at \SI{5}{K} in the \ce{SD_2}. Hydrogen impurities have typically been found in the form of HD and not para-\ce{H_2}. However, since this was the only library available and impurity concentrations in \ce{D_2} are usually around 0.2\%~\cite{morris_measurements_2002}, those data were considered to give a reasonable approximation of the absorption problem. The impact of the impurities was assessed at the exit of the twister where statistical uncertainties were lower. The results are plotted in \cref{fig:sensitivity_solid-pH2}. For concentrations ranging up to 1\%, an average drop in neutron counts of roughly 0.35\% was highlighted for $\lambda > \SI{40}{\angstrom}$ for every 0.01\% increase of the hydrogen content. For the scattering instruments opening, this linear approximation is less accurate, showing a relatively larger loss, especially for hydrogen concentrations higher than 0.5\%. An impurity level of 0.2\% is achievable in \ce{D_2}. Considering more pure \ce{D_2} reaching 0.05\%, these effects would hardly downgrade the gains observed considering the ideal crystal.

Similarly, the addition of 1\% para-\ce{D_2} to the material mix did not lead to significant effects exceeding the statistical uncertainty at the exit of the twister. However, due to recombination following the radiolysis induced by fast neutrons and gammas, it could be expected that the para-\ce{D_2} concentration will increase as a function of the operation time. After the radiolysis of an ortho-\ce{D_2} molecule, according to the natural proportion of the two spin isomeric forms, the recombination will produce a para-\ce{D_2} molecule 33\% of the time \cite{iverson_radiolysis}. Preliminary calculations performed using a solid para-\ce{D_2} thermal scattering library at \SI{5}{K} showed small losses (less than 5\%) for neutrons with $\lambda>\SI{40}{\angstrom}$ when adding 30\%  para-\ce{D_2}. As the two spin isomers' cross sections are similar in the VCN energy range, this last result is not surprising. Even though the neutronic performance at \SI{5}{K} is reassuring, a high content of para-\ce{D_2} would drastically decrease the thermal conductivity of the crystal, leading to difficulties to keep it at \SI{5}{K}. The neutronic effects of para-\ce{D_2} at higher temperatures should then be further investigated in a dedicated study.

\subsubsection{Crystal Imperfections}
Growing a perfectly transparent \ce{SD_2} crystal is clearly known as a challenge \cite{anghel_solid_2018,morris_measurements_2002,atchison_measured_2005,thomasBrys,IR_thesis,korobkina_growing_2022}. The goal of perfect optical transparency is hindered by the crystal defects due to thermal stress. If high-purity ortho-\ce{D_2} is achievable, cracks and frost buildup on the surface seem inevitable. The former are produced during the cooling phase and also during temperature cycling, while the latter is a snow-like layer on the \ce{SD_2} surface created by a small amount of \ce{D_2} undergoing sublimation during a proton pulse.

As they are strongly affecting the UCN yield, those defects have been extensively studied but there is limited evidence of the interaction with higher energy neutrons in the VCN range. In this section, we address this point based on the available knowledge, keeping in mind that a more detailed study would be necessary for further applications.

The impact of cracks on the \ce{SD_2} transparency has been studied in \cite{morris_measurements_2002,thomasBrys}. The presence of cracks induces a sharp change in potential $\Delta V_F = \SI{104}{neV}$ at the boundaries between \ce{D_2} and vacuum, resulting in elastic scattering at the interface, an  effect that has been observed and quantified in a SANS experiment with cold neutrons \cite{thomasBrys}. UCN can be totally reflected due to their low energies, thus reducing the transmitted intensity. The same work also shows how the transmission of a directional VCN beam is heavily impacted by a deteriorated crystal. The small-angle elastic scattering at the boundaries explains the cross section increase but the effect that a a smaller mean-free path could have on the moderation and isotropic emission of VCN is less clear.

One possibility is that assuming a smaller mean-free path could ultimately lead to the calculation of sub-optimal dimensions for the moderator. Using the method adopted for ND in \cite{rizzi_SANSND}, SANS could be added to the \ce{SD_2} thermal scattering library in order to model cracks. The experimental Porod exponents taken from \cite{thomasBrys,IR_thesis} could be used as starting point and would allow modelling a real crystal under different cooling conditions. In any case, less damage to the crystal was reported when the cooling from the liquid phase was slow and the temperature cycling was kept varying between \SI{5}{K} and \SI{10}{K}.

Surface degradation has been extensively studied from both theoretical and experimental perspectives in \cite{anghel_solid_2018}. The creation of a frost layer at the \ce{SD_2}-vacuum interface is due to \ce{D_2} sublimation. This frost layer is then poorly cooled and poorly attached to the \ce{SD_2} surface. As the number of layers increases with the number of proton pulses, the optical opacity of the surface increases. A procedure called conditioning allows sublimating and resublimating the frost as better-quality \ce{SD_2}, while recovering the UCN yield. The physics behind the neutron loss is resulting from an alteration of the optical potential shape (from a stepped to a smooth profile), specular and non-specular scattering (due to roughness \cite{atchison_diffuse_2010}), and diffusion from randomly oriented crystal facets. Each of these mechanisms prevents the extraction of UCN, but are energy-dependent processes described by the optical laws of reflection \cite{anghel_solid_2018}. In particular, the probabilities for reflection $P_R$ and transmission $P_T$ an interface are defined by:
\begin{eqnarray}
    P_R &=& \left|\frac{\sqrt{E_\perp}-\sqrt{E_\perp  - \Delta V_F}}{\sqrt{E_\perp}+\sqrt{E_\perp  - \Delta V_F}}\right|^2\\
    P_T &=& 1- P_R
\end{eqnarray}
where $\Delta V_F$ is the change in neutron optical potential at the interface \ce{D_2}-vacuum and $E_\perp$ is the component of kinetic energy perpendicular to the boundary surface before the interaction. For VCN energies of the order of \SI{e-5}{eV}, it is clear that $P_T$ is almost equal to 1 for almost all emitted neutrons. Experimental results reported in \cite{anghel_solid_2018} show that already the fastest UCN (\SI{320}{neV}) are less impacted by the frost layers, leading to a harder spectrum after continuous pulse operations.
\subsubsection{\ce{SD_2} Source Summary}
%taken from the conclusion of the paper
Simulations for a full-\ce{SD_2} source predict order-of-magnitude gains for $\lambda> \SI{40}{\angstrom}$ neutrons, along with gains exceeding a factor of two in the $\num{10}<\lambda<\SI{40}{\angstrom}$ range, in comparison with a conventional \ce{LD_2} moderator design of similar volume and following the same engineering constraints required by the existing ESS layout. A significant fraction of these gains are owing to an ND reflecting layer encasing the source, which have a high albedo for VCNs.
 
Although it is clear that both cracks and surface degradation will need a dedicated study which takes into account the specific conditions and constraints at ESS, it is likely that imperfections in \ce{SD_2} -- which are generally considered detrimental for UCN production -- should not significantly affect higher energies in the VCN range.

Novel heat-extraction solutions are required to make the proposed design feasible, with preliminary studies indicating that embedding a foam structure of beryllium or graphite into the moderator volume would be viable with the ESS initial operating proton beam power of \SI{2}{MW}. Engineering such a cooling system for the design power of \SI{5}{MW} may present greater difficulties, and further study would be required.

Implementation of a such VCN moderator can be expected to open new avenues of research in neutron scattering and fundamental physics studies with neutrons. 

\subsection{Combined option}
\label{sec:combined}
The idea of a VCN source combining an \ce{SD_2}-based VCN converter with an existing CN source was  proposed  at the first HighNESS workshop on UCN and VCN Sources at ESS~\cite{nesvizhevsky_why_2022}. The rationale behind the concept is to maximize the VCN flux to the NNBAR experiment. A dedicated VCN source could overcome the well-known low phase-space density (the neutron flux decreases as the square of longitudinal neutron velocity in a cold source), while retaining the sensitivity increase as the square of the free flight time ($\approx \lambda^2$). The net gain factor would then be simply proportional to the increase in the phase-space density over the cold source. The original concept is shown in \cref{fig:original_valery} and it can be described as follows:
 \begin{figure}[tbp!]
    \centering
    \includegraphics[width=0.7\columnwidth]{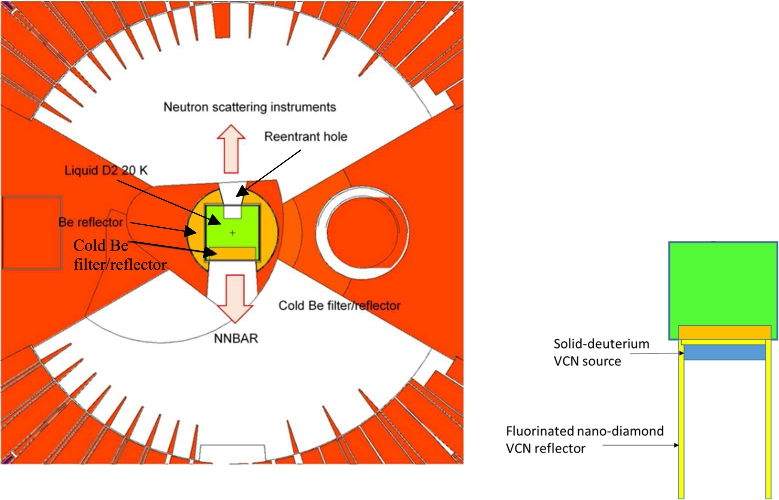}
    \caption[Original concept of the \ce{SD_2} converter]{Original concept of the \ce{SD_2} converter as presented in \cite{nesvizhevsky_why_2022}}.
    \label{fig:original_valery}
\end{figure}
\vspace{0.2cm}
\begin{itemize}
    \item[-]  An \ce{SD_2} VCN converter at a temperature of {$\approx$}4\,K is placed in front of the \qtyproduct{24x40}{cm} NNBAR opening.
    \item[-] A long channel filled with fluorinated detonated nanodiamonds (F-DNDs) is placed at the exit as extraction system, increasing the total flux of VCN at \SI{2}{m} due to multiple reflections from the walls and enhanced directional extraction to the experiment.
    \item[-] A thin F-DND layer is placed between the \ce{LD_2} source and the \ce{SD_2} converter to increase the chance of moderation inside the converter volume. At the same time, such a thin layer is nearly transparent for CNs, hence the incident CNs are virtually unaffected.
\end{itemize}
\vspace{0.2cm}

Contrary to UCN extraction with \ce{SD_2}, VCNs penetrate through the inhomogeneities in the \ce{SD_2} bulk, and are thus more easily extracted with minimal losses. %This should not be taken to mean the a hybrid design concept would not work with other converter materials.

The source flexibility and the ease of implementation within the current \ce{LD_2} cold source make this an interesting design for a general-purpose VCN beamline. Therefore, the performance of a possible converter on the WP7 has also been studied.

\subsubsection{NNBAR side}
\begin{figure}[!tb]
    \centering
    \includegraphics[width=0.6\textwidth]{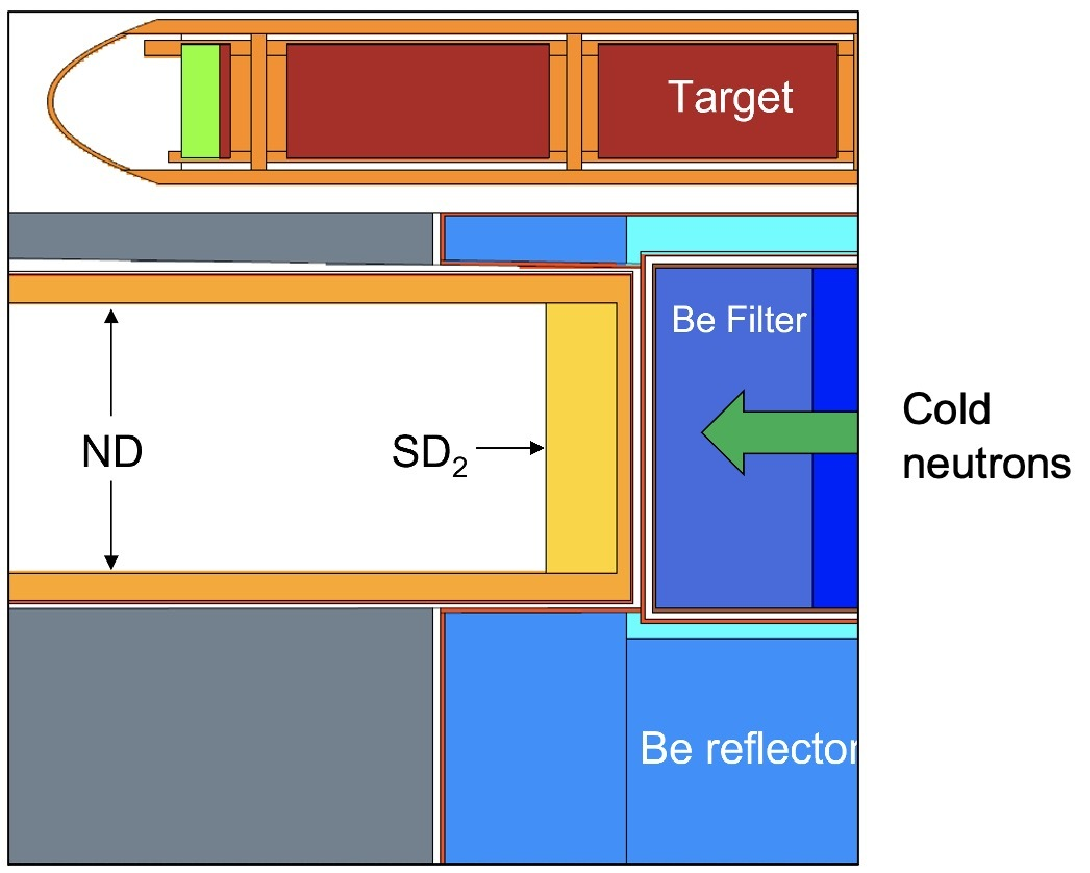}
    \bigskip
    \raisebox{.4\height}{\begin{tabular}[b]{l l}
    \toprule
      \multicolumn{2}{c}{Specifications} \\ 
      \midrule
      Converter thickness & \SI{5}{cm} \\
      Converter vessel & \SI{0}{mm} \\
      ND reflector (sides) & \SI{2}{cm} \\
      ND reflector (back) & \SI{1}{cm} \\
      External Al case & \SI{2}{mm} \\
      Internal Al walls & \SI{0}{mm} \\
      \bottomrule
    \end{tabular}}
    \captionlistentry[table]{A table beside a figure}
    \captionsetup{labelformat=andtable}
    \caption[MCNP model of the preliminary test run]{(Left) MCNP model of the preliminary test run to check the separate and combined effect of ND and the \ce{SD_2} block. (Right) Parameter table for this model.}
    \label{fig:preliminary_spec}
  \end{figure}

As a starting point, the simplest study one can imagine is to measure the effect of a thin converter on the VCN flux at the beam port, which is located \SI{2}{m} from the center of the moderator. For this test, we put \SI{5}{cm} of \ce{SD_2} in a 170-cm-long ND-filled channel with 2-cm reflector thickness on the sides and \SI{1}{cm} in the direction of the cold neutron beam. The reason for having a thick ND layer is to increase the diffusive transport of VCNs. The very cold energy range is far from the quasi-specular regime; thus, the intense albedo from the powder is exploited instead. The first tests did not have any structure around the converter and for the inner walls. This is clearly an oversimplification, since in reality both the \ce{D_2} (gas) and the NDs (powder) must be stored in a container, but it allows for a verification of the predicted enhancing effect, neglecting the engineering complications. The model used is shown in detail in \cref{fig:preliminary_spec}.

The performance of the converter is measured with a current tally (F1) through a fixed surface placed at the beam port with several energy-grouped resolutions and coarse angle bins. The effect on the neutron counts was studied separately for each component of the system. The  \cref{tab:preliminary_results} summarizes these preliminary results at all angles.

\begin{table}[tbp]
  \centering
  \caption[Preliminary converter results at all angles for NNBAR]{Preliminary results at all angles of the \ce{SD_2} converter design on the NNBAR side. The tally is measuring the counts at all angles through a fixed surface placed at the beam port. Gains are measured over the third optimized baseline (see~\cref{sec:3iter}).}
    \begin{tabular}{c cccc}
    \toprule
     Gains & \multicolumn{1}{l}{ Al casing} & \multicolumn{1}{l}{\ce{SD_2} only} & \multicolumn{1}{l}{NDs only} & \multicolumn{1}{l}{\ce{SD_2} + NDs} \\
     \midrule
    \phantom{2}$\SI{4}{\angstrom} < \lambda < \SI{10}{\angstrom}$ & 0.98 & 0.54 & 1.71 & 0.87 \\
    $\SI{20}{\angstrom} < \lambda < \SI{40}{\angstrom}$ & 0.95 & 0.71 & 0.84 & 1.76 \\
    $\lambda > \SI{40}{\angstrom}$ & 0.86 & 0.93 & 0.44 & 16.9 \\
    \bottomrule
    \end{tabular}%
  \label{tab:preliminary_results}%
\end{table}%

First, the Al case was added in the opening, which produced overall small losses in the energy ranges studied. Then, only the \ce{SD_2} crystal was inserted, without the surrounding NDs. As a result, cold neutron (\SIrange{4}{10}{\angstrom}) and VCNs from \SIrange{20}{40}{\angstrom} are less, while an increase in the coldest neutron counts ($\lambda > \SI{40}{\angstrom}$) is the first sign of enhanced VCN production over the \ce{LD_2} baseline.

Similarly, results for ND reflector but no \ce{SD_2} are shown. In this third case, it is possible to observe an increase in the cold counts for the ND guide only, which is due to the quasi-specular reflection. On the other hand, huge losses are observed in the VCN energy region. The difference with the previous model, and cause of the losses, is in the additional layer at the emission surface, which is back-scattering and diffusing the VCNs coming from the \ce{LD_2} moderator.

Finally, the two pieces are coupled, \ce{SD_2} and NDs in the Al case, and the VCN counts at the beam port almost double from \SIrange{20}{40}{\angstrom} and increase by a factor 17 above \SI{40}{\angstrom}. In the next section, this tentative simple case is improved by an optimization of the parameters.  

\subsubsection{Optimization of the ideal case}
\begin{figure}
    \centering
    \includegraphics[height=0.32\textheight]{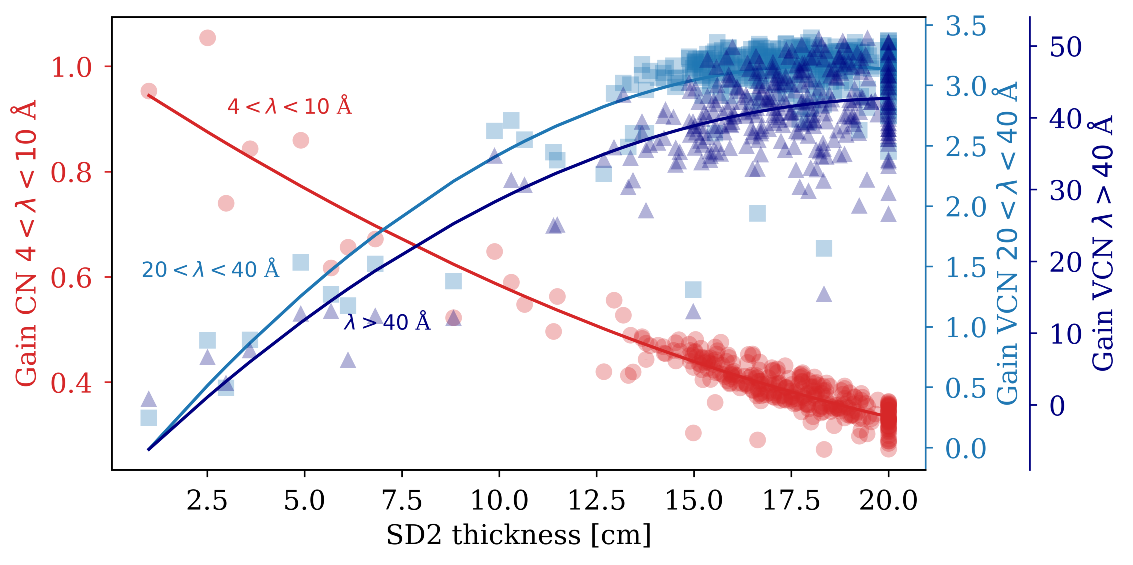}
    \includegraphics[height=0.32\textheight]{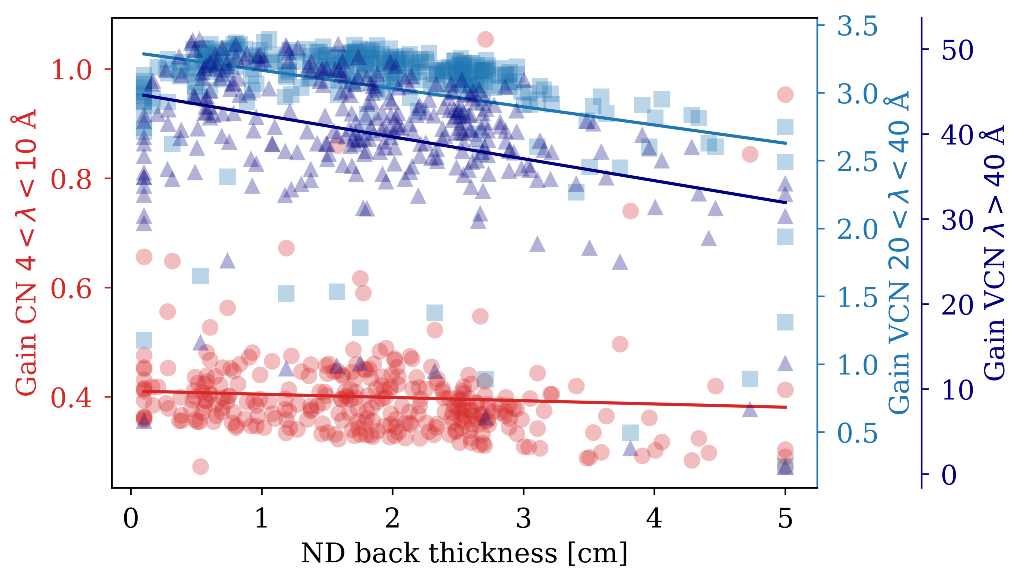}
    \includegraphics[height=0.32\textheight]{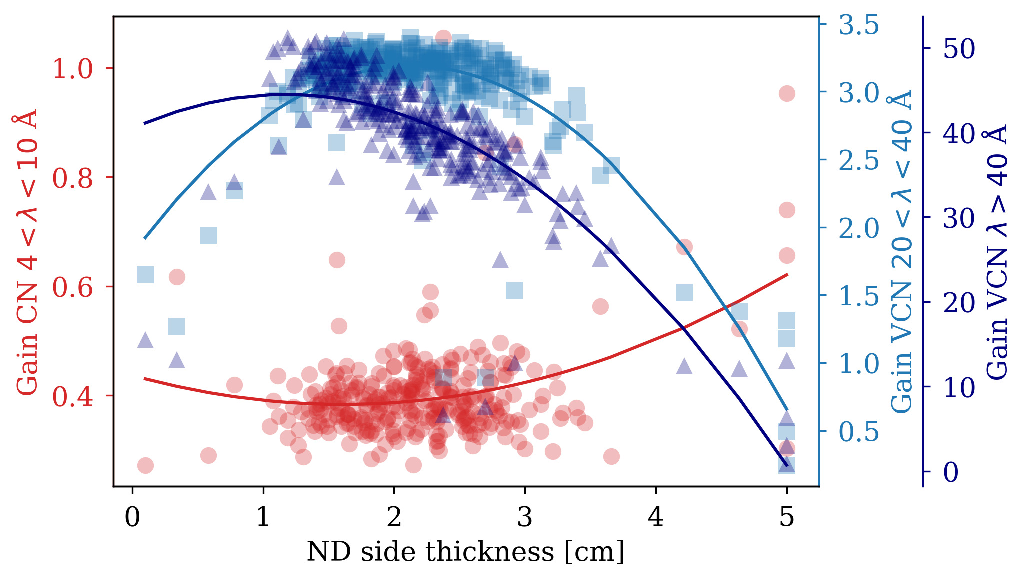}
    \caption[Results of the multi-dimensional optimization.]{Results of the multi-dimensional optimization. Each run of the optimization is represented by three vertically aligned points, one for each energy-group of interest and each one with its own scale for the gains over the \ce{LD_2} baseline. Trend lines are shown with the sole purpose of guiding the reader's eye.}
    \label{fig:optimization_converter}
\end{figure}

For a robust and methodical approach to the optimization task, Dakota~\cite{Dakota_6.18} and the EGO algorithm were used. The optimization parameters were the thickness of the \ce{SD_2} block and the thickness of the ND layer both on the sides of the guide-like system and behind the \ce{SD_2} (in front of the emission source) designated as side and back thicknesses, respectively. The FOM chosen to drive the algorithm is the VCN counts at the beam port between 20 and \SI{40}{\angstrom} and at all angles. In fact, it is more convenient to use this tally since it converges faster while following approximately the same behavior as the VCNs above \SI{40}{\angstrom}. The results of this multi-dimensional optimization are shown in \cref{fig:optimization_converter}.

Before discussing the results, the content of the plots in~\cref{fig:optimization_converter} should be clarified:
\vspace{0.2cm}
\begin{itemize}[leftmargin=1cm]
    \item[-] Each run of the optimization is represented by three vertically aligned points, one for each energy-group of interest and each one with its own scale for the gains over the \ce{LD_2} baseline.
    \item[-] Since we are showing one parameter at a time, the presence of different counts values for the same parameter value is to be interpreted as a run where one (or both) of the other parameters was changed.
    \item[-] The lines shown are mere trend lines with the sole purpose of guiding the reader's eye and not fitted models.
    \item[-] The EGO algorithm does not sample uniformly the parameter space, hence the larger clusters of points visible in the plots are formed to better sample the region around the maximum of the FOM. This is also why this approach can be much faster than a simpler evaluation of a parameter grid.
\end{itemize}     
\vspace{0.3cm}

The first important result of the optimization is that large blocks of \ce{SD_2} produce the highest VCN yields. Here \SI{20}{cm} was the upper boundary given to Dakota for the thickness, but it is reasonable to assume that there would be higher gains for larger crystals. However, with approximately \SI{14}{L} of \ce{SD_2}, this moderator has already unprecedented dimensions for a crystal at \SI{5}{K}, with non-negligible issues pertaining to cooling and engineering (the estimated heat deposition from prompt neutron and gammas is \SI{2.5}{kW}). The cost for the high gains in the VCN energy range are considerable losses in the cold range around \SI{4}{\angstrom}. A possible explanation for these losses is in the large \ce{SD_2} elastic cross section (see \cref{fig:xs_sod}) that decreases the mean free path, effectively making the cold source appear further away from the beam port and increasing the chances of absorption or leakage.

Additionally, the thickness of the crystal seems to be the most important parameter by looking at the range of the results, while the thickness of the back ND layer is the least crucial for an optimal design. The latter is due to the relatively high transmission fraction of cold neutron through thick ND layers. Meanwhile, once the neutrons are moderated to VCN energy, that layer, even when few millimeters thin, suddenly becomes opaque, reducing the chances of leakage.

\begin{figure}[!tb]
    \centering
    \includegraphics[width=0.4\textwidth]{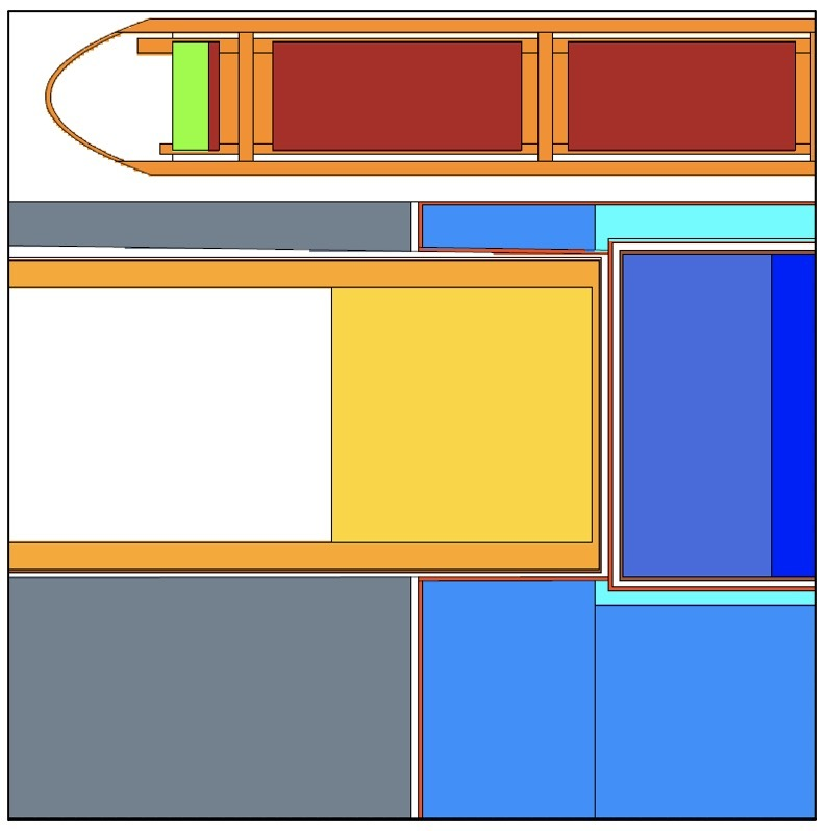}
    \qquad
    \raisebox{0.1\height}{\begin{tabular}[b]{l l}
    \toprule
      \multicolumn{2}{c}{Specifications} \\ 
      \midrule
      \textbf{Converter thickness} & \SI{20}{cm} \\
      Converter vessel & \SI{0}{mm} \\
      \textbf{ND reflector (sides) }& \SI{2}{cm} \\
      \textbf{ND reflector (back)} & \SI{0.55}{cm} \\
      External Al case & \SI{2}{mm} \\
      Internal Al walls & \SI{0}{mm} \\
      Prompt heat-load & \SI{2.5}{kW} \\
      \bottomrule
    \end{tabular}}
    \captionlistentry[table]{A table beside a figure}
    \captionsetup{labelformat=andtable}
    \caption[MCNP model of the optimized converter]{(Left) MCNP model of the optimized converter. (Right) Parameter table for this model. The optimized parameter are in bold.}
    \label{fig:optimized_spec}
  \end{figure}

Conversely, the thickness of the ND layer on the sides seems to have a more important role. In particular, both too-thick and too-thin values are penalizing for VCNs. In the first case because there is not enough reflecting material, while in the second case, too much moderating material is lost to accommodate NDs. Also, the emission surface is considerably shaded by the reflector layer, inevitably reducing the neutron counts at the beam port. The optimized model is shown in~\cref{fig:optimized_spec}. Compared to the \ce{LD_2} baseline, this model is able to deliver 47 times more neutrons above \SI{40}{\angstrom} and 3 times more neutrons in the range from \SIrange{20}{40}{\angstrom}, while retaining only 40\% of the cold neutrons between 4 and \SI{10}{\angstrom}.

\subsubsection{Engineering details}
\label{subsec:engineering_details}
\begin{figure}[!tb]
    \centering
    \includegraphics[width=0.4\textwidth]{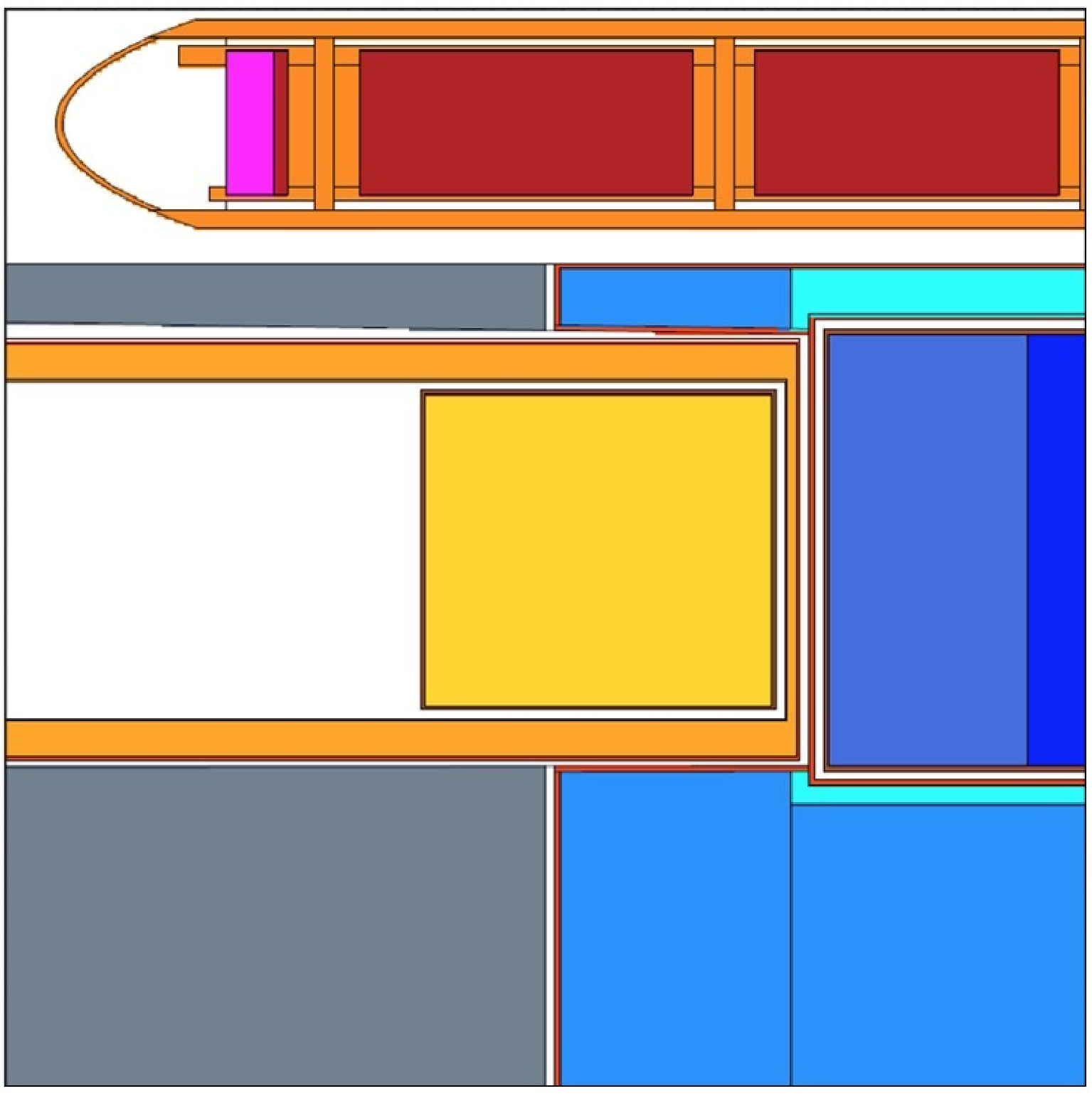}
    \qquad
    \raisebox{0.1\height}{\begin{tabular}[b]{l l}
    \toprule
      \multicolumn{2}{c}{Specifications} \\ 
      \midrule
      Converter thickness & \SI{20}{cm} \\
      Converter vessel & \SI{2}{mm} \\
      ND reflector (sides) & \SI{2}{cm} \\
      ND reflector (back) & \SI{0.55}{cm} \\
      External Al case & \SI{2}{mm} \\
      Internal Al walls & \SI{1}{mm} \\
      Prompt heat-load & \SI{2.5}{kW} \\
      \bottomrule
    \end{tabular}}
    \captionlistentry[table]{A table beside a figure}
    \captionsetup{labelformat=andtable}
    \caption[MCNP model of the optimized converter with engineering details]{(Left) MCNP model of the optimized converter with engineering details. (Right) Parameter table for this model. }
    \label{fig:engineering_wBeFilter}
  \end{figure}

In order to make the model more realistic, we added an aluminum vessel \SI{2}{mm} thick at \SI{5}{K} around the \ce{SD_2} crystal. An additional \SI{5}{mm} vacuum gap separates the converter vessel from the ND reflector surrounding it. Finally, we add \SI{1}{mm} of aluminum as internal cladding for the reflector. We have already mentioned that the model without inner Al structure was a simplified test to study the validity of the idea. For a ND guide, this is quite a critical feature, especially in the VCN range. This model is shown in \cref{fig:engineering_wBeFilter}.

\indent Compared to the ideal case, this model performs much worse. We estimated that, in terms of gain over the baseline, it yields only 2.6 times more neutrons above \SI{40}{\angstrom} and 10\% more neutrons in the range from \SIrange{20}{40}{\angstrom}. While \SI{1}{mm} is already a reasonable thickness, one way to improve the model is to use thinner Al cladding for the extraction. Another way to improve performance would be using other metal foils, like magnesium~\cite{chernyavsky_enhanced_2022}. Preliminary calculations with \SI{1}{mm} of pure Mg showed gains over the baseline as high as a factor of 12 for neutrons above \SI{40}{\angstrom} and a factor of 2  in the range from \SIrange{20}{40}{\angstrom}. Here, it is probably most noteworthy that the flexibility of the design could be used to improve the performances, keeping in mind that reducing the thickness of the inner walls or changing the material tends to be beneficial. 

\subsubsection{Improvements to the design}
\begin{figure}[!bt]
    \centering
    \includegraphics[width=0.4\textwidth]{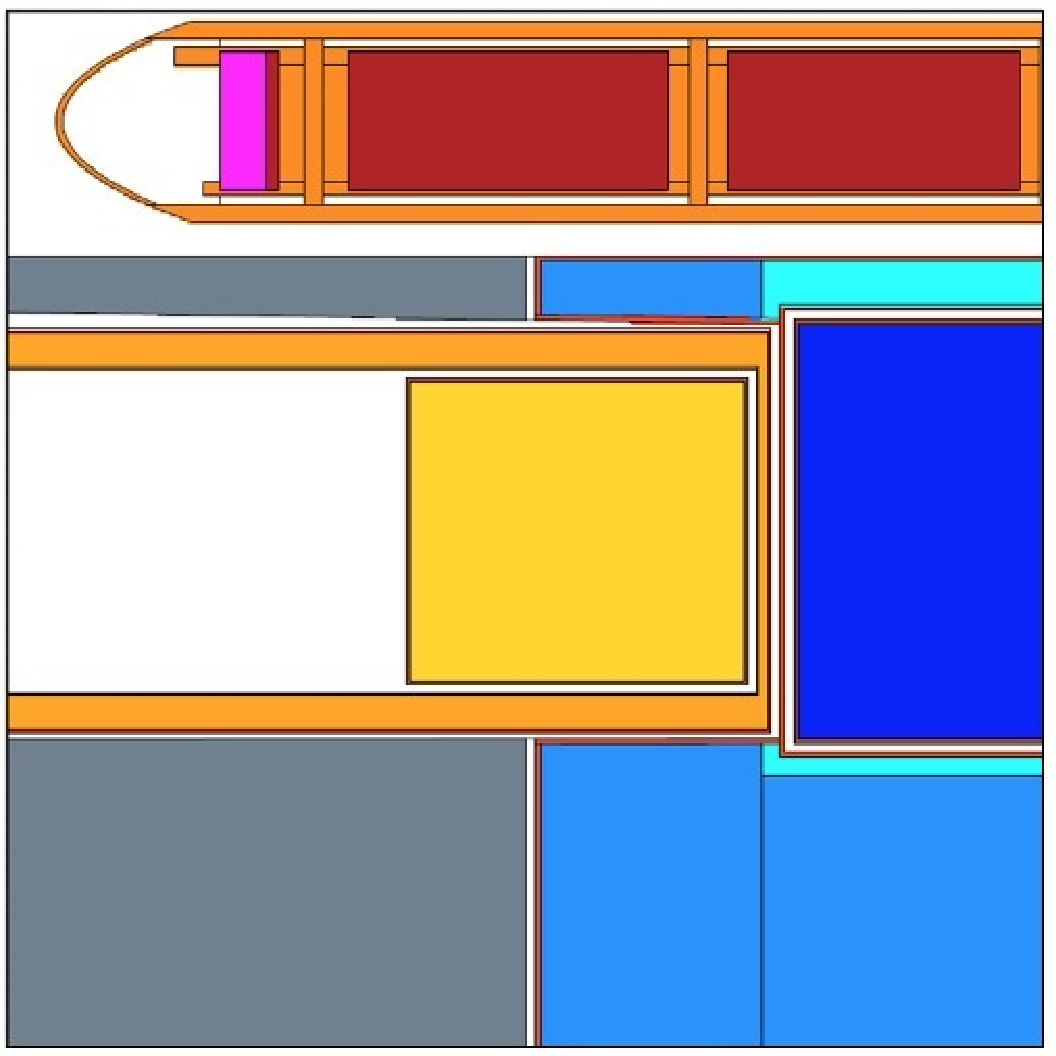}
    \qquad
    \raisebox{0.1\height}{\begin{tabular}[b]{l l}
    \toprule
      \multicolumn{2}{c}{Specifications} \\ 
      \midrule
      Converter thickness & \SI{20}{cm} \\
      Converter vessel & \SI{2}{mm} \\
      ND reflector (sides) & \SI{2}{cm} \\
      ND reflector (back) & \SI{0.55}{cm} \\
      External Al case & \SI{2}{mm} \\
      Internal Al walls & \SI{1}{mm} \\
      Prompt heat-load & \SI{3}{kW} \\
      \bottomrule
    \end{tabular}}
    \captionlistentry[table]{A table beside a figure}
    \captionsetup{labelformat=andtable}
    \caption[MCNP model of the optimized converter without Be filter]{(Left) MCNP model of the optimized converter with engineering details and without Be filter in the cold source. (Right) Parameter table for this model.}
    \label{fig:engineering_woutBeFilter}
  \end{figure}

The first change we can make to the model that will predictably improve the performance is removing the cold beryllium filter from the \ce{LD_2} vessel. The reason for this is that the beryllium at \SI{20}{K} reflects any neutrons with wavelengths below \SI{4}{\angstrom} back to the moderator; whereas for longer wavelength neutrons, the crystal is almost transparent. Although this is beneficial for NNBAR, whose FOM is proportional to $\lambda^\text{2}$, it is detrimental for a VCN converter, given that neutrons in the range from \SI{2}{\angstrom} to \SI{4}{\angstrom} contribute significantly to the VCN production in \ce{SD_2}. The model without the Be filter is shown in~\cref{fig:engineering_woutBeFilter}. Removing the filter to add more moderating \ce{LD_2} has the effect of increasing the gains by $\approx \text{20}\%$ over the case with the filter (\cref{fig:engineering_wBeFilter}). Another important result is in the increase to \SI{3}{kW} of the heat deposition in the moderator volume due to the higher neutron and gamma flux.

Since the thermal and cold neutron flux have an important role in VCN production when with \ce{SD_2}, a further improvement to the design should be obtained by moving the converter closer to the source. In other words, it is worth studying the effect of embedding the converter inside the \ce{LD_2} vessel, regardless of its feasibility. To this end, we inserted the converter \SI{11}{cm} deep inside the cold source. Due to the major shift in the design, we chose to optimize this model separately. The parameters are, within the statistical uncertainty, identical to the previous optimization, except for the thickness of the back ND layer which is \SI{1.2}{cm} (still a low-sensitivity  parameter).

\begin{figure}[!bt]
    \centering
    \includegraphics[width=0.4\textwidth]{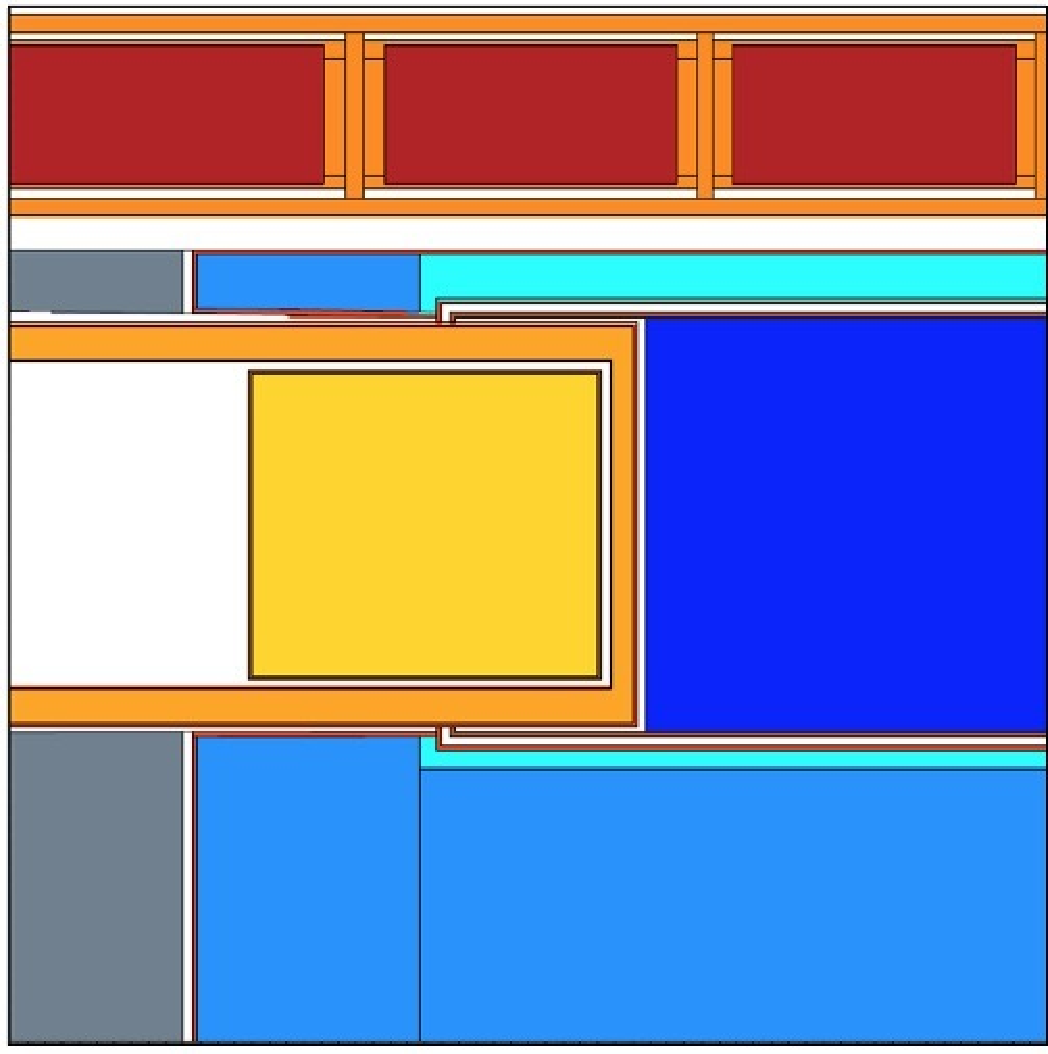}
    \qquad
    \raisebox{0.1\height}{\begin{tabular}[b]{l l}
    \toprule
      \multicolumn{2}{c}{Specifications} \\ 
      \midrule
      \textbf{Converter thickness} & \SI{20}{cm} \\
      Converter vessel & \SI{2}{mm} \\
      \textbf{ND reflector (sides)} & \SI{2}{cm} \\
      \textbf{ND reflector (back)} & \SI{1.25}{cm} \\
      External Al case & \SI{2}{mm} \\
      Internal Al walls & \SI{1}{mm} \\
      Prompt heat-load & \SI{5.8}{kW} \\
      \bottomrule
    \end{tabular}}
    \captionlistentry[table]{A table beside a figure}
    \captionsetup{labelformat=andtable}
    \caption[MCNP model of the optimized converter embedded in the \ce{LD_2 vessel}]{(Left) MCNP model of the optimized converter embedded in the \ce{LD_2} cold source. (Right) Parameter table for this model. The optimized parameters are in bold.}
    \label{fig:inside_LD2}
  \end{figure}

The model implementing this idea is shown in \cref{fig:inside_LD2}. In terms of performances, this solution provides an average factor of two improvement for all energies over the case with the Be filter (\cref{fig:engineering_wBeFilter}). If we compare the neutron counts over the baseline without a VCN converter, this model is able to deliver 5.7 times more neutrons above \SI{40}{\angstrom} and 2.5 times more neutrons in the range from \SIrange{20}{40}{\angstrom}, while retaining 60\% of the cold neutrons between 4 and \SI{10}{\angstrom}. However, once again, the price to pay for the increased flux is a significant increase in heat-load. We estimate that just moving the converter \SI{11}{cm} inside the moderator nearly doubles the heat deposition to a value of \SI{5.8}{kW}.

\subsubsection{Converter for neutron scattering instruments}
\begin{figure}[!tb]
    \centering
    \includegraphics[width=0.4\textwidth]{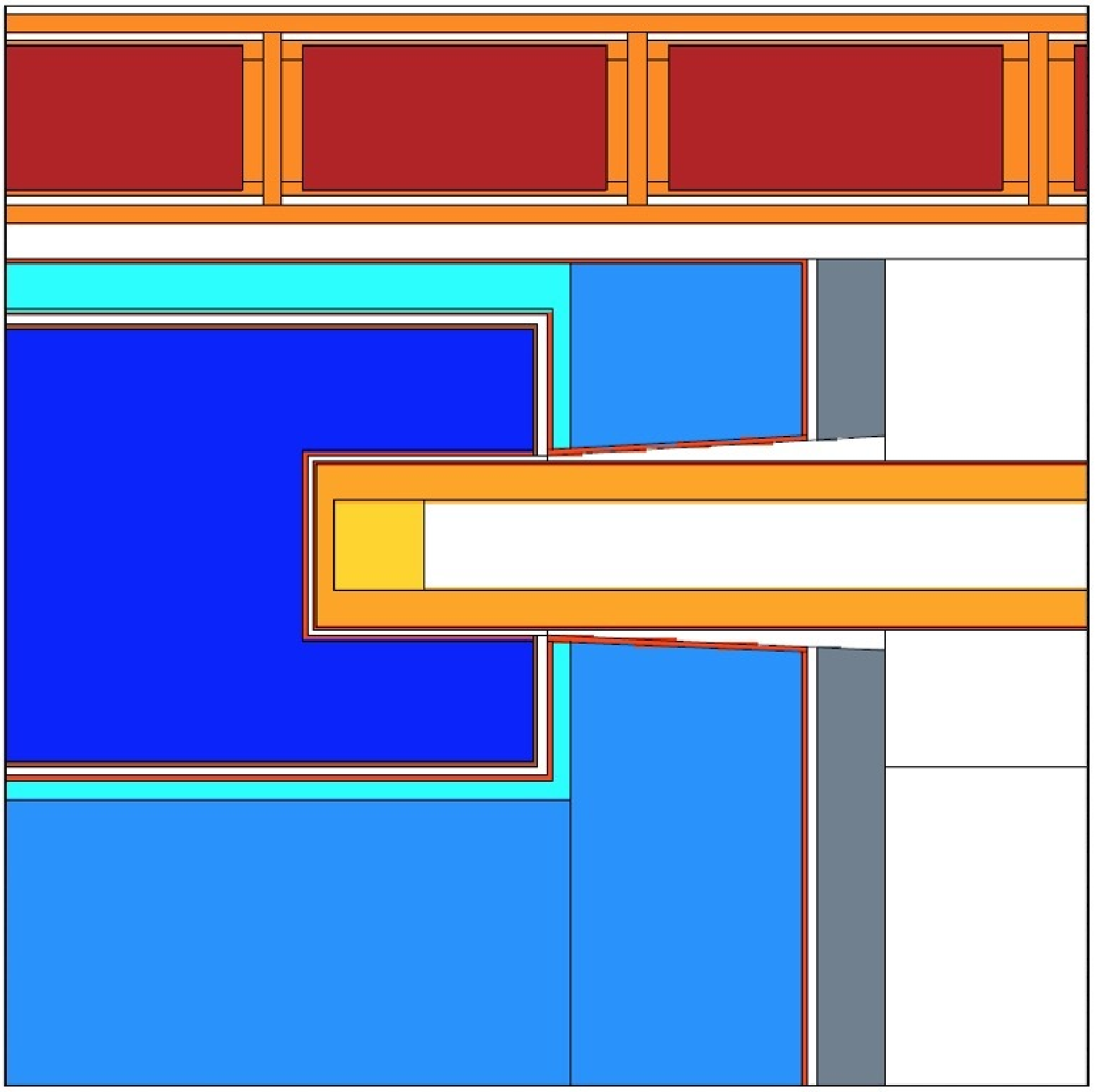}
    \qquad
    \raisebox{0.1\height}{\begin{tabular}[b]{l c}
    \toprule
      \multicolumn{2}{c}{Specifications} \\ 
      \midrule
      Opening size & \qtyproduct{10x10}{} \si{cm^2}  \\
      Converter thickness & \SI{5}{cm} \\
      Converter vessel & \SI{0}{mm} \\
      ND reflector (sides) & \SI{2}{cm} \\
      ND reflector (back) & \SI{1}{cm} \\
      External Al case & \SI{2}{mm} \\
      Internal Al walls & \SI{0}{mm} \\
      \bottomrule
    \end{tabular}}
    \captionlistentry[table]{A table beside a figure}
    \captionsetup{labelformat=andtable}
    \caption{(Left) MCNP model of the converter embedded on the WP7 side for the preliminary test run to check the separate and combined effect of NDs and the \ce{SD_2} block. (Right) Parameter table for this model.}
    \label{fig:preliminary_WP7}
  \end{figure}

Similarly to the tests on the NNBAR side, the effects on VCN flux were studied for a thin converter on the WP7 side. The F1 tally measures the current at a fixed surface placed at the beam port, \SI{2}{m} from the center of the moderator, with the same energy-angle resolution. For this test, the opening is \qtyproduct{10x10}{cm} and the \ce{SD_2} crystal is \SI{5}{cm} thick.  The ND guide is \SI{190}{cm} long with 2-cm-thick walls on the sides and \SI{1}{cm} in the direction of the cold neutron beam.  No structure is present around the converter and for the inner walls. The model used is shown in detail in \cref{fig:preliminary_WP7}.

\begin{figure}[bt!]      
    \begin{subfigure}[b]{0.48\textwidth}
        \centering
        \includegraphics[width=\textwidth]{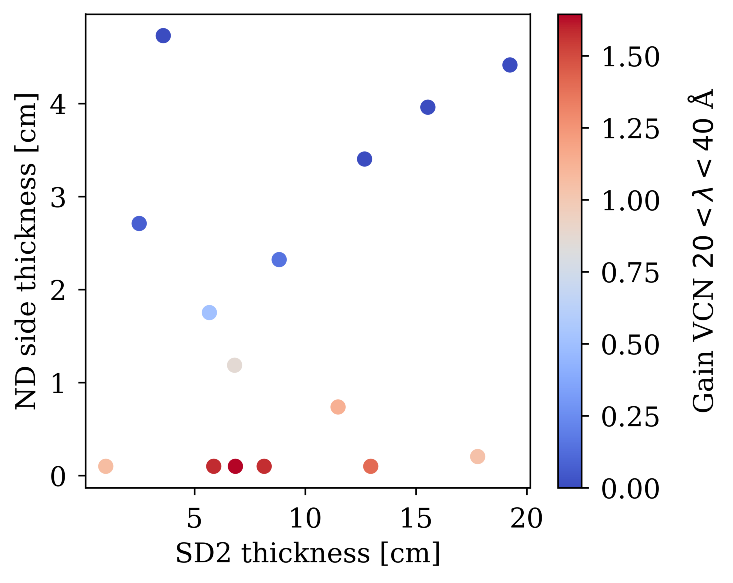}
        \subcaption{}
        \label{fig:Dakota_10x10_side}
    \end{subfigure}
    \hfill
    \begin{subfigure}[b]{0.48\textwidth}
        \centering        
        \includegraphics[width=\textwidth]{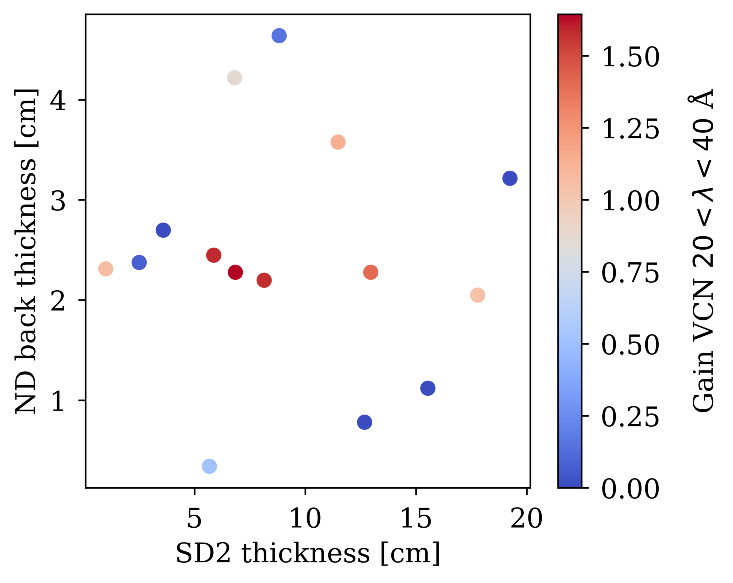}
        \subcaption{}
        \label{fig:Dakota_10x10_back}
    \end{subfigure}
\caption[Dakota optimization with a \qtyproduct{10x10}{cm} opening]{Results of the multi‐dimensional Dakota optimization of the \qtyproduct{10x10}{cm} opening for WP7. Each run of the optimization is represented by a point. Only gains for the VCN energy‐group from \SIrange{20}{40}{\angstrom} are shown. The gains are calculated over the baseline with the same opening. (a) \ce{SD_2} thickness vs ND layer side thickness (b) \ce{SD_2} thickness vs ND layer back thickness.}
\label{fig:Dakota_10x10}
\end{figure}

The preliminary test with this tentative setup did not show, compared to the baseline, any gain in using the converter independently from the energy-group considered. To be sure that the bad performances observed are not strictly bound to a particular configuration, the most important parameters of the geometry were optimized with Dakota. Namely, the EGO algorithm found the best values for the \ce{SD_2} thickness, the ND layer thickness on both the sides and the back. Similarly to the NNBAR case, the figure of merit of the optimization was the VCN counts between \SI{20}{\angstrom} and \SI{40}{\angstrom} and the minimum value set as boundary for the ND layers is \SI{0.1}{cm}. Despite the precision being lower than the NNBAR case, especially at very long wavelength, the results suggest with a good degree of convergence that the most favorable configuration has only \SI{1}{mm} of NDs on the sides~(\cref{fig:Dakota_10x10}).

The fact that gains are observed over the baseline in the VCN range suggests that inserting a block of \ce{SD_2} has the intended effect of further moderating the cold spectrum coming from the \ce{LD_2}. Also, as expected, the system is not very sensitive, within few centimeters, to the thickness of the NDs in the back due to the high neutron transmission in the cold range.

However, the configuration that maximizes the gains between \SI{20}{\angstrom} and \SI{40}{\angstrom} is found at the minimum of the ND thickness on the sides. The different behavior observed between WP7 and NNBAR in the optimization of the ND side layer thickness is explained by a poor parameterization of the model. As already mentioned, the increase of the reflector thickness on the side has the effect of reducing the amount of \ce{SD_2} and occluding the emission surface. Despite being present also on the NNBAR side, this effect seems to be crippling only on the WP7 side.

\begin{figure}[!tb]
    \centering
    \includegraphics[width=0.4\textwidth]{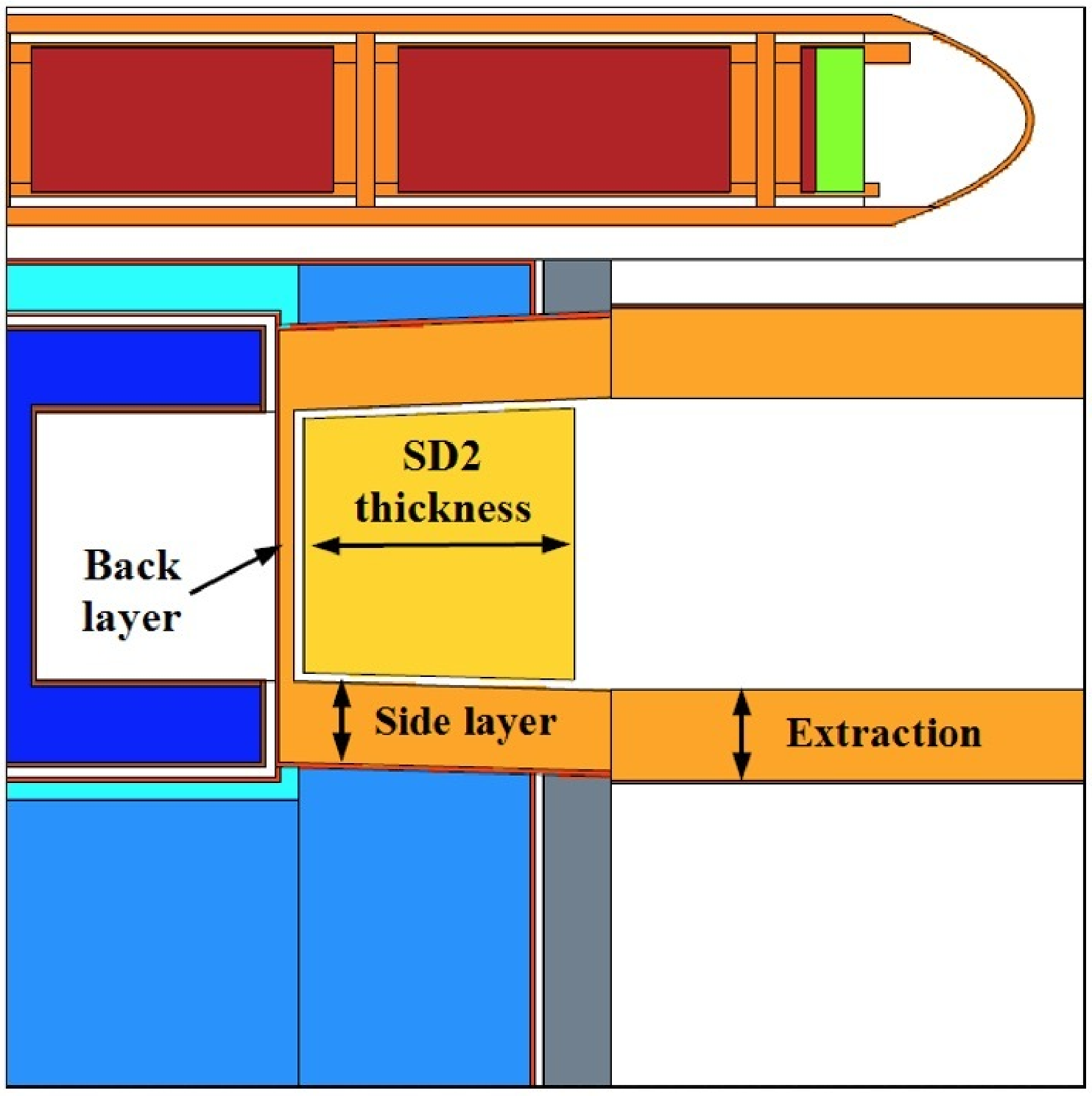}
    \quad
    \raisebox{0.1\height}{\begin{tabular}[b]{l c}
    \toprule
      \multicolumn{2}{c}{Specifications} \\ 
      \midrule
      Opening size & \qtyproduct{15x15}{} \si{cm^2}  \\
      \textbf{Converter thickness} & \SI{15.6}{cm} \\
      Converter vessel & \SI{0}{mm} \\
      \textbf{ND reflector (sides)} & \SI{4.5}{cm} \\
    \textbf{ND reflector (back)} & \SI{0.8}{cm} \\
     \textbf{ND reflector (extraction)} & \SI{5}{cm} \\
      External Al case & \SI{2}{mm} \\
      Internal Al walls & \SI{0}{mm} \\
      Prompt heat-load & \SI{860}{W} \\
      \bottomrule
    \end{tabular}}
    \captionlistentry[table]{A table beside a figure}
    \captionsetup{labelformat=andtable}
    \caption{(Left) MCNP model of the converter for the WP7 side with improved parameterization (Right) Parameter table for this model. The optimized parameters are
in bold.}
    \label{fig:better_WP7}
  \end{figure}

\begin{figure}[btp!]  
    \begin{subfigure}[b]{\textwidth}
        \centering
        \includegraphics[width=\textwidth]{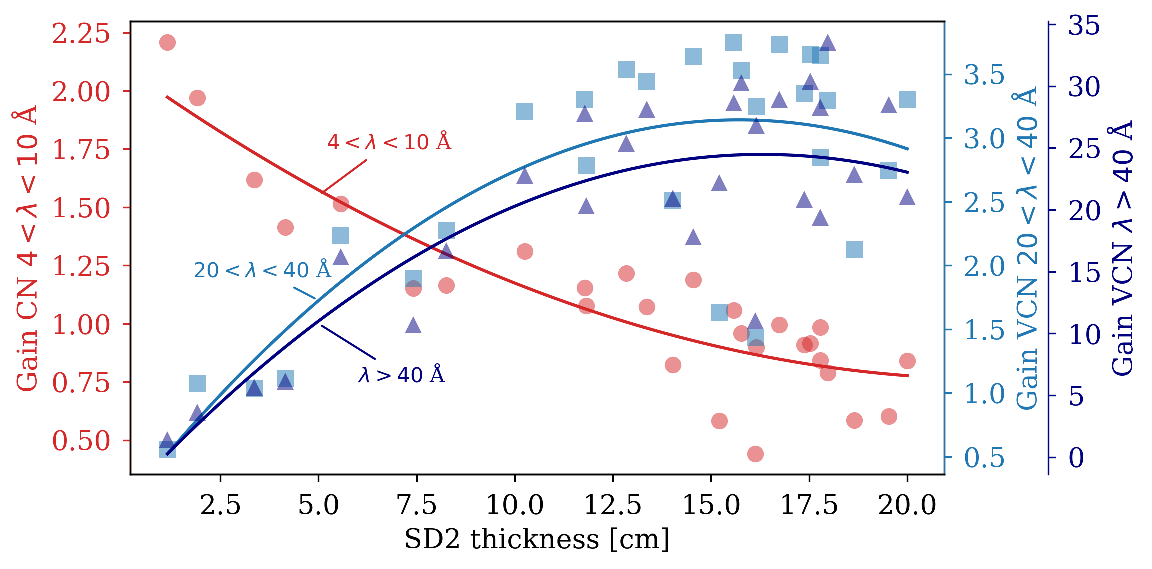}
        \subcaption{}
        \label{fig:SD2_thickness_WP7_15x15}
    \end{subfigure}
    \begin{subfigure}[b]{0.48\textwidth}
        \centering
        \includegraphics[width=\textwidth]{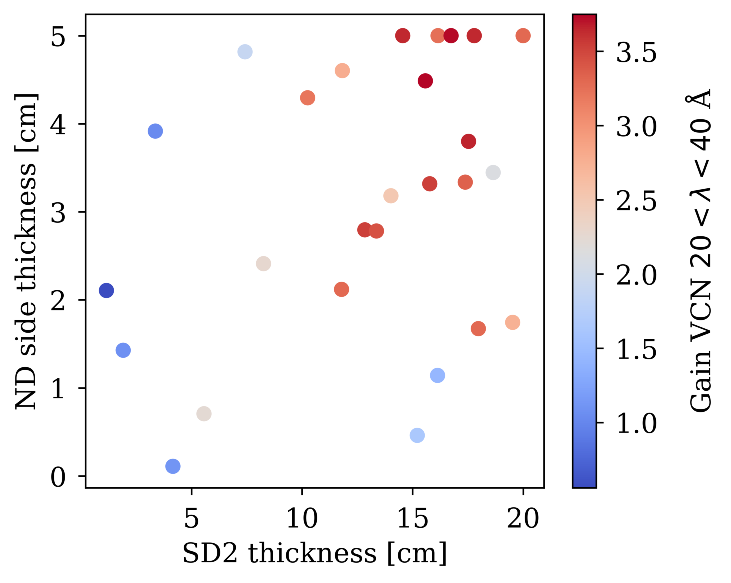}
        \subcaption{}
        \label{fig:SD2vsSide_WP7_15x15}
    \end{subfigure}
    \hfill
    \begin{subfigure}[b]{0.48\textwidth}
        \centering
        \includegraphics[width=\textwidth]{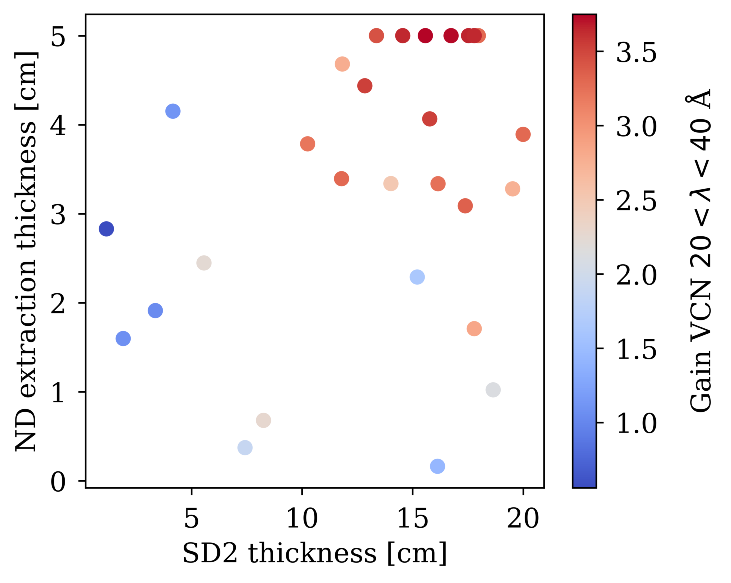}
        \subcaption{}
        \label{fig:SD2vsExtraction_WP7_15x15}
    \end{subfigure}
\caption{Results of the multi‐dimensional Dakota optimization of the \qtyproduct{15x15}{cm} opening for WP7. The gains are calculated at all angles over the baseline with the same opening. (a) each run of the optimization is represented by three vertically aligned points, one for each energy-group of interest and each one with its own scale for the gains. Trend lines are shown with the sole purpose of guiding the readers' eye. (b) each run of the optimization is represented by a point. Only gains for the VCN energy‐group from \SIrange{20}{40}{\angstrom} are shown. The two parameters plotted are: \ce{SD_2} thickness vs ND layer side thickness (c) \ce{SD_2} thickness vs. ND layer thickness in the extraction.}
\label{fig:optimization_WP7_improved}
\end{figure}

One way to fix this is to move the \ce{SD_2} block out of the reentrant hole and by changing the thickness of the ND layer outward at the cost of the reflector, not the emission surface. In this setup, the \ce{SD_2} block has the same lateral dimensions as the opening and it is decoupled from the NDs. This allows for an unrestricted increase to the ND layers on the sides to reduce the leakage, without occluding the opening. Finally, the thickness of the reflector layer outside the twister (i.e. the guide-like part) is also decoupled and optimized separately. The cost in terms of performance for this setup is a lower cold flux due to the longer distance from the center of the cold source. The optimized model for a \qtyproduct{15x15}{cm} opening is shown in \cref{fig:better_WP7}, while in \cref{fig:optimization_WP7_improved} an overview of the results of the optimization is presented.


From the analysis of the parameter space, the behavior of the improved model is more similar to what was expected by such system. In \cref{fig:SD2_thickness_WP7_15x15}, the different trends of the gains for different energy groups as a function of the \ce{SD_2} thickness are highlighted. It is clear that, for small amounts of \ce{SD_2} in the opening, the overall effect of the setup is to increase the CN transport by exploiting the quasi-specular reflection on the ND layers on the sides. Notably, the factor 2 increase in the flux is compatible with the results in \cref{sec:NDextraction} at all angles. As the thickness of the converter increases, the shift at longer wavelengths becomes more and more pronounced. Around 15-\SI{20}{cm}, the converter yields an average gain factor of more than 3 for VCNs between \SI{20}{\angstrom} and \SI{40}{\angstrom} and 30 for VCNs at longer wavelengths and at all angles. In \cref{fig:SD2vsSide_WP7_15x15} and \cref{fig:SD2vsExtraction_WP7_15x15} the gains from \SIrange{20}{40}{\angstrom} are plotted as function of the \ce{SD_2} thickness and ND layer on the sides and in the extraction, respectively. In both cases, a thicker reflector is correlated with higher gains. In the case of the NDs on the sides, the removal of standard reflector does not seem to produce a significant loss.

It should be noted at this point that each step of the optimization is computationally more expensive and tally precision is lower than for the NNBAR opening. This limits the exploration of the parameter space and the extraction of key insights about the system. Despite the lower cold flux on the converter caused by the longer distance from the center of the moderator, the system is able to deliver order-of-magnitudes gains in the lower VCN range, with potentially smaller impact on the cold flux at the beam port and a smaller heat load to remove (cfr. \cref{fig:better_WP7}). Finally, in \cref{subsec:engineering_details} the losses caused by adding the engineering details were discussed for the setup in the NNBAR opening. Similar observations are reasonably valid for the WP7 opening.   

\subsubsection{Gains at small angles}
\begin{figure}[btp!]  
    \begin{subfigure}[b]{\textwidth}
        \centering
        \includegraphics[width=\textwidth]{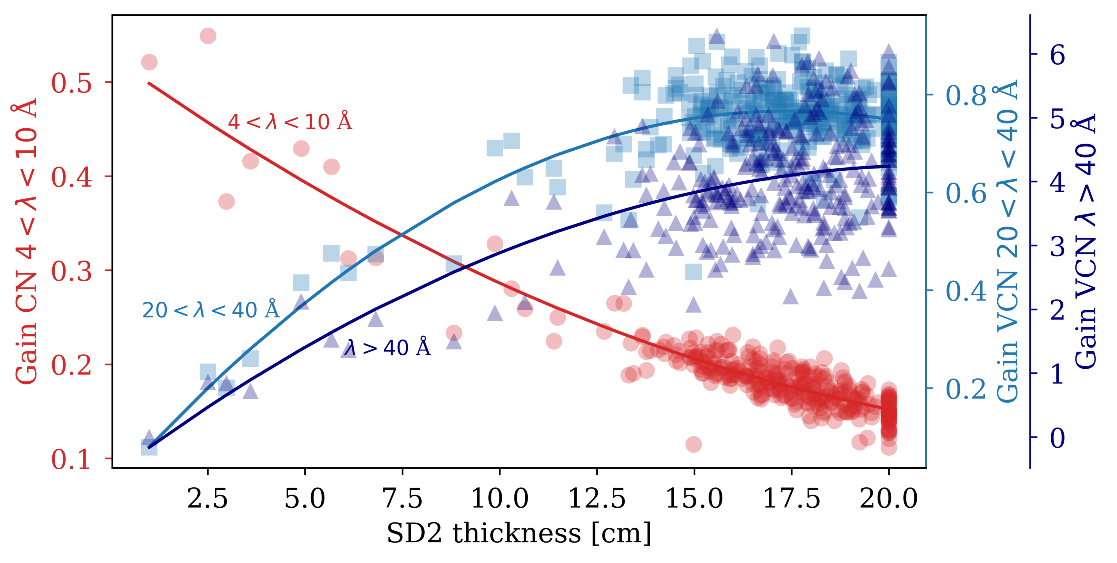}
        \subcaption{}
        \label{fig:SD2_thickness_NNBAR_0}
    \end{subfigure}
    \begin{subfigure}[b]{\textwidth}
        \centering
        \includegraphics[width=\textwidth]{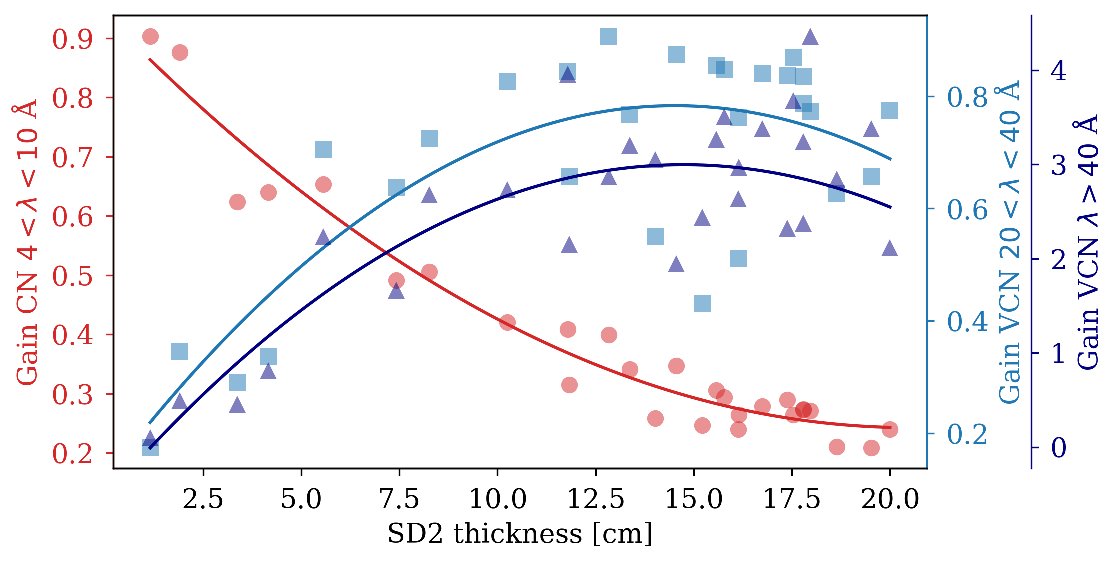}
        \subcaption{}
        \label{fig:SD2_thickness_WP7_15x15_small}
    \end{subfigure}
\caption{Results of the multi‐dimensional Dakota optimization of the (a) NNBAR opening and (b) the \qtyproduct{15x15}{cm} opening for WP7. The gains are calculated at small angle over the baseline with the same opening. Each run of the optimization is represented by three vertically aligned points, one for each energy-group of interest and each one with its own scale for the gains. Only the \ce{SD_2} thickness is reported. Trend lines are shown with the sole purpose of guiding the readers' eye.}
\label{fig:optimization_both_small}
\end{figure}

The results discussed so far are for neutrons reaching the beam port at any angle. The presence of a ND channel allows neutrons with a large divergence to reach the recording surface. However, for most conventional neutron scattering instruments, neutrons with large divergence are collimated along the way and do not contribute to the flux at the sample. The gains for the neutrons with vertical divergence between \SI{0}{\degree} and \SI{2}{\degree} are shown in \cref{fig:SD2_thickness_NNBAR_0} for the NNBAR opening, and \cref{fig:SD2_thickness_WP7_15x15_small} for the WP7 opening.

The gains are reported solely as a function of the \ce{SD_2} thickness, as the strongest predictor. While the trends are similar to the all-angles cases, the scale of the VCN gains is smaller. Up to \SI{20}{\angstrom}, there is no advantage in using the converter. The initial VCNs produced inside the \ce{LD_2} cannot reach the beam port due to the thin ND layer between the main source and the converter. Since ND channels have proven to be effective in improving the neutron flux at small divergence, the reason for the lower gains can be attributed only to a lesser production in the \ce{SD_2}. The neutron flux out of the \ce{SD_2} at small divergence does not compensate for the initial VCN flux, yielding an overall loss between 20--30\%. Above \SI{40}{\angstrom}, the neutron production by the \ce{SD_2} crystal over the \ce{LD_2} baseline is enough to achieve a net gain. Due to the wide applications of low-divergence neutrons, this effect should be studied in detail, in order to tailor the source to the specific needs of the instruments placed at the end of the extraction system. 

\FloatBarrier
\subsection{Beam Extraction with nanodiamonds}
\label{sec:NDextraction}
\begin{figure}[bt]      
    \begin{subfigure}[b]{0.46\textwidth}
        \centering
        \includegraphics[width=0.9\textwidth]{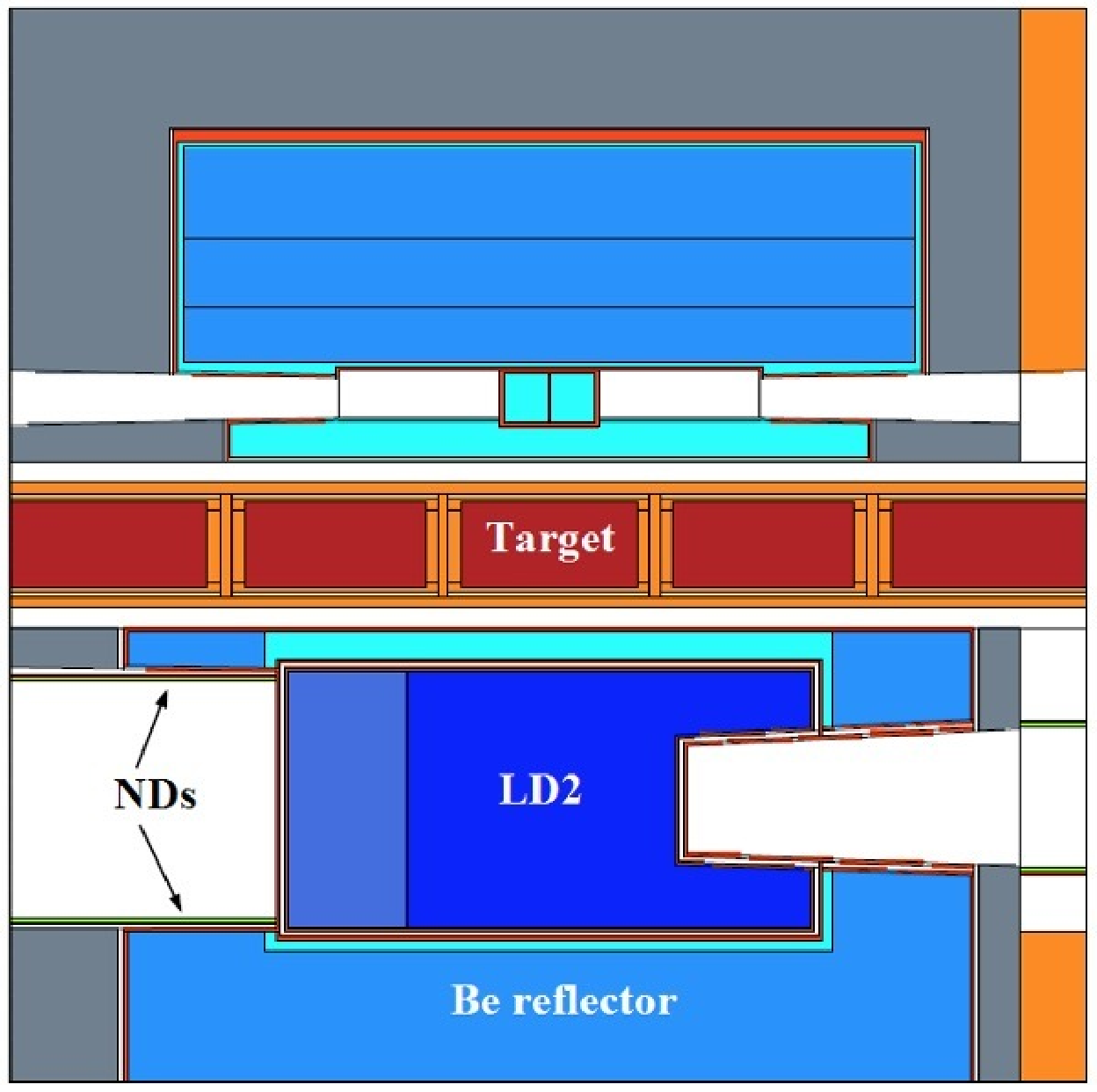}
        \subcaption{}
        \label{fig:LD2_baseline_XY}
    \end{subfigure}
    \hfill
    \begin{subfigure}[b]{0.46\textwidth}
        \centering        
        \includegraphics[width=0.9\textwidth]{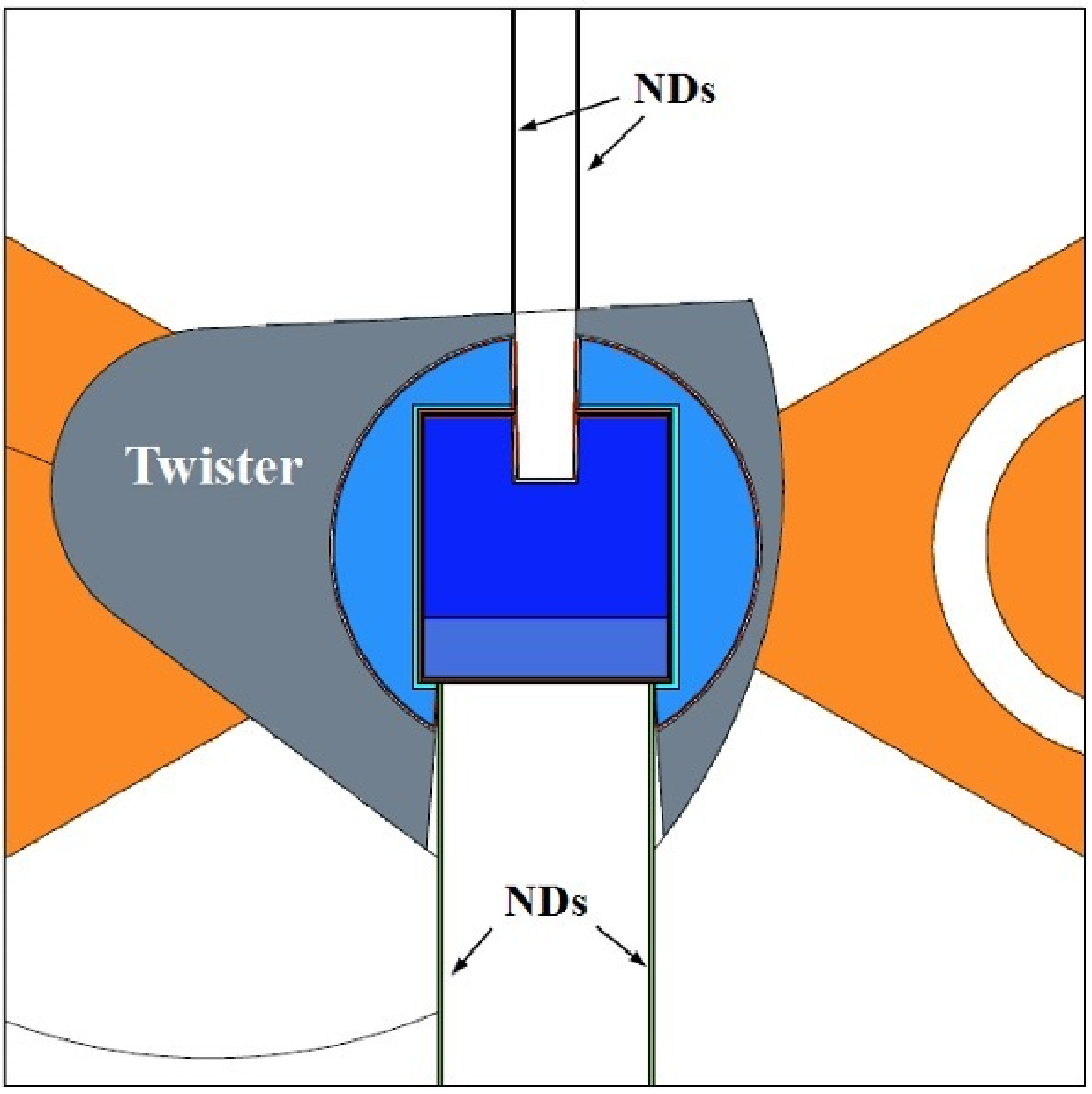}
        \subcaption{}
        \label{fig:LD2_wND_zoom}
    \end{subfigure}
\caption[Baseline with ND extraction system]{MCNP model of the \ce{LD_2} baseline with ND extraction system (a) vertical cross section, perpendicular to the proton beam direction. (b) Cross section parallel to the target plane with the proton beam impinging from the left.}
\label{fig:LD2_wND}
\end{figure}

An additional option for improving VCN yield regardless of the source configuration is to use a dedicated guide-like system to more efficiently extract the VCN tail from the \ce{LD_2} spectrum. In a dedicated study, since the quasi-specular reflection on thin ND powder samples has been experimentally observed \cite{cubitt_quasi-specular_2010, aleksenskii_clustering_2021, nesvizhevsky_reflection_2008, nesvizhevsky_fluorinated_2018}, a straight ND guide with walls thickness of \SI{5}{mm} was considered adequate to observe the improvement of extracting CNs and VCNs. Preliminary simulations were conducted on both sides of the \ce{LD_2}. The ND powder had an effective density of 0.6 \si{g/cm^3}, so it was compressed inside the hollow space created by two Al walls. The inner wall was a \SI{0.1}{mm} thick Al foil, and the outer Al wall had a thickness of \SI{2}{mm}. The geometry is shown in \cref{fig:LD2_wND}.

\begin{figure}[htb]
    \centering
    \includegraphics[width=0.9\textwidth]{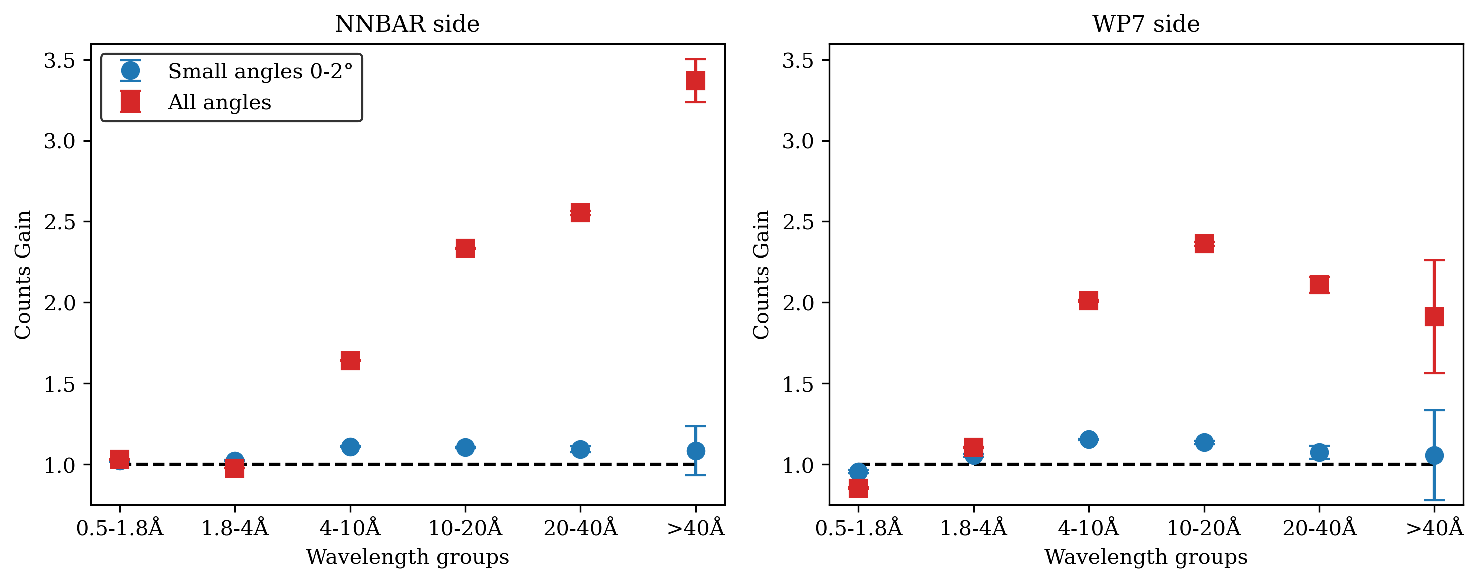}
    \caption[Gains with ND guide as function of neutron wavelength groups]{Gains with ND guide \SI{2}{m} away from the center of the moderator (beam port position) as a function of neutron wavelength groups for NNBAR (left) and WP7 (right) openings.}
    \label{fig:gains_ND_guide}
\end{figure}

\begin{figure}[htb]
    \centering
    \includegraphics[width=0.9\textwidth]{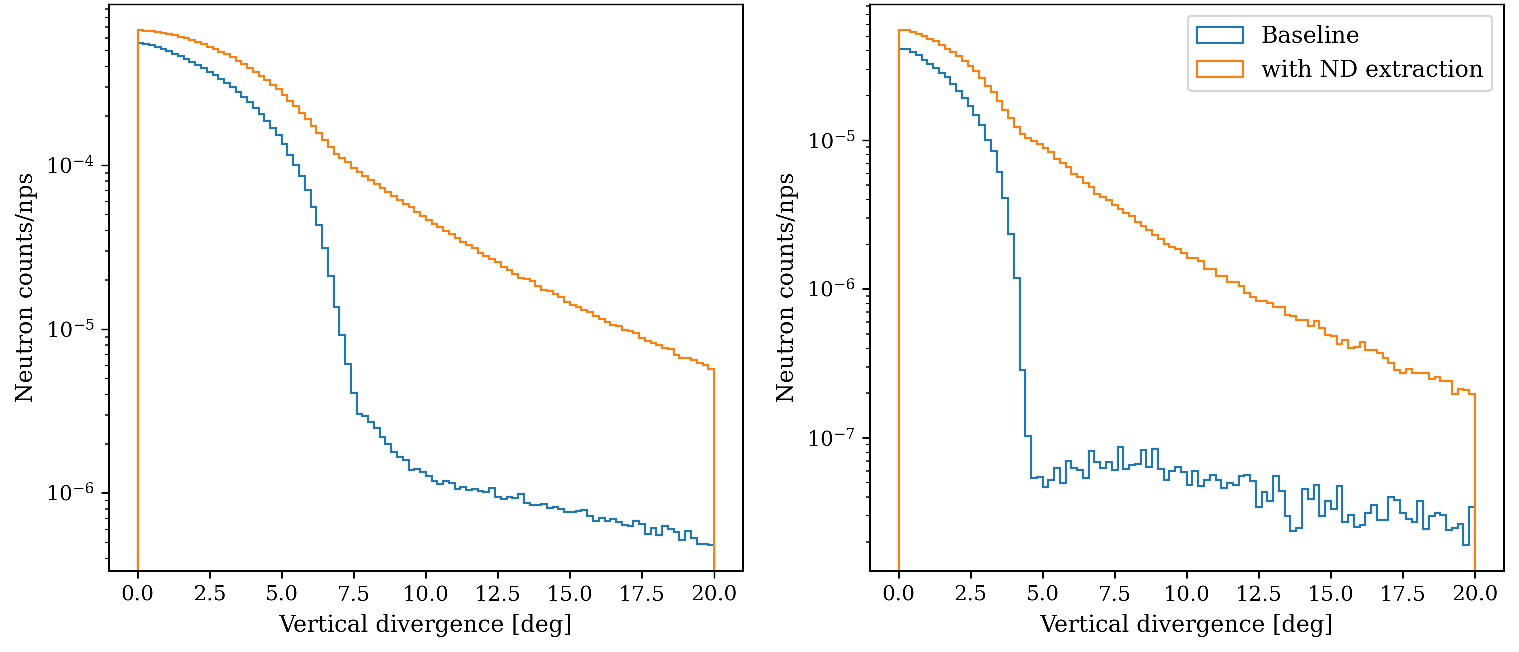}
    \caption[Divergence at the beam port for baseline without and with ND guide]{Divergence of cold neutrons at the beam port ($\lambda >\SI{4}{\angstrom}$) for the baseline without (blue) and with (orange) ND guide. Left for the NNBAR opening, right for the WP7 opening. The vertical axis is neutron counts/nps. }
    \label{fig:div_LD2_4AA}
\end{figure}

The result of the study are summarized in~\cref{fig:gains_ND_guide,fig:div_LD2_4AA}. As expected, the ND guide can transport cold neutrons and can provide on average a factor of 2 gain at \SI{2}{m} over the baseline. However, most of these gains are obtained for neutrons reaching the beam port with vertical divergence greater than \SI{2}{\degree}, that is, the more pronounced effect is observed in the region of the phase space where fewer neutrons (or none at all) would be allowed to reach the beam port without the guide. The gain provided by the ND guide reached the maximum at around \SI{10}{\angstrom} on the WP7 side, and then decrease in the VCN energy range. Here, due to the multiple large‐angle interactions, the reflections on the ND layer very quickly produce an isotropic distribution. The large NNBAR beam port is more likely to accept those large divergence neutrons, while in the smaller guide on the neutron scattering side, they are easily lost (absorbed, back-scattered, and so on) after multiple interactions with the ND and the Al walls.
The idea of a ND guide between the emission surface and the beam port as a VCN-enhancing system is further investigated in the following section.

\FloatBarrier
\subsection{Comparison of the different options}
\label{sec:comparison}
\begin{figure}[bt!]      
    \begin{subfigure}[b]{0.48\textwidth}
        \centering
        \includegraphics[width=\textwidth]{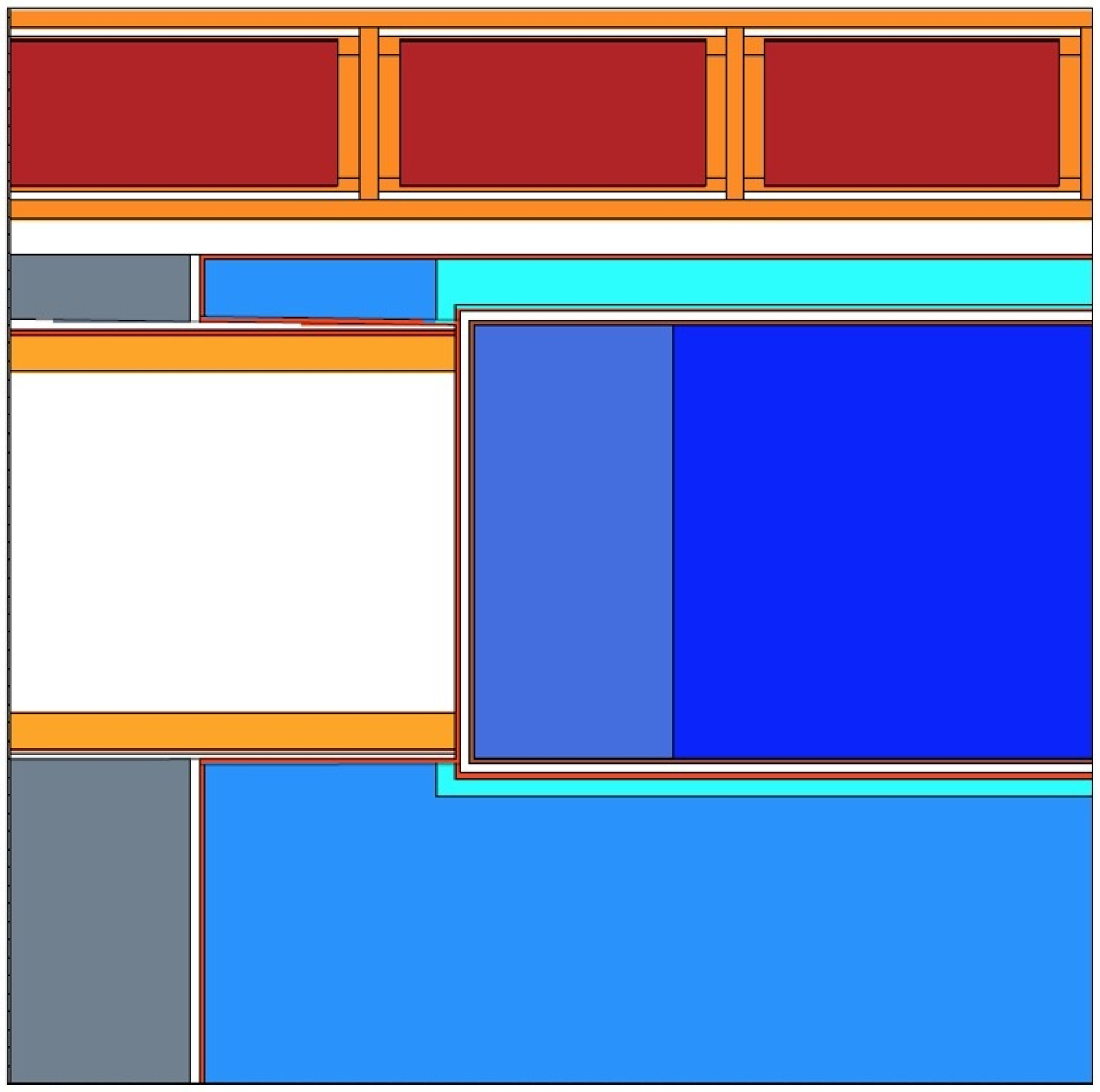}
        \subcaption{}
        \label{fig:baseline_2wND_XY}
    \end{subfigure}
    \hfill
    \begin{subfigure}[b]{0.48\textwidth}
        \centering
        \includegraphics[width=\textwidth]{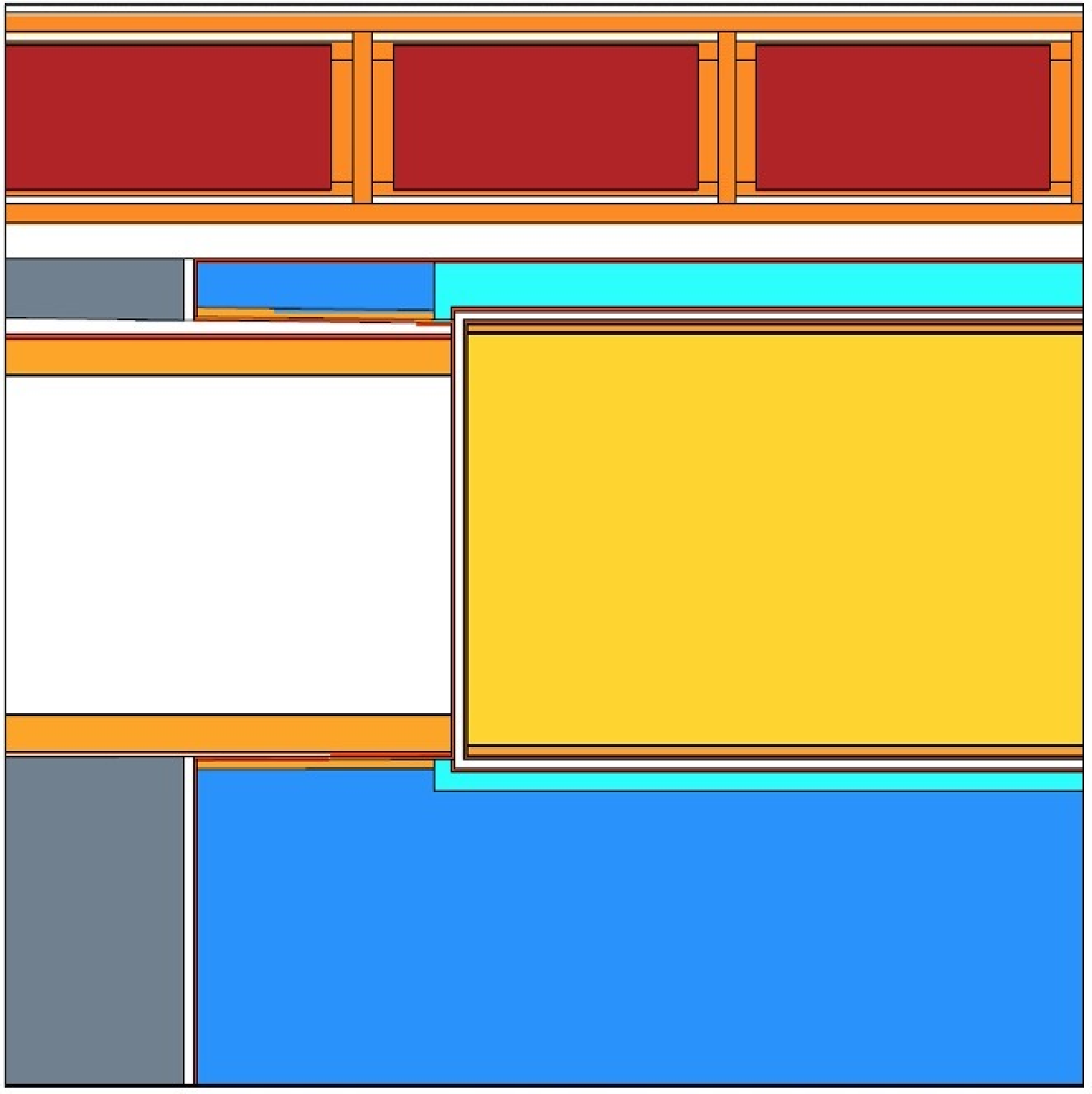}
        \subcaption{}
        \label{fig:sd2_comparison_XY}
    \end{subfigure}
    \begin{subfigure}[b]{0.48\textwidth}
        \centering
        \includegraphics[width=\textwidth]{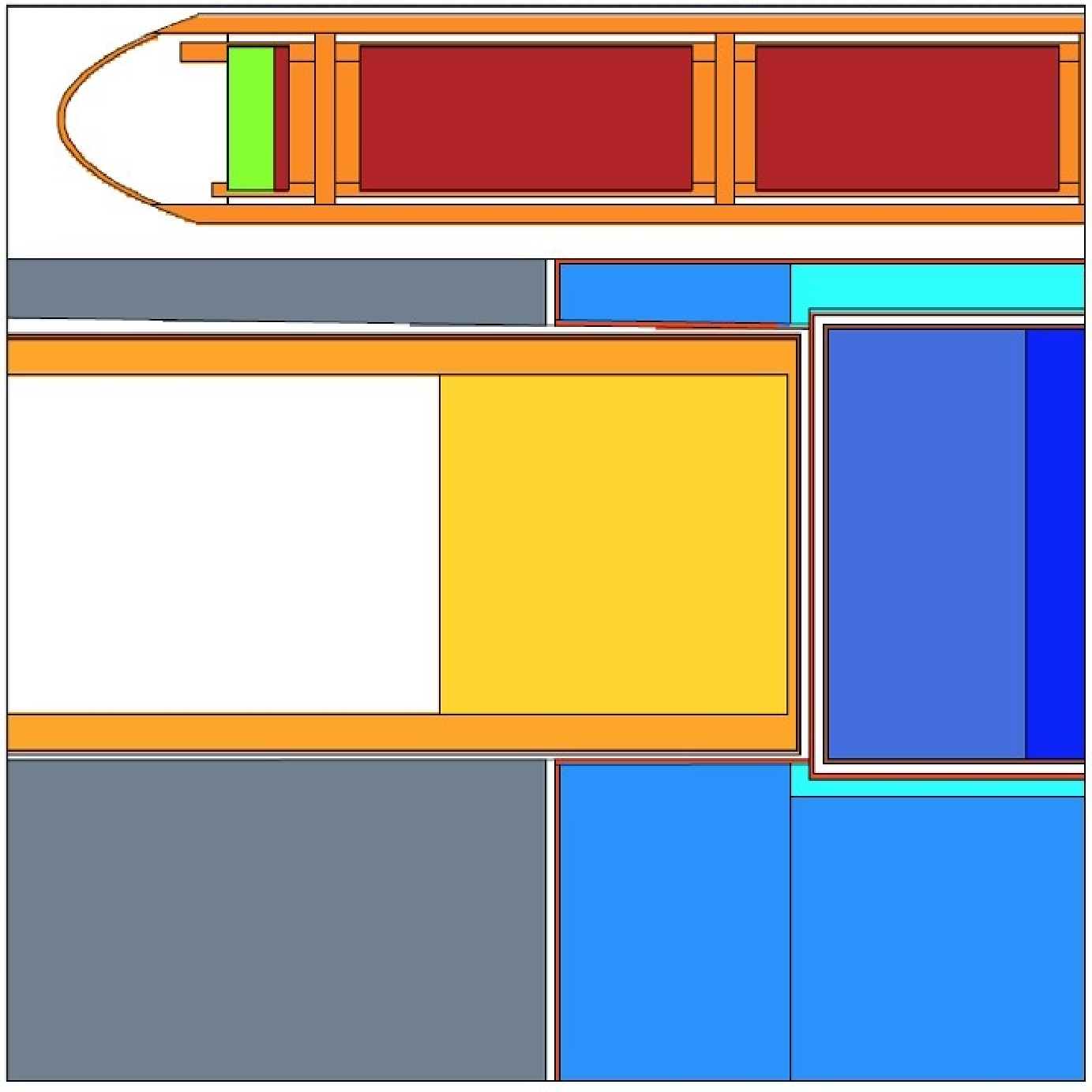}
        \subcaption{}
        \label{fig:converter_ideal_XY}
    \end{subfigure}
    \hfill  
    \begin{subfigure}[b]{0.48\textwidth}
        \centering
        \includegraphics[width=\textwidth]{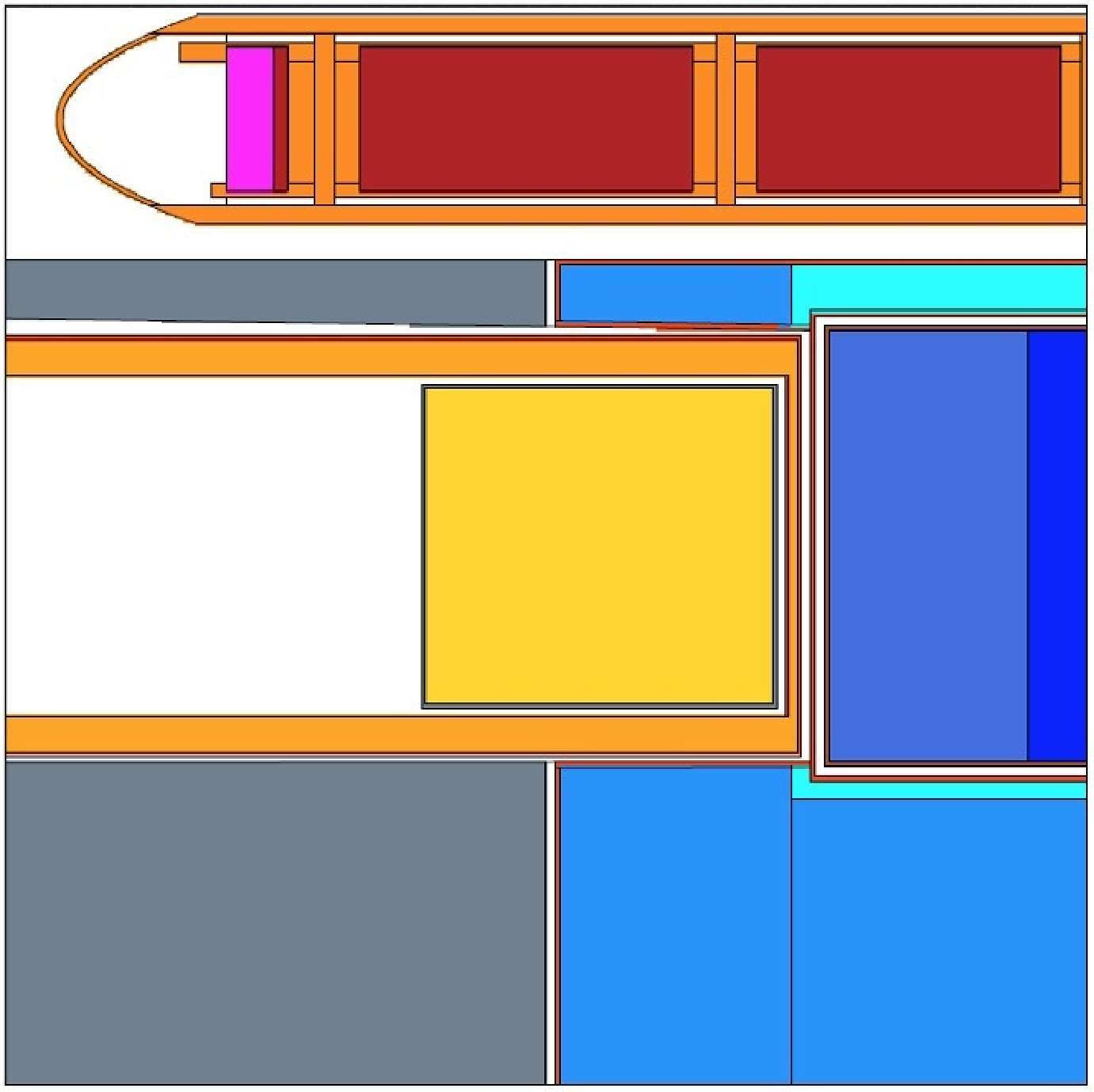}
        \subcaption{}
        \label{fig:converter_real_XY}
    \end{subfigure}
    \hfill
\caption{MCNP models of the VCN sources compared in this work: (a) standard \ce{LD_2} with advanced ND extraction system (b) dedicated \ce{SD_2} with advanced ND extraction system (c) combined \ce{LD_2} and \ce{SD_2} source with ND extraction (ideal case) (d) combined \ce{LD_2} and \ce{SD_2} source with ND extraction (Al vessel and void gap).}
\label{fig:comparison_geometries}
\end{figure}
In this section, we compare the performance of the VCN options presented thus far, except for the clathrates option because it is still at a conceptual level. The models compared here are shown in~\cref{fig:comparison_geometries}. In order for the comparison to be meaningful, it is important to account for the differences that are not at the core of the design (e.g. the size of the recording surface, or the thickness of the Al walls). Here, we focus only on the NNBAR opening since its size is very similar between the cases, and it is more reliable in terms of statistical convergence. Below is a description of each model that is part of the comparison:

\vspace{0.2cm}
\begin{itemize}
    \item[-] The baseline used is an \ce{LD_2} \qtyproduct{24x45x48.5}{cm} box at \SI{22}{K} with a reentrant hole on the neutron scattering side of approximately \qtyproduct{10x10}{cm} and \SI{12.5}{cm} and straight walls in the direction perpendicular to the target. The NNBAR side is \qtyproduct{24x40}{cm} and a 11-cm cold Be filter at the exit. This baseline was initially conceived to accommodate several equally-illuminated instruments on a plane parallel to the target. The only adjustment to be made to the model in view of the comparison is the dimension of the recording surface at \SI{2}{m} from the center of the moderator. This is set for all the cases at \qtyproduct{19x35}{cm}.
    \item[-] The ND extraction discussed in \cref{sec:NDextraction} has a thin ND layer on the sides. Here it is increased to \SI{2}{cm}, which should increase the effect of the enhanced VCN transport at the expense of the cold neutrons. The inner Al wall that contains the ND is \SI{1}{mm} thick. The model is presented in \cref{fig:baseline_2wND_XY}.
    \item[-] For the dedicated full \ce{SD_2} source, a ND extraction system identical to the previous case was added (\cref{fig:sd2_comparison_XY}). The model for this comparison assumes ideal \ce{SD_2} at \SI{5}{K} with no cooling structure inside the moderating volume. As seen in \cref{sec:full_SD2}, this is far from being a realistic description of the moderator, but it is based on the same sets of assumptions that define the baseline (i.e. ideal \ce{LD_2} at \SI{22}{K} with no flow channels inside).
    \item[-] In the combined option for the NNBAR side, the \ce{SD_2} block is \SI{20}{cm} thick and the ND layer between the source and the \ce{SD_2} block is \SI{0.55}{cm}. The ND extraction system is the same as the previous cases, as well as the recording surface. The difference between the ideal case in \cref{fig:converter_ideal_XY} and the one in \cref{fig:converter_real_XY} is only the \SI{2}{mm} Al vessel and the \SI{5}{mm} vacuum gap surrounding the \ce{SD_2}. The only change made for the comparison is adjusting the position and the dimensions of the ND guide to match the previous cases. 
\end{itemize}
\vspace{0.2cm}

\begin{figure}[bt!]      
    \begin{subfigure}[b]{0.85\textwidth}
        \centering
        \includegraphics[width=0.85\textwidth]{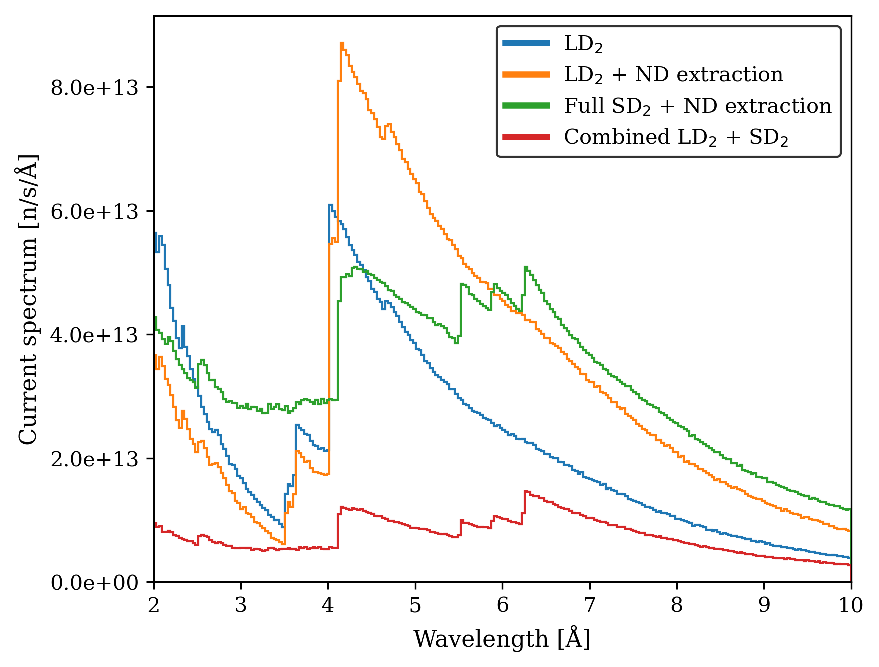}
        \subcaption{}
        \label{fig:comparison_spectra_lin}
    \end{subfigure}
    \hfill
    \begin{subfigure}[b]{0.85\textwidth}
        \centering        
        \includegraphics[width=0.85\textwidth]{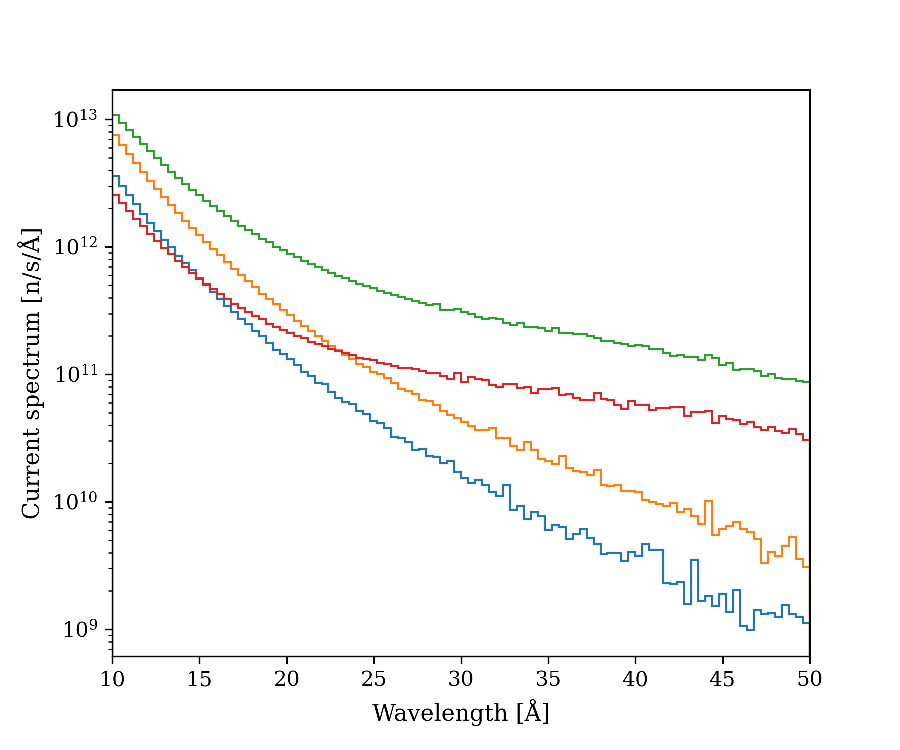}
        \subcaption{}
        \label{fig:comparison_spectra_log}
    \end{subfigure}
\caption[]{Current at all angles [\si{n/s/\angstrom}] at \SI{2}{m} from the moderator center for all the VCN options studied in this section. (a) Bragg peaks from Be Filter, ND and \ce{D_2} crystal are visible in the range between 2 and \SI{10}{\angstrom}. (b) VCN tails are highlighted in the range between 10 and \SI{50}{\angstrom}.}
\label{fig:comparison_spectra}
\end{figure}

For each of these cases, we calculated the current in \si{n/s} and the brightness in \si{n/s/cm^2/sr/\angstrom}. The calculation of the absolute brightness for this comparison is similar to what has been done in \cref{sec:full_SD2}. First, we recorded the raw counts per spallation proton through the surface at \SI{2}{m} from the center of the moderator. Then, we divided the counts at $\pm\,\SI{2}{\degree}$ from the surface normal, by the emission surface, by the solid angle in the \SI{2}{\degree} cone, and by the wavelength range of the group, then we multiplied the result by the expected proton current on the ESS target at \SI{5}{MW}, as in \cref{eq:brightness}. All the results in the appropriate units are presented in \cref{tab:comparison_VCN}. To facilitate the comparison, in \cref{tab:comparison_VCN_gains} we calculated the gains over the baseline for each option, while in \cref{fig:comparison_spectra} the current spectra at all angles highlights the effect of ND, \ce{SD_2}, and the combination of the two in two wavelength regions.

The best option for both the figure of merits studied is undoubtedly the dedicated full \ce{SD_2} moderator at \SI{5}{K}. From the neutronic point of view, order-of-magnitude VCN gains are complemented by a high CN yield. This makes such a moderator an ideal candidate to replace the \ce{LD_2} cold source. However, as already highlighted in \cref{sec:full_SD2}, this solution comes with great challenges from the engineering point of view. 

On the other hand, the easiest solution is using the \ce{LD_2} moderator as it is would be to place a guide-like system made of a thick layer of ND and enhance the VCN tail of the cold spectrum. The simulations show that the longer the wavelength, the higher the gain factor for the current; but when it comes to brightness (low angles) the gains are more limited. The interactions with dedicated VCN instruments need to be investigated for a complete overview of the impact of such device. There is, nevertheless, room for further improvement in the design both in terms of materials (e.g. metal walls, packing of the powder etc.) and configuration (e.g. position of the guide) that should be further investigated in a new project. 

The option of combining \ce{SD_2} and \ce{LD_2} is a midpoint between the previous two cases. Even though the possibility of cooling a 5-K \ce{D_2} block so close to the spallation target is yet to be determined, this case promises to be much easier to handle if compared to the full \ce{SD_2}. At the same time, other challenges would arise in terms of the engineering development (e.g. finding room for the cooling infrastructure in the limited space of the opening, replacing components etc.) that do not need to be faced in the case of a simpler ND guide. In terms of neutronic performance, this option can potentially provide order-of-magnitude gains in the VCN current, without structural changes to the CN baseline. As with the other cases, the optimization of the materials used for the extraction and the containment of the \ce{D_2} is crucial to preserve the ideal gains. Lastly, this option takes a significant toll on the CNs, which are drastically reduced at \SI{2}{m} with the insertion of the \ce{SD_2} block. A possible explanation for this effect is the decrease of the neutron mean free path due to the elastic scattering inside the \ce{D_2} crystal, which increases the effective distance between the moderator surface and the recording surface. 
\FloatBarrier
%%%%%%% BIG TABLE %%%%%%%%%%%%%%%%%%%%%%%%%%
\begin{table}[!bp]
  \centering
\rotatebox{90}{%
    \begin{minipage}{0.9\textheight}
      \def\arraystretch{1.2}
      \centering
      \setlength{\tabcolsep}{1.5mm}
      \small
      \centering
      \captionof{table}{Neutron current and brightness in wavelength groups for the five VCN sources used in the comparison (\cref{fig:comparison_geometries}), calculated for the NNBAR opening. The statistical relative error is also reported.}
\label{tab:comparison_VCN}
    \begin{tabular}{l c c c c c c c c c c}
    \toprule
	 \multirow{3}{*}{}& \multicolumn{10}{c}{Current at all angles [\si{n/s}] } \\
     \cmidrule{2-11}
	& \multicolumn{2}{c}{$\SI{1.8}{\angstrom}<\lambda<\SI{4}{\angstrom}$}     &	\multicolumn{2}{c}{$\SI{4}{\angstrom}<\lambda<\SI{10}{\angstrom}$} 
 &  \multicolumn{2}{c}{$\SI{10}{\angstrom}<\lambda<\SI{20}{\angstrom}$}
 &	\multicolumn{2}{c}{$\SI{20}{\angstrom}<\lambda<\SI{40}{\angstrom}$}
 &	\multicolumn{2}{c}{$\SI{40}{\angstrom}<\lambda<\SI{90}{\angstrom}$} \\	
\cmidrule{2-11}
 &	Value &	\% Rel. Err. &	Value &	\% Rel. Err..&	Value&	\% Rel. Err.&	Value&	\% Rel. Err.	& Value &	\% Rel. Err. \\
\midrule
Baseline & \num{6.20E+13}  &	0.1	& \num{1.26E+14}  &	0.1	& \num{9.73E+12}  &	0.1	& \num{6.17E+11}  &	0.5	& \num{3.19E+10}  &	3.7 \\
\makecell[l]{Baseline \\(ND extraction)} & \num{4.42E+13} &	0.2	& \num{2.19E+14} &	0.1	& \num{2.08E+13} &	0.1	& \num{1.50E+12} &	0.3	& \num{1.10E+11} &	2.0	 \\
\makecell[l]{Full \ce{SD_2}\\(ND extraction)} & \num{7.15E+13}  &	0.2	& \num{1.99E+14}  &	0.1	& \num{3.63E+13}  &	0.1	& \num{7.44E+12}  &	0.2	& \num{2.65E+12}  &	0.6 \\
\makecell[l]{Combined option\\(ideal)}  & \num{1.45E+13} & 	0.3	& \num{4.76E+13} & 	0.1	& \num{8.30E+12} & 	0.1	& \num{2.10E+12} & 	0.4	& \num{1.14E+12} & 	0.7 \\ 
\makecell[l]{Combined option\\(Al vessel)} & \num{1.52E+13} &	0.3 &	\num{5.09E+13} &	0.1 &	\num{7.22E+12} &	0.1 &	\num{1.28E+12} &	0.4 &	\num{3.30E+11} &	1.3 \\
\midrule
 & \multicolumn{10}{c}{Brightness [\si{n/s/cm^2/sr/\angstrom}]} \\
\midrule
Baseline     & \num{7.12E+11}     &	0.4	& \num{7.67E+11}  &	0.3	& \num{3.73E+10}   &	0.3	& \num{1.18E+09}    &	1.3	& \num{2.25E+07}     &	10 \\
\makecell[l]{Baseline \\(ND extraction)}  & \num{9.14E+11} &	0.4 &	\num{1.10E+12} &	0.3 &	\num{5.29E+10} &	0.3 &	\num{1.67E+09} &	1.3 &	\num{2.86E+07} &	9.1 \\
\makecell[l]{Full \ce{SD_2}\\(ND extraction)} & \num{1.46E+12} &	0.5	& \num{9.27E+11} &	0.2	& \num{8.66E+10} &	0.2	& \num{6.71E+09} &	1.0	& \num{5.64E+08} &	3.5 \\
\makecell[l]{Combined option\\(ideal)} & \num{2.72E+11} &	0.8 	& \num{1.86E+11} &	0.3 	& \num{1.71E+10} &	0.3 	& \num{1.40E+09} &	1.7 	& \num{1.80E+08} &	5.1 \\
\makecell[l]{Combined option\\ (Al vessel)} & \num{3.35E+11} &	0.8	& \num{2.35E+11} &	0.3	& \num{1.72E+10} &	0.3	& \num{1.00E+09} &	1.8	& \num{5.79E+07} &	9.6  \\
\bottomrule
    \end{tabular}
    \end{minipage}}
\end{table}
%%%%%%% BIG TABLE Gains %%%%%%%%%%%%%%%%%%%%%%%%%%
\begin{table}[!bp]
  \centering
\rotatebox{90}{%
    \begin{minipage}{0.95\textheight}
      \def\arraystretch{1.2}
      \centering
      \setlength{\tabcolsep}{1.5mm}
      \centering
      \captionof{table}{Neutron current and brightness gains over Baseline in wavelength groups calculated from \cref{tab:comparison_VCN}. Propagated statistical error is also reported.}
\label{tab:comparison_VCN_gains}
    \begin{tabular}{l g w g w g w g w g w}
    \toprule
	 \multirow{3}{*}{}& \multicolumn{10}{c}{Gain on current at all angles} \\
     \cmidrule{2-11}
	& \multicolumn{2}{c}{$\SI{1.8}{\angstrom}<\lambda<\SI{4}{\angstrom}$}     &	\multicolumn{2}{c}{$\SI{4}{\angstrom}<\lambda<\SI{10}{\angstrom}$} 
 &  \multicolumn{2}{c}{$\SI{10}{\angstrom}<\lambda<\SI{20}{\angstrom}$}
 &	\multicolumn{2}{c}{$\SI{20}{\angstrom}<\lambda<\SI{40}{\angstrom}$}
 &	\multicolumn{2}{c}{$\SI{40}{\angstrom}<\lambda<\SI{90}{\angstrom}$} \\	
\cmidrule{2-11}
 &	Value &	\% Rel. Err. &	Value &	\% Rel. Err..&	Value&	\% Rel. Err.&	Value&	\% Rel. Err.	& Value &	\% Rel. Err. \\
\midrule
\makecell[l]{Baseline \\(ND extraction)} 	 & 0.7  &	0.4 &	1.7	& 0.1   &	2.1	&     0.2 &	2.4	& 1.0 &	3.4	& 7.4\\
\makecell[l]{Full \ce{SD_2}\\(ND extraction)}  & 1.2    &	0.3 &	1.6	& 0.1   &	3.7	&     0.2 &	12 & 	0.8 &	83& 	6.0\\
\makecell[l]{Combined option\\(ideal)}   & 0.2    &	0.3 &	0.4	& 0.1   &	0.9	&     0.2 &	3.4	& 0.9 &	36& 	6.1\\
\makecell[l]{Combined option\\(Al vessel)}  & 0.2 &	0.3 &	0.4	& 0.1   &	0.7	&     0.2 &	2.1	& 0.9 &	10& 	6.2\\
\midrule
 & \multicolumn{10}{c}{Gain on Brightness } \\
\midrule
\makecell[l]{Baseline \\(ND extraction)}      & 1.3 &	0.6 &	1.4   &	0.2   &	1.4	& 0.4 &	1.4	& 1.9 &	1.3	& 14 \\
\makecell[l]{Full \ce{SD_2}\\(ND extraction)}   & 2.1 &	0.6 &	1.2   &	0.2   &	2.3	& 0.4 &	5.7	& 1.6 &	25	& 11 \\
\makecell[l]{Combined option\\(ideal)}            & 0.4 &	0.9 &	0.2   &	0.3   &	0.5	& 0.4 &	1.2	& 2.1 &	9.5	& 11 \\
\makecell[l]{Combined option \\(Al vessel)}        & 0.5 &	0.9 &	0.3   &	0.3   &	0.5	& 0.4 &	0.9	& 2.3 &	3.1	& 14 \\
\bottomrule
    \end{tabular}
    \end{minipage}}
\end{table}
\FloatBarrier

\section{Deuterated clathrates hydrates} 
\label{sec:dch}
% \begin{figure}[hbt!]
%    \begin{subfigure}[b]{0.48\textwidth}
%        \centering
%        \includegraphics[width=.75\textwidth]{vcn_fig/SphereWithPureClathrate_Blahoslav.eps}
%        \subcaption{}
%        \label{fig:SphereWithPureClathrate_Blahoslav.eps}
%    \end{subfigure}
%    \hfill
%    \begin{subfigure}[b]{0.48\textwidth}
%        \centering        
%        \includegraphics[width=\textwidth]{vcn_fig/ClathrateInSphere_NeutronFlux_eV_Blahoslav.eps}
%        \subcaption{}
%        \label{fig:ClathrateInSphere_NeutronFlux_eV_Blahoslav.eps}
%    \end{subfigure}
%    \caption{Measuring neutron flux in a spherical moderator to compare the oxygen clathrate hydrate, LD$_2$ and SD$_2$. (a) Model of the spherical moderator  (b) Neutron flux in the moderator.}
%        \label{fig:VCN_Sphere_1}
%    \end{figure}
%
%\begin{figure}[hbt!]
%\begin{center}
%\includegraphics[width=0.75\textwidth]{vcn_fig/Clathrate_Delivery_Box_80x80x80_Blahoslav.eps}
%\caption{Toy model with the volume of moderator box of 80 x 80 x 80 $cm^3$.}
%\label{fig:Clathrate_Delivery_Box_80x80x80_Blahoslav.eps}
%\end{center}
%\end{figure}
%
%\begin{figure}[hbt!]
%\begin{center}
%\includegraphics[width=0.75\textwidth]{vcn_fig/SpectrumAt50cm_ToyModel_80x80x80_Blahoslav}
%\caption{Neutron energy spectrum measured 50 cm away from the emission surface, with size of moderator box of 80 $\times$ 80 $\times$ 80 \si{cm^3}.}
%\label{fig:SpectrumAt50cm_ToyModel_80x80x80_Blahoslav}
%\end{center}
%\end{figure}

Another set of materials investigated within the HighNESS project are deuterated clathrate hydrates \cite{sloan_clathrate_2008} (DCH), which are water-based solids with large crystallographic unit cells. The moderation potential of clathrate hydrates lies in their low energy modes, which are a consequence of the ability of these so-called inclusion compounds to host guest molecules in cages that are formed by networks of hydrogen-bonded water molecules.

While moderation from collective excitations, such as phonons, in a moderator material are kinematically restricted; localized, dispersion-free excitations, such as displacement of confined molecules -- often referred to as Einstein modes --, molecular rotations, librations or paramagnetic excitations do not suffer from this limitation and allow for an efficient neutron slowdown even at lowest neutron temperatures~\cite{Czamler_2023} \cite{zimmer2016neutron}. Clathrate hydrates present numerous localized modes, making them a promising choice for VCN moderators. However, their actual moderation capability for the VCN range still needs to be demonstrated experimentally.
 
Three hydrate systems were investigated within the HighNESS project. The first one hosting deuterated tetrahydrofuran (THF-d) as a guest molecule (molecular formula: \ce{17D_2O.C_4D_8O}) \footnote{This system was studied before and has shown a broad band of low-energy modes, as experimentally demonstrated by \cite{Celli_2012} and \cite{conrad_inelastic_2004}.}, the second one hosting molecular oxygen (\ce{O_2}) as a guest molecule  (molecular formula: \ce{17D_2O.}$\sim$ \ce{3O_2})\footnote{The oxygen abundance is, unlike the THF-d , non-stoichiometric and strongly depends on the sample preparation. This is represented by "$\sim$".} and the third one hosting both THF-d and \ce{O_2} (molecular formula: \ce{17D_2O.C_4D_8O} $\sim$ \ce{3O_2}). Therein, \ce{O_2} occupies the small cages, which are twice as abundant as the bigger cages occupied by the THF-d.

The biggest advantage of the first system is that it can be manufactured with high yields by freezing a stoichiometric mixture of THF-d and heavy-water (\ce{D_2O}). Introducing \ce{O_2} to the hydrate allows for an additional channel of neutron slow-down. The magnetic triplet ground state of molecular oxygen with its zero-field splitting of \SI{0.4}{meV} allows for moderation via a cooling cascade mechanism, as described in~\cite{zimmer2016neutron}. The notably low excitation energy, in contrast to modes in the lattice and THF with energies in the range of several meV, is expected to considerably improve the moderation efficiency towards VCN, with the inelastic magnetic scattering providing a final cooling decrement. In this section we present results of the experimental characterization of THF-hydrates (\cref{sec:exp_clathrate}), as well as first results from neutronic simulations using moderators of \si{O_2}-clathrate (\cref{sec:simulation_hydrate}).

\subsection{Experimental characterization of clathrate hydrates} \label{sec:exp_clathrate}

A necessary characteristic to assess the suitability of a material for moderation purposes is the dynamic structure factor $S(\textbf{q}, \omega)$, which accounts for its structure and excitation spectrum and enters the cross section for VCN production. To determine this quantity in absolute units for the THF-d clathrate hydrate, inelastic scattering experiments using thermal- and cold-neutron time-of-flight instruments at ILL have recently been performed in addition to diffraction experiments for sample characterization. Results of these measurements have been published in~\cite{Czamler_2023}, the key findings are summarized below. Additional experiments on VCN transmission can provide important complementary information, which determines the maximum depth from which VCNs can be extracted from a moderator to a beam.

\subsubsection{Manufacturing of THF-hydrates}  \label{sec_manufacturing}
A starting point of this experimental campaign is establishing a reliable and scalable technique, that allows production of relatively large quantities of hydrates with minimal amounts of residual ice. This technique is described in detail in ~\cite{Czamler_2023} for THF-hydrates having different protonated and deuterated components. The obtained structures were studied by neutron diffraction at the high-intensity two-axis diffractometer D20.
%An exemplary powder pattern and details of the associated Rietveld-refinement are given in \cref{sec_diff_collection}. The structure analysis is followed by a study of the low-energy dynamics of different THF-hydrate samples, as described in \cref{sec_spectroscopy}.

\begin{figure}[h!]
\centering
\hspace{-0.25cm}\includegraphics[width=1.\linewidth]{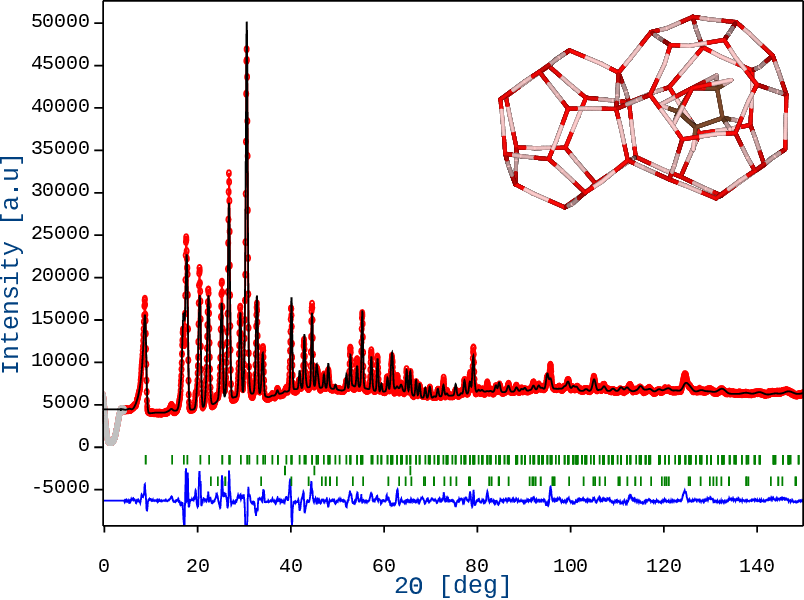}
\caption[Diffraction pattern of fully deuterated THF-hydrates ($\mathrm{THF-d} \cdot \mathrm{D_2O}$).]{Diffraction pattern of fully deuterated THF-hydrates ($\mathrm{THF-d} \cdot \mathrm{D_2O}$) formed in situ taken on D20 with $\lambda=$ \SI{1.546}{\angstrom} at a temperature of 230 K (red) and the corresponding multi-phase Rietveld-refinement (black). The blue line depicts the difference between the data and the fit. The calculated peak positions (green ticks) correspond to phases considered for the refinement, which are, besides the THF-d-hydrate, the aluminum sample environment and residual hexagonal ice. The upper right corner shows 1 of the 8 large cages ($5^{12} \, 6^4$) within the unit cell hosting a THF-d molecule, as well as 1 of the 16 empty small cages  ($5^{12}$). The experimental conditions are described in detail in ~\cite{Czamler_2023}. The data is available under ~\cite{data_Exp_1_42}.}
\label{fig_diffraction_pattern}
\end{figure}

THF (molecular formula \ce{C_4H_8O}) is an organic compound which has a ring structure. A great advantage of THF-hydrates, compared to most other clathrate hydrate compounds, is that they form at ambient pressure from a stoichiometric mixture of its two liquid components, water ($\mathrm{H_2 O}$) and THF ($\mathrm{C_4 H_8 O}$), in a ratio of 17 to 1~\cite{KIDA2021104284} at \SI{280}{K} and below. After carefully weighing and mixing the two liquids with a teflon-coated magnetic stirrer, a cool-down of the solution results in solidification in the CS-II structure (see \cref{tab_sII}). The diffractogram that verifies this structure is shown in Fig. \ref{fig_diffraction_pattern}.
 %The THF-hydrate forms the CSII structure (see \cref{tab_sII}) under ambient pressure -- once the individual components are carefully weighted and mixed with a teflon-coated magnetic stirrer -- by quenching the mixture with liquid nitrogen.
%The CS-II structure is the most common among clathrate hydrates and its ideal unit cell contains 136 ($\mathrm{H_2 O}$) molecules forming 16 small and 8 large cages. This structure has been extensively studied previously (see, for example,~ \cite{gough_composition_1971-1}). The study presented here focuses on the yields of the CS-II structure and quantifying the residual ice, for two different methods to cool down the sample.
\begin{table}[h]
\caption{Characteristics of the CS-II hydrate crystal cell structure. In the case of THF-hydrates the unit cell formula reduces to $8 \, \mathrm{THF}$:$136 \, \mathrm{H_2O}$, with the small cages remaining empty. Table adapted from \cite[p. 60]{sloan_clathrate_2008}.}
\centering
\begin{tabular}{l|l}
Crystal system          & Cubic                                                                                                                                                                                             \\
Space group             & Fd3m (N$^\circ$227)                                                                                                                                                                                    \\
Lattice description     & Face centered                                                                                                                                                                                     \\F
Lattice parameters      & \begin{tabular}[c]{@{}l@{}}a = \SI{17.1}{\angstrom}-\SI{17.33}{\angstrom}\\ $\alpha = \beta = \gamma = 90^\circ$\end{tabular} \\
Number of cages         & 8 large ($5^{12} \, 6^4$), 16 small ($5^{12}$)                                                                                                                                                  \\
Ideal unit cell formula & $8 (5^{12} \, 6^4) \cdot 16 (5^{12}) \cdot 136 \mathrm{H_2O} $                                                                            
\end{tabular}

\label{tab_sII}
\end{table}
Depending on the sample, a clathrate weight percentage of  $95.1 \pm 1.5 \; \%$ to $98.4 \pm 1.6  \; \%$ was reached for the samples investigated, with the uncertainty being dominated by the quality of the fit.
The details of the refinement are given in \cref{tab_refinement}. The fit is significantly affected by the crystallite size, attributed to the texture arising during in situ formation. Despite this it provides evidence that the clathrate structure is reliably produced with high purity from the stoichiometric liquid mixture.

\begin{table}[]
\caption{Refinement details of the diffraction pattern depicted in \cref{fig_diffraction_pattern}. The Bragg R-factor ($\mathrm{R}_\mathrm{B}$) and the weighted profile R-factor ($\mathrm{R}_\mathrm{wp}$) indicate an acceptable fit.}
\centering
\begin{tabular}{l|ll}
Phase           & $\mathrm{THF-d}\cdot\mathrm{D_2O}$                              & Ice $\mathrm{I}_\mathrm{h}$                                                                   \\ \hline
Crystal System  & cubic                                     & hexagonal                                                                \\
Space Group     & F d -3 m  & P 63/m m c                              \\
Cell parameters & a = 17.22(7)                            & \begin{tabular}[c]{@{}l@{}}a = b = 4.50(7)\\ c = 7.35(4)\end{tabular} \\
$\mathrm{R}_\mathrm{B}$  & 10.49                                     & 27.92                                                                    \\
$\mathrm{R}_\mathrm{wp}$          & 4.89                                      & 4.89        
\end{tabular}

\label{tab_refinement}
\end{table}

\subsection{Low-Energy Excitations of THF-Hydrates}

After verifying the formation of the CS-II structure in the samples subjected to both solidification through quenching and slower cooldown, we undertook an investigation of their low-energy excitations. For this purpose, experiments were carried out using the ILL time-of-flight (TOF) spectrometers, specifically IN5 \cite{IN5_layout} and Panther \cite{Panther_layout}. These experiments aimed to measure the dynamic structure function $S(q, \omega)$ across a substantial portion of the $(q, \omega)$-space. Details can again be found in  ~\cite{Czamler_2023}. The data obtained from these measurements play a crucial role in benchmarking density functional theory (DFT) and molecular dynamics (MD) simulations, as elaborated in \cite{Shuqi_ECNS_23} and in this report. The four samples with different combinations of protonated and deuterated components as described in \cref{tab_samples} were measured at a temperature of \SI{1.5}{K}. Data reduction, analysis and visualization were carried out using the Mantid software package \cite{mantid_software} \cite{ARNOLD2014156}.

 \begin{table}[]
 \caption{Differently deuterated and protonated samples prepared for spectroscopy experiments. The deuteration of either the guest molecule or the host lattice allows the highlighting of different parts of the sample in the scattering signal. Note that the full or partial deuteration is capable of slightly changing the dynamics of the sample, while the structure remains the one described in \cref{tab_sII}.}
 \centering
 \begin{tabular}{l|lll}
           & Abbrev. & Host    & Guest       \\ \hline
 Fully protonated & $\mathrm{THF \cdot H_2O}$ & $136 \, \mathrm{H_2O}$ & $8 \, \mathrm{C_4 H_8 O}$ \\
 Deuterated guest & $\mathrm{THF-d \cdot H_2O}$ & $136 \, \mathrm{H_2O}$ & $8 \, \mathrm{C_4 D_8 O}$ \\
 Deuterated cage  &$\mathrm{THF \cdot D_2O}$ & $136 \, \mathrm{D_2O}$ & $8 \, \mathrm{C_4 H_8 O}$ \\
 Fully deuterated &$\mathrm{THF-d \cdot D_2O}$ & $136 \, \mathrm{D_2O}$ & $8 \, \mathrm{C_4 D_8 O}$
 \end{tabular}
 \label{tab_samples}
 \end{table}

\begin{figure}[h!]
\centering
\begin{subfigure}{.5\textwidth}
  \centering
   \includegraphics[width=1\linewidth]{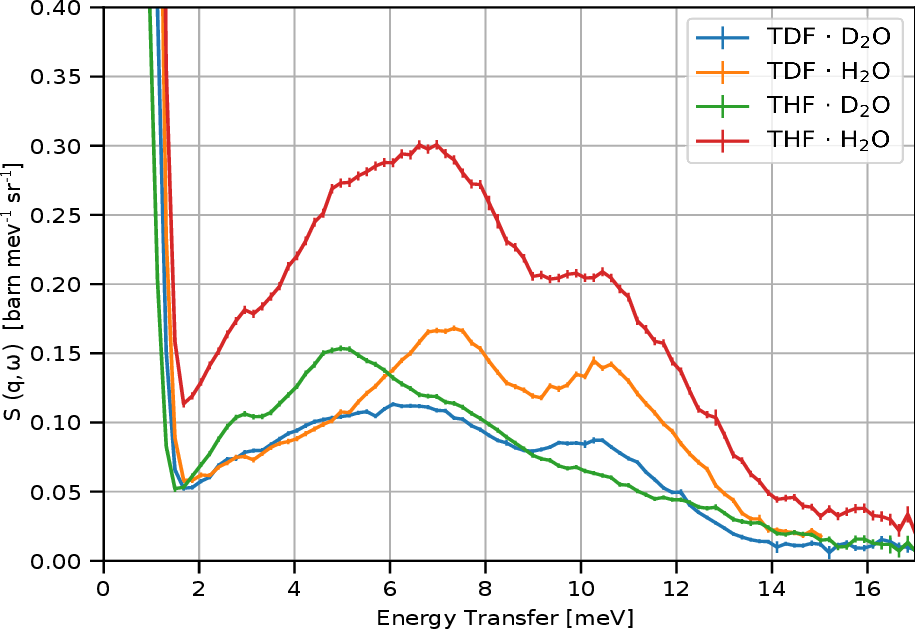}
  \caption{}
  \label{fig_spec_panther}
\end{subfigure}%
\begin{subfigure}{.5\textwidth}
  \centering
  \includegraphics[width=1\linewidth]{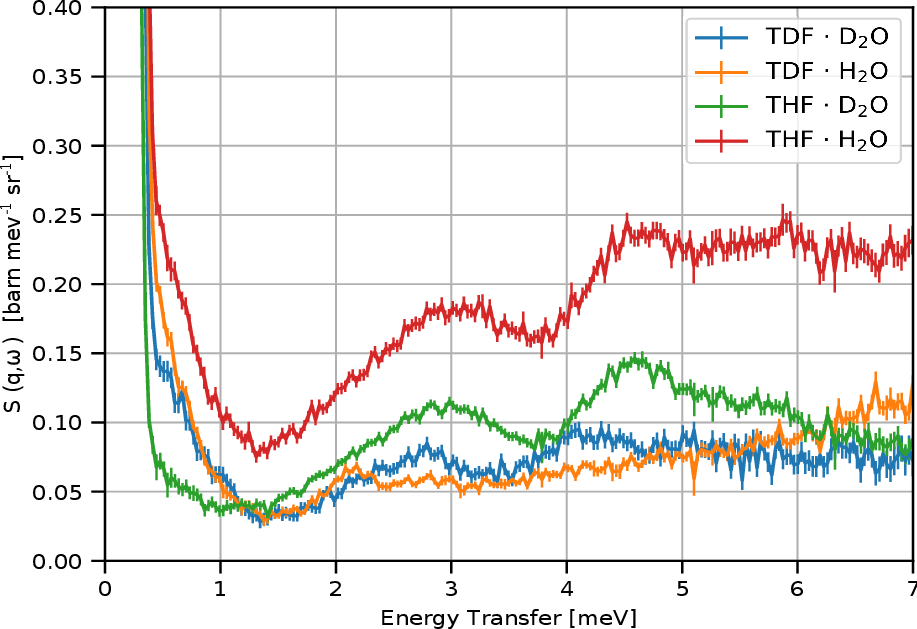}
  \caption{}
  \label{fig_spec_IN5}
\end{subfigure}
\caption{(a) Constant $q$-slice at $q = (4  \pm 1)\; \SI{}{\angstrom^{-1}}$ through $S(q, \omega)$ for different deuterations and protonations of the THF-hydrate measured at \textbf{Panther} with an incident energy $E_i=\SI{19}{meV}$ at a temperature of $T=\SI{1.5}{K}$. The characteristic peaks at \SI{7}{meV} and \SI{10.5}{meV} of CS-II can be well discerned. Preliminary results, data is available under \cite{data_Exp_1_49}. (b) Constant $q$-slice at $q =(3.5 \pm 0.75) \;  \SI{}{\angstrom^{-1}}$ through $S(q, \omega)$ at \textbf{IN5} with an incident energy $E_i =\SI{9}{meV}$ at a temperature of $T=\SI{1.5}{K}$. The observed peaks are due to localized excitations of the THF molecules (see text). The shoulder of the elastic peak is not a feature of the sample but back scattering of the sample environment. The experimental conditions are described in detail in ~\cite{Czamler_2023}. The data data is available under \cite{data_Exp_1_42}.}
\label{fig:tof_spectroscopy}
\vspace{-0.1cm}
\end{figure}

\cref{fig_spec_panther,fig_spec_IN5} show two spectra for different THF hydrate samples measured at Panther and IN5, respectively. These spectra are obtained by integration over a given $q$-range, providing an average of the coherent signal in this range.

%\begin{figure}[h]
%  \centering
%  \includegraphics[width=.7\linewidth]{vcn_fig/Panther_wide_units.eps}
%  \caption[Constant $q$-slice at $q = (4  \pm 1)\; \SI{}{\angstrom^{-1}}$ through $S(q, \omega)$ for different deuterations and protonations of the THF-hydrate measured at \textbf{Panther}.]{Constant $q$-slice at $q = (4  \pm 1)\; \AA^{-1}$ through $S(q, \omega)$ for different deuterations and protonations of the THF-hydrate measured at \textbf{Panther} with an incident energy $E_i=\SI{19}{meV}$ at a temperature of $T=\SI{1.5}{K}$. The characteristic peaks at \SI{7}{meV} and \SI{10.5}{meV} of CS-II can be well discerned. Preliminary results, data is available under \cite{data_Exp_1_49}.}
%  \label{fig_spec_panther}
%\end{figure}
In the investigated energy range, the dynamics of the host structure are primarily determined by the translational modes of the $\mathrm{H_2O}$ or $\mathrm{D_2O}$ molecule. This results in two distinct peaks at approximately \SI{7}{meV} and \SI{10.5}{meV} for each CS-II hydrate structure (see, e.g.  \cite{Chazallon_2002},\cite{schober2003coupling},\cite{conrad_inelastic_2004},\cite{Celli_2012},\cite{broseta_gas_2017}). This can also be observed in \cref{fig_spec_panther} for both the protonated and the deuterated host lattice. The most pronounced peaks can be observed in the $\mathrm{THF-d \cdot H_2O}$ sample (orange), as the influence of the deuterated guest molecule is reduced in the scattering signal.
%\begin{figure}[]
%\centering
%  \includegraphics[width=.7\linewidth]{vcn_fig/IN5_units.eps}
%  \caption[Constant $q$-slice  at $q =(3.5 \pm 0.75) \; \AA^{-1}$ through $S(q, \omega)$ for different deuterations and protonations of the THF-hydrate measured at \textbf{IN5}.]{Constant $q$-slice  at $q =(3.5 \pm 0.75) \; \AA^{-1}$ through $S(q, \omega)$ for different deuterations and protonations of the THF-hydrate measured at \textbf{IN5} with an incident energy $E_i =\SI{9}{meV}$ at a temperature of $T=\SI{1.5}{K}$. The observed peaks are due to localized excitations of the THF molecules (see text). The shoulder of the elastic peak is not a feature of the sample but back scattering of the sample environment. Preliminary result, data is available under Ref. \cite{data_Exp_1_42}.}
%  \label{fig_spec_IN5}
%\end{figure}
The vibrations of the guest molecule occur at lower energy levels and exhibit clear peaks around \SI{2.9}{meV} and \SI{4.7}{meV}, as highlighted in the $\mathrm{THF \cdot D_2O}$ sample (depicted in green) in \cref{fig_spec_IN5}. These localized excitations hold significant potential for moderation into the VCN range.
Substituting hydrogen with deuterium results in an enhanced mass and consequently a greater moment of inertia. This applies to both the translational modes of the host lattice and the excitations of the THF molecule, resulting to a shift towards lower energies. This observation aligns with the phonon density of states (PDOS) computed within the collaboration \cite{Shuqi_ECNS_23}.

%The clathrate hydrates are water-based solids with large crystallographic unit cells. They are capable of hosting guest molecules in cages that are formed by a grid of hydrogen-bonded water molecules. The low energy modes presented in the guest molecules such as tetrahydrofuran (THF) are able to cool down neutrons to the energy range of VCNs.

%The magnetic ground state of oxygen as a hosted molecule with its zero-field splitting of $\SI{0.4}{meV}$ provides with a possibility of an efficient moderation to energy range of VCNs (see \cite{zimmer2016neutron}) that is supported by the inelastic magnetic scattering~\cite{Czamler_2023}. In this section, we present first results from neutronic simulations using moderators of \si{O_2}-clathrate.

\subsection{Simulation of a clathrate hydrate moderator} \label{sec:simulation_hydrate}

\begin{figure}[!htb]
    \begin{subfigure}[b]{0.48\textwidth}
        \centering
        \includegraphics[width=\textwidth]{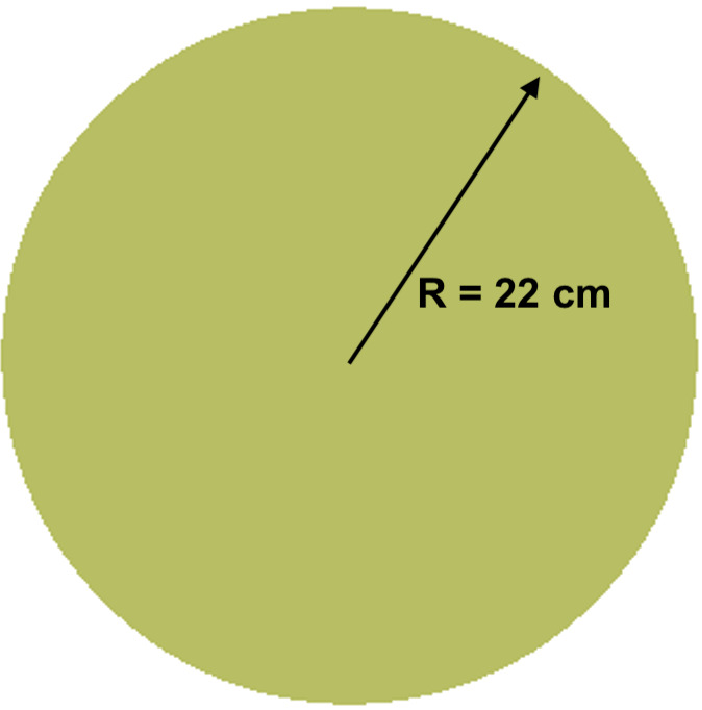}
        \subcaption{}
        \label{fig:SphereWithPureClathrate_Blahoslav}
    \end{subfigure}
    \hfill
    \begin{subfigure}[b]{0.48\textwidth}
        \centering        
        \includegraphics[width=\textwidth]{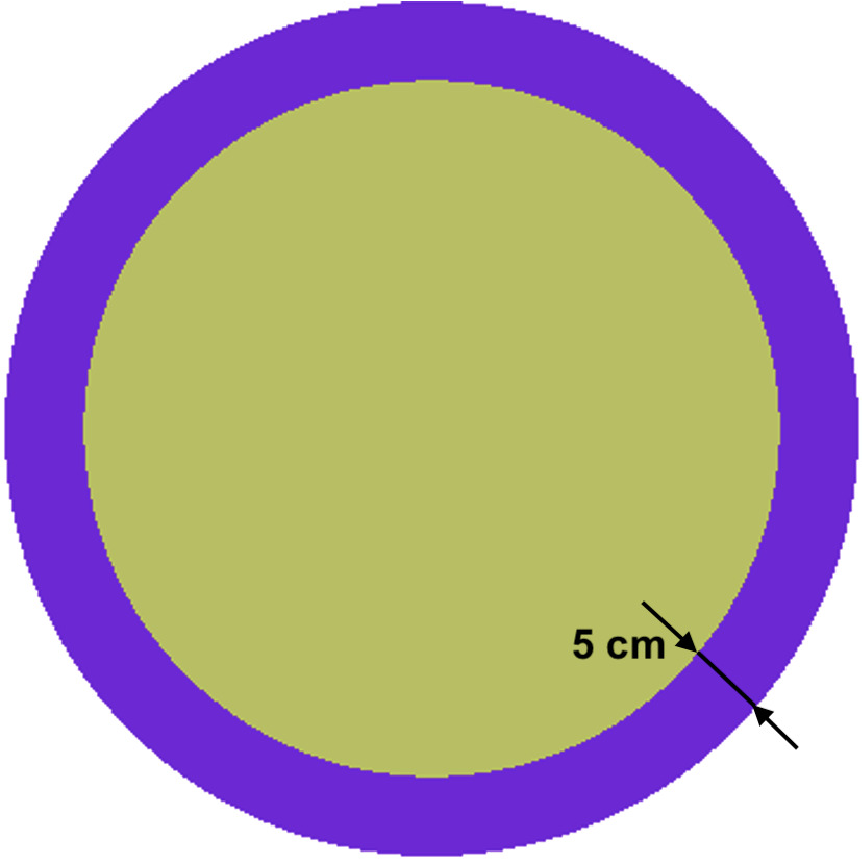}
        \subcaption{}
        \label{fig:SphereWithClathrateAndMgH2Reflector_Blahoslav}
    \end{subfigure}
\caption{An initial model of the clathrate VCN source. (a) Initial model with sphere of radius 22\,cm filled with clathrate hydrate. (b) Second model with additional 5\,cm thick layer of \ce{MgH2} cold neutron reflector.}
\label{fig:preliminary}
\end{figure}

To study the use of clathrate hydrate as a VCN moderator, neutron flux in a simple spherical model was initially calculated in OpenMC~\cite{ROMANO201590}. In this model, a sphere with radius of 22 cm was filled with the clathrate hydrate at 2\,K (see \cref{fig:SphereWithPureClathrate_Blahoslav}) with a thermal (1\,eV) isotropic point source of neutrons located in the center of sphere. The radius of the sphere was chosen so that it has a similar volume as the baseline \ce{LD2} cold moderator. The sphere was subsequently filled with \ce{LD2} and \ce{SD2} for comparison with clathrate.

\begin{figure}[!htb]
    \centering
    \includegraphics[width=0.96\columnwidth]{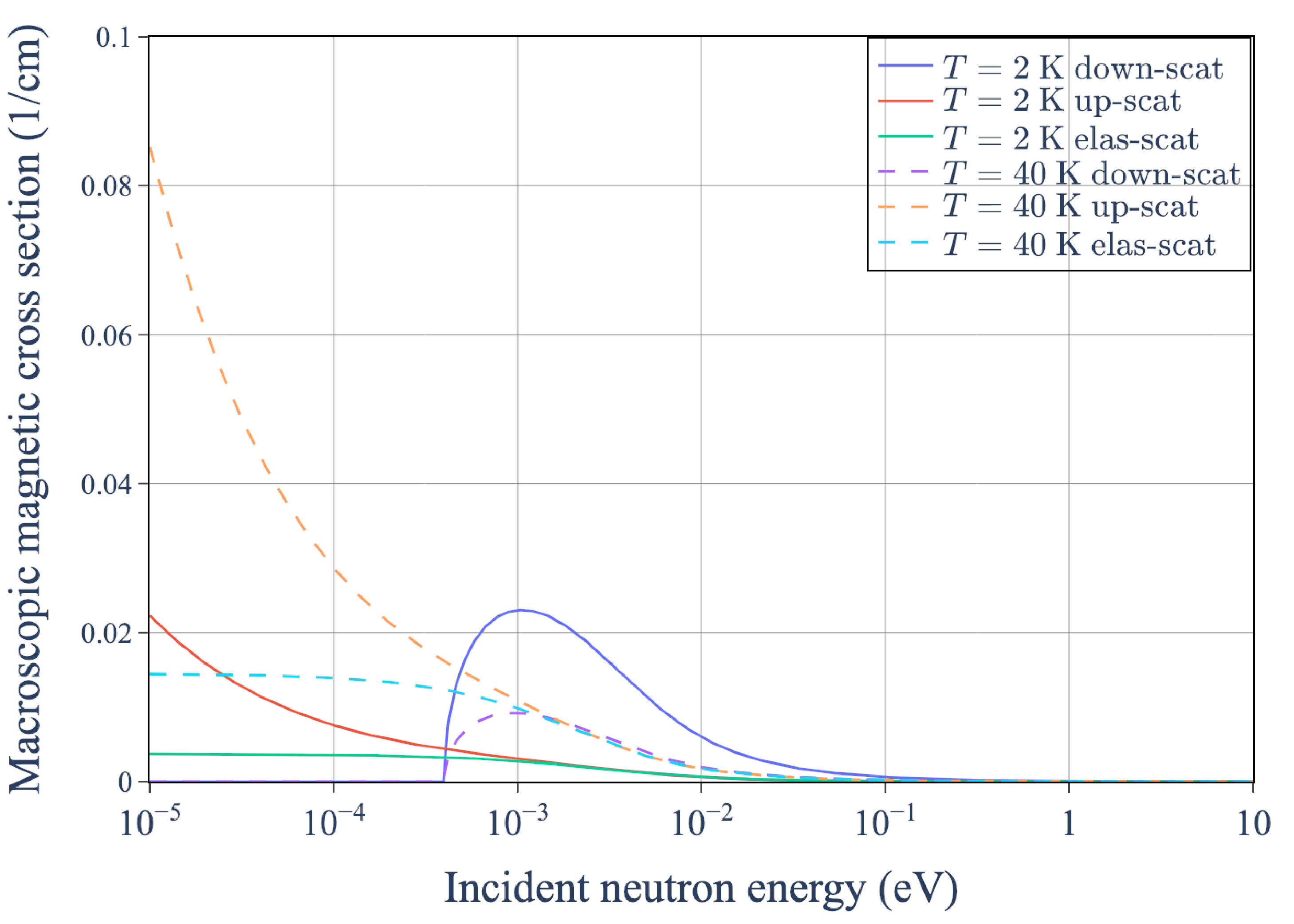}
    \caption{Cross-section for the magnetic scattering in 
    \si{O_2}-clathrate hydrate with 136 \ce{D_2O} and 24 \ce{O_2} in each unit cell. Generated using the tool ncplugin-MagScat presented in Ref.~\cite{xu2023magnetic}.}
    \label{fig:ClathrateCrossSection_Blahoslav}
\end{figure}

\begin{figure}[!htb]      
    \begin{subfigure}[b]{0.48\textwidth}
        \centering
        \includegraphics[width=\textwidth]{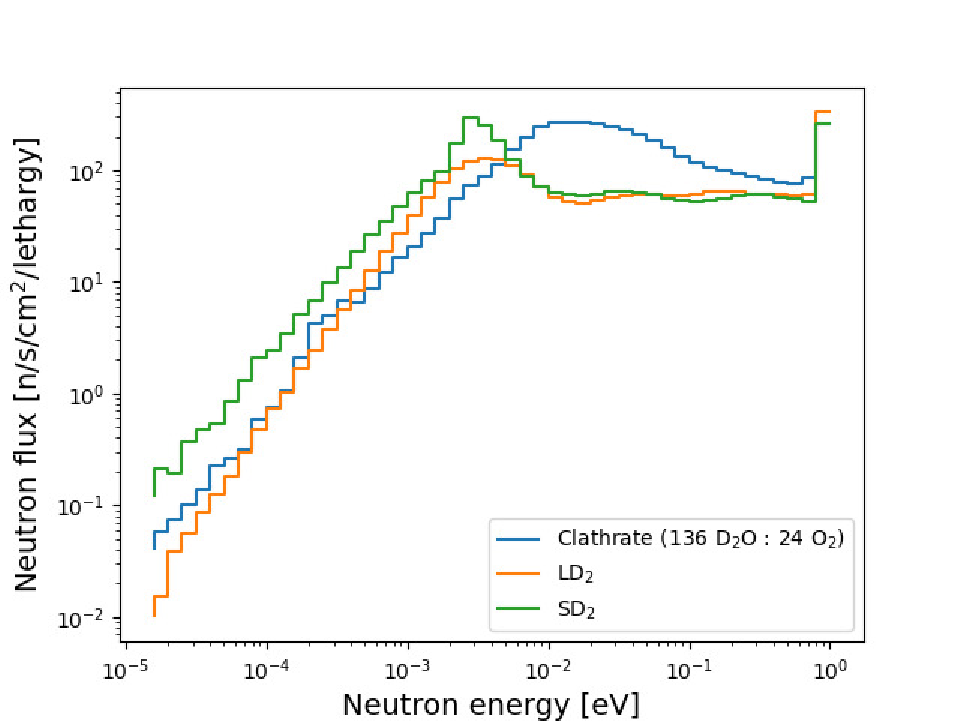}
        \subcaption{}
        \label{fig:ClathrateInSphere_NeutronFlux_eV_Blahoslav}
    \end{subfigure}
    \hfill
    \begin{subfigure}[b]{0.48\textwidth}
        \centering        
        \includegraphics[width=\textwidth]{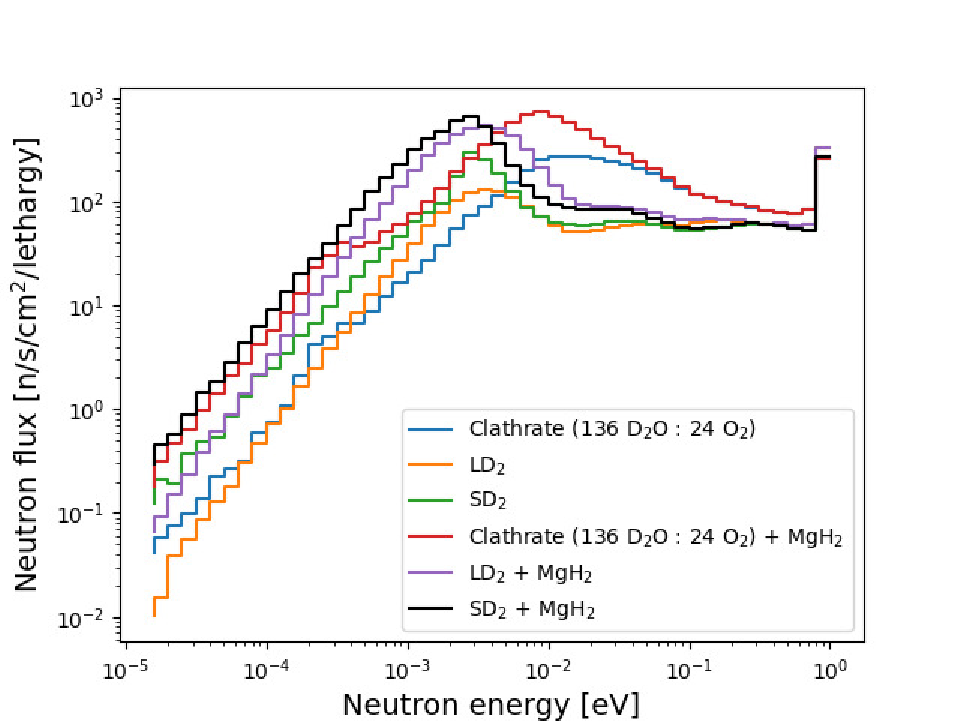}
        \subcaption{}
        \label{fig:ClathrateInSphereAndMgH2_NeutronSpectrum_Blahoslav}
    \end{subfigure}
\caption{Calculated neutron flux inside a sphere filled by \si{O_2}-clathrate, \ce{LD2} or \ce{SD2}. (a) Model without the cold neutron reflector. (b) Comparison of the model with and without the cold neutron reflector, which was is \ce{MgH2} at 20\,K.}
%\label{fig:preliminary_2}
\end{figure}

\begin{figure}[!htb]     
    \begin{subfigure}[b]{0.48\textwidth}
        \centering
        \includegraphics[width=\textwidth]{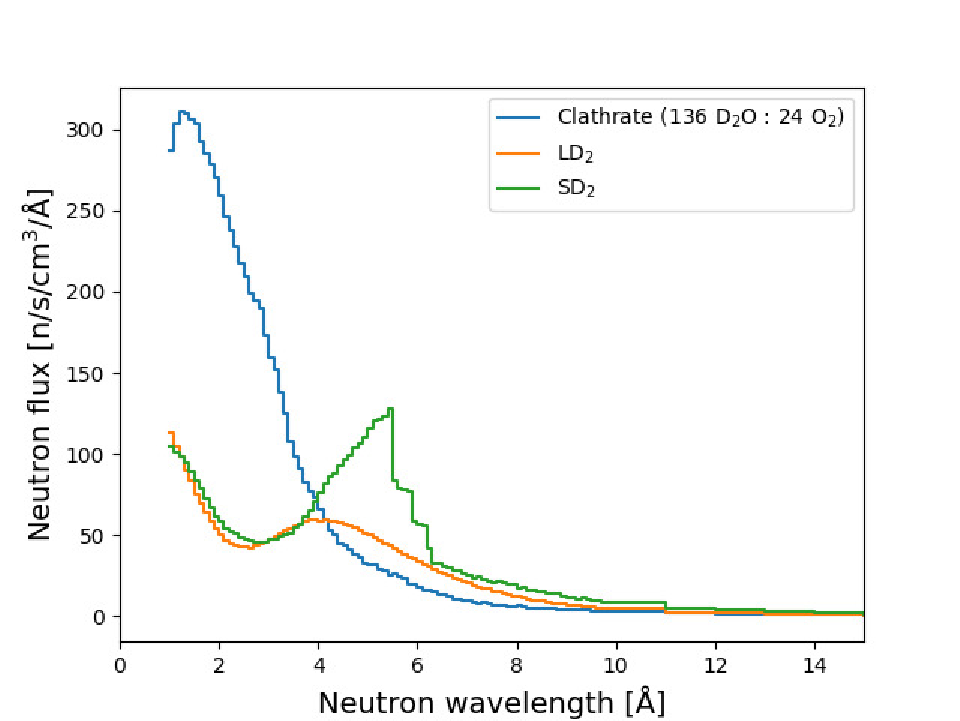}
        \subcaption{}
        \label{fig:ClathrateInSphere_NeutronFlux_AA_-1_Blahoslav}
    \end{subfigure}
    \hfill
    \begin{subfigure}[b]{0.48\textwidth}
        \centering        
        \includegraphics[width=\textwidth]{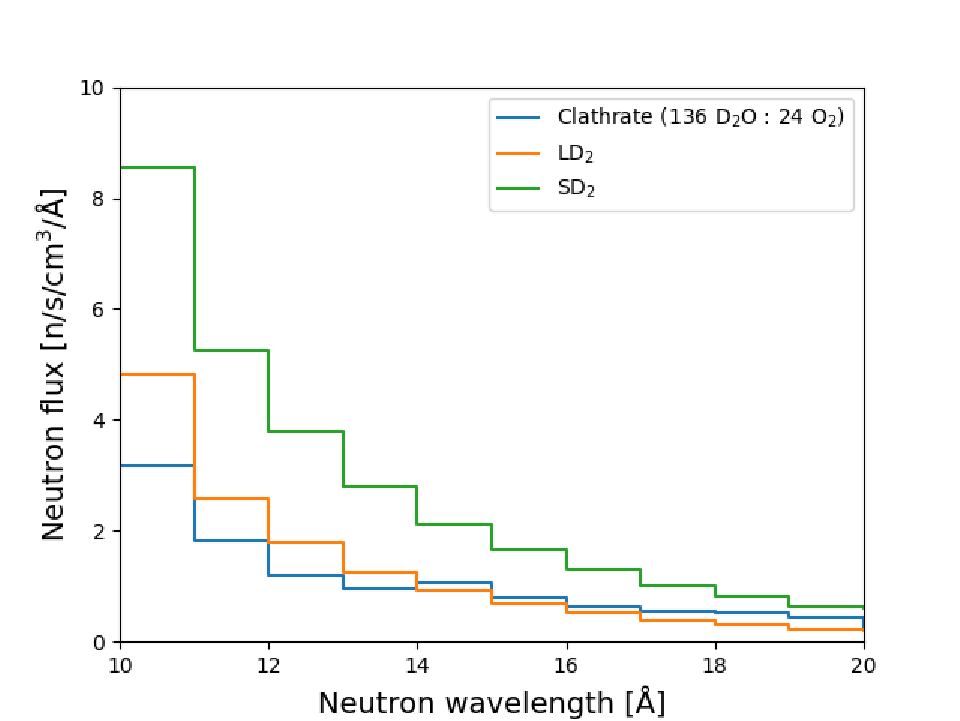}
        \subcaption{}
        \label{fig:ClathrateInSphere_NeutronFlux_AA_-2_Blahoslav}
    \end{subfigure}
\caption{Neutron spectrum computed in the sphere filled by \si{O_2}-clathrate, \ce{LD2} or \ce{SD2}. (a) Spectrum between 1 and 15 \AA. (b) Spectrum between 10 and 20 \AA.}
%\label{fig:preliminary_2}
\end{figure}

\begin{figure}[!htbp]
    \begin{subfigure}[b]{0.48\textwidth}
        \centering
        \includegraphics[width=\textwidth]{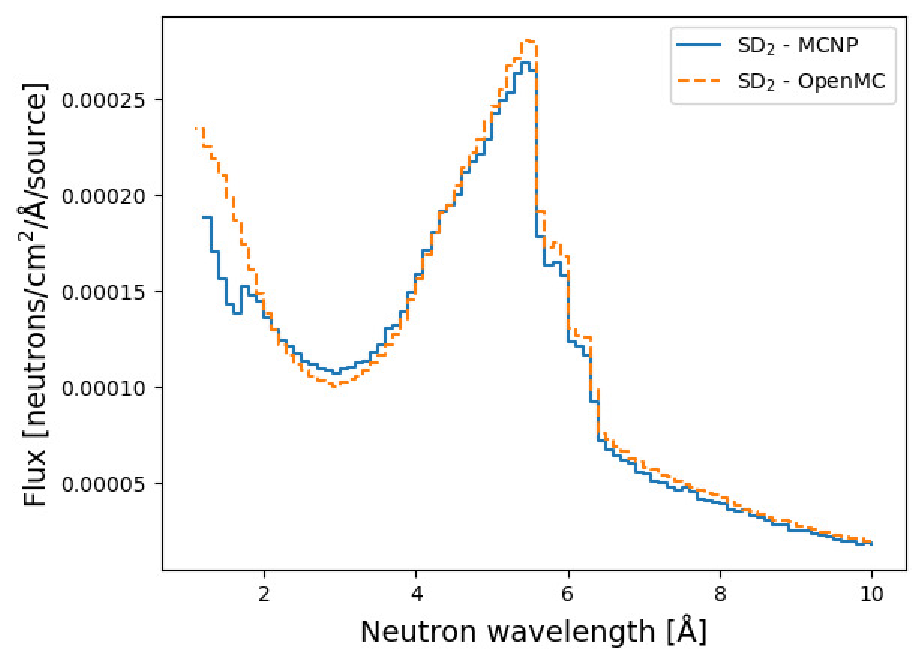}
        \subcaption{}
        \label{fig:SD2Benchmark_MCNP_OpenMC_Blahoslav}
    \end{subfigure}
    \hfill
    \begin{subfigure}[b]{0.48\textwidth}
        \centering        
        \includegraphics[width=\textwidth]{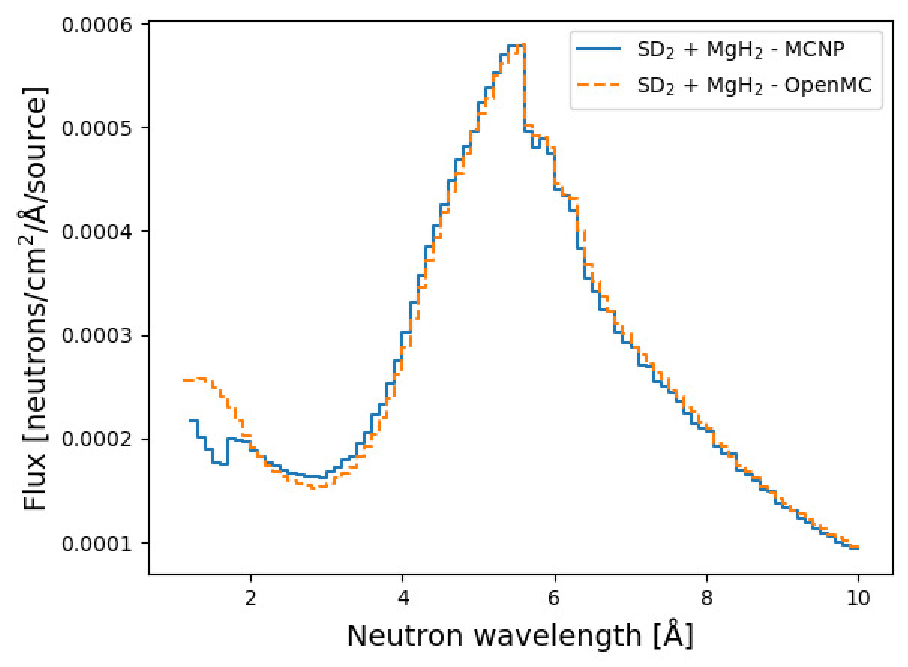}
        \subcaption{}
        \label{fig:SD2+MgH2_Benchmark_MCNP_OpenMC_Blahoslav}
    \end{subfigure}
\caption{Comparison of the neutron wavelength spectra for the spherical moderator filled with \ce{SD2} calculated in MCNP and OpenMC. (a) Spectra without the \ce{MgH2} reflector at 20 K. (b) Spectra with the \ce{MgH2} reflector at 20 K.}
\label{fig:preliminary_25}
\end{figure}

The neutron energy spectrum for each case can be seen in \cref{fig:ClathrateInSphere_NeutronFlux_eV_Blahoslav}, and the neutron wavelength spectrum for each case can be seen in \cref{fig:ClathrateInSphere_NeutronFlux_AA_-1_Blahoslav} and \cref{fig:ClathrateInSphere_NeutronFlux_AA_-2_Blahoslav}. Here, the \ce{SD2} outperforms the clathrate for both the CN ($\SI{2}{\AA}<\lambda<\SI{10}{\AA}$) and VCN range ($\lambda>\SI{10}{\AA}$), while the clathrate was only competitive with \ce{LD2} in the VCN range above $\SI{14}{\AA}$. The reason for the sudden increase in the neutron flux at $\SI{14}{\AA}$ for the clathrate-filled moderator is that the cross section for the magnetic down-scattering is maximal at that wavelength (see \cref{fig:ClathrateCrossSection_Blahoslav}). The wavelength spectrum in \cref{fig:ClathrateInSphere_NeutronFlux_AA_-1_Blahoslav} shows a peak for \ce{SD2} at about $\SI{5}{\AA}$, which appears due to a sharp peak in the elastic cross section for \ce{SD2} at this wavelength. These results from an OpenMC simulation are consistent with those from an identical simulation in MCNP (see \cref{fig:SD2Benchmark_MCNP_OpenMC_Blahoslav}).

\begin{figure}[!htbp]
    \begin{subfigure}[b]{0.48\textwidth}
        \centering
        \includegraphics[width=\textwidth]{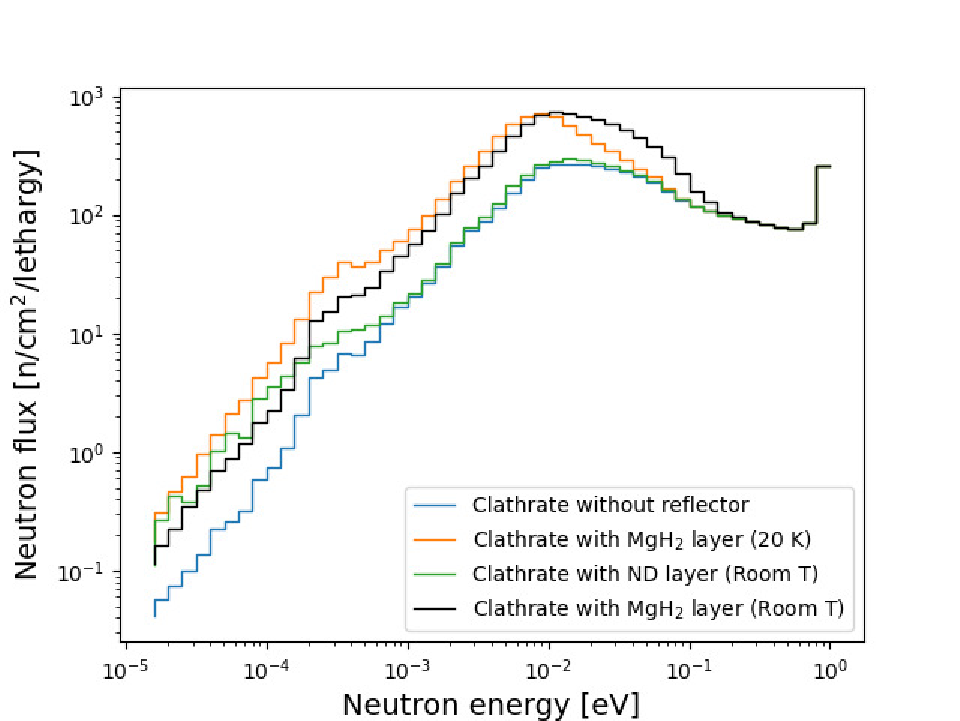}
        \subcaption{}
        \label{fig:TestingColdNeutronReflectorAroundClathrate_eV_Blahoslav}
    \end{subfigure}
    \hfill
    \begin{subfigure}[b]{0.48\textwidth}
        \centering        
        \includegraphics[width=\textwidth]{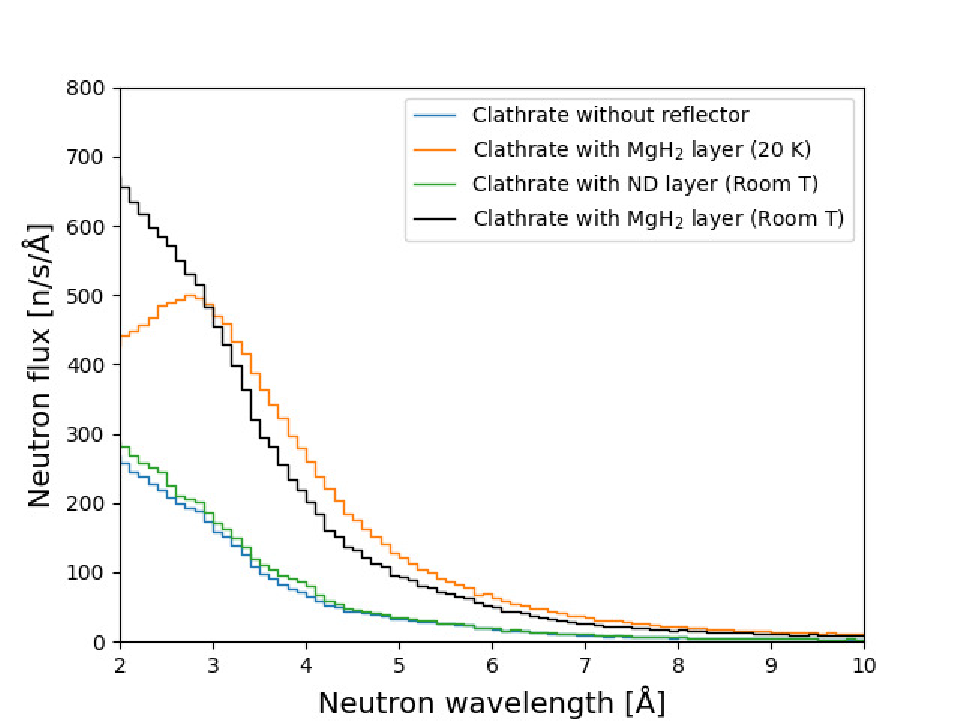}
        \subcaption{}
        \label{fig:TestingColdNeutronReflectorAroundClathrate_AA_Blahoslav}
    \end{subfigure}
\caption{Neutron spectra computed in the sphere filled by \si{O_2}-clathrate and different neutron reflector materials around the moderator. (a) Spectrum in energy. (b) Spectrum in wavelength.}
%\label{fig:preliminary_2}
\end{figure}

A limiting factor of the selected clathrate at 2\,K is the large mean free path of cold neutrons, about 30\,cm, for magnetic down scattering that can be deduced from \cref{fig:ClathrateCrossSection_Blahoslav}. The cross sections are computed using the tool ncplugin-MagScat in which the physics of inelastic neutron magnetic scattering~\cite{zimmer2016neutron} is implemented~\cite{xu2023magnetic}. To mitigate this deficiency, a 5\,cm thick shell of material to reflect cold neutrons was designed around the sphere to allow for more interactions between cold neutrons and the clathrate (see \cref{fig:SphereWithClathrateAndMgH2Reflector_Blahoslav}). This shell was filled with \ce{MgH2} because of its promising properties as a cold neutron reflector, as reported by Granada et al.~\cite{granada2020studies} and based on a study of several neutron reflector materials. This study was conducted using the model in~\cref{fig:SphereWithClathrateAndMgH2Reflector_Blahoslav} where the 5\,cm thick outer shell was filled with several possible cold neutron reflectors (results shown in \cref{fig:TestingColdNeutronReflectorAroundClathrate_eV_Blahoslav,fig:TestingColdNeutronReflectorAroundClathrate_AA_Blahoslav}). In this case, the presence of \ce{MgH2} at 20\,K maximized the neutron flux for the cold and very cold regions.

% The presence of MgH2 or polyethyelene at 20 K maximized the neutron flux for the cold and very cold regions; MgH2 might be preferable on account of possible  degradation of polyethylene when exposed to high neutron fluxes. 

\begin{figure}[!htbp]     
    \begin{subfigure}[b]{0.48\textwidth}
        \centering
        \includegraphics[width=\textwidth]{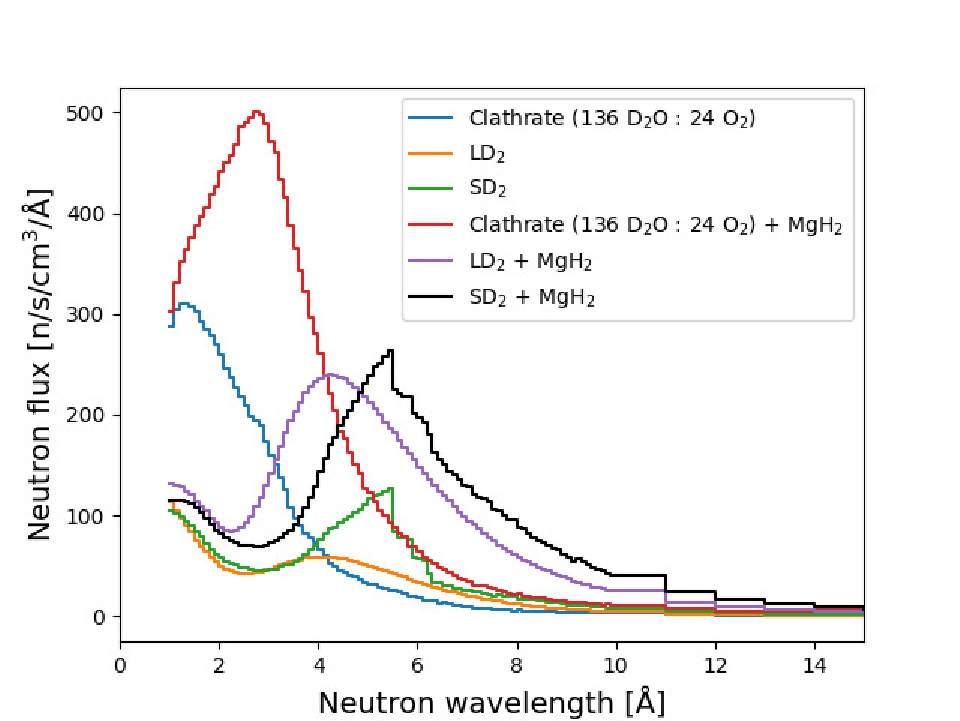}
        \subcaption{}
        \label{fig:ClathrateInSphere_NeutronFlux_AA_0_Blahoslav}
    \end{subfigure}
    \hfill
    \begin{subfigure}[b]{0.48\textwidth}
        \centering        
        \includegraphics[width=\textwidth]{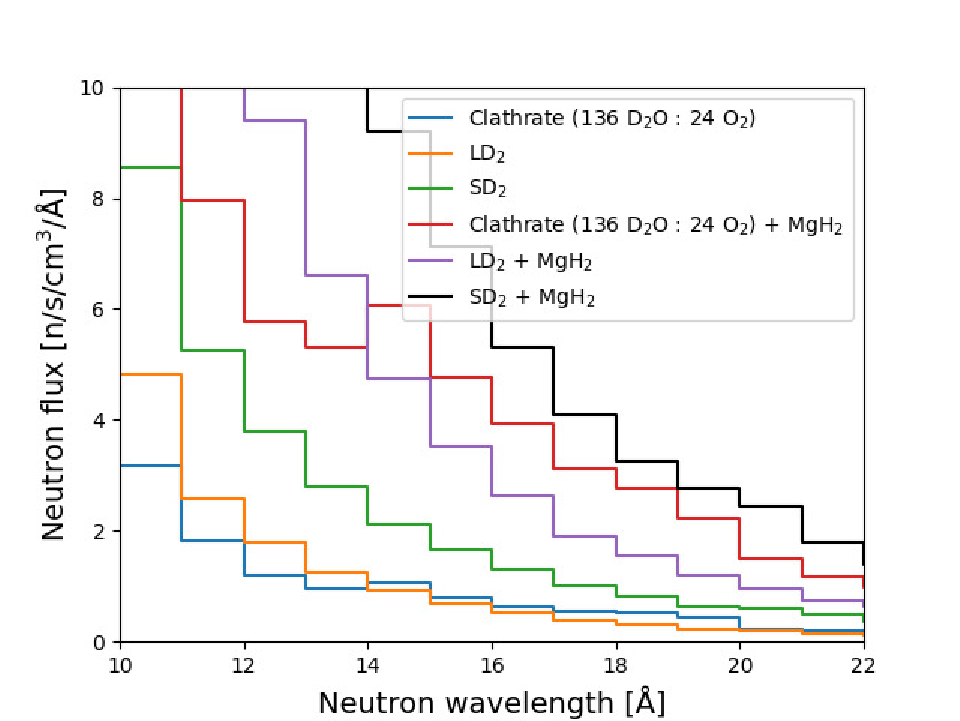}
        \subcaption{}
        \label{fig:ClathrateInSphere_NeutronFlux_AA_1_Blahoslav}
    \end{subfigure}
\caption{Neutron spectrum computed in a sphere filled by \si{O_2}-clathrate, \ce{LD2}, or \ce{SD2}; with or without the \ce{MgH2} reflector at 20\,K. (a) Spectrum between 1 and 15\,\AA. (b) Spectrum between 10 and 22\,\AA.}
%\label{fig:preliminary_2}
\end{figure}

The resulting neutron energy and wavelength spectra for the spherical moderator surrounded by the cold neutron reflector is shown in \cref{fig:ClathrateInSphereAndMgH2_NeutronSpectrum_Blahoslav,fig:ClathrateInSphere_NeutronFlux_AA_0_Blahoslav,fig:ClathrateInSphere_NeutronFlux_AA_1_Blahoslav}. It is clear that the presence of the \ce{MgH_2} reflector improves the efficiency of the clathrate as a VCN moderator much more than for the case of \ce{SD2} and \ce{LD2}. The performance of clathrate with an \ce{MgH2} reflector nearly reached that of \ce{SD2} with a \ce{MgH2} reflector in this scenario. The peak for the \ce{SD2} spectrum at about \SI{5}{\AA} appears in the case with the \ce{MgH2} reflector as well. This result from OpenMC is again consistent with the results from MCNP (see \cref{fig:SD2+MgH2_Benchmark_MCNP_OpenMC_Blahoslav}).

\begin{figure}[!htbp]
    \centering
    \includegraphics[width=0.65\columnwidth]{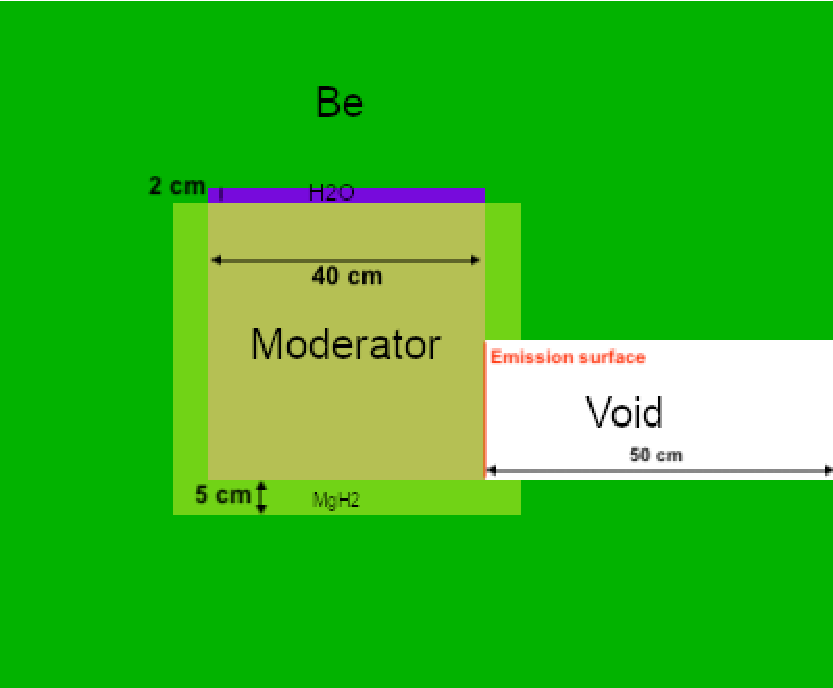}
    \caption{Drawing of the OpenMC toy model of a clathrate moderator, with size of moderator box of 40 $\times$ 40 $\times$ 40 \si{cm^3}.}
    \label{fig:ToyModel_Box40x40x40_Blahoslav}
\end{figure}

\begin{figure}[!htbp]
    \centering
    \includegraphics[width=.9\textwidth]{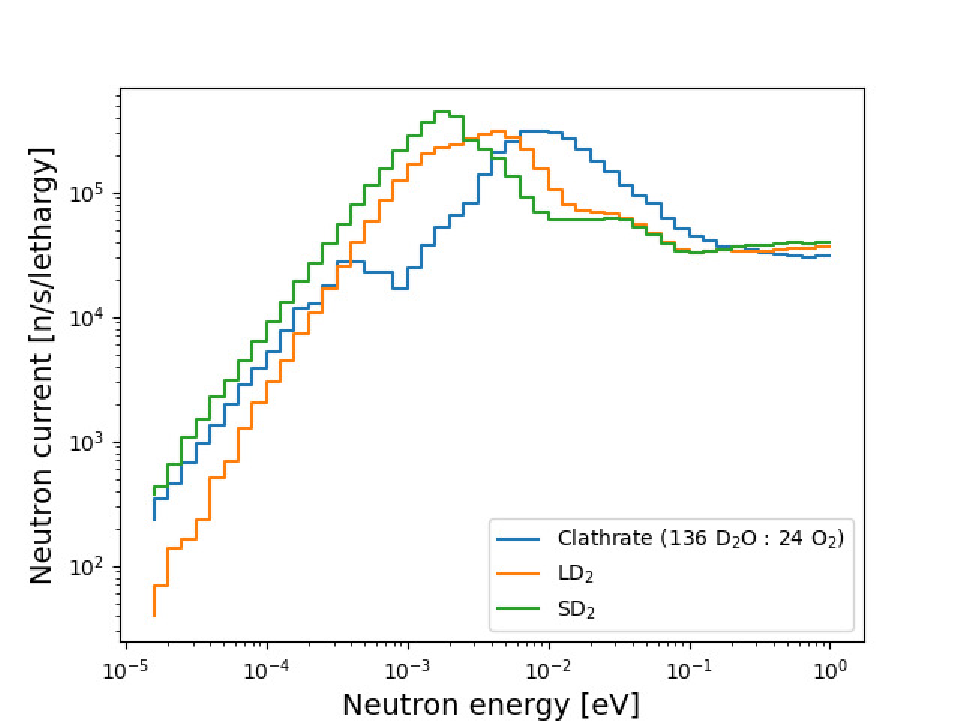}
    \caption{Neutron energy spectrum computed 50 cm away from the emission surface with size of moderator box of 40 $\times$ 40 $\times$ 40 \si{cm^3}.}
    \label{fig:preliminary_4}
\end{figure}

\begin{figure}[!htbp]      
    \begin{subfigure}[b]{0.48\textwidth}
        \centering
        \includegraphics[width=\textwidth]{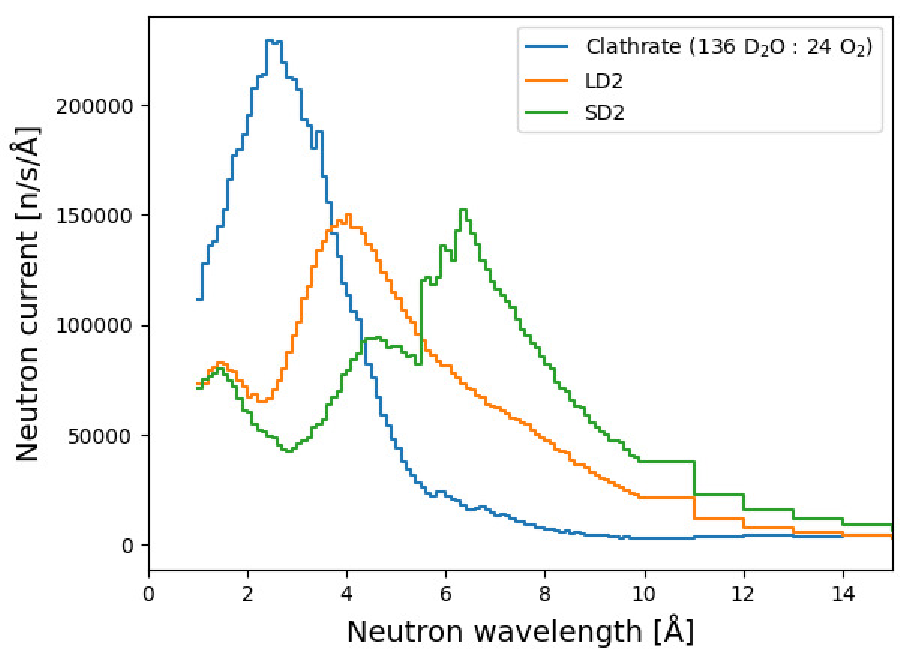}
        \subcaption{}
    \label{fig:SpectrumAt50cm_ToyModel_40x40x40_AA_0_Blahoslav}
    \end{subfigure}
    \hfill
    \begin{subfigure}[b]{0.48\textwidth}
        \centering        
        \includegraphics[width=\textwidth]{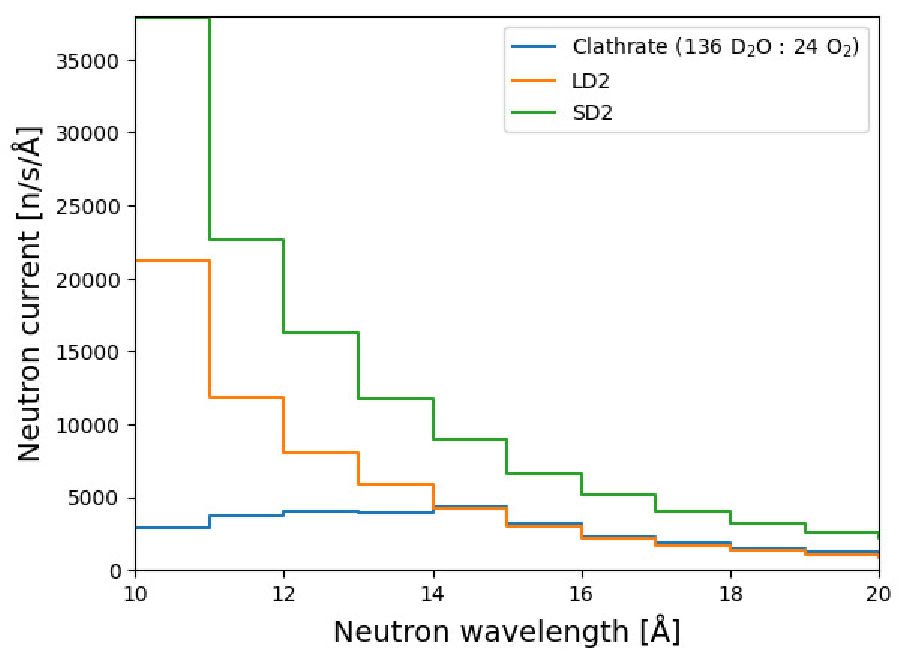}
        \subcaption{}
        \label{fig:SpectrumAt50cm_ToyModel_40x40x40_AA_1_Blahoslav}
    \end{subfigure}
\caption{Neutron wavelength spectra computed 50\,cm away from the emission surface for the model with moderator volume of 40 $\times$ 40 $\times$ 40 \si{cm^3}. (a) Spectra from 1 to 15 \AA. (b) Spectra from 10 to 20 \AA.}
\label{fig:preliminary_8}
\end{figure}

%\begin{figure}[!htbp]      
%    \begin{subfigure}[b]{0.48\textwidth}
%        \centering
%        \includegraphics[width=\textwidth]{vcn_fig/SpectrumAt50cm_ToyModel_80x80x80_AA_0_Blahoslav.eps}
%        \subcaption{}
%\label{fig:SpectrumAt50cm_ToyModel_80x80x80_AA_0_Blahoslav}
%    \end{subfigure}
%    \hfill
%    \begin{subfigure}[b]{0.48\textwidth}
%        \centering        
%        \includegraphics[width=\textwidth]{vcn_fig/SpectrumAt50cm_ToyModel_80x80x80_AA_1_Blahoslav.eps}
%        \subcaption{} \label{fig:SpectrumAt50cm_ToyModel_80x80x80_AA_1_Blahoslav}
%    \end{subfigure}
%\caption{Neutron wavelength spectrum measured 50 cm away from the emission surface for the model with moderator volume of 80 $\times$ 80 $\times$ 80 \si{cm^3}. (a) The spectrum from 1 to 15 \AA. (b) The spectrum from 10 to 20 \AA.}
%\label{fig:preliminary_9}
%\end{figure}

A toy model was designed after finishing the initial study to simulate more realistic conditions for a source made of clathrate hydrate inserted below the spallation target at the ESS. This model contains a cell with a beryllium thermal reflector, a 2-cm thick thermal premoderator filled with \ce{H_2O}, and a cubic-shaped moderator filled either with the clathrate hydrate, \ce{LD2}, or \ce{SD2} surrounded by a 5-cm thick \ce{MgH_2} cold reflector filled. The emission surface has an  area of 20 $\times$ 20 \si{cm^2}, designed together with a 50-cm vacuum pipe for neutron extraction. A surface source with area of 40 $\times$ 40 \si{cm^2} was located at the top of the \ce{H2O} premoderator. The volume of the moderator cell was fixed at 40 $\times$ 40 $\times$ 40 \si{cm^3} (see \cref{fig:ToyModel_Box40x40x40_Blahoslav}). 

The neutron energy spectrum was calculated (see \cref{fig:preliminary_4}) 50\,cm away from the emission surface in the extraction pipe. The neutron wavelength spectra in \cref{fig:SpectrumAt50cm_ToyModel_40x40x40_AA_0_Blahoslav,fig:SpectrumAt50cm_ToyModel_40x40x40_AA_1_Blahoslav} show that the clathrate hydrate becomes a competitive moderator material with \ce{LD2} above about $\SI{14} \AA$. This result is consistent with the results from the spherical moderator and \ce{MgH_2} reflector model. 
%The peak of the \ce{SD2} at about 7 Å that is caused by the elastic part of the cross section not only appeared in the simulation in OpenMC, but was also confirmed by the simulation performed in MCNP as can be seen on \cref{fig:40x40x40ToyModel_SD2_OpenMC_MCNP_Comparison}. 
%The volume of the box moderator was increased to 80 $\times$ 80 $\times$ 80 \si{cm^3} while keeping the dimensions of cold reflector, premoderator and extraction pipe aligned to the bottom of moderator. The results are shown in Ref.~\cite{xu2023magnetic}.
\FloatBarrier

%% file: ucnsource.tex
%\section{Introductory Concepts for Ultra-Cold Neutrons (UCN) (Luca and WP4 team)}
\section{Design of the ESS  Ultracold Neutron source}
\label{sec:UCN_intro}

\subsection{UCN scientific case and general overview of the field } \text

The extremely low energies of ultracold neutrons, in the order of several hundred neV and below, enable them to exhibit unique properties. Neutrons are massive, neutral, nuclear particles and are affected by the gravitational, magnetic, week and strong interaction. Usually, the UCN’s energy is defined as being smaller than the effective strong interaction potential $V_\mathrm{F}$ (Fermi potential) of Beryllium, which is 252 neV. The reflecting properties, together with gravity and magnetic fields led to the idea of building traps for UCNs \cite{Zeldovich_59,Vladimirskii_1961}, in which they can be stored for a long time limited by neutron $\beta$-decay. This characteristic makes them highly responsive to subtle phenomena. In many cases, despite their significantly lower fluxes, UCNs prove to be more sensitive probes compared to thermal and cold neutrons.

%The extremely low energy levels of ultracold neutrons, often on the order of several \SI{}{neV}, enable them to exhibit unique properties. They can be reflected from various materials at all angles of incidence, depending on the energy range associated with the material.  
%Consequently, UCNs can be stored in material traps\cite{Zeldovich_59}, magnetic traps\cite{Vladimirskii_1961}, or even be subjected to gravitational confinement. UCNs have the potential to remain within experimental setups for extended periods, sometimes spanning several hundred seconds. 
%This characteristic makes them highly responsive to subtle phenomena. In many cases, despite their significantly lower fluxes, UCNs prove to be more sensitive probes compared to thermal and cold neutrons.

It was F.L. Shapiro, who pointed out the possibility of precession measurements with UCN \cite{Shapiro_1968}, followed by the first reported observations of UCNs \cite{Lushchikov_1969} \cite{STEYERL196933}. 
This led to rapid developments in the field during the 1970s, resulting in substantial enhancements in precision in measuring the neutron lifetime \cite{history_plot}. They also played a crucial role in setting constraints on the electric dipole moment (EDM) of the neutron \cite{Golub_Pendlebury_1972}, and later in the discovery of gravitational eigenstates of UCN above a flat mirror \cite{Valery_2002}.

Increasing the quantity of ultracold neutrons introduced into experiments remains a valuable approach for improving experimental precision. In cases where the free neutron lifetime is not the main limiting factor for losses, as observed in neutron EDM measurements, significant enhancements can also be achieved by extending the storage duration\footnote{It's important to note that certain experiments depend on a substantial transmitted flux of UCNs rather than a dense storage.}. However, it is important to highlight that regardless of the scenario, existing experiments with ultracold neutrons are primarily restricted by counting statistics. For a comprehensive overview of the scientific capabilities of UCNs and the specific needs of respective experiments the reader can refer to \cite{ABELE20231}.

Currently, there are only a handful of operational UCN sources worldwide, including those at ILL, PSI, TRIGA-Mainz, and LANL. PNPI in Russia has operated several UCN sources in the past, and a new one is currently under construction. Additionally, various UCN sources are either in the construction phase or have been proposed at prominent facilities like FRM-II in Germany, TRIUMF in Canada, LANL and SNS in the USA. For a comprehensive overview of the global UCN landscape, refer to \cite{serebrovworkshop2022}.

The first UCN source developed for user applications is PF2 at the ILL. It is based on the vertical extraction of VCNs from the tail of the Maxwellian distribution of a cold source, followed by final slowdown to UCNs using the \textit{Steyerl-turbine} \cite{pf2_VCN}. All the other sources, operating, under design or under construction, are  based on neutron conversion, using either SD$_\text{2}$ at about 5 K, or superfluid $^\text{4}$He (He-II) at about 1 K or less. While other materials have been considered and are still studied (see \cite{zimmerworkshop2022} and references therein), these two materials have been identified as particularly suitable for UCN production. The transition from a cold neutron  to an ultracold neutron occurs in both of these materials through a single scattering event. Consequently, the UCN do not reach thermal equilibrium with the medium, leading to the classification of such sources as ``superthermal". It is imperative to maintain the medium at extremely low temperatures to minimize the loss of UCN due to up-scattering.

One of the primary objectives of HighNESS is to design a UCN source at ESS. Extensive studies have been conducted, drawing from the two materials mentioned above. This exploration includes various concepts for UCN sources, ranging from those positioned in close proximity to the neutron-production target (referred to as ``in-pile" options) to sources located further away and supplied with a cold neutron beam (referred to as ``in-beam" options). You can find a comprehensive list of concepts and potential locations for possible UCN sources at ESS in \cite{zaniniworkshop2022}.
The findings presented herein are based on the assumption of ESS operating with a proton beam power of \SI{5}{MW}, equivalent to a beam energy of \SI{2}{GeV} and an average current of \SI{2.5}{mA}. As discussed in \cref{timeline}, ESS will start its operation with the reduced power of 2 MW  corresponding to an \SI{800}{MeV} beam energy and maintaining an average current of \SI{2.5}{mA}. Given the linear increase of the neutron spallation yield, the calculated UCN performance and heat loads can be directly scaled to the \SI{2}{MW} operational scenario.

\subsubsection{VCN to Feed UCN Production}
The parallel development of a VCN source in HighNESS has a positive side effect of improving the flux in the wavelength ranges relevant to UCN production (which are discussed in detail in \cref{sec:MethodsForCalculatingUCN}). Thus, the VCN design options proposed in \cref{sec:vcn} may be highly beneficial in terms of UCN production.

For example, as highlighted in \cite{GOLUB1975133}, superfluid He-II can be utilized as a converter in the production of UCN through a superthermal cooling process of CN or VCN (see \cref{sec:HePUCNcalculation}). This is made possible by the crossing dispersion relations between superfluid He-II and the free neutron, allowing neutrons with wavelengths of approximately \SI{8.9}{\angstrom} (equivalent to a kinetic energy of 1.0\,meV) to be scattered down to the ultracold regime emitting a single phonon. An enhancement of the flux of neutrons with wavelengths around \SI{8.9}{\angstrom} would be a valuable asset for in-beam UCN-sources exploiting this mechanism. This includes variants with in-situ UCN production and detection approaches, as proposed in \cite{degenkolb_approaches_2023}.

\subsection{In-pile and in-beam UCN sources}
\label{sec:IntroInPileInBeam}

\begin{figure}[hbtp!]
\begin{center}
\includegraphics[width=0.8\textwidth]{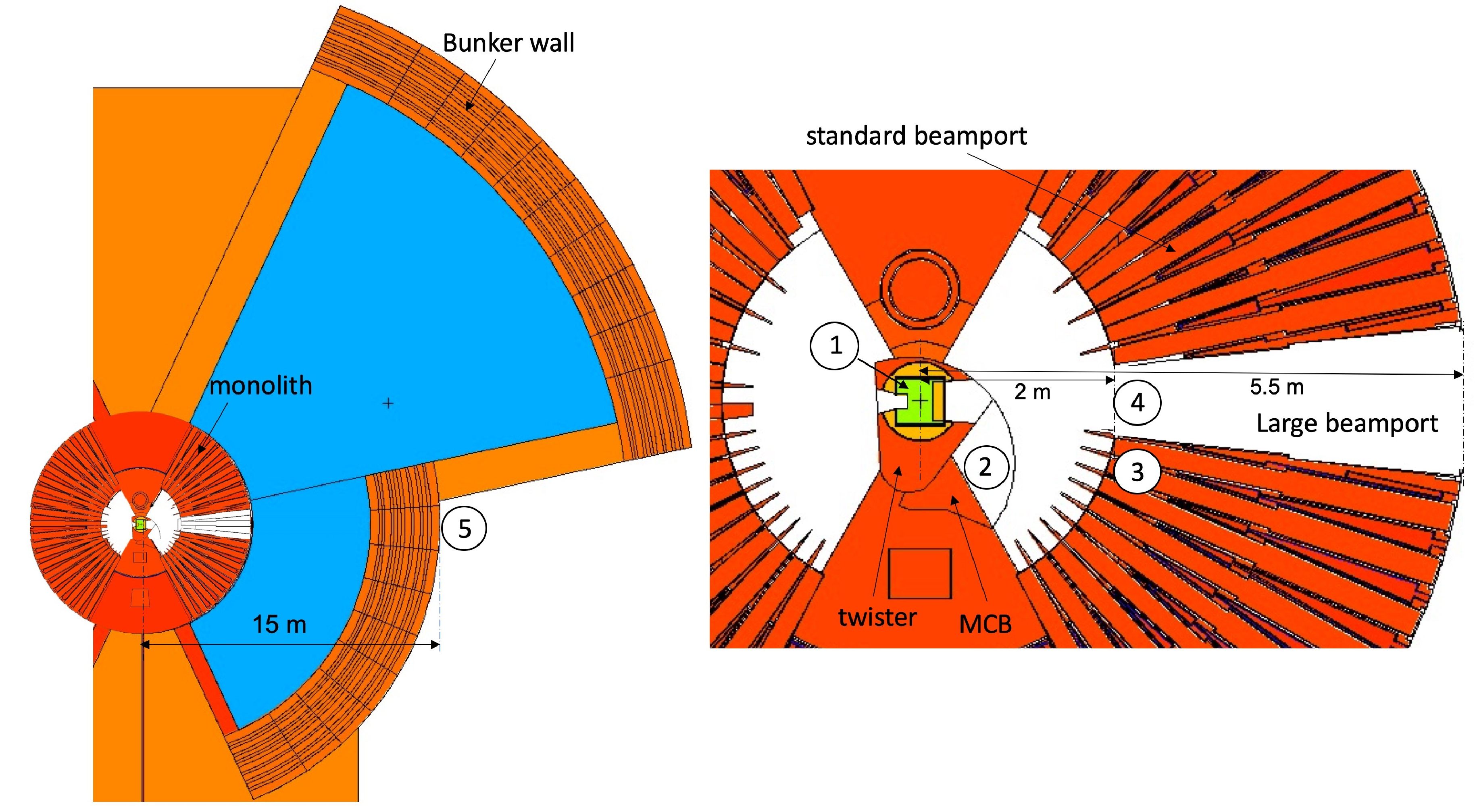}
\caption{
A horizontal cut through the target region, at the height of the LD$_2$ moderator (shown in green) situated below the spallation target (not
visible). The cylindrical region of radius 5.5 m around the center represents the shielding monolith (shown in red). The right figure is a zoom of
the central part of the left figure. About half of its 42 standard beamports are visible in the cut plane. The possible locations of UCN sources, as
studied within the HighNESS project are: (1) inside the “twister”; (2) inside the moderator cooling block; (3) in a standard beamport; (4) in the
large beamport (shown as a white segment in the monolith); (5) outside the “bunker”, a heavy concrete shielding structure (shown in orange)
placed around the monolith; the minimum distance of this location from the moderator is 15 m. See \cite{zaniniworkshop2022} for details and explanation of options
for the various source positions. Reprinted from \cite{zaniniworkshop2022}.
}
\label{fig:oldfig4}
\end{center}
\end{figure}

We have divided the possible implementations of UCN sources at ESS into two groups: {\it in-beam} and {\it in-pile} UCN sources. \\
\indent An in-pile source is placed near the primary source of neutrons, i.e., a reactor core, or the  target in a spallation source. Thus, an in-pile source has  the advantage of receiving a high neutron flux, with the possibility of delivering high UCN yields. This comes however with the challenges  related to positioning such a source near the target, and in particular to keep the source at the desired low temperature (about 5 K for an SD$_\text{2}$ source, and below 2 K for a He-II source). Conversely, an in-beam source is placed at some distance (typically several tens of meters) from the primary cold source. Such an in-beam source, fed by a cold neutron beam, will have lower UCN production rate than for an in-pile source, but there are several advantages, such as: easier engineering (for example, more freedom about the placement of the source, more available space); much easier cooling of the source to sub-K temperature; easier access for maintenance. \\
\indent  The list of the identified in-pile and in-beam options is given in \cite{zaniniworkshop2022} and shown in \cref{fig:oldfig4}. We have adopted the convention to label a source located inside the ESS monolith, i.e., within 5.5 m from the center of the monolith (corresponding to the center of the main moderator) as {\it in-pile}. A source located at larger distance is considered  {\it in-beam} (cf. Figure 4 in \cite{zaniniworkshop2022}).
In HighNESS, we have intestigated only
 one in-beam concept, which has been published in \cite{zimmerworkshop2022}. This in-beam concept exploits the larger field of view offered by the large beamport, and adopts novel nested mirror optics \cite{zimmer2016_MultimirrorImaging}. Alternatively, a regular beamline, i.e. from one of the available beamports of ESS, could be used for an in-beam source similar to the ILL SuperSUN \cite{Chanel}. 
  Performance estimates are given in \cite{skyler2022}. 
  
\indent An alternative approach for UCN studies is to perform the intended experiment in-situ\cite{skyler2022}. Such experiments have previously been realized and are proposed for ESS and other facilities. In this approach, the measurements are performed inside the source, which is based on He-II. The proposed experiments are EDM searches, thus requiring polarized neutrons. The detector is also placed inside the source and can be based on 
$^3$He. The major advantage of in-situ measurements is the fact that transport losses are avoided altogether.
There are some challenges, in particular related to the guiding of cold neutrons in the source, in-situ detection of UCN, and balancing UCN accumulation time with measurement time.
In-situ measurement might be possible for the in-beam option \cite{zimmerworkshop2022} that we propose which uses a He-II source placed at a distance from the ESS cold source, and nested mirror optics to transport the neutrons from the LBP to the UCN source. Additionally, an in-situ configuration could also be possible for a He-II source placed directly in the LBP at about 2 m from the cold source. \\
\indent Concerning  the possible converter materials for in-beam and in-pile UCN sources, we have adopted the following: in the case of in-pile sources, both SD$_\text{2}$ and He-II are possible candidates; therefore both materials have been studied. For an in-beam source, we considered only the option of He-II. The reason for this choice is that we expect higher total UCN production in a large He-II converter placed in a cold neutron beam than in a SD$_\text{2}$ converter. The latter, even when placed in beam, would have to be of relatively small volume, to overcome the issue of poor UCN extraction from the bulk of the SD$_\text{2}$ converter. Because of that, SD$_\text{2}$ is considered a valid candidate for a UCN source only in the in-pile option. In fact, all the UCN sources in operation (LANL, PSI) or under construction (FRM-II), based on SD$_\text{2}$, are  in-pile designs.

In summary, the list of possible UCN sources which have been analyzed within the HighNESS project is the following:

\begin{itemize}
    \item[-] He-II in twister;
    \item[-] \ce{SD_2} in twister;
    \item[-] He-II in MCB;
    \item[-] \ce{SD_2}  in MCB;
    \item[-] He-II in LBP;
    \item[-] He-II in beam.
\end{itemize}

Regarding the last case, He-II in-beam, we have only studied the source's performance using the LBP where thanks to the large available solid angle, special optics can be employed. Performance of a generic in-beam He-II source at the ESS using a regular beamport, can be estimated rescaling the results obtained from this case. 

\subsection{UCN production}
\label{sec:production}

There are several quantities  that can be used to characterize the performance of a UCN source:

\begin{itemize}

    \item [-] UCN production rate density ${P_\text{UCN}}$  [cm$^\text{-3}$ s$^\text{-1}$]. This is the rate of production of UCN inside the source, per unit volume of the source.

    \item[-] UCN production rate $\dot{N}_\text{UCN}$ [s$^\text{-1}$]. Following from integration of ${P_\text{UCN}}$ over the volume of the UCN source, it characterizes the total production rate.
    
    \item[-] UCN density in the source $\rho_\text{UCN}$ [cm$^\text{-3}$]. This quantity is in general time-dependent but here to be understood as the number of UCN per unit volume in the source in steady-state operation. It is sometimes also called saturated UCN density and given by the product of ${P_\text{UCN}}$ and the lifetime $\tau$ of UCNs in the source. $\tau$ is strongly dependent on the material and on its temperature. For SD$_\text{2}$ at 5 K, it is about 40 ms \cite{brys2007extraction,doegethesis}. For He-II, assuming UCN accumulation in a converter vessel with material walls, it is given by the expression \cite{zimmer2014}

    \begin{equation}
    \tau^{-1}=\tau^{-1}_\beta + \tau^{-1}_\text{up}+\tau^{-1}_\text{He3}+\tau^{-1}_\text{wall}\ . 
    \end{equation}

In this expression, $\tau_\beta$ is the mean lifetime of the neutron (880 s); $\tau_\text{up}$ is the up-scattering rate, which for a temperature below 1 K is given by \cite{golub1979,golub1983} $\tau_\text{up} \approx (T[K])^7/100 [s]$; $\tau_\text{He3}$ is the rate of UCN absorption by $^3$He; and $\tau_\text{wall}$ is the rate of UCN loss due to interaction with the walls of the vessel.

\item[-] Total UCN number (or saturated UCN number) in the source $N_\text{UCN}$. It is given by the production rate $\dot{N}_\text{UCN}$ times the UCN lifetime $\tau$ in the source.

\item[-] UCN density in the storage vessel $\rho_\text{UCN, EXP}$[cm$^\text{-3}$]. For an experiment with UCN, the quantity to be optimized is the sensitivity to the effect to be measured, rather than the UCN density in the source. A large UCN density $\rho_\text{UCN,EXP}$ and a large total number of UCNs trapped in an experimental storage vessel, $N_\text{UCN,EXP}$, are often numbers to be considered and optimized in designing an experiment. To estimate these quantities for a specific UCN source, it is necessary to first have at least a preliminary design of a UCN experiment and then calculate the fraction of UCN losses attributed to various factors, such as extraction losses, transport losses, and losses inside the storage vessel. In general, experiments involving a large vessel, such as some neutron lifetime experiments, often aim at maximizing $N_\text{UCN,EXP}$. Certain other experiments relying on extreme suppression of systematic effects, such as nEDM experiments, are preferably done with smaller vessels, in which a high value of $\rho_\text{UCN,EXP}$ is then crucial \cite{zimmer2014}.

\item[-] Total (saturated) number of UCNs in the storage vessel $N_\text{UCN,EXP}$. For this quantity, the same considerations as discussed in the previous bullet point apply.
    
\end{itemize}

Given the multitude of quantities involved, it becomes evident that there is no single, unambiguous figure of merit for optimizing a UCN source. In our studies, our primary focus was on achieving high $\dot{N}$ and $P$, with our analysis primarily centered on the source itself. Assessments of UCN experiment performance connected to a source should be conducted in a subsequent stage, during the design of the UCN experiments.\\
\indent For the same reason, the performance of SD$_\text{2}$-based and He-II based sources are not directly comparable: the UCN lifetimes in the two materials are very different, of the order of 40 $\mu$s for SD$_\text{2}$, and of the order of hundreds of seconds for He-II below 1 K. The scopes of these two source types are thus different as well. SD$_\text{2}$ close to a primary cold source can produce a large UCN flux which is suitable for experiments in flow-through mode or if large vessels need to be filled, whereas He-II enables UCN accumulation to high saturated UCN densities, which can be advantageous for small storage vessel with a long storage time constant. Finally a prerequisite for in-pile source variants, where the UCN converter medium is installed close to a primary cold source, is the capability to transport UCNs with low losses over large distances of several tens of meters.

\subsection{Methods for calculating the UCN production rate density $\text{P}_\text{UCN}$} \label{sec:MethodsForCalculatingUCN}
\subsubsection{Calculating $\text{P}_\text{UCN}$ in SD$_\text{2}$ }
\label{sec:prod_rate_density}
\begin{figure}[tbh!]      
    \centering        
    \includegraphics[width=0.85\textwidth]{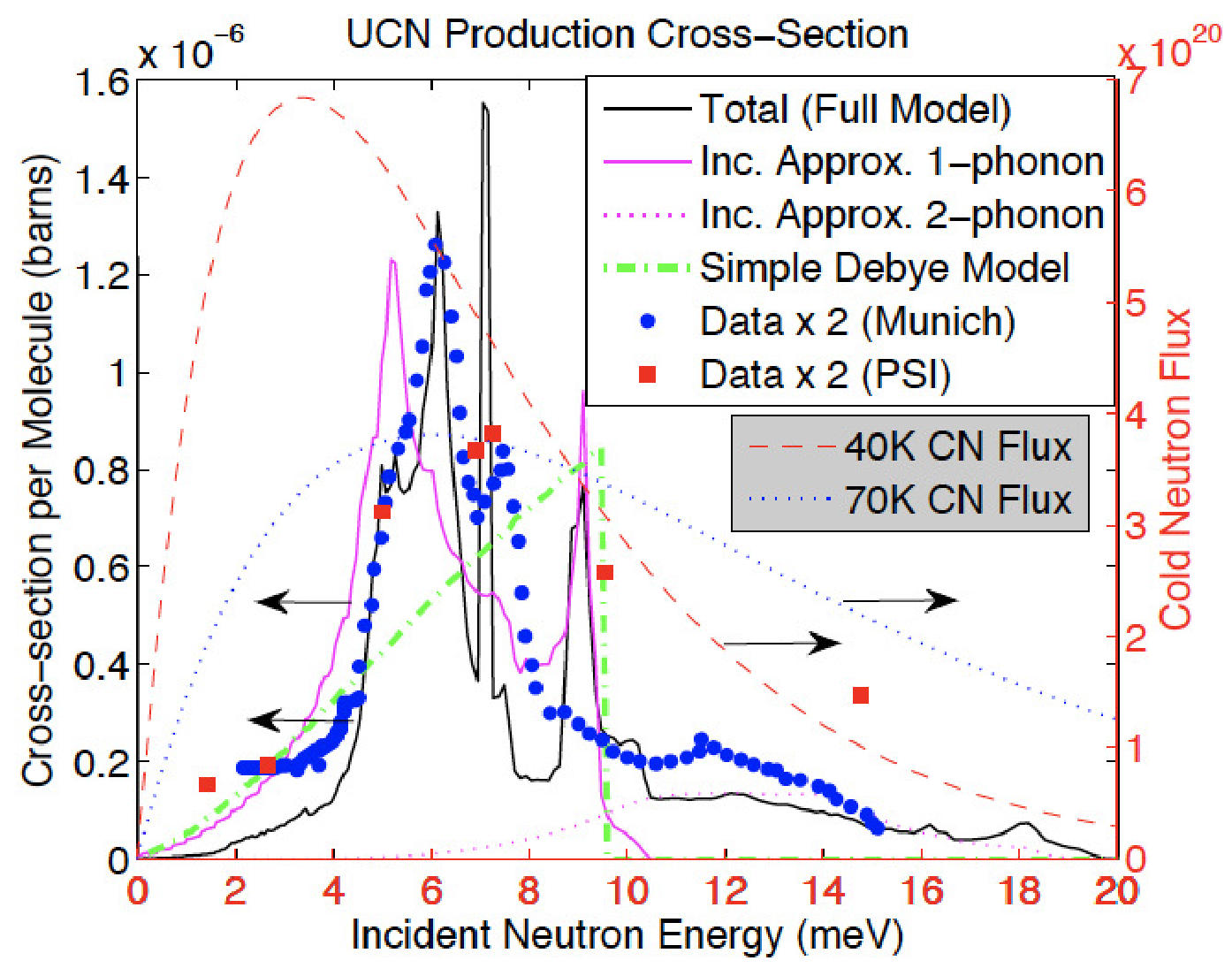}
\caption{The comparison of UCN production cross sections in SD$_\text{2}$. Adapted from Liu et al. \cite{liu2010coherent}. The red squares depict the cross section measured at PSI \cite{atchison2007cold}, whereas the blue circles depict the cross section calculated at TUM in Munich \cite{frei2011understanding}. The calculation of cross section at TUM was based on experimental data from neutron time-of-flight experiment at IN4 (ILL).}
\label{fig:UCNSD2CrossSection_TUMPSIComparison}
\end{figure}

$P_\text{UCN}$ cannot be estimated directly in an MCNP simulation by the conventional method of measuring a neutron flux tally, since MCNP does not transport neutrons in the energy range of UCNs. For converters made of SD$_\text{2}$, $P_\text{UCN}$ was therefore estimated by calculating the cold neutron flux and multiplying it by the UCN production cross-section. The accuracy of such calculation depends on the accuracy of the thermal scattering libraries used in the MCNP calculation, as well as on the accuracy of the UCN production cross-section in SD$_\text{2}$. Several SD$_\text{2}$ UCN cross-sections have been published. We adopted the cross section calculated by Frei et al. \cite{frei2011understanding} at the Technical University of  Munich. This cross-section was calculated from a dynamical scattering function extracted from the IN4 experiment conducted at ILL with the SD$_\text{2}$ sample kept at 4 K and an incident neutron energy of 67 meV (see \cref{fig:UCNSD2CrossSection_TUMPSIComparison}). This cross-section  differs slightly from the one measured in an experiment at the FUNSPIN beamline at PSI with a sample kept at 8 K (see \cref{fig:UCNSD2CrossSection_TUMPSIComparison}). The observed discrepancies between these two cross sections can be, for example, explained by a different crystal orientation, sample temperature, sample purity and crystal quality.

\subsubsection{Calculating $\text{P}_\text{UCN}$ in He-II  } \label{sec:HePUCNcalculation}

\begin{figure}[tbh!]
\centering
\includegraphics[width=0.85\textwidth]{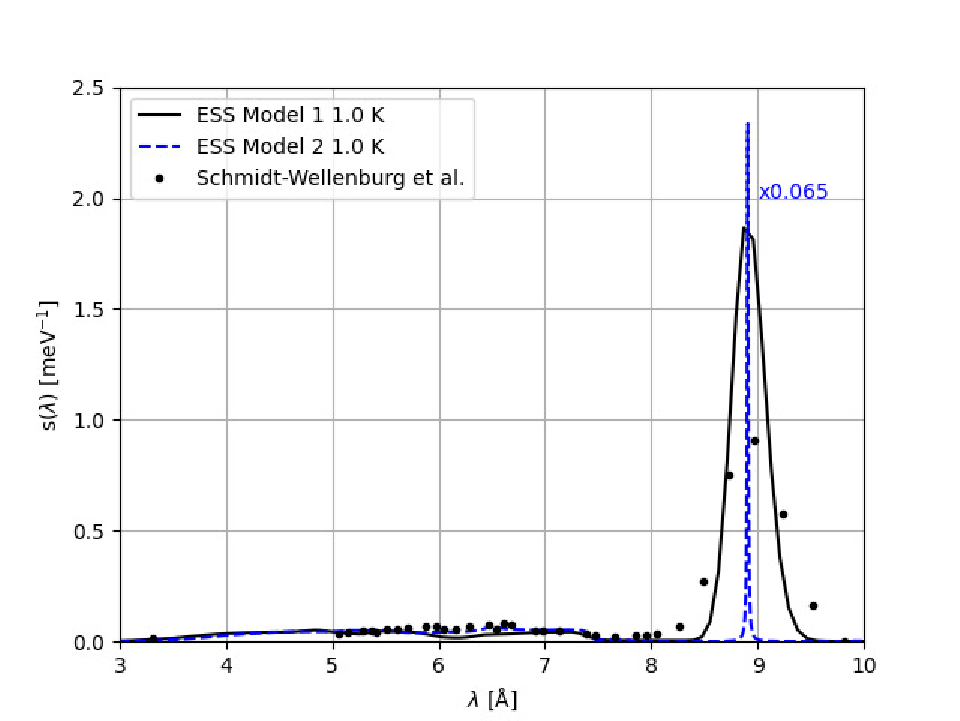}
\caption{The scattering function, $s(\lambda)$, for UCN production in He-II. The curve from Schmidt-Wellenburg et al. \cite{Schmidt_2009} has been determined experimentally. The ESS model 1 curve was used for this work and is based on an early version of the scattering library \cite{ICNS_he4}, while ESS model 2 is an improved version, which can be used for future work \cite{GRANADA2023168284}. The single-phonon peak for ESS Model 2 has been multiplied by a factor of 0.065 for visualization purposes.}
\label{fig:s_lambda}
\end{figure}

%The interaction processes of neutrons in He-II differs from the previously described SD$_\text{2}$.
The scattering function for He-II, $s(\lambda)$, which the UCN production calculations are based on, is shown in \cref{fig:s_lambda}. The model used for the library generation is based on the separation of the single-phonon excitations, described by the phonon-roton dispersion relation, and the multi-phonon excitations, which have been included using a weighted frequency spectrum with the phonon-expansion approach in the Gaussian approximation \cite{D4.4}.

The scattering function can then be derived from this model and has also been determined experimentally \cite{Schmidt_2009}. Both functions are shown in \cref{fig:s_lambda}. ESS Model 1 refers to the version used for this work, while ESS Model 2 is based on a revised methodology \cite{GRANADA2023168284}, which can be used in future work.
It can be seen in the figure that the main contribution to the UCN production is given by a narrow wavelength band around 8.9~\AA~,but also higher energetic neutrons between 2~\AA~and 6~\AA~contribute to UCN production (via multi-phonon processes).
%Another major difference of He-II compared to SD$_\text{2}$ is the vanishing absorption cross section that leads to a lifetime of the converted UCN in He-II in the order of the half life of the unbounded neutrons ($\approx$ 10 min) compared to SD$_\text{2}$ ($\approx$40ms). 
%(THIS PART WILL BE IMPROVED AND EXPANDED) \\

Following \cite{Schmidt_2009}, $P_\text{UCN}$ is given by

\begin{equation}
P_\text{UCN} = N \sigma V_c \frac{k_c}{3 \pi}
\int_{0}^{\infty}\frac{d \phi}{d \lambda}
s(\lambda) \lambda d\lambda,
\label{eq:He_II_P_UCN}
\end{equation}

\noindent where $N$ is the He number density, $\sigma$=1.34 b is the bound neutron scattering cross section for $^\text{4}$He; 
$\hbar k_c = \sqrt{2 m_n V_c}$;  $V_c$ is the wall potential of the converter vessel with respect to the Fermi potential of He-II (18.5 neV at SVP); we consider beryllium as wall material, with an optical potential of 252 neV; therefore  $V_c$= 233 neV;
$s(\lambda)$ is the UCN scattering function as a function of the neutron wavelength $\lambda$; $\frac{d\phi}{d \lambda}$ is the differential incident flux.

\subsubsection{UCN production rates as a function of the distance from the \ce{LD_2} moderator}

\begin{figure}[tbh!]      
    \centering        
    \includegraphics[width=0.85\textwidth]{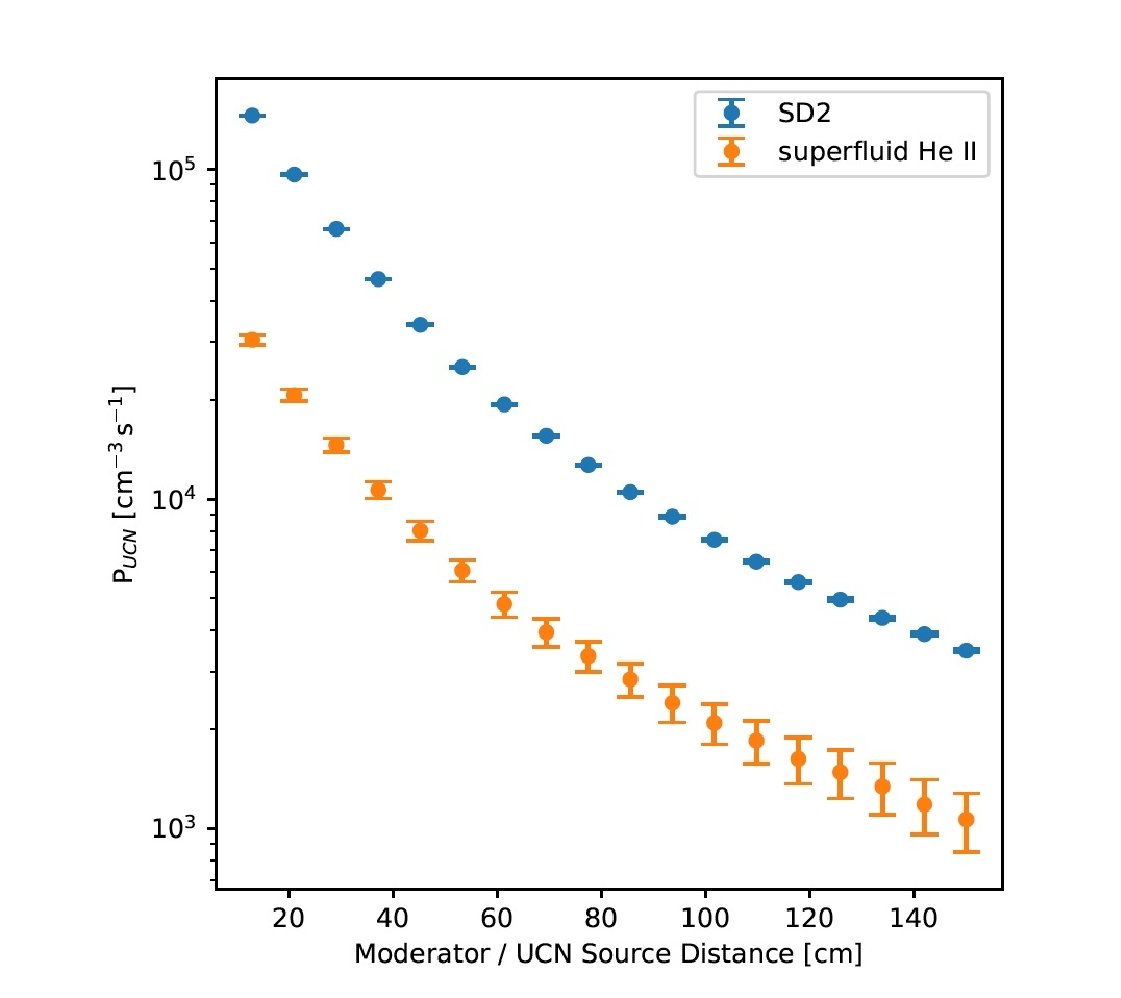}
\caption{Calculated production rate densities for He-II \cite{Schmidt_2009} and SD$_\text{2}$ \cite{liu2010coherent} as a function of the
distance from the LD$_2$ moderator. The  distances are from the center of the UCN source to
the edge of the beryllium filter on the NNBAR side of the lower moderator. The UCN converters are not modelled.}
\label{fig:distance}
\end{figure}

The calculated production rate density in He-II and SD$_\text{2}$, as a function of the distance from the cold moderator, ranging from the moderator surface to 150 cm away, is shown in \cref{fig:distance}.
These estimates are obtained following the procedure explained in the sections above.
Neutron fluxes are calculated at a given distance from the moderator surface; the SD$_\text{2}$ or He-II converters are not modelled. These curves
are therefore representative of the production in He-II or thin
films of SD$_\text{2}$, i.e., for configurations where the perturbation of the incoming cold 
flux, from the converter
itself, is negligible.

\subsection{\ce{SD_2} source in twister}
\label{sec:sd2_intwister}

The first option that we consider is to place the UCN source as close as possible to the cold source inside the lower moderator plug in the ESS twister (see Section
\ref{sec:fac}). 
This option corresponds to {\it option 1} in \cref{fig:oldfig4}.
%For an explanation  of the geometry and the purpose of the twister, 
The rationale behind this concept is  to maximize the cold flux delivered to the UCN converter and therefore the UCN production, in agreement with the results shown in \cref{fig:distance}. This solution requires either to design a dedicated second-generation cold moderator or to find a design that complements the first-generation cold source, without impacting the needs of the NNBAR experiment and of the neutron scattering instruments.  
As indicated in \cite{zaniniworkshop2022}, 
a possible design which, thanks to the reduced amount of SD$_\text{2}$ inside the twister, might be possible to cool and operate, consists of
a thin slab of SD$_\text{2}$ placed close to the main cold source. We  explore several ideas using a thin SD$_\text{2}$ converter in the twister in \cref{sec:thinslab,sec:hole,sec:cyl}.
A larger volume of SD$_\text{2}$, if coolable at the high power levels of ESS, could be a very promising source of VCNs, as well as UCNs.
This is discussed in \cref{sec:prim}.

%, in view of the known difficulty in extracting UCNs from the bulk of a large SD$_\text{2}$ volume, we decided to explore more in depth the option where the amount of SD$_\text{2}$ inside the twister is limited. 
%(\textcolor{red}{we should however add the case of UCN extracted from the last 2 cm of a full SD$_\text{2}$ moderator which I think has been studied}). \\
%With both \ce{LD_2} dedicated or not 
%\indent The case of a  He-II converter inside the twister, instead of SD$_2$ would most likely be of only academic interest, i.e., to determine the maximum UCN production rate in the case where a large He-II converter fills most of the available place inside the twister. Such a source would most likely be impossible to cool at the desired temperatures; additionally, due to the need of a large He-II volume, it would most likely fill most of the available space in the twister; this would prevent the placement of a highly performing CN or VCN source below the spallation target, hence strongly limiting the scientific possibilities of instruments looking at the lower moderator systems. Nevertheless, even if purely for academic reasons, such a case should be studied in the future.

\subsubsection{Thin-slab external converter}
\label{sec:thinslab}
\begin{figure}[tb!]      
    \begin{subfigure}[b]{0.48\textwidth}
        \centering
        \includegraphics[width=\textwidth]{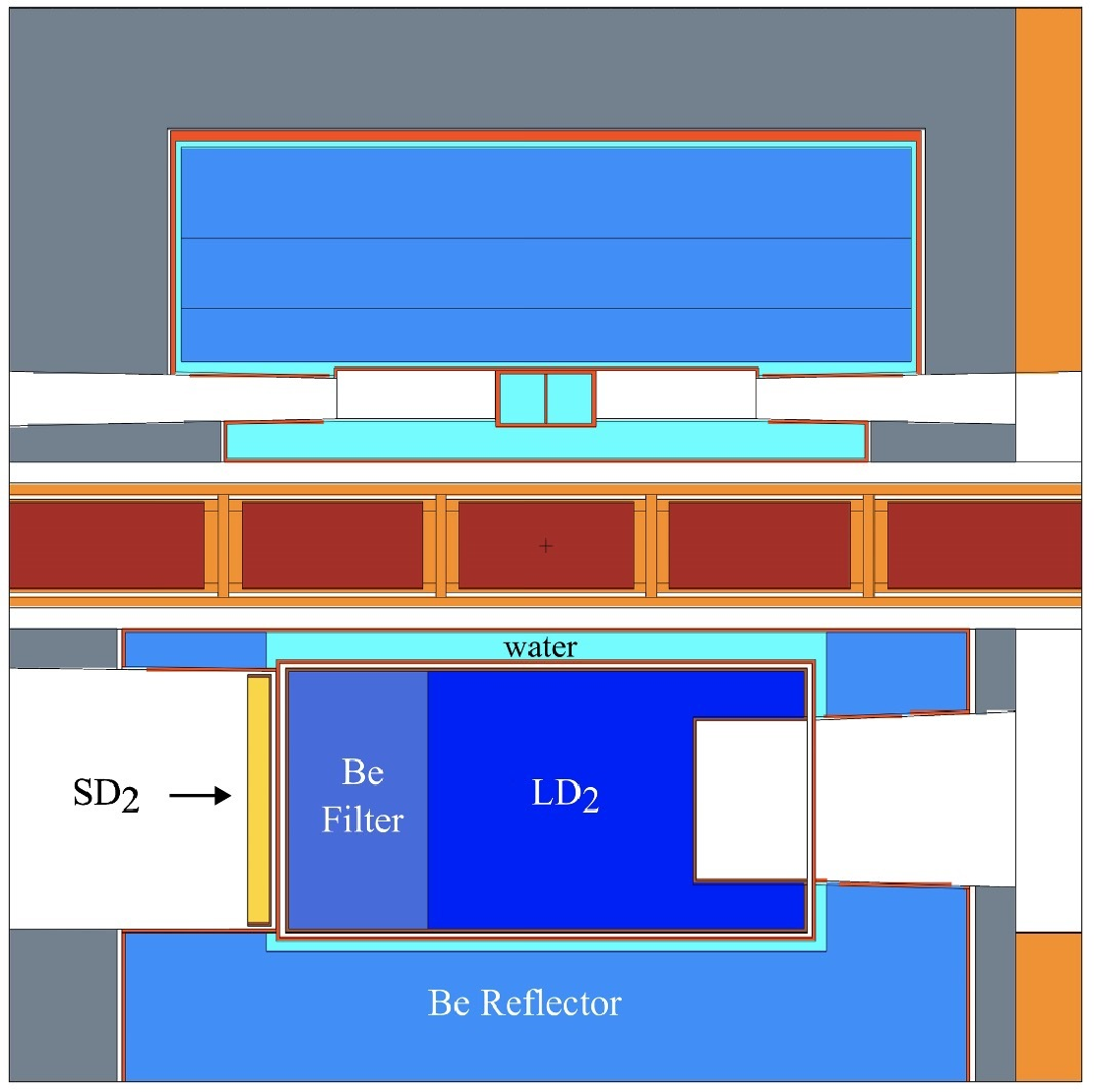}
        \subcaption{}
        \label{fig:baseline_XY}
    \end{subfigure}
    \hfill
    \begin{subfigure}[b]{0.48\textwidth}
        \centering
        \includegraphics[width=\textwidth]{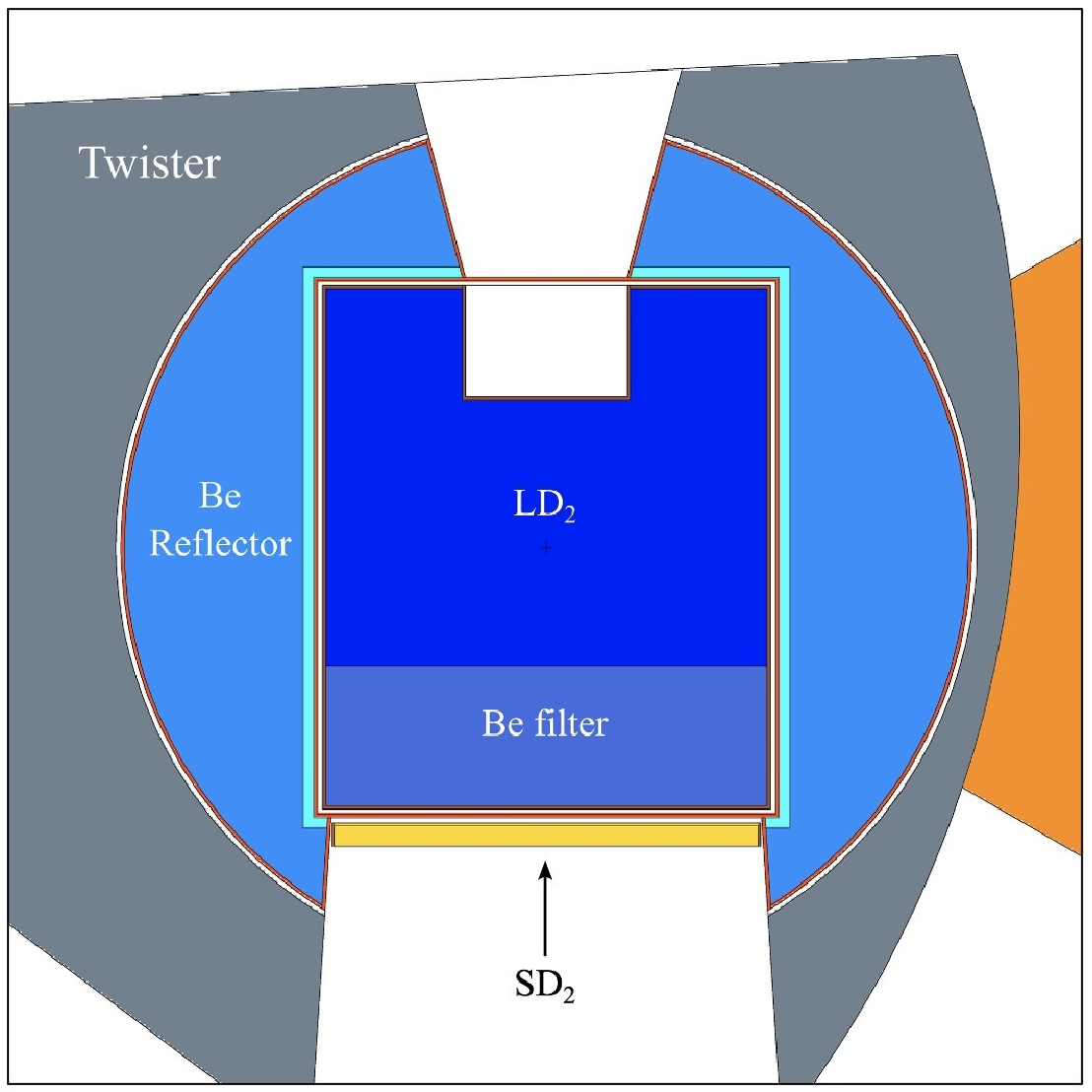}
        \subcaption{}
        \label{fig:baseline_XZ}
    \end{subfigure}
\caption{MCNP model of a 2-cm-thick SD$_\text{2}$ UCN source complementing the LD$_\text{2}$ baseline for UCN production. (a) vertical cut, perpendicular to the proton beam direction. (b) cut parallel to the target plane with the proton beam impinging from the left. }
\label{fig:baseline_geom}
\end{figure}

The most straightforward implementation of a UCN source in the twister consists of adding a UCN converter to the current optimized design of the cold source.  We added a 2-cm-thick slab of SD$_\text{2}$ at \SI{5}{K} covering the NNBAR opening completely. The SD$_\text{2}$ volume is embedded in an \ce{Al} case \SI{2}{mm} thick, but there is no \ce{Al} on the extraction side. In a real case, there would be a very thin window.
A view of the MCNP model is shown in \cref{fig:baseline_geom}.

\begin{figure}[tb!]      
    \begin{subfigure}[b]{0.48\textwidth}
        \centering
        \includegraphics[width=\textwidth]{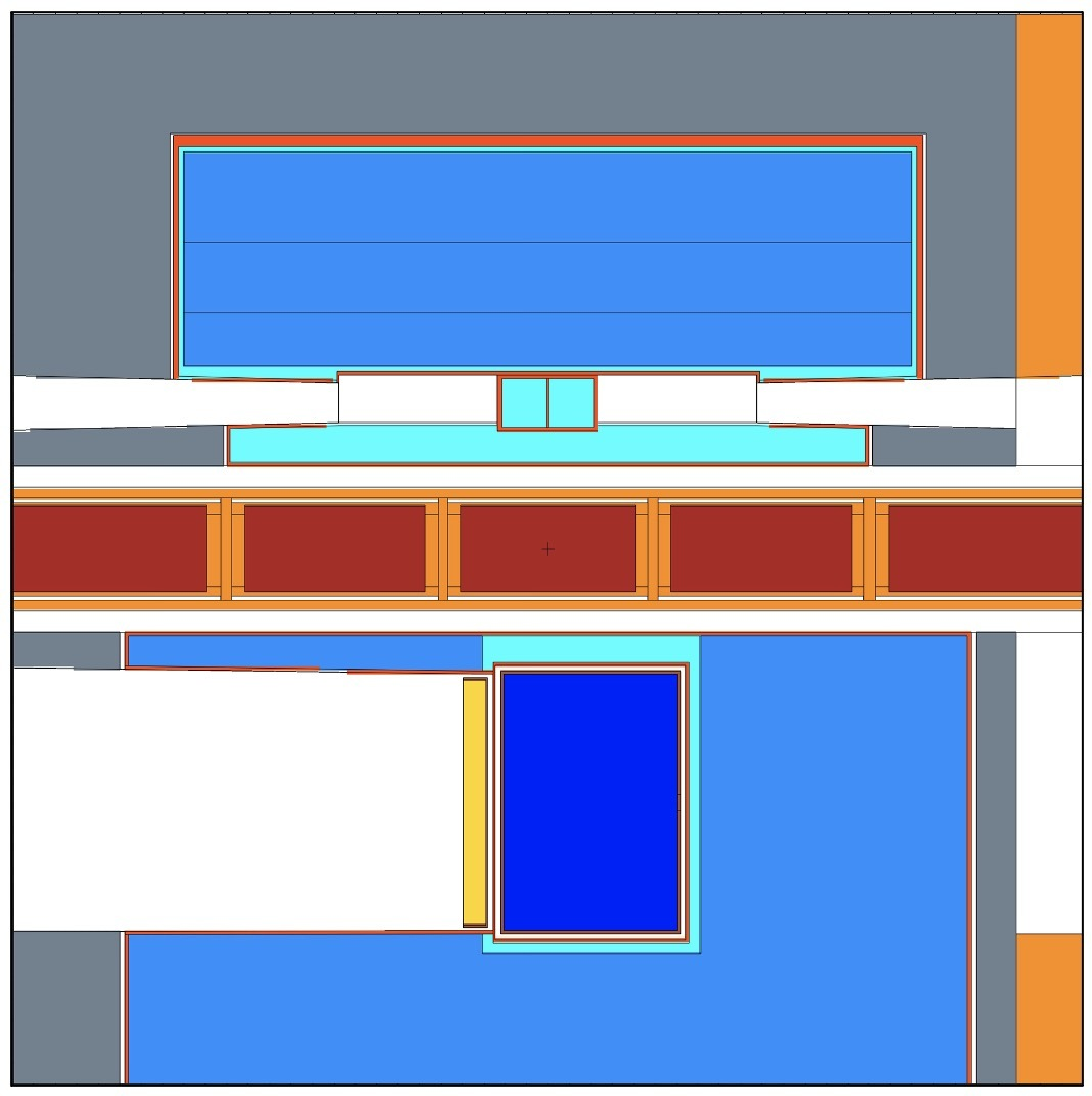}
        \subcaption{}
        \label{fig:baseline_opt_XY1}
    \end{subfigure}
    \hfill
    \begin{subfigure}[b]{0.48\textwidth}
        \centering        
        \includegraphics[width=\textwidth]{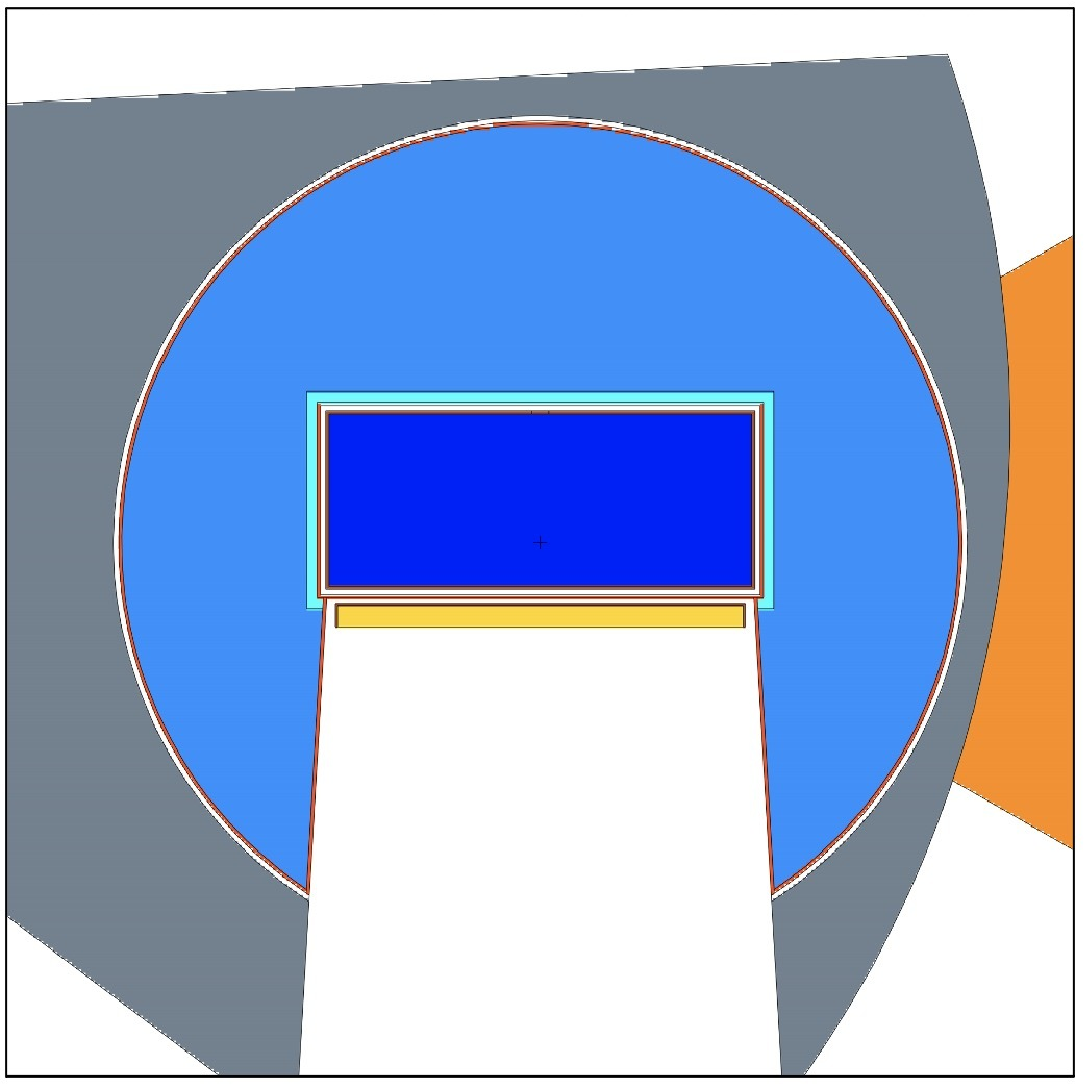}
        \subcaption{}
        \label{fig:baseline_opt_XZ1}
    \end{subfigure}
\caption{Fixed 2-cm SD$_\text{2}$ UCN source with optimized cold LD$_\text{2}$ moderator. (a) vertical cut, perpendicular to the proton beam direction. (b) cut parallel to the target plane with the proton beam impinging from the left. The figure of merit for the optimization was the mean UCN production rate density inside the SD$_\text{2}$ converter.}
\label{fig:baseline_opt_geom}
\end{figure}

\indent The choice of the SD$_\text{2}$ thickness comes from  evidence reported in literature \cite{brys2007extraction,doegethesis} on the effective production depth of extracted UCN. In particular, it was estimated that roughly 50\% of the UCN produced in the last \SI{2}{cm} of a 16-cm thick converter are extracted. This percentage drops to 30\% between 2 and \SI{4}{cm} in depth, while making it increasingly more inefficient to keep the \ce{SD_2} at \SI{5}{K}.
In our concept, the volume of the converter is 1.8 liters. The UCN production rate density from the thin slab, calculated with the method highlighted in \cref{sec:prod_rate_density}, was estimated to be \SI{3.07e5}{n/s/cm^3}. The corresponding $\dot{N}_\text{UCN}$ is then \SI{5.56e8}{n/s}. The estimated prompt heat-load on the converter (both SD$_\text{2}$ and \ce{Al} vessel) is \SI{760}{W}. The $\beta$ and $\gamma$ contribution from \ce{^{28}Al} and the surrounded activated material has  been treated separately in \cref{sec:activation}. \\
\indent A first improvement to the design may come from the removal of the cold \ce{Be} filter. As a matter of fact, the gains for $\lambda > \SI{4}{\angstrom}$ produced by the filter come at the expense of neutrons in the range \SIrange{2}{4}{\angstrom}, i.e., between 5 meV and 20 meV. 
%, which are likely to be reflected back in the LD$_\text{2}$ volume. 
While the increase at longer wavelengthts is advantageous for NNBAR, whose FOM is proportional to $\lambda^\text{2}$, it is detrimental for a UCN converter, since neutrons in the range from 
 \SI{2}{\angstrom} to \SI{4}{\angstrom}  contribute significantly to the UCN production in SD$_\text{2}$ (cfr. \cref{fig:UCNSD2CrossSection_TUMPSIComparison}).
 Removing the filter to add more moderating LD$_\text{2}$ has the effect of increasing the UCN production rate density up to \SI{4.70e5}{n/s/cm^3} (a 53\% gain) and the prompt heat-load to \SI{1}{kW}, while decreasing the neutron yield on the neutron scattering instruments side by 5\%.
Adopting this solution at ESS would entail achieving a high-intensity UCN source by  complementing the pre-existing design without having to design a new moderator. Such a solution could be implemented after the NNBAR experiment. The drawbacks are a higher heat-load on the converter, a possible lack of space for cooling infrastructure and lower performance on the dedicated WP7 opening in case of removing the Be filter.  \\
\indent It is also of interest to study the performance of the thin-slab converter in the ideal case where the lower moderator would be 
optimized for UCN production, as opposed to the previous case, where a thin SD$_\text{2}$ layer was added to an existing design. Such a source would still emit CNs and VCNs, but would be optimized for UCNs.
This  case represents the \textit{non plus ultra} for an in-pile UCN source based on SD$_\text{2}$ (at least when a thin-slab is used), and therefore is a term of comparison for the other cases.
%which acts both as an upper limit and as a yardstick for the other cases. 
We assume the cold moderator to be completely filled with ortho-LD$_\text{2}$ at \SI{22}{K} (with 2.5 wt\% Al) and with only one fixed-dimensions opening \qtyproduct{24x40}{cm}. The vertical dimension was also kept fixed to \SI{24}{cm}, while the other two (parallel and transverse to the proton beam axis) were varied. The center of the moderator-converter system along the transverse-to-proton-beam axis was also a variable in the Dakota optimization (more information on Dakota and the optimization algorithm in \cite{Dakota_6.18}). The optimization algorithm found the best $P_\text{UCN}$ for a \qtyproduct{40x16x24}{cm} cold moderator with the center shifted \SI{4}{cm} away from the opening, placing the converter closer to the region of highest cold intensity. The MCNP model of the optimum is shown in \cref{fig:baseline_opt_geom}. The estimated $P_\text{UCN}$ and $\dot{N}_\text{UCN}$ are \SI{7.72e5}{n/s/cm^3} and \SI{1.35e9}{n/s}, respectively, for a total prompt heat-load of \SI{2.91}{kW}. It should be noted that since the converter's volume does not vary much (it shrinks slightly to fit the divergent opening as it goes deeper), we can consider this case to be also optimal for the production rate $\dot{N}_\text{UCN}$ since it differs from $P_\text{UCN}$ only by a (constant) volume.
After the optimization results we can  conclude that an ideal thin-slab SD$_\text{2}$-based UCN source placed inside the twister does not need large volumes of cold moderator material as long as it is kept close to the hot spot of cold neutron production. Clearly, this higher production rate comes at the expense of a higher heat-load to remove from the converter.
\subsubsection{Delayed heat deposition from activation}
\label{sec:activation}
A correct estimation of the heat deposition on the UCN converter is essential for the \textit{in-pile} solution, where the high radiation flux could make it challenging to keep the converter at \SI{5}{K}. Additional heat-load, which is not taken into account by the F6 tally in MCNP (which only calculated prompt energy deposition), comes from the decay of activation products.
Considering the  use of Al in the moderator vessels, we expect the most important contributor to be \ce{^{28}Al}:
\begin{equation}
 \ce{^{27}Al + n -> ^{28}Al -> ^{28}Si^{*} + {\beta} +\overline{\nu} -> ^{28}Si + {\gamma} }
 \end{equation}
The half-life of \ce{^{28}Al} is \SI{2.245}{min}, it decays to a metastable state of \ce{^{28}Si} with the emission of an electron (average energy \SI{1.24}{MeV}), which in turn decays to stable \ce{^{28}Si} emitting a photon (one line at \SI{1.79}{MeV}). In MCNP it is possible to estimate the saturation activity of \ce{^{28}Al} from the incident neutron flux using a tally multiplier (FM card). The FM card is used to apply a multiplicative response function from the MCNP6 cross-section libraries. In this case, we used the radiative capture library \ce{(n,\gamma)}, identified with the ENDF/B (MT) reaction number 102. One tally with the respective FM card was defined for each cell containing aluminum. The volume of the cell is normalized to 1 so that the value obtained is in \si{Bq}, instead of the default \si{Bq/cm^3}. The next step is to calculate the energy deposition from the decay electron and $\gamma$. The particles are generated uniformly in the cell with the intensity equal to the saturation activity. This step is repeated for each particle and for each aluminum cell. The electrons are generated with a continuous $\beta$-decay spectrum taken from \cite{eckerman_availability_1994}. The results for the model in \cref{fig:baseline_geom} obtained with this method are reported in \cref{tab:delayed_heatload}.
\begin{table}[tbp]
\centering
\def\arraystretch{1.5}
    \setlength\tabcolsep{0pt}
    \caption{Delayed heat deposition from \ce{^{28}Al} decay. The columns correspond to the source cells. For each cell, the scored heat-load in both the SD$_\text{2}$ and the Al vessel is reported. Two different tables are presented for electrons and photons.}
    \begin{tabular*}{0.95\textwidth}{@{\extracolsep{\fill}} r c c c c c c }
    \toprule
     & \multicolumn{6}{c}{Delayed electron heat-load [W]} \\
    \cmidrule{2-7}
    To\textbackslash From & Flow channels & LD$_\text{2}$ vessel & REH vessel & Vacuum jacket & SD$_\text{2}$ vessel & Total \\
    \midrule
    SD$_\text{2}$ & \num{6.51E-04} & \num{1.95E-02} & 1.63  & 1.14  & 10.9  & 12.1 \\
    SD$_\text{2}$ vessel & \num{2.93E-03} & \num{2.24E-01} & 7.33  & 22.7  & 47.6  & 70.5 \\
    \midrule
    Total  & \num{3.59E-03} & \num{2.44E-01} & 8.96  & 23.8  & 58.5  & 82.6 \\
    \midrule
    & \multicolumn{6}{c}{Delayed $\gamma$ heat-load [W]} \\
    \cmidrule{2-7}
    To\textbackslash From & Flow channels & LD$_\text{2}$ vessel & REH vessel & Vacuum jacket & SD$_\text{2}$ vessel & Total \\
    \midrule
    SD$_\text{2}$  & \num{5.38e-1} & 3.38 & \num{6.12e-02}  & 3.89  & 1.14  & 9.01 \\
    SD$_\text{2}$ vessel & \num{1.00} & \num{6.62} & \num{1.17e-01}  & 8.17  & 3.36  & 19.3 \\
    \midrule
    Total  & \num{1.54} & \num{9.99} & \num{1.78e-1}  & 12.0  & 4.50  &  28.3\\
    \bottomrule
    \end{tabular*}%
  \label{tab:delayed_heatload}%
\end{table}%
The short range of electrons in aluminum, compared to the thickness of the containers, means that the largest contribution to the decay heat-load in the SD$_\text{2}$ Al vessel comes from the $\beta$-decay in the vessel itself, while the second largest contribution comes from the $\beta$-decay in the Al vacuum jacket few millimeters away. In the SD$_\text{2}$ cell, the single most important source of decay heat is the electron from the Al vessel, while, in total, photons and electrons have a similar impact on the heat-load. 
The heat deposition from the decay products can be compared with the deposition from the prompt radiation. The prompt heat-load estimated with the F6 tally is \SI{483}{W} for the SD$_\text{2}$ cell and \SI{275}{W} for the Al vessel. These estimations point out that the heat deposition from delayed decay products accounts for almost 30\% of the heat in the Al vessel, hence it should not be overlooked. For SD$_2$, this contribution, while not negligible, is less significant (about 4\% increase).

\subsubsection{Reentrant hole design}
\label{sec:hole}
\begin{figure}[tb!]      
    \begin{subfigure}[b]{0.48\textwidth}
        \centering
        \includegraphics[width=\textwidth]{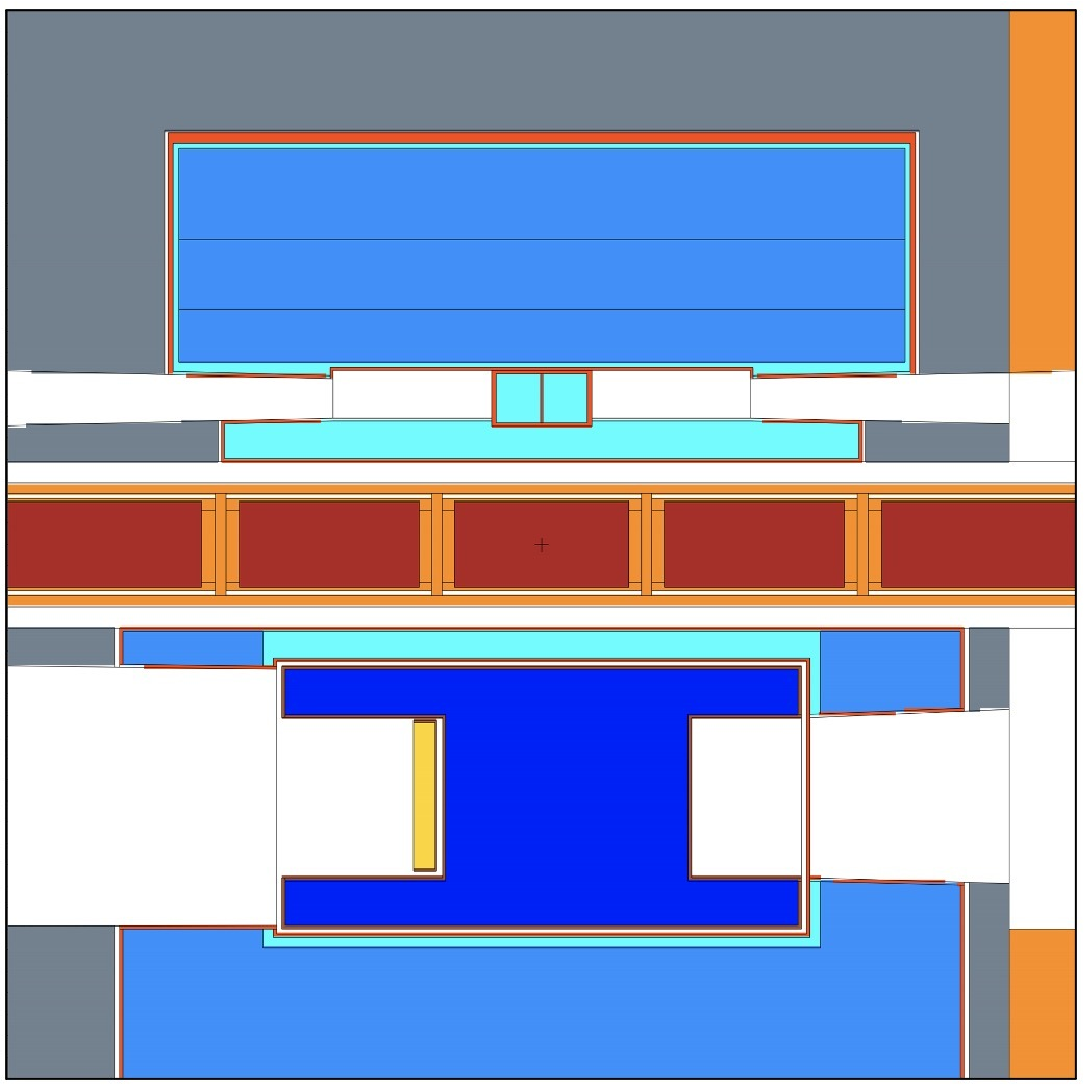}
        \subcaption{}
        \label{fig:REH_XZ}
    \end{subfigure}
    \hfill
    \begin{subfigure}[b]{0.48\textwidth}
        \centering        
        \includegraphics[width=\textwidth]{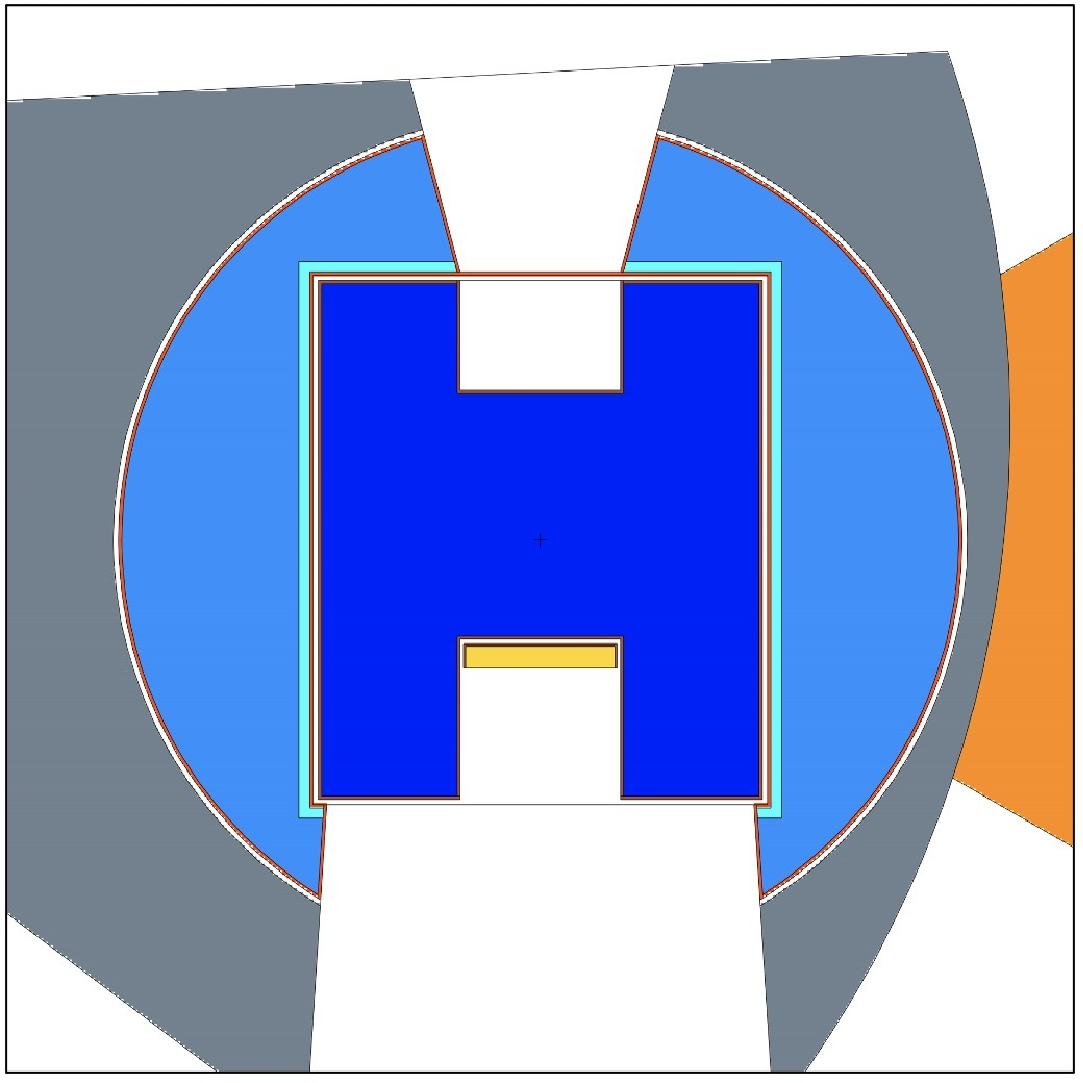}
        \subcaption{}
        \label{fig:REH_XY}
    \end{subfigure}
\caption{Fixed 2-cm SD$_\text{2}$ UCN source in a reentrant hole inside the LD$_\text{2}$ moderator, hence closer to the hot-spot of cold neutrons production. (a) vertical cut, perpendicular to the proton beam direction. (b) cut parallel to the target plane with the proton beam impinging from the left. The figure of merit for the optimization was the UCN production rate density inside the SD$_\text{2}$ converter.}
\label{fig:REH_geom}
\end{figure}
In light of the results of the ideal case, we explored also a design that would possibly reconcile the needs of both a neutron scattering instruments opening delivering CN on one side and a UCN source on the other; such design could be adopted  after the NNBAR experiment is completed. The two openings compete for the high CN intensity produced at the center of the large LD$_\text{2}$ moderator. For the first tentative design, we added a 2-cm-thick SD$_\text{2}$ converter at the bottom of a \qtyproduct{15x15x15}{cm} reentrant hole on the NNBAR side. Hence, this converter has a smaller volume of \SI{384}{cm^3}. It is important for the UCN extraction that no \ce{Al} layer is interposed between the converter and the channel. Hence the \ce{Al} window (\SI{3}{mm}) separating in the previous models the vacuum jacket and extraction channel (also vacuum) is removed. The geometry is shown in \cref{fig:REH_geom}.  
The estimated $P_\text{UCN}$ thus obtained is \SI{1.31e6}{n/s/cm^3} with a total prompt heat-load of \SI{0.56}{kW}. The smaller dimension of this converter is compensated by the high production rate density achieved thanks to the proximity to the LD$_\text{2}$ core. However, this corresponds to  $\dot{N}_\text{UCN}=\SI{5.03e8}{n/s}$ which is lower than simply putting the converter outside the LD$_\text{2}$ vessel. 

\subsubsection{Cylindrical 3-opening design}
\label{sec:cyl}
\begin{figure}[tb!]      
        \centering
        \includegraphics[width=0.5\textwidth]{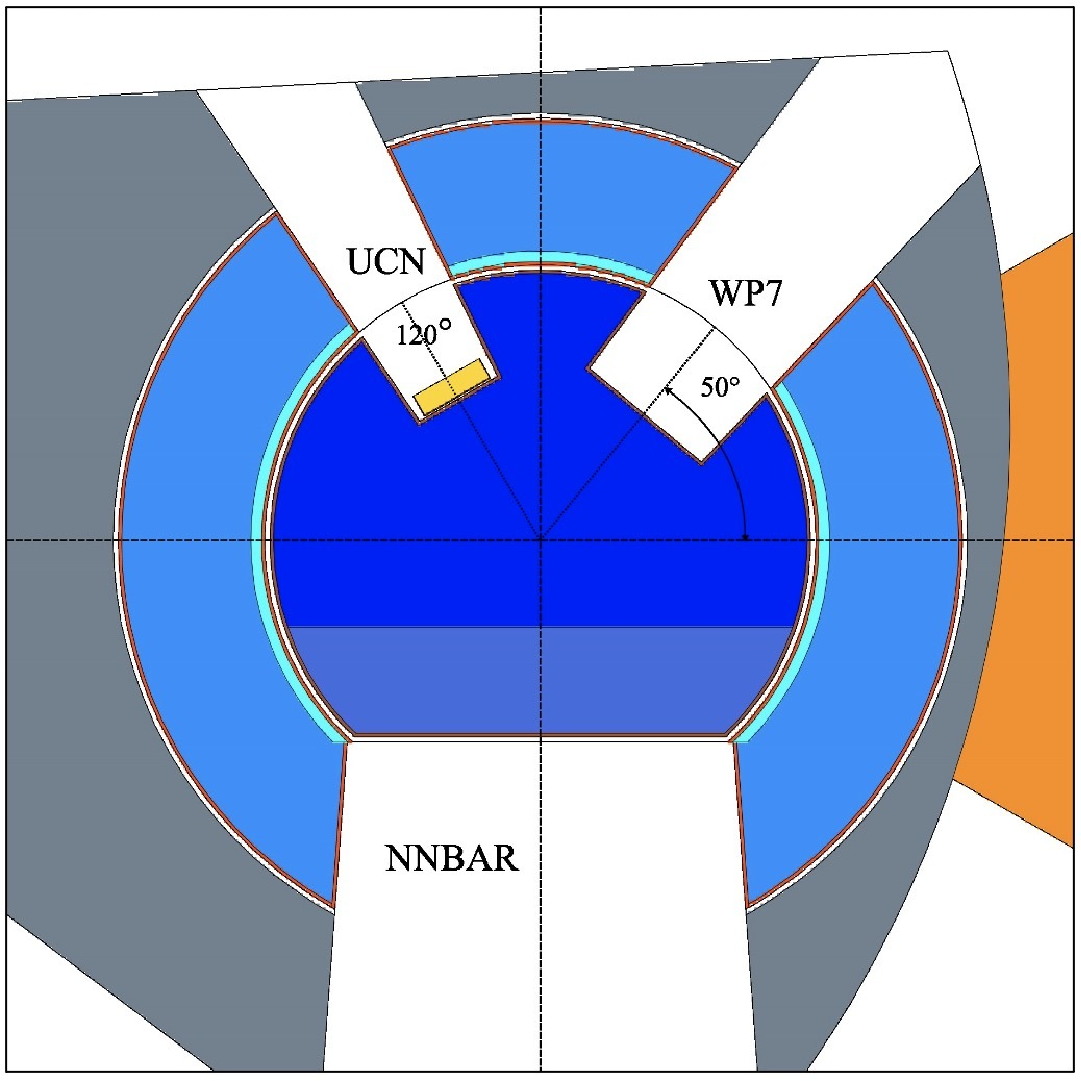}
\caption{Cylindrical cold moderator (\SI{45}{cm} diameter) with three openings for NNBAR, UCN and neutron scattering experiments. The SD$_\text{2}$ converter has a fixed thickness of \SI{2}{cm}. The cut is parallel to the target plane with the proton beam impinging from the left.}
\label{fig:3O_BigCyl_sub}
\end{figure}

We studied the possibility to have a UCN in-pile source together with an opening for NNBAR and one for standard scattering experiment. We deemed not possible to accommodate a third opening in the box geometry presented so far, so we switched the shape of the cold moderator to a cylinder with a \SI{45}{cm} diameter and \SI{24}{cm} height. The model is shown in \cref{fig:3O_BigCyl_sub}. 
The NNBAR opening is of the same size and in the same position as before, since it is constrained by the position of the large beamport at \SI{2}{m}. As in the developing of the lower moderator it was found that the cylindrical emission surface in the NNBAR opening was detrimental to the performance, the cylinder was cut to obtain a flat rectangular emission surface. A \SI{10}{cm} thick cold Be filter was also included. The WP7 opening is a \qtyproduct{15x15}{cm} square surface \SI{10}{cm} deep in the moderator, placed at \SI{50}{\degree} with respect to the proton beam direction (as a comparison, the NNBAR opening is placed at \SI{270}{\degree} in this coordinate system). Finally, the UCN converter is placed in a \qtyproduct{10x10}{cm} opening at \SI{120}{\degree}, also \SI{10}{cm} deep in the moderator. It has again a fixed thickness of \SI{2}{cm} for a total SD$_\text{2}$ volume of \SI{128}{cm^3}. At this exploratory stage, these dimensions are tentative, but it should be noted that we can apply the same principle of competition for the hot-spot described for the opposing openings. Hence, we opted for a symmetric configuration. The estimated $P_\text{UCN}$  is \SI{1.74e6}{n/s/cm^3} while the production rate is \SI{2.22e8}{n/s}. The prompt heat-load on this converter is \SI{520}{kW}. The estimated FOMs for the NNBAR and WP7 openings are \SI{2.33e17}{n\cdot\angstrom^2/s/sr} and \SI{2.84e15}{n/s/sr}. These values should be compared with the ones reported in \cref{tab:performance_table_iteration1} for the original LD$_\text{2}$ baseline. We can see that the two original openings would lose 7\% and 12\%, respectively, in this configuration.  For a more fair comparison, one should consider the baseline model without the \SI{3}{mm} Al window at the interface between the vacuum jacket and the opening, since this is not considered in this model. However, regardless of the absolute value of the loss, at this stage we are more interested in showing the concept and its feasibility.

It is clear that this is a limited study and a wider optimization effort, which takes into account all the parameters including the NNBAR and WP7 FOMs, could find one or more better compromises to accommodate the three applications.

\subsubsection{SD$_\text{2}$ as primary moderator}
\label{sec:prim}
\begin{figure}[tb!]      
    \begin{subfigure}[b]{0.555\textwidth}
        \centering
        \includegraphics[width=\textwidth]{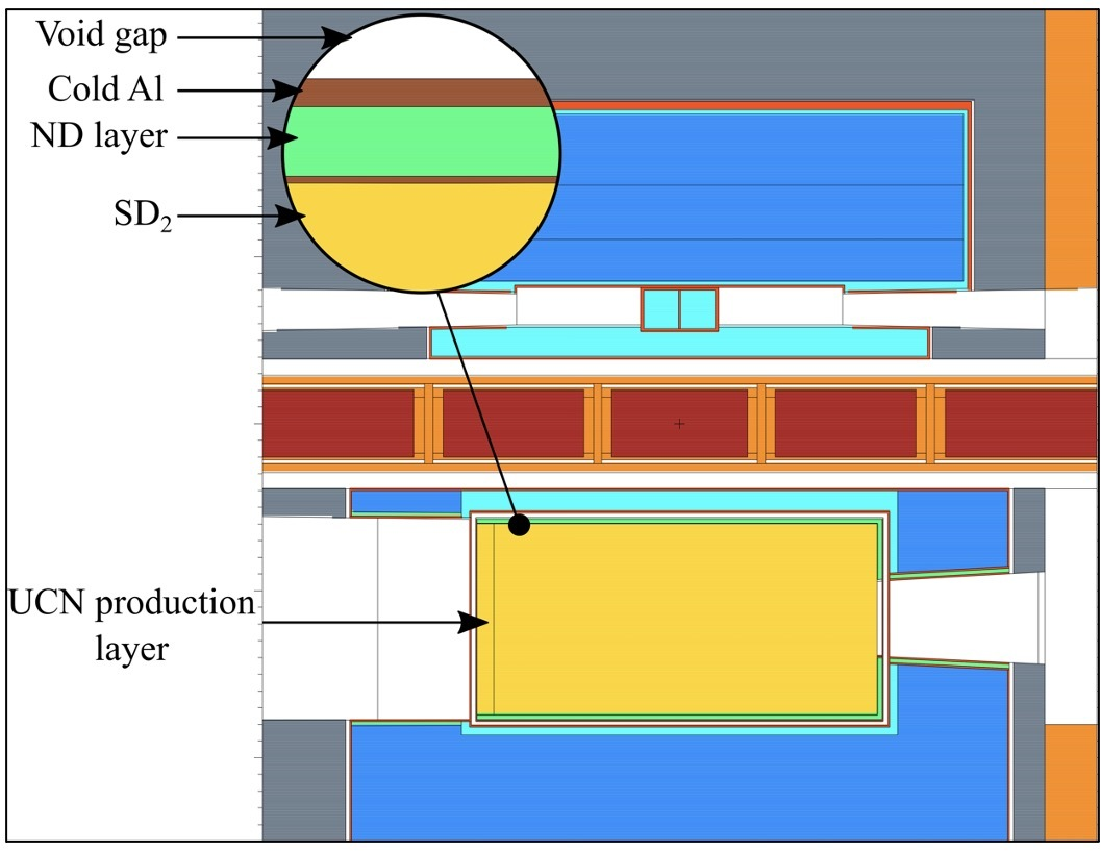}
    \subcaption{}
    \label{fig:fullSD2a}
    \end{subfigure}
    \hfill
    \begin{subfigure}[b]{0.435\textwidth}
        \centering        
        \includegraphics[width=\textwidth]{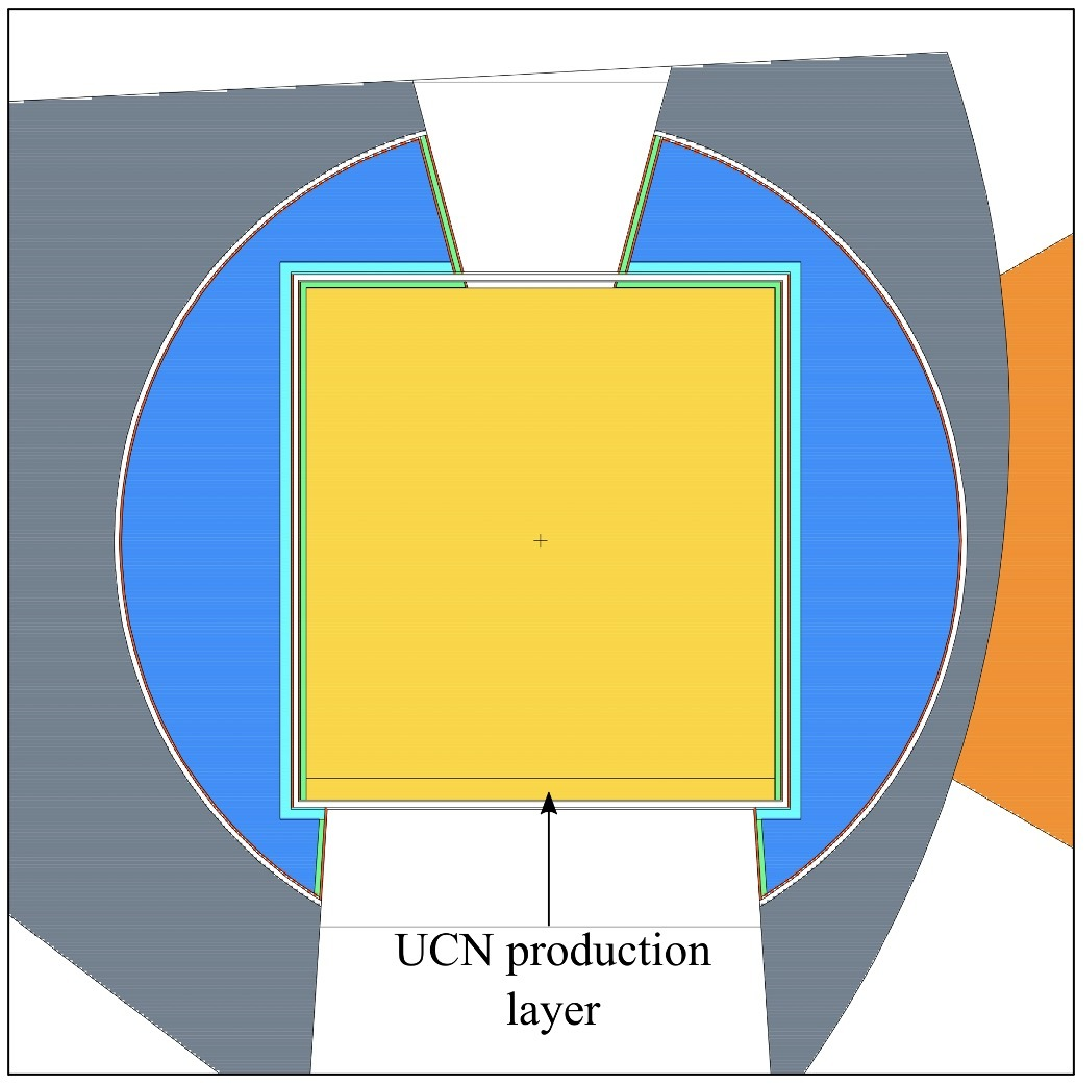}
        \subcaption{}
        \label{fig:fullSD2b}
    \end{subfigure}
\caption{MCNP model of a \qtyproduct{41 x 48 x 24}{\centi\metre} SD$_\text{2}$ moderator for VCN and UCN production. (a) vertical cut, perpendicular to the proton beam direction. The inset zooms on the 5-mm ND reflector layer and 
 its aluminum case (b) cut parallel to the target plane with the proton beam impinging from the left.}
\label{fig:fullSD2}
\end{figure}

The last in-twister design is based on the observation that the high-intensity \ce{SD_2}-based VCN source not only gave the highest VCN yield, but also produced UCN, even though only the ones produced in the last few \si{cm} of the crystal can be extracted.
%\textcolor{red}{CHECK NEXT SENTENCE ALREADY MENTIONED} These calculations could be carried out for the first time thanks to the recent development of thermal scattering libraries for both solid deuterium (\textcolor{red}{REF in CDR}) and nanodiamonds in \textsc{MCNP\,6.2}.
The geometry is shown in \cref{fig:fullSD2}.

An extensive description of the model, the method and the results of the VCN performance of this multi-purpose source, which would be the first of its kind, is provided in \cref{sec:full_SD2}. Here, we only report the relevant values concerning the UCN production. The total volume of SD$_\text{2}$ is \SI{48.2}{L}, but only the last \SI{2}{L} (corresponding to the last \SI{2}{cm} on the NNBAR side) are considered for the UCN production. The $P_\text{UCN}$ in  this smaller volume is \SI{6.56e5}{n/s/cm^3}, which means a UCN production rate of \SI{1.32e9}{n/s}. Concerning the heatload, the main challenge of this design is to keep a large volume of ortho-D$_\text{2}$ in the solid state at \SI{5}{K}. We estimated that the heat-load on the moderator would be approximately \SI{40}{kW} at 5 MW operation, but this number is expected to increase once the decay heat-load from the surrounding aluminum is taken into account. Preliminary simulations of embedded foam-like aluminum and aluminum-beryllium alloy cooling structures have shown promising results in improving the moderator thermal conductivity. In conjunction with a standard liquid-helium cooling pipe, these foams should allow for a rate of heat extraction capable of keeping the solid deuterium well below its melting point, at least for the \SI{2}{MW} scenario at ESS. In particular, the use of beryllium alloys or graphite in such cooling structures appears to be more favorable in both terms of cooling (due to reduced self-heating) and neutronics. Despite the obvious challenges, this idea opens the door to a completely new future for ESS, where VCN and UCN are produced with an intensity never seen before, allowing scientists to explore new instrument concepts and, ultimately, generating new science. For further discussion on this design see \cref{sec:full_SD2}. % this reference is in the VCN section

Estimated values for each design in this section are listed in \cref{tab:comparison_twister} for comparison.

\begin{table}[tbp!]
\centering
\def\arraystretch{1.5}
    \setlength\tabcolsep{0pt}
\caption{Summary table of the calculations for the in-twister options. The optimization was driven by maximising $P_\text{UCN}$, but since the volume is approximately constant, the same result would be obtained by choosing $\dot{N}_\text{UCN}$ instead. The reported heat-load is only the prompt radiation heat-load.} 
\begin{tabular*}{0.95\textwidth}{@{\extracolsep{\fill}} r c c c c c c }
\toprule
 & \makecell{SD$_\text{2}$ Volume \\ $[\si{L}]$}  & \makecell{$P_\text{UCN}$ \\
 $\left[\si{n/s/cm^3}\right]$}& \makecell{$\dot{N}_\text{UCN}$ \\
 $\left[\si{n/s}\right]$} & \makecell{Heat-load\\ $[\si{W}]$}& \makecell{ WP7 FOM \\ $[\si{n/s/sr}]$} & \makecell{ NNBAR FOM \\ $[\si{nÅ^2/s/sr}]$}\\
\midrule
\textbf{Baseline + UCN} &      \num{1.81} &\num{3.07e5} & \num{5.56e8} & 760 & \num{3.23e15}& - \\
\textbf{No Be filter + UCN} &  \num{1.81} &\num{4.70e5} & \num{8.51e8} & 1000 & \num{3.06e15} & -\\
\makecell[rr]{\textbf{Optimized}\\\textbf{UCN-only}} &\num{1.75}  &\num{7.72e5} &  \num{1.35e9} & 2910 & - & -\\
\midrule
\textbf{Reentrant Hole}  & \num{0.38} & \num{1.31e6} &  \num{5.03e8} &560 & \num{2.81e15} & -\\
\midrule
\textbf{3-openings cylinder}  & \num{0.13} & \num{1.74e6} &  \num{2.22e8} & 520 & \num{2.84e+15} & \num{2.33e+17} \\
\midrule
\textbf{Full SD$_\text{2}$ moderator}  & \num{48.2} & \num{6.56e+05} &  \num{1.32e9} & 39886 & - & - \\
\bottomrule
\end{tabular*}
\label{tab:comparison_twister}
\end{table}

\subsection{In-Twister He-II UCN Source}
\label{sec:lHe_intwister}

Similar to the models discussed in \cref{sec:sd2_intwister}, placing a UCN source within the twister has been investigated for cases having He-II as the UCN production material. This solution imposes the most difficulty in terms of heating, while also presenting considerable engineering constraints.

\begin{figure}[tbh!]      
    \centering        
    \includegraphics[width=0.65\textwidth]{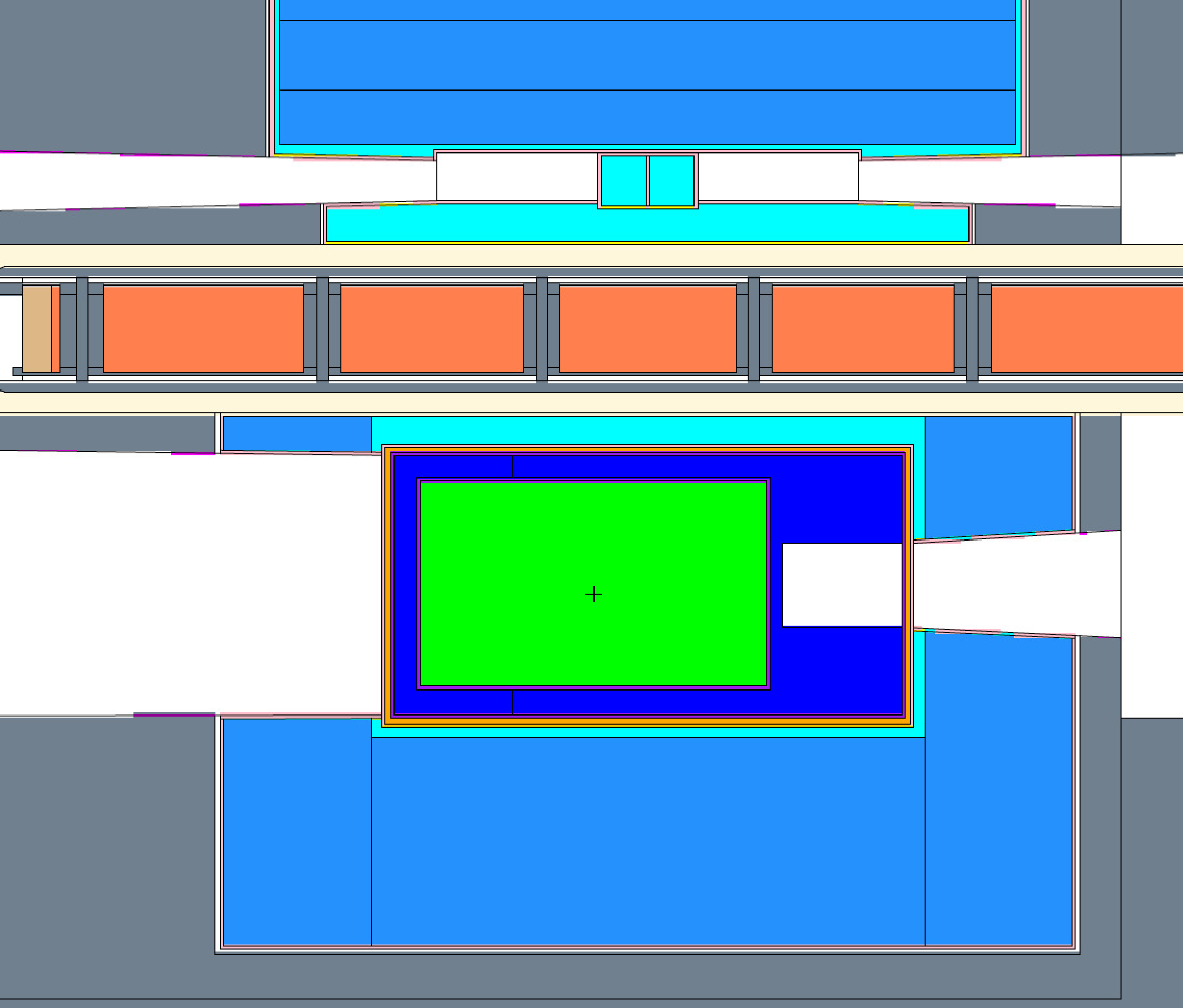}
\caption{In-pile UCN source with He-II (light green) nearly fully diplacing the \ce{LD_2} (dark blue) moderator volume.}
\label{fig:lHe_inpile_full_geom}
\end{figure}

A preliminary model is shown in \cref{fig:lHe_inpile_full_geom}, where the majority of the \ce{LD_2} volume is displaced by He-II at a temperature of \SI{1.2}{K}.  Although engineering such an insert to the baseline \ce{LD_2} model could be relatively straightforward, the heatload reported in this model in \cref{tab:lhe_intwister} is nearly 10\,kW. While a kW-scale heatload may be feasible for \ce{SD_2} when considering its optimal temperature range of roughly 4--\SI{10}{K}, He-II is ideally kept below \SI{1}{K}~\cite{golub1979,golub1983} and heatloads exceeding  \SI{100}{W} are likely unfeasible. Despite this model being impractical, it should be observed that the total neutron production rate exceeds that of $\dot{N}_\text{UCN}$ for the Full \ce{SD_2} model by roughly a factor of four (and with the Full \ce{SD_2} having roughly twice the volume). The two results for this model in \cref{tab:lhe_intwister} are for trials with and without a MgH$_2$ reflective layer enclosing the full moderator volume, with an improvement exceeding a factor of two for the trial with added MgH$_2$.

\begin{figure}[tb!]      
    \begin{subfigure}[b]{0.5\textwidth}
        \centering
        \includegraphics[width=\textwidth]{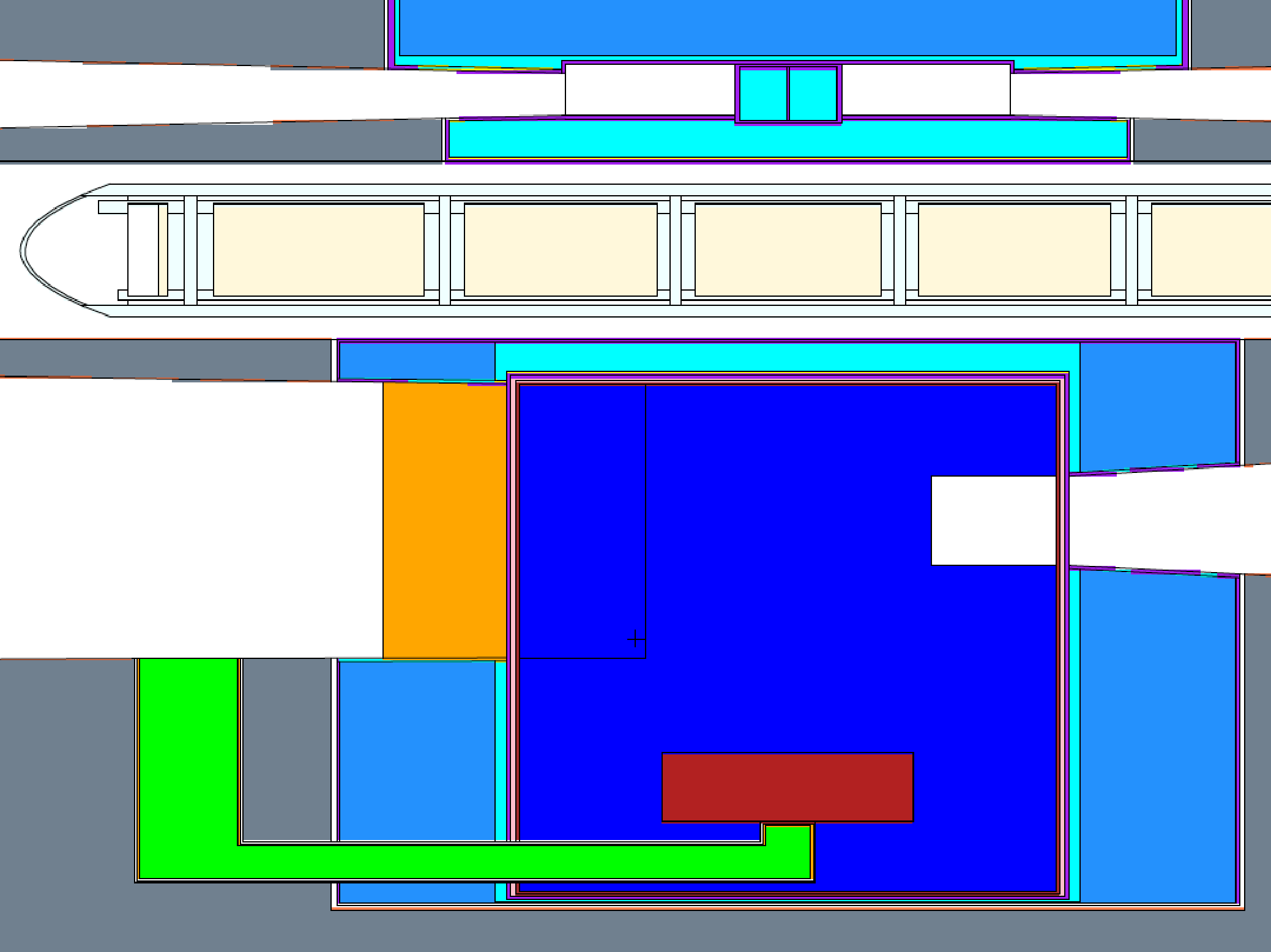}
        \subcaption{}
        \label{fig:lHe_inpile_extract_XY}
    \end{subfigure}\\
    \hfill
    \begin{subfigure}[b]{0.5\textwidth}
        \centering        
        \includegraphics[width=\textwidth]{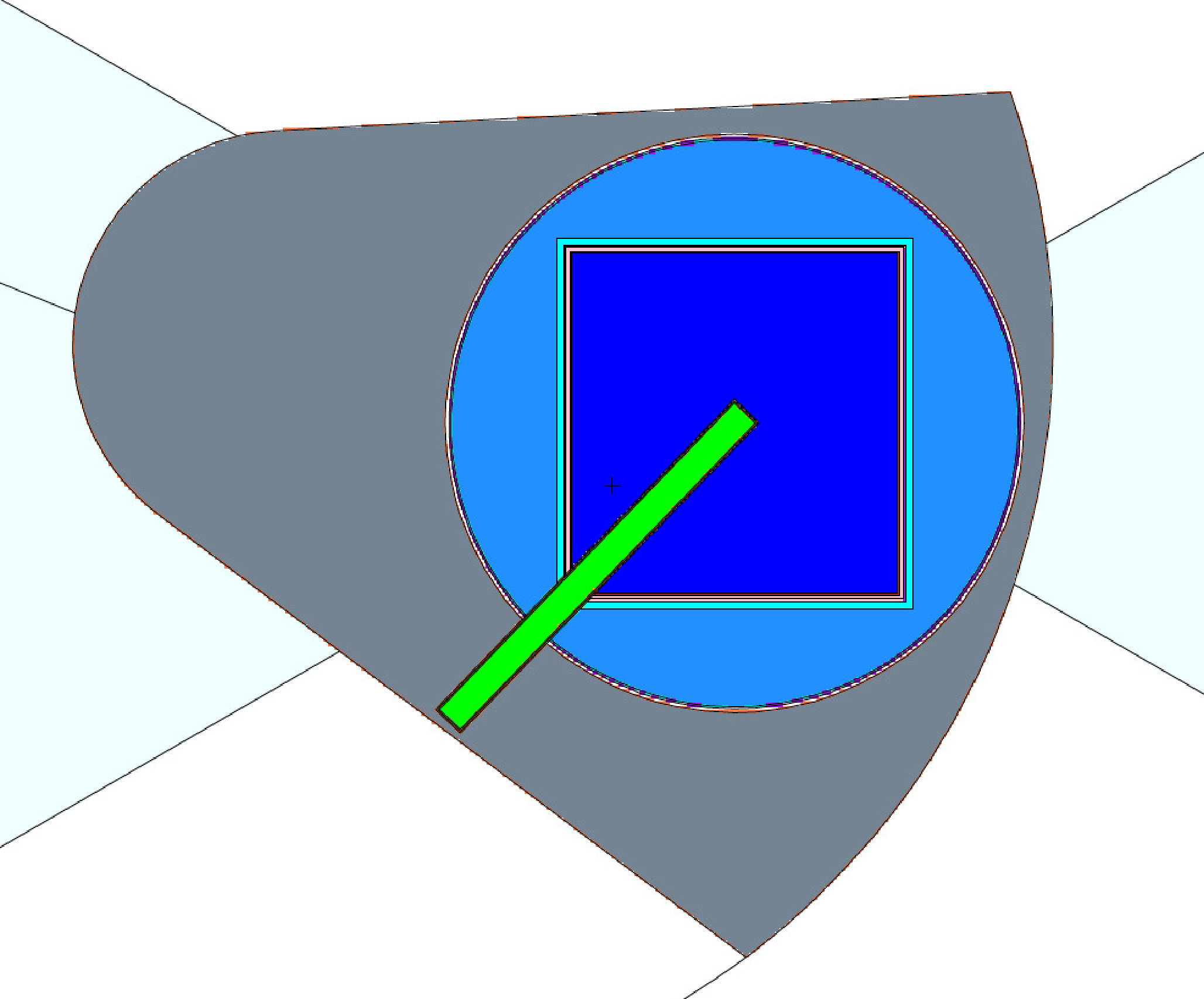}
        \subcaption{}
        \label{fig:lHe_inpile_extract_XZ}
    \end{subfigure}
\caption{In-pile liquid helium UCN source (light green) with extraction channel; vertically extended \ce{LD_2} moderator (dark blue); and single-crystalline bismuth gamma shield (red). (a) Cross-section of the extraction channel. (b) Top-down view of the moderator and extraction channel within the twister.}
\label{fig:lHe_inpile_extract_geom}
\end{figure}

A more realistic model is shown in \cref{fig:lHe_inpile_extract_geom}, where the moderator is extended downward with a small 64\,cm$^3$ primary UCN production volume of He-II and an extraction line extending to a beamport adjacent the NNBAR opening. In this case, the single-crystalline bismuth shielding reduces photon heating while being relatively transparent to neutrons in the critical UCN production range for He-II near 9\,{\AA} (see \cref{fig:s_lambda})~\cite{adib_2003_bi_filter}. Polycrystalline bismuth was also tested and is an effective gamma-shield, but shows a penalty in UCN production of roughly 4\%. The best available thermal scattering kernel for single-crystalline bismuth was at a temperature of 77~K. For further studies a library at a temperature of 22~K would be necessary to match that of the \ce{LD_2} moderator; it is expected that this would increase the He-II UCN production by further reducing the bismuth cross-section near 9{\,\AA}. (See also: \cref{sec:bi_gamma_shielding}.)

From \cref{tab:lhe_intwister}, one can see that the heatload drops by roughly two orders of magnitude, while the UCN production rate per cm$^3$ is only reduced by approximately 75\%. This reduction can be attributed to the increased distance from the He-II volume to the target, with a small penalty due to the bismuth gamma shield.

\begin{figure}[tb!]      
    \begin{subfigure}[b]{0.5\textwidth}
        \centering
        \includegraphics[width=\textwidth]{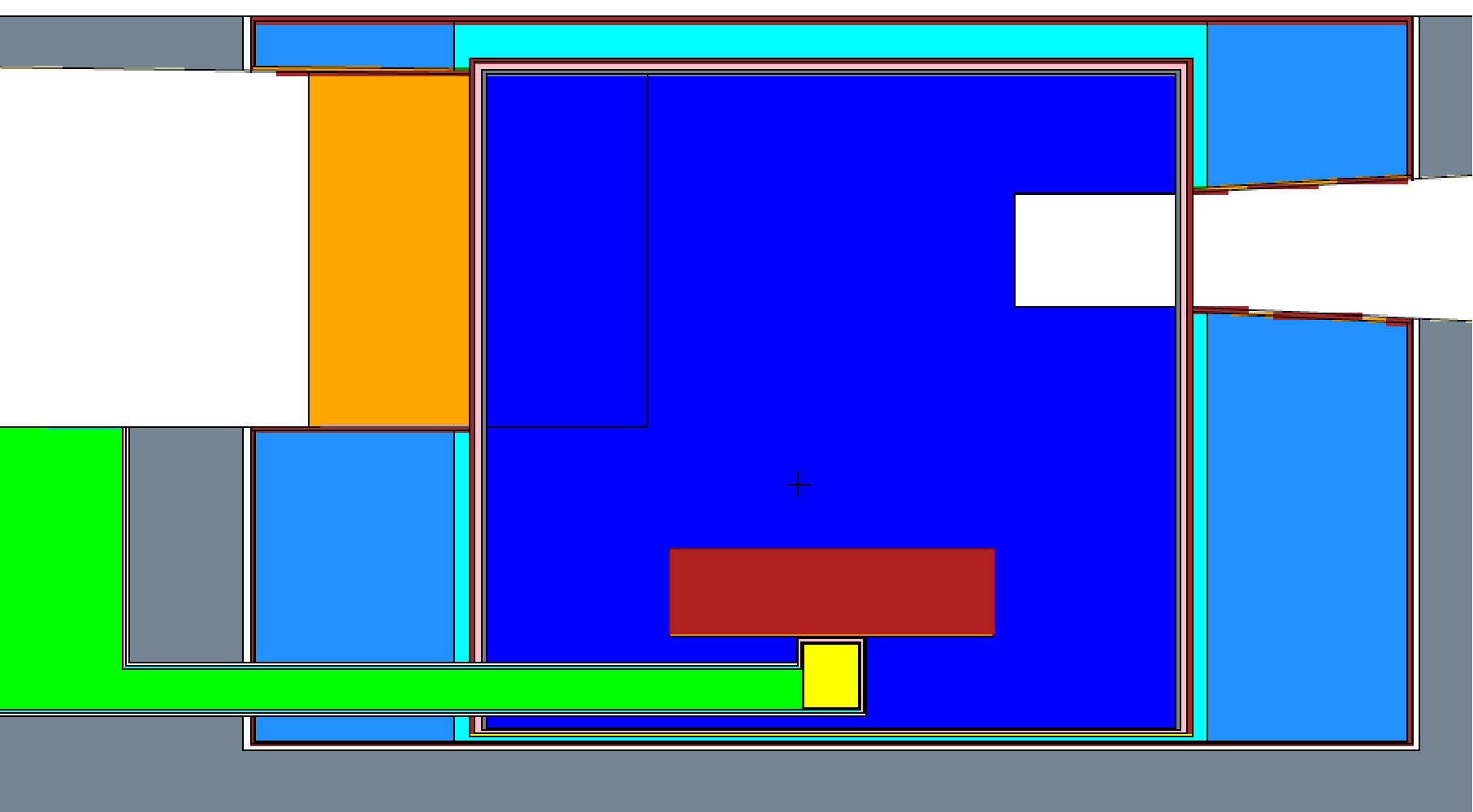}
        \subcaption{}
        \label{fig:lHe_SD2_hyb_inpile_extract_XY}
    \end{subfigure}\\
    \hfill
    \begin{subfigure}[b]{0.5\textwidth}
        \centering        
        \includegraphics[width=\textwidth]{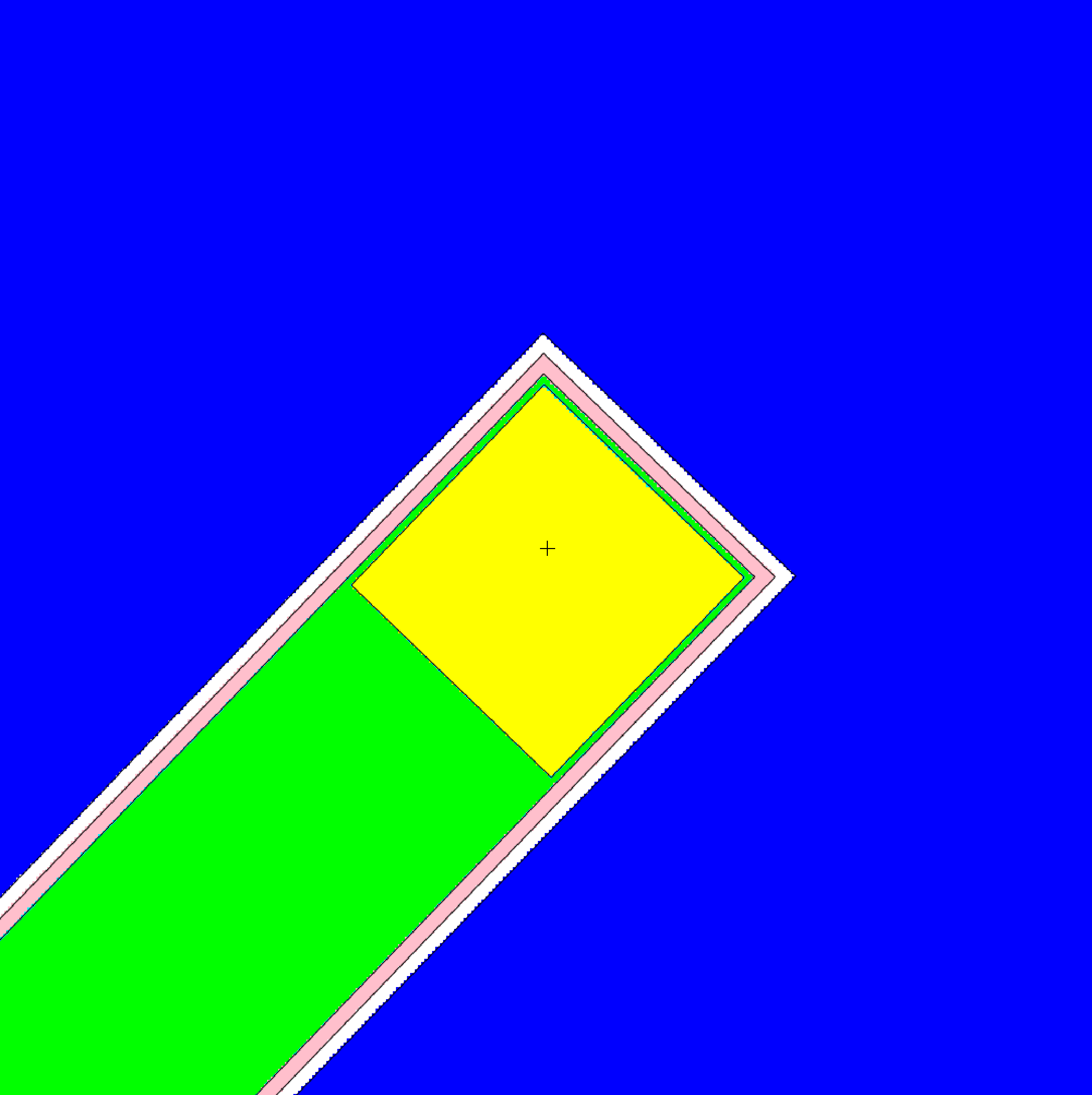}
        \subcaption{}
        \label{fig:lHe_SD2_hyb_inpile_extract_XZ}
    \end{subfigure}
\caption{In-pile liquid helium UCN source and extraction channel (light green) with added thin-slab \ce{SD_2} UCN source (yellow); a vertically extended \ce{LD_2} moderator (dark blue); and single-crystalline bismuth gamma shield (red). (a) Cross-section of the hybrid source and extraction channel. (b) Top-down view of the primary UCN production volume.}
\label{fig:lHe_SD2_hybrid_inpile_extract_geom}
\end{figure}

A hybrid model is then shown in \cref{fig:lHe_SD2_hybrid_inpile_extract_geom}, where a thin slab of \ce{SD_2} is added in the critical UCN production volume centered directly below the impact point of protons on target (the proton beam direction is pointing out of the page). For this case, the He-II and \ce{SD_2} volumes contribute roughly equal amounts to the total UCN yield in neutrons per second, with the \ce{SD_2} being better suited for absorbing the higher heatload and producing orders of magnitude more neutrons per cm$^3$, and the longer lifetime of UCNs in He-II allowing for a larger volume.

\begin{table}[tbp!]
\centering
\def\arraystretch{1.5}
    \setlength\tabcolsep{0pt}
\caption{Calculations for liquid helium UCN sources placed in-pile. The reported heat-load is only the prompt radiation heat-load.} 
\begin{tabular*}{0.95\textwidth}{@{\extracolsep{\fill}} r c c c c c}
\toprule
 & \makecell{He-II/{\ce{SD_2}} Volume \\ $[\si{L}]$}  & \makecell{$P_\text{UCN}$ \\
 $\left[\si{n/s/cm^3}\right]$}& \makecell{$\dot{N}_\text{UCN}$ \\
 $\left[\si{n/s}\right]$} & \makecell{Heat-load\\ $[\si{W}]$}\\
\midrule
\textbf{Full He-II Moderator + MgH$_2$} (\cref{fig:lHe_inpile_full_geom}) &      \num{21.9} &\num{2.58e5} & \num{5.56e9} & 9933\\
\textbf{Full He-II Moderator + No MgH$_2$} &      \num{21.9} &\num{1.23e5} & \num{2.69e9} & 9915\\
\midrule
\textbf{Small He-II w/extraction} (\cref{fig:lHe_inpile_extract_geom}) &      \num{1.25} &\num{6.3e4} & \num{7.8e7} & 113\\
\midrule
\textbf{Small He-II + \ce{SD_2} slab w/extraction} (\cref{fig:lHe_SD2_hybrid_inpile_extract_geom}) & & & & \\
   \ce{SD_2} slab only &
  \num{0.06} &\num{1.22e6} & \num{7.7e7} & 10\\
   He-II extraction line only&
  \num{1.19} &\num{5.14e4} & \num{6.1e7} & 99\\
  \textbf{Total} &      \num{1.25} & -- & \num{1.34e8} & 109\\
\bottomrule
\end{tabular*}
\label{tab:lhe_intwister}
\end{table}

\subsection{\ce{SD_2}  in MCB}
\begin{figure}[tbh!]      
    \begin{subfigure}[b]{0.48\textwidth}
        \centering
        \includegraphics[width=\textwidth]{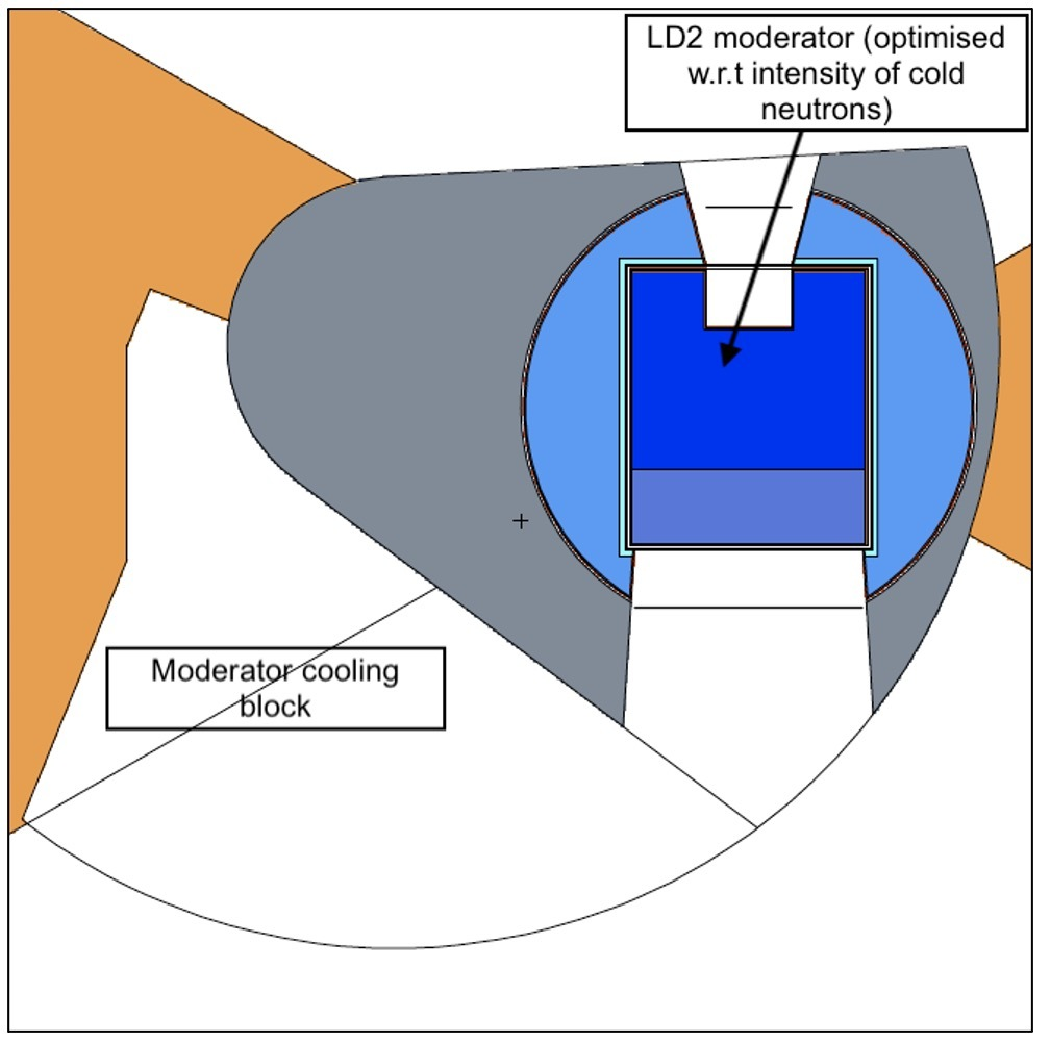}
        \subcaption{}
        \label{fig:InitialBaselineModelOfLD2AndMCB_Blahoslav}
    \end{subfigure}
    \hfill
    \begin{subfigure}[b]{0.48\textwidth}
        \centering        
        \includegraphics[width=\textwidth]{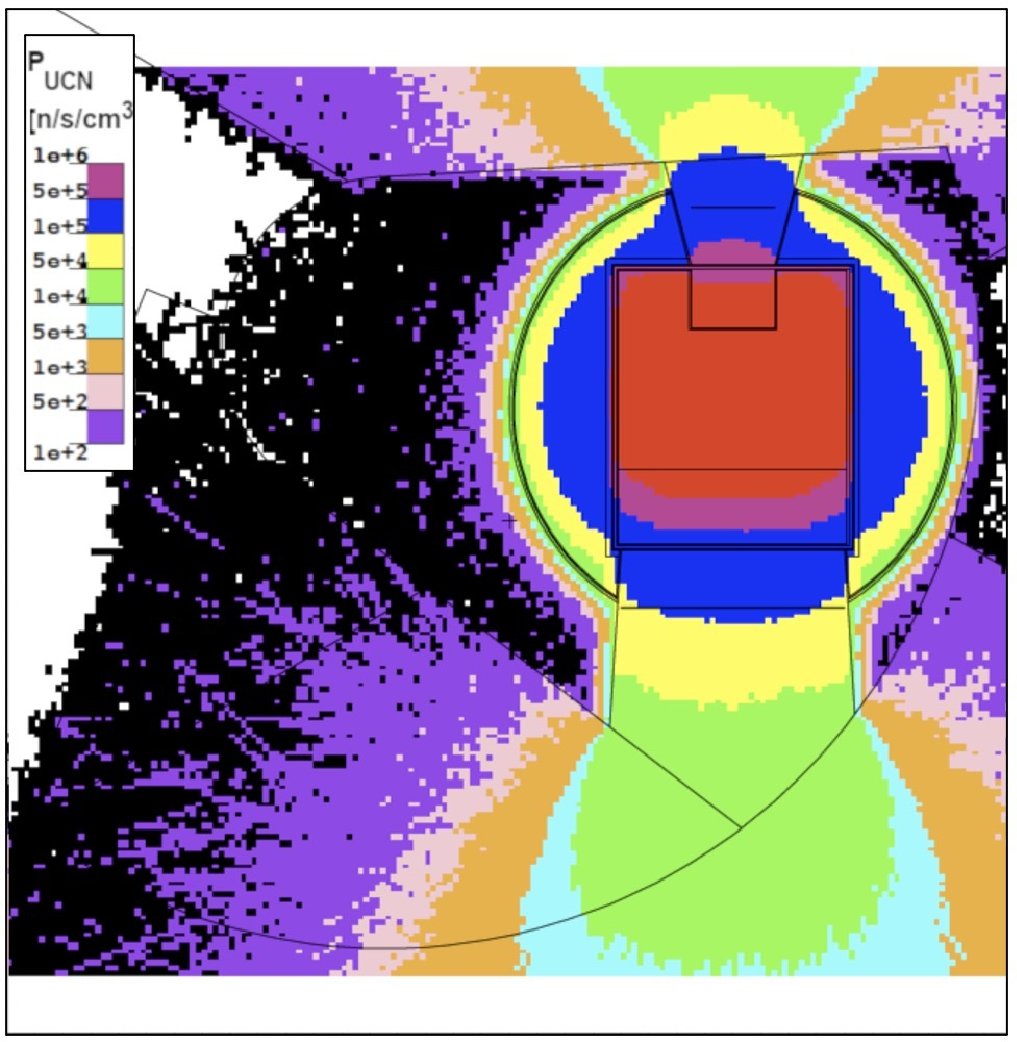}
        \subcaption{}
        \label{fig:P_UCNMap_BaselineModel}
    \end{subfigure}
\caption{(a) The baseline geometry for LD$_2$ cold neutron moderator and MCB. (b) Map of $P_\text{UCN}$.}
%\label{fig:SphericalSD2}
\end{figure}

\begin{figure}[tbh!]      
    \centering        
    \includegraphics[width=0.48\textwidth]{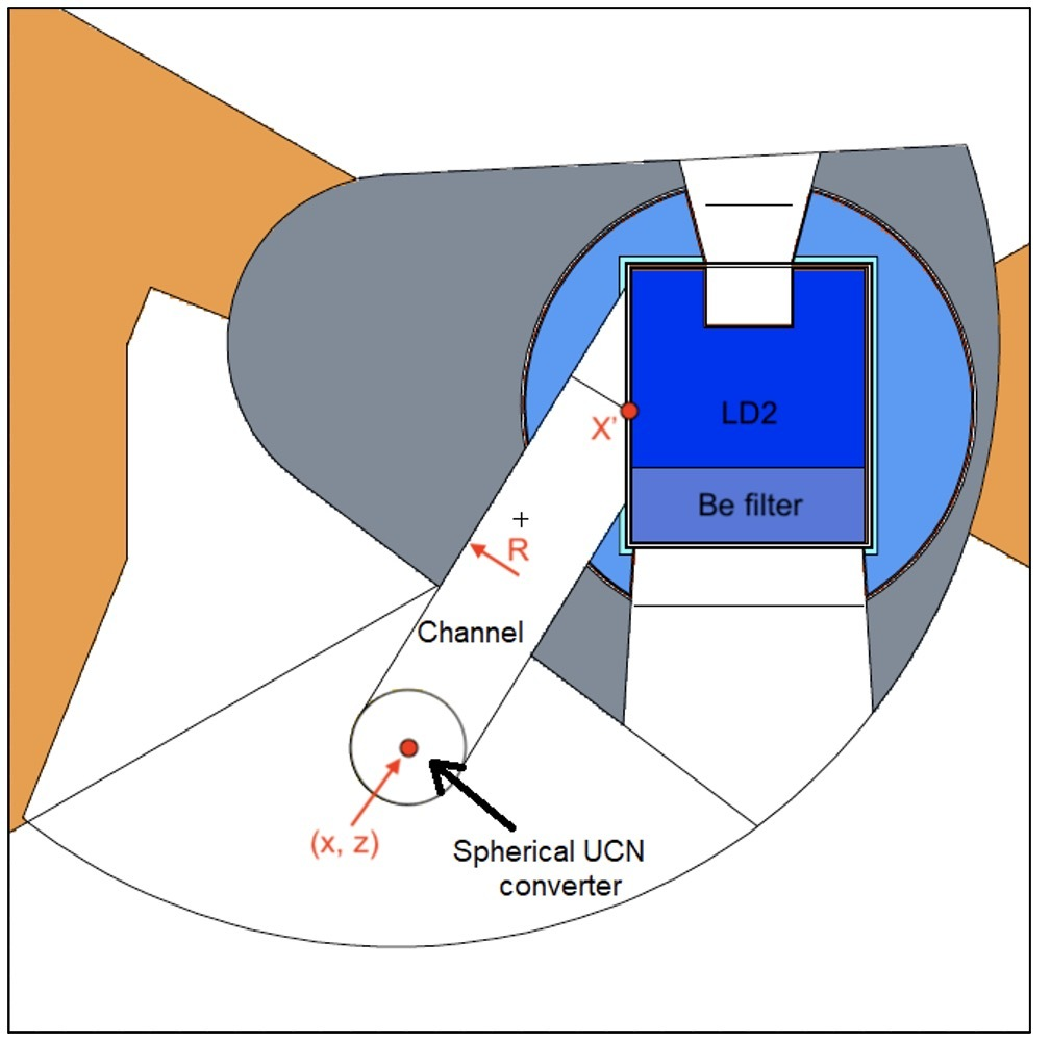}
\caption{Parameterised model of the channel and UCN converter.}
\label{fig:SphericalSD2AndCylindricChannel_Parameters}
\end{figure}

The possibility of placing the UCN SD$_\text{2}$ converter in the moderator cooling block was studied.
The MCB is a piece made of steel located in the target vessel with the sole purpose of shielding (see Section 3.2 in \cite{zaniniworkshop2022}).
A part of the MCB that surrounds the twister and the LD$_2$ moderator (see \cref{fig:InitialBaselineModelOfLD2AndMCB_Blahoslav}) can be potentially modified to fit the UCN converter inside \cite{zaniniworkshop2022}. In this section, we describe the analysis of a SD$_\text{2}$ converter, while a He-II converter in the MCB is described in \cref{subsec:he-in-mcb}.
The basic concept is to place the SD$_\text{2}$ UCN converter into MCB near the cold source (see position 2 of Figure 4 in \cite{zaniniworkshop2022}), which would make use of cold neutrons streaming from the LD$_2$ moderator to produce UCNs.
However, as it is clear from \cref{fig:P_UCNMap_BaselineModel},
in the current baseline model few cold neutrons can reach the MCB location, 
leading to small $P_\text{UCN}$. This can be solved by upgrading the baseline model and, in particular, the Be reflector and steel surrounding the LD$_2$ moderator, by designing a channel connecting the LD$_2$ moderator and the MCB. \\
% $P_\text{UCN}$
% UCNCrossSectionSD2_Munich.png
\indent The initial model that was studied comprised of a cylindrical channel to transport cold neutrons from the surface of the LD$_2$ moderator to the MCB and spherical UCN converter in which $P_\text{UCN}$ was calculated (see \cref{fig:SphericalSD2AndCylindricChannel_Parameters}). In this case, the UCN converter was defined as a void cell in the geometry definition in MCNP, so the calculated $P_\text{UCN}$ was underestimated since the effect of scattering of cold neutrons within the SD$_\text{2}$ material was not taken into account. However, comparing relative values of $P_\text{UCN}$ that were calculated for each set of parameters in the model were sufficient for this initial study.  Both the cylindrical channel and the UCN converter were aligned with the center of LD$_2$ moderator on the plane perpendicular to the y axis. \\
\indent Subsequently, the model was parametrized such that the horizontal position of the channel's opening at the LD$_2$ moderator surface $X'$, the channel radius $R$ and the SD$_\text{2}$ spherical converter position in $x$ and $z$ could be varied. First, a general parametric study was performed. Namely, the position of SD$_\text{2}$ converter was varied at a fixed radius of the channel (R = 6, 8, and 10 cm) and a fixed alignment of the opening of the channel with the surface of box-shaped LD$_2$ moderator. The alignment was either with the LD$_2$ moderator centre or Be filter. The parametric study revealed that a larger channel radius gave rise to higher $P_\text{UCN}$, but at the cost of a drop in the NNBAR FOM. \\
\indent Moreover, $P_\text{UCN}$ was at a maximum at a given channel radius when the length of the channel was minimized. Therefore, the channel opening aligned with the Be filter, together with the UCN converter placed in the MCB with a slight offset in $x$ and $z$ w.r.t. $X'$ to minimize the length of the channel, resulted in the highest observed $P_\text{UCN}$. \\
\indent This finding was also independently confirmed by the Efficient Global Optimization algorithm that searched for maximal $P_\text{UCN}$ by varying all the parameters: $x$, $z$, $X'$ and $R$. Although the Be filter suppressed the spectrum of neutrons below 4 Å, the models with channel opening aligned with the Be filter still performed the best w.r.t. $P_\text{UCN}$. \\
\indent Simultaneously, the decrease in the NNBAR FOM was minimal. This was unexpected, since cold neutrons with a wavelength between 2 and 4 Å partially contribute to $P_\text{UCN}$. However this loss was compensated by the gain from designing a shorter channel. These results together with high $P_\text{UCN}$ observed around the NNBAR emission surface (see \cref{fig:P_UCNMap_BaselineModel}) motivated a further study of SD$_\text{2}$ converter located in the MCB, but with a direct view of the NNBAR emission surface. \\
\indent The maximal acceptable decrease in NNBAR FOM was chosen to be 5~\% for any model of a UCN converter in the MCB. The NNBAR FOM of the baseline model without the UCN converter and transport channel was 2.53 $\times$ 10$^\text{17}$, the lowest acceptable FOM therefore 2.40 $\times$ 10$^\text{17}$. Of the models fulfilling this requirement, the highest observed $P_\text{UCN}$ was obtained for a cylindrical channel with a radius of 6 cm aligned with the Be filter. This design was therefore selected as the reference model. The shape of the UCN converter was redesigned to reflect that UCNs can only be extracted from a very thin of layer SD$_\text{2}$. This resulted in a disc-shaped UCN converter with 2 cm thickness. Filling this UCN converter by SD$_\text{2}$ increased the measured $P_\text{UCN}$ to 1.1 $\times$ 10$^\text{4}$.  

% \begin{figure}[tbh!]      
%    \begin{subfigure}[b]{0.48\textwidth}
%        \centering
%        \includegraphics[width=\textwidth]{ucn_fig/SphericalSD2AndCylindricChannel.eps}
%        \subcaption{}
%        \label{fig:SphericalSD2AndCylindricChannel}
%    \end{subfigure}
%    \hfill
%    \begin{subfigure}[b]{0.48\textwidth}
%        \centering        
%        \includegraphics[width=\textwidth]{ucn_fig/SphericalSD2AndCylindricChannel_Parameters.eps}
%        \subcaption{}
%        \label{fig:SphericalSD2AndCylindricChannel_Parameters}
%    \end{subfigure}
% \caption{(a) Concept of extraction channel in the twister to feed an UCN converter placed in MCB. (b) Parameterised model of the channel and UCN converter.\textcolor{red}{remove figure}}
% \label{fig:SphericalSD2}
% \end{figure}
% \begin{table}[tbp!]
% \caption{Maximal observed $P_\text{UCN}$ in SD$_\text{2}$ converter for fixed radius of cylindrical channel.}

\begin{figure}[tbh!]   
    \begin{subfigure}[b]{0.50\textwidth}
        \centering        
        \includegraphics[width=\textwidth]{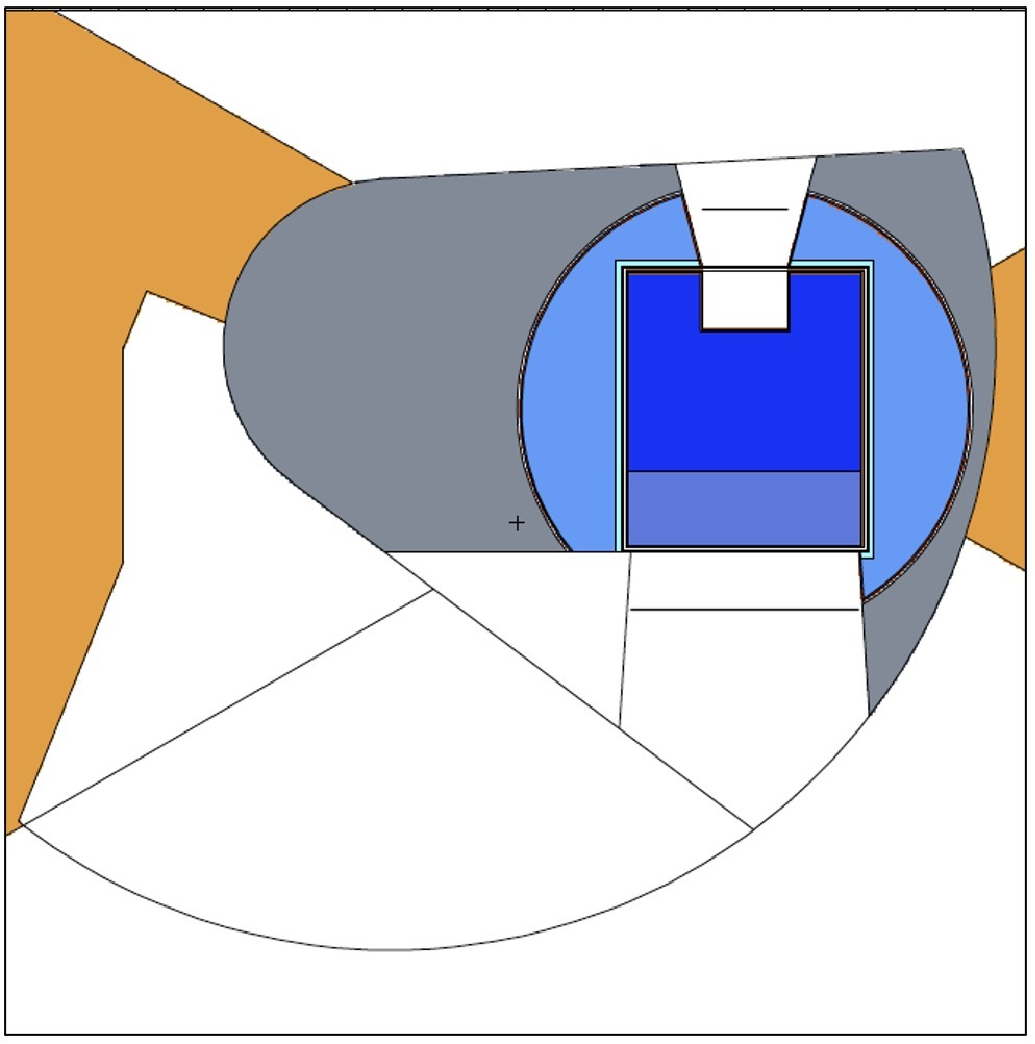}
        \subcaption{}
        \label{fig:HorizontalCutInTwister_UCNSourceInMCB_Blahoslav}
    \end{subfigure}
    \hfill
    \begin{subfigure}[b]{0.48\textwidth}
        \centering
        \includegraphics[width=\textwidth]{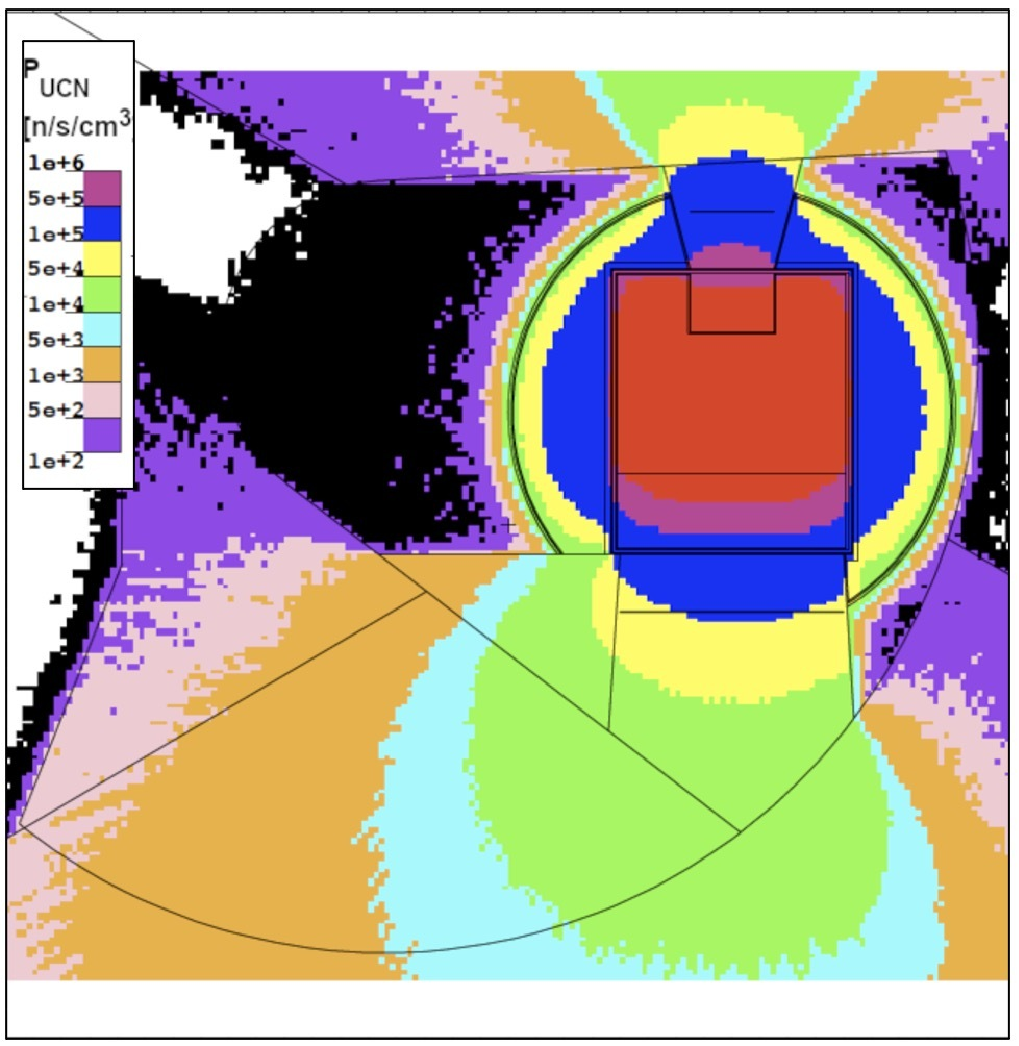}
        \subcaption{}
        \label{fig:P_UCNMap_CutByHorizontalPlane_SD2InMCB}
    \end{subfigure}
\caption{A model with material removed in twister below a horizontal plane to allow for a direct view between the SD$_\text{2}$ converter in MCB and NNBAR emission surface (a) Drawing of the model.  (b) Map of $P_\text{UCN}$.}
\label{fig:HorizontalCutStudy_0}
\end{figure}

\begin{figure}[!htb]  
\begin{subfigure}[b]{0.48\textwidth}
        \centering        
        \includegraphics[width=\textwidth]{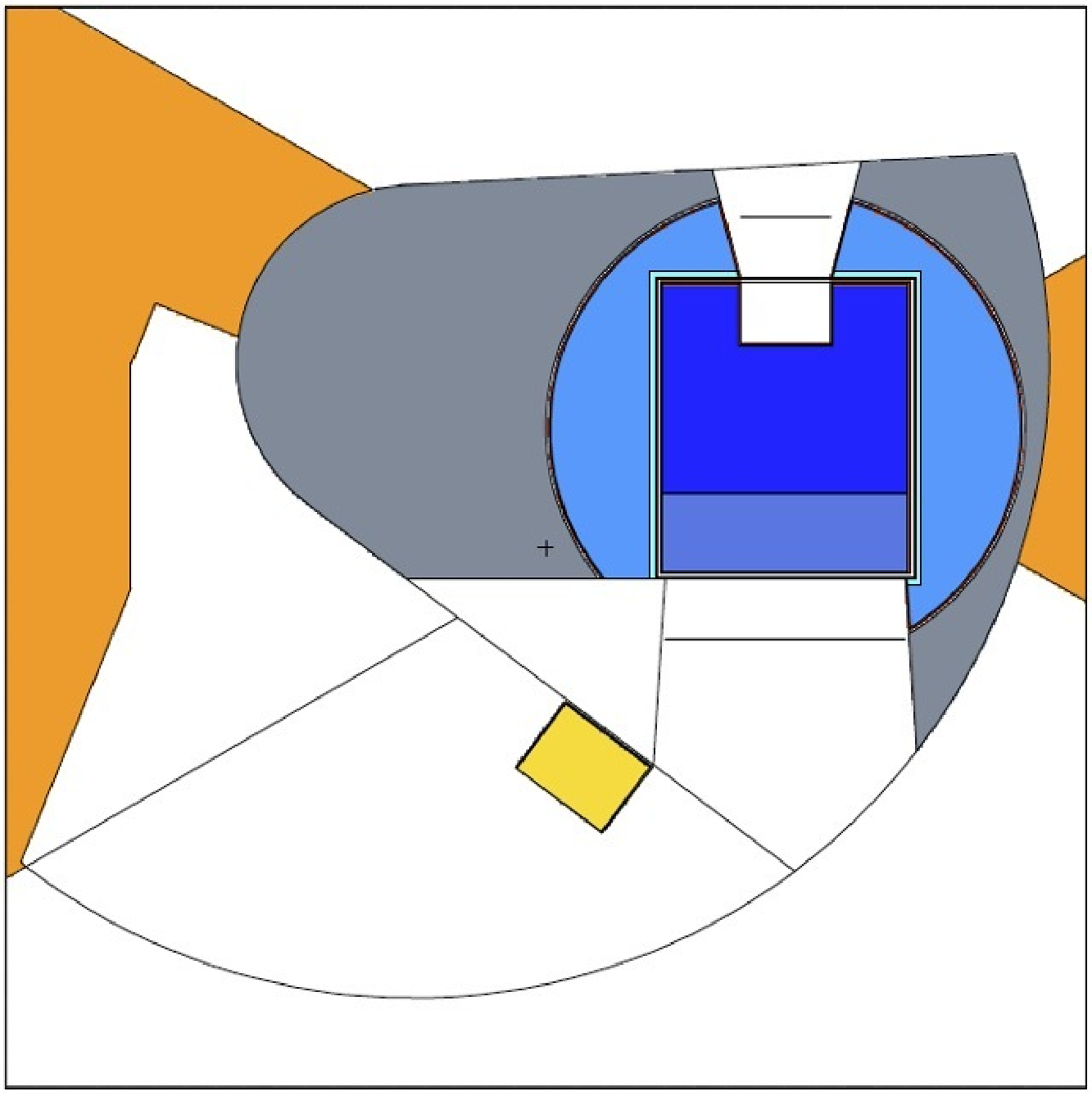}
        \subcaption{}
        \label{fig:SD2ConverterInMCB_Thick_CutByHorizontalPlane}
    \end{subfigure}
    \hfill
    \begin{subfigure}[b]{0.505\textwidth}
        \centering
        \includegraphics[width=\textwidth]{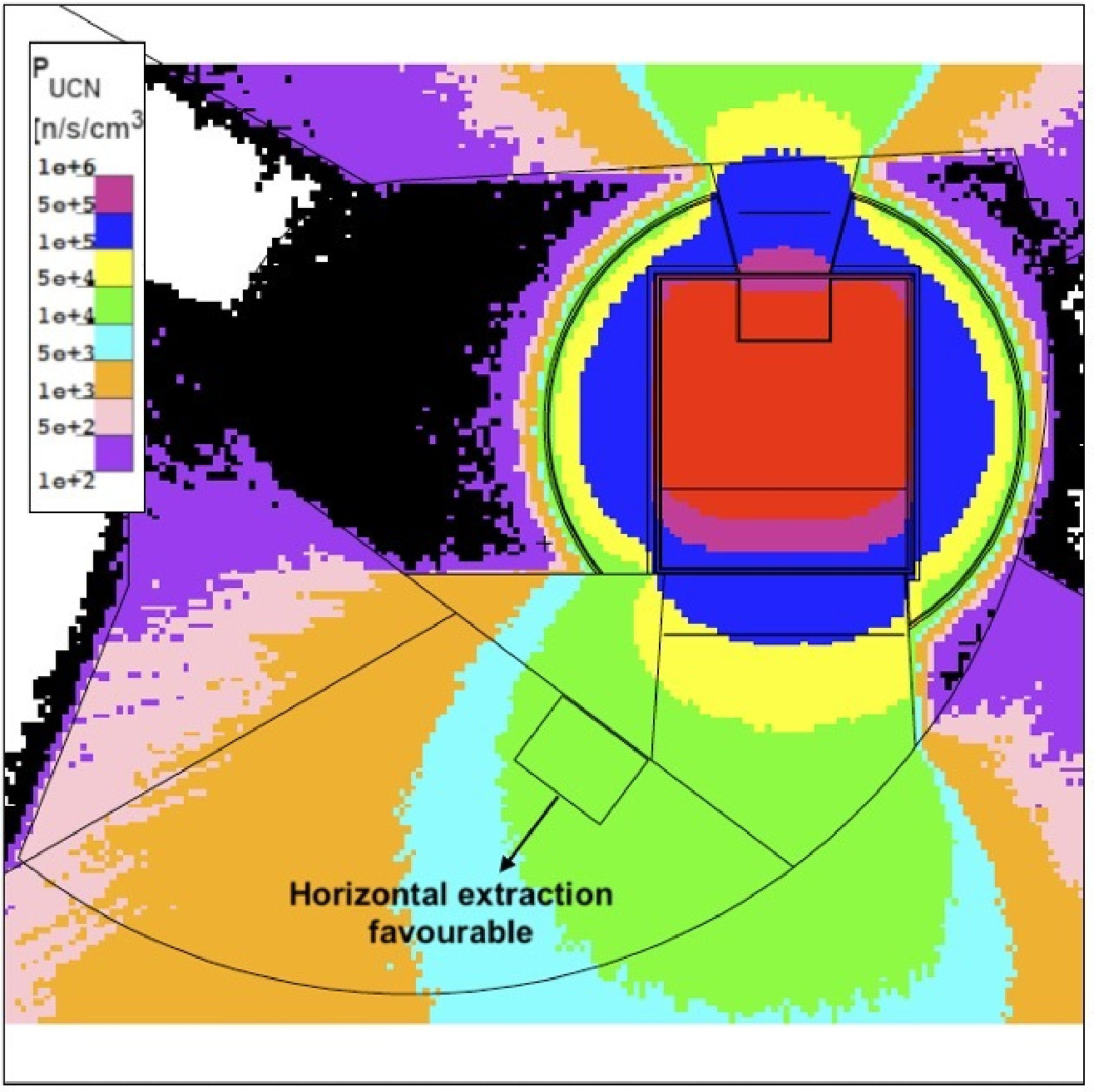}
        \subcaption{}
        \label{fig:P_UCNMap_CutByHorizontalPlane_SD2InMCB_PlacingSD2Converter}
    \end{subfigure}

\caption{(a) The thick SD$_\text{2}$ UCN converter in the MCB for which $P_\text{UCN}$ and heat-load was estimated. (b) The thick SD$_\text{2}$ converter placed within the region with $P_\text{UCN}$ of the order of $10^\text{4}s$.}
\label{fig:SphericalSD4}
\end{figure}

\begin{figure}[tbh!]      
    \begin{subfigure}[b]{0.5\textwidth}
        \centering
        \includegraphics[width=\textwidth]{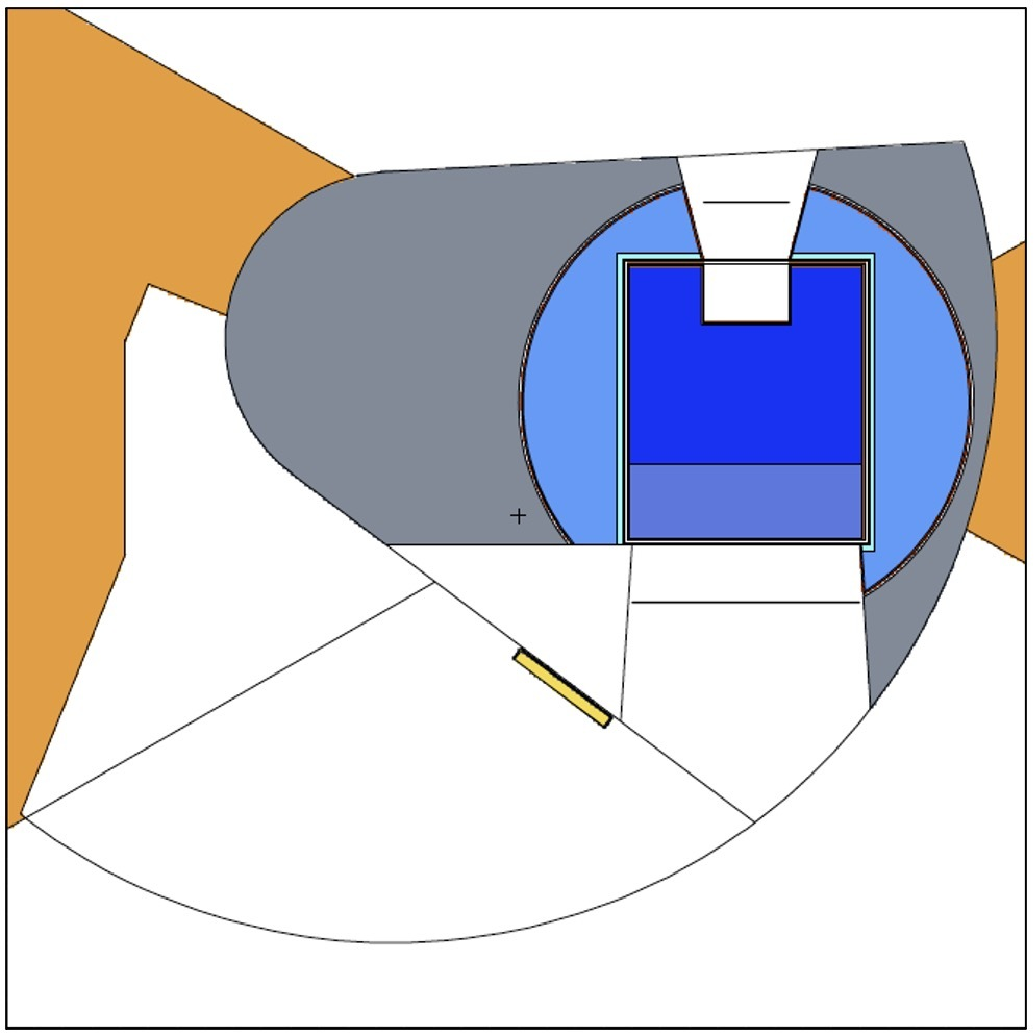}
        \subcaption{}
        \label{fig:SD2ConverterInMCB_Thin_CutByHorizontalPlanel}
    \end{subfigure}
    \hfill
    \begin{subfigure}[b]{0.51\textwidth}
        \centering        
        \includegraphics[width=\textwidth]{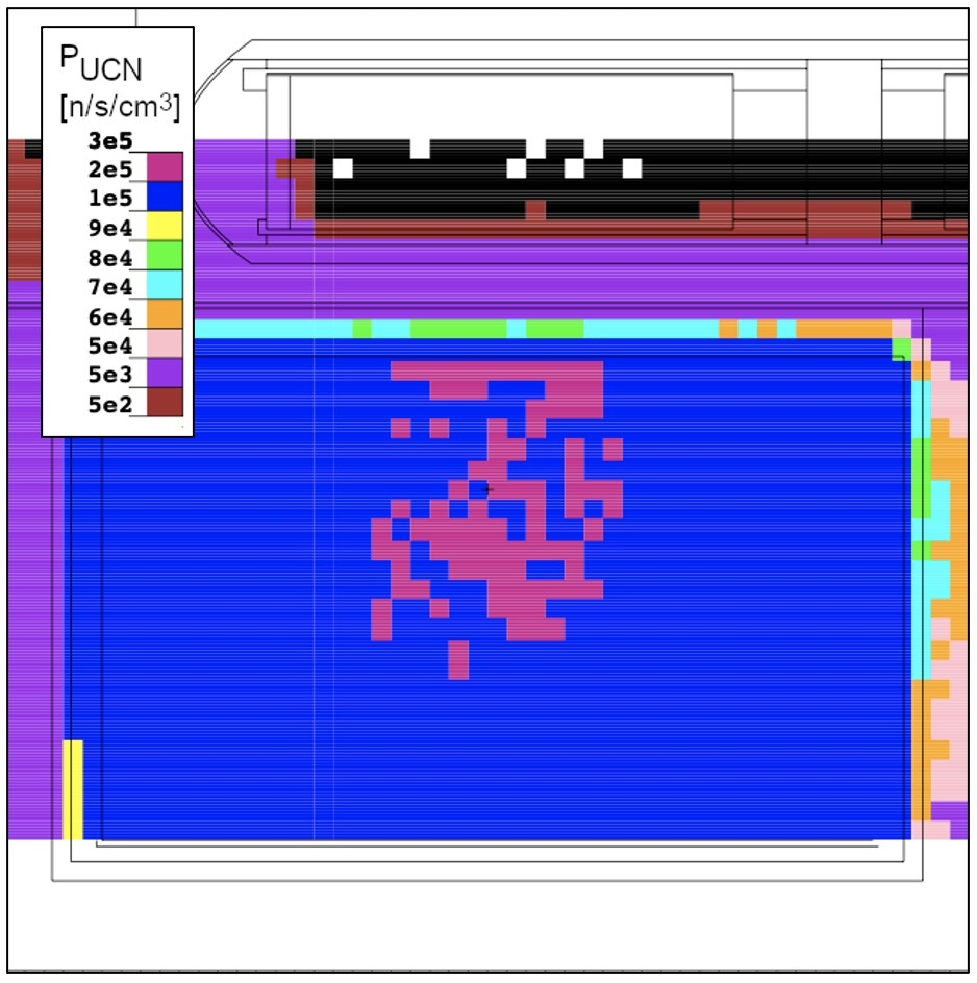}
        \subcaption{}
        \label{fig:P_UCNMap_NNBAREmissionSurface_Blahoslav}
    \end{subfigure}
\caption{{(a) Thin box-shaped SD$_\text{2}$ UCN converter placed in the MCB. (b) $P_\text{UCN}$ map in the $yz$-plane measured at the NNBAR emission surface.}}
%\label{fig:SphericalSD2}
\end{figure}

\begin{figure}[!htb]      
    \begin{subfigure}[b]{0.45\textwidth}
        \centering
        \includegraphics[width=\textwidth]{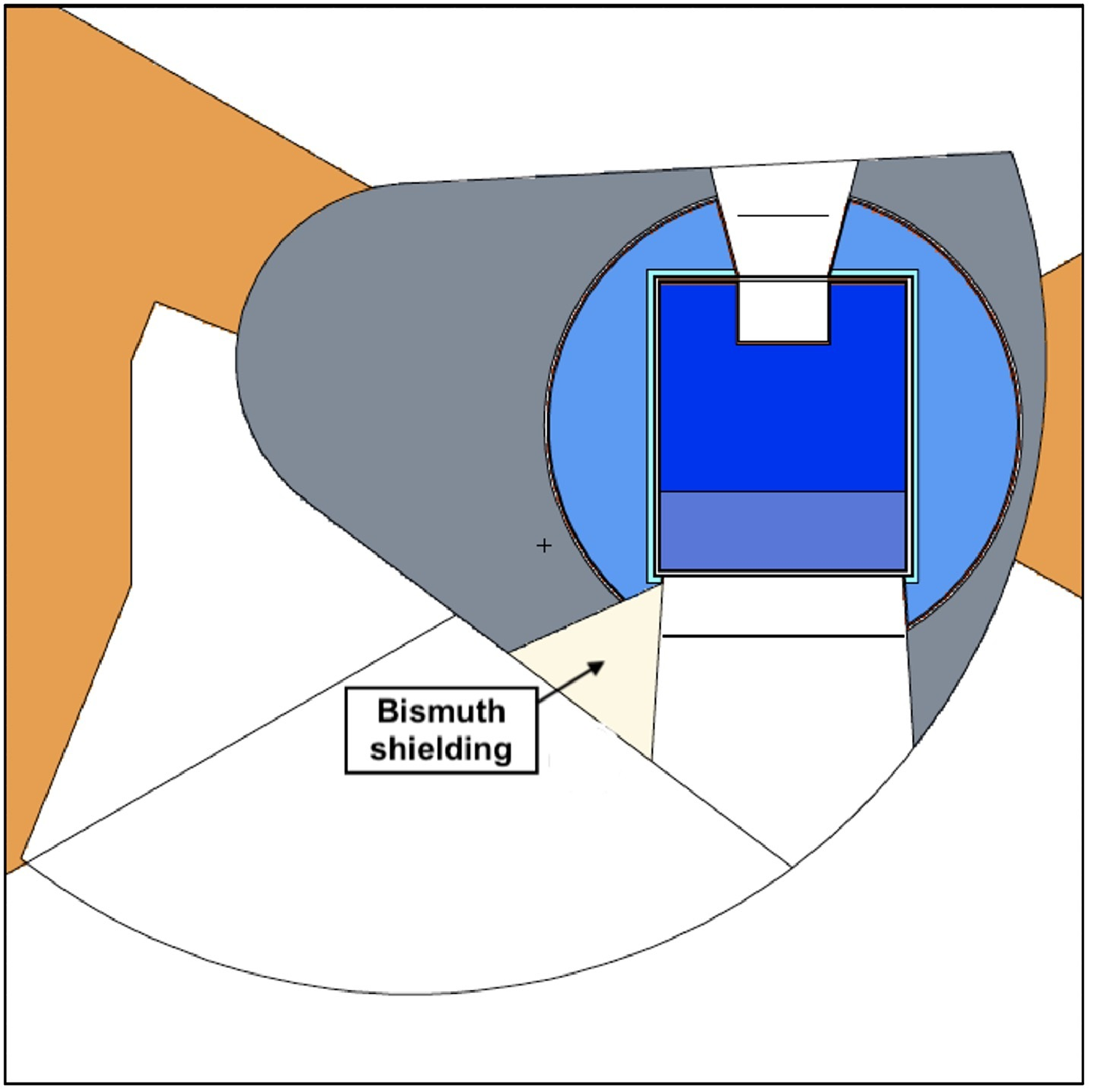}
        \subcaption{}
        \label{fig:Model_GammaShieldingByBismuthForMCB_Blahoslav}
    \end{subfigure}
    \hfill
    \begin{subfigure}[b]{0.45\textwidth}
        \centering        
        \includegraphics[width=\textwidth]{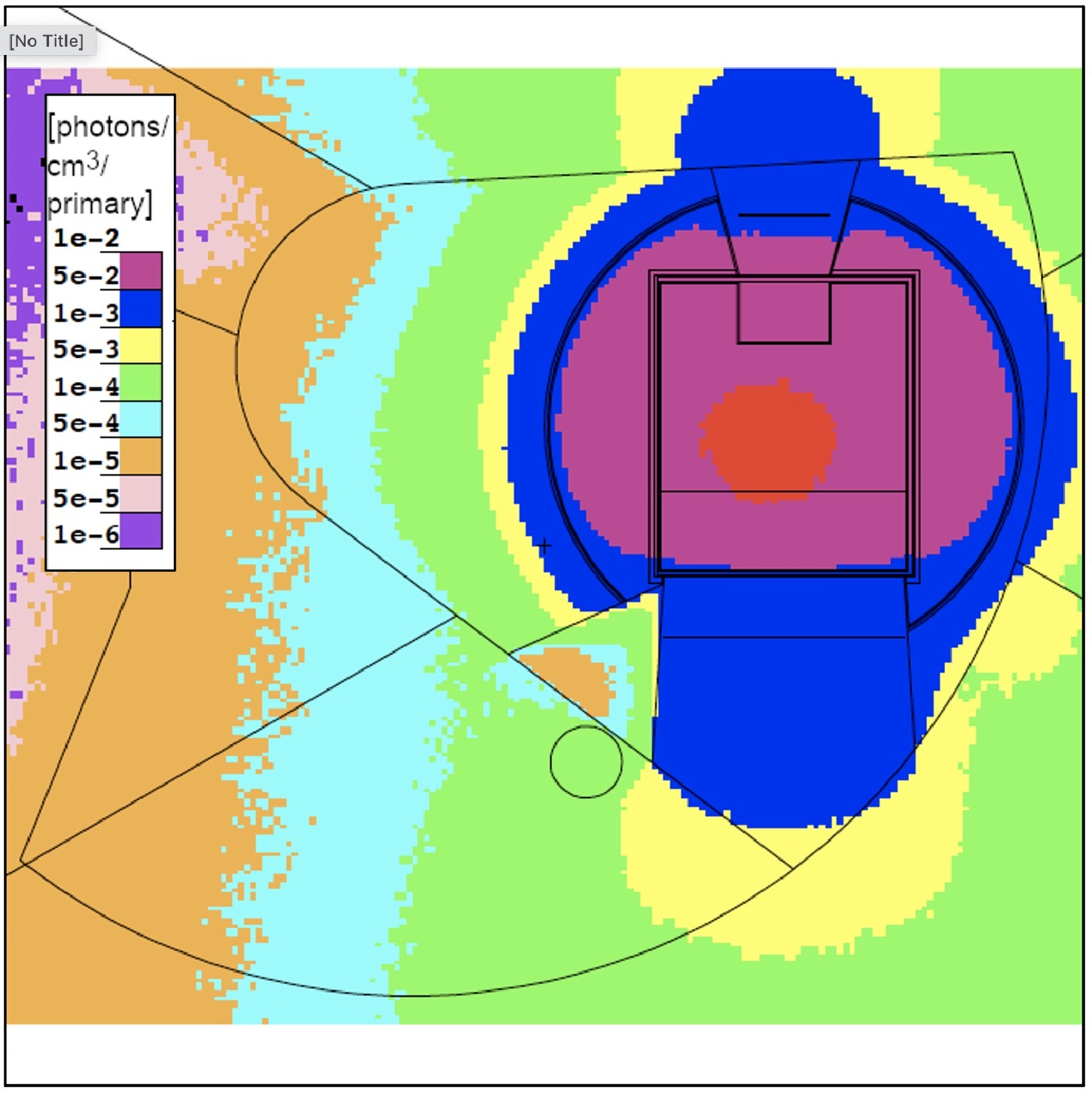}
        \subcaption{}
        \label{fig:GammaShieldingByBismuthForMCBTest}
    \end{subfigure}
\caption{(a) Inserting the bismuth shielding into the twister in order to suppress photons before they reach MCB. (b) bismuth shielding in the twister reduced the flux of photons in MCB significantly.}
% \label{fig:SphericalSD2_3}
% \end{figure}
% \begin{figure}[!htb]      
    \begin{subfigure}[b]{0.48\textwidth}
        \centering
        \includegraphics[width=\textwidth]{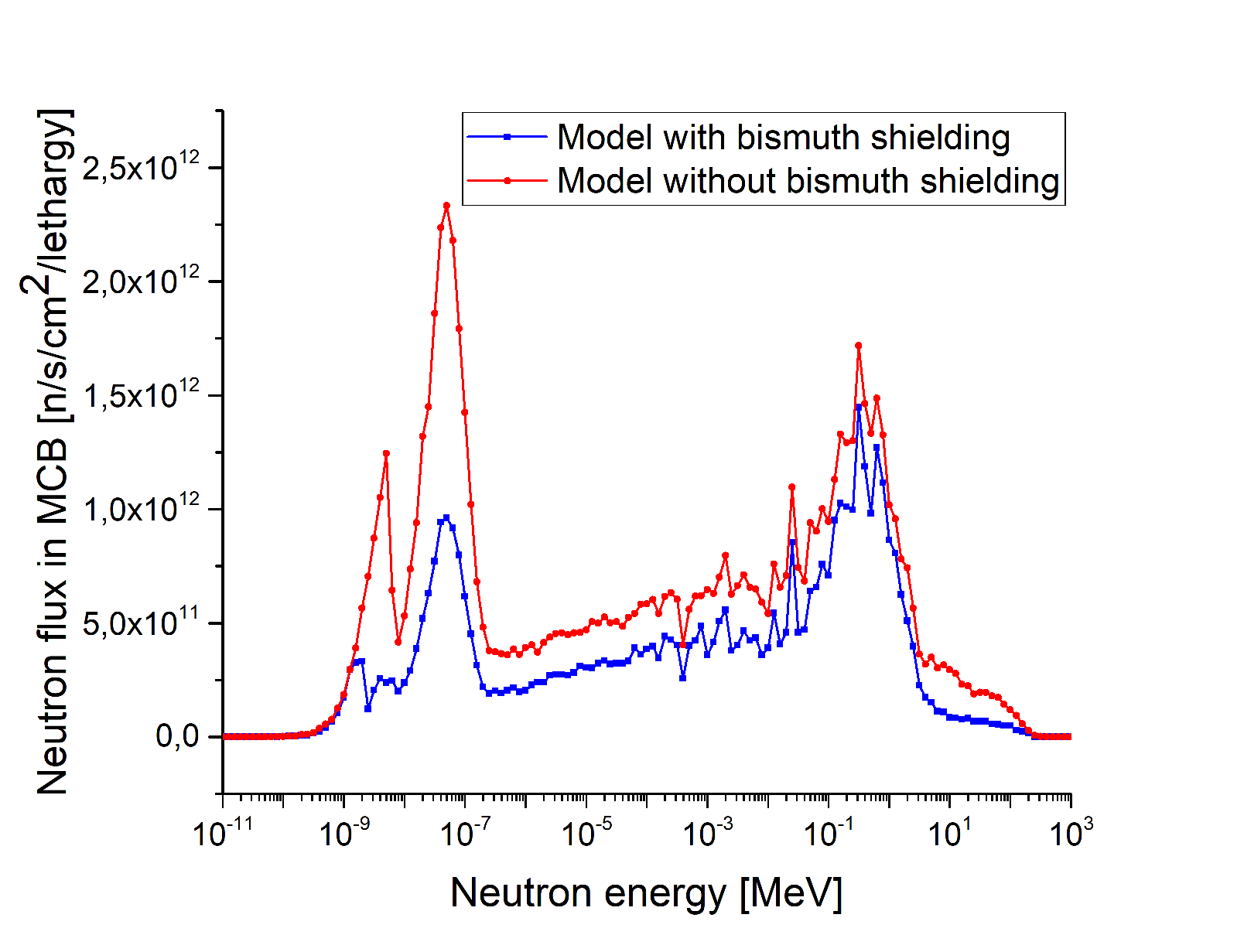}
        \subcaption{}
        \label{fig:MeVBiEffect_Blahoslav}
    \end{subfigure}
    \hfill
    \begin{subfigure}[b]{0.48\textwidth}
        \centering        
        \includegraphics[width=\textwidth]{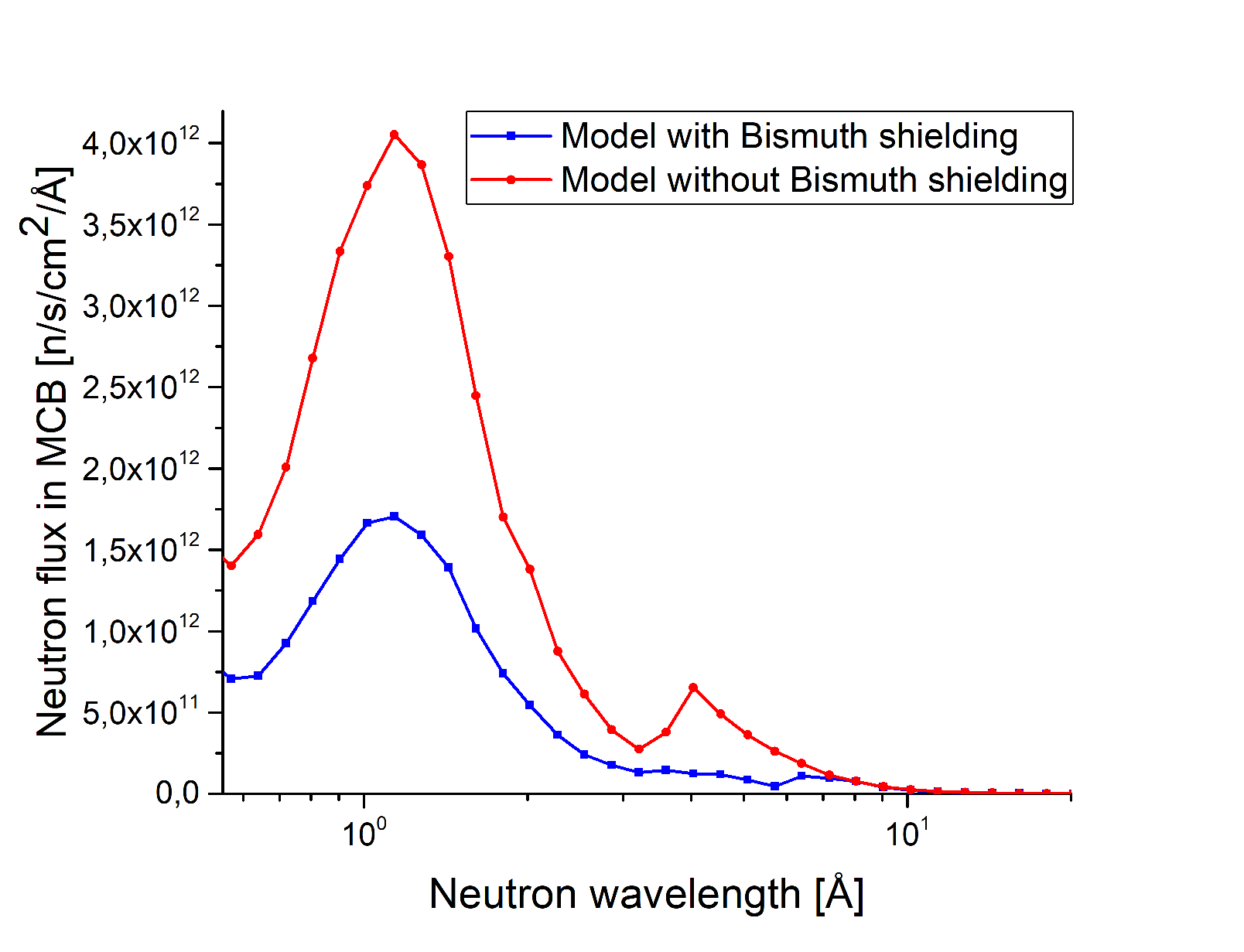}
        \subcaption{}
        \label{fig:ABiEffect_Blahoslav}
    \end{subfigure}
\caption{(a) The neutron spectrum in the MCB. The red line depicts the spectrum without bismuth shielding in the twister, and the blue line depicts the spectrum with bismuth shielding in the twister. (b) The flux of cold and thermal neutrons with wavelengths $>$ 1 Å was clearly suppressed in the MCB by the bismuth shielding.} 
\end{figure}

\indent Subsequently, an SD$_\text{2}$ converter in the MCB with direct view of the NNBAR emission surface was studied. This design was modelled by placing the converter further down in MCB with respect to the $x$ axis, and defining a plane in the twister below which all material was removed. Consequently, cold neutrons emitted from the NNBAR emission surface could reach the SD$_\text{2}$ converter in the MCB without obstacles, as shown in \cref{fig:HorizontalCutInTwister_UCNSourceInMCB_Blahoslav}. Based on \cref{fig:P_UCNMap_CutByHorizontalPlane_SD2InMCB}, the SD$_\text{2}$ converter in the MCB could reach $P_\text{UCN}$ of the order of 10$^\text{4}$ cm$^\text{-3}$s$^\text{-1}$. Subsequently, the model of the SD$_\text{2}$ converter was designed and placed into region with the highest expected $P_\text{UCN}$ in the MCB, i. e. within the green shaded region in \cref{fig:P_UCNMap_CutByHorizontalPlane_SD2InMCB_PlacingSD2Converter}. The resulting geometry of a thick box-shaped SD$_\text{2}$ UCN converter with a thickness of 13.4 cm can be seen in \cref{fig:SD2ConverterInMCB_Thick_CutByHorizontalPlane}. This scenario implies a horizontal extraction of UCNs. The Al container was designed around the SD$_\text{2}$ UCN converter with a wall thickness of 3 mm that was reduced to 0.5 mm on the emission surface of the converter. $P_\text{UCN}$ calculated for this converter was 4.5 $\times$ 10$^\text{4}$ cm$^\text{-3}$s$^\text{-1}$. 
% ucn_fig/InitialCutInTwister_UCNSourceInMCB_Blahoslav.eps
% fig:InitialCutInTwister_UCNSourceInMCB_Blahoslav
% (a) Material removed below a tilted plane.
\indent The initial thick box-shaped SD$_\text{2}$ UCN converter was redesigned since previous experiments~\cite{brys2007extraction} showed that it is much easier to extract UCNs from a thin slab of SD$_\text{2}$ converter. This constraint together with the fact that $P_\text{UCN}$ of the order of 10$^\text{4}$ cm$^\text{-3}$s$^\text{-1}$ could be reached close to the border between the MCB and the twister, i. e. at the shortest possible distance to the NNBAR emission surface, favoured a following design of the SD$_\text{2}$ converter (see \cref{fig:SD2ConverterInMCB_Thin_CutByHorizontalPlanel}).
%\cref{ucn_fig/SD2ConverterInMCB_Thin_CutByHorizontalPlane}). 
$P_\text{UCN}$ measured for this thin box-shaped SD$_\text{2}$ UCN converter was 3.8 $\times$ 10$^\text{4}$ cm$^\text{-3}$s$^\text{-1}$. \\
\indent Estimating the heat-load on the SD$_\text{2}$ converter and surrounding Al vessel was also an integral part of the study. The prompt heat-load for the thick box-shaped SD$_\text{2}$ UCN converter was 440 W, where the heat deposited by neutrons was 220 W, and the heat deposited by photons was 200 W. However, this estimation did not take into account a contribution due to the neutron activation of isotope $^\text{27}$Al with subsequent $\beta$-decay of $^\text{28}$Al  where both an electron and a $\gamma$-ray were emitted. The decay heat for the thick UCN SD$_\text{2}$ converter was 18 W. Similarly, the prompt heat-load for a thin box-shaped SD$_\text{2}$ UCN converter was 150 W, where the heat deposited by neutrons was 60 W, and the heat deposited by photons was 80 W. The decay heat of $^\text{28}$Al  was 9 W. \\
% ]{ucn_fig/P_UCNMap_CutByPlane_SD2InMCB.eps
% fig:InitialCutInTwister_UCNSourceInMCB_Blahoslav
% (a) Material removed below a tilted plane.
% {fig:P_UCNMap_CutByPlane_SD2InMCB}
% (a) material removed below a tilted plane
%\begin{figure}[tbh!]      
%    \begin{subfigure}[b]{0.48\textwidth}
%        \centering
%        \includegraphics[width=\textwidth]{ucn_fig/P_UCNMap_CutByPlane_SD2InMCB.eps}
%        \subcaption{NNBAR FOM 2.52e+17}
%        \label{fig:P_UCNMap_CutByPlane_SD2InMCB}
%    \end{subfigure}
%    \hfill
%    \begin{subfigure}[b]{0.48\textwidth}
%        \centering        
%        \includegraphics[width=\textwidth]{ucn_fig/P_UCNMap_CutByHorizontalPlane_SD2InMCB.eps}
%        \subcaption{NNBAR FOM 2.51e+17}
%        \label{fig:P_UCNMap_CutByHorizontalPlane_SD2InMCB}
%    \end{subfigure}
% \caption{Map of $P_\text{UCN}$ for the model with (a) material removed below a tilted plane and (b) material removed below a horizontal plane. \textcolor{red}{remove figure, combine two right figures into a single one }}
%\label{fig:SphericalSD2}
%\end{figure}
\indent All the calculated variations of $P_\text{UCN}$ over a grid in the $xz$-plane were done with the center of LD$_\text{2}$ moderator lying at $y$ = -23.7 cm in our model. Therefore, it was also important to explore the $P_\text{UCN}$ variation in the $yz$-plane over the NNBAR emission surface (see \cref{fig:P_UCNMap_NNBAREmissionSurface_Blahoslav}). It is apparent that a slightly higher $P_\text{UCN}$ could be reached by placing the SD$_\text{2}$ UCN converter slightly closer to the water premoderator, where the converter would benefit from neutrons with wavelengths closer to the thermal region. 
Finally, a preliminary study of the effect of bismuth on reducing the heat-load of photons on the SD$_\text{2}$ UCN converter was performed. For this purpose, a shielding block made of bismuth was inserted into the twister (see \cref{fig:Model_GammaShieldingByBismuthForMCB_Blahoslav}). The flux of photons was significantly reduced in the MCB by the effect of the bismuth shielding (see \cref{fig:GammaShieldingByBismuthForMCBTest}). However, this came at the cost of a significant reduction of $P_\text{UCN}$ within the SD$_\text{2}$ converter. The reason for this is illustrated in \cref{fig:MeVBiEffect_Blahoslav} which shows the neutron energy spectrum obtained in the spherical volume just outside the Bi shielding visible in \cref{fig:GammaShieldingByBismuthForMCBTest}. This close to both the moderator and the spallation target, the spectrum consists of a mixture of moderated and fast neutrons, extending up into the MeV range. However, it is clear that especially the cold part of the spectrum is strongly suppressed by the bismuth shielding. This is further illustrated in \cref{fig:ABiEffect_Blahoslav}, showing the wavelength distributions for the cold spectrum with and without bismuth.
In fact, bismuth has a Bragg cutoff wavelength larger than 6.6~\AA. Therefore, it is probably not well suited for a SD$_2$ source, which requires neutrons with such wavelengths. For He-II on the other hand, for which neutrons with $\lambda=8.9$~\AA~are most effective for UCN production, the Bi filter is very interesting, as discussed in \cite{serebrovworkshop2022}.
\\

\subsection{He-II in MCB} \label{subsec:he-in-mcb}

%In deliverable \textcolor{red}{insert reference} a draft engineering concept for the implementation of a UCN source into the existing ESS framework has been rendered. The results of the MCNP particle transport simulations define the expected heat loads which have to be handled. To realise this an approach is shown which combines the implementation in the position with the highest neutron flux with a combined concept to guide the produced UCN out of the source while removing the heat load from it via a He-II based heat pipe system. \textcolor{red}{update the text (guide part has been removed?}

%\subsubsection{UCN production investigation by MCNP particle transport simulations} \label{ch:2-1}

%The implementation of an UCN moderator system has been investigated by WP4 with focus on 5 different positions respectively converter media. In the table shown on the right side of \cref{fig:2-1} is the result of the performed particle transport simulations by MCNP regarding the UCN production rate as figure of merit concluded. Attention should be paid that option 1 has the highest UCN production rate, but is replacing the VCN moderator. 
The option of a UCN moderator placed in the moderator cooling block based on helium in phase-state II (He-II) is investigated in this section.
This option is very promising
 if CN, VCN and UCN sources are intended to be installed and operated in parallel.
 %, simultaneously this is also the option with the highest expected heat load. 
%Therefore, aspects related to the further engineering phase have been regarded for this position of highest engineering demands in terms of heat load handling combined with benefits given by the location. 
Therefore, preliminary engineering studies have been performed for this option. 
The following engineering study is unconstrained by the final decision for an implementation concept in this location.

\begin{figure}[hbt!]
\begin{center}
\includegraphics[width=1\textwidth]{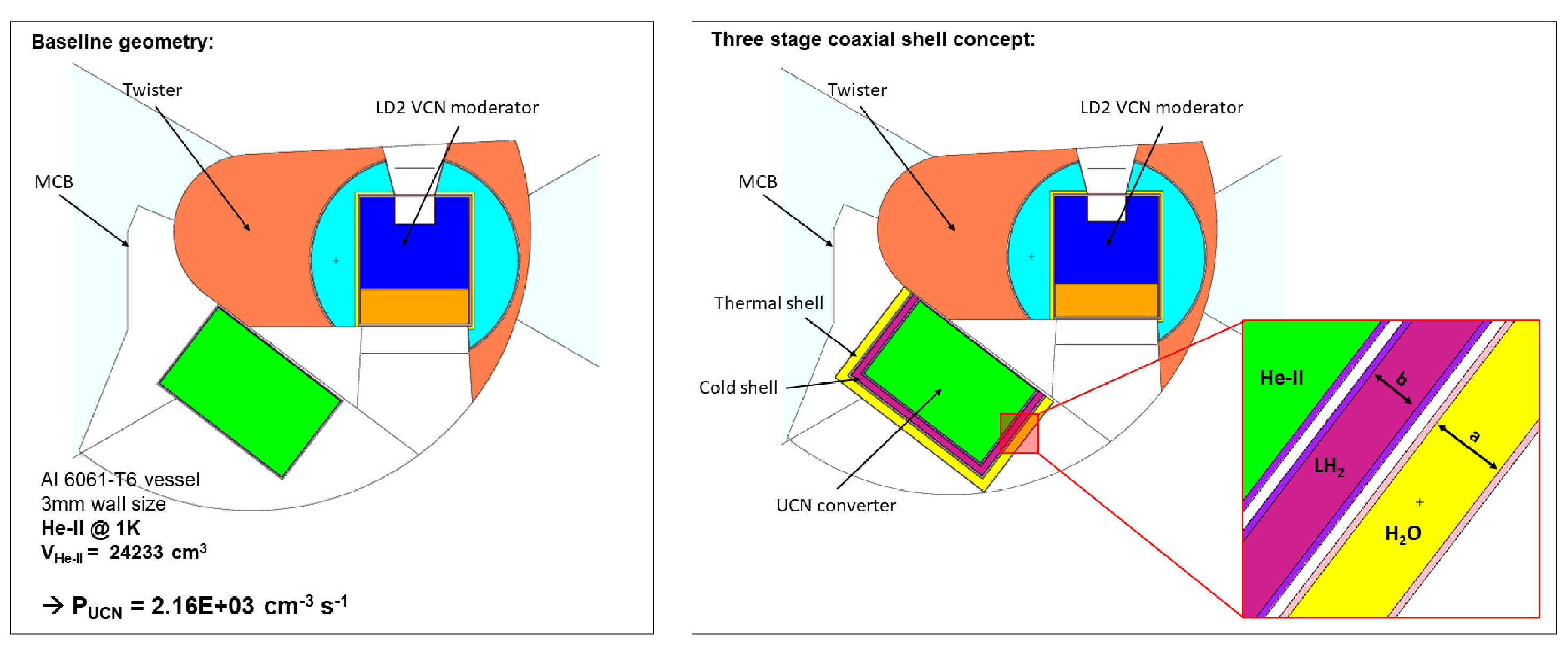}
\caption{MCNP simulation of coaxial-moderator shell design. Horizontal cut through the MCNP simulated geometry. Left: baseline geometry with He-II filled vessel. Right: Coaxial-moderator shell geometry. The moderator shell thickness a \& b has been varied in a parameter study.}
\label{fig:MCB_geometry}
\end{center}
\end{figure}

%\begin{figure}[hbt!]
%\begin{center}
%\includegraphics[width=1\textwidth]{ucn_fig/WP5/2-1.eps}
%\caption{\textcolor{red}{remove this figure} Overview of investigated UCN source implementation positions. \\ Left: horizontal cuts through the target bunker shows the different options for the implementation of the UCN source. Right: Table with the results of the MCNP particle transport simulations. Location 3 is equal to 4 in terms of radiation fields and therefore not explicitly shown.}
%\label{fig:2-1}
%\end{center}
%\end{figure}
 A three stage coaxial shell moderator concept and a channel option to lead cold neutrons from the LD$_\text{2}$ moderator to the UCN source has been studied
 in an attempt to increase the UCN production rate \cite{D4.4}.  In the channel case (\cref{fig:SphericalSD2AndCylindricChannel_Parameters}) the result was a loss in the UCN production rate for every channel diameter, reflector material and wall size. 
 Removing parts of the twister as depicted in \cref{fig:MCB_geometry} leads to a gain in the UCN production rate and is therefore the preferred option.

A coaxial-moderator shell design has been evaluated in a parametric study. The baseline model is shown in \cref{fig:MCB_geometry} and consists of an aluminum T6061-T6 vessel with a constant wall size of \SI{3}{mm} filled with He-II at \SI{1}{K}. The volume of the He-II content is about 24 L. The shell model consists of two different premoderators encapsulating the He-II vessel. The first premoderator, with thickness $a$, is a H$_{2}$O layer at room temperature and is
supposed to thermalize fast neutrons.
Inside of it is an additional
 layer with thickness $b$ of parahydrogen at \SI{20}{K}, to moderate thermal neutrons into a cold neutron regime (8.9\AA) in which the probability of direct conversion in He-II to the desired UCNs is highly increased (see \cref{fig:s_lambda}).

%\begin{figure}[hbt!]
%\begin{center}
%\includegraphics[width=0.7\textwidth]{ucn_fig/WP5/2-3.eps}
%\caption{Cross section curve of neutrons with He-II at \SI{1}{K} \textcolor{red}{This figure is duplicate - see Figure 3!}}
%\label{fig:2-3}
%\end{center}
%\end{figure}

The moderator media are encased by \SI{3}{mm} Al T6061-T6 spaced with a \SI{5}{mm} vacuum gap between the vessels as depicted in \cref{fig:MCB_geometry}. The variation in the media thicknesses varies between $a$: \SIrange{15}{35}{mm} for the thermal and $b$: \SIrange{5}{25}{mm} for the cold moderator. 

%\begin{figure}[hbt!]
%\begin{center}
%\includegraphics[width=0.7\textwidth]{ucn_fig/WP5/2-4.eps}
%\caption{Bubble plot of the coaxial moderator shell design parameter study. The figure inside the bubbles gives the UCN production rate relative to the baseline model for the parameter sets were green indicates gain and red loss.}
%\label{fig:2-4}
%\end{center}
%\end{figure}

%The outcome is shown in the bubble plot \cref{fig:2-4}. The figures in the bubbles indicates the relative gain (green), respectively loss (red) to the baseline model.

It can be observed that the maximum gain of $\approx$2\% is reached for a case of thick thermal and thin cold moderator layers. The result is that the shell concept is not highly performing compared to the baseline model, but can be followed up in further simulations with different geometries and moderator media even if an outstanding gain in performance is unlikely. Therefore, the baseline model is the option that will be considered in the further engineering process.\\
%The at first surprising conclusion that the simple baseline model has already a performance competitive to the other UCN source implementation options will be analyzed in the following. \\ \\
\cref{tab:UCNpos5heatload} shows the expected heat loads on the aluminum structure and on the He-II filling, separated for neutron and gamma radiation. It can be observed that the highest heat load is caused by gamma radiation absorbed in the aluminum structure. 

% insert Tab 2.1!!!%
\begin{table}[tbp]
\begin{center}
\begin{tabular}{ l c c c}
     \toprule
        & Al & He-II & \textbf{$\Sigma$} \\
    \midrule
	\textbf{$\gamma$} & \SI{173}{W} & \SI{84}{W} & \SI{256}{W}  \\
	n & \SI{11}{W} & \SI{61}{W} & \SI{71}{W}  \\

	\bottomrule
 &  &  & \textbf{\SI{328}{W}} \\
 \end{tabular}
    \caption{Heat load on the He-II vessel separated  by material and type of radiation.}
    \label{tab:UCNpos5heatload}
\end{center}
\end{table}

%\end{table}{ l c c c}
%        & \textbf{Al} & \textbf{He-II} & \textbf{$\Sigma$} \\
%    \midrule
%	\textbf{$\gamma$} & \SI{173}{W} & \SI{84}{W} & \SI{256}{W}  \\
%	\textbf{n} & \SI{11}{W} & \SI{61}{W} & \SI{71}{W}  \\
%	 &  &  & \textbf{\SI{328}{W}} \\
%	\bottomrule

%    \caption{Heat load on the He-II vessel separated  by material and type of radiation.}
%	\label{tab:UCNpos5heatload}

The neutron absorption cross section in most materials generally increases towards lower neutron energies. Therefore, the gain in UCN conversion by He-II is overcome by the loss due to absorption of the premoderated neutrons in aluminum structures needed to add moderation stages; simultaneously the heat load in the system is increased. \cref{fig:s_lambda} shows that for lower neutron wavelength then the narrow \SI{8.9}{\AA} band of high interaction probability with He-II, the interaction probability is significantly lower but still present. In this regime the energy transfer from the neutron to the He-II is mediated in several steps by phonon interactions. Due to He-II's extremely low absorption cross section, higher-energy neutrons are either converted to UCNs as desired or pass through He-II without interaction.
The conclusion for the further engineering process is that the He-II vessel of the UCN source should be designed under the premise to minimize the aluminum content to the needed for fulfilling the demands on structural integrity of the nuclear code for pressure vessels while maximizing the encased He-II volume, without additional moderation devices. If in the following engineering process the heat load is above a critical level that can not be handled by available He cryostates, additional gamma filter shells can be considered, which will most likely imply some loss of the UCN performance of the source.\\
It can also be considered to manufacture the vessel from beryllium or from a beryllium\slash aluminum alloy which would considerably reduce the heat load on the vessel. These materials are currently not qualified by the RCC-MRx nuclear code for pressure vessels although the material parameters are promising. Therefore an accreditation for the use of these materials can be taken into account. \\
The mounting position in the moderator cooling block offers the possibility to integrate the UCN source in the existing ESS infrastructure by replacing and rebuilding a minimum of existing components. The moderator cooling block consists of staked and actively cooled stainless steel shielding blocks which can be partially replaced or overworked. Another benefit of the location is the already existing supply infrastructure with cooling water and cold gaseous as well as liquefied He.

In conclusion, the MCB location has, besides a high UCN production rate, different advantages. The UCN source can be implemented by changing and overworking a small number of existing components while most of the existing ESS structure can be left untouched. 
%The guiding and extracting of the UCN can be done in a vertical direction which offers the possibility to combine the UCN guiding with the cooling of the moderator by a He-II based heat pipe concept. 
Furthermore, the location offers existing supply infrastructure with cooling water and helium on different temperature levels.

\subsubsection{Recommended further engineering investigations} \label{ch:2-1-3}
\begin{itemize}
    \item Engineering on UCN source concept:
    \begin{itemize}
        \item Guiding of UCN and interfacing to experiments
        \item Inner wall coating ($^{58}$Ni)
        \item Purification of He-II
    \end{itemize}
\end{itemize}    
\begin{itemize}
    \item Experimental investigations on He-II regarding:
    \begin{itemize}
        \item Phase stability
        \item Heat conductivity
        \item Experimental simulation of ESS conditions based on MCNP simulation results (expected heat load, pulse structure)
    \end{itemize}
\end{itemize}
\begin{itemize}
\item Limits for cross section area for heat transport through He-II
\end{itemize}
%Depending on experimental results: \\ VALENTINA ASK MATTHIAS 
%\begin{itemize}
 %   \item Is the heat load of concept 2 (passive heat pipe) manageable or is the use of filters needed (trade-off between heat load $\leftrightarrow$ neutronic performance)
%\end{itemize}

\subsection{He-II in LBP}
\indent Another option for creating a UCNs source at ESS is based on He-II placed at the LBP, which is described extensively in Section~3.4 of \cite{zaniniworkshop2022}. A He-II based UCN source can benefit from the extended lifetime of the UCNs within the neutron absorption-free medium, allowing for storage and therefore, high density of UCNs. Additionally, this location can result in lower heat loads compared to the other options discussed so far. While the LBP is initially intended for use by the NNBAR experiment, it could potentially be repurposed for UCN production at a later stage.

\begin{figure}[tbh!]
\centering
\includegraphics[width=0.9\textwidth]{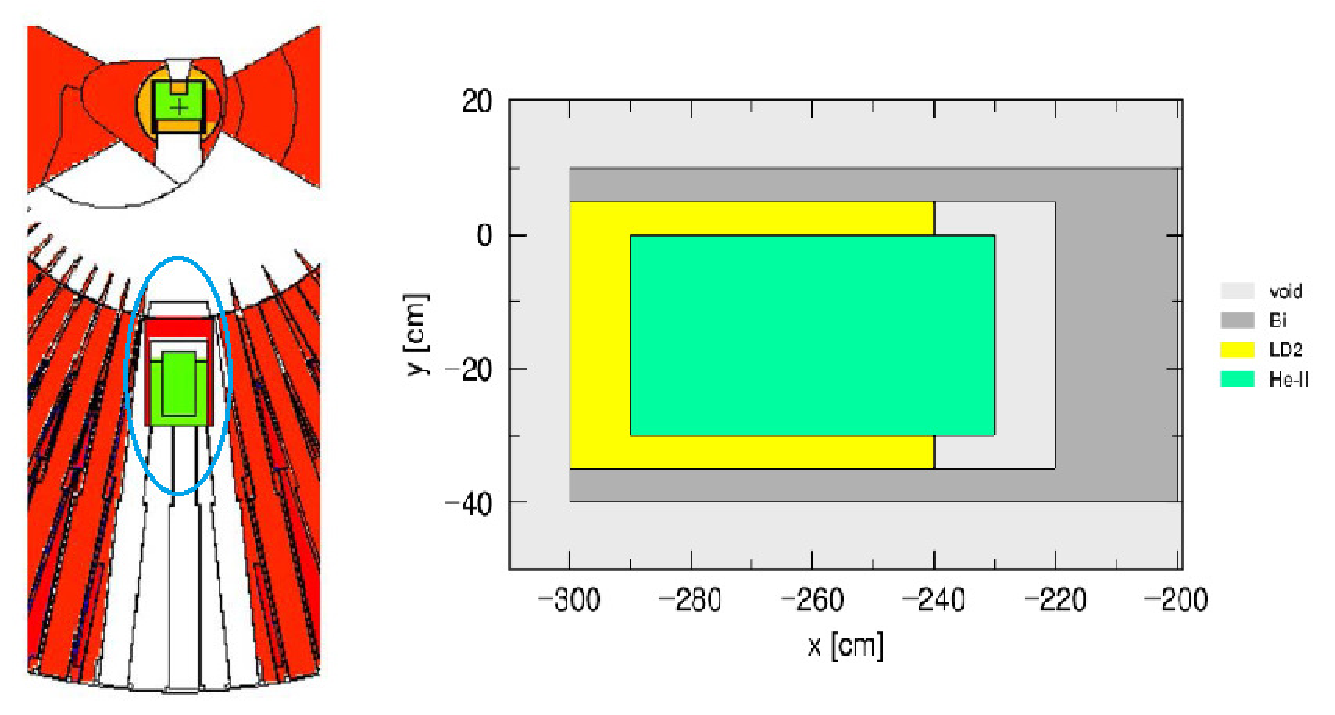}
\caption{\textit{Left}: MCNP geometry showing the He-II source backed by a LD$_\text{2}$ reflector in the large beamport, concept of Serebrov and Lyamkin \cite{serebrovworkshop2022}.
\textit{Right}: The geometry and the materials used in the UCN source located at LBP plotted by PHITS 3.27 \cite{phits}. }\label{fig:LBP_HeII_geo}
\end{figure}

Two different concepts of UCN source that could use the LBP location were identified. In the first option, which is studied in this section, 
the LBP can host a He-II converter for UCN production (represented as number 4 in \cref{fig:oldfig4}), placed inside the monolith, with the tip of the source at a distance above approximately 2 m from the center of the lower moderator. As it is described in \cref{sec:IntroInPileInBeam}, we still consider this option as ``in-pile", even though it is a larger distance from the primary source compared to the options in the twister or in the MCB, discussed in the previous sections. This classification is justified by the fact that at that location, a He-II converter will receive a neutron spectrum with a large component of fast neutrons, unlike a pure in-beam converter which will receive mainly cold neutrons and a smaller fraction of background neutrons. In the second option (location 5 in \cref{fig:oldfig4}, considered as ``in-beam", we still use the LBP, but the source is placed further away, outside the neutron bunker. This option is described in \cref{sec:InBeam}.\\
%The idea of using He-II source as an ``in beam" solution has been implemented in PIK reactor in Russia \cite{serebrovworkshop2022} and in SuperSUN at  ILL \cite{chanel2022concept}.  The UCN density achievable in an ``in-beam" configuration is expected to  be lower than that of an ``in-pile" configuration. This is because of the distance from the primary neutron source. However, as we get closer to the primary source, i.e. creating an ``in-pile" source, the heat-load will increase and therefore a powerful cryogenic equipment is indispensable which can make the ``in-pile" solution costly. Thus, it is worthwhile to investigate the use of UCN in an ``in-beam" configuration at ESS. 
\indent For the in-pile option under study in this section, we followed the proposal by Serebrov and Lyamkin \cite{serebrovworkshop2022}, consisting of a 58-liter vessel (\qtyproduct{60x30x32}{cm}) filled with He-II, surrounded by a reflector made of LD$_2$ and bismuth shielding. The  geometry is depicted in \cref{fig:LBP_HeII_geo}, and it will be referred to as the baseline geometry henceforth. \\
%As mentioned previously, the bismuth filter serves for protection against gamma radiation, while being highly transmissive for 8.9 \AA~neutrons that induce the one-phonon process of UCN production. \\
%The bismuth reflector is for gamma protection and to suppress losses which may occur at the walls of the converter vessel and therefore, decrease the UCN lifetime \cite{teshigawara2019measurement}. \\
\indent \cref{tab:LBP} presents a summary of the UCN production for the baseline geometry \cite{zaniniworkshop2022}. For this study, the He-II thermal scattering library mentioned at \cref{sec:HePUCNcalculation} is used in MCNP. By implementing the scattering function from ESS model 1, \cref{fig:s_lambda} into \cref{eq:He_II_P_UCN}, the UCN production rate density is calculated as 368.5 $\si{n/s/cm^3}$. The total prompt heat-load only on the He-II converter at the specified source location is 8.2 W. To further enhance this design, it is essential to understand how each component of the UCN source impacts UCN production and heat load.

\begin{table}[tbp!]
\caption{Performance of baseline geometry of He-II UCN source in the LBP.}
\centering
\def\arraystretch{1.5}
   \setlength\tabcolsep{0pt}
    \begin{tabular*}{0.95\textwidth}{@{\extracolsep{\fill}}  r c c c c}
 \toprule
 & \makecell{\ce{He-II} Volume \\ $[\si{L}]$}  &  \makecell{$P_\text{UCN}$ \\
 $\left[\si{n/s/cm^3}\right]$}& \makecell{$\dot{N}_\text{UCN}$ \\
 $\left[\si{n/s}\right]$}  &  \makecell{Heat-load\\  $[\si{W}]$}   \\
\midrule
\textbf{He-II in LBP (Baseline)} &      \num{57.6} &  \num{368.5} &  \num{2.1e7} & 8.2 \\

\bottomrule
\end{tabular*}
 \label{tab:LBP}
\end{table}
The main objectives are twofold: first, to maximize the neutron flux within the energy range of 1-10 meV within the He-II converter. Second, to effectively manage the total heat load on the He-II converter. This heat load encompasses contributions from both neutrons and photons on the He-II converter as well as the aluminum shell surrounding the He-II. Additionally, there is a decay heat load resulting from aluminum beta activation, which accounts for about 24 \% of the total heat load originating from neutrons and photons \cite{serebrov_development_2023}. Throughout the optimization process, we maintained the He-II volume constant at the preliminary design value of 57.6 L for comparability. Subsequently, we investigated the impact of each component on the performance of the UCN source.

\subsubsection{Bismuth Filter for Gamma Shielding}
\label{sec:bi_gamma_shielding}
\begin{figure}[tbh!]
\centering
\includegraphics[width=1.0\textwidth]{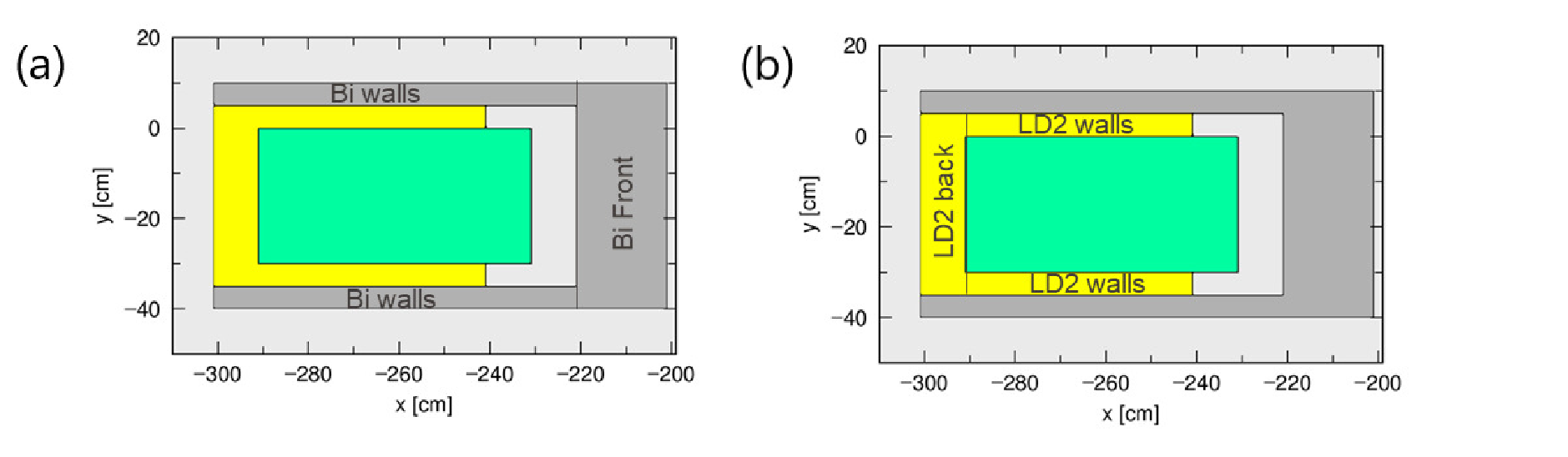}
\caption{Studying the baseline geometry to assess the impact of individual components: (a) Investigation of bismuth shielding by the front section and walls; (b) Examination of the effect of LD$_\text{2}$ by the back section and walls.}
\label{fig:optimization}
\end{figure}

\indent To investigate the impact of bismuth on the resulting UCN density and heat load, we segregated this part of the geometry into the front section and the section encompassing the walls, as depicted in \cref{fig:optimization} (a). The outcomes, outlined in \cref{tab:bismuth-effect}, indicate that the presence of Bi in the walls does not markedly influence the heat load or the UCN production rate density when compared to the front part. Removing Bi from the front portion results in a doubling of UCN gain, albeit at the cost of increasing the heat load by approximately a factor 6.

\begin{table}[tbp!]
\caption{Effect of Bi in the front and in the walls on the relative gain in UCN production rate density and relative heat load compared to the baseline geometry.}
\centering
\def\arraystretch{1.5}
   \setlength\tabcolsep{0pt}
    \begin{tabular*}{0.95\textwidth}{@{\extracolsep{\fill}}  r c c}
 \toprule
 & \makecell{Relative gain \\ $P_\text{UCN}$}  &  \makecell{Relative heat load \\
He-II, n + $\gamma$} \\
\midrule
Baseline &      \num{1} &  \num{1}  \\
Baseline-Absence of Bi-walls &      \num{0.85} &  \num{1.3} \\

Baseline-Absence of Bi-Front &      \num{2.37} &  \num{5.7}\\

Baseline-Absence of total Bismuth &      \num{1.95} &  \num{6.7}\\

\bottomrule
\end{tabular*}
 \label{tab:bismuth-effect}
\end{table}

\begin{figure}[tbh!]
\centering
\includegraphics[width=0.6\textwidth]{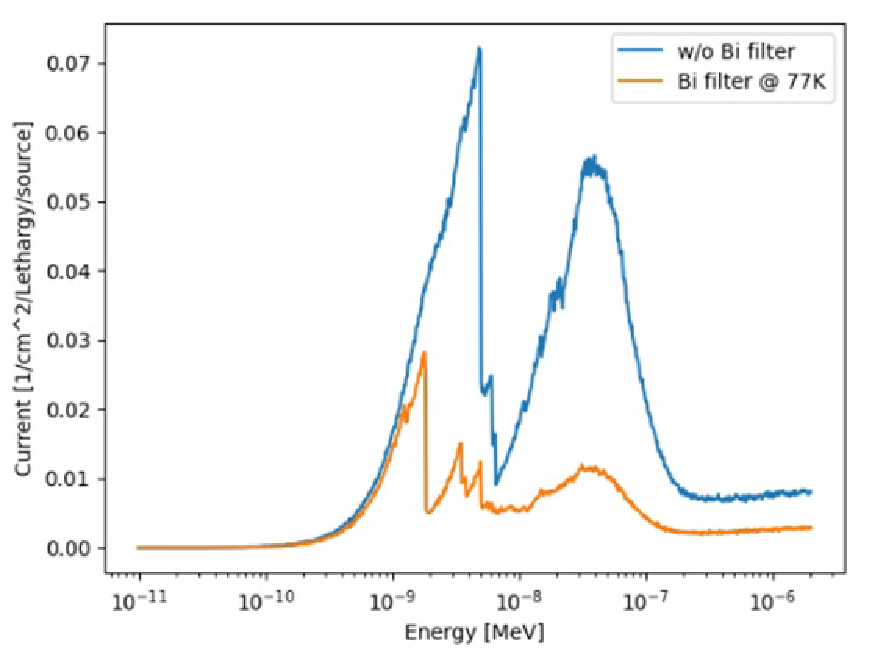}
\caption{The cold source spectrum when there is no bismuth filter and at the presence of bismuth filter at 77 K as a function of energy}
\label{fig:TransmitSpectrumBi}
\end{figure}

The significant impact of removing Bi on UCN gain can be understood by examining the transmission spectrum with and without Bi in the front section, as illustrated in \cref{fig:TransmitSpectrumBi}. This comparison reveals a substantial drop within the 1-10 meV range, which is crucial for UCN production.

\begin{figure}[tbh!]
\centering
\includegraphics[width=1.0\textwidth]{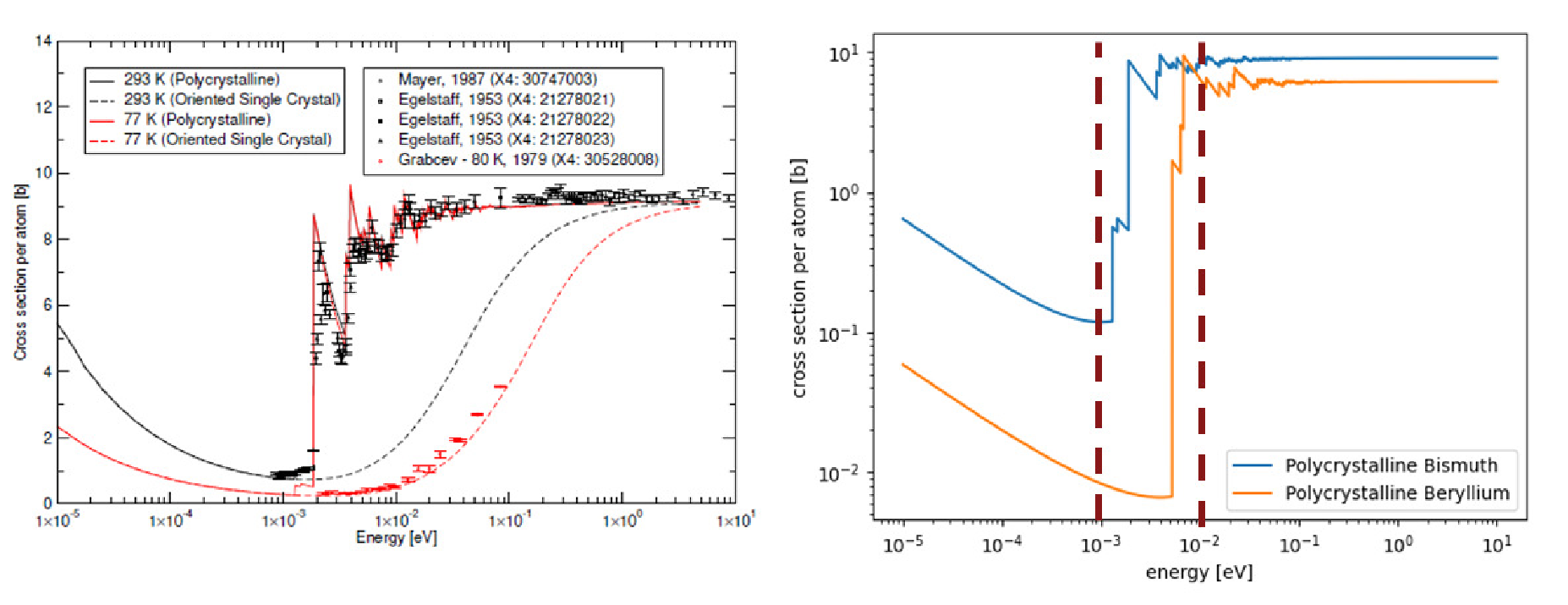}
\caption{Left: Total neutron interaction cross section of neutron for polycrystalline and oriented single crystalline bismuth; Right: Total cross section of neutrons for polycrystalline bismuth and beryllium.}
\label{fig:BiScPcBe}
\end{figure}

This effect arises from the presence of Bragg peaks in the cross-section of polycrystalline bismuth, as shown in \cref{fig:BiScPcBe} (left). To compensate for this drop, we need to identify a material that scatters neutrons less effectively within the 1-10 meV range. Looking again at \cref{fig:BiScPcBe}, oriented single crystal bismuth at 77 K exhibits a minimal scattering cross section within this range. Additionally, beryllium, shown on the right side of the figure, also has a lower scattering cross section.

The results of replacing polycrystalline bismuth in the front section with single crystalline bismuth and beryllium are detailed in \cref{tab:frontFilter}. Single crystalline bismuth leads to a 1.75-fold increase in gain compared to the baseline, with only a 30 \% increase in heat load. On the other hand, beryllium shows lower performance in both UCN gain (1.6-fold increase) and heat load reduction (1.9-fold increase).

\begin{table}[tb]
\caption{Effect of different front filter materials on the UCN production rate density and the heat load on He-II converter. PC: polycrystalline ; SC: single crystalline}
%\begin{adjustbox}{width=0.7\columnwidth,center}
\label{tab:frontFilter}
\centering
\begin{tabular}{ccc}
\toprule
Front Filter & Relative $P_\text{UCN}$ & He-II heat load (n+$\gamma$)\\ 
\midrule
void & 2.3 & 5.2 \\
Bi PC @ 77K & 1  & 1 \\
Bi SC @ 77K & 1.75 & 1.3 \\
Be PC @ 77K & 1.6 & 1.9 \\
\bottomrule 
\end{tabular}
%\end{adjustbox}
\end{table}

\begin{figure}[tbh!]
\centering
\includegraphics[width=1.0\textwidth]{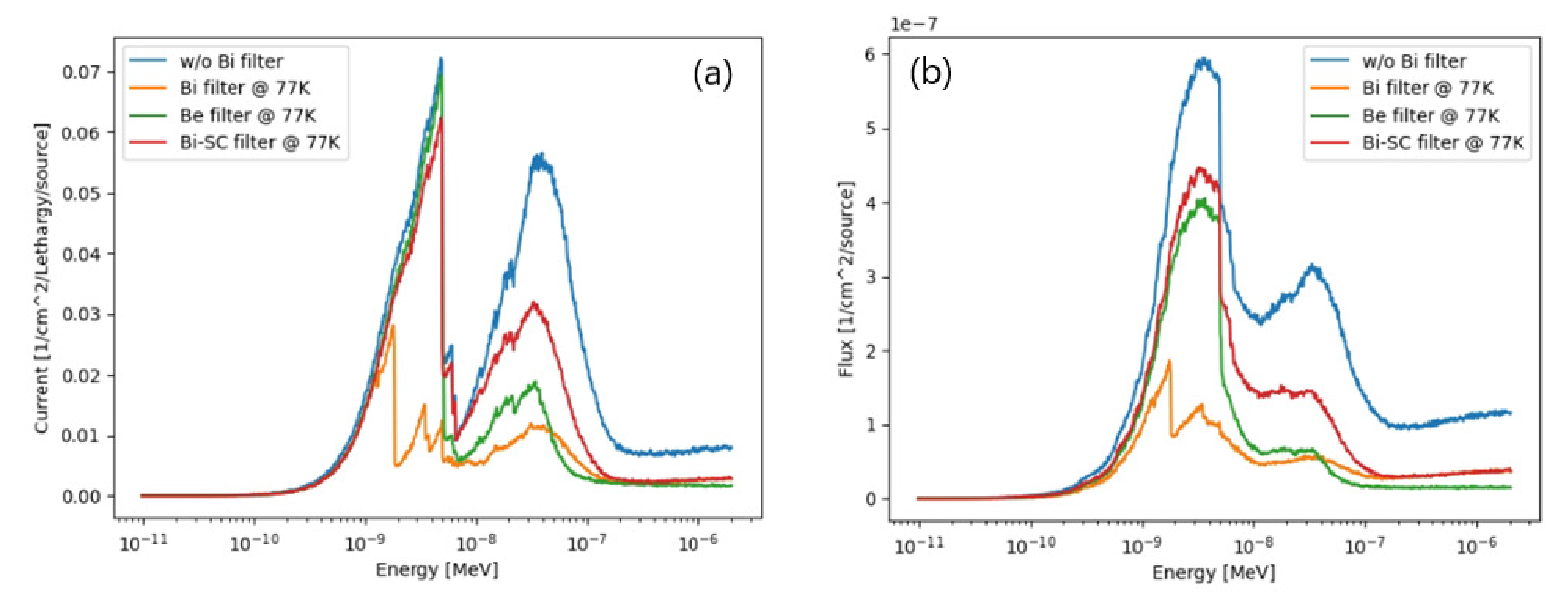}
\caption{(a) Transmittance spectrum of incoming cold neutrons for different materials; (b) Flux of neutrons inside the He-II converter.}
\label{fig:FrontFilterSpec}
\end{figure}

\cref{fig:FrontFilterSpec}(a) displays the transmittance spectrum of various filters, demonstrating the superior transmission of neutrons within the 1-10 meV range for single crystalline bismuth and beryllium. Additionally, \cref{fig:FrontFilterSpec}(b) illustrates the neutron spectrum inside the He-II, revealing a greater flux for single crystalline bismuth and beryllium when compared to polycrystalline bismuth. This observation aligns with the findings presented in the table.

\subsubsection{LD$_\text{2}$ as a CN reflector}

\indent Following the same methodology employed for the bismuth studies, we divided the LD$_\text{2}$ into two parts: the back section and the walls, as depicted in \cref{fig:optimization}(b). The impact of each part of the LD$_\text{2}$ reflector is outlined in \cref{tab:LD2reflector}. In total, the presence of LD$_\text{2}$ leads to a 42 \% gain in UCN production rate density and has a favorable effect on reducing heat load. This is primarily attributed to the implementation of the inverted pre-moderator scheme, allowing cold neutrons a second opportunity for conversion to UCNs upon reflection back into the He-II. The reduction in heat load arises from the fact that a portion of the heat is absorbed by LD$_\text{2}$, thereby reducing the amount absorbed by the He-II converter.

\begin{table}[tb]
\caption{Effect of different parts of LD$_\text{2}$ reflector on the UCN production rate density $P_\text{UCN}$ and total heat load on He-II converter}
%\begin{adjustbox}{width=0.7\columnwidth,center}
\label{tab:LD2reflector}
\centering
\begin{tabular}{ccc}
\toprule
 & Relative gain $P_\text{UCN}$ & He-II heat load (n+$\gamma$)\\ 
\midrule
baseline & 1 & 1 \\
Absence of \ce{LD_2}-walls & 0.65  & 1.2 \\
Absence of \ce{LD_2}-back & 0.87 & 1.01 \\
Absence of \ce{LD_2}-total & 0.58 & 1.21 \\
\bottomrule 
\end{tabular}
%\end{adjustbox}
\end{table}

\begin{figure}[tbh!]
\centering
\includegraphics[width=1.0\textwidth]{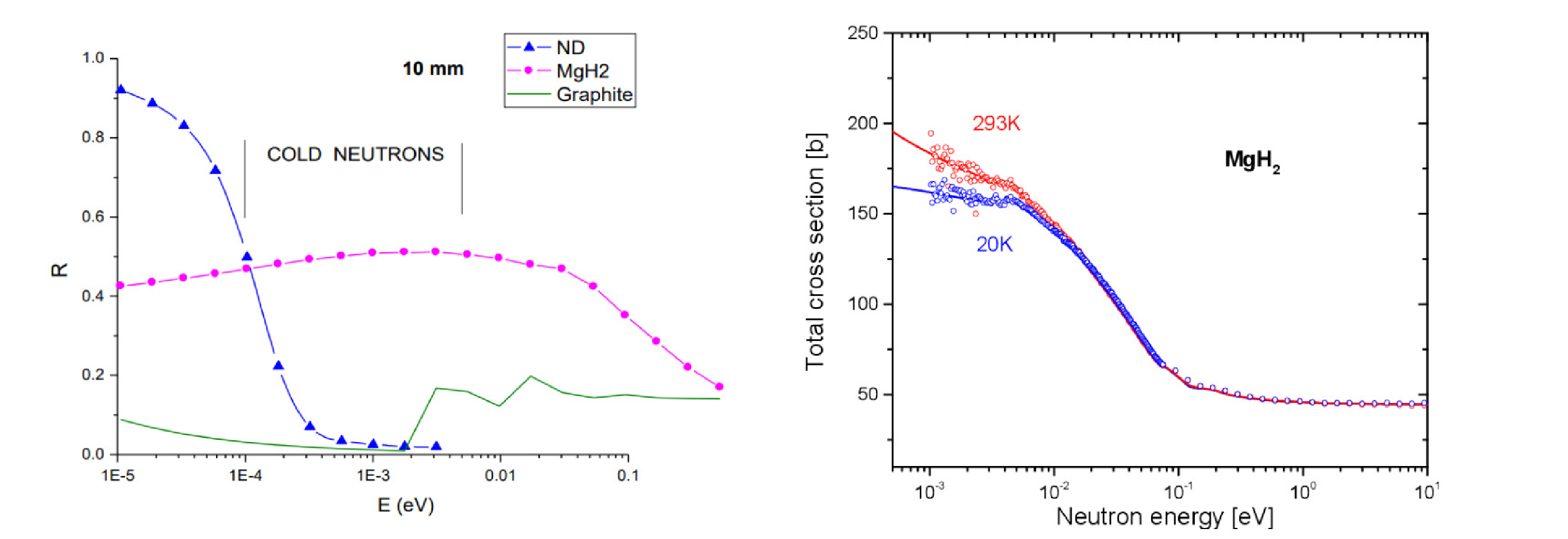}
\caption{Left: Reflectivity of nanodiamonds, graphite and MgH$_\text{2}$ as a function of neutron energy \cite{granada2020studies}; Right: Total neutron interaction cross section of MgH$_\text{2}$ at 20K and 293K \cite{Granada2021}.}
\label{fig:MgH2}
\end{figure}

Apart from using LD$_\text{2}$ as a cold reflector, there are other materials that hold potential as effective reflectors for the cold regime, such as nanodiamonds and MgH$_\text{2}$. \cref{fig:MgH2} displays the reflectivity of these materials, distinctly highlighting that MgH$_\text{2}$ exhibits better reflection compared to the other two, specifically for neutrons within the 1-10 meV range. Additionally, maintaining MgH$_\text{2}$ at 20K results in reduced scattering, as illustrated on the right-hand side of the same figure.

\cref{tab:MgH2} provides a comparison of the relative UCN production rate density in scenarios without bismuth walls and with the substitution of LD$_\text{2}$, resulting in a 39 \% increase in gain. This gain factor could see another 41 percent improvement with the addition of a 1-cm layer of MgH$_\text{2}$ surrounding the He-II converter. However, increasing the temperature of MgH$_\text{2}$ leads to a loss in performance, almost equivalent to the baseline geometry. This is likely attributed to up-scattering events, which diminish the flux of cold neutrons within the He-II medium, and underscores the significance of keeping the source at low temperatures.

\begin{table}[tb]
\caption{Effect of MgH$_\text{2}$ layer on the UCN production rate density $P_\text{UCN}$ in He-II converter}
%\begin{adjustbox}{width=0.7\columnwidth,center}
\label{tab:MgH2}
\centering
\begin{tabular}{cc}
\toprule
Geometry & Relative gain $P_\text{UCN}$ \\ 
\midrule
baseline & 1 \\
Absence of Bi walls & 0.85 \\
Bi walls replaced by LD$_\text{2}$@20K & 1.24 \\
1 cm MgH$_\text{2}$@20K + 9 cm LD$_\text{2}$@20K & 1.65 \\
1 cm MgH$_\text{2}$@293K + 9 cm LD$_\text{2}$@20K & 1.04 \\
\bottomrule 
\end{tabular}
%\end{adjustbox}
\end{table}

\subsubsection{Adding engineering details}
\begin{figure}[tbh!]
\centering
\includegraphics[width=1.0\textwidth]{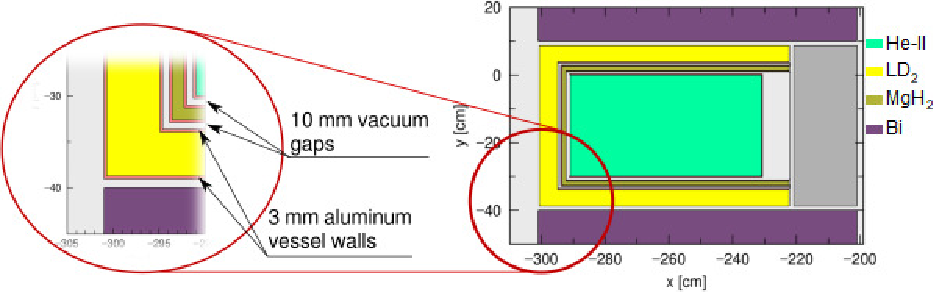}
\caption{Adding engineering details, including aluminum vessel and vacuum gaps to the source design. A close-up view is provided for better clarity. The yellow region }
\label{fig:engineering}
\end{figure}

\indent Until now, our discussion has been based on an idealized design concept. In reality, maintaining He-II at 1K, MgH$_\text{2}$ at 20K, and Bi at 77 K requires proper isolation of each component through the use of vacuum insulation. Furthermore, both He-II and LD$_\text{2}$ are fluids and necessitate the presence of a vessel, along with considerations for the vessel's material, and the inclusion of vacuum gaps within the design. As per the technical designs, a 3 mm aluminum vessel encompassing all components and a 10 mm vacuum gap can adequately fulfill these requirements, as illustrated in \cref{fig:engineering}.

\begin{table}[tb]
\caption{Effect of adding engineering details on the UCN production rate density $P_\text{UCN}$ in He-II converter}
%\begin{adjustbox}{width=1.0\columnwidth,center}
\label{tab:engineering}
\centering
\begin{tabular*}{1.0\textwidth}{cccccc}
\toprule
Geometry & \makecell{Relative gain \\ $P_\text{UCN}$} & \makecell{Heat load [W]\\ He-II (n+$\gamma$)}& \makecell{Heat load [W]\\ Al (n+$\gamma$)}& \makecell{Heat load [W]\\ Al ($\beta$)}&  \makecell{Total \\Heat load [W]}\\ 
\midrule
Baseline (Ideal) & 1 & 8.0 & - & - & 8.0 \\
\hline
Baseline & 0.76 & 9.3 & 3.05 & 2.26 & 14.6 \\
\hline
\makecell{Optimized \\(MgH$_\text{2}$+LD$_\text{2}$+ Bi PC)} & 1.02 & 10.6 & 7.0 & 5.2 & 22.8 \\
\hline
\makecell{Optimized \\(MgH$_\text{2}$+LD$_\text{2}$+ Bi SC)} & 1.6 & 13.5 & 10.7 & 8.0 & 32.2 \\
\hline
\makecell{Optimized \\(MgH$_\text{2}$+LD$_\text{2}$+ Be)} & 1.52 & 19.5 & 17.8 & 13.2 & 50.5 \\
\bottomrule 
\end{tabular*}
%\end{adjustbox}
\end{table}

\subsubsection{Outlook}
According to the calculations, employing single crystalline bismuth at 77 K and MgH$_\text{2}$ at 20 K presents an opportunity to enhance UCN gain while still maintaining the heat load below 100 W. While this design did not account for the UCN extraction scheme, it is worth noting that being situated within a large beam port provides enough space for implementing extraction channels. Additionally, considering the specific instrument and UCN experiment, there may be room for optimization in the volume of He-II. \cref{tab:LBP-final} shows the final values for the optimized geometry of the UCN source at LBP considering the engineering details.

\begin{table}[tbp!]
\caption{Performance of optimized geometry of He-II UCN source in the LBP.}
\centering
\def\arraystretch{1.5}
   \setlength\tabcolsep{0pt}
    \begin{tabular*}{0.95\textwidth}{@{\extracolsep{\fill}}  r c c c c}
 \toprule
 & \makecell{\ce{He-II} Volume \\ $[\si{Liters}]$}  &  \makecell{$P_\text{UCN}$ \\
 $\left[\si{n/s/cm^3}\right]$}& \makecell{$\dot{N}_\text{UCN}$ \\
 $\left[\si{n/s}\right]$}  &  \makecell{Heat-load\\  $[\si{W}]$}   \\
\midrule
\textbf{He-II in LBP (Final design)} &      \num{57.6} &  \num{590} &  \num{3.4e7} & 32.2 \\

\bottomrule
\end{tabular*}
 \label{tab:LBP-final}
\end{table}

%\section{He-II in MCB (Mathias)} %\label{ch:He-IIinMCB}

\subsection{In-beam UCN source}
\label{sec:InBeam}

In this section the design and possible performance of an in-beam UCN source for the ESS are described. 
More details can be found in a recent publication  \cite{zimmerworkshop2022}). 

The central component of in-beam UCN source is a neutron delivery system 
consisting of nested-mirror optics (NMO). The bright neutron emission surface of the large liquid-deuterium moderator has to be imaged onto the UCN source, a remotely located superfluid-helium converter.
% reference to production section
This approach allows the converter to be placed far away from the high-radiation area in the ESS shielding bunker. The focused beam illuminates the converter under a solid angle that may significantly exceed the solid angle transportable by a straight guide equipped with the most advanced supermirror coating. In this way a large gain in flux density and hence UCN production rate density is achieved.

%Neutron transport simulations performed with the design suggest that such a UCN source would offer saturated UCN densities at the top of the range of other current projects

%The paper represents an important conceptual step forward, towards the realization of an in-beam UCN source at the ESS.

\subsubsection{Basic principles of an in-beam UCN source}

The three main components to an in-beam UCN source are: a cold neutron moderator, a neutron optical delivery system
(NODS), and a converter vessel, filled with superfluid Helium (He-II). 
The problem of optimizing the UCN production according
to a particular set of desired characteristics requires a careful consideration of not only the many parameters describing these three components, but also the interplay between them. 

%The present section is intended to serve as a
% point of entry into this problem, and will focus on a simplified picture of the in-beam source in order to discuss the
% concepts that are key to UCN production. Here, we will mainly consider the case in which one desires a high density of
% UCN in the source. The discussion is ordered proceeding in the upstream direction, that is, starting from the converter
% surface upon which cold neutrons are incident and working back toward the moderator. 

\begin{figure}[htbp!]       
\centering
\includegraphics[width=\textwidth]{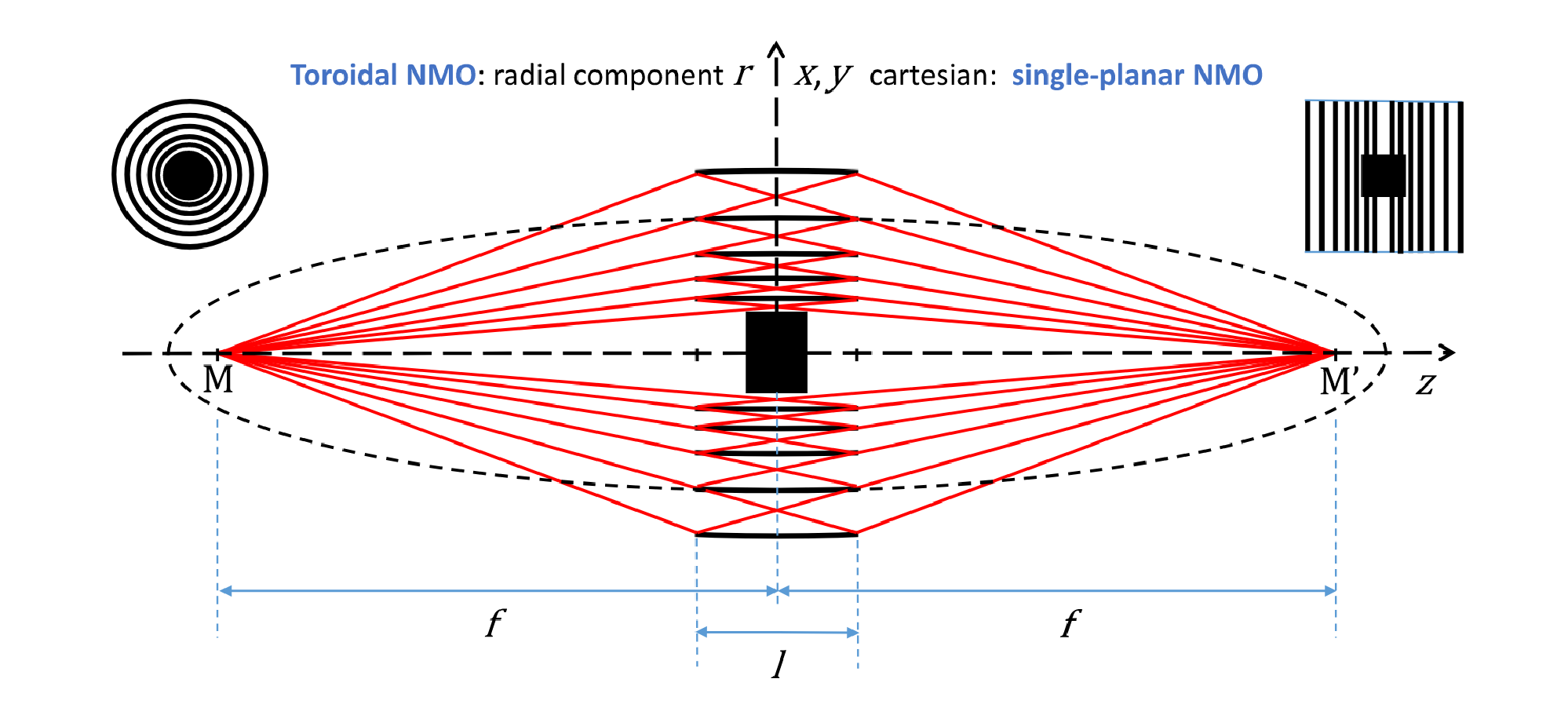} 
\caption{Schematic of an elliptic nested mirror optical (NMO) system. The foci M and M' are separated by the distance $2f$ and are common for a set of ellipses. One of them is indicated by the dashed line. The surfaces of the mirrors are formed by truncating the ellipses to a common length $l$. 
The two types of symmetry about the optical axis of the NMO are indicated in the upper corners; the views show the NMO surfaces projected onto a surface normal to the optical axis $z$. Image from \cite{zimmerworkshop2022}.}
\label{fig_inbeam_principle}
\end{figure}

%idea and mathematics of nested optics generation 
A possible architecture for the transport system for neutrons diverging from a source to a target region is an elliptical guide, with its focal points coinciding with the center of the source and the detector, respectively \cite{zimmer2016_MultimirrorImaging,zimmer2019_ImagingNestedmirror}.
%An ellipse has the property to reflect a beam that emanates from one of its focal points directly to the other one. 
The layers of several guides can then be nested to build up a spatial tight component (see \cref{fig_inbeam_principle}) to form a NMO system. 
These devices accept neutrons within a wide, geometrically defined angular range. 
Two main implementations of NMO can be considered, either with a toroidal or with a planar symmetry (see inlays in \cref{fig_inbeam_principle}). In the former case, the device is rotationally symmetric about the optical axis; each mirror surface is a section of an ellipsoid of revolution, and a single reflection transports a neutron from source to target. In the planar case, the mirrors have a local translational symmetry in a direction transverse to the optical axis. Refocusing the beam in both transverse dimensions then requires a combination of two planar NMO rotated by $\SI{90}{\degree}$ about the optical axis. Such a system images the beam by two reflections, one for each transverse dimension.

\begin{figure}[htbp!]     
\centering
\includegraphics[width=\textwidth]{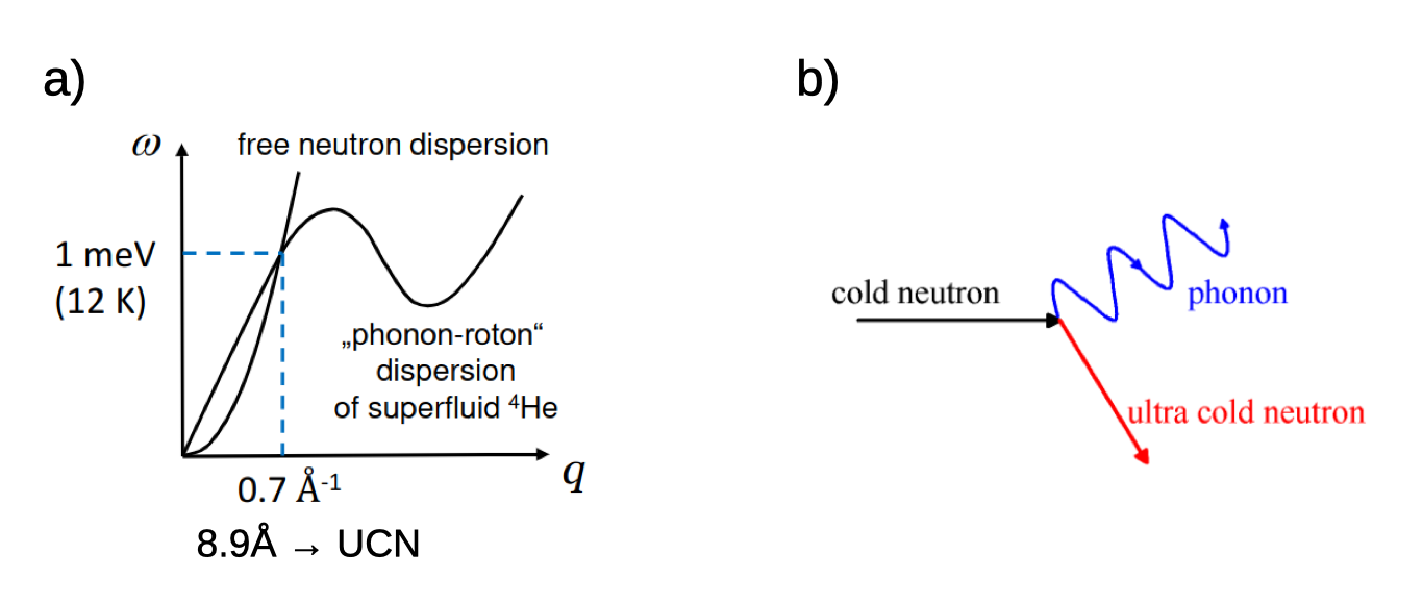} 
\caption{a) Dispersion curves of free neutrons and phonons in superfluid helium $^4 \rm He$. b) Elastic scattering process. A cold neutron excites a phonon and becomes ultracold}
\label{fig_inbeam_process}
\end{figure}

%Superthermal
The in-beam UCN source for this report is proposed to be of the ``superthermal” type.
The mechanism was proposed 1975 by Golub and Pendlebury \cite{golub1975}
and in contrast to UCN production by neutron moderation, cold neutrons are ``converted” to UCN by imparting nearly their entire kinetic energy to elementary excitations -- most commonly phonons -- of the source medium in single scattering events. 
In \cref{fig_inbeam_process} the process for 
superfluid $^{4}\rm He$ (He-II) is depicted. 
The dispersion relations of the free neutron and the phonon-roton of the superfluid helium cross at an energy $E^{\ast }{\approx}1.04~\text{meV}$, corresponding to a wavelength  
$\lambda ^{\ast}{\approx}8.9~\text{Å}$. A cold neutron with in an energy resp. wavelength centered around this value can lose almost all its energy to the converter medium and become ultracold by exciting a phonon.   
Besides this dominant “single-phonon” process a 
second contribution to UCN production accounts for inelastic scattering of  cold
neutrons that involves several phonons. Such “multi-phonon” processes occur over a wider range of the
neutron spectrum, but, for typical cold beams delivered by neutron guides, contribute less than  $30 \%$ \ to the total UCN production \cite{korobkina2002, schmidt-wellenburg2015}.
The simulations and estimates for the studies in this section did only take the single-phonon contribution into consideration.
The production rate for UCNs up to a maximum energy  $V_{\mathrm{c}}=233(2)$~neV, defined by the wall potential of the converter vessel is \cite{Schmidt_2009}:
\begin{equation}
P_{\mathrm{I}}\left(V_{\mathrm{c}}\right)=\left.4.97(38) \cdot 10^{-8} \frac{\rm \AA}{\mathrm{cm}} \frac{\mathrm{d} \phi}{\mathrm{d} \lambda}\right|_{\lambda^*}
\label{inbeam_prod_rate}
\end{equation}
where $\frac{\mathrm{d} \phi}{\mathrm{d} \lambda} |_{\lambda^*}$ is the flux at the converter evaluated at the conversion wavelength $\lambda ^{\ast}$ of $8.9~\text{Å}$.

This makes clear that for such an in-beam UCN device the goal becomes to maximize the average of 
$\frac{\mathrm{d} \phi}{\mathrm{d} \lambda} |_{\lambda^*}$  
at the converter.

\subsubsection{In-beam UCN source implementation at the ESS}
\begin{figure}[htbp!]     
\centering
\includegraphics[width=0.8\textwidth]{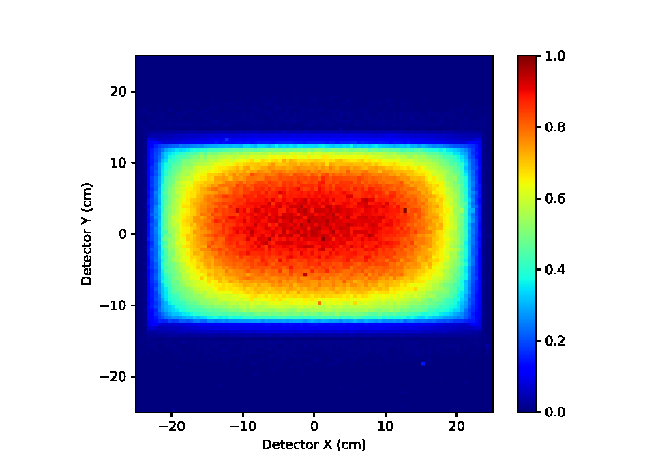} 
\caption{
Depiction of a (simulated) intensity map on the ESS \ce{LD_2} moderator surface for neutrons with wavelengths around $8.9\text{Å}$. The data represents neutrons detected at a distance of 4 meters from the emission surface, passing through a mid-point "pinhole" between the detector and moderator and portray the moderator surface as seen through the LBP. Note that the image had to be inverted (up to down and left to right) to compensate for the "camera obscura" effect. The peak intensity is observed closer to the upper edge  positioned below the target wheel.
}
\label{fig_inbeam_Brilliance}
\end{figure}

\begin{figure}[htbp!] 
\begin{center}
 \includegraphics[width=\textwidth]
{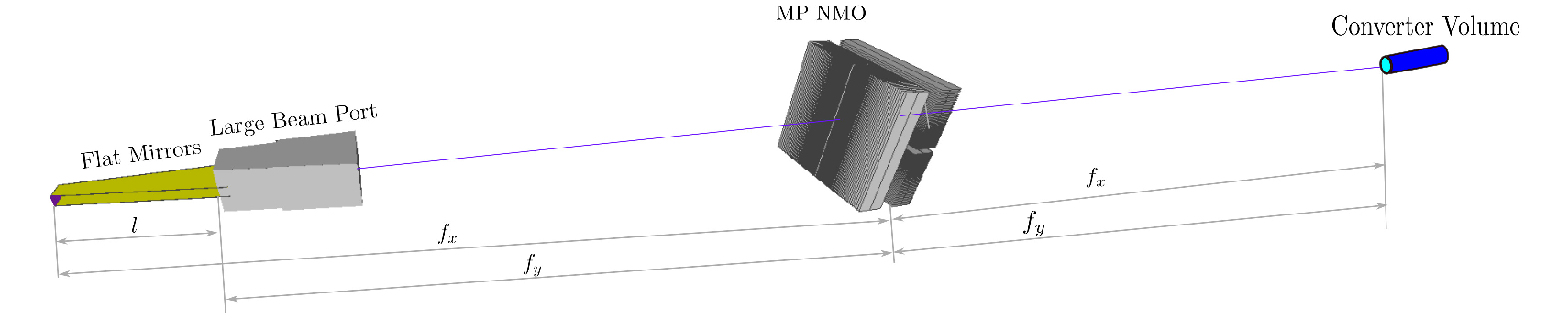} 
\caption{ Schematic of the proposed Neutron Optical Delivery System for the Large Beam Port at the ESS. 
The parameters quoted below have been used for the McStas simulations.
The distance between the moderator, at left, and the converter entrance, at right, is  $\SI{35}{m}$. Neutrons are emitted at the \ce{LD_2} moderator surface
from an area  $w_{\mathit{mod}}\times h_{\mathit{mod}}$, where the physical height $h_{\mathit{mod}}=\SI{24}{cm}$.
and width $w_{\mathit{mod}}=\SI{40}{cm}$, respectively. Neutrons are initially guided in the
vertical direction by two parallel, flat mirrors that end at the LBP at $\SI{2.67}{m}$ The beam is imaged onto the entry of the converter by two planar elliptic NMO that consist of mirrors of lengths  $L_{\mathit{mir}}=\SI{0.5}{m}$. The semi-minor axis of the outermost mirror of both devices is  $b_{0,x}=b_{0,y}=\SI{1.85}{m}$.
The vertically focusing NMO has a focal length of  $f_y=\SI{16.165}{m}$ \ and is placed halfway between converter and 
the end of the flat mirrors, whereas the horizontally focusing NMO ($f_x=\SI{17.5}{m}$) is centred between 
the converter  and the moderator. All mirrors of the NODS are coated with 
the same broadband  $m=6$ supermirror. 
%The mirrors of the NMO are deposited on  $0.5\mathit{mm}$ \ thick
%silicon substrates, 
%
%%and the simulation accounts for neutron refraction and absorption by these substrates. 
%A beam stop
%of area  %$w_b^2=0.25m^2$ \ placed in front of the first %NMO blocks the direct view of moderator at the converter. 
}
\label{fig_inbeam_NDOS}
\end{center}
\end{figure}

The concepts pointed out in the previous section
shall now be applied to an implementation at the ESS in order to estimate the UCN densities that may be possible using NMO to transport cold
neutrons to a He-II converter vessel. 
The ideal location for such a UCN source would make use 
of the LBP, through which moderators in the vicinity of the spallation target wheel can be viewed with large solid angle. 
The emission parameters of the \ce{LD_2} moderator are
\begin{equation}
A_{LD2}=40_x\times 24_y cm^2 \ \ \ \ and \ \ \  b_{LD2}^{\ast}=3.4\times
10^{11}s^{-1} {cm}^{-2} {sr}^{-1} \ \ \ \ {at} \ \ \SI{5}{MW},
\end{equation}
with $b_{LD2}^{\ast}$ being the the mean brilliance at $\SI{8.9}{\angstrom}$ over the moderator surface $A_{LD2}$.
In \cref{fig_inbeam_Brilliance} a simulated intensity map of neutrons with wavelengths near $\SI{8.9}{\angstrom}$ emitted by the surface of
\ce{LD_2} moderator is shown. 
%The large area combined with and efficient transport of the cold neutrons leaving the LD2 moderator to the converter
%surface would result in a large total UCN production rate in comparision to a smaller but brighter moderator.  
%
An efficient extraction of the divergent beam  
at  $8.9~\text{Å}$ \ from the \ce{LD_2} moderator by an elliptic NMO
requires covering a large solid angle. 
%Although this is possible through the LBP, the rather “anisotropic” access geometry there
%makes this task less than straightforward. 
%We note, however, that no horizontally reflecting mirrors close to the
%moderator are necessary, keeping the view of the moderator surface at neighboring beamlines unobstructed and thus
%available for other purposes. Let us first consider the horizontal direction. The maximal angle  $\theta _x$ \ at which
%an NMO can view a surface element within a central region of the moderator depends on the width  $w_{\mathit{mod}}$
%\ of this region. Taking  $w_{\mathit{mod}}=22\mathit{cm}$, for example, a choice that is motivated further below,
%would correspond to  $\theta _x=5.6{}^{\circ}$, and would require an  $m=6.4$ \ coating on the outermost NMO mirrors. 
%
A double-planar elliptic NMO would be best adapted to image the beam, whose horizontal and vertical extents are defined at the moderator surface and guide end, respectively, 
onto the He-II converter. 
In the vertical direction, access is limited from the bottom by a  $\SI{2}{\meter}$ \ long horizontal shielding plate, which, is flat rather than tapered and cuts into the field of view; see Figure 5 in \cite{zaniniworkshop2022}.  % There was also a reference to "the lower left of \cref{ch:2} which is missing"
In order to minimize losses of neutrons emitted downward, the bottom plate should be covered by a mirror starting close to
the moderator. If one then places a second mirror, parallel to this, one obtains a vertical, “one-dimensional” guide of the height of the moderator and its length extending to the LBP (see \cref{fig_inbeam_NDOS} for a schematic). 
The planar elliptic NMO with horizontal mirrors can then extract
neutrons from the “virtual source” defined by the end of this guide. It would have a different focal length than its vertical counterpart.
The He-II converter would be installed at a distance of $\SI{35}{m}$. 
 
% Using the neutron transport efficiency of the NODS shown in Figure~5, estimated from the simulations shown in Figure~6 to be
%  $\epsilon _{\mathit{tr}}=40\mathit{}$, we can finally give estimates for UCN production induced by the single-phonon
% process in a He-II converter installed at a distance of  $35m$ \ from the LD2 moderator. The production rate density in

%To minimize the fraction of neutrons transported
%by multiple reflections in the guide,  $h_g$ \ should be maximized, which is also in line with the aim of maximizing
%the viewable portion of the moderator surface. Choosing  $h_g=22\mathit{cm}$, which is slightly less than the height of
%the moderator emission surface ( $24\mathit{cm}$), would permit full vertical illumination of the guide when there is a
%gap  $d_g$ \ of several centimeters, likely necessary for technical reasons, between its entrance and the moderator
%surface.
%

% result mcstas simulations and flux estimates

\begin{figure}[htbp!]     
\centering
\includegraphics[width=\textwidth]{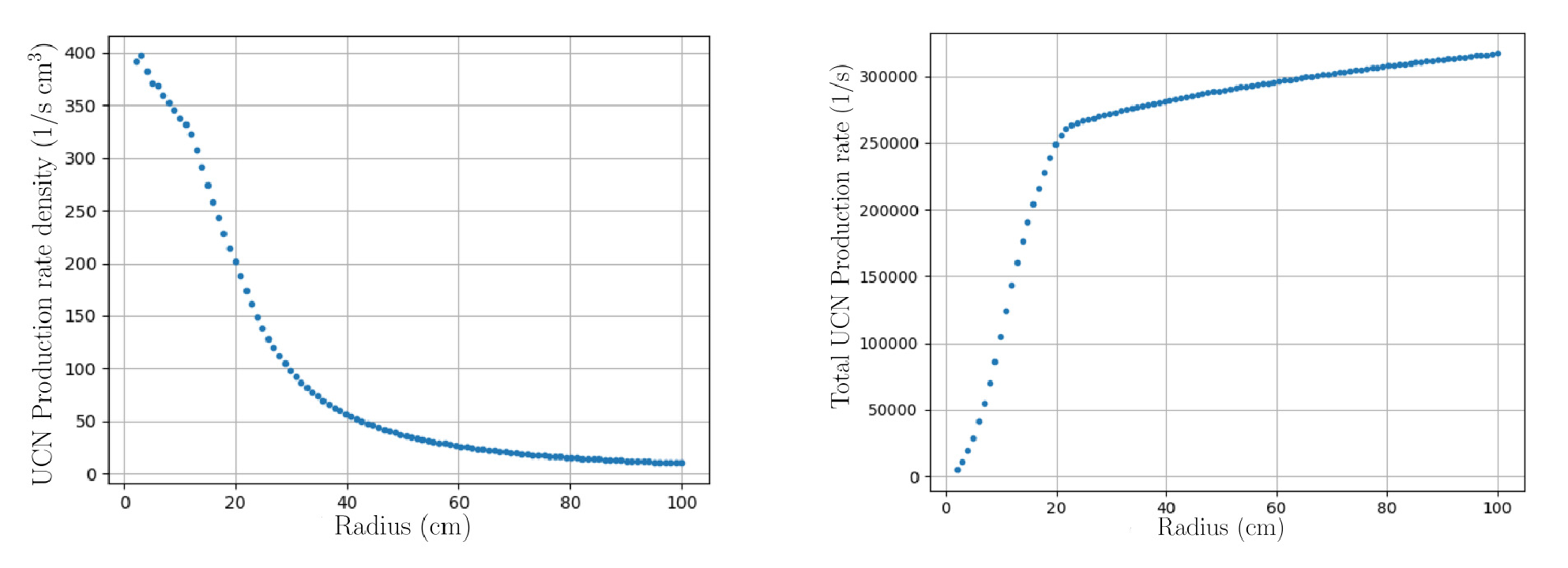}
\caption{Relative (left) and  total (right) production rates at the UCN converter volume as a function of the radius of the converter.}
\label{fig_inbeam_P_rates}
\end{figure}

To determine its performance, the proposed system has been simulated in McStas. The NMO have been realized using the library developed for the NNBAR reflector (see Section 7 of the HighNESS Conceptual Design Report Volume II ). The recorded flux at the converter area is converted to a production rate using \cref{inbeam_prod_rate}.
In \cref{fig_inbeam_P_rates} the plots of the relative and total productions rates at the converter and the total UCN production rate for converter volumes of different radii are shown.   
For a converter with diameter $\SI{22}{cm}$ and taking into account losses due to the thickness of the NMO layers of $\SI{22}{\percent}$ we obtain a production rate of:

\begin{equation}
p_{\mathrm{I}} = \SI{269}{\per\second\per\cubic\centi\meter} 
\label{inbeam_P_result}
\end{equation}
Assuming the UCN storage time constant of the converter to be  $\tau =300s$, the corresponding saturated UCN density will be (using $\rho = \tau \times p_{\mathrm{I}}$):

\begin{equation}
\rho _{\mathit{sat}} = \SI{8.06e4}{\per\cubic\centi\meter} 
\label{inbeam_density_result}
\end{equation}
 %$\rho _{\mathit{sat}}=8.06 \times 10^4\mathit{cm}^{-3}$\ \ \ \ \ \ \ \ (19)
%
in a short converter. 

It can be concluded that the parameters of the UCN in-beam type source described in this section propose a very attractive
option since it offers a saturated UCN density at the top of the range of other current projects, which rely on UCN production in vicinity of a strong primary cold neutron source. 
This kind of UCN source is complementary to in-pile type proposals, which generally possess a much higher UCN production rate, but have to overcome cooling and extraction difficulties. It's also important to consider the practical advantages of having the converter located remotely from the primary source. By being far removed from strong radiation fields, the required cooling power is significantly reduced, and the UCN source is easily accessible for troubleshooting. Moreover, UCN transportation is limited to short distances, and nuclear licensing procedures are less likely to pose direct obstacles to the project.

%VALENTINA COMMENTED 
%It should be mentioned that the implementation of a vertical guide close to the ESS spallation target might be complicated by the high flux of fast neutrons there.
%Even for supermirrors on metallic substrates, developed for in-pile guide sections, sufficientradiation hardness still needs to be experimentally demonstrated. If such a guide is not feasible, beam extraction would have to be accomplished solely by NMO, in which case, the aforementioned geometrical restrictions and partial obstruction of the moderator view would reduce the possible UCN yield.

% Maximal UCN production in the converter would be obtained by matching its entrance shape to that of the rectangular beam
% provided by the NODS, which is shown in Figure~6 with a circular converter entrance. Placing an  $m=6.4$ \ supermirror
% neutron guide at the converter walls would then transport the provided neutron beam through an elongated source, within
% which the volume-averaged UCN production density would be decreased by imperfect reflectivity and by transmission
% losses, Eq. 12. In a source with a cylindrical supermirror guide, the transported neutron flux would be reduced by a
% factor  ${\approx}\pi /4$, cf. Eq. 14. For such a situation, with a converter having diameter  $22\mathit{cm}$ \ and
% length  $3m$, and assuming a loss of UCN production on the order of  $20$ \ due to imperfect mirror reflectivity, the
% total UCN production rate and total saturated UCN number  $N_{\mathit{sat}}$ \ in the converter would be

\subsection{Comparison of the UCN sources }

In \cref{table:summarytable} the performances of the different concepts are compared.
Overall, we have analyzed several solutions with SD$_\text{2}$ in the twister and in the MCB, and different solutions using He-II in the twister, MCB, LBP and in-beam with a NMO system.
Some general conclusions are:

\vspace{0.3cm}

\begin{itemize}
    \item[-] $P_\text{UCN}$ is higher in SD$_\text{2}$ than in He-II, in agreement with previous studies. The most direct comparison can be done for location 2, where we tested both SD$_2$ and He-II sources.

    \item[-] High $\dot{N}_\text{UCN}$  values can be reached by all the  concepts. For the case of He-II, higher values are reached thanks to the larger volumes of the converters.
    
    \item[-] Heat-loads are much lower for the He-II placed further away from the spallation target, compared to the other options. This is a clear advantage for a solution based on a He-II source placed at least 2\,m away from the spallation target, or in-beam outside the ESS neutron bunker.
    
    \item[-] In-source $\rho_\text{UCN}$ values can be obtained by multiplying $P_\text{UCN}$ by the UCN lifetime in the medium. As previously discussed, this is about 40 ms in SD$_\text{2}$, while it is of hundreds of seconds for He-II. However, even the shorter lifetime of UCNs in SD$_\text{2}$ allows for a large release fraction (of roughly 50\%) for a thin SD$_\text{2}$ converter of 2\,cm or less thickness \cite{brys2007extraction}. A comparison of in-source $\rho_\text{UCN}$ between He-II and SD$_\text{2}$ is not necessarily conclusive because the two sources are typically used in different ways, as discussed in \cref{sec:production}.

\end{itemize}

\begin{table}[]
\caption{Summary table with the performance of different UCN source concepts. For details, see discussions in the corresponding sections. The source locations are the ones from \cref{fig:oldfig4}. The beam values from \cite{D4.3} have been corrected with a 10 \% reduction to account for differences in the MCNP models used.}
\begin{center}
\begin{tabular}{ccccc}
\toprule
  Option & Volume  & $P_\text{UCN}$   & $\dot{N}_\text{UCN}$   & Heat\\ 
   &  [L] &   [cm$^\text{-3}$ s$^\text{-1}$] &  [s$^\text{-1}$]  & [Watt] \\ \hline
\multicolumn{5}{c}{SD$_\text{2}$ thin slab in twister - location 1} \\  
\cref{fig:baseline_geom} & 1.81 & 3.1 $\times$ 10$^5$ & 5.6 $\times$ 10$^8$ &  760 \\
\cref{fig:baseline_opt_geom} & 1.75 & 7.7 $\times$ 10$^5$ & 1.4 $\times$ 10$^9$ &  2910 \\ 
\cref{fig:REH_geom} & 0.38 & 1.3 $\times$ 10$^6$ & 5.0$\times$ 10$^8$ &  560 \\
\cref {fig:3O_BigCyl_sub} & 0.13 & 1.7 $\times$ 10$^6$ & 2.2$\times$ 10$^8$ &  520 \\ \hline
\multicolumn{5}{c}{Full He-II  in twister - location 1} \\
\cref{fig:lHe_inpile_full_geom}& 21.9 &\num{2.58e5} & \num{5.65e9} & 9933\\
\hline
\multicolumn{5}{c}{Thin He-II in twister w/extraction - location 1} \\
\cref{fig:lHe_inpile_extract_geom}& 1.25 &\num{6.3e4} & \num{7.8e7} & 113\\
\hline
\multicolumn{5}{c}{Thin SD2 + He-II in twister w/extraction - location 1} \\
\cref{fig:lHe_SD2_hybrid_inpile_extract_geom} & 1.25 & -- & \num{1.34e8} & 109\\
(\ce{SD_2} Only) & 0.06 &\num{1.22e6} & \num{7.7e7} & 10\\
(\ce{He-II} Only) & 1.19 &\num{5.14e4} & \num{6.1e7} & 99\\
\hline
\multicolumn{5}{c}{Full SD$_\text{2}$ in twister - location 1} \\  
\cref{fig:fullSD2} & 48.2 & \num{6.56e+05} & \num{1.32e9} & 39886 \\ \hline
\multicolumn{5}{c}{SD$_\text{2}$ thin slab in MCB - location 2} \\ 
\cref{fig:SD2ConverterInMCB_Thin_CutByHorizontalPlanel} & \num{0.91}  & \num{3.8e4} & \num{3.4e7} &  159 \\ \hline
\multicolumn{5}{c}{He-II in MCB - location 2} \\ 
\cref{fig:MCB_geometry} & 24.3 & 2160 & \num{5.23e7} &  328 \\ \hline
\multicolumn{5}{c}{He-II in MCB + 77\,K Bi Shield - location 2} \\ 
\cref{fig:MCB_geometry} & 24.3 & 1435 & \num{3.49e7} &  153 \\ \hline
\multicolumn{5}{c}{He-II in MCB + 293.6\,K Bi Shield - location 2} \\ 
\cref{fig:MCB_geometry} & 24.3 & 1150 & \num{2.79e7} & 145 \\ \hline
\multicolumn{5}{c}{He-II in LBP - location 4} \\ 
\cref{fig:LBP_HeII_geo}& 58 & 590 & \num{3.4e7} &  32 \\ \hline
\multicolumn{5}{c}{He-II in beam - location 5} \\ 
\cref{fig_inbeam_NDOS}  &114 & 234 & \num{1.53e7} & \\ \hline
\bottomrule
\end{tabular}
\label{table:summarytable}
\end{center}
\end{table}

\section{Integration of the CN, VCN and UCN sources }
\label{sourceintegretion}

\subsection{Summary of the main results for CN, VCN and UCN sources}

As presented in the previous sections, the HighNESS project explored various potential solutions for CN, VCN, and UCN sources. It's important to recall that HighNESS had two primary objectives: designing intense neutron sources and shifting the neutron spectrum towards colder energies. These goals can be achieved through several approaches.

In the following discussion, we will first summarize and discuss the results for each type of source individually. Then, we will explore how these different sources can be integrated most efficiently.

%\subsubsection{CN moderator}

The second ESS source is centered around the CN moderator designed in HighNESS. The reference design, utilizing LD$_2$ and a Be filter/reflector, achieves an order-of-magnitude increase in intensity compared to the upper moderator. This source is expected to serve both the NNBAR experiment and various neutron scattering studies. It's worth noting that this CN source is the only one that has been fully designed. This is not surprising, as it utilizes LD$_2$, a well-established high-intensity cold source with decades of proven experience.

There is, however, an alternative to this design, namely a \ce{SD_2} moderator, that was discussed in the context of VCN and UCN sources. While preliminary engineering design considerations seem promising for operating such a moderator at least up to 2 MW, further in-depth R\&D would be necessary to realize and operate this source in a high-temperature, strong-radiation environment. As a result, implementing this source would only be possible after years of research, therefore the most likely scenario is that this source will be installed at a later stage, likely after the completion of the NNBAR experiment. It can then be concluded that the ESS second source should be based on the LD$_2$ moderator for the first several years of operation.

%\subsubsection{VCN moderator}
Unlike the CN source, the VCN moderator presented several promising options for investigation. Our study followed two distinct but interconnected paths: one involved the use of dedicated VCN materials, while the other explored the use of advanced reflectors to enhance VCN source performance. In the dedicated materials category, we considered SD$_2$ and deuterated clathrate hydrates. Additionally, we explored advanced reflectors such as nanodiamonds and MgH$_2$, which could be employed not only with dedicated VCN materials but also with LD$_2$.

Furthermore, a variety of hybrid configurations using both LD$_2$ and SD$_2$ were analyzed, thanks also to the availability of accurate thermal scattering libraries (several of which were calculated in the framework of the HighNESS project). Among the various options, it seems that the most promising ones in the foreseeable future are based on the use of SD$_2$ in conjunction with nanodiamond reflectors, where \ce{LD_2} is fully displaced as the moderating medium. Deuterated clathrate hydrates have been explored for the first time for the development of a practical VCN source. The first results in HighNESS indicate that larger volumes of such moderating materials would be necessary, compared to \ce{SD_2}. This would probably make them impractical for ESS as primary VCN-source options; nevertheless the potential for the use of this material is established, and  investigations should be extended to their use as reflectors, or  in conjunction with a primary source.

A full SD$_2$ source is expected to deliver an order-of-magnitude increase in VCN brightness above 40\,\AA. This result was not anticipated at the outset of the HighNESS project: there were no prior works with VCN sources placed so near a spallation target, and reliable thermal scattering libraries needed development or improvement. This result is therefore very important and could be groundbreaking in the field of neutron scattering, if such a source was realized.

%\subsubsection{UCN source}

%\label{table:summarytable}
Regarding UCN sources, HighNESS explored a wider range of concepts compared to VCN sources. The project identified various in-pile options both in the initial proposal and during its course. Furthermore, the availability of two proven materials, SD$_2$ and He-II, expanded the array of potential options. To provide a comprehensive overview, the in-pile performances are summarized in \cref{table:summarytable}. Additionally, the project conducted an in-depth study of an in-beam option.

\subsection{Integration options based on an \ce{LD_2} primary source}
This option was considered the most promising in the HighNESS proposal. The rationale behind this concept involves having a primary high-intensity cold source that can also serve as a source to feed secondary VCN and UCN sources. Consequently, the project dedicated significant effort to explore this possibility, which culminated in a detailed design of the LD$_2$ moderator. This design underwent several phases of neutronic and engineering optimization.

\begin{figure}[htbp!]     
\centering
\includegraphics[width=0.7\textwidth]{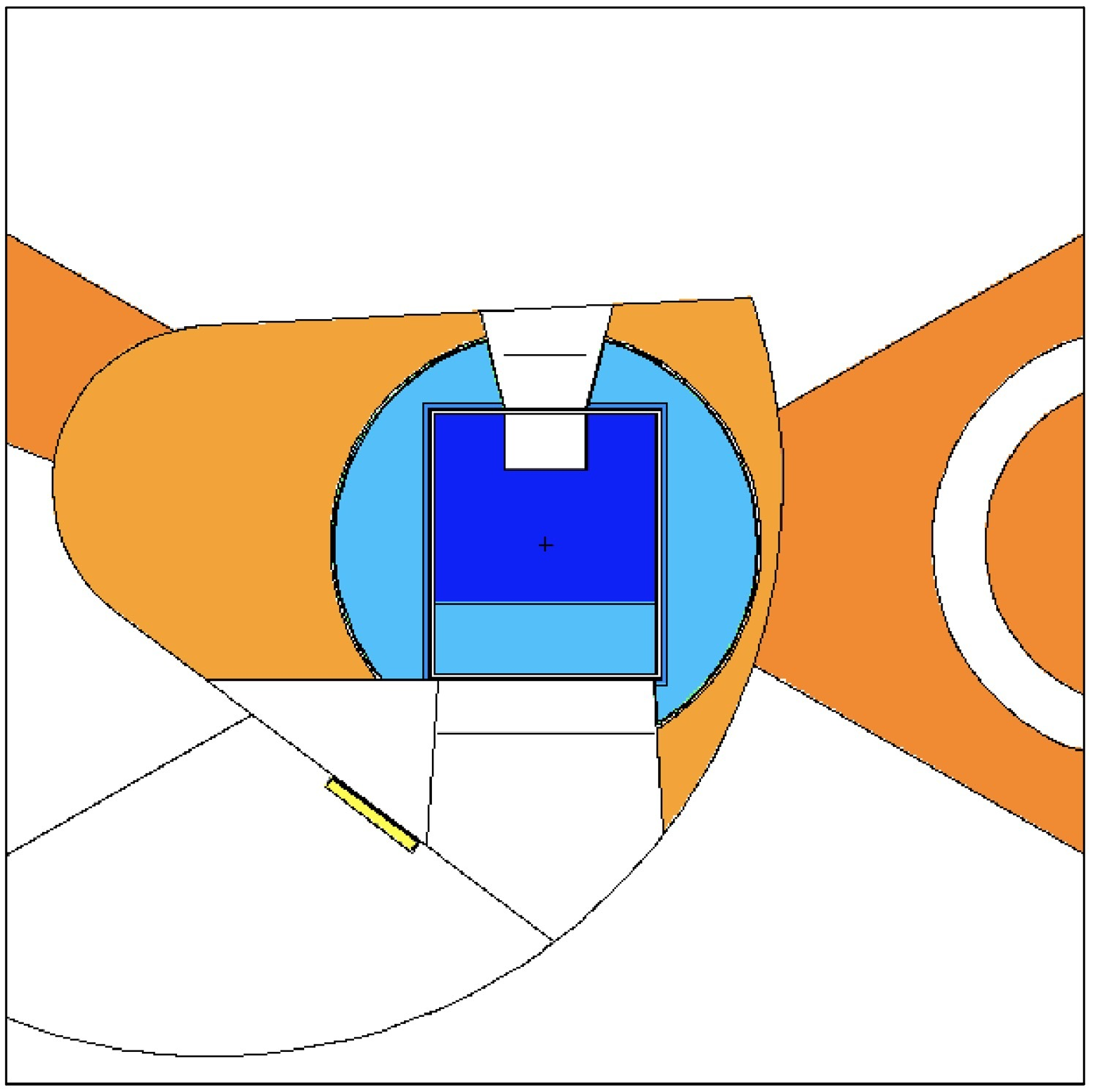} 
\caption{An example of integration of primary cold neutron source made of \ce{LD_2} and secondary UCN source made of \ce{SD_2}. \ce{LD_2} is depicted by dark blue color and \ce{SD_2} is depicted by yellow color. The Be filter on NNBAR opening is depicted by the light blue color.}
\label{fig:MCBIntegration}
\end{figure}

For this option we explored two possibilities for the placement of the secondary source: the MCB as can be seen on Figure \ref{fig:MCBIntegration}, and the LBP.

\subsubsection{\ce{LD_2} in twister, UCN in the MCB}

For this option, from day one ESS has an intense CN source and a UCN source in the MCB, which has no interference with the operation of NNBAR. Neutronic design of this option has been investigated for UCN sources both based on SD$_2$ and He-II, showing great potential for a high-performance source.

\subsubsection{\ce{LD_2} in twister, followed by UCN in LBP after NNBAR}
\label{sere}

\begin{figure}[tb!]      
    \begin{subfigure}[b]{0.48\textwidth}
        \centering
        \includegraphics[width=\textwidth]{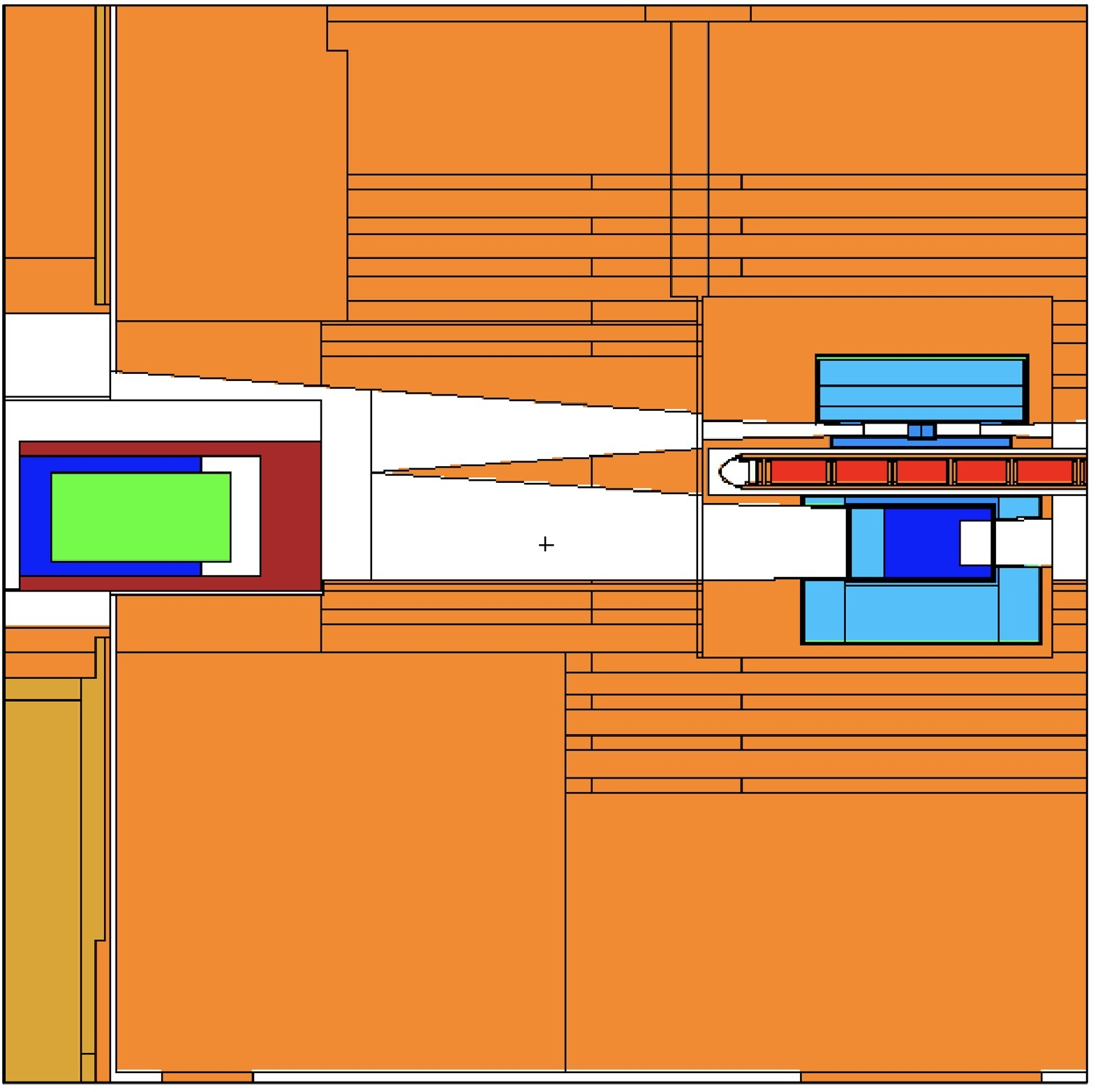}
        \subcaption{}
        \label{fig:baseline_opt_XY2}
    \end{subfigure}
    \hfill
    \begin{subfigure}[b]{0.48\textwidth}
        \centering        
        \includegraphics[width=\textwidth]{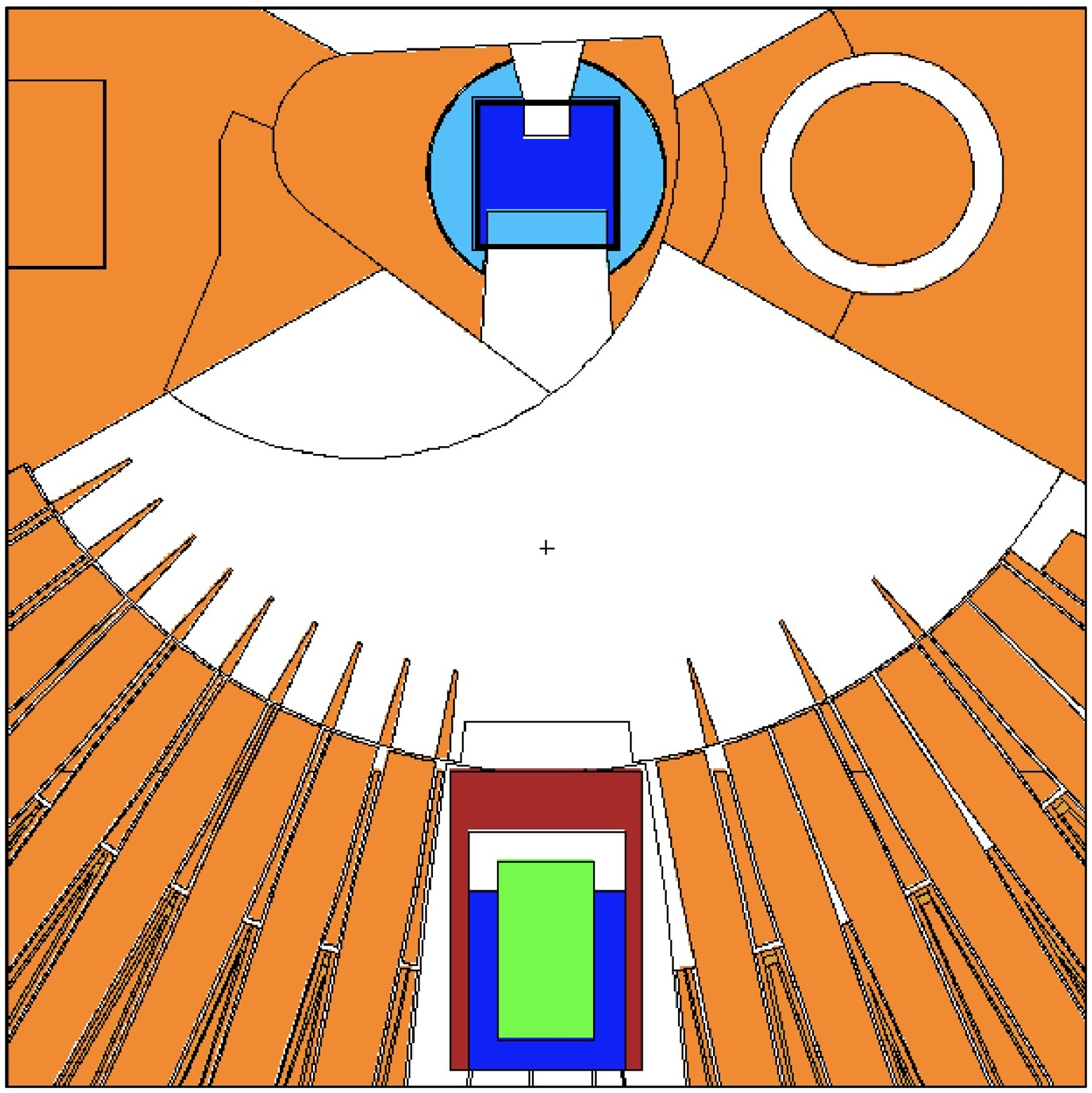}
        \subcaption{}
        \label{fig:baseline_opt_XZ2}
    \end{subfigure}
\caption{Integration of the primary cold source made of \ce{LD_2} depicted by the dark blue color located below the target and secondary UCN source made of He-II depicted by the green color located in the NNBAR beamport. The model is based on the proposal from Serebrov. The red color depicts the ESS target made of tungsten. The dark red color around He-II converter depicts the shielding made of bismuth.}
\label{fig:SerebrovIntegration}
\end{figure}

In this scenario, ESS initially uses the LD$_2$ moderator alone, and the NNBAR experiment is conducted for several years. Once the NNBAR experiment is completed, the LBP can be repurposed for other applications. We have conducted detailed studies of two options for this case: the in-pile option (see Figure \ref{fig:SerebrovIntegration}) and the in-beam option using nested mirror optics.

\subsubsection{\ce{LD_2} in twister,  UCN in regular beamport}

This configuration represents a third option for a secondary UCN source fed by the \ce{LD_2} moderator, wherein a regular beamport is utilized in order to avoid interfering with the NNBAR experiment. Additionally, in this case the UCN source can be in-pile, inside the monolith, or in-beam. We did not perform a neutronic study of this in-beam option, however its performance can be estimated by either scaling the results for the LBP in-pile solution see \cref{sere}, or from estimates made at the HighNESS workshop~\cite{chanelConceptStrategySuperSUN2022}.

\subsection{Integration option based on an \ce{LD_2} primary source, upgraded to deliver also VCN and UCN}

The options based on using the LD$_2$ moderator as primary source described above do not have a dedicated VCN source. This is due to the fact that, for a VCN source to surpass the performance of an LD$_2$ moderator, in the wavelength range of the VCNs, such a source must necessarily be placed inside the twister, close to the hot spot of neutron production. The only other possibility for a VCN source without using a dedicated material that has been explored in HighNESS is the use of advanced reflectors (NDs) to more efficiently transport the coldest part of the spectrum to an instrument or guide entrance. However, it was found to only have a significant effect at large divergence angles, which is not immediately relevant for the intended applications of the second source. An upgraded LD$_2$ cold source, with the addition of a SD$_2$ slab on one of its extraction sides -- or in one of the more complex hybrid options detailed in \cref{sec:sd2_intwister} -- could be a valid alternative for a combined CN/VCN source. 

\subsection{A full-\ce{SD_2} source for CN, VCN and UCN?}
A VCN source based on a full-SD$_2$ moderator at 5\,K was also designed as part of HighNESS. With respect to the LD$_2$ source, an SD$_2$ moderator delivers more neutrons for wavelengths above about 6\,\AA, while the performance below this range is lower. However, it would provide a slight increase in the FOM of NNBAR, which would make it very interesting as it would be a source of CN, VCN and UCN. 

In this scenario, ESS would start with a \ce{LD_2} moderator located below the target; later the \ce{LD_2} moderator would be replaced with a \ce{SD_2} moderator with the same dimensions. 
%The \ce{SD_2} allows extraction of CN, VCN and UCN.
This scenario  would be possible only if the big cooling challenges to operate a solid deuterium source at high power would be solved;  by the time ESS  will reach the 5 MW average power, a compromise design solution should probably be adopted, where the moderator is part \ce{LD_2} and part \ce{SD_2}.

%This scenario is the most attractive because it pushes for the development of a working SD2 source which will allow VCN.

%It is also a gradual development, starting with the proven \ce{LD_2} technology and upgrading it later on with \ce{SD_2} (with advanced cooling techniques, use of foams to increase thermal conductivity, and other improvements)

\subsection{Comparison of the different scenarios}

Based on the above considerations, it is evident that there are several competitive possibilities for providing a second source that complements the upper high-brightness bi-spectral moderator. To select the most promising option among these, several factors need to be taken into account, particularly focusing on performance, operability, and technical feasibility.

 \begin{table}[h!]
\caption{The different integration scenarios.}
\label{tab:compscenarios}
\begin{center}
\begin{tabular}{ c c  c c }
\toprule
%\texttt{iel} & Description \\ 
\midrule
option & CN source & VCN source & UCN source \\ \hline
1 & LD$_2$ & none & none  \\
2 & LD$_2$ & none & SD$_2$ or He-II in MCB \\
3$^a$ & LD$_2$ & none & He-II in beam in regular beamport \\
4$^b$ & LD$_2$ & none & He-II in LBP  \\
5$^c$ & LD$_2$ & none & He-II in beam in LBP (NMO) \\
6$^d$ & Hybrid LD$_2$/SD$_2$ & SD$_2$+ND & SD$_2$   \\
7$$ & Hybrid LD$_2$/SD$_2$ & SD$_2$+ND & He-II in beam  \\
8$^e$ & SD$_2$ & SD$_2$ & SD$_2$  \\
\bottomrule
\end{tabular}
\end{center}
{\footnotesize
$a$:  not  studied in HighNESS, performance can be estimated from SuperSUN at ILL. \\
$b$: installed after NNBAR is completed. \\
$c$: UCN in beam, use of NMO; installed after NNBAR is completed. \\
$d$: UCN extracted from SD$_2$. \\
$e$: full SD$_2$ option. \\
}
\end{table}

The list of possible options for an integrated source is in \cref{tab:compscenarios}.

The eight rows in \cref{tab:compscenarios} could be loosely considered as a possible sequence of possible configurations for the lower moderator plug. Starting from having only the LD$_2$ moderator in the twister, it would be possible to add in parallel a UCN source in the MCB, or a UCN source in beam using a regular beamport other than the LBP.

Different options for UCN sources could be implemented after NNBAR is completed and the LBP is available (options 3 and 4).

To establish a high-intensity VCN source, we have determined that the use of a dedicated material in the twister is essential. However, the development of technology to operate a 5 K moderator in close proximity to the spallation target is required.  Additionally, the need to commence operation of the lower moderator with a high-intensity cold moderator suggests that the implementation of a strong VCN source would likely occur at a later stage, potentially after the NNBAR experiment is completed. In terms of performance, a full SD$_2$ moderator is comparable to an LD$_2$ moderator, but it delivers an order of magnitude or more VCNs and is also a proven material for UCN. Therefore, option 8 could potentially represent the final stage of the HighNESS source, where a single SD$_2$ moderator can offer competitive CN fluxes, unprecedented VCN fluxes, and a world-leading UCN source.

%\textcolor{red}{fill in}

%% file: neutronscattering.tex
\section{Neutron scattering instruments for the ESS second source  } 
\label{sec:neutronscattering}

% This has been exmplained, reduce
During the late stages of the design of the European Spallation Source it was realized that reducing the height of the cold hydrogen moderator could significantly increase its brightness, presenting a new opportunity for both moderator and instrument design \cite{butterfly_optimization}. The increase in brightness did not fully compensate for the decrease in height, meaning the height would directly control the trade-off between intensity and brightness. The optics for all instruments, which will be served by the newly designed ``butterfly'' moderator \cite{ESS_moderators}, were reoptimized for a range of moderator heights \cite{butterfly_optimization}, and a choice was made to reduce the height from 10\,cm to 3\,cm, as this benefited almost all instruments, and in some cases increased the simulated flux on sample with over 150\,\%.
%Originally the target monolith had been designed to accommodate two identical moderators, each illuminating half of the beamports, yet due to the changes in engineering constraints from the smaller size of the redesigned moderator, it was possible to illuminate all beamlines using a single moderator slot. The redesigned moderator was named the butterfly and placed in the top moderator slot, leaving the lower slot free for future upgrades. The purpose of the HighNESS project is to investigates options for this lower moderator slot with the aim of designing a moderator that complements the butterfly moderator.

There were however a few science cases that proved to be more difficult on the smaller moderator, for example imaging, which benefits from a larger moderator due to better homogeneity and field of view (FoV). Some types of instruments can tolerate a large divergence, for example backscattering instruments or time-of-flight spectrometers, and their performance is more closely related to the intensity of the moderator than its brightness. For these reasons it was natural for the HighNESS project to investigate larger moderators to complement the ``butterfly'' moderator with its small height. The ``butterfly'' moderator has a bispectral design with both a thermal water part and cold hydrogen part \cite{ESS_moderators}. On the other hand, the HighNESS, larger moderator, focuses on the cold side, and provides a colder spectrum which can potentially benefit small-angle neutron scattering (SANS), imaging and spin-echo instruments.

In order to optimize the moderator proposed by HighNESS, a number of instrument concepts that would utilize this new moderator were considered along with suitable figure-of-merit (FOM) expressions. This allowed an iterative approach for moderator design. Each new design could be evaluated using these instrument concepts to assess the performance of the moderator, considering not only the flux at its surface but also its impact on the performance of the instruments reliant on the moderator.

In this section we share the final design of the instruments along with their performance for each of the 4 proposed moderator sizes, providing the necessary data to choose the best of the 4 options. The considered instruments were two SANS instruments, one using conventional optics and the other using focusing optics, as well as an imaging instrument. The work on a spin-echo instrument using focusing optics was not finished within the time constraints of the project but shows promising results and should be further developed in future studies. Each conceptual instrument is compared with the relevant ESS instruments under construction to ensure they add new capabilities and avoid replicating existing capabilities in the instrument suite, to as high a degree as possible. 

As described previously, the HighNESS project has also investigated a solid deuterium moderator which would have significantly better performance in the VCN wavelength range. When considering instrumentation for such a source, it is important to have optical tools for counteracting the effects of gravity. The last part of this section investigates the use of prisms in combination with focusing optics to create achromatic optics, which will be an essential tool for future VCN-neutron scattering instruments.

\subsection{Moderator alternatives}
The moderator optimization process arrived at a large liquid deuterium vessel with different sizes of exits towards the condensed matter instruments investigated here. The final four candidates were the following sizes,

\begin{align*}
\text{3$\times$3\,cm$^2$, 5$\times$\,5\,cm$^2$, 10$\times$\,10\,cm$^2$, and 15$\times$\,15\,cm$^2$.}
\end{align*}

\begin{figure}
    \centering
    \includegraphics[width=0.55\columnwidth]{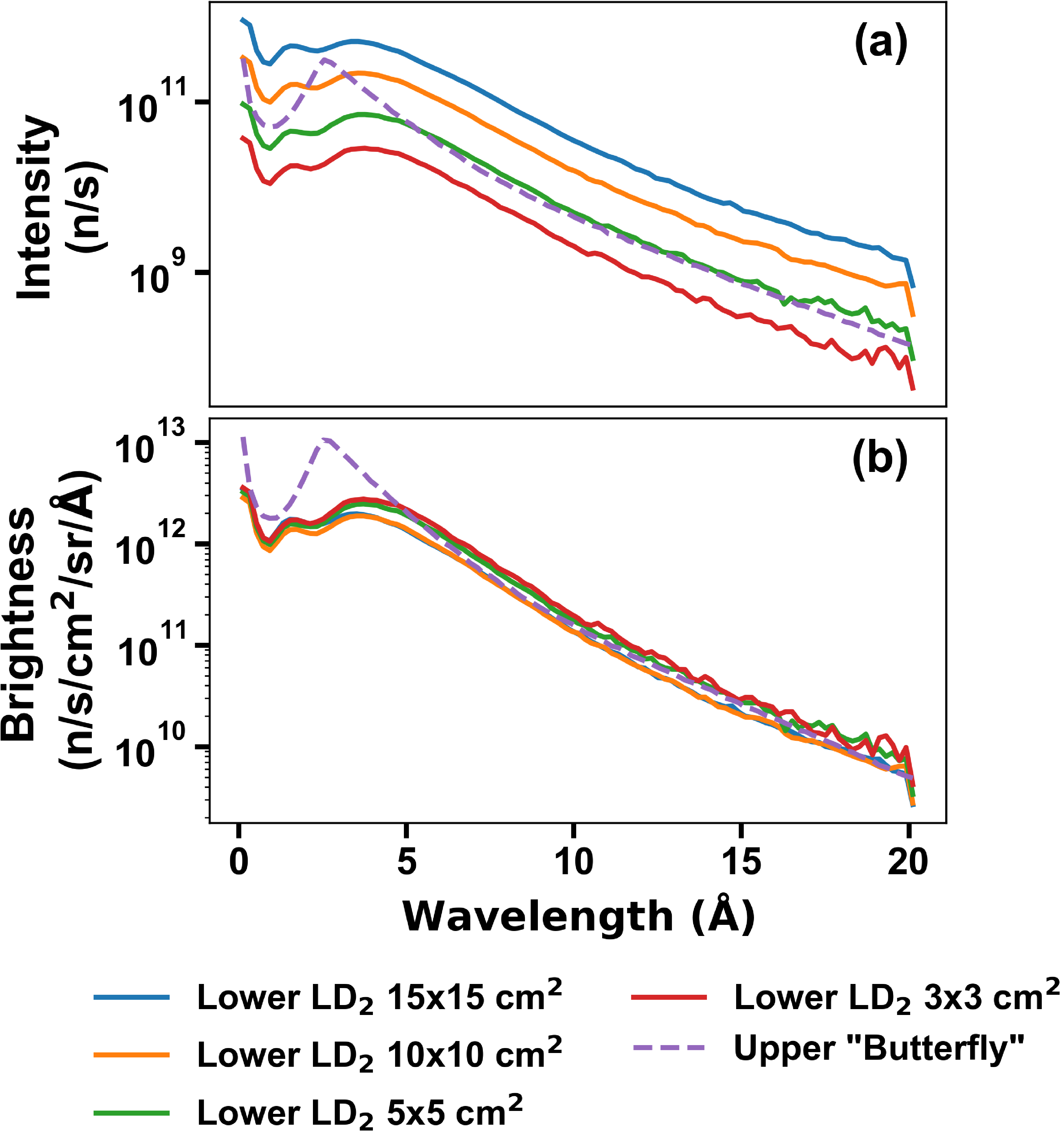}
    \caption{Spectra of the proposed lower liquid deuterium moderators, in terms of (a) neutron intensity and (b) neutron brightness, measured at 2\,m distance, over the beam-port. The corresponding cold spectrum of the upper ``butterfly'' moderator is plotted over for comparison.}
    \label{fig:Spectra}
\end{figure}

In \cref{fig:Spectra} the intensity and brightness of these four moderators are plotted along with the corresponding values for the cold part of the upper ``butterfly'' moderator. While the smaller openings do have a higher brightness, this effect is much smaller for the deuterium moderator than the hydrogen-based ``butterfly'', meaning that differences in brightness are less pronounced. At wavelengths larger than 5\,\AA\ the brightness of the larger moderators is very similar to that of the ``butterfly'', while the two smaller moderators have slightly higher brightness than the upper moderator. With the similar brightness, the larger moderators ($>$\,5\,$\times$\,5\,cm$^2$) have a significantly larger neutron intensity when compared to the ``butterfly'' moderator.

\subsection{Optimization of small-angle neutron scattering instruments}
The SANS science case for a larger moderator would naturally focus on larger samples, though the large source can also be utilized by focusing optics. As such, two different designs were investigated, a conventional SANS named ConvSANS and a Wolter optics-based SANS instrument named WOF-SANS. There are two SANS instruments under construction at the ESS, the short-length LoKI and the medium-length SKADI. For ConvSANS to be complementary to these instruments, it is chosen to increase its length to achieve better resolution and smaller minimum $Q$. The WOF-SANS instrument uses focusing optics to benefit from the large moderator area and due to the focusing properties can avoid a long collimation section, resulting in a shorter instrument with greater bandwidth, while maintaining high resolution for even smaller $Q$ values.

\subsubsection{Conventional SANS}
\begin{figure}[tbh!]
    \centering
    \includegraphics[width=0.95\columnwidth]{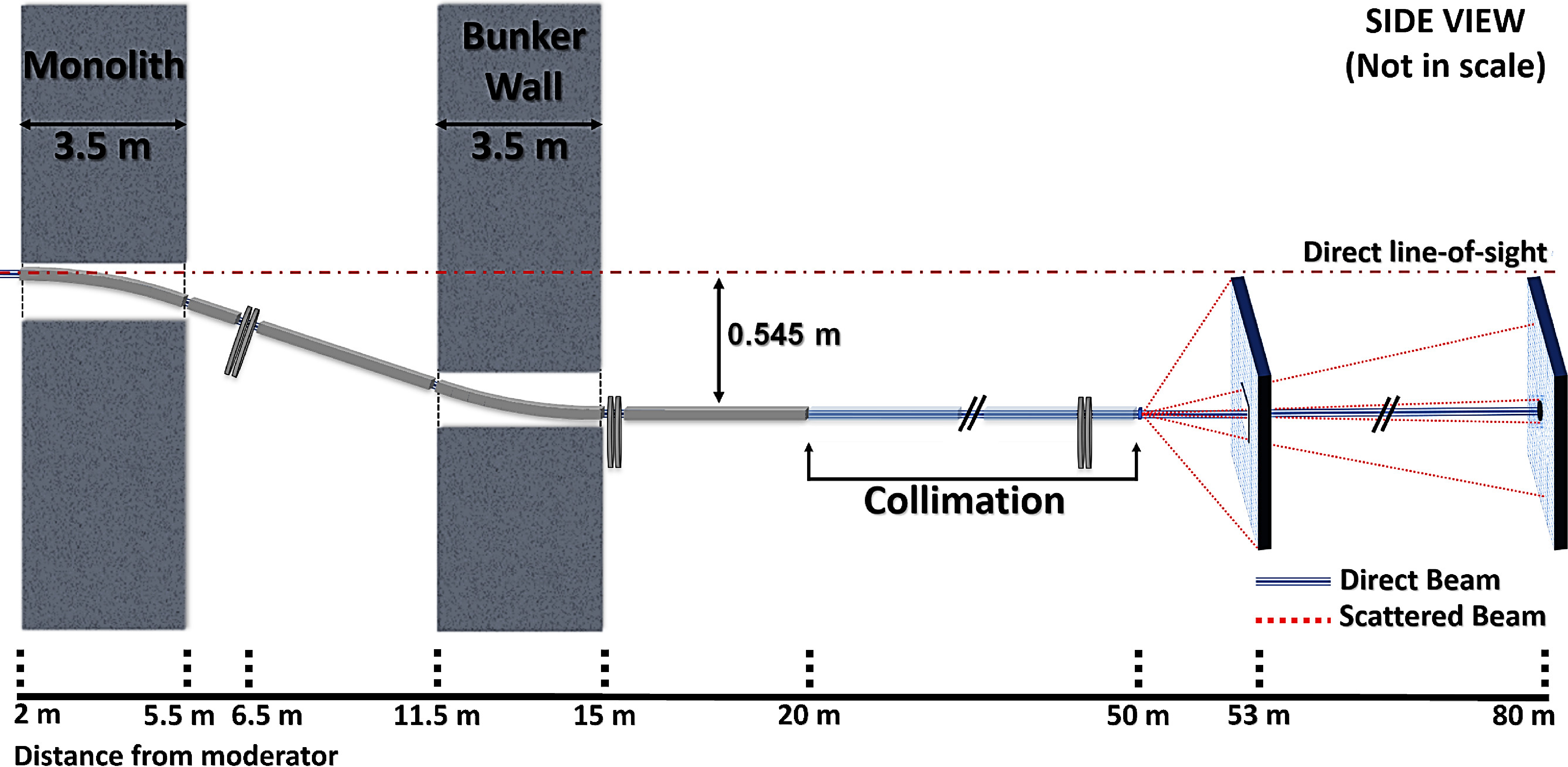}
    \caption{Overview of ConvSANS instrument geometry.}
    \label{fig:ConvSANS_CAD}
\end{figure}

The Conventional SANS (ConvSANS) conceptual design presented here is based on a typical pinhole collimation geometry. A schematic of the instrument is given in \cref{fig:ConvSANS_CAD}. 
The current design consists of a single block 30\,m long collimation, thus having, for initial assessment, a constant collimation distance. In a practical realization the collimation will consist of sections to enable to relax resolution and adapt to different sample sizes. Currently, two fixed-position detectors are placed at 3\,m (wide angles) and 30\,m (small angles) from the sample, respectively. The combination of the long collimation with the two detector positions, will allow for a large $Q$-range coverage, allowing at the same time to maintain high $Q$ resolution for large sample sizes. The relatively long total instrument length, set at 80\,m, leads to good wavelength resolution, while the long collimation provides high angular resolution, which can be traded for large samples and thus high intensity. The total length of the instrument limits the wavelength bandwidth to about 3.5\,\AA\, when operating in the standard 14\,Hz mode, however, this can be extended to about 7.0\,\AA\ in the pulse skipping 7\,Hz mode of operation.

The present instrument design assumes a constant guide size of 6\,$\times$\,6\,cm$^2$ throughout the instrument. The start of the neutron guide is at 2\,m from the moderator surface, at the beginning of the monolith wall. The instrument makes use of a pair of benders located in the monolith and the bunker wall, respectively, that are used to avoid direct line-of-sight to the moderator and act as a short wavelength cut-off filter. The benders are designed with a radius of curvature of 61.25\,m and a length of 3.5\,m. The use of the bender pair provides for twice out of line-of-sight curvature to minimize the intrinsic background, and offsets the beamline down vertically by about 0.55\,m. In the in-between space of the monolith wall and the bunker wall, two tilted straight neutron guides, 1\,m and 5\,m in length, respectively, connect the two benders, with a small gap between them reserved for chopper installation. After the second bender, a straight guide of 5\,m in length brings the beam back into horizontal position and connects the bender with the collimation system. After initial Monte Carlo investigations, all neutron guides and benders are coated with $m$\,=\,4 supermirrors, however, this should be subjected to further optimization. The 30\,m collimator starts at 20\,m from the moderator, where a first aperture, hereafter referred to as source aperture ($D1$), is located, and ends at 50\,m from the moderator, where the sample aperture ($D2$) is located. 

Wavelength selection and frame overlap prevention is performed using two double-disc choppers. A bandwidth double-disk chopper is placed in the in-bunker section of the instrument, between the two tilted neutron guides, at 6.5\,m from the moderator. An additional pair of choppers is placed right after the bunker wall, at 15\,m, downstream of the second bender, to suppress frame overlap. A third pair of choppers can also be added further downstream, at 40\,m within the collimation section, to further suppress frame overlap.

The detector configuration employed for the current set-up is based on a "window-frame" design (see \cref{fig:ConvSANS_CAD}). The front detector is at a fixed position, 53\,m from the moderator and 3\,m from the sample position. It has a 3\,$\times$\,3\,m$^2$ surface area with a 0.327\,$\times$\,0.327\,m$^2$ window opening at the center. The rear detector, also at fixed position, is located at 80\,m from the moderator and 30\,m from the sample position, with a 3\,$\times$\,3\,m$^2$ surface area. In the current design, there is no specific detector technology considered (e.g., $^3$He gas detector or $^{10}$B-based detector).

\begin{figure}[tbh!]
    \centering
    \includegraphics[width=0.6\columnwidth]{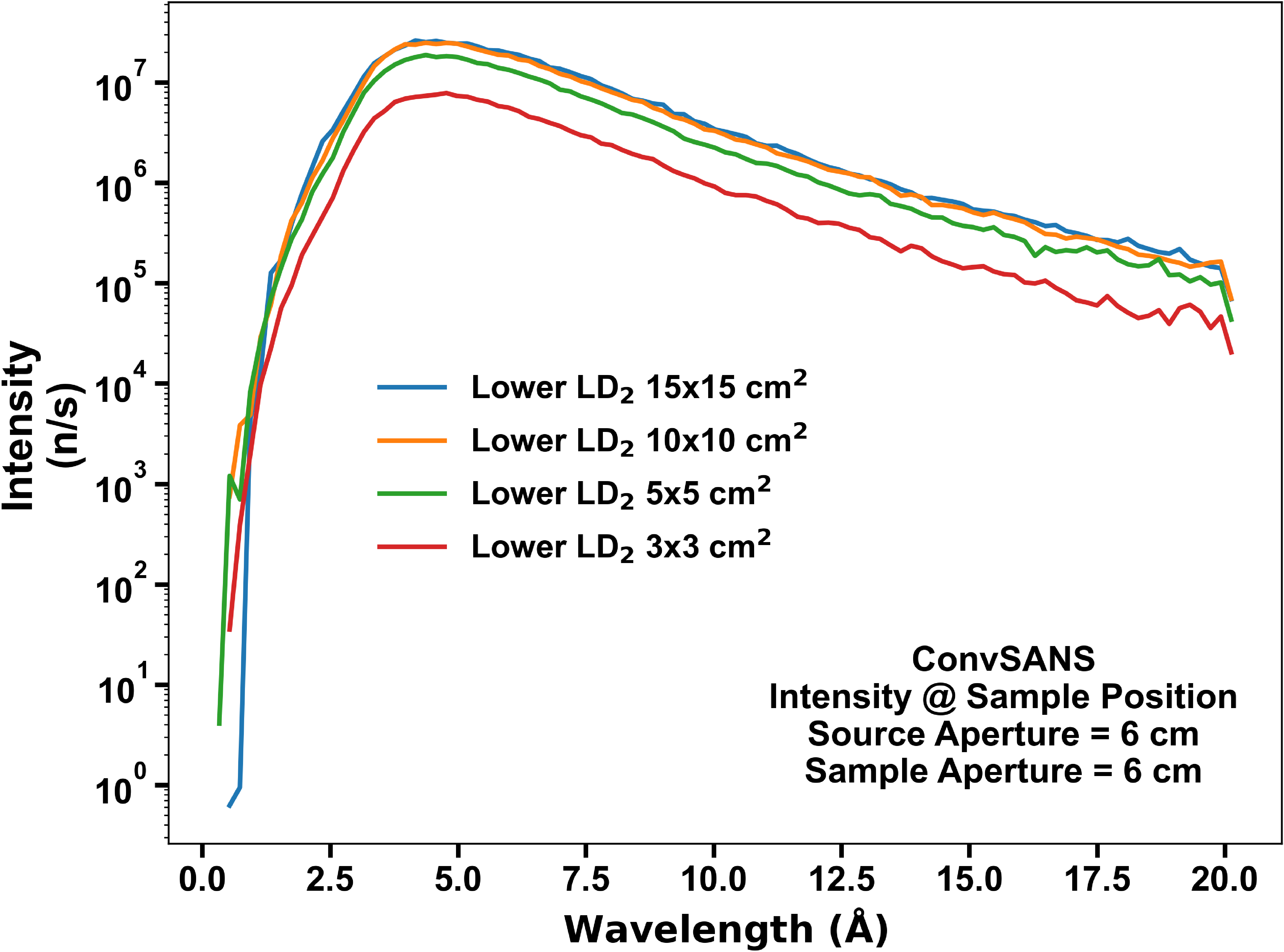}
    \caption{Intensity on sample for ConvSANS instrument for each proposed moderator size measured over the full wavelength range.}
    \label{fig:ConvSANS_intensity}
\end{figure}

While the instrument has flexible collimation configurations for different samples, it was chosen to investigate the neutron intensity on sample for the configuration that accommodates the largest possible sample, here a source aperture of 6\,cm and a sample aperture of 6\,cm. The spectrum on the sample with the choppers open (full spectrum) can be seen in \cref{fig:ConvSANS_intensity}. In this configuration the performance is similar for all moderators, with only the smallest delivering a distinctly lower intensity. Thus, for configurations intended for larger samples and moderate resolution, the benefit from the larger moderators increases, though it is important to note that even for higher resolution setups, the largest moderator still achieves the highest intensity.

\subsubsection{SANS with Wolter focusing optics}
\begin{figure}[tbh!]
    \centering
    \includegraphics[width=0.95\columnwidth]{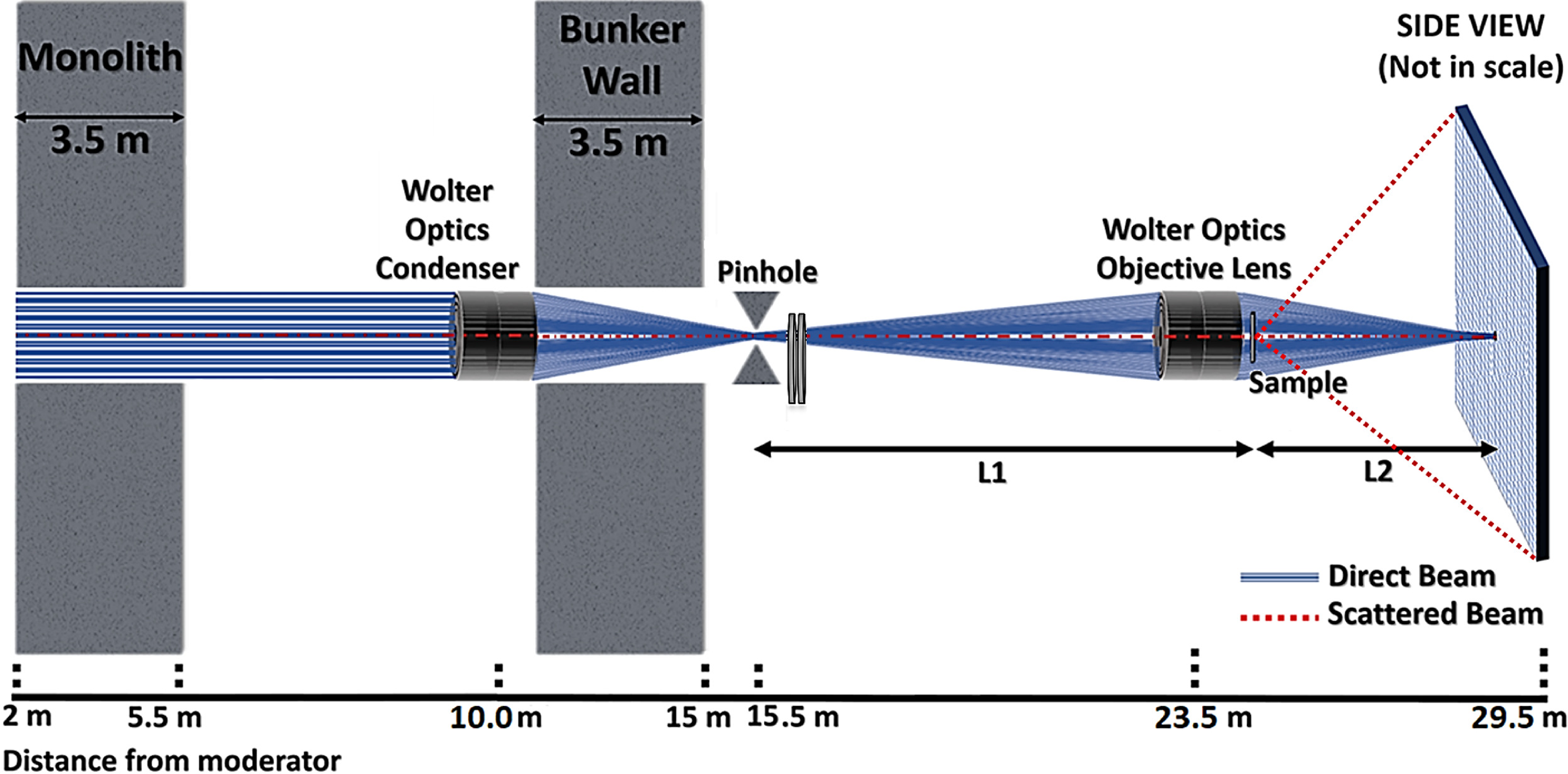}
    \caption{Overview of WOF-SANS instrument geometry. L1 is the distance from the pinhole to the sample, L2 the one from the sample to the detector.}
    \label{fig:WOFSANS_CAD}
\end{figure}

The WOF-SANS design makes use of a pair of reflective Type I Wolter optics \cite{Wolter_early} to take advantage of the large moderator surface and thus increase the neutron intensity at the sample position. The concept of such a focusing SANS instrument is described in \cite{focused_SANS_resolution}. A schematic of the current instrument design can be seen in \cref{fig:WOFSANS_CAD}. The total length of the instrument is 29.5\,m defining a wavelength bandwidth of 9.5\,\AA\ when operating with the 14 Hz source frequency and can be extended to about 19\,\AA\ at pulse skipping mode.

The first Wolter optics system is used as a condenser lens, and it is located in the in-bunker section of the instrument, 10\,m from the moderator (considering its middle point). It is designed accordingly to have the highest possible throughput. For an initial assessment, it consists of 28 nested paired parabolic and hyperbolic (P-H) sections, currently 1.3\,m and 1.0\,m in length, respectively (2.3\,m total length), with a total radius of 7.5\,cm. The innermost radius of 1.0\,cm is not covered and might be filled with an absorber. The focal length of the condenser is 5.5\,m thus its focal point is 15.5\,m from the moderator. The use of parabolic mirrors as the first point of contact for the incoming beam, implies that predominantly the lowest divergence component of the phase space will be focused. The mirrors are coated, for initial assessment, with supermirror coating of $m$\,=\,3.0. 
Lower coating quality is anticipated to suffice, as highly divergent portions of the beam, similar to the conventional SANS approach, will not be taken into account. Instead, a larger area of the moderator will be considered.
An aperture of 4\,mm in diameter is placed downstream of the bunker wall, at the focal point of the condenser, to suppress any out-of-focus rays and any neutrons that pass through the condenser without reflecting. This position will be heavily shielded and the aperture will be designed to deal with such a situation and to shield the large diameter beam of background radiation.

The second Wolter optics system is used as an objective lens and it is located 23.5\,m from the moderator (considering its middle point). Its focal lengths, aperture-to-objective and objective-to-detector, are 8\,m and 6\,m, respectively, resulting in a (de)magnification $M$\,=\,0.75. As such, the beam spot size on the detector center is expected to be 3\,mm in diameter. The objective lens consists of 25 nested paired elliptical and hyperbolic (E-H) sections, 0.9\,m and 0.82\,m in length, respectively (1.72\,m total length). Its maximum radius is 10\,cm, allowing it to collect the full divergence coming from the aperture (i.e., from the condenser). The E-H mirrors are also coated supermirror of $m$\,=\,3.0, however, this is subject to optimization. The sample is placed between the objective lens and the detector. Different positions are possible with corresponding impact on illuminated sample size (beam size), $Q$-range, and resolution. 
A bandwidth double-disk chopper is placed right after the bunker wall and the slit of the focusing position, at 16 m. An additional pair of choppers can be placed before the objective lens to suppress frame overlap. The position sensitive detector is considered to have a surface area of 3\,$\times$\,3\,m$^2$ and is placed at a fixed position matching the focal point of the E-H optics, at 29.5\,m from the moderator. As in the case of ConvSANS there is no assumption of specific detector technology.

\begin{figure}[tbh!]
    \centering
    \includegraphics[width=0.6\columnwidth]{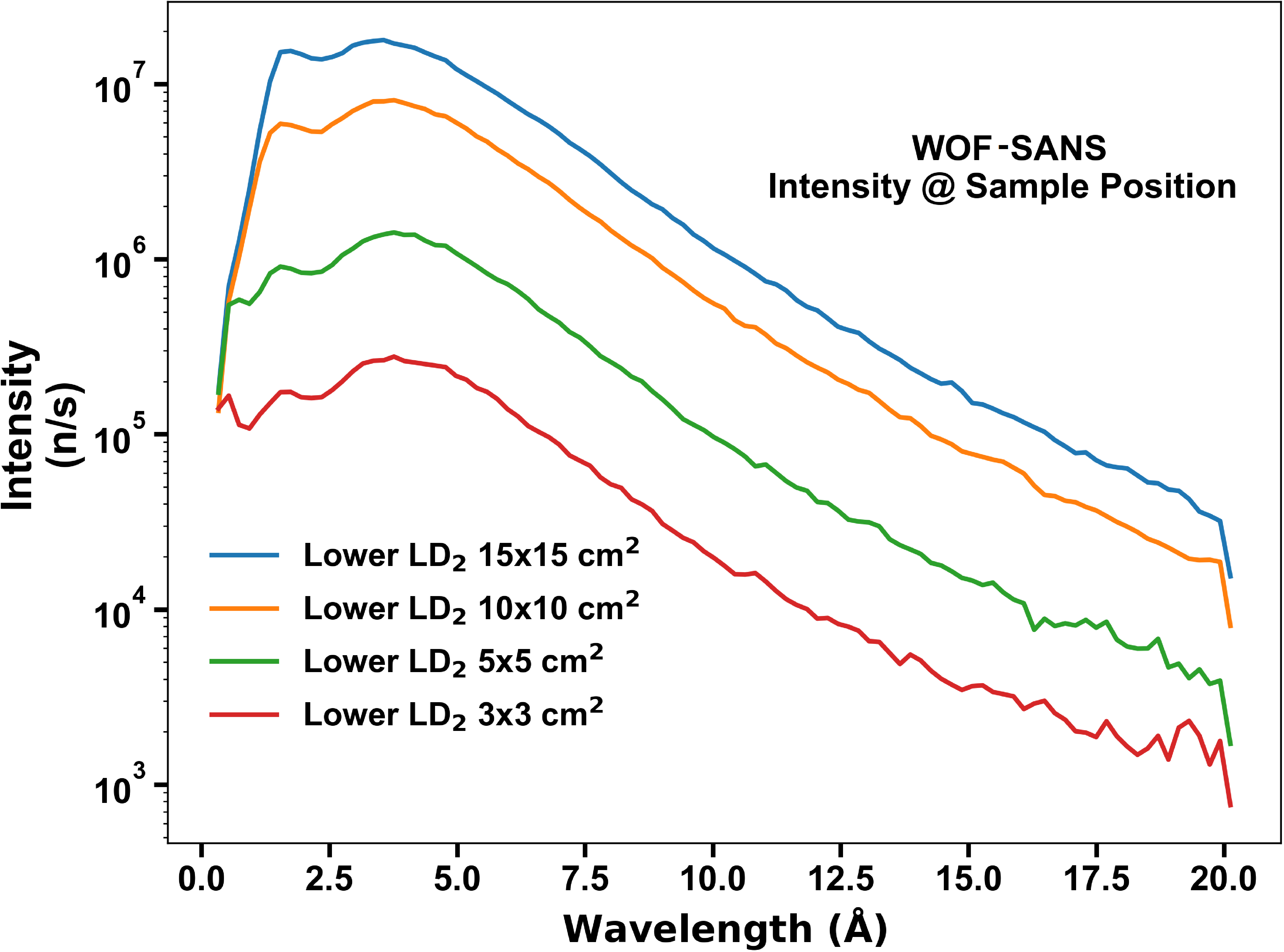}
    \caption{Intensity on sample for WOF-SANS instrument for each proposed moderator size, measured over the full cross section of the objective lens and over the full wavelength range.}
    \label{fig:WOFSANS_intensity}
\end{figure}

The spectrum on the detector position with all choppers open can be seen in \cref{fig:WOFSANS_intensity}. Due to the focusing of the instrument, this is equivalent to the intensity at the sample regardless of position (when neglecting losses from air scattering). The WOF-SANS instrument shows higher sensitivity to the size of the moderator, attributed to the use of a parabolic section that performs better with larger sources.

\subsubsection{FOM Definition}
In general, determining the FOM of a SANS instrument can be complex since there are many parameters coming into play and different sets of samples and/or configurations need to be taken into account. For our SANS instruments, at this early stage of conceptual design, the useful parameters that can be considered are the intensities at the sample position ($I_{sample}$), as calculated from the simulations, the $Q_{min}$ and $Q_{max}$ values, as well as $\sigma^2_Q(Q_{min})$. Similar to the FOM defined by information theory, we can define the following simple FOM formula:

\begin{equation}
FOM = \frac{I_{sample}}{\sigma^2_Q(Q_{min})}\cdot ln\left (\frac{Q_{max}}{Q_{min}}\right ). 
\end{equation}

With the above definition we see that the FOM is proportional to the intensity, as well as the ratio of $Q_{max}$ and $Q_{min}$. The $Q$-resolution variance is inversely proportional to the FOM since lower values (improved resolution) will increase the FOM. 

\subsubsection{Performance of conceptual SANS instruments}
\begin{figure}[tbh!]
    \centering
    \includegraphics[width=0.55\columnwidth]{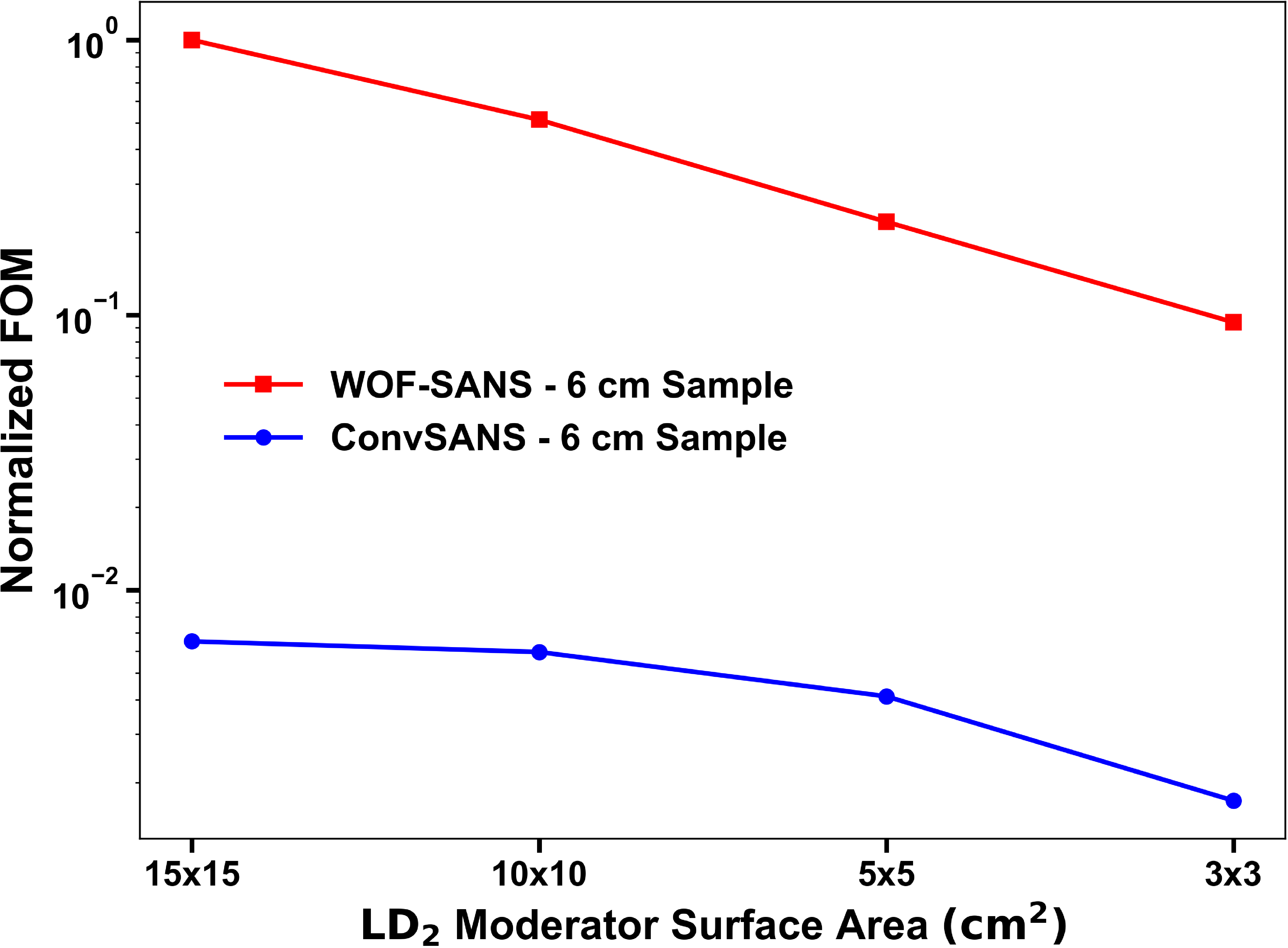}
    \caption{FOM at fixed sample size (maximum considered) as function of moderator size.}
    \label{fig:SANS_FOM}
\end{figure}

With this definition of the FOM, both the SANS instruments and moderator candidates can now be directly compared. 
% In \cref{fig:SANS_FOM_Q} the figure of merit is shown for both instruments and all moderators as a function of Q, showing the WOF-SANS generally has higher performance than the conventional instrument.
% \begin{figure}[tbh!]
%     \centering
%     \includegraphics[width=0.75\columnwidth]{wp7_figures/deliverable_figures/figure6.png}
%     \caption{Figure of merit as function of scattering vector magnitude for both SANS instrument concepts and all proposed moderators.}
%     \label{fig:SANS_FOM_Q}
% \end{figure}
To illustrate the performance of the instruments as a function of moderator size, the FOM at the same maximum usable sample size of 6\,cm is plotted for each moderator in \cref{fig:SANS_FOM}, which clearly shows that both SANS instrument concepts perform best at the largest source.

\newpage
\subsubsection{Performance comparison to ESS instruments}
\begin{figure}[!h]%[!htbp]
  \centering
  \includegraphics[width=0.7\columnwidth]{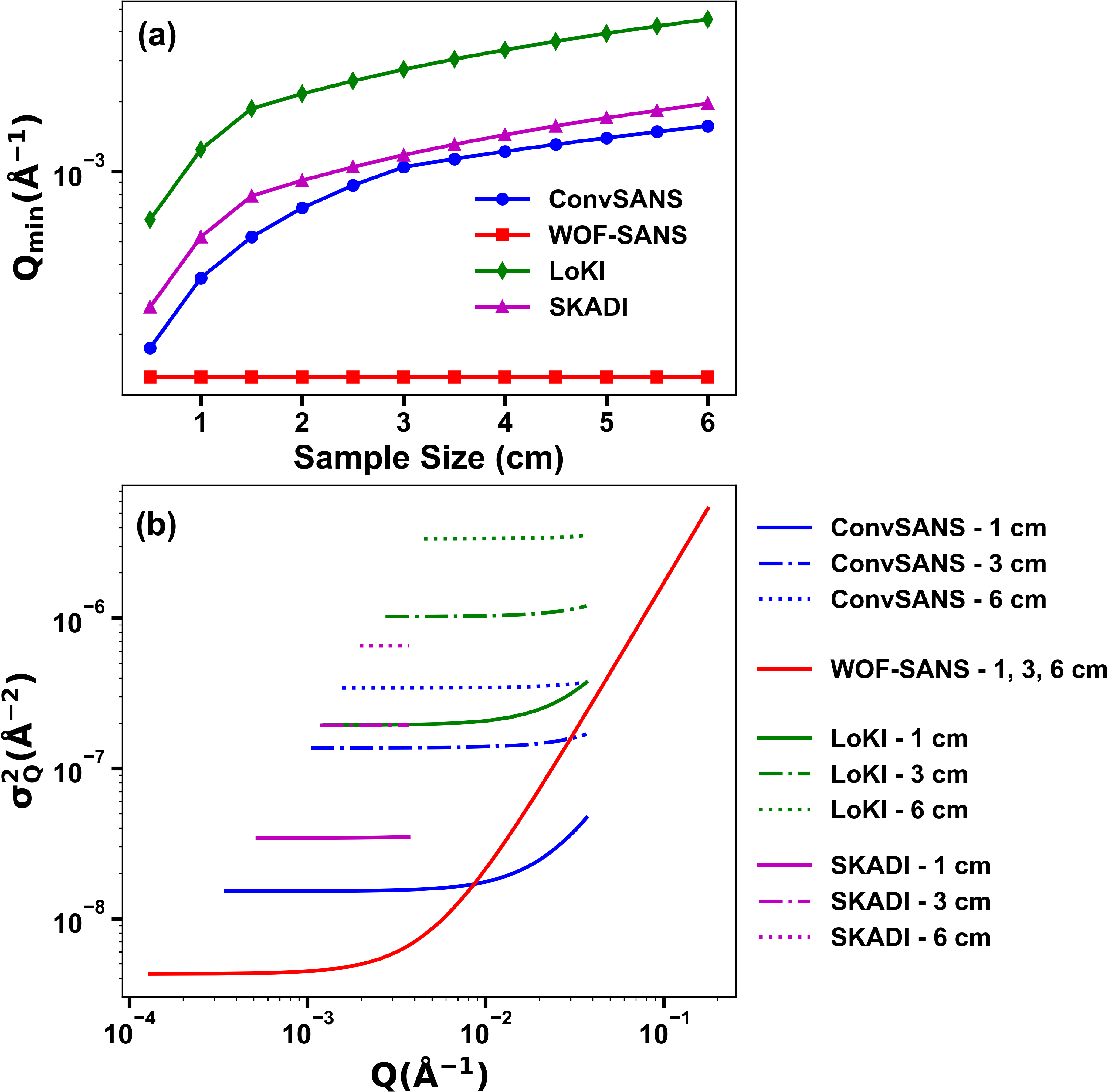}
  \caption{(a) Calculated $Q_{min}$ values for ConvSANS, WOF-SANS, LoKI, and SKADI as a function of sample size. (b) $Q$ resolution variance, $\sigma^2_Q$, as a function of scattering vector $Q$ for ConvSANS and WOF-SANS in comparison to those of LoKI and SKADI. All resolutions were calculated for $\lambda_{max}$\,=\,12\,\AA\ and for the longest configuration of each instrument, on the rearmost detector (single detector for WOF-SANS).}
  \label{fig:Qmin&resolution}
\end{figure}

In addition to comparing the two conceptual SANS instruments with each other for each proposed moderator, it is important to view their performance in the context of the SANS instruments already under construction at the ESS. In this section, the 15\,$\times$\,15\,cm$^2$ moderator has been selected, while LoKI and SKADI were simulated using the upper ``butterfly'' moderator and their latest McStas instrument files. Fig.\,\ref{fig:Qmin&resolution}a depicts $Q_{min}$ values as a function of sample size for ConvSANS, WOF-SANS, LoKi, and SKADI. Fig.\,\ref{fig:Qmin&resolution}b contains plots of $\sigma^2_Q$ as a function of the scattering vector $Q$ for the different instruments calculated on their rearmost detector and for three chosen sample sizes: 1\,cm, 2\,cm, and 3\,cm. The LoKI and SKADI detector surface areas used for the calculations were 1.0\,$\times$\,1.0\,m$^2$ and 0.2\,$\times$\,0.2\,m$^2$, respectively.

From our calculations, the WOF-SANS instrument exhibits the lowest $Q_{min}$ value among the instruments, which due to its focusing geometry is independent of sample size. When we consider the three pinhole collimation geometry instruments (ConvSANS, LoKI, and SKADI), it is seen that the instrument featuring the tightest collimation yields lower $Q_{min}$ values, as expected. Regarding the resolution, WOF-SANS emerges as the top performer across a significant portion of its $Q$-range, followed by ConvSANS, SKADI, and LoKI, in that respective order. WOF-SANS dominates at the low and middle $Q$-ranges while it has lower resolution at the higher $Q$ values. This is due to the wavelength resolution contribution dominating at high $Q$; WOF-SANS, despite its superior geometrical resolution (apparent at low $Q$), has only moderate wavelength resolution as a consequence of its short length. Furthermore, we see that the WOF-SANS instrument offers the largest dynamic $Q$-range, independent of sample size, an important attribute given its utilization of a single detector at a fixed position. This is a result of the unique focusing geometry of the instrument and a substantial wavelength bandwidth.

\begin{figure}[!h]%[!htbp]
  \centering
    \includegraphics[width=0.55\columnwidth]{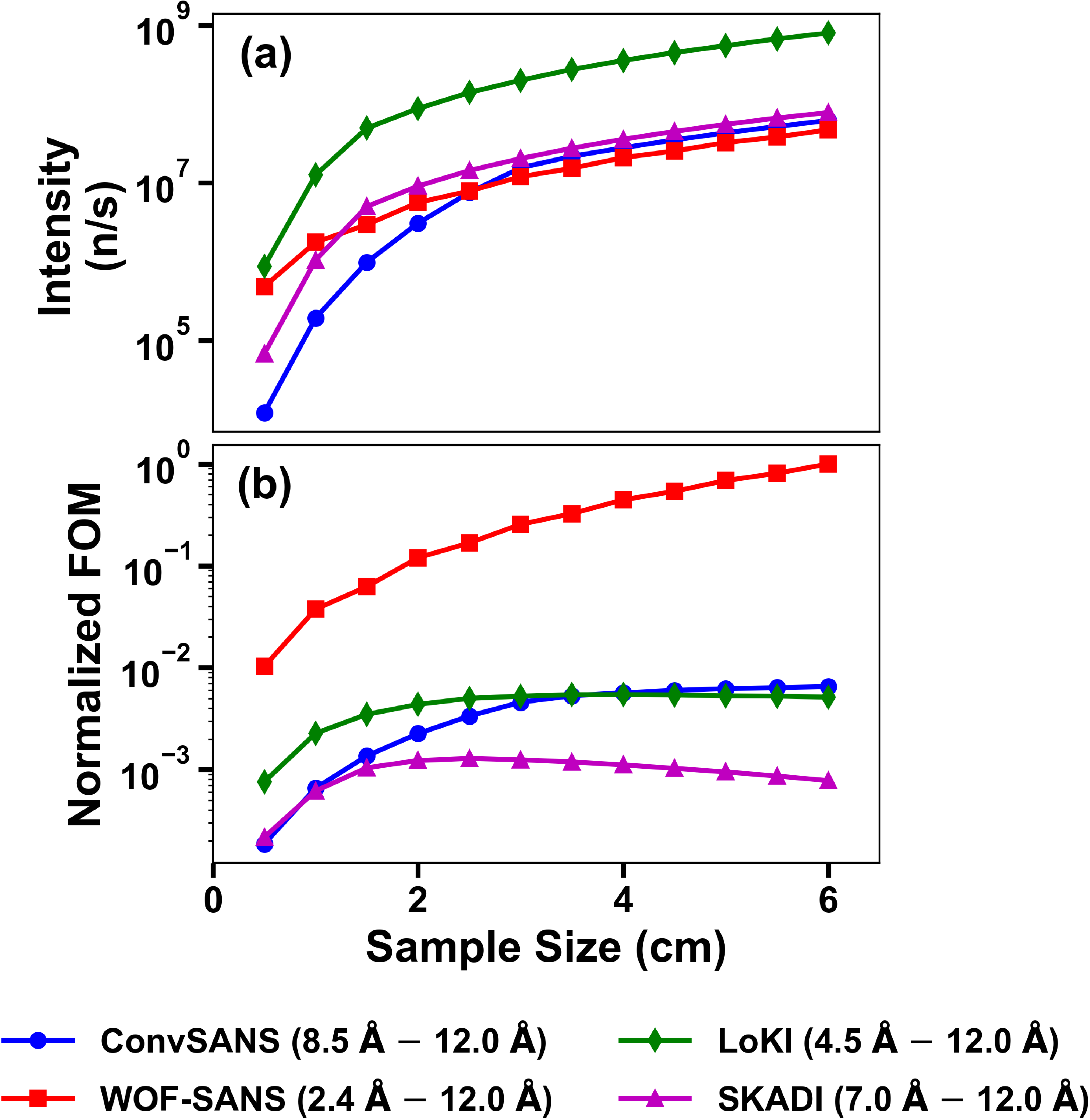}
  \caption{(a) Total neutron intensity at the sample position as a function of sample size for ConvSANS, WOF-SANS, LoKI, and SKADI. (b) Calculated normalized FOM of WOF-SANS, ConvSANS, LoKI, and SKADI as a function of sample size.}
  \label{fig:SANSspecra&FOM}
\end{figure}

The instruments are also compared in terms of neutron intensity on sample and FOM when considering their rearmost detector configuration, with the results given in \cref{fig:SANSspecra&FOM}. Compared to the ESS instruments LOKI and SKADI, the ConvSANS instrument sacrifices some flux especially at the smallest sample sizes to achieve the better resolution characteristics already discussed. The WOF-SANS instrument, overall offers relatively high intensity when considering the full optics cross-section. However, to maintain such intensity, the samples need to move along the beam with smaller samples positioned closer to the detector deteriorating both $Q_{min}$ and resolution. As such, it was decided that it is better to maintain high resolution and low $Q_{min}$ values by keeping the sample close to the optics and sacrificing intensity with the use of sample slits. The defined FOM is used to calculate a FOM curve as a function of sample size for all instruments. This comparison can be seen on \cref{fig:SANSspecra&FOM}b. From the results it is clear the WOF-SANS is significantly different from the three conventional instruments and in general has superior performance due to its high resolution and low $Q_{min}$.

\subsubsection{SANS Conclusion}
It was shown that both conceptual SANS instruments, ConvSANS and WOF-SANS, perform best on the largest 15\,$\times$\,15\,cm$^2$ moderator. Using this moderator, the ConvSANS instrument was able to fit into the suite of ESS SANS instruments by extending the minimum $Q$ and providing a slightly better resolution at the cost of some flux. The WOF-SANS with its different mode of operation, when also considering $Q$ ranges and corresponding resolution, was shown to perform significantly better than ESS instruments under construction.

\subsection{Optimization of imaging instrument}
\begin{figure}[tbh!]
    \centering
    \includegraphics[width=0.95\columnwidth]{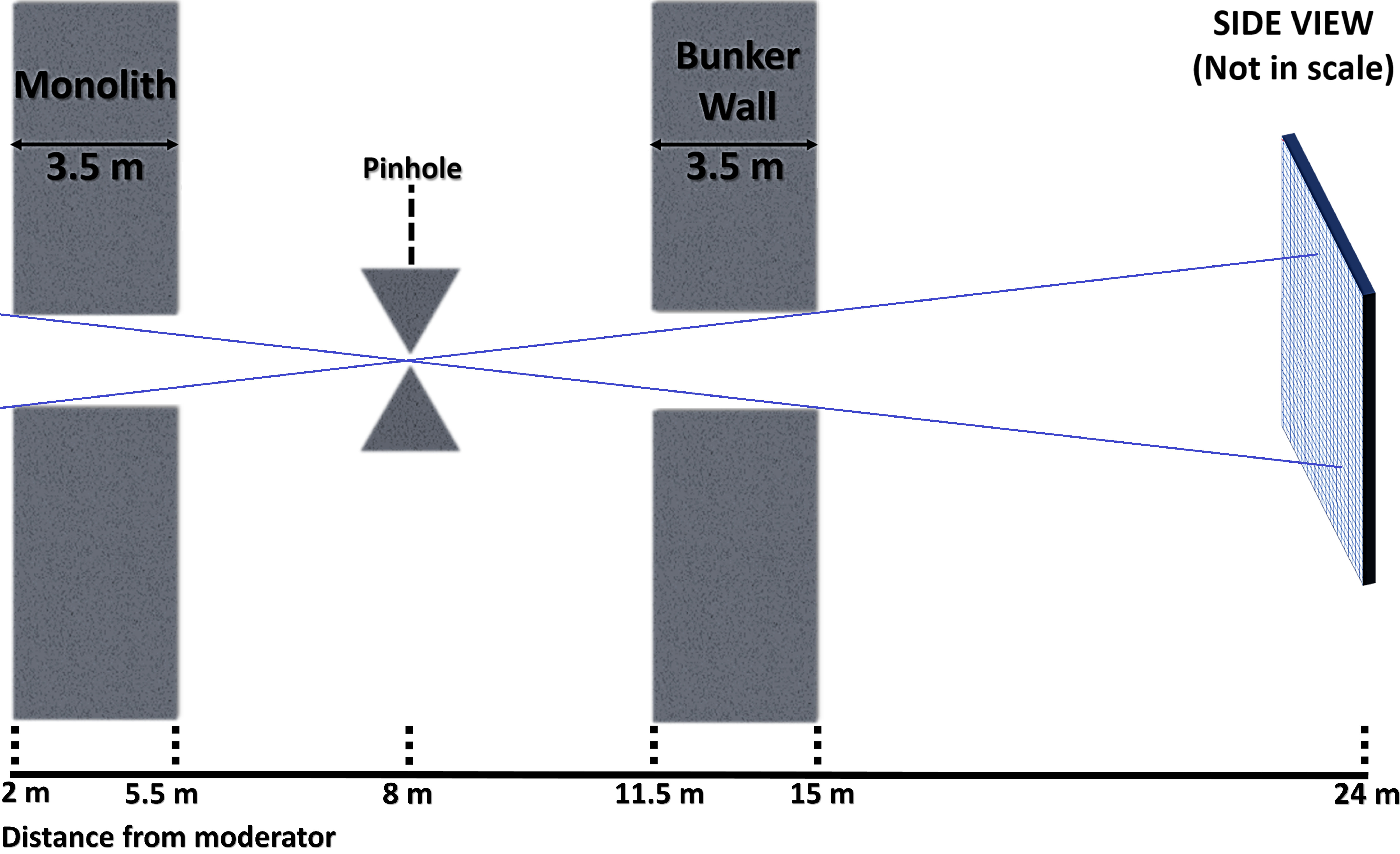}
    \caption{Overview of the imaging instrument geometry.}
    \label{fig:Imaging_CAD}
\end{figure}

The ESS instrument suite already has an imaging instrument under construction, namely ODIN \cite{andersen}, which uses a complex chopper system and a neutron guide to achieve enhanced wavelength resolutions necessary for Bragg edge analyses and other wavelength resolved imaging methods. Not all experiments will require these capabilities, and thus ODIN would be well complimented by a simple pinhole based imaging instrument that would exploit the larger and more homogeneous FoV from a larger moderator, and e.g., very moderate wavelength resolution. Given the known positions of the monolith and bunker, as well as the requirement for a wide FoV, it was determined that the pinhole should be located 8 meters from the moderator, with the detector situated up to 24 meters away. The geometry can be seen in \cref{fig:Imaging_CAD}. With the maximum distance between the detector and the pinhole being twice that between the pinhole and the moderator, the anticipated maximum FoV will be twice the width and height of the moderator.

\begin{figure}[tbh!]
    \centering
    \includegraphics[width=0.55\columnwidth]{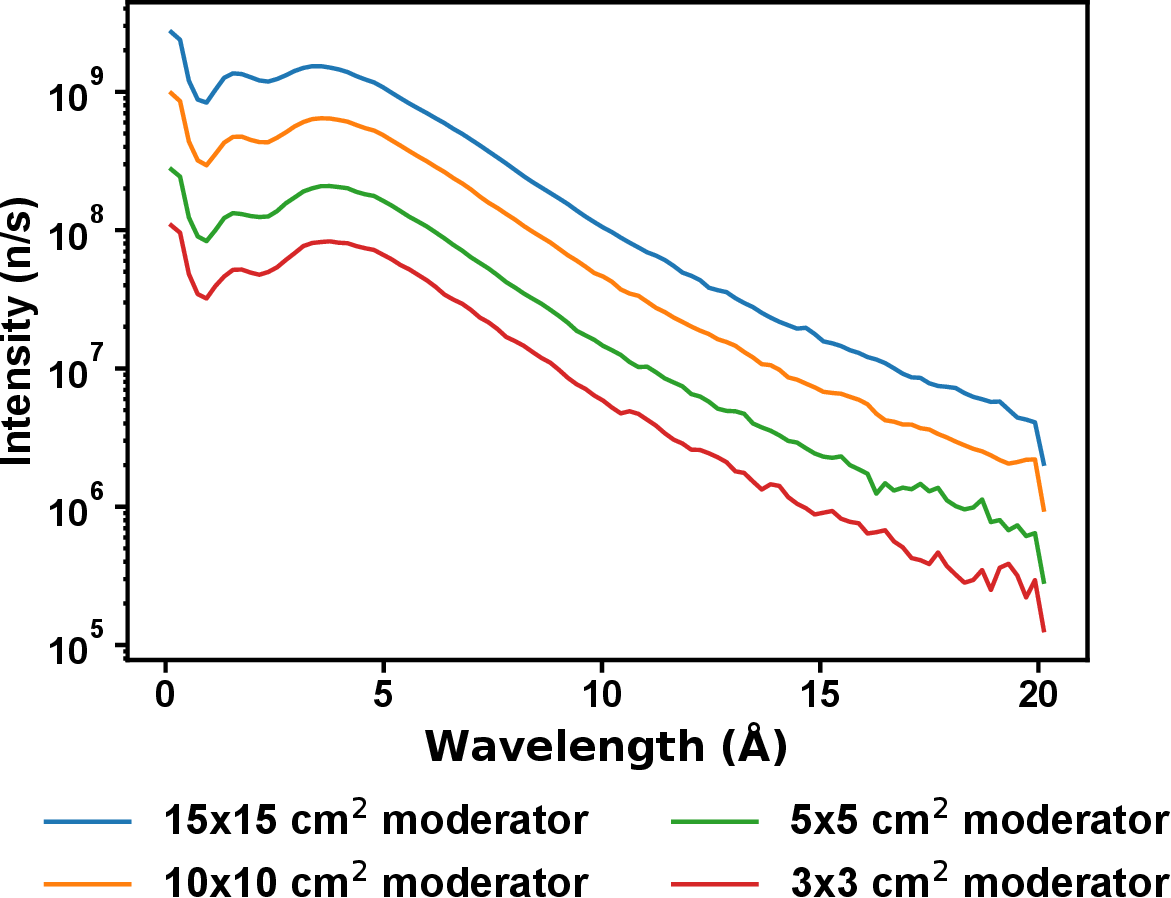}
    \caption{Intensity on detector for the imaging instrument when using a 2\,$\times$\,2\,cm$^2$ pinhole with each of the different proposed moderators}
    \label{fig:Imaging_intensity}
\end{figure}

The spectrum reaching the detector with a pinhole of 2\,$\times$\,2\,cm$^2$ can be seen in \cref{fig:Imaging_intensity} and scales with the moderator intensity as expected. In this work square pinholes are used over the usual circular counterpart in order to allow comparisons when using the moderator itself as a pinhole.

\subsubsection{FOM Definition}
For an imaging instrument there are several factors that should contribute to its FOM. In the following, intensity refers to the intensity on the detector, $L$ to the distance from pinhole to detector and $D$ to the size of the pinhole.

\begin{itemize}
    \item Intensity: The FOM should increase when the required exposure time is decreased, which is here determined by the intensity in the pixel on the detector with the lowest intensity within the defined FoV.
    \item FoV: the FOM would be higher for an instrument that can provide a larger FoV, so the FOM is multiplied by the FoV area. Unless otherwise specified, the FoV is found as the largest square area centered on the detector where no pixel is less than 75\% of the maximum intensity.
    \item Collimation: the FOM should be normalized to the collimation provided by the instrument, changing the $L/D$ ratio results in a $(D/L)^2$ change in the intensity. To normalize for the increase in intensity from relaxed resolution, the FOM is multiplied with $(L/D)^2$.
    \item Evenness: The FOM should decrease as the image becomes less flat. Here, we assess flatness using the standard deviation of intensity in the detector pixels within the defined FoV. This assessment is represented as the square root of the inverse of the standard deviation.

\end{itemize}
\vspace{3mm}
\noindent The full expression for the FOM of an imaging instrument used in this work can thus be written as,
\begin{equation}
 FOM = I_{minimum}\cdot FOV \cdot \left(L/D\right)^2 \cdot\sqrt{\frac{1}{std(I)}}. 
\end{equation}

\subsubsection{Imaging instrument performance}
The imaging instrument can adjust its collimation through changes to the pinhole and the position of the sample/detector along the beam outside the bunker (on an optical bench). For the purpose of this report we have investigated four different configurations, all of which at the outmost measurement position at 24\,m, as shown in \cref{table:NIparams1}.

\begin{table}[h]
\caption{Investigated distance and pinhole configurations for imaging instrument.}
\label{table:NIparams1}
\centering
\begin{tabular}{ccc}
\hline
L (cm) Pinhole detector distance & D (cm) pinhole size & L/D \\
\hline
1600~cm & 1~cm & 1600 \\
1600~cm & 2~cm & 800 \\
1600~cm & 3~cm & 533 \\
1600~cm & 5~cm & 320 \\
\hline
\end{tabular}
\label{table:pinhole}
\end{table}

\begin{figure}[tbh!]
    \centering
    \includegraphics[width=0.55\columnwidth]{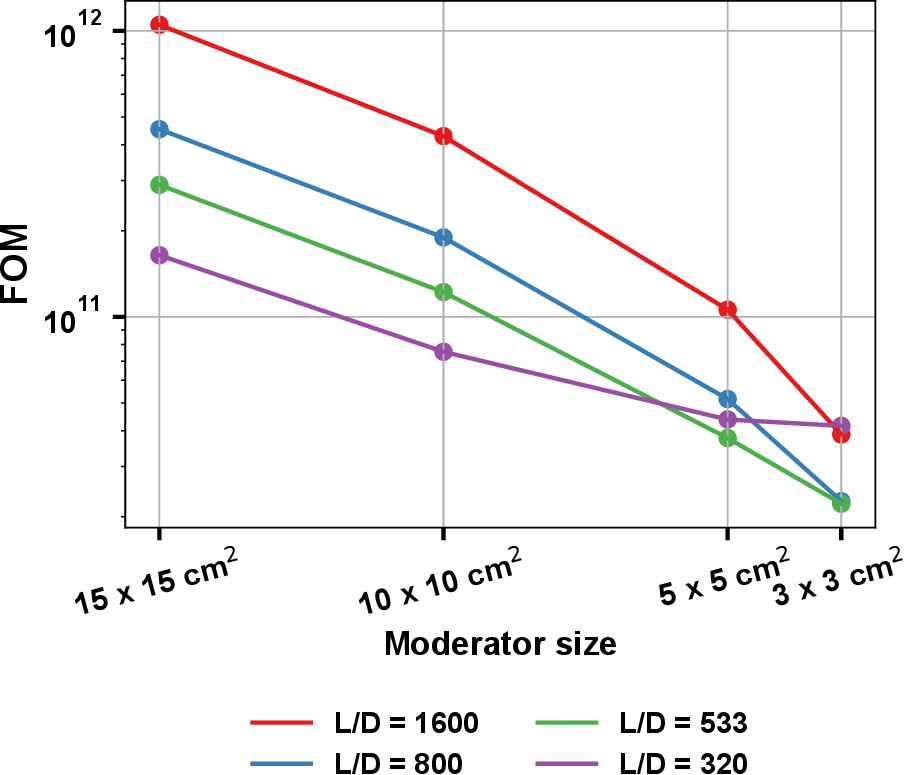}
    \caption{FOM for the imaging instrument with different pinhole sizes resulting in different $L/D$ values for all proposed moderators.}
    \label{fig:Imaging_FOM_fixed_view}
\end{figure}

If we assume that the optimal FoV for each configuration is twice the dimensions of the viewed moderator, we can evaluate the FOM for each moderator at each configuration. The result is shown in \cref{fig:Imaging_FOM_fixed_view}. Generally, the larger moderators perform better, and this trend is stronger at high collimation ratios than at low.

Since the two smallest moderators correspond directly to the pinholes used for the two lowest $L/D$ values, it is possible to reconfigure the instrument to not use a pinhole for these configurations as seen in \cref{table:NIparams2}.

\begin{table*}[!h]
\centering
\caption{Configurations of imaging instrument that use the moderator itself as pinhole definition when possible.}
\label{table:NIparams2}
\begin{tabular}{|c|cccc|}
\hline
\begin{tabular}[c]{@{}c@{}}Moderator\\ Size (cm$^2$)\end{tabular} & \multicolumn{1}{c|}{\begin{tabular}[c]{@{}c@{}}L/D\,=\,320\\ D\,=\,5\,$\times$\,5\,cm$^2$\end{tabular}} & \multicolumn{1}{c|}{\begin{tabular}[c]{@{}c@{}}L/D\,=\,533\\ D\,=\,3\,$\times$\,3\,cm$^2$\end{tabular}} & \multicolumn{1}{c|}{\begin{tabular}[c]{@{}c@{}}L/D\,=\,800\\ D\,=\,2\,$\times$\,2\,cm$^2$\end{tabular}} & \begin{tabular}[c]{@{}c@{}}L/D\,=\,1600\\ D\,=\,1\,$\times$\,1\,cm$^2$\end{tabular}\\ 
%\hline
\hline
15\,$\times$\,15 & & & \multicolumn{2}{c|}{\multirow{3}{*}{\begin{tabular}[c]{@{}c@{}}Pinhole 8\,m from moderator \\ Detector 24\,m from moderator\\ L\,=\,16\,m\end{tabular}}}\\ \cline{1-1}
10\,$\times$\,10 & & & \multicolumn{2}{c|}{}\\ \cline{1-2}
5\,$\times$\,5 & \multicolumn{1}{c|}{\begin{tabular}[c]{@{}c@{}}No pinhole\\ Detector 16\,m from moderator\\ L\,=\,16\,m\end{tabular}} & & \multicolumn{2}{c|}{}\\ \cline{1-3}
3\,$\times$\,3 & \multicolumn{1}{c|}{\begin{tabular}[c]{@{}c@{}}No pinhole, D\,=\,3\,$\times$\,3\,cm$^2$\\ Detector 9.6\,m from moderator\\ L\,=\,9.6\,m\end{tabular}} & \multicolumn{1}{c|}{\begin{tabular}[c]{@{}c@{}}No pinhole\\ Detector 16\,m from moderator\\ L\,=\,16\,m\end{tabular}} & &\\ \hline
\end{tabular}
\end{table*}

\begin{figure}[tbh!]
    \centering
    \includegraphics[width=0.55\columnwidth]{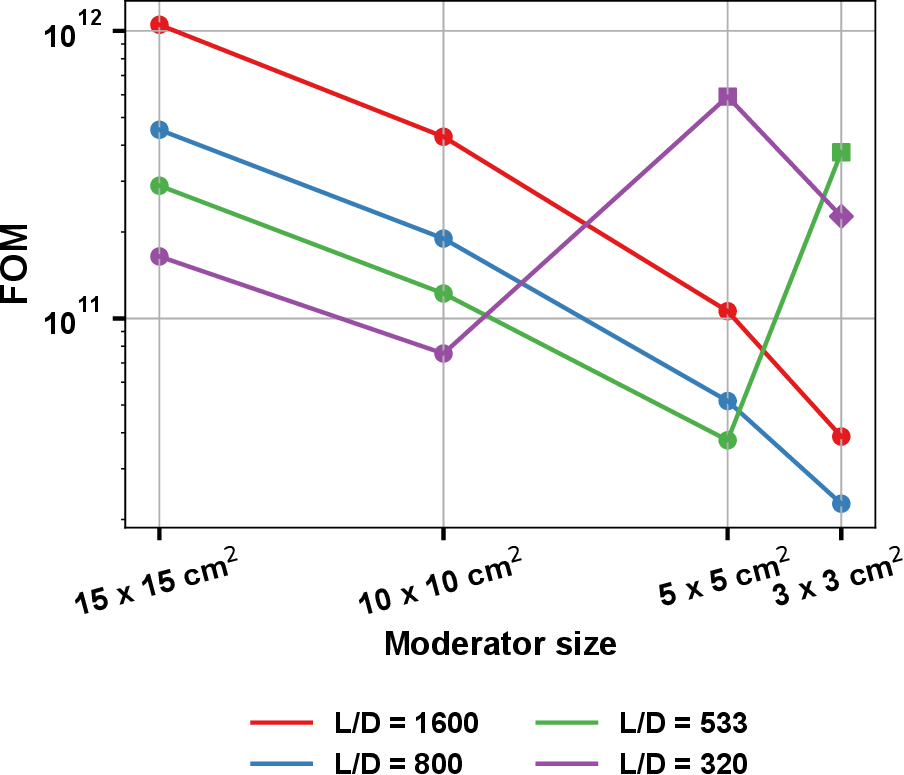}
    \caption{FOM for the imaging instrument with different pinhole sizes resulting in different $L/D$ for all proposed moderators. The moderator was used as pinhole when feasible, these cases are shown as square points. The single diamond configuration represents the case where the distance had to be reduced to 9.6\,m to achieve desired $L/D$.}
    \label{fig:Imaging_no_slit}
\end{figure}

The configurations using a 16\,m distance from the moderator would be limited by restrictions in allowed size of gaps in the bunker wall, which ends 15\,m from the moderator. Here we assume that these gaps can be increased by additional shielding around the instrument cave. The lowest $L/D$ of 320 for the 3\,$\times$\,3\,cm$^2$ moderator can only be achieved by moving the detector position into the bunker. Even though this is most likely not feasible, we included it for completeness.
Assuming the FoVs would be twice the one of the moderator, these configurations result in the FOM shown in \cref{fig:Imaging_no_slit}. Note that the low collimation options for the smallest moderators now achieve much higher FOM for some lower collimation configurations, but struggle with high collimation.

The feasibility of using the moderator as pinhole is doubtful as neutrons could scatter close to it and still reach the sample, blurring the image. The required opening in the shielding would be much larger than the other configurations, resulting in severely increased background. The configurations using the moderator as pinhole were studied for completeness, yet it was decided not to continue the investigation despite their impressive low collimation performance.

\begin{figure}[tbh!]
    \centering
    \includegraphics[width=0.99\columnwidth]{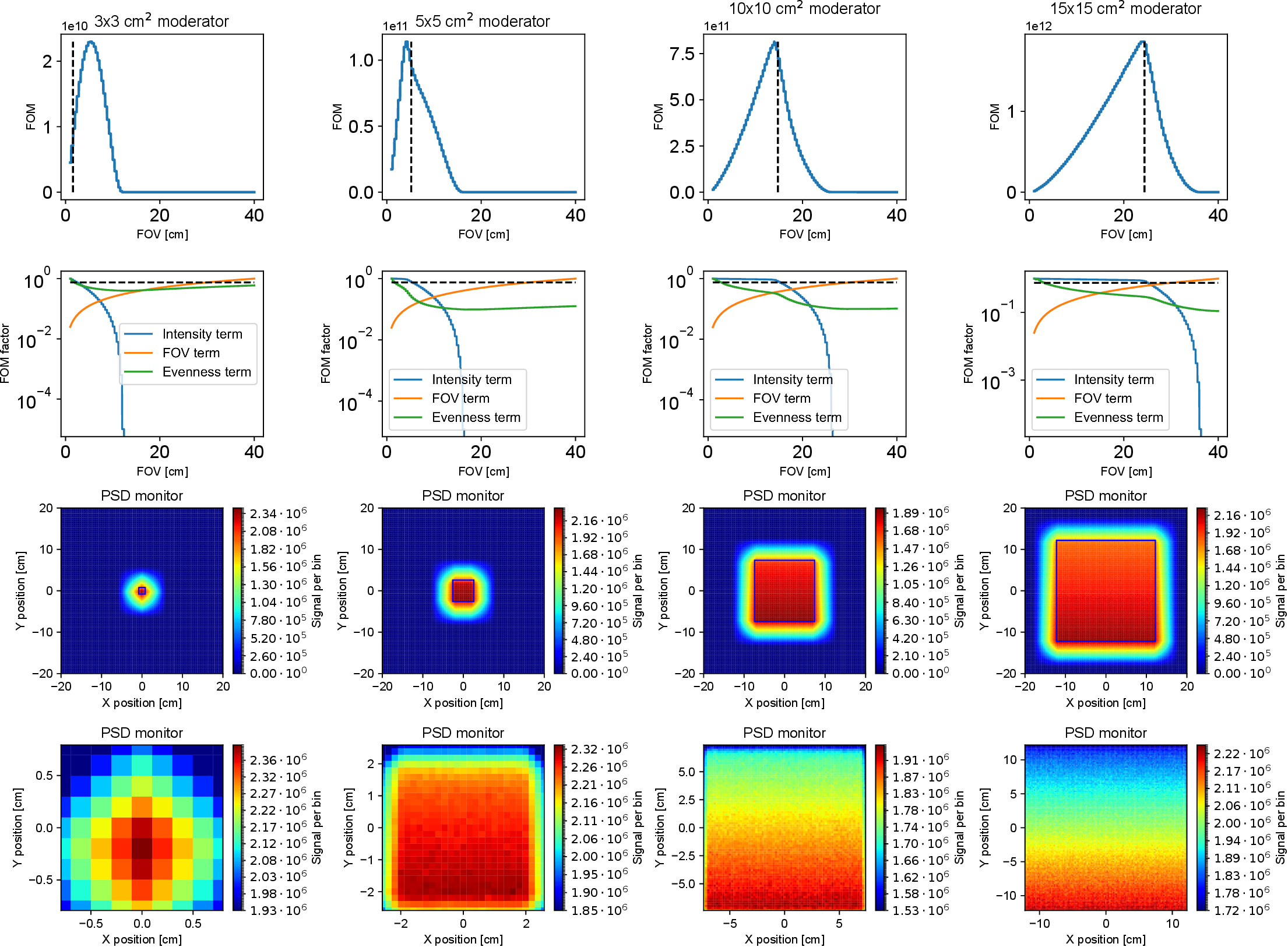}    
    \caption{Example of FOM calculation for all moderators with 2\,$\times$\,2\,cm$^2$ pinhole corresponding to a $L/D$ of 800. Each column corresponds to a moderator size. The top graph is the FOM as function of used FOV with the dashed line showing the used FOV where falloff of 75\,\% of maximum intensity is used. The second row is a decomposition of the factors included in the FOM along with the ratio between lowest and highest intensity. The third row is the full detector with a black box showing the FOV. The last row is the FOV cut from the detector showing the homogeneity, notice the limited color scale.}
    \label{fig:Imagine_analysis}
\end{figure}

So far the FoV has been assumed to be twice that of the moderator. The beam for each configuration is now examined to find the FoV corresponding to the intensity at the edge falling to no less than 75\% of the maximum intensity. An example of such an investigation is seen in \cref{fig:Imagine_analysis}. The smallest moderator is almost the same size as the pinhole, and thus suffers in terms of evenness, yet the larger moderators each have a successively larger homogenous area. They also have increased brightness in the area nearest the target at top of the moderator, which after inversion from the pinhole camera correspond to a larger intensity near the bottom of the detector, detracting from the flatness of the distribution. The same procedure is followed for all configurations.

\begin{figure}[tbh!]
    \centering
    \includegraphics[width=0.85\columnwidth]{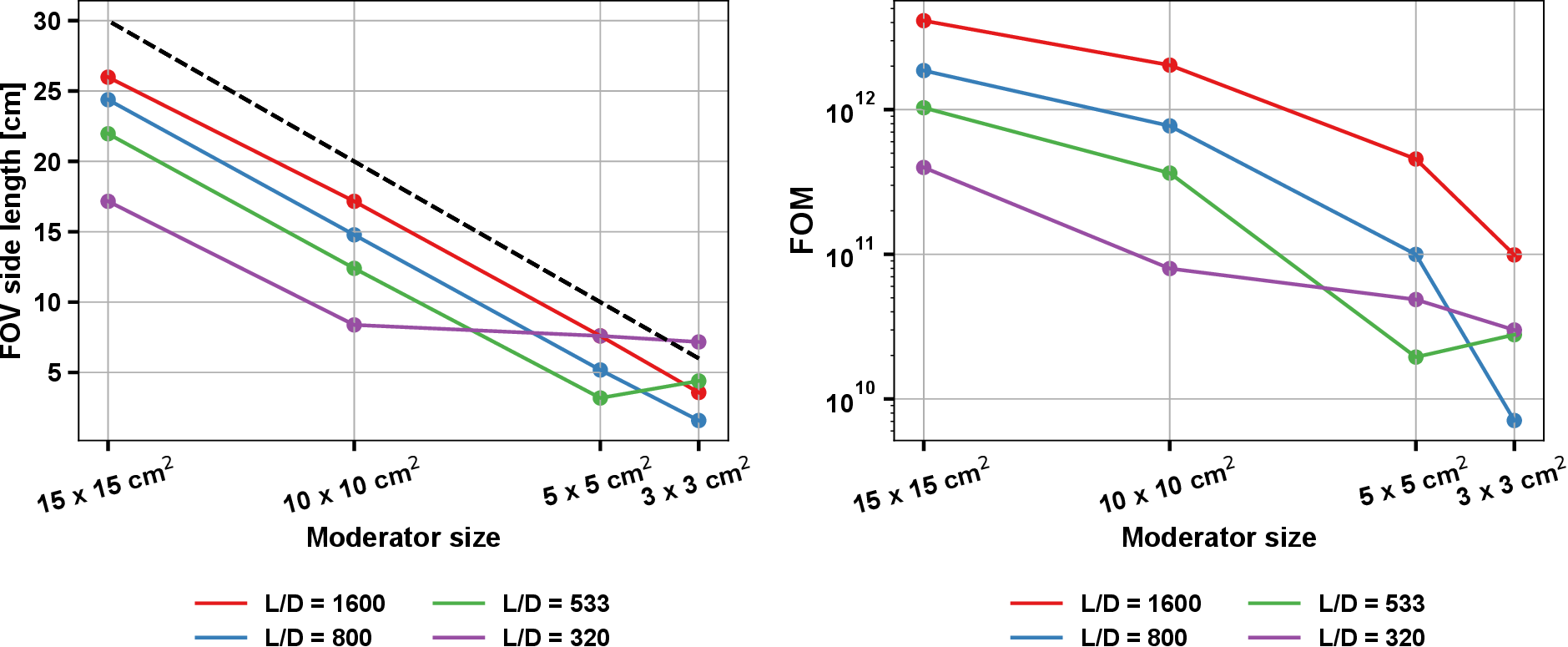}
    \caption{Left: FoV where the minimum intensity is 75\,\% of the maximum for all configurations of moderators and pinholes. The dashed line is twice the moderator size for comparison. Right: The corresponding FOM for all configurations of moderators and pinholes.}
    \label{fig:Imaging_variable_FOM}
\end{figure}

If we plot the achieved FoV and corresponding FOM for each configuration of pinhole and moderator size (\cref{fig:Imaging_variable_FOM}), it can be seen, that the found FoV is less than twice the moderator size, except for a single configuration of the smallest moderator.
The overall trend in the FOM is similar to the fixed fields of view with pinholes in all configurations, and thus we conclude that the fixed field version is sufficient for further analysis.

\subsubsection{Performance comparison to ESS instrument}
\begin{figure}[!h]%[!htbp]
  \centering
    \includegraphics[width=0.55\columnwidth]{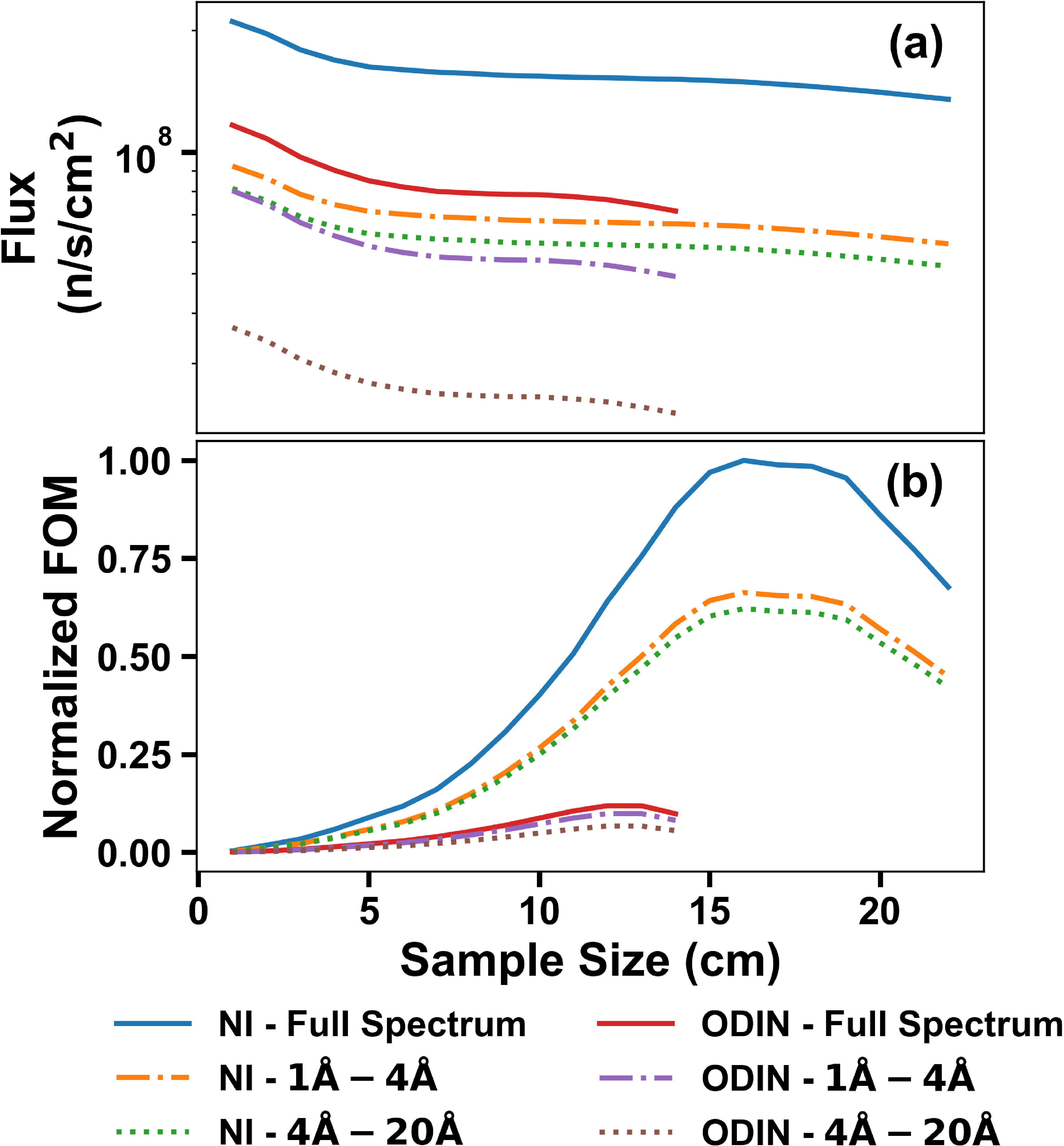}
  \caption{(a) Total neutron flux at the sample position of NI and ODIN as a function of sample size. (b) FOM of NI and ODIN as a function of sample size. All simulations and  calculations were done for $L/D$\,=\,300 and for three spectral regions: full spectrum, 1\,\AA\,--\,4\,\AA, 4\,\AA\,--\,20\,\AA.}
  \label{fig:NIresults}
\end{figure}

In order to investigate whether the conceptual imaging instrument is relevant for ESS it is compared to the ODIN imaging instrument already under construction. Here it is assumed, that the 15\,$\times$\,15\,cm$^2$ moderator is chosen for the conceptual instrument, while the performance of ODIN is calculated using the bispectral extraction of the ``butterfly'' moderator with same proton power on target. For this comparison it is important to note that ODIN uses a guide to extend the instrument length and thus significantly improves energy resolution, along with a comprehensive chopper setup that allows it to customize this resolution on demand. Such options are much more limited for the conceptual instrument without a guide. A comparison of flux as a function of sample size/FoV for a nominal $L/D$\,=\,300 and for three spectral regions can be seen in \cref{fig:NIresults}a from which it is clear that the conceptual instrument achieves a higher flux when comparing same spectral regions. The maximum FoV/sample size for each instrument is chosen where the flux is at about 75\,\% peak flux in the penumbra. 

The FOMs used in this section for the two instruments are also compared as a function of sample size, and the results are shown in \cref{fig:NIresults}b. The conceptual imaging instrument achieves a higher FOM over the FoV/sample size range, which is primarily caused by the lack of energy resolution in the FOM. This is however justified by the requirement of optimizing an instrument that should compliment ODIN. While the conceptual "simple" imaging instrument can handle the presumably large subset of experiments that do not require energy resolution with significantly shorter measurement times, ODIN could focus on experiments, where customizable energy resolution is necessary.

\newpage
\newpage
\subsection{Achromatic optics for long wavelengths}
\label{prismasec}

Typically, condensed matter instrumentation for neutron scattering can ignore the effect of gravity. When going to the VCN region, though, this becomes a serious issue. While they can be transported with traditional guides, any long collimation section or focusing can be problematic. A part of the HighNESS project was studying how prisms could be used to counteract gravity when used in conjunction with focusing optics.

\begin{figure}[tb!]
    \centering
    \includegraphics[width=0.7\columnwidth]{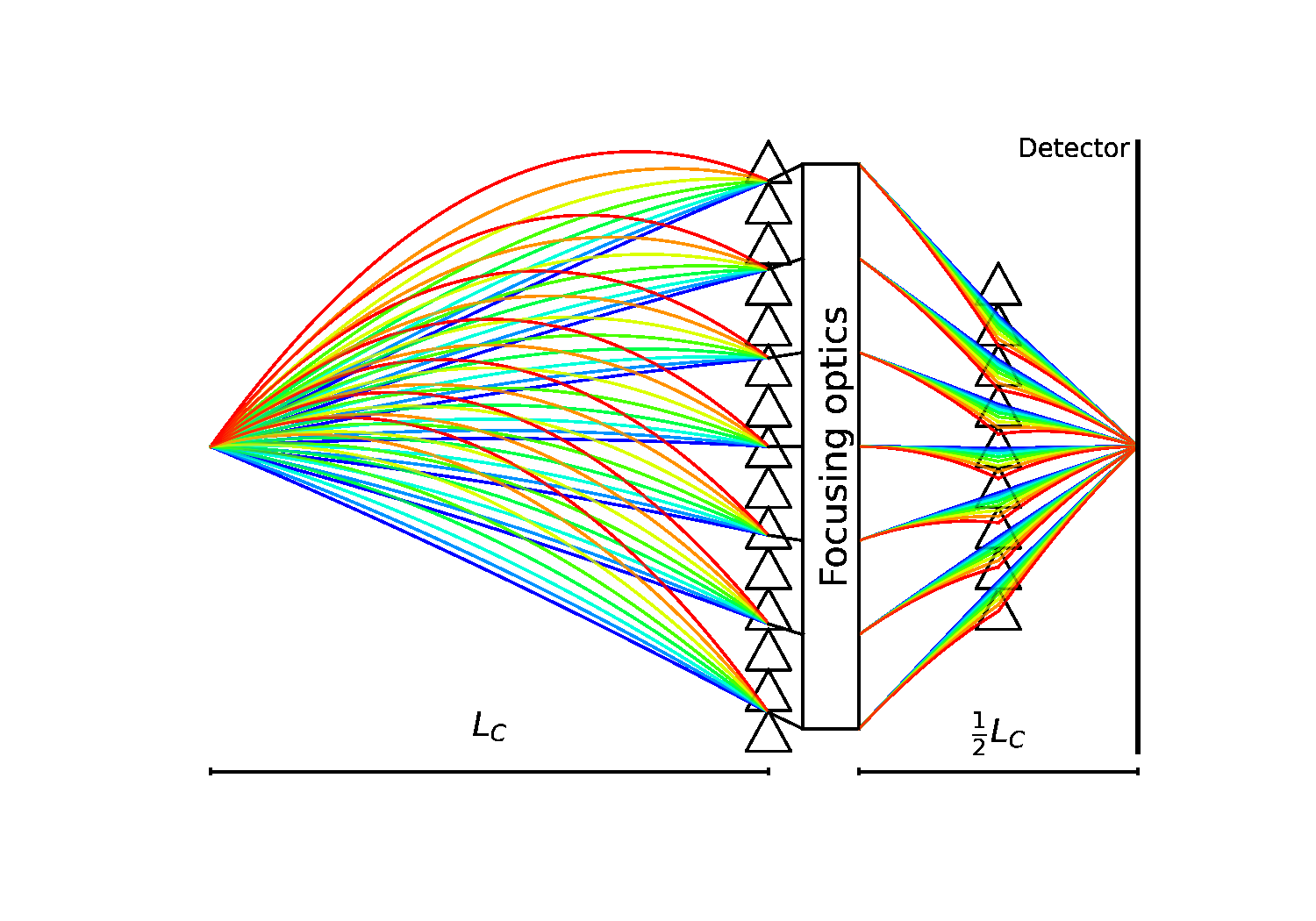}
    \caption{Diagram of two prisms canceling gravity between them. The distance has to correspond to the characteristic length of the prisms, which depend both on the material and shape. (Not to scale and not illustrating the actual path within the prisms.)}
    \label{fig:characteristic_length}
\end{figure}

\begin{figure}[tb!]
    \centering
    \includegraphics[width=0.7\columnwidth]{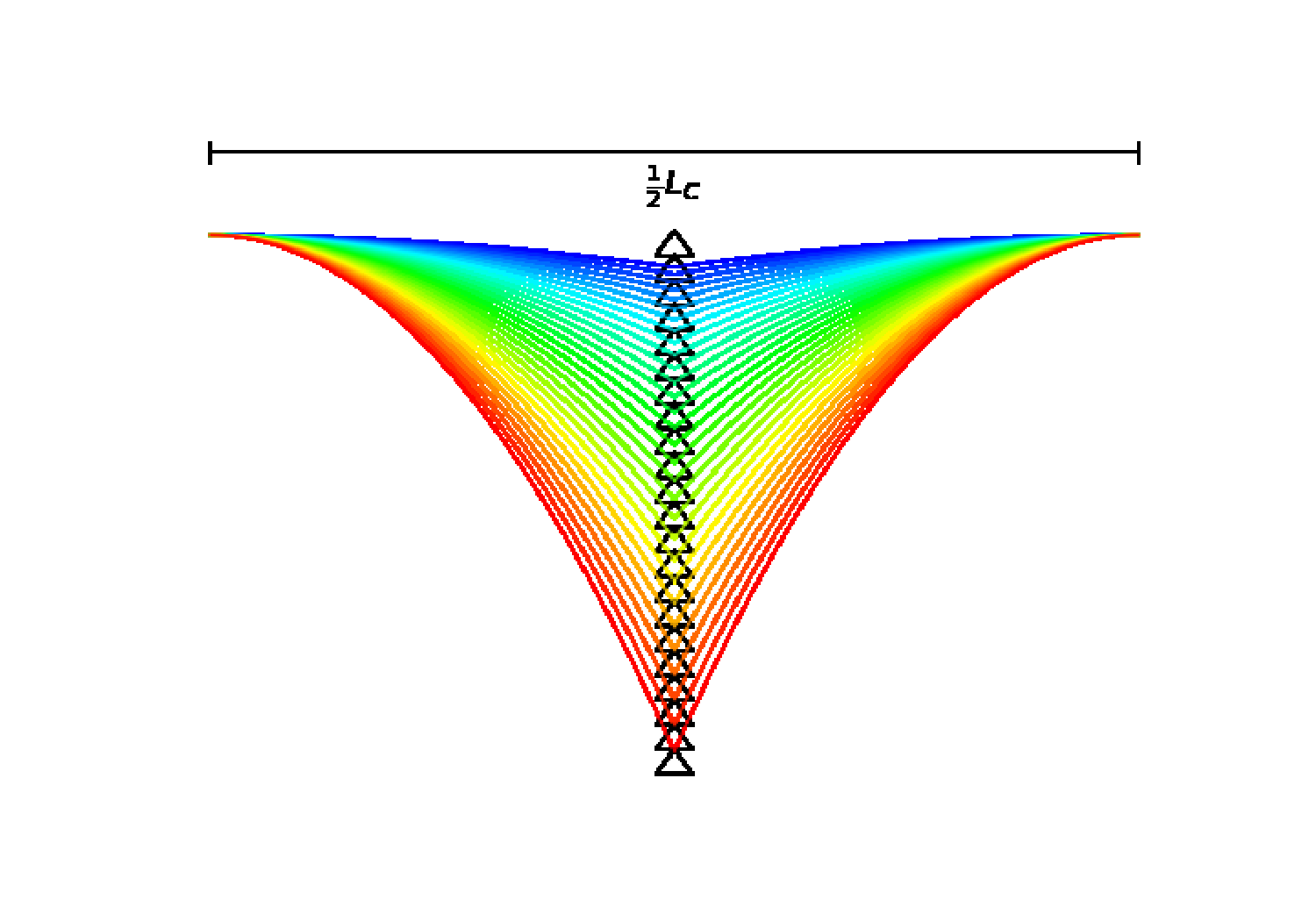}
    \caption{Diagram of prism stack placed centrally cancelling the effect of gravity over half a characteristic length of the prisms used (Not to scale and not illustrating the actual path within the prism.)}
    \label{fig:half_characteristic_length}
\end{figure}

A well-collimated white beam incident on a prism will be split according to wavelength such that all the trajectories meet at a later point. The distance to this point corresponds to the properties of the prism, both its geometry and material, and is here called the critical distance for a prism. At this point the beam will be divergent, but can be recombined with a second prism, as seen in \cref{fig:characteristic_length}. For some applications it is unwanted to have a prism close to the destination of a beam. Here an alternative geometry can be used with a stack of prism halfway between the origin and target, though the distance covered in this configuration is only half the characteristic length, see \cref{fig:half_characteristic_length}.

These configurations have been tested with the Monte Carlo ray-tracing simulation tool McStas and work as expected, even for beams with some divergence, which act as if there was no gravity in the corrected region.

\begin{figure}[tb!]
    \centering
    \includegraphics[trim = 0mm 25mm 0mm 0mm, width=0.8\columnwidth]{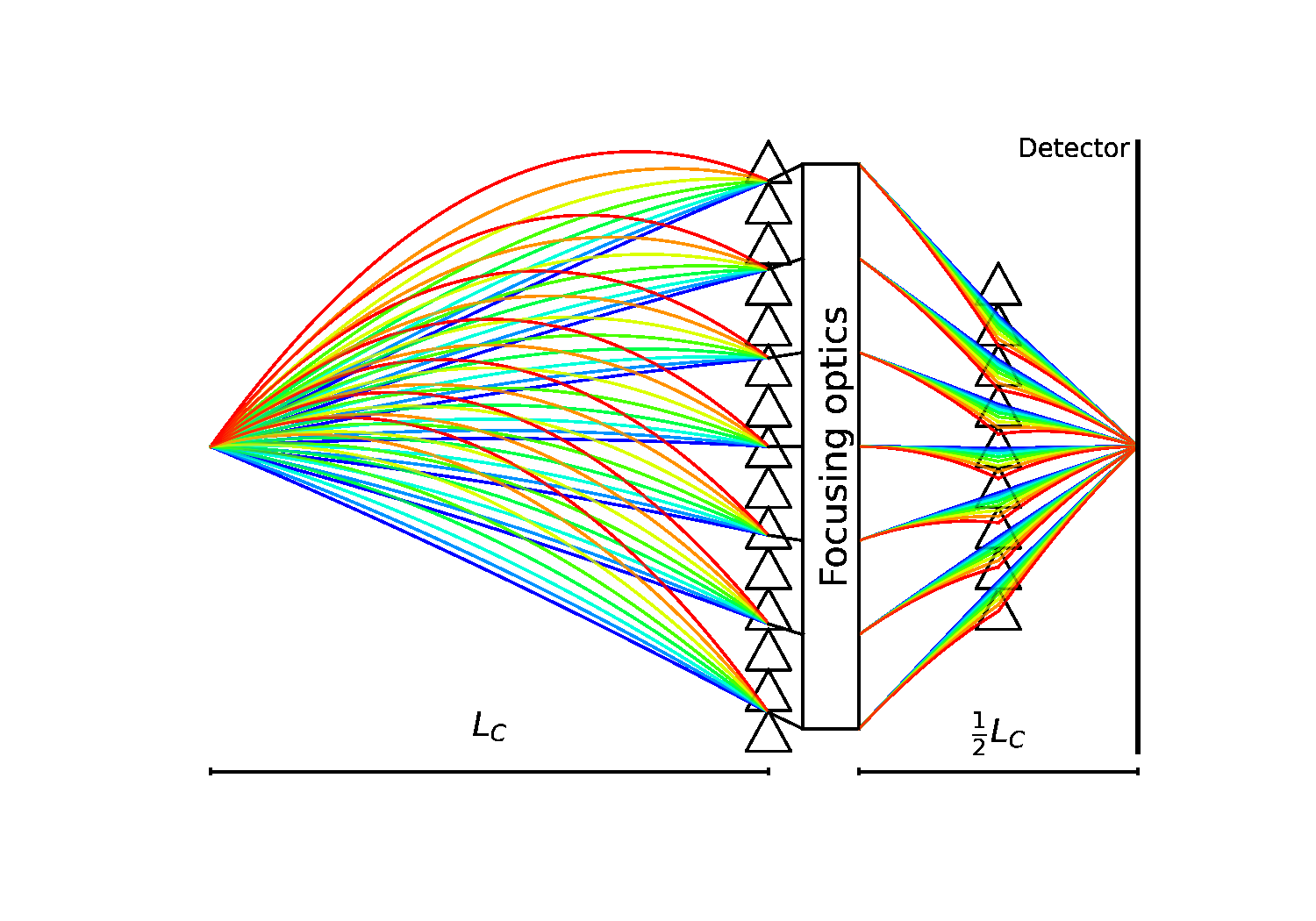}
    \caption{Illustration of Wolter optics with correcting prisms. Blue paths are low wavelength with little gravity impact and the red color has the highest wavelength and large impact from gravity.}
    \label{fig:prism_optics}
\end{figure}

To demonstrate this further we combine these techniques with focusing Wolter optics. As they have limited depth, the majority of the flight path would be in free space where the prism systems can counteract the vertical displacement due to gravity. This system is shown in \cref{fig:prism_optics}. The instrument starts with a point source and is followed by a stack of Be prisms after 4.88~m which correspond to their characteristic length. The focusing optics are placed right after the prisms. Gravity correction of the focused beam is performed with a stack at the halfway point, again using Be prisms. The focal points of the focusing optics are thus on the source and 2.44~m after the end of the focusing optic.

\begin{figure}[tb!]
    \centering
    \begin{subfigure}[b]{0.45\textwidth}
    \includegraphics[width=0.96\columnwidth]{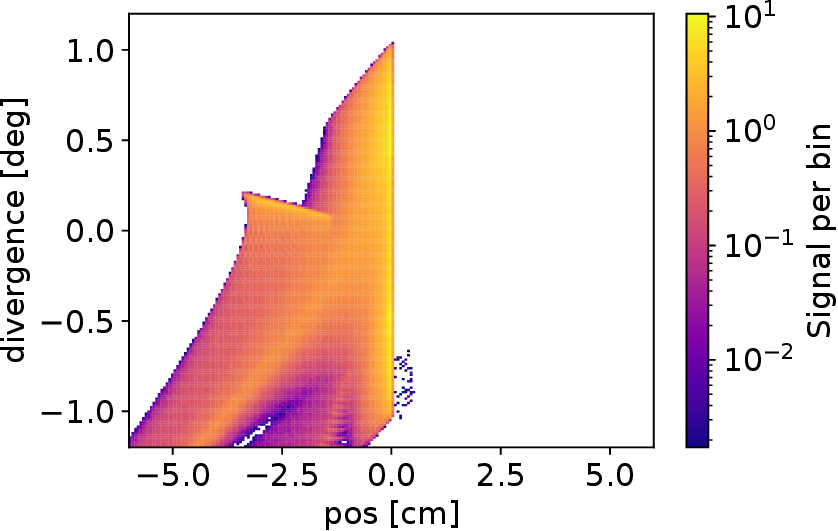}
    \caption{Without correcting prisms.}
    \label{fig:normal_acceptance_beam}
    \end{subfigure}
    \hfill
    \begin{subfigure}[b]{0.45\textwidth}
    \includegraphics[width=0.96\columnwidth]{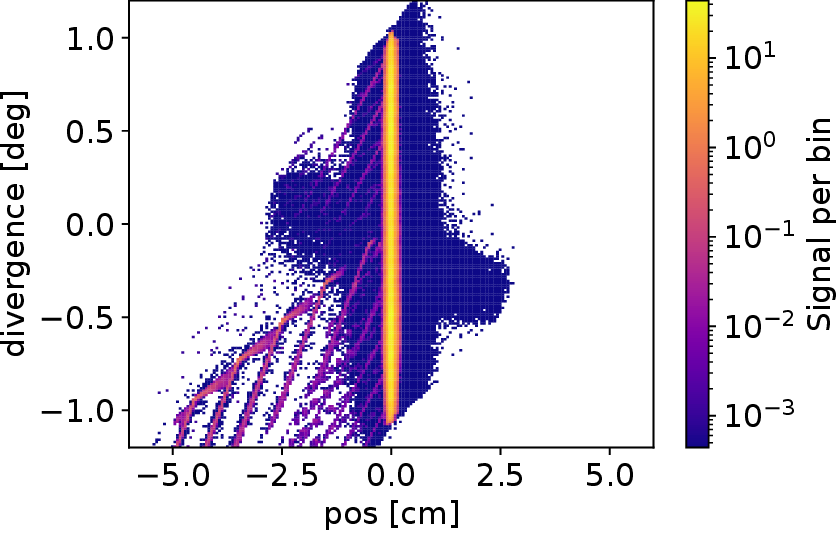}
    \caption{With correcting prisms.}
    \label{fig:prism_acceptance_beam}
    \end{subfigure}
    \caption{Acceptance diagram of vertical divergence and position at the detector with a white beam 10--70\,\AA{}}
\end{figure}

\begin{figure}[tb!]
    \centering
    \begin{subfigure}[b]{0.45\textwidth}
    
    \includegraphics[width=0.96\columnwidth]{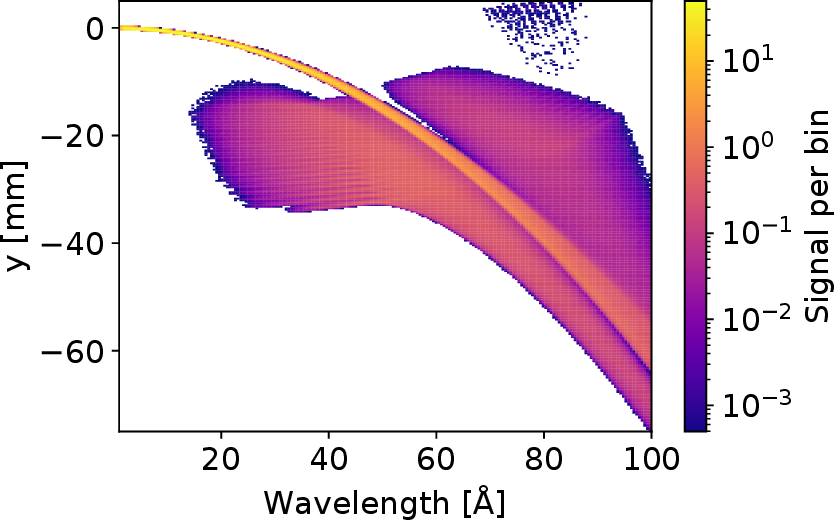}
    \caption{Without correcting prisms.}
    \label{fig:normal_wavelength_drop}
    \end{subfigure}
    \hfill
    \begin{subfigure}[b]{0.45\textwidth}
    \includegraphics[width=0.96\columnwidth]{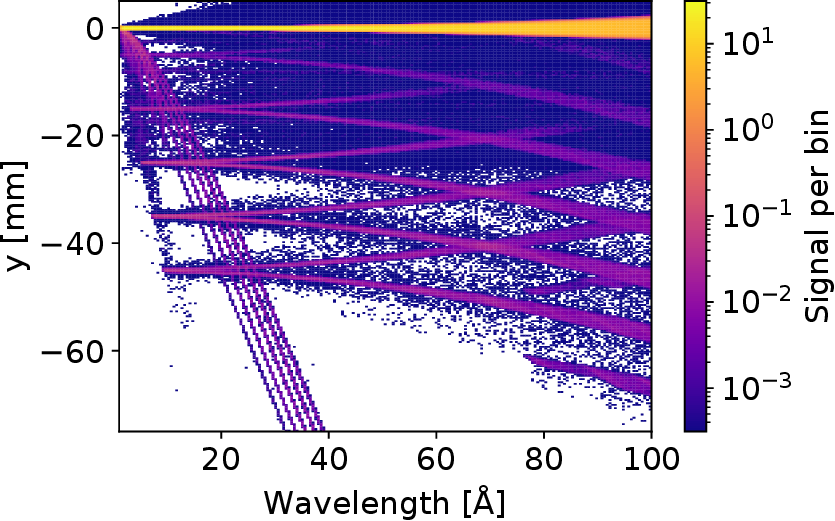}
    \caption{With correcting prisms.}
    \label{fig:prism_wavelength_drop}
    \end{subfigure}
    \caption{Height distribution of the focused beam as a function of wavelength.}
\end{figure}

In \cref{fig:normal_acceptance_beam} and \cref{fig:prism_acceptance_beam} the beam quality of this system is compared to a version without the correcting prims. The difference in beam quality is stark, with the uncorrected version having a drop and complex correlations between position and divergence. The corrected instrument does have artifacts, these are however at intensities 3 orders of magnitude lower than the intended beam. The beam drop as a function of wavelength is investigated in the same manner in \cref{fig:normal_wavelength_drop} and \cref{fig:prism_wavelength_drop}. Again, it becomes evident that even over such small distances, correcting measures are necessary for the VCN wavelength range.

Such an optical system can be a part of the solution for creating efficient and accurate instrumentation that functions over a very broad wavelength range. The system could for example be used to create a SANS instrument, where additional length could be achieved with multiple columns of prisms. These preliminary studies are promising and warrant further investigation in future research.

\subsection{Neutron Scattering instrumentation summary and cost estimation}
\begin{figure}[tbh!]
    \centering
    \includegraphics[width=0.55\columnwidth]{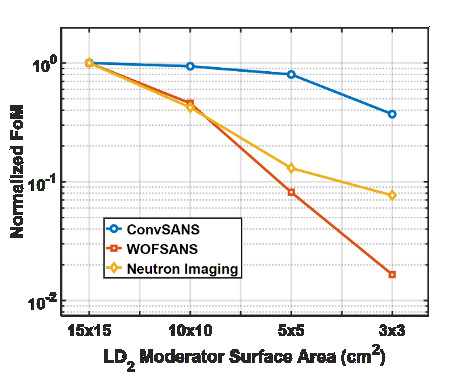}
    \caption{Normalized FOM as function of moderator size for the three instruments, all normalized to the 15\,$\times$\,15\,cm$^2$ moderator. The SANS instruments are for $Q=1.0\cdot10^{-4}$ Å$^{-1}$ while the imaging instrument is fixed FoV, L/D=533 and with pinholes for all moderator sizes.}
    \label{fig:FOM}
\end{figure}

This section investigated the performance of three conceptual instruments, a conventional SANS instrument, a Wolter optics-based SANS instrument and an imaging instrument for four moderator candidates of varying size. The smaller moderators had a higher brightness, but in contrast to the similar survey for the ``butterfly'' moderator, all investigated instruments preferred the largest moderator as seen \cref{fig:FOM}. The trade-off between brightness and size for the investigated liquid deuterium moderator was simply smaller than for the hydrogen-based ``butterfly'' moderator, resulting in the opposite outcome. The large liquid deuterium moderator thus complements the ``butterfly'' moderator well as it provides a large surface with a similar brightness at wavelengths over 5 Å, and is especially suited to instruments with performance scaling driven by moderator size or neutron intensity.

It was shown that when using the proposed 15x15~cm$^2$ moderator, the conceptual instruments are all highly competitive with the ESS instruments under construction and are complementary as they avoid covering exactly the same usecases while offering new capabilities. A rough estimate of the cost for each instrument can be found in \cref{table:cost}

\begin{table}[h]
    \centering
        \caption{Cost estimates of conceptual instruments.}
    \begin{tabular}{lc} 
        \hline 
        Instrument & Cost estimate [M\EUR{}] \\
        \hline
        ConvSANS & 20-30 \\
        WOFSANS & 20-30 \\
        Imaging instrument & 15-20 \\        
        \hline
    \end{tabular}
    \label{table:cost}
\end{table}

Furthermore, it was investigated how to counteract gravity when transporting neutron beams with a large wavelength range; here, prisms were shown to work well in conjunction with Wolter focusing optics.

%% file: tsl.tex
\section{Thermal Scattering Libraries}
\label{sec:tsl}
\subsection{Introduction}
Here we present the development of simulation software for describing neutron interactions in novel moderator and reflector materials considered of interest for the HighNESS project. The main interest is the development of software in order to describe neutron interactions in NDs, magnesium hydride, graphitic compounds with extended Bragg-edges compared to normal graphite, and the clathrate hydrates. To support this work, several methods were developed to include new physics processes, such as small-angle neutron scattering and magnetic scattering, into the Monte-Carlo simulation process.

Traditionally, thermal neutron scattering data for Monte-Carlo simulations are usually distributed as part of the thermal scattering sublibrary (TSL) of major evaluated nuclear data libraries (ENDF/B \cite{ref_endf8}, JEFF \cite{ref_jeff} and JENDL \cite{ref_jendl}). In the past decade, these libraries have seen an increase in activity related to the development of the TSL, in part due to the work of the Subgroup 48 of the Working Party for Evaluation Cooperation \cite{SG42}. Nevertheless, the coverage of materials, and particularly of materials of interest in the development of cold neutron sources, is limited. 

In addition, these libraries are also restricted by the ENDF format \cite{Herman2009}. In its current version, elastic scattering can be either represented using a coherent or incoherent model. In December 2020, a format change was proposed \cite{Zerkle2020} to include both components in what is called \emph{mixed elastic} scattering. To make use of this format, the tool that is currently freely available for the generation of thermal scattering libraries, the LEAPR module of NJOY\cite{Macfarlane2017}, must be modified and recompiled to support the calculation of coherent scattering in new materials. In addition, there is no simple way to include additional physics processes such as small-angle neutron scattering and magnetic scattering through this approach.

To overcome these limitations, we investigated a number of approaches for including improved thermal neutron scattering data in Monte-Carlo simulations. 
These include approaches based on extended thermal neutron scattering libraries in a modified ENDF format in addition to calling the code NCrystal \cite{Cai2020,CAI2019400,kittelmann2021elastic} on-the-fly during a Monte-Carlo simulation. 

Each of the above mentioned developments used input based on state-of-the art molecular modelling systems. This included the usage of both density-functional-theory (DFT) techniques in addition to classical molecular dynamics. 

An important part of the process of development of new models and tools is benchmarking against experimental data. Data for this purpose is in particular very limited for materials of interest to very cold and ultracold neutron applications. For this reason, we also carried out an experimental campaign at the BOA instrument at the Paul Scherrer Institute to explore its possible use as a neutron beamline for this purpose and also to investigate in particular extend Bragg-edge graphitic compounds.

In the following, we first present the molecular modelling simulations, followed by advancements in nuclear libraries and extension of NCrystal to include new physics, and finally work on benchmarking and investigations with the BOA instrument.

\subsection{Molecular Modelling}
To support the development of new neutron scattering models, extensive molecular modelling calculations were carried out using a variety of techniques. The main outcomes of such calculations are the structure, in the form of the atomic positions, and the phonon frequency distribution. These two quantities can be used as input to the neutron scattering models. 

\subsubsection{Molecular modelling for MgH$_2$ / MgD$_2$}
\label{sec:md-mgh2-mgd2}
\begin{figure}[h!]
	\centering
	\includegraphics[width=0.6\columnwidth]{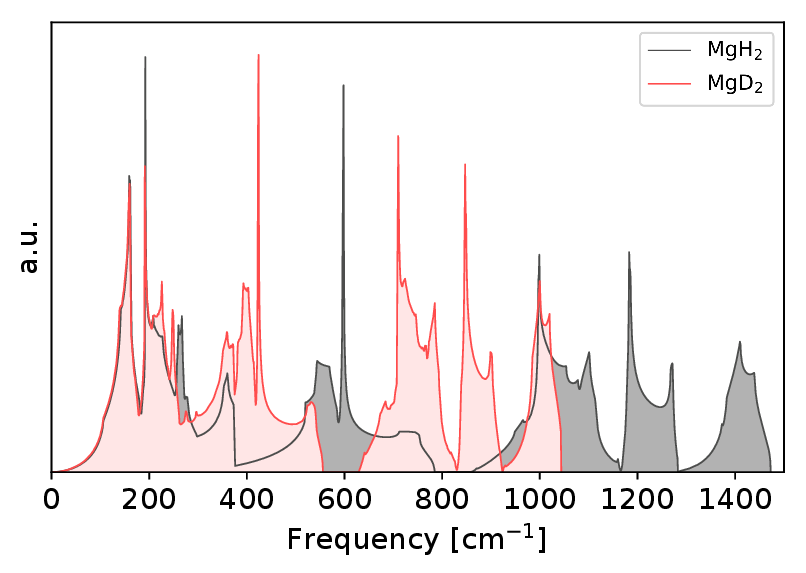}
	\caption{Comparison of the theoretical phonon DOS for MgH$_2$ and MgD$_2$ (from Ref. \cite{ramic2021njoyncrystal}). 
 }
	\label{fig:phdos}
\end{figure}

% Davide and Marco
%
%- Summary of the work
The phonon spectra inputs for the MgH$_2$, and its deuterated variant MgD$_2$, libraries were generated using Density Functional Perturbation Theory (DFPT) \cite{Baroni} with the Perdew-Burke-Ernzerhof (PBE) \cite{pbe} approximation for the exchange and correlation energy functional. The calculations also employed ultrasoft pseudopotentials \cite{uspseudo,Dalcorso2014} and plane wave expansions of Kohn-Sham orbitals as given in the Quantum-ESPRESSO (QE) package \cite{QMEspresso}. The  phonon density of states (PDOS) of MgH$_2$ and MgD$_2$ are shown in \cref{fig:phdos} \cite{ramic2021njoyncrystal}.

The results for MgH$_2$ are very similar to previous ab-initio works \cite{Ohba,Schimmel}. In particular, in \cite{Ohba} the overall validity of the isotropic approximation of the mean square displacement is assessed at different temperatures. The high-frequency phonons, due to the motion of hydrogen, are almost exactly scaled by a factor $\sqrt{2}$ by isotopic substitution while the acoustic part of the spectrum is essentially unchanged. 

\subsubsection{Nanodiamonds}\label{sec:nanodiamonds_finite_spectra}
NDs of interest as reflector materials for cold neutrons are synthesized by detonation techniques \cite{teshigawara2019measurement}. The resulting samples consist of nanoparticles with a distribution in diameter centered at about 5\,nm \cite{teshigawara2019measurement}. Transmission electron microscopy (TEM) 
measurements of NDs synthesized under different conditions revealed a mostly diamond-like structure of the nanoparticles with, however, a partial surface graphitization whose extent depends on the preparation conditions, thermal history, and purification process \cite{Mochalin2011}.
Full transformation of NDs into carbon onions was in fact observed by heating at about 1670\,K \cite{Qiao2006}.
Impurities such as H, N and O are also present, resulting from the purification methods or from oxidation in air. A fraction of about 5 atomic $\%$ of H was measured, for instance, in commercial samples of detonation NDs used by Vasiliev et al. in ref. \cite{Vasiliev2015} to measure the heat capacity as a function of temperature. This latter work shows that the heat capacity of 5\,nm NDs is notably larger than that of bulk diamond up to 300 K. 

This result suggests a sizable difference in the PDOS of the NDs with respect to the bulk which is confirmed by the inelastic neutron scattering data reported by Shiryaev et al. in Ref.~\protect\cite{Shiryaev2020}.

Since the PDOS is expected to affect both the elastic (via the Debye-Waller factor) and the inelastic contributions to the neutron scattering cross section, it is important to have reliable structural models of the NDs to provide both atomic positions and the phonon DOS for the calculation of the total neutron scattering cross section.

To this aim, we generated models of NDs  by molecular dynamics (MD) simulations with the classical interatomic potential AIREBO \cite{Stuart2000}. This potential is suitable for generating reliable structural models of NDs as shown by several previous works, including the formation of a graphitic shell upon annealing \cite{Matsubara2020}. However, the classical potential is less reliable in reproducing the PDOS, which is the key property needed to compute the neutron scattering cross section. Our strategy thus consists of using the classical potential to investigate the dependence of the structural properties on the annealing process and then to use the NDs models generated with the AIREBO potential to compute the PDOS with a more reliable, but much more computationally demanding scheme. To this aim we chose the  Gaussian Approximation Potential (GAP) from Ref. \cite{Rowe2020}.

We have investigated  NDs with diameters of 1.4, 2.5 and 5 nm, containing about 200, 750, and 5900 atoms, respectively. The initial configuration of the ND was built in two different ways, either as a spherical shape or by using the Wulff construction that minimizes the overall surface energy. To this aim we used the theoretical surface energies of the surfaces of diamonds from Ref. \cite{Tran2013}, obtained from Density Functional Theory (DFT). 

\begin{figure*}[t!] 
  \centering
\includegraphics[width=0.30\textwidth]{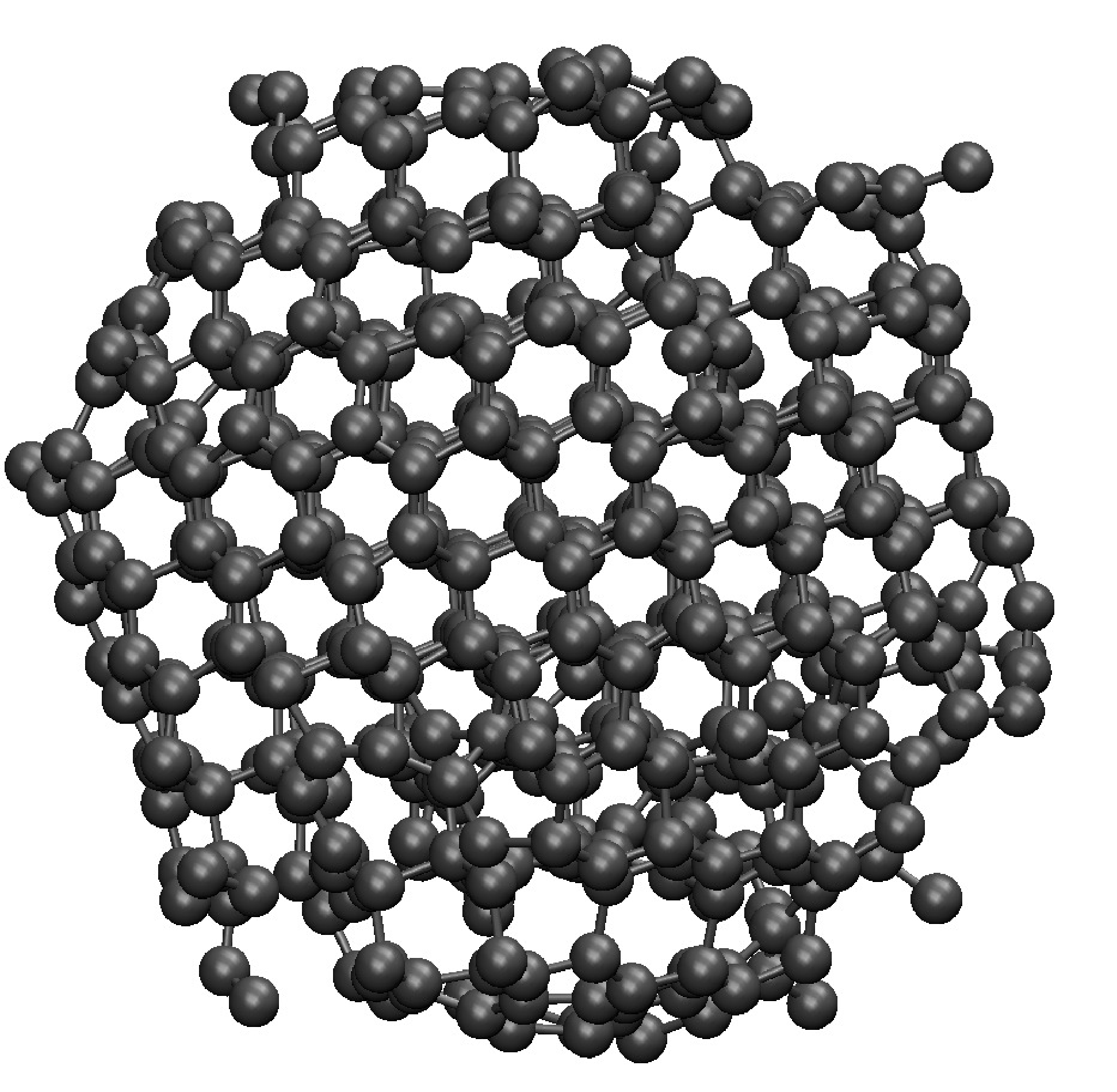}
\includegraphics[width=0.30\textwidth]{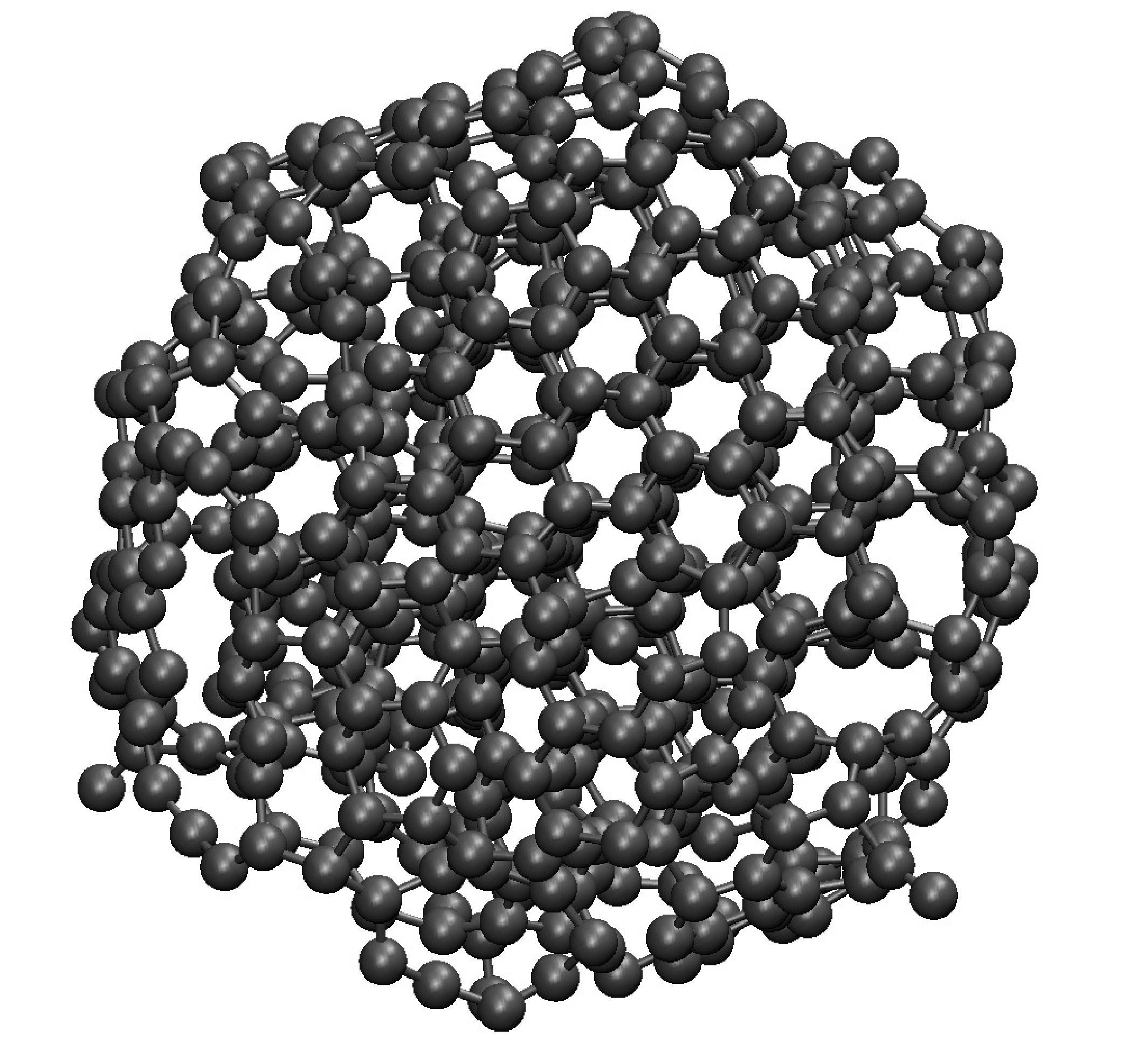}
\includegraphics[width=0.30\textwidth]{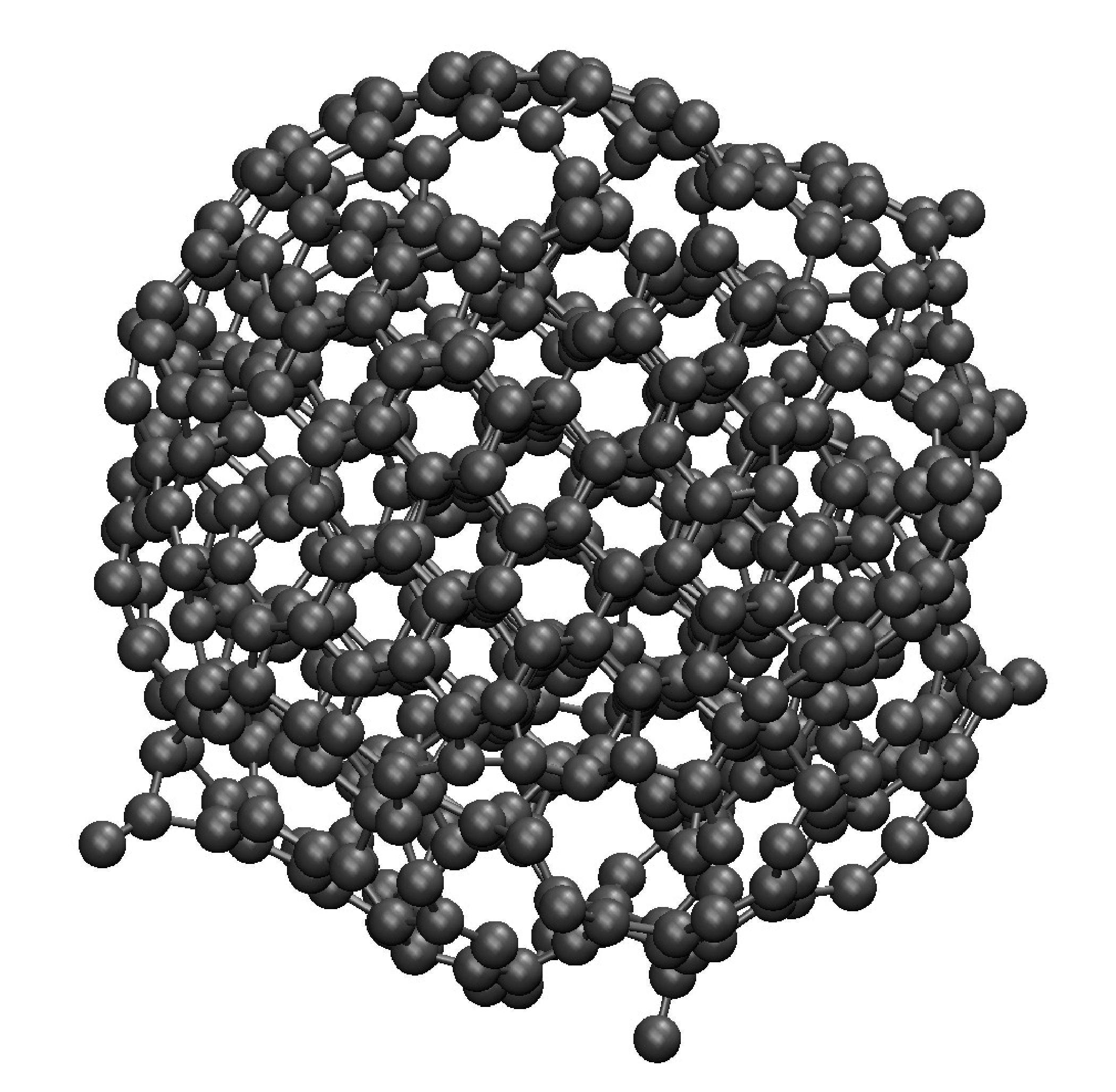}
  \caption{Snapshots of the evolution of the 2.5 nm ND annealed at  2000 K with the AIREBO potential extracted at (left) 10 ps, (central) 500 ps  and (right) 1 ns. The initial configuration was built by the Wulff construction.}
  \label{fig:graphitization}
\end{figure*}

We then annealed the models at several different temperatures in the range 500-3000 K for 1 ns. Snapshots of the evolution of the 2.5 nm ND as a function of time at  2000 K are shown in \cref{fig:graphitization}. 
A partial graphitization of the outermost layers is clearly visible.

Preliminary calculations of PDOS with the AIREBO potential show that the PDOS is more similar to that of the bulk diamond for annealing at the lower temperatures while sizable changes occur at and above 2000 K due to partial graphitization with a reduction of the optical peak and an increase of the PDOS due to acoustic phonons at low frequency. This effect is reduced by increasing the size of the NDs as also shown by the results of Ref. \cite{Matsubara2020} on  1442-atom NDs.

\begin{figure*}[t!] 
  \centering

 \includegraphics[width=0.7\textwidth]{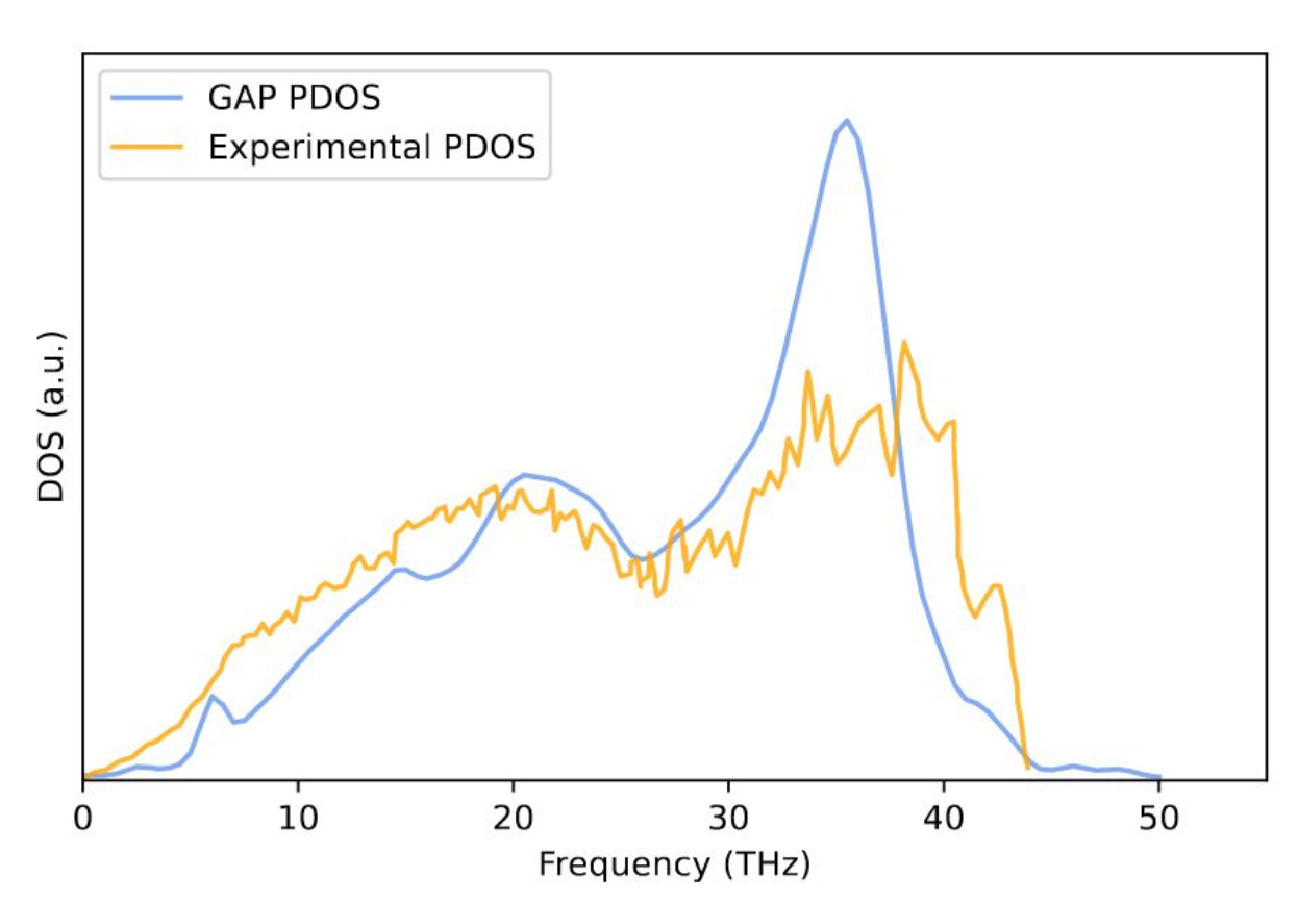}

  \caption{Phonon density of states computed with the GAP potential for the 5.0 nm ND annealed at 500 K (from Ref. \cite{dijulio2023_refl}), compared with experimental data from inelastic neutron scattering from Ref. \protect\cite{Shiryaev2020}.}
  \label{PDOS_New}
\end{figure*}

To compare with experimental data  we resort, however, to the more accurate GAP potential. In \cref{PDOS_New} we compare the theoretical PDOS of the largest  5 nm ND annealed at 500 K  with INS data from Ref. \protect\cite{Shiryaev2020}. The PDOS was obtained by Fourier transforming the velocity-velocity correlation function extracted from simulations 20 ps long. The agreement with experimental data is overall good and it improves for the models annealed at higher temperatures which corresponds to a higher degree of graphitization.
These theoretical PDOS have been used to compute the neutron scattering cross section presented in the sections below.

\subsubsection{Graphitic compounds}\label{sec:molmod}
We have explored several graphitic compounds through DFT calculations. This includes graphite oxide and intercalated graphite oxide, also in deuterated forms, and fullerite. We present below a selection of these calculations, further details can be found in \cite{xu2023magnetic} for fullerite.

Graphite oxide (GO) is made of graphitic layers with a fraction of carbon atoms bound to oxygen atoms either in an epoxy group  (C-O-C)  or within a hydroxyl (C-OH). This leads to a wrinkled hexagonal structure with a typical 6-7 {\AA } distance between the layers \cite{Mermoux,Lerf,Szabo}, which is also ruled by the presence of hydrogen bonds among hydroxyls in adjacent layers.  Intercalated water molecules bound to the hydrophilic hydroxyl groups have a marginal effect on the structure of the individual layers, but they lead to a further expansion of the interlayer distance \cite{Szabo}. The C/O and C/H ratio of graphite oxide depends on the preparation conditions with reported compositions ranging from C$_8$O(OH)$_2$ to  C$_{12}$O(OH)  (see Ref. \cite{Boukhvalov} for a review). 

To provide structural and phononic data for the calculation of the neutron scattering cross sections, we  optimized the geometry of different models of GO by means of DFT calculations. We first considered the model with composition C$_8$O(OH)$_2$ proposed in Ref. \cite{Boukhvalov}. The supercell contains two graphitic planes each made of 2x2 unit cells of a single graphite layer (8 carbon atom per layer) with one epoxy group and two hydroxyl groups per layer pointing in opposite directions with respect to the graphitic plane. The two graphitic planes in the unit cell have oxygen atoms and hydroxyls in the same positions but the planes can slide one with respect to the other during the geometry optimization. We then considered a second larger model with a supercell containing two graphitic planes each made of 3x3 unit cells of a single graphite layer (18 carbon atom per layer) with one epoxy group and two hydroxyl groups per layer, still pointing in opposite directions with respect to the graphitic plane, 
which corresponds to a composition C$_{18}$O(OH)$_2$.  

Finally, we considered a hydrated form of the small model by adding a water molecule in the van der Waals gap between adjacent graphitic layers which corresponds to a composition C$_8$O(OH)$_2$-H$_2$O.
The geometry of these models have been optimized by DFT calculations
with the QE suite of programs \cite{QMEspresso}, the PBE \cite{pbe} approximation for the exchange and correlation energy functional, ultrasoft pseudopotentials \cite{uspseudo} and a plane wave expansion of Kohn-Sham orbitals. We also included semiempirical van der Waals interactions \'a la Grimme (D2) \cite{D2}.   

\begin{figure}[h!]
\centering
\hspace{1 truecm}
%\begin{subfigure}{
 %{\includegraphics[height=5cm]{tsl_fig/graphite-oxide-high-cutoff-side-view-all-free.eps}}
 {\includegraphics[height=5cm]{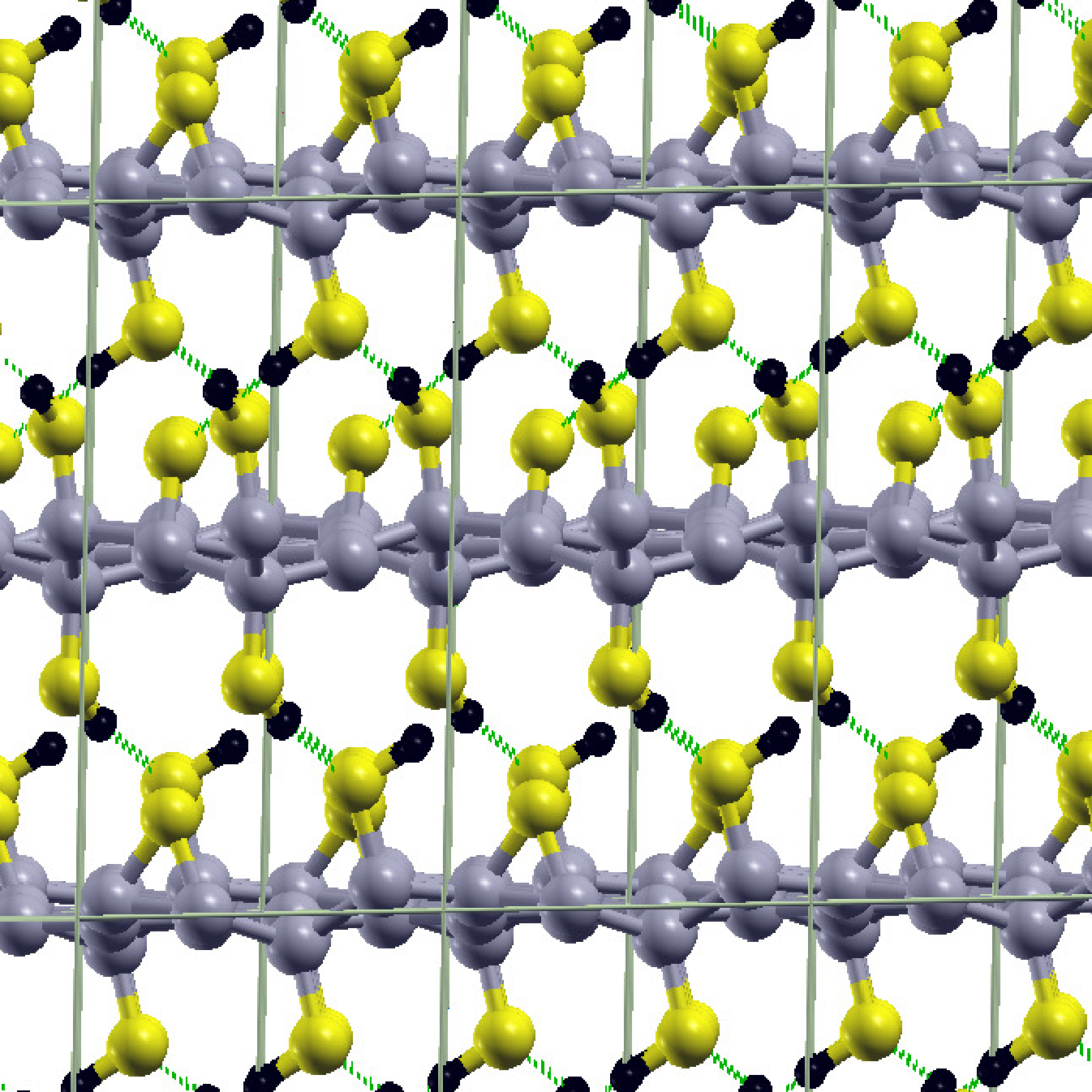}}
 \hspace{1 truecm}
%{\includegraphics[height=5cm]{tsl_fig/graphite-oxide-lower-coverage-all-free-side-view.eps}}
{\includegraphics[height=5cm]{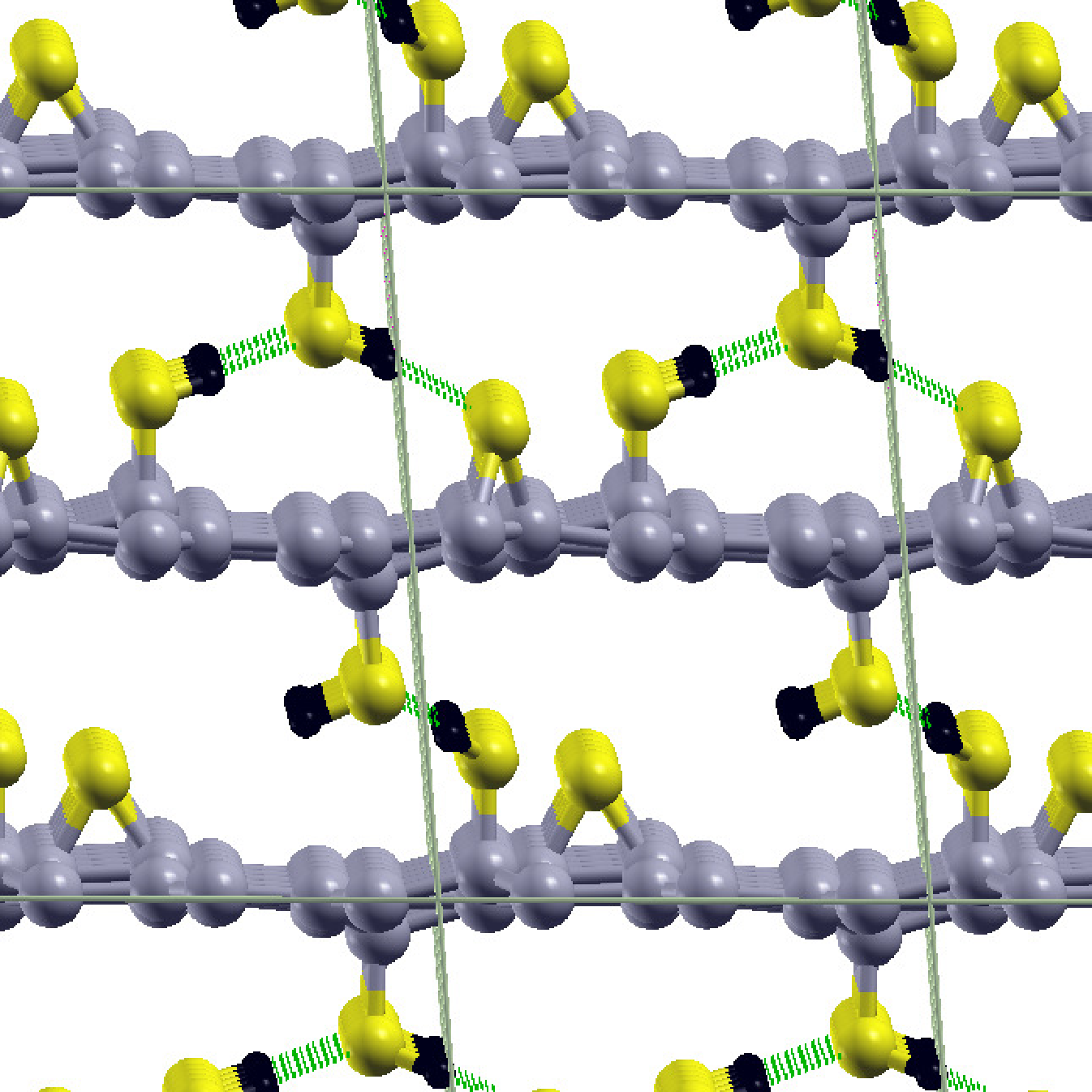}}
% }
%\end{subfigure}
\caption{ Side  views of  (right) the small model of GO (composition C$_8$O(OH)$_2$) and (left) the large model of GO (composition C$_{18}$O(OH)$_2$). Several replicas of the supercell are shown. Hydrogen bonds are depicted by green lines. Carbon, oxygen and hydrogen atoms are depicted by gray, yellow and black spheres.}
\label{geosmall}
\end{figure}

The side views of the final configuration of the small and large model of GO are shown in \cref{geosmall}. The sliding of the two graphitic planes leads to the formation of H-bonds among the hydroxyls and between one hydroxyl and the epoxy group. There are two different interplanar distances corresponding to 5.244 and 4.525 {\AA } in the small model and to 4.616 and 4.699 {\AA } in the large one. In the hydrated model, with composition C$_8$O(OH)$_2$-H$_2$O, the water molecules induce a swelling of the structure with larger interlayer distances of 6.131 and 6.141 \AA.

\begin{figure}[h!]
\centering
\vspace{ 1 cm}
% {\includegraphics[height=6.0cm]{tsl_fig/graphene-oxide-high-cutoff-PDOS-no-symmetry.eps}}
{\includegraphics[height=6.0cm]{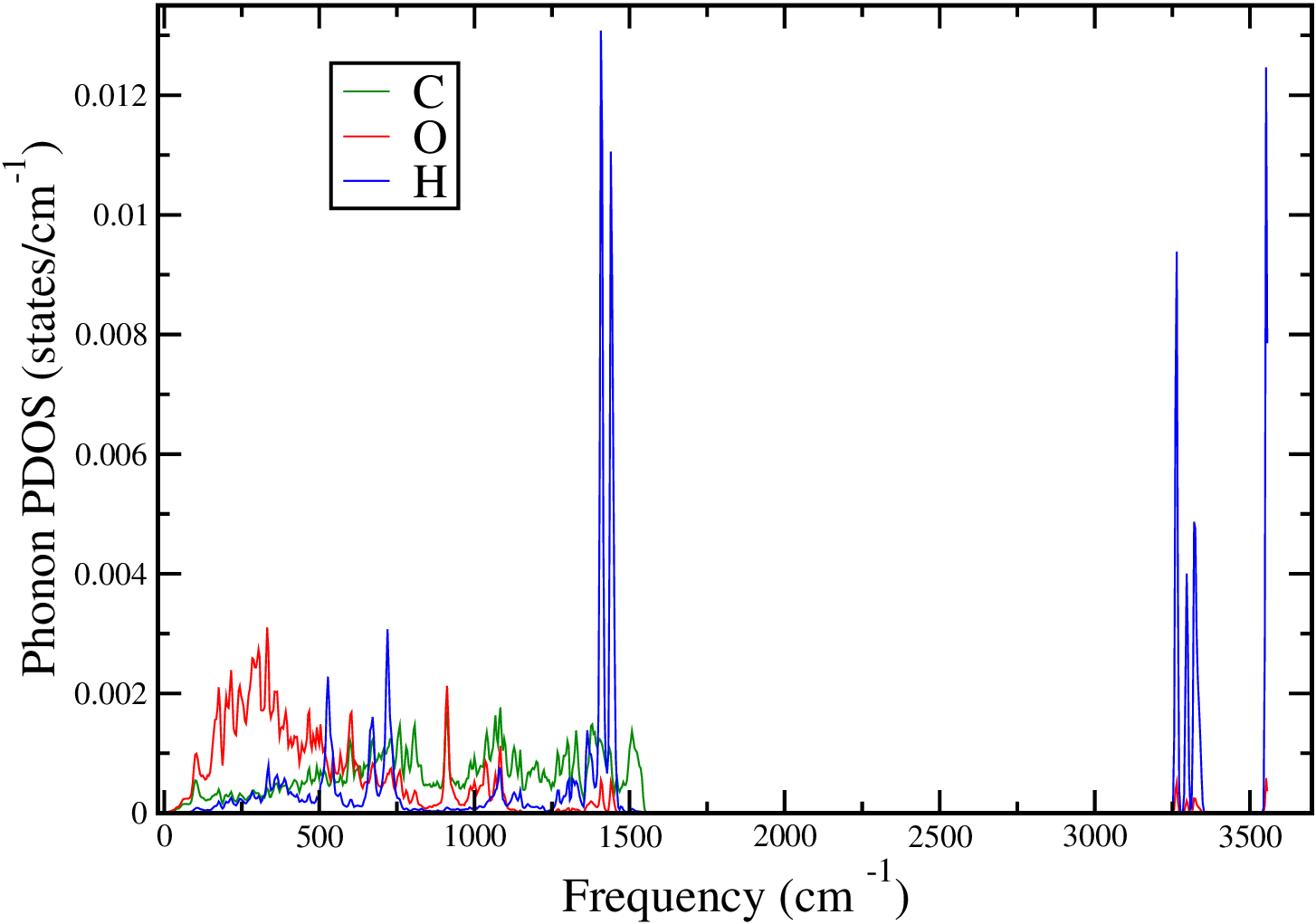}}

{\includegraphics[height=6.0cm]{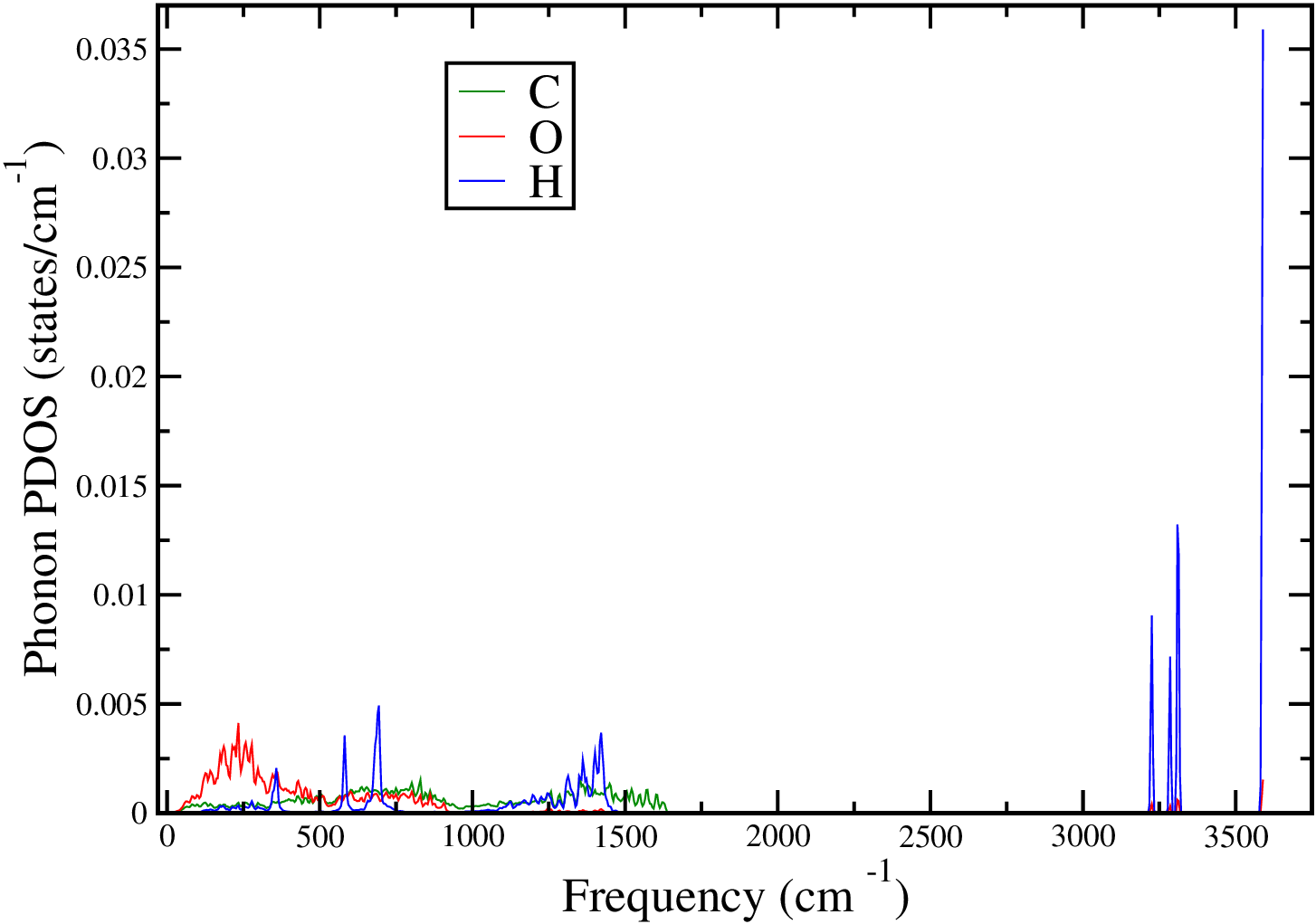}}
% {\includegraphics[height=6.0cm]{tsl_fig/graphene-oxide-lower-coverage-PDOS-no-symmetry.eps}}
\caption{Phonon density of states (PDOS) projected on different atomic species, each normalized to one for (top panel) the small model with composition C$_8$O(OH)$_2$ and (bottom panel) in the large model with  composition C$_{18}$O(OH)$_2$.}
\label{DOSsmall}
\end{figure}

We then computed the PDOS for all models within DFPT \cite{Baroni} as implemented in the QE code.  We also computed the PDOS for deuterated models. Examples of the PDOS projected on different atomic species, each normalized to one, are shown in \cref{DOSsmall} for the small and large models with no intercalated water. Atomic positions and PDOS of the models discussed above have been used as input to the neutron scattering models.

% %\end{figure}
\subsubsection{Clathrate hydrates}\label{sec:molmodclathrate}

\begin{figure}[!t]
    \centering
	\includegraphics[width=.75\textwidth]{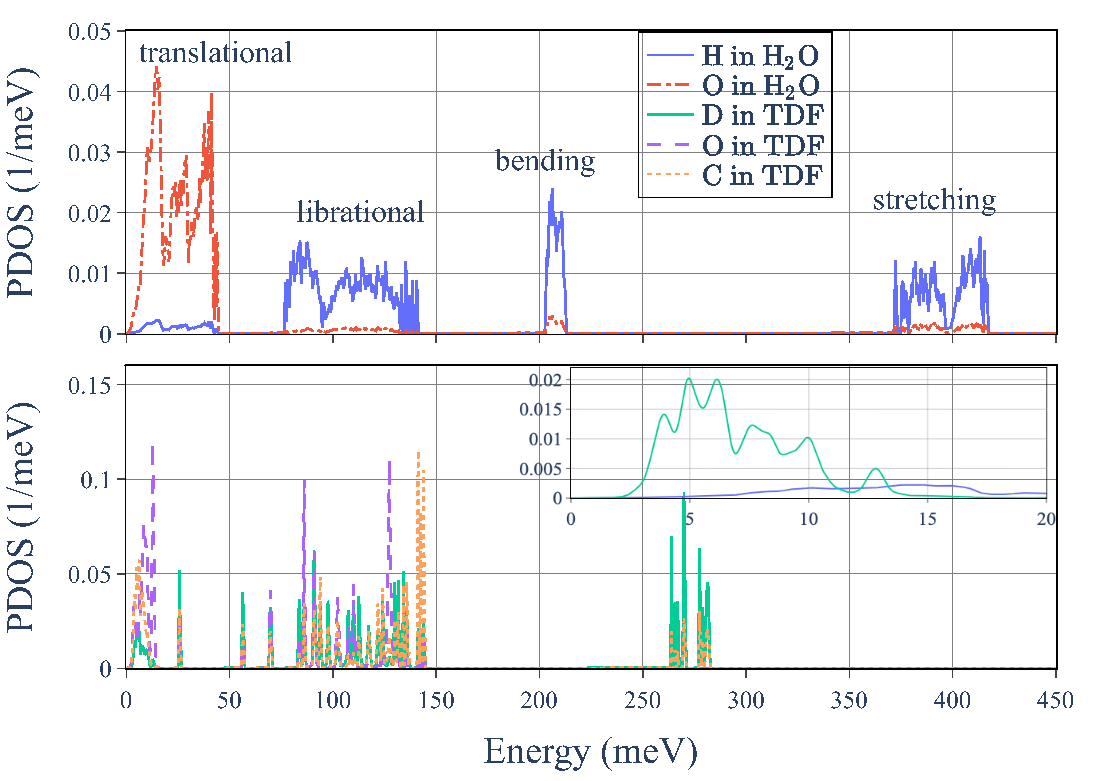}
	\caption{Normalised PDOS of TDF-containing hydrogenated clathrate hydrate projected on different atomic species from DFT calculations. The global PDOS can be obtained by summing the projected PDOS multiplied by the number of corresponding atoms in the unit cell.}
    \label{fig:tdf-h2o-pdos}
\end{figure}

The initial configuration of the THF-containing hydrogenated clathrate hydrates in structure II was taken from Ref.~\cite{lenz_structures_2011}. The unit cell of the face centered cubic lattice contains 34 \ce{H2O} molecules and two THF molecules inserted in the large cages. The unit cell was replicated to build the conventional cubic cell with 136 \ce{H2O} and 8 THF molecules. The initial position of the THF molecules was selected by performing classical MD simulations with GROMACS~\cite{Gromacs} using the SPC/E potential for water~\cite{SPC} and the general Amber force field for THF and THF-water interactions~\cite{GAFF}. The initial structure was optimised by means of DFT calculations with CP2k~\cite{CP2k}. We employed the PBE approximation for the exchange and correlation functional~\cite{pbe} and norm conserving pseudopotentials~\cite{GTH1}. The Kohn-Sham orbitals were expanded in a triple-zeta-valence plus polarisation basis set while the electronic density was expanded in plane waves  as implemented in CP2k. Semiempirical van der Waals interactions were included according to Grimme (D3)~\cite{D3}.
The phonons were computed using CP2k combined with phonopy~\cite{TOGO20151} at the experimental lattice parameter.
The force constant matrix was computed from forces at finite atomic displacements in the conventional cubic cell (136 water molecules). The Coulombic long range contribution was computed analytically from effective charges and the dielectric constant obtained in turn from ionic forces and polarisation at finite electric field within the Berry phase approach~\cite{Berry} implemented in CP2k.

\indent The PDOS was computed for all possible four combinations of TDF/THF and hydrated/deuterated clathrate.
As an example, we show in \cref{fig:tdf-h2o-pdos} the PDOS
projected on different atomic species for the TDF-H$_2$O clathrate. At low energy, the translational modes of the \ce{H2O} molecules dominate. In this spectral region, the experimental neutron weighted PDOS of type II clathrate hydrates has two characteristic peaks at about 7 and 10.5 meV ~\cite{schober2003coupling,celli2012pdos}. This double peak feature is less evident in the theoretical PDOS projected on the water molecules. The position of the two peaks is also blue-shifted to about 10 and 14.5 meV. This double peak feature is instead better reproduced by classical simulations with rigid water molecules as reported for instance in Refs.~\cite{english,celli2012pdos}. A possible improvement of the theoretical PDOS would then result from the use of the DFT-PDOS for the intramolecular  modes and the classical result for the  PDOS in the translational low frequency region as we will discuss later on. The band from \SI{70}{meV} to \SI{150}{meV} is due to molecular librations. At higher energy, intramolecular modes are dominant with \ce{O}-\ce{H} bending and \ce{O}-\ce{H} stretching  at around \SI{200}{meV} and \SI{400}{meV}, respectively. The PDOS of atoms bound in the TDF molecules are illustrated in the lower panel of \cref{fig:tdf-h2o-pdos}. The PDOS of the THF-H$_2$O and TDF-D$_2$O can be found in Ref.
\cite{Shuqi_ECNS_23}. The PDOS of the cage can be transferred to other clathrates, based on weak guest-host coupling assumption as discussed in Ref.~\cite{schober2003coupling}.

We  considered two other type II clathrates, which include the clathrate with O$_2$ molecules in the large and small cages \cite{chazallon2002anharmonicity} and a mixed clathrate with THF in the large cages and O$_2$ molecules in the small cages. For these structures, we performed classical MD simulations at 10 K with the same force-field with rigid water molecules and partially rigid THF used for the generation of the initial structure of the THF-clathrate mentioned previously. We considered a supercell made of 2$\times$2$\times$2 conventional cubic cells each containing 136 water molecules. The PDOS were computed by Fourier transforming the velocity-velocity autocorrelation function extracted from  MD trajectories. 

To include in the PDOS the contribution from intramolecular modes of the water molecule, we used the DFT results obtained for the type II clathrate containing THF discussed above. In the approximation of weak coupling between water and the guest molecules in the cage, we expect that the intramolecular modes of H$_2$O would be little affected by the replacement of THF by O$_2$ at least for the purpose of the calculation of the neutron scattering cross section. 

\begin{figure}
    \centering
    \includegraphics[width=0.75\textwidth]{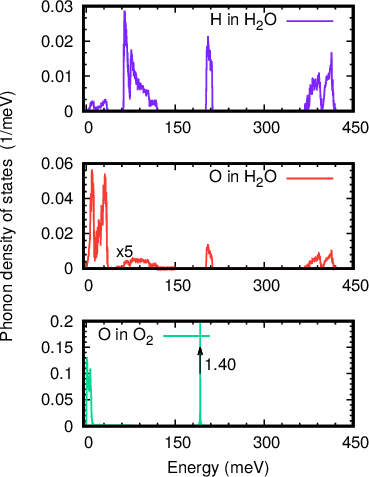}
	\caption{The phonon density of states of O$_2$-H$_2$O clathrate projected on the different atomic species obtained from atomistic simulations (see text). Each projected PDOS is normalized to one. The total PDOS can be obtained by summing the projected PDOS multiplied by the number of corresponding atoms in the unit cell. 
 For the sake of clarity, the projected PDOSs are multiplied by five  above 50 meV for O in water.
 }
   \label{fig:O2-H2O-sp}
\end{figure}

To describe both the intermolecular and intramolecular modes we then built a PDOS by merging the classical PDOS in the low frequency region of translational and librational modes with the DFT PDOS in the high frequency region for the intramolecular modes of water. The resulting PDOS projected on the different atomic species are shown in
shown in \cref{fig:O2-H2O-sp} for the O$_2$-clathrate hydrate. The simulated PDOS serve to compare with the measurements presented in \cref{sec:vcn}.

%\clearpage
\subsection{Advancements in Nuclear Data Libraries}
\label{sec:NJOY+NCrystal}

The design of reactors, neutron sources, and nuclear systems requires calculations of neutron distributions using radiation transport codes. A major source for nuclear data used in radiation transport codes, such as MCNP, OpenMC, and PHITS, are the ENDF files. These files include sub-libraries for different kinds of particles, in particular thermal neutron scattering data in form of TSL and written in ENDF-6 format \cite{Herman2009}. In ENDF format, File 7 stores the thermal neutron scattering data in two sections: one for elastic and one for inelastic scattering. In general, the scattering cross section of neutrons can be decomposed into four parts: coherent elastic, incoherent elastic, coherent inelastic, and incoherent inelastic scattering. Elastic scattering represents the scattering of a neutron without the exchange of energy with the target material in the laboratory frame of reference, while inelastic scattering represents the scattering of a neutron where it either gains or loses energy. Coherent scattering contains the interference terms of the scattering and therefore is sensitive to the structure of the material, whereas incoherent elastic scattering depends on the self correlation for each atom at different times. 

At the moment, in the ENDF-6 format, the elastic section stores either the coherent elastic or incoherent elastic cross section only. In the inelastic section, either the incoherent inelastic cross section is stored in the incoherent approximation or the inelastic is stored as a sum of incoherent and coherent inelastic parts calculated in the incoherent approximation. To overcome the limitation in the elastic section, a change was recently proposed \cite{Zerkle2020} for the thermal scattering format, which will in this paper be referred to as the ``mixed-elastic format''. Implementation of the new format makes it possible to store both the coherent and incoherent elastic cross sections in the elastic section, one after another one, in the same format in which they were stored before individually. This is important for materials like MgD$_2$ which have a significant coherent and incoherent elastic component. Without the new format, thermalization of neutrons couldn't be properly calculated for such a material using the current format.

Thermal scattering files can be produced with the LEAPR module of NJOY, which is available under a free software license but its support for coherent elastic scattering is limited. On the other hand, the NCrystal library \cite{Cai2020, CAI2019400,kittelmann2021elastic} has an extensive treatment of the calculation of thermal neutron scattering cross sections directly but cannot generate ENDF-6 formatted files. In order to remedy this, we have combined these two tools, using NCrystal to generate the microscopic data that is later used by NJOY to produce thermal scattering libraries. Additionally, we have modified NJOY to support the proposed mixed-elastic format and produce thermal scattering .ACE files with coherent and incoherent elastic scattering. The Monte-Carlo code OpenMC was modified as well to support this new format. This development is described in the NJOY+NCrystal paper \cite{ramic2022njoyncrystal} in great detail, however a summary is given below.

Since MgH$_2$ and MgD$_2$ have significant incoherent and coherent elastic components, the main motivation for developing NJOY+NCrystal was to enable accurate calculations of thermalization and moderation of neutrons in these materials, in both the current and proposed format. Additionally, NJOY+NCrystal was motivated by a need to implement the mixed-elastic format proposal in an open source code, such as NJOY. In essence, NJOY+NCrystal is a customized version of NJOY that relies on NCrystal to provide the necessary information for the calculations of the coherent and incoherent elastic components. Additionally, both NJOY and NCrystal utilize CMake and GitHub repositories for the installation, hence making the integration smoother. 

The changes to NJOY can be split into two parts. The first part is to provide an interface between LEAPR and NCrystal so that LEAPR can calculate coherent and incoherent elastic cross section components for any crystalline material. The calculated elastic components are then stored in the tsl-ENDF file for both the current and proposed mixed-elastic format. The second part consists of implementing changes to the THERMR and ACER modules of NJOY so that the new mixed-elastic format can be read and handled properly in order to produce .ACE files used by Monte-Carlo particle transport codes to sample neutron scattering events. A summary of the generated libraries are given below.
%\newpage
%\subsubsection{Ace libraries for MgH$_2$ / MgD$_2$ and other materials}
%\label{sec:ace-libraries}
\subsubsection{MgH$_2$ / MgD$_2$}
% Kemal
% 
% - Description of the library
% - Description of the repository

We computed the total scattering neutron cross sections of polycrystalline MgH$_2$ and its deuterated variant MgD$_2$ using the NJOY~+~NCrystal tool and input from the DFT simulations described above. MgH$_2$ recently emerged as a good candidate material for cold neutron reflectors \cite{Granada2021},  while MgD$_2$ is of interest because of an even lower absorption cross section. As mentioned before, calculating the total cross section of a material is one of the most important validation tests of the newly created libraries.

\begin{figure}[h!] % replace 't' with 'b' to force it to be on the bottom
  \centering
  \includegraphics[width=0.6\columnwidth]{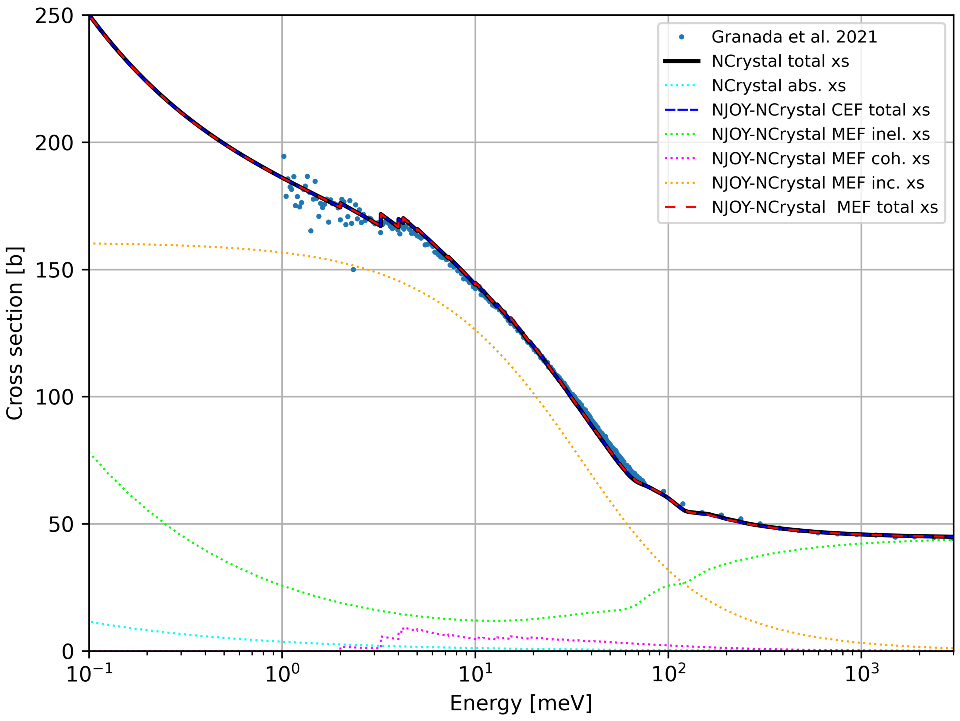}
  \caption{Total scattering cross section comparison for MgH$_2$. From \cite{ramic2022njoyncrystal}.}
  \label{fig:mgh2_total_xs}
\end{figure}

From the calculated phonon spectra, tsl-ENDF files were created for both MgH$_2$ and MgD$_2$, using the NJOY~+~NCrystal tool. In \cref{fig:mgh2_total_xs} it can be seen that the agreement between the measured total cross section data from Granada \cite{Granada2021} and the calculated curves (which include an absorption cross section ($\sigma_{2200}^\text{H} = 0.3326$ b, $\sigma_{2200}^\text{Mg} = 0.063$ b) is excellent. Since MgH$_2$ is an incoherent scatterer, due to the large incoherent cross section of hydrogen, if MEF option cannot be utilized, the CEF option is a good approximation as well. MEF stands for mixed-elastic format, while CEF stands for current ENDF format. The latter includes approximations within the ENDF format and can be used in standard versions of Monte-Carlo software, as described in more detailed in \cite{ramic2022njoyncrystal}.

\begin{figure}[h!] % replace 't' with 'b' to force it to be on the bottom
  \centering
  \includegraphics[width=0.6\columnwidth]{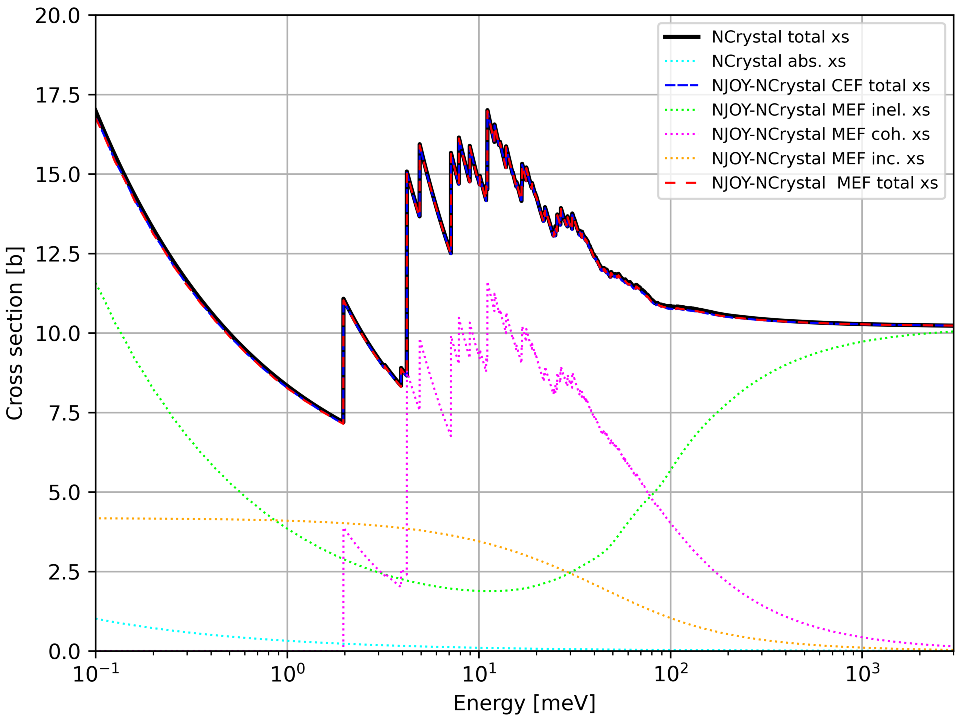}
  \caption{Total scattering cross section comparison for MgD$_2$. From \cite{ramic2022njoyncrystal}.}
  \label{fig:mgd2_total_xs}
\end{figure}

\cref{fig:mgd2_total_xs} shows a comparison between the NCrystal and the NJOY+NCrystal calculated total cross sections for MgD$_2$, as well as different scattering components as calculated in NJOY+NCrystal and NCrystal. It can be seen that the total cross section for MgD$_2$ is an order of magnitude lower than for MgH$_2$. This arises from the incoherent hydrogen, which has a cross section of 82.02 barns, and is replaced with coherent deuterium with a cross section of 7.64 barns. Although the CEF option is not recommended for deuterated materials due to the presence of both coherent and incoherent cross sections, the approximation for compounds works well in terms of total cross section because the coherent component is assigned to the Mg(MgD$_2$) library and the incoherent component to the D(MgD$_2$) library.

\subsubsection{Other materials}\label{sec:other_materials}
In addition to including the MgH$_2$ and MgD$_2$ evaluations, the NJOY+NCrystal library contains 213 tsl-ENDF evaluations, created for 112 new or updated materials, in both the current ENDF-6 format and in the new proposed mixed-elastic format. A summary of the NJOY+NCrystal library is given in the NJOY+NCrystal paper \cite{ramic2022njoyncrystal}.

Different methodologies were employed to create the final tsl-ENDF files. Some of the materials were created from existing .ncmat files in current and previous versions of NCrystal. For some of the materials, mostly monoatomic metals, the VDOS curves in existing .ncmat files were updated, mostly using different sources from the literature. For the rest of the materials, phonon curves were obtained by utilizing ab-initio calculations from a phonon database at Kyoto University \cite{wwwphonodb,MaterialsProject,ONG2013314,ONG2015209}. In summary, the files from the database were used with Phonopy \cite{TOGO20151} to calculate phonon eigenvalues and eigenfrequencies, which were then utilized with oClimax \cite{ref_oclimax} to extract partial phonon spectra. A detailed explanation of how the phonon spectrum and crystal structure was obtained for each material is provided in the comments section of the tsl-ENDF files. 

The experimental data for the validation of the tsl-ENDF files is scarce. Wherever experimental data were available, tsl-ENDF files were validated against total cross-section measurements and diffraction data, while all materials were validated against specific heat capacity curves as a minimum standard for acceptance. The link to the exact files used for each material at the Kyoto University database can be found inside the tsl-ENDF files in the HighNESS Github repository. For validation, the correct crystal structure was a starting point (a reference for each crystal structure can be found inside tsl-ENDF files as well), followed by comparison with the experimental specific heat capacity and if available, total cross section measurements. For $\alpha$ and $\beta$ SiC, diffraction data was used as well as a means of validation of the libraries.

\subsubsection{Libraries for the THF- and TDF-clathrates}\label{subsec:theo_xs}

 \begin{figure}[!t]
 	\centering
 	\includegraphics[width=.75\textwidth]{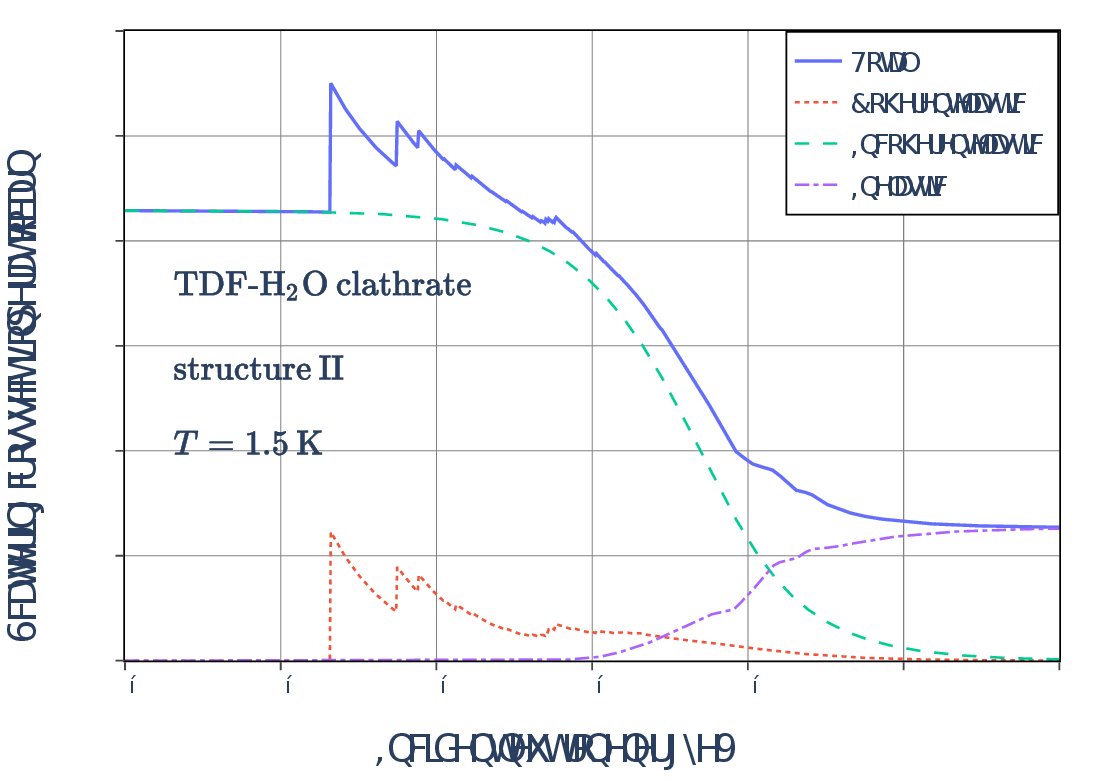}
 	\caption{Neutron scattering cross sections of type II TDF-containing hydrogenated clathrate hydrate.}
     \label{fig:tdf-h2o-xs}
 \end{figure}

 The lattice parameters, atomic positions of the crystalline structure, and the PDOS from the molecular modelling calculations were converted into input files for the NCrystal~toolkit. \cref{fig:tdf-h2o-xs} shows the total scattering cross section of the THF-containing deuterated clathrate hydrate, which is the sum of coherent elastic, incoherent elastic and inelastic cross sections. Incoherent elastic scattering is dominant for neutron energy below a few meV because of the large incoherent cross section of hydrogen~\cite{dawidowski2013neutron}. It should be pointed out that these scattering cross sections are calculated based on \cref{fig:tdf-h2o-pdos}. 

 The large Bragg cutoff (around \SI{0.2}{meV} or $20$~\AA) is an advantage for cold neutron moderation because neutrons can be reflected within the cage structure thus increasing the interactions with the guest molecules. The cross sections of the totally hydrogenated and deuterated THF-clathrates are presented in Ref.~\cite{Shuqi_ECNS_23}. These theoretical neutron scattering cross sections serve to compare against the measurements performed at the ILL as shown in \cref{sec:vcn}.

 \subsubsection{Libraries for graphitic compounds}
 \begin{figure}[!htbp]
     \centering
     \includegraphics[width=.75\textwidth]{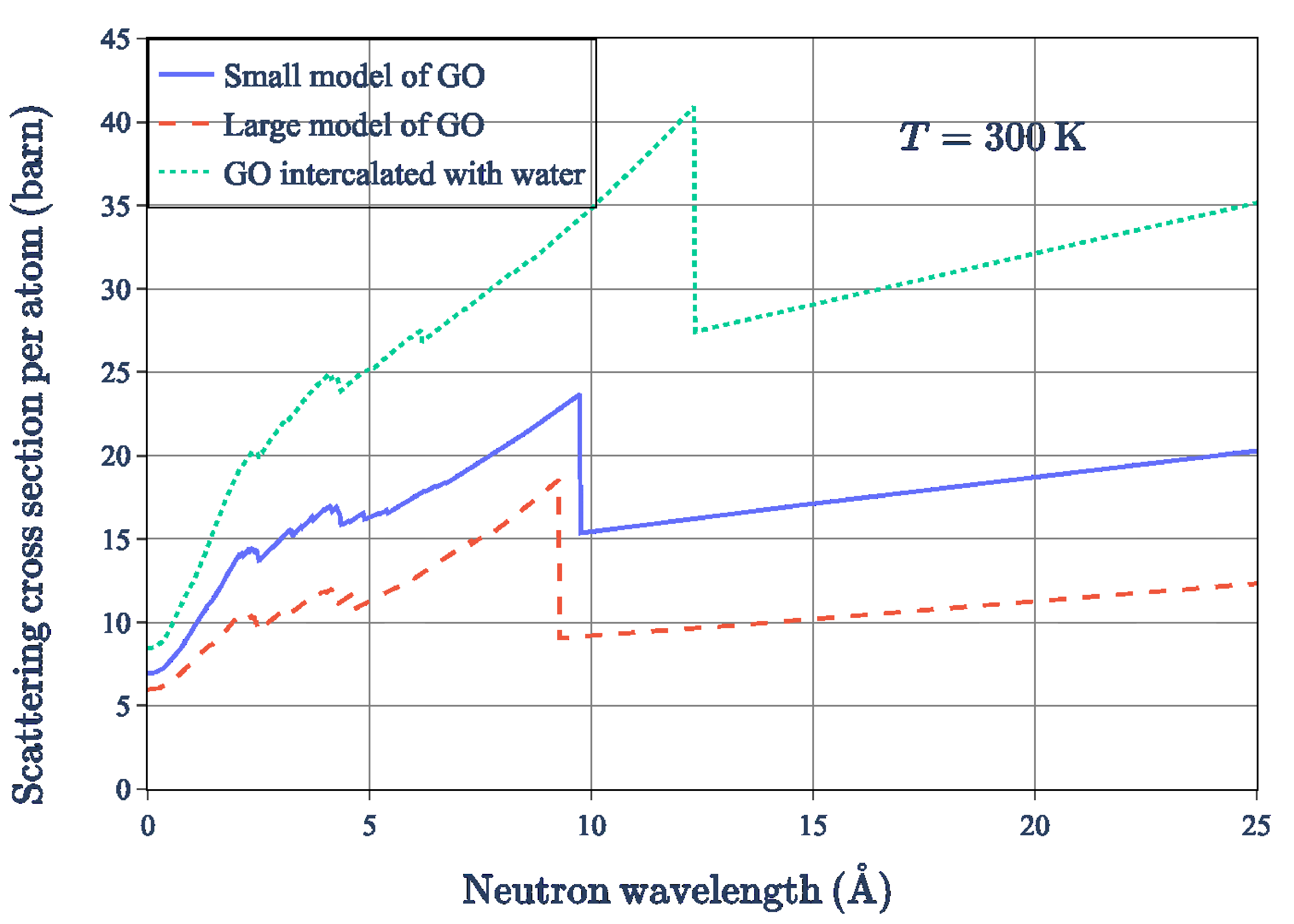}
     \caption{Neutron scattering cross sections of the small and large models of hydrogenated GO, and hydrogenated GO intercalated with water molecule.}
     \label{fig:GO_xs_H}
 \end{figure}

 \begin{figure}[!htbp]
     \centering
     \includegraphics[width=.75\textwidth]{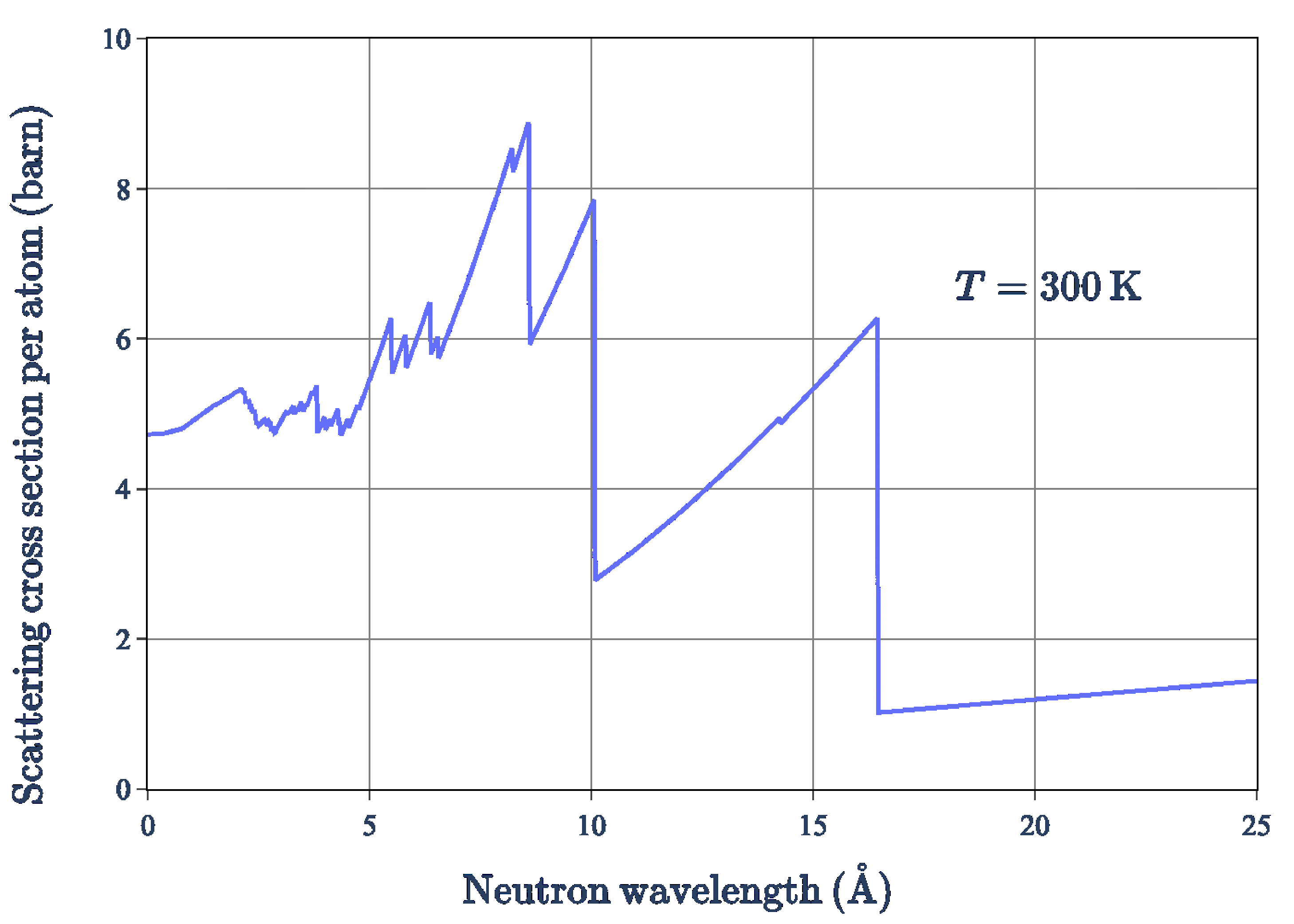}
     \caption{Neutron scattering cross section of fullerite, calculated based on the PDOS and crystalline structure presented in Ref.~\cite{xu2023magnetic}.}
     \label{fig:C60_xs}
 \end{figure}
 For the graphitic material presented in \cref{sec:molmod}, the crystalline structure and the corresponding PDOS are merged into a .ncmat file by using the NCrystal internal tool. Examples of the theoretical neutron scattering cross sections are presented in \cref{fig:GO_xs_H,fig:C60_xs}. 

 \cref{fig:GO_xs_H} show the results for GO for the two different simulation models in addition to GO including water intercalation. It can be seen that the addition of the water extends the Bragg-cutoff to higher wavelengths. The scattering cross section per atom of the small model (\ce{C8O(OH)2}) is larger than that of the large model (\ce{C18O(OH)2}), because the atomic density of hydrogen is higher in the small model. The effect of deuteration is to the lower the cross-sections, due to the exchange of H to D, which results in a lower incoherent elastic scattering in the material.
 
\cref{fig:C60_xs} shows the results for the fullerite model, which exhibits an extension of the Bragg-cutoff to even higher wavelengths, compared to the graphite oxide models. The neutron scattering cross section is obtained using the crystalline structure and the DFT-calculated PDOS which are detailed in Ref.~\cite{xu2023magnetic}.

In addition to the above, the .ncmat files for all graphite compounds investigated can be found in the HighNESS Github repository. This includes deuterated graphite oxide, deuterated intercalated graphite oxide, and a couple other intercalated graphites.

\clearpage
\newpage
\subsection{Extensions of NCrystal to include new physics}
A unique feature for the code NCrystal is that it can be called on-the-fly during a Monte-Carlo simulation. This makes it possible to avoid limitations in the ENDF nuclear data format and include new physics that were not possible before through usage of plugins.

The support for plugins was introduced in NCrystal release versions 2.2.x-2.4.x and they are to be developed according to the descriptions given on the wiki pages \cite{NCrystalPlugins1,NCrystalPlugins2}. The development process includes picking a name for the plugin, forking the plugin template library and developing the plugin to combine new physics models with the existing models in NCrystal. A template repository and example infrastructure to facilitate validation, debugging and bench-marking of the new plugin are provided to make it easier for new users to get started. After the plugin is developed, it can be compiled as a shared library or directly into a given NCrystal installation and all plugins developed should appear as forks of the ncplugin-template repository \cite{NCrystalForks}. In addition, NCrystal release version 2.1.0 includes the ability to define custom data in the NCrystal material data files, which can be used by the new physics models developed in the plugin. These developments have made it possible to include small-angle neutron scattering, magnetic scattering, and texture effects directly into NCrystal and will be described below.

\subsubsection{Small-angle neutron scattering plugin}
\label{sec:nanodiamond_sans_model}
The design and optimization of reflector systems for very‐cold and ultracold neutron sources based on NDs critically relies on an efficient and validated implementation of SANS. For this reason, we investigated different approaches for integrating in Monte-Carlo codes a pre-existing empirical model for SANS in NDs powder. The first possibility consists of extending the NCrystal poly-crystalline kernel by adding the SANS process through a plugin which is interfaced through NCrystal with other radiation transport codes, e.g. McStas, Geant4 or OpenMC. The other possibility is to implement the ND calculation directly in the Monte-Carlo code and use a modified .ACE SANS file, as was done in \cite{granada2020studies}. This work has been presented in earlier papers \cite{rizzi_SANSND,MarquezDamian2021}.

The empirical model that serves as a starting point for both implementations is based on the work by Granada et. al. \cite{granada2020studies}\cite{Granada2021} and was implemented prior to the beginning of this project \cite{MarquezDamian2021}. The idea is to fit a simple model to the SANS structure factor measured by Teshigawara et. al. \cite{teshigawara2019measurement}. This approach takes into consideration the nanoparticle size distribution, because it affects the experimental SANS structure factor.

\begin{figure}[h!]
      \includegraphics[width=0.9\columnwidth]{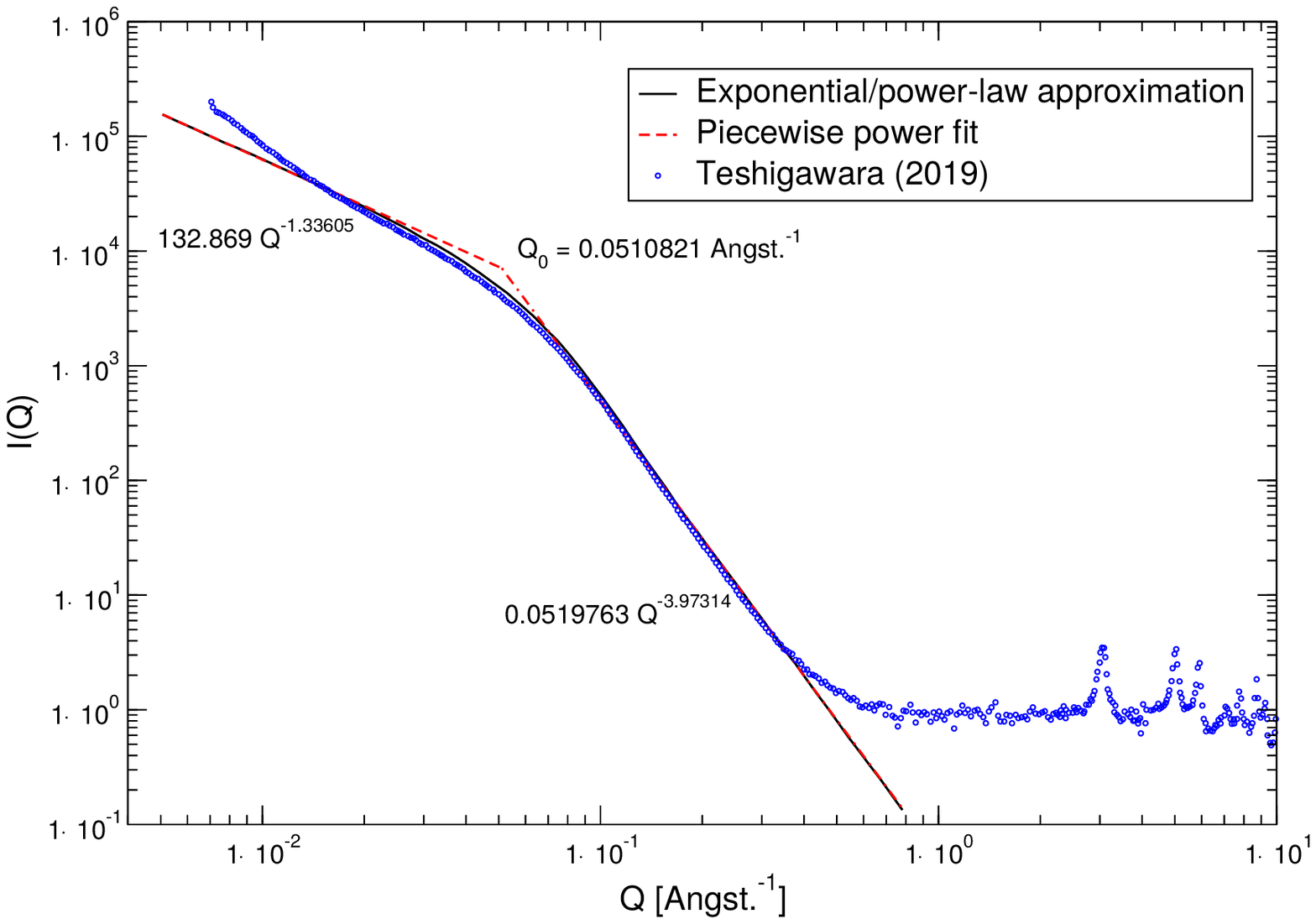}
\caption{Structure factor $I(Q)$ in the Teshigawara paper fitted by a power-exponential law and then further simplified into a piecewise power law fit.}
\label{fig:iofq_ND_teshi}
\end{figure}

In the Teshigawara work, the experimental results were fit using the unified exponential/power-law approximation, as expressed in Eq.~(3) in \cite{Avdeev2013}, while this implementation further simplifies the fit function in a piecewise power law:

\begin{equation}
   I(Q)=\left\lbrace \begin{array}{ll}
                  A_1 Q^{b_1} & Q< Q_0\\
                  A_2 Q^{b_2} & Q> Q_0
                \end{array} \right. 
\end{equation}
The results, along with the initial $I(Q)$ data are shown in \cref{fig:iofq_ND_teshi}.
By using the relation between the SANS structure factor and the SANS cross section,  and considering that for neutron energies $E_n > \SI{0.01}{meV}$ (lower limit in many Monte Carlo simulation software) it is always true that the incident wavenumber $k_0$ satisfies $ k_0 > 6.95\times10^{-2}$~\AA$^{-1}> Q_0 = 5.11\times10^{-2}$~\AA$^{-1}$, %ForJIMDorDDJ not sure what is needed here, comment by Valentina. FIXED. Added definition of k0
then it is possible to use the following expression for the total SANS cross section:
\begin{equation}
    \begin{split}\sigma(k_0) = \dfrac{\sigma_0}{2k_0^2}\int_0^{2k_0} q I(q) \mathrm{d} q = & \dfrac{\sigma_0}{2k_0^2}\left (\int_0^{Q_0} q I(q) \mathrm{d} q + \int_{Q_0}^{2k_0} q I(q) \mathrm{d} q \right)\\
    = & \dfrac{\sigma_0}{2k_0^2} \left[\dfrac{A_1}{b_1+2}Q_0^{b_1+2}+\dfrac{A_2}{b_2+2}(2 k_0)^{b_2+2} -\dfrac{A_2}{b_2+2}Q_0^{b_2+2} \right] \end{split}desc
\end{equation}
This cross section as a function of the wavevector $k_0$ is used by Monte-Carlo codes during run-time to compute the macroscopic total cross section and sample the distance to next collision. If there is a collision, the code randomly picks the reaction based on the ratio of its cross section to the total one. For different types of scattering, the outgoing energy and direction are sampled according to the underlying physics. In the case of SANS, there is no energy exchange, but the direction still needs to be determined. The cumulative probability distribution for the scattering vector $Q$ is given by:
\begin{equation}
CDF(k_0, Q) = \left\lbrace \begin{array}{ll}\dfrac{\dfrac{\sigma_0}{2k_0^2}}{\sigma(k_0)} \dfrac{A_1}{b_1+2}Q^{b_1+2} & Q<Q_0\\ \dfrac{\dfrac{\sigma_0}{2k_0^2}}{\sigma(k_0)}\left[ \dfrac{A_1}{b_1+2}Q_0^{b_1+2}+\dfrac{A_2}{b_2+2}Q^{b_2+2} -\dfrac{A_2}{b_2+2}Q_0^{b_2+2} \right]& Q>Q_0\\\end{array}\right.
\label{eq:cdf}
\end{equation}
The value of the outgoing $Q$ is sampled by first inverting \cref{eq:cdf} analytically and then evaluating $CDF^{-1}(\xi)$ at a random number $\xi \sim U(0,1)$. The scattering cosine is finally computed as $\mu = 1 - \dfrac{1}{2} \left[\dfrac{Q}{k_0}\right]^2$. \\
This model has been implemented as a plugin in NCrystal and has been extensively benchmarked in \cite{rizzi_SANSND}.

\begin{figure}[h]
\begin{center}
    \includegraphics[width=0.7\columnwidth]{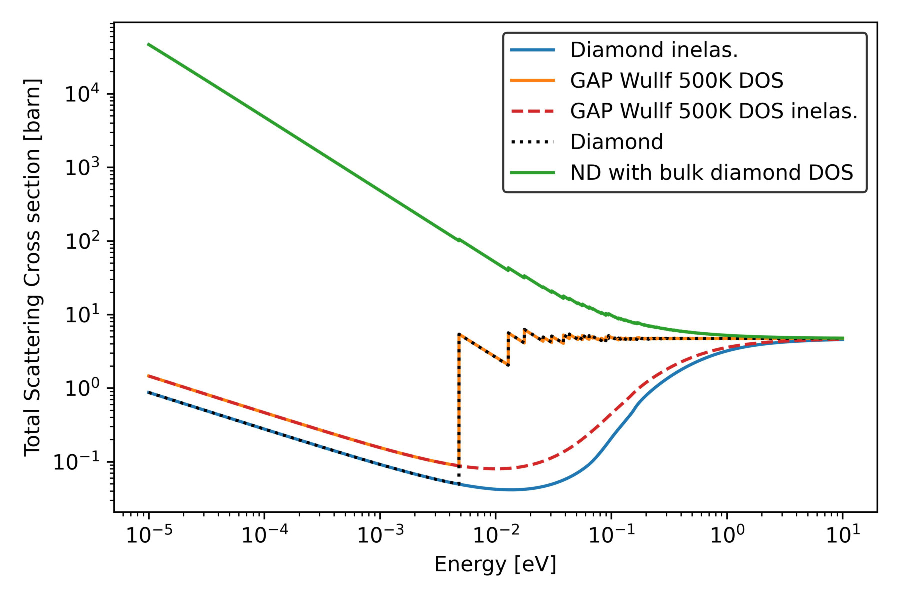}
\caption{Calculated scattering cross-sections and components
for NDs using the phonon frequency distribution of
perfect diamond and the phonon frequency spectrum based on
a finite-size of the ND particles. From \cite{dijulio2023_refl}}
\label{fig:ncrystal_tsx_diamond_vs_nanodiamonds}
\end{center}
\end{figure}

In \cref{fig:ncrystal_tsx_diamond_vs_nanodiamonds} we can see the calculated total scattering cross-sections using the phonon frequency distribution for perfect diamond and the one based on the finite size of the ND particles using the NCrystal plugin. It can be seen that the scattering cross-section is dominated by small-angle neutron scattering at low-energies, while the finite size of the NDs impacts the inelastic component of the cross-section.

\subsubsection{Magnetic scattering plugin}

Clathrate hydrates hosting oxygen are considered as promising candidates due to neutron inelastic magnetic scattering with oxygen~\cite{zimmer2016neutron}. A tiny energy around \SI{0.4}{meV}, which corresponds to the zero-field splitting constant, can be removed from the neutrons due to the magnetic interactions with the electrons of oxygen molecules in paramagnetic states. Based on the theory derived by Zimmer~\cite{zimmer2016neutron}, the equations of the magnetic scattering kernels are implemented as a plugin named ncplugin-MagScat in NCrystal. 

At low temperatures, molecules in crystalline oxygen appear to be antiferromagnetically ordered~\cite{freiman2004solido2,liu2004afmo2}. Oxygen molecules are in paramagnetic states when they are kept sufficiently far away to avoid magnetic ordering, e.g., enclathrated in clathrate hydrates~\cite{zimmer2016neutron}. Paramagnetic molecular oxygen possesses a spin triplet ground state characterising the spin state projection along the symmetry axis of the molecule. The transition of the magnetic levels or spin orientation results in changing the neutron kinetic energy with \SI{0.4}{meV} due to the molecular zero-field splitting. The physics of neutron scattering with paramagnetic oxygen is derived in Ref.~\cite{zimmer2016neutron}. Nevertheless, only the inelastic magnetic cross sections are given, the derivation of the theoretical outgoing neutron distribution or the scattering kernels required by Monte-Carlo simulations remains undone. Scattering kernels or $S(\overrightarrow{Q},\omega)$ are tabulated data giving the probability of neutron scattering with target nuclei or electron system as a function of neutron wave vector transfer $\overrightarrow{Q}=\overrightarrow{k}-\overrightarrow{k^\prime}$ and energy transfer $\hbar\omega=E-E^\prime$. The approximation of neglecting the Debye-Waller factor in the calculations of paramagnetic scattering cross sections in Ref.~\cite{zimmer2016neutron} can also be improved. The physics of neutron nucleus scattering can be included to take into account the impacts of atoms presented in the cage structure.

The aim is thus to include and validate the physics of neutron scattering with paramagnetic oxygen in the Monte-Carlo neutron transport simulations. To this end, we use the open source software package NCrystal to calculate the neutron nucleus scattering kernels and benefit from its flexibility and versatility to include the magnetic physics in a plugin named ncplugin-MagScat. NCrystal~can be used as a backend for the Monte-Carlo particle transport code OpenMC, which provides a possibility to validate the implemented magnetic scattering model by comparing with the experimental measurements reported in the literature, performed by Chazallon et al. on \ce{O2} clathrate hydrate~\cite{chazallon2002anharmonicity} and Renker et al. on \ce{O2}-\ce{C60}~\cite{renker2001intercalation}, respectively.

Based on Ref.~\cite{zimmer2016neutron}, the neutron magnetic scattering kernels or dynamic structure factors $S_{\textrm{mag}}(Q,\omega)$ are derived. In this section we only present the final form of the equations. Without external magnetic fields, the double differential magnetic scattering cross section for unpolarised neutrons is given by

\begin{equation}
	\dfrac{\Diff 2\sigma_{\textrm{mag}}}{\diff \Omega \diff {E^\prime}}=b_{\textrm{m}}^2\left(\sqrt{\dfrac{E^\prime}{E}}S_{\textrm{mag},\pm}(Q,\omega)+S_{\textrm{mag},0}(Q,\omega)\right),
	%\label{eq:ddxsmSQw1}
\end{equation}

where $b_{\textrm{m}}=\SI{5.404}{fm}$ is the magnetic scattering length, $Q$ is the neutron wavenumber transfer, $\hbar\omega=E-E^\prime$ is the neutron energy transfer, $S_{\textrm{mag},\pm}(Q,\omega)$ and $S_{\textrm{mag},0}(Q,\omega)$ represent respectively the magnetic inelastic ($+$ for up-scattering and $-$ for down-scattering) and elastic scattering kernels and are given by

\begin{equation}
	S_{\textrm{mag},\pm}(Q,\omega)=\exp\left(-(\langle u^2\rangle+\dfrac{\ln(2)}{\Gamma_{\textrm{mag}}^2})Q_{\pm}^{2}\right)g_{\pm}(T)\delta(\hbar\omega\pm D),
	%\label{eq:SQwpm1}
\end{equation}

and

\begin{equation}
	S_{\textrm{mag},0}(Q,\omega)=\exp\left(-(\langle u^2\rangle+\dfrac{\ln(2)}{\Gamma_{\textrm{mag}}^2})Q_{0}^{2}\right)g_{0}(T)\delta(\hbar\omega).
	%\label{eq:SQw01}
\end{equation}

$\langle u^2\rangle$ is the mean-squared displacement (MSD) which is temperature-dependent and can be calculated from the PDOS. $\Gamma_{\textrm{mag}}=\SI{15}{nm^{-1}}$ represents the half width at half maximum (HWHM) of the magnetic form factor which is approximated by a Gaussian function~\cite{kleiner1955magnetic}. $D$ is the zero-field splitting constant and $g(T)$ are functions which contain thermal average of spin matrix elements.

\begin{figure}[htbp]
	\centering
\begin{subfigure}[b]{0.445\textwidth}
	\centering
    \includegraphics[width=\textwidth]{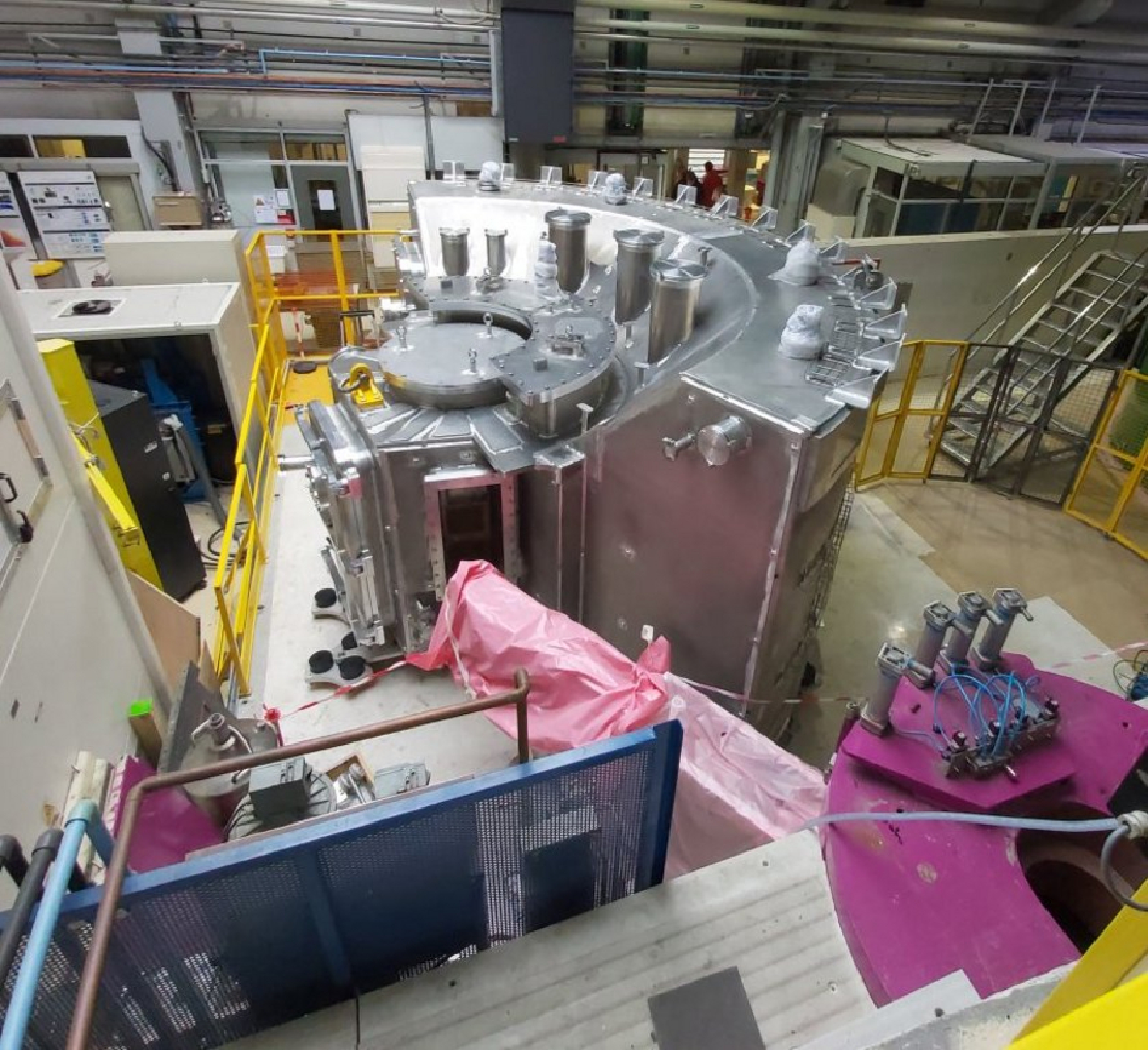}
    \caption{IN6}
    \label{fig:in6_view}
\end{subfigure}
\hfill
\begin{subfigure}[b]{0.545\textwidth}
    \centering
    \includegraphics[width=\textwidth]{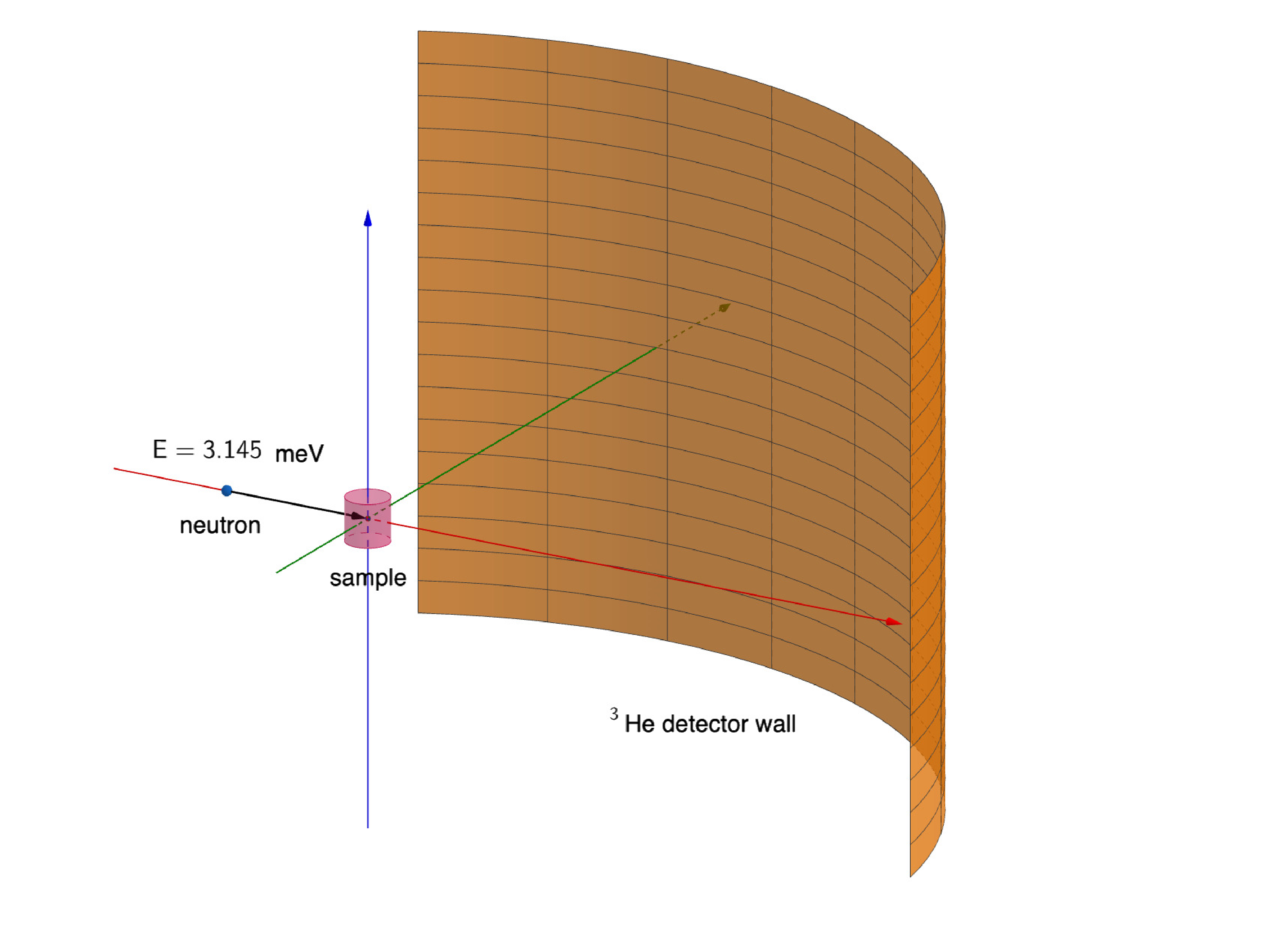}
    \caption{Simplified model}
    \label{fig:in6_simplified}
\end{subfigure}
	\caption{Time-of-flight spectrometer IN6 at ILL.}
    \label{fig:in6_model}
\end{figure}

\begin{figure}[htbp]
	\centering
	\includegraphics[width=\textwidth]{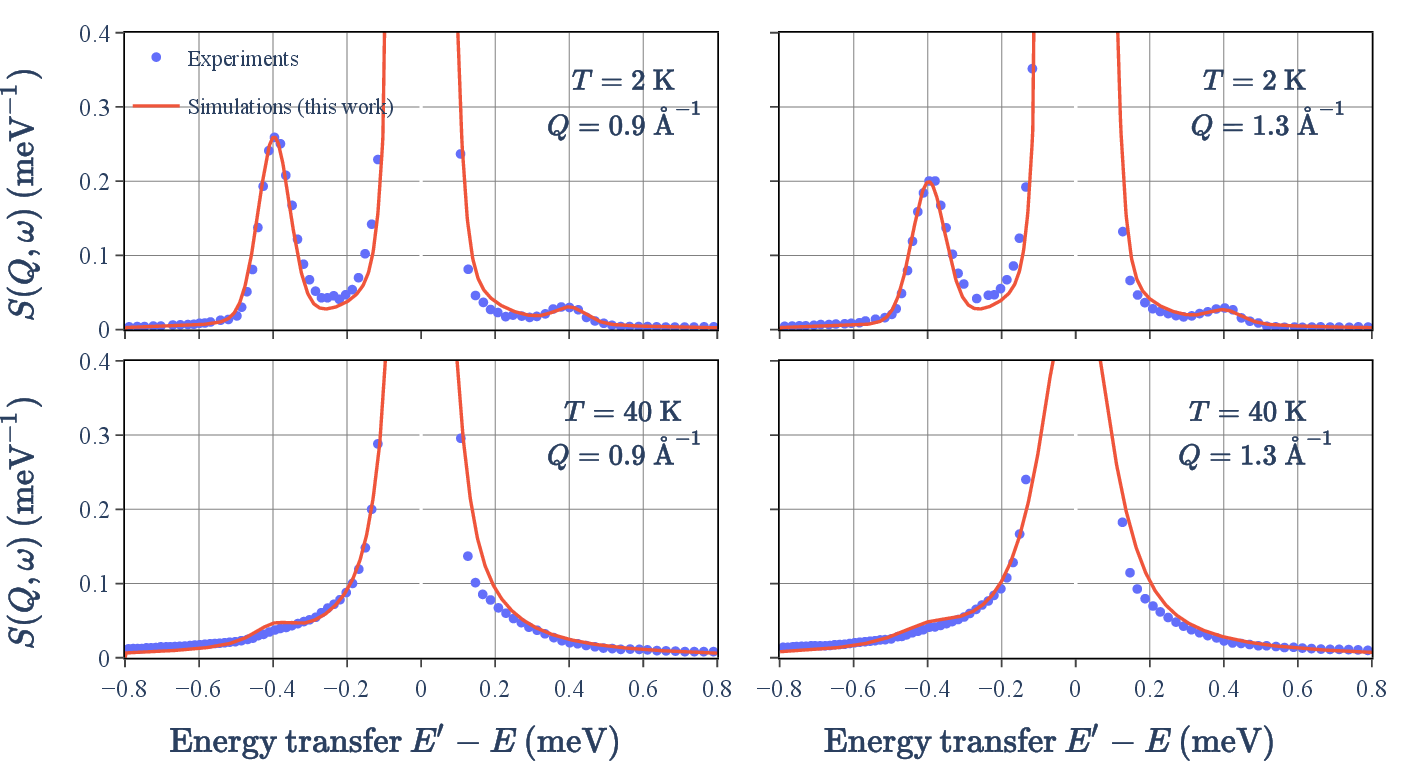}
	\caption{Comparisons between simulated and experimental $S(Q,\omega)$ on \ce{O2}-clathrate~\cite{chazallon2002anharmonicity} at \SI{2}{K} and \SI{40}{K}.}
	\label{fig:O2-D2O-sim-exp}
\end{figure}

The above model has been integrated into a NCrystal plugin and more details are listed in Ref.~\cite{xu2023magnetic}. It is pointed out in Ref.~\cite{zimmer2016neutron} that the inelastic neutron scattering due to magnetic effects was observed in the experiments performed by Chazallon et al.~\cite{chazallon2002anharmonicity} on \ce{O2}-clathrate and Renker et al.~\cite{renker2001intercalation} on \ce{O2}-\ce{C60}. Both of the measurements were performed on the time-of-flight spectrometer IN6 (\cref{fig:in6_view}) at ILL. The incident neutron wavelength was set to be $5.1$~\AA~(\SI{3.145}{meV}) for these two measurements. The incident wavelength is larger than the Bragg cutoff of aluminum (\ce{Al}) ($\approx4.6$~\AA), which has the advantage to avoid possible contamination from \ce{Al}. In our work, we investigate the magnetic effects through Monte-Carlo simulations. Similar to the process developed for simulating neutron diffraction measurements on uranium dioxide~\cite{xu2021uo2diffraction}, the calculation scheme is composed of MD and DFT calculations, Monte-Carlo simulations, and experimental correction.

MD and DFT calculations are detailed in Ref.~\cite{xu2023magnetic}. The simulations were performed by using the Monte-Carlo particle transport code OpenMC. From version 0.13.3, OpenMC enables users to define materials through the NCrystal~.ncmat file and call the corresponding neutron scattering cross sections on-the-fly. The neutron magnetic scattering kernels calculated by the developed plugin ncplugin-MagScat~and the neutron nucleus scattering kernels generated from PDOS can thus be taken into account in the Monte-Carlo simulations. 

Based on detailed characteristics found on the website~\cite{ill2023in6}, a simplified model of IN6 spectrometer (\cref{fig:in6_simplified}) was used in the OpenMC simulations. The distance between the sample position and the cylindrical \ce{^{3}He} detector wall is $L=\SI{2.48}{m}$. The detector wall of IN6 covers a large azimuthal angular range from \SI{10}{\degree} to \SI{115}{\degree} and a vertical angular range near to \SI{\pm 15}{\degree}. The detector wall is composed of 337$\times$109 grids along the azimuthal and vertical angular ranges, respectively. The spatial resolution is thus equal to $\SI{1.3}{cm}\times\SI{1.2}{cm}$. 

The experimental corrections were performed to take into account the resolution of the detectors and the motions of the guest oxygen molecules which are temperature-dependant. Good agreement is obtained between the simulated and experimental $S(Q,\omega)$ at \SI{2}{K} and \SI{40}{K} (see \cref{fig:O2-D2O-sim-exp}), confirming the magnetic scattering physics implemented in the developed plugin. The results on \ce{O2}-\ce{C60} are detailed in Ref.~\cite{xu2023magnetic}. Preliminary investigations of the magnetic model in the VCN production are presented in \cref{sec:vcn}.

\subsubsection{Texture plugin}

A polycrystalline material, whose grains are randomly oriented, is referred to as a powder. The calculation of the coherent elastic scattering cross section $\sigma_{\textrm{coh}}^{\textrm{el}}(\lambda)$ for powder in NCrystal is given by~\cite{Cai2020,kittelmann2021elastic}:
\begin{equation}
    \sigma_{\textrm{coh}}^{\textrm{el}}(\lambda)=\dfrac{\lambda^2}{2V_{\textrm{uc}}}\sum_{hkl}d_{hkl}|F_{hkl}|^2,
    \label{eq:md}
\end{equation}
where $V_{\textrm{uc}}$ is the volume of the unit cell of the crystal, $d_{hkl}$ represents the distance between adjacent atomic plans (hkl), and $|F_{hkl}|^2$ is the form factor depending on the Debye-Waller coefficient or mean squared displacement of the atoms.

Nevertheless, grains having a preferred orientation or texture are observed in polycrystalline materials. The presence of texture can have a significant impact on $\sigma_{\textrm{coh}}^{\textrm{el}}(\lambda)$. The texture can be described using the orientation distribution function (ODF) which gives the probability of finding grains with a specific orientation in a polycrystalline material~\cite{wright2005texture}. Several models which incorporate the ODF in the calculation or correction of $\sigma_{\textrm{coh}}^{\textrm{el}}(\lambda) $ exist~\cite{laliena2020texture,santisteban2012texture,malamud2014texture,dessieux2019texture}. In our work, we implemented the March-Dollase model as a plugin in NCrystal to investigate the texture effects in graphitic materials in transmission measurements.

The March-Dollase model assumes an axially symmetric orientation distribution around the beam direction~\cite{sato2011rietveld}. \cref{eq:md} is corrected by multiplying the following term $P_{hkl}(\lambda,d_{hkl})$:
\begin{equation}
	P_{hkl}(\lambda,d_{hkl})=\dfrac{1}{2\pi}\int_{0}^{2\pi}\left(R^2B_{hkl}^2+\dfrac{1-B_{hkl}^2}{R}\right)^{-\frac{3}{2}}\textrm{d}\phi,
\end{equation}
where
\begin{equation}
	B_{hkl}=\cos(A_{hkl})\sin(\theta_{hkl})+\sin(A_{hkl})\cos(\theta_{hkl})\sin(\phi),
\end{equation}
\begin{equation}
	A_{hkl}=\arccos\left(\dfrac{hH+kK+lL}{\sqrt{h^2+k^2+l^2}\sqrt{H^2+K^2+L^2}}\right),
\end{equation}
\begin{equation}
	\theta_{hkl}=\arcsin(\dfrac{\lambda}{2d_{hkl}}),
\end{equation}
and $(HKL)$ is the preferred orientation axis parallel to the beam direction. $R$ is a coefficient representing the degree of crystallographic anisotropy. A polycrystalline material can have more than one preferred orientation. We use the fraction $f_{hkl}$ and $R_{hkl}$ associated to each preferred orientation $hkl$ with $\sum f_{hkl}=1$.

\begin{figure}[htbp]
    \centering
    \includegraphics[width=.7\textwidth]{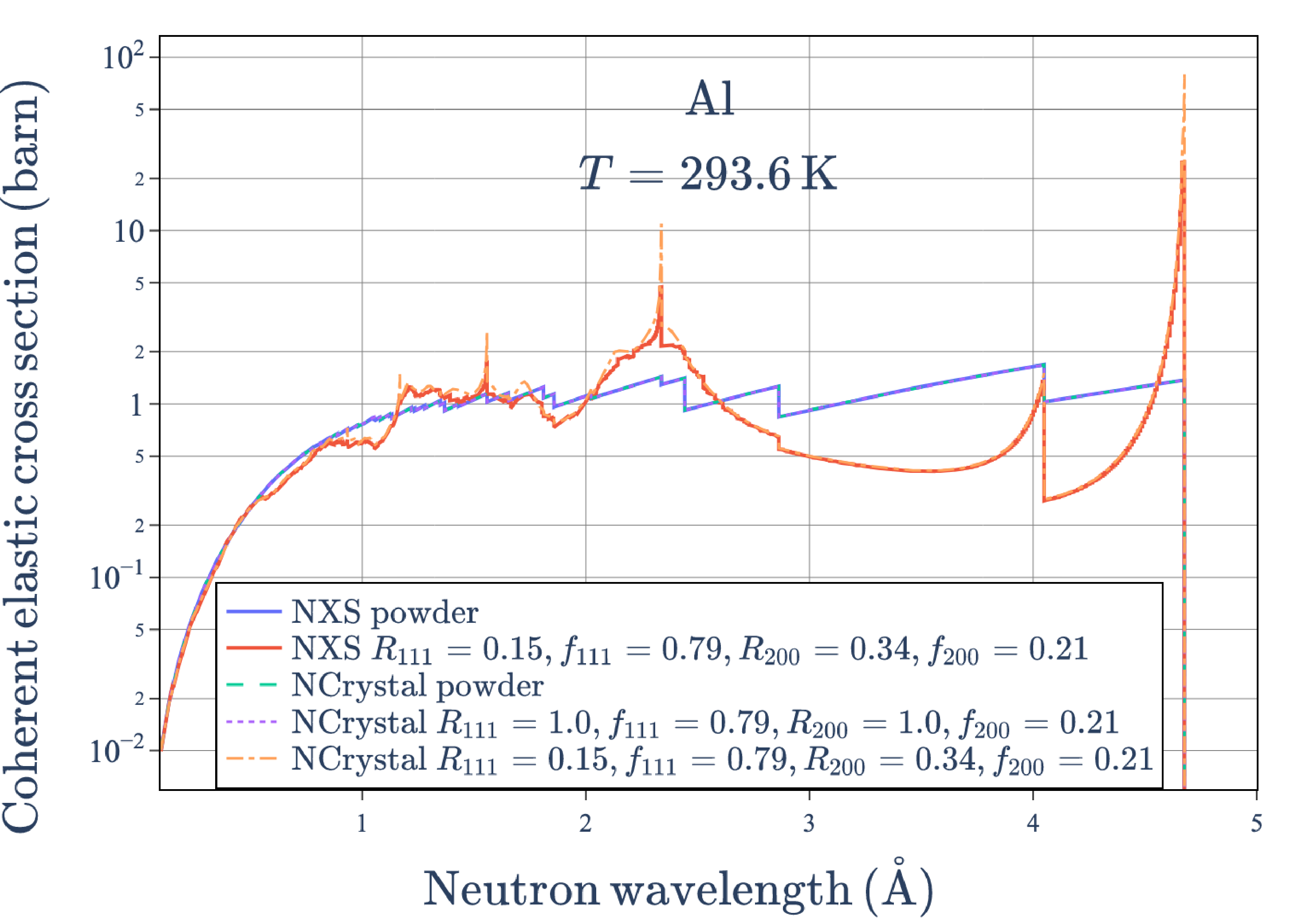}
    \caption{Comparison of coherent elastic scattering cross section calculated by the NXS code and the NCrystal plugin for aluminum with and without texture.}
    \label{fig:Al_tex}
\end{figure}

\begin{figure}
    \centering
    \includegraphics[width=0.9\textwidth]{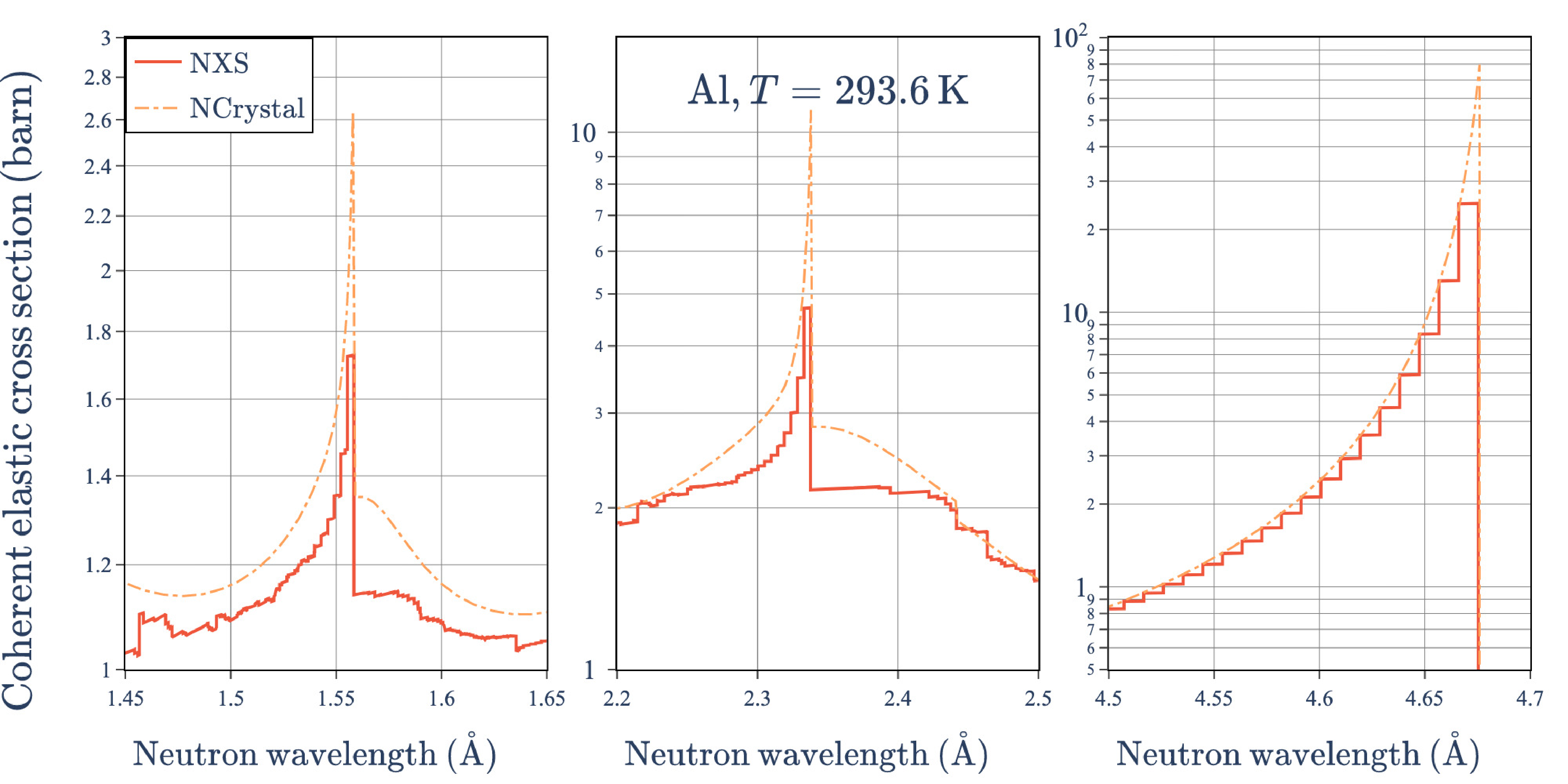}
    \caption{Zoom of cross section for aluminum with texture shown in \cref{fig:Al_tex}.}
    \label{fig:Al_tex_zoom}
\end{figure}

The March-Dollase model has been implemented in the NXS code~\cite{boin2012nxs}. The coherent elastic cross sections for aluminum calculated by NXS and our NCrystal plugin are compared in \cref{fig:Al_tex}. We obtain an excellent agreement for powder aluminum. In \cref{eq:md}, when $R=1$, $P_{hkl}(\lambda,d_{hkl})$ becomes a constant equal to 1, representing the random orientation of grains. The identical cross section obtained by putting $R=1$ serves to verify our implementation for the texture model. For $R$ different from 1, the texture effects on the coherent elastic cross sections are clearly observed. We obtained a good agreement with the NXS code. A zoomed up image of the three peaks is presented in \cref{fig:Al_tex_zoom}. Our calculations show finer peaks compared to the NXS code because we use a finer neutron wavelength discretization. The comparison of cross section with NXS serves to verify our implementation of the March-Dollase model in the developed plugin.

\subsection{Experimental investigations of graphitic compounds}

To support the work on the graphitec compounds, we carried out a series of measurements at the BOA \cite{MORGANO201446} beamline at the Paul Scherrer Institute. These measurements were deemed necessary due to the lack of information of neutronic properties of expanded Bragg-edge graphitic compounds, and as a way to understand the structural and dynamical properties of the material better. 

For the measurements, we prepared several different graphitic compound samples in flake form. These included normal graphite, graphite oxide, deuterated graphite oxide, graphite oxide intercalated with water and deuterated graphite oxide intercalated with heavy water. 

\begin{figure}[htb!]
\centering
\vspace{ 1 cm}
{\includegraphics[height=8.0cm]{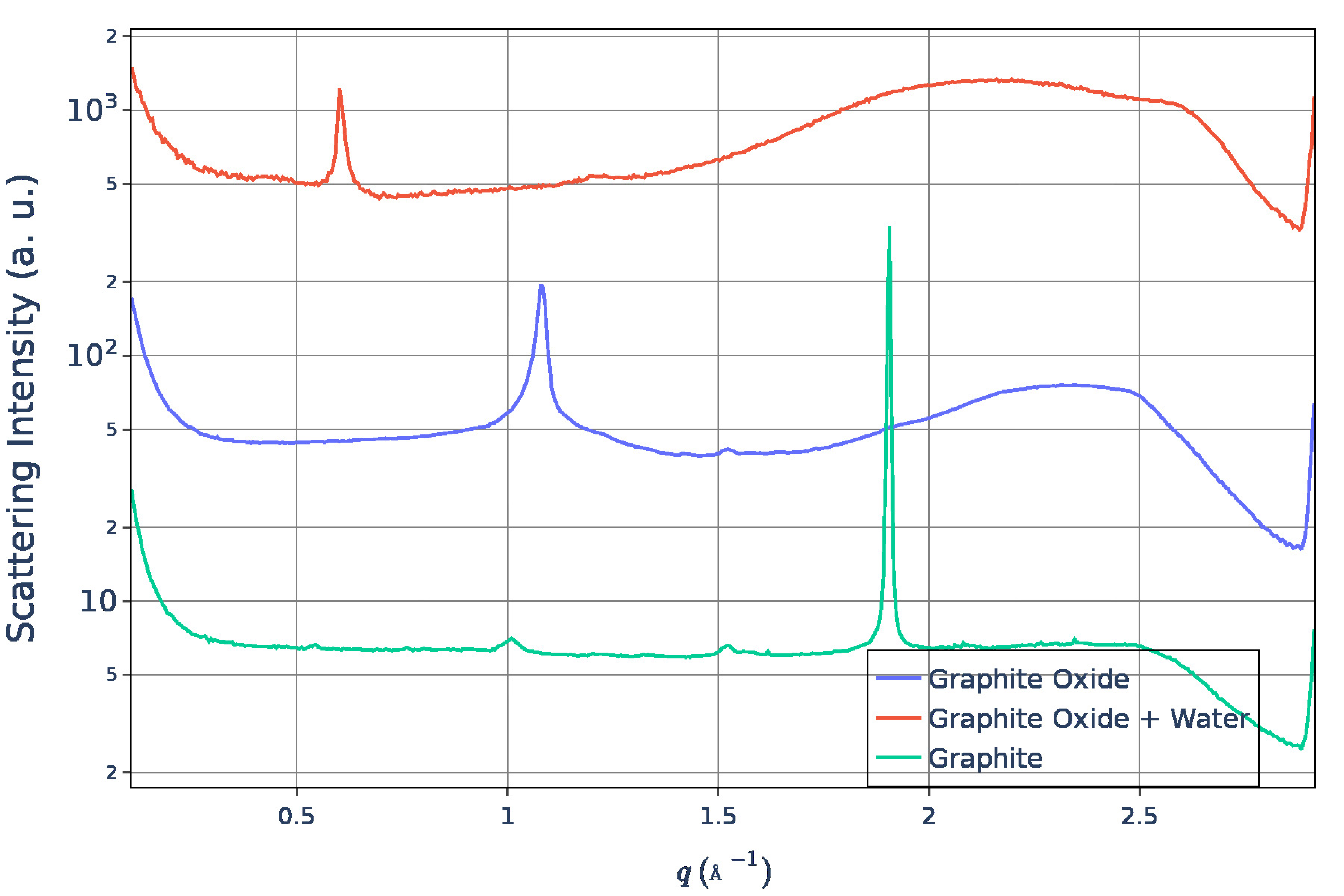}}
\caption{Examples of diffraction data collected at CoSAXS.}
\label{maxiv-diffraction}
\end{figure}

\cref{maxiv-diffraction} shows examples of (wide-angle X-ray scattering) WAXS data collected at the CoSAXS beamline at MAX-IV for graphite, graphite oxide, and graphite oxide mixed with water. The plot clearly shows a decreasing value of $q$ for the narrow peak, which indicates an increased interplanar spacing and thus a larger Bragg cutoff. The shift in the peak between graphite and graphite oxide is due to the oxidation of the graphite, while the shift in peak between graphite oxide and the graphite oxide mixed with water sample is due to the intercalation of water molecules between the planes of the graphite oxide. 

\begin{figure}[t!]
\centering
\vspace{ 1 cm}
{\includegraphics[height=8.0cm]{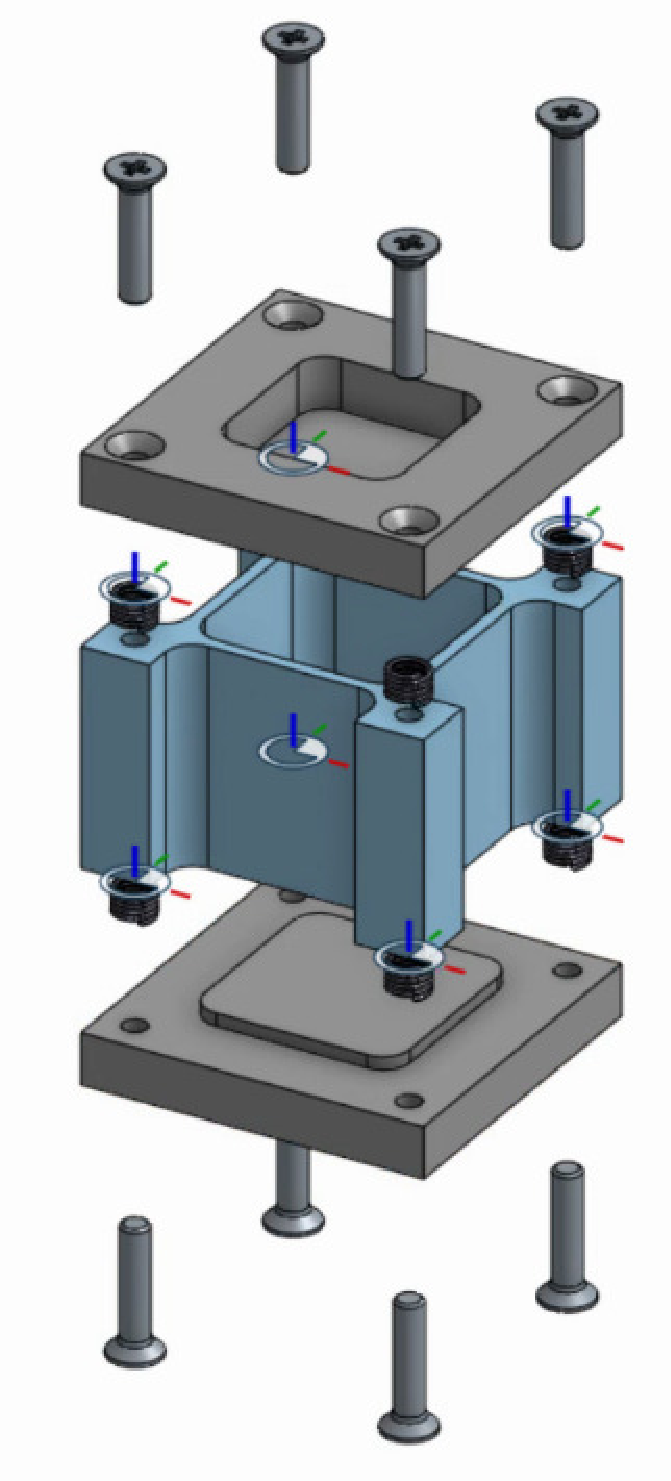}}\; {\includegraphics[height=6.0cm]{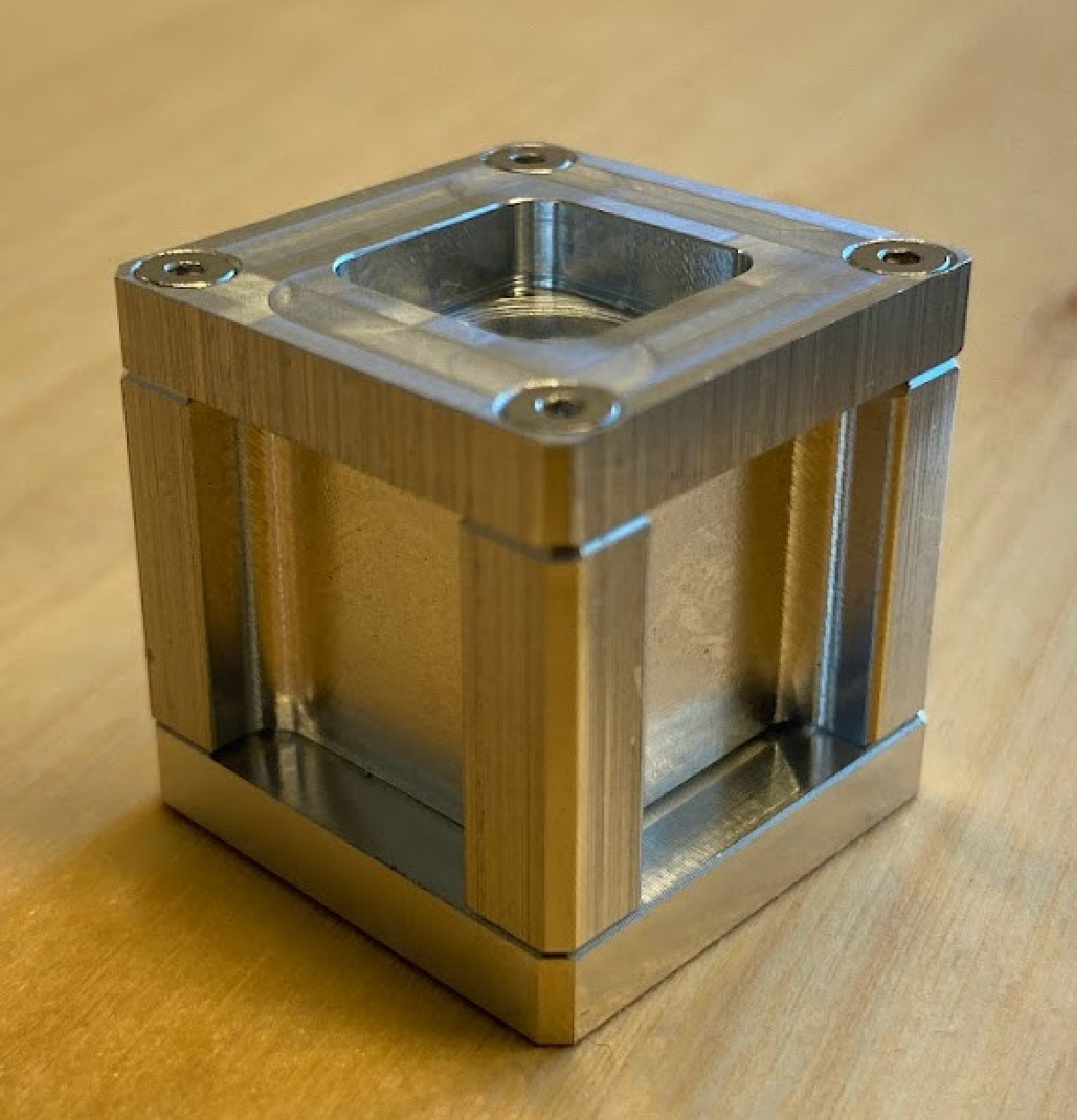}}
\caption{Sample holder for triaxial transmission experiments.}
\label{sample_holders}
\end{figure}

For the experimental campaign at BOA, we also developed a triaxial transmission sample holder (\cref{sample_holders}), which provides the same optical depth and the same center for the three directions. This was used to explore texture effects in the samples.

The measurements were carried out over three different beamtimes at BOA. In the first experiment we used the imaging setup of the instrument, however it was found that the background levels above 8 {\AA} were too high in order to resolve Bragg-edges above this limit. 

\begin{figure}[htb!]
\centering
\vspace{ 1 cm}
%{\includegraphics[height=8.0cm]{tsl_fig/spectracompare.eps}}
{\includegraphics[height=8.0cm]{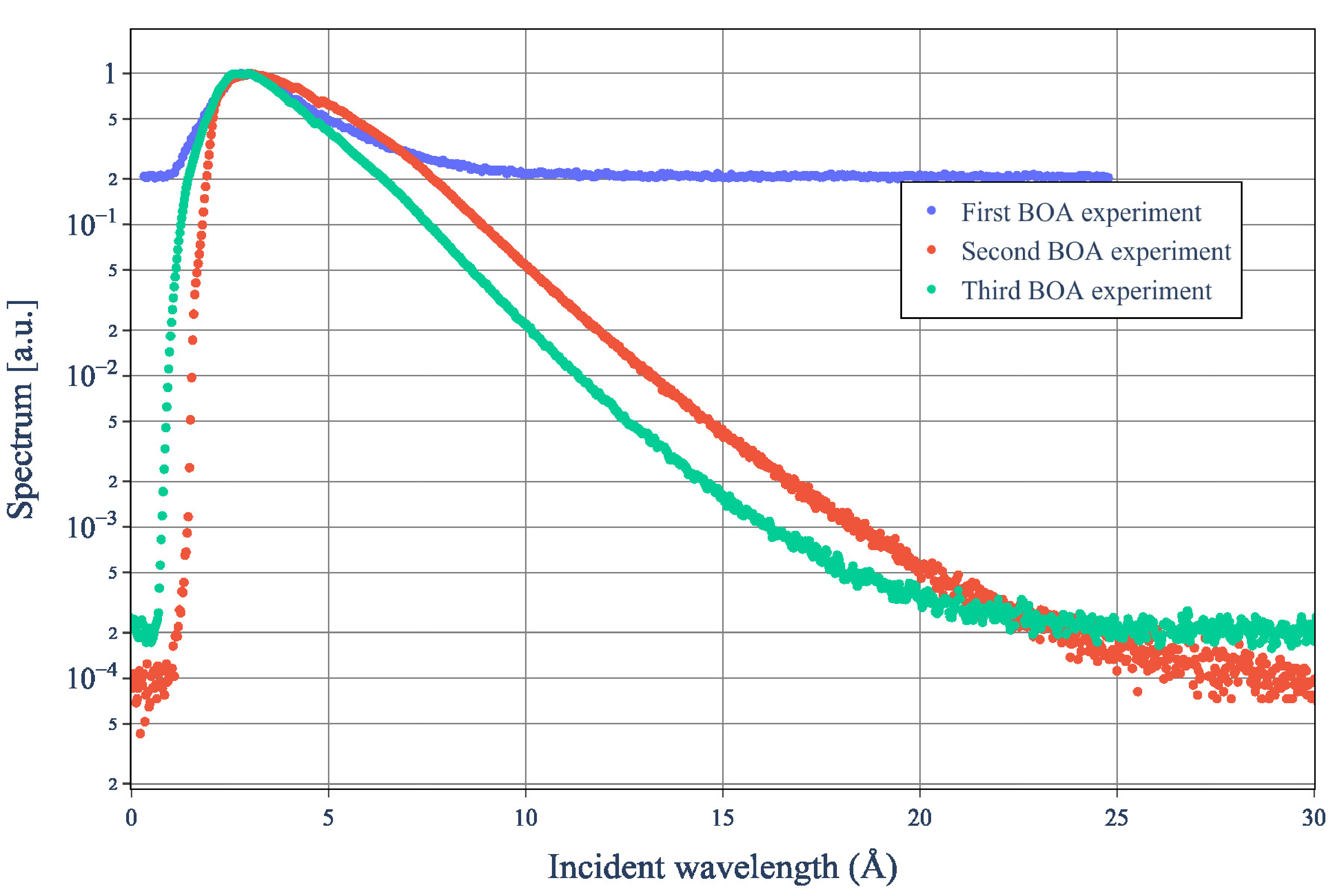}}
\caption{Measured spectra for the three experiments at BOA in PSI. From the first to the second experiment, background was significantly reduced, which resulted in an increase of the usable wavelength range. From the third experiment, the neutron mirror was removed, which made it possible to measure in the range $1-2\,\AA$, at the expense of the long wavelength region.}
\label{spectracompare}

\centering
\vspace{ 1 cm}
%{\includegraphics[height=8.0cm]{tsl_fig/graphite.eps}}
{\includegraphics[height=8.0cm]{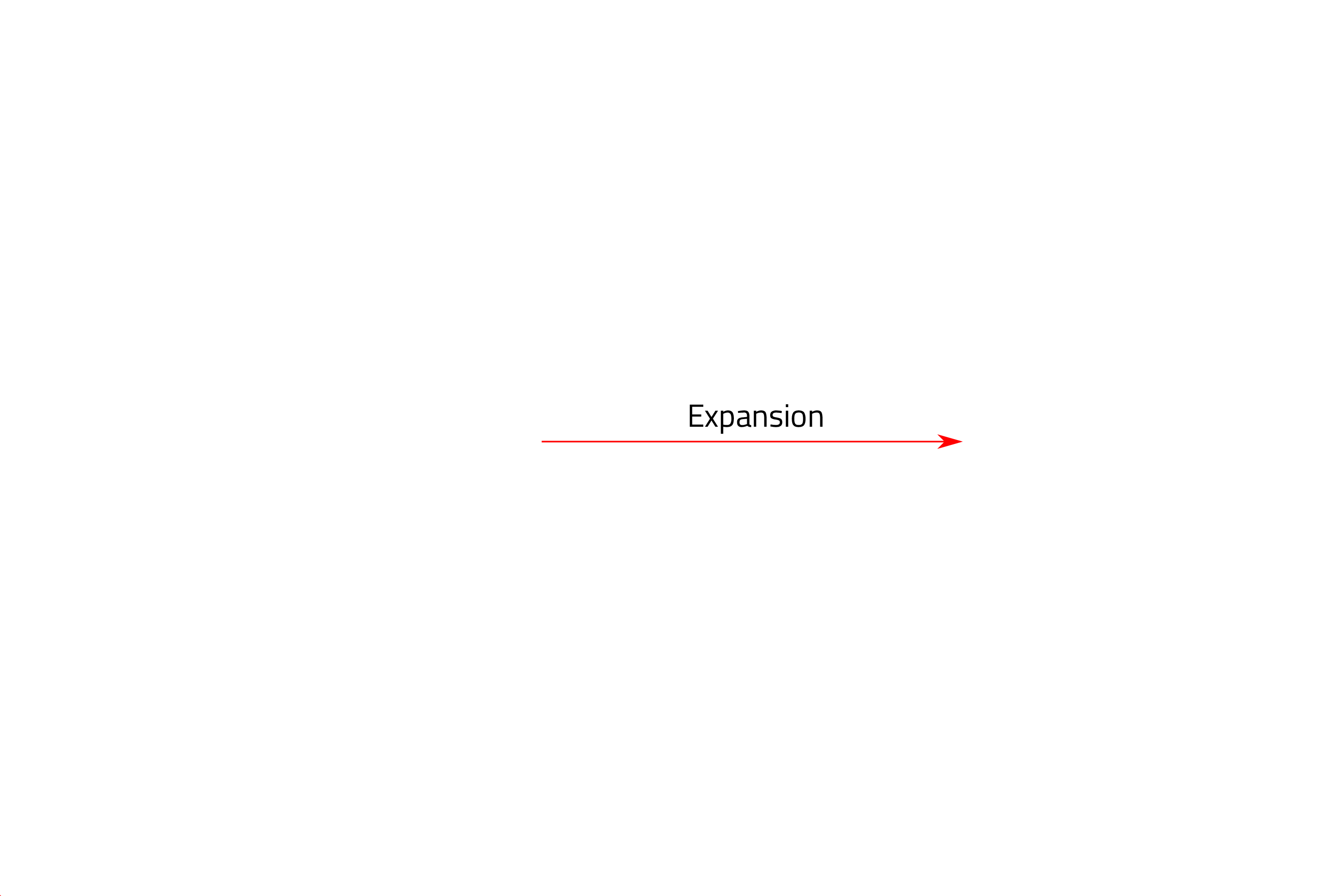}}
\caption{Measured total cross-sections for graphite and graphite oxide in the second PSI experiment.}

\label{exp2}
\end{figure}

For the second measurement, we made major revisions to the setup in order to reduce the background above about 8 $\AA$. This included the usage of a neutron collimation system from the PSI Neutron Optics and Computing Group, in combination with a flat, 8.5 mm thick multi-tube He-3 detector. We also used a flat $m=4$ mirror for a better separation of the cold neutron beam from the fast beam component, which we identified as a contribution to the background in the first run. A main goal of this experiment was to determine the best geometry of the optics, shielding and detector that maximizes the signal-to-background ratio. The resulting spectrum from the revised setup compared to the spectrum from the first measurement is shown in \cref{spectracompare}. As seen in the figure, the reduction in the background level was dramatic, resulting in a spectrum reaching up to around 25 {\AA } in wavelength. \cref{exp2} shows the resulting total cross-section for graphite oxide using the revised setup, where it is now possible to see clearly a Bragg edge appearing around 12 \AA. 

The main aim of the third measurement at BOA was to investigate texture effects in the graphite oxide samples. The spectrum obtained from this measurement is also shown in \cref{spectracompare}, where the differences at lower wavelength appear as the mirror was not used in the latter measurements. 

\begin{figure}[t!]
\centering
\vspace{ 1 cm}
{\includegraphics[height=8.0cm]{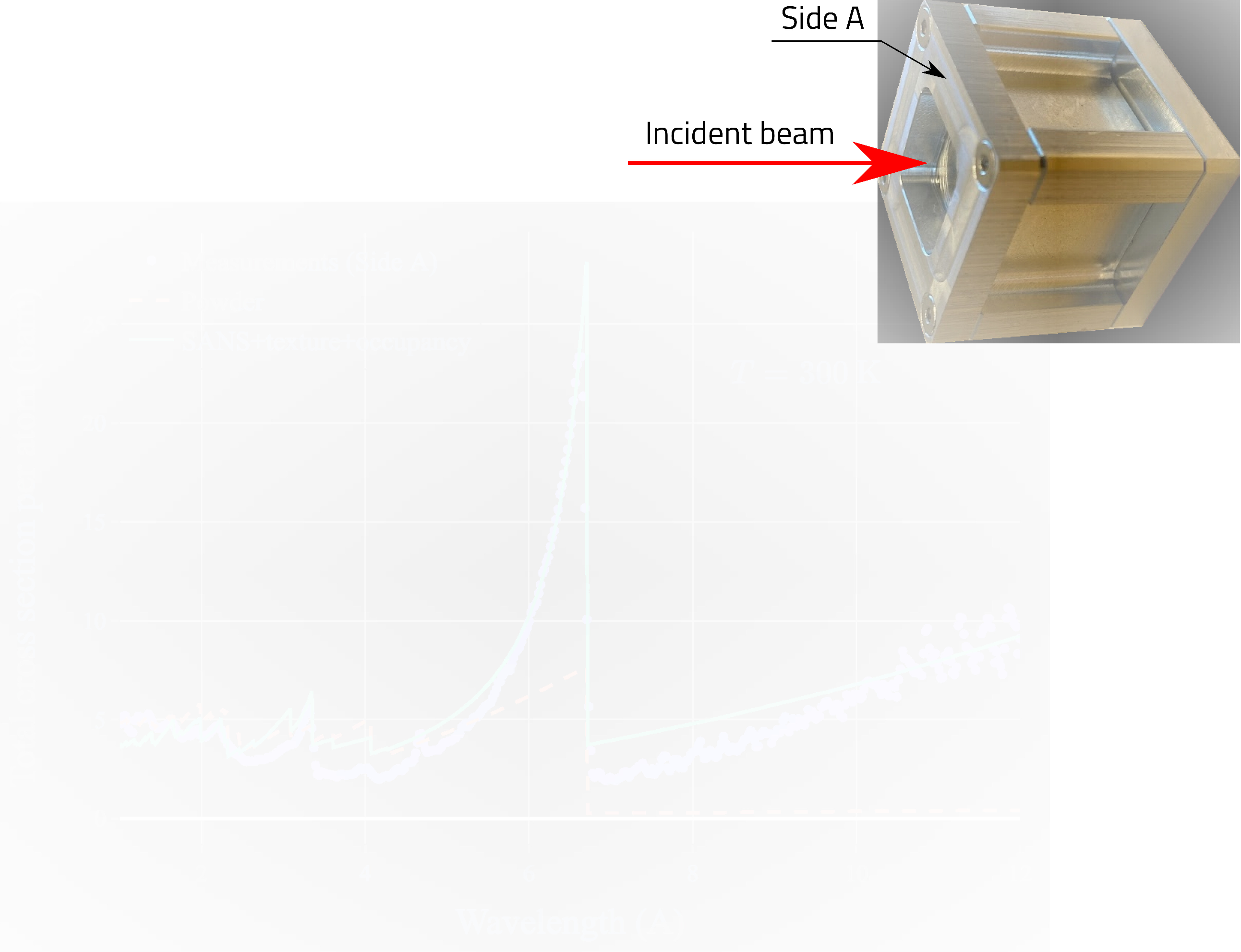}}
\caption{Total cross section for graphite flakes, measured and computed in the direction of compression.}
\label{C_sideA_XS}
\end{figure}

A preliminary analysis of the results confirms the presence of texture in the samples, which can be appropriately represented using the March-Dollase texture implemented in NCrystal. \cref{C_sideA_XS} shows the total cross section in the direction of compression for graphite flakes. It can be well represented by a combination of a texture model, to include the orientation of the flakes during the filling of the sample holder, and a SANS model to include the microscopic porosity of the graphites. This combination of models will also be used for further analysis of the measured spectra for other graphitic materials.

\subsection{Advances on NCrystal}\label{sec:advances_ncrystal}
In addition to the above mentioned developments, NCrystal has undergone extensive upgrades over the course of the HighNESS project. This goes back to NCrystal release version 2.1.0. These releases include developments motivated not only by the HighNESS project, but also by outside projects and cover a wide range of areas such as improvements to the code and data format, extending and introducing new features and data, and bug fixes, just to name a few. A limited selection of specific examples are highlighted below. More information can also be found in the change logs \cite{NCrystal_Changelog}, covering developments not only related to this work but also outside projects. Several of these developments are listed below.

NCrystal version 2.5.0 was a major technical upgrade to NCrystal which included changes to more than 200 files amounting to around 15,000 lines of code. The main motivation for these improvements was to bring NCrystal code up to modern C++ standards, improve caching strategies, provide support for multi-threading, and to prepare for future improvements in NCrystal, for example for better support for integration with Monte-Carlo codes, such as Geant4 and OpenMC, and future possibilities for describing thermal neutron scattering in multi-phase materials. Of particular interest is the support for multi-threading. This allows NCrystal to be run safely in the multi-thread mode of a given Monte-Carlo code, such as OpenMC, thus making it possible to take advantage of the gains in efficiency due to multithreading. 

Release version 2.6.x includes improvements to handling observed artifacts in calculated inelastic cross-sections due to low-granularity beta-grids and phonon density of states curves in addition to the possibility to estimate atomic mean-squared-displacements from the phonon density of states. The latter contribution improves the reliability of the Debye-Waller factors at low-temperatures in elastic scattering components.

Release version 2.7.0 \cite{NCrystal_270} of NCrystal includes a massive addition of 64 new crystalline materials to the NCrystal library. Many of the new materials are of wide interest to the nuclear community, but serve as an important validation for NCrystal. Included in this library are also the data generated in this work for magnesium hydride and magnesium deuteride. The release also contains the useful scripts, ncrystal\_verifyatompos and ncrystal\_onlinedb2ncmat, for verification of input of the crystal structure of a material and also against multiple online databases (such as materialprojects.org or the Crystallography Open Database), respectively.

NCrystal was also updated to include improved sampling of scattering kernels for very cold and ultracold neutron producing materials. This was included in the release 3.1.0 of NCrystal. One practical challenge, facing in particular UCN moderator design studies, is that only a tiny fraction of scattered neutrons will normally be left with an energy at the desired UCN scale (hundreds of neV). Thus, in order to get sufficient statistics, while keeping computational requirements reasonable, it is necessary to employ some sort of biasing or variance reduction scheme. This is facilitated by making it possible to split a scattering process into two components: a down-scatter-to-UCN-regime process and a process with the rest of the physics, by respectively appending \texttt{";comp=inelas;ucnmode=only"} and \texttt{";ucnmode=remove"} to a given NCrystal material cfg-string. Furthermore, the dedicated UCN process is implemented with a special improved model which is believed to be completely free of any modelling artifacts. A new configuration parameter, ucnmode is used to control this UCN process split-up (cf. CfgRefDoc for details). Note that in order to divide neutrons into UCN and non-UCN regimes, the code also needs the definition of a UCN threshold energy. If not set explicitly, this will default to 300~neV ($\approx 522$ Å). 

In order to properly implement and validate the work described above, several internal utilities were added related to scattering kernels. This includes code for easily evaluating scattering kernels at any alpha and beta point, as well as utilities for what is essentially exact sampling or integration for cross sections. These utilities are useful as they allow us to validate scattering kernel models better, are used directly in the new UCN-production models, and will hopefully facilitate future improvements. 

NCrystal was further improved in order to facilitate the HighNESS thermal neutron scattering school held in May 2023 at the ESS. In preparation for this school, new tools were added, including for example better material creation via a python API.

\subsection{Contents of the HighNESS repository}

\paragraph{}

The software described here is freely available in the Github repository of HighNESS and is hosted in the Github platform, following the directives of the HighNESS Data Management Plan. The following is a list of the major software developments within the project.\\

\begin{itemize}
    \item OpenMC+Ncrystal:\\
    \url{https://github.com/highness-eu/openmc/tree/mixed_ncrystal}
    \item NCrystal NDs pĺugin\footnote{Plugin support requires installing NCrystal version 2.2.0 or later: \url{https://github.com/mctools/ncrystal}
}:\\
    \url{https://github.com/highness-eu/ncplugin-SANSND}
    \item NJOY+Ncrystal:\\
    \url{https://github.com/highness-eu/NJOY2016/tree/njoy_ncrystal}\\
    \url{https://github.com/highness-eu/ncrystal/tree/njoy_ncrystal}
    \item NJOY+Ncrystal Library (MgH$_2$, MfD$_2$ and additional materials):\\
    \url{https://github.com/highness-eu/NJOY-NCrystal-Library}
    \item OpenMC with modifications to support mixed-elastic scattering:\\
    \url{https://github.com/highness-eu/openmc/tree/njoy_ncrystal}
    \item Texture Plugin:\\
    \texttt{\url{https://github.com/highness-eu/ncplugin-CrysText} }
    \item NCMAT files for graphitic materials:\\
    \texttt{\url{https://github.com/highness-eu/ncmat-graphitic} }
    \item Magnetic Scattering Plugin:\\
    \texttt{\url{https://github.com/highness-eu/ncplugin-MagScat}
    }
    \item NCMAT files for clathrates:\\
    \texttt{\url{https://github.com/highness-eu/ncmat-clathrates} }
\end{itemize}

\subsection{Conclusions}

In this section we have presented developments of simulation software for describing neutron interactions in novel moderator and reflector materials considered of interest for the HighNESS project. The main focus has been the development of software to describe interactions in ND particles, magnesium hydride, graphitic compounds with extended Bragg-edges compared to normal graphite, and the clathrate hydrates. 

For this purpose we have investigated two main approaches for including thermal neutron scattering data in Monte-Carlo simulations. In the first case we developed the tool NJOY+NCrystal to support a number of options for creating nuclear data in the format for Monte-Carlo simulations. The tool supports the creation of libraries for polycrystalline materials in a format that can be directly used with these codes, or in an improved mixed-elastic format which can be used with modified Monte-Carlo codes. 

This approach is however still limited to polycrystalline materials with standard elastic and inelastic scattering components. Thus, to include effects like small-angle neutron scattering and magnetic scattering, we developed a plugin feature for NCrystal. Through this approach Monte-Carlo codes can access NCrystal on-the-fly, instead of accessing a nuclear data library, making it possible to extend the types of physics process included in the simulation.

Additionally, we have carried out extensive molecular modelling simulations in order to provide updated and new data for input to the neutron scattering models. For example, we investigated finite size effects on the phonon frequency spectrum of ND particles in addition to producing new data for magnesium hydride, graphitic compounds and the clathrate hydrates.

Finally, we carried out an experimental campaign at the BOA beamline of PSI in order to aid in the benchmarking of the graphitic compound models and to investigate the possibility of using this beamline for measurements of very cold neutron scattering cross-sections. We demonstrated through these experimental efforts that the background on the instrument could be reduced to a level where it was possible to resolve the Bragg-edges around 12 \AA\ in our graphitic compound samples.

%\subsection{Acknowledgements} 
%We would like to acknowledge the Paul Scherrer Institute for access to the beam time at the facility under proposal numbers 20212829, 20221100 and 20222657. We acknowledge MAX IV Laboratory for time on Beamline COSAXS under Proposal 20221553. Research conducted at MAX IV, a Swedish national user facility, is supported by the Swedish Research council under contract 2018-07152, the Swedish Governmental Agency for Innovation Systems under contract 2018-04969, and Formas under contract 2019-02496.
%We would also like to thank the ESS Manufacturing Workshop and the ESS sample environment group for machining the triaxial and cylindrical sample holders. 

%% file: adreflectors.tex
\section{Experimental campaign with advanced reflectors}
\label{sec:adref}
In this section, the experimental efforts to
test the advanced reflector configurations highlighted during the HighNESS project are presented. 
In line with the plan described in the HighNESS proposal, the optimal beam extraction configuration for a mock-up experiment is identified in \cref{sec:design_experiment}. The host institute chosen for this experiment is the Budapest Neutron Center (BNC), where a cold moderator test facility (CMTF) has been constructed~\cite{ICNS2022_source}. The purpose of a CMTF is to provide users with a facility designed to test moderators, both for high-power and compact sources. In this setup, an out-of-pile moderator and reflector system is fed by the fast and thermal neutrons coming from the \SI{10}{MW} BNC research reactor. 
The setup of the HighNESS experiment is designed to test the effect of advanced reflectors both around the cold source and in the beam extraction, taking into account the constraints imposed from the facility.

After having approximated the appropriate dimensions for the reflector layer and the diffusive extraction channel, this information was transferred to the engineering team, and served as starting point for the engineering design. Several iterations between neutronic and engineering design were done before the prototype was manufactured.
First neutron and gamma dose measurements were performed at 1\,MW reactor power at the channel outlet (near reflector) and outside the bunker on June 5$^\text{th}$ 2023. The low level of the background allowed for the beamline to also be opened at 10 MW. The outside bunker measurements were found to be within the local regulation limits; some parts of the bunker, however, needed to be reinforced. 

In the context of a separate experiment planned by the Mirrotron company, the CMTF installation was completed in the reactor hall on June 23$^\text{rd}$ 2023, making it possible to obtain a 2D image from the moderator using a pin-hole system. Due to delays linked to reactor operation, however, the HighNESS measurement campaign was postponed until September 18$^\text{th}$ through the 29$^\text{th}$.   

After the preliminary experiment in Budapest Neutron Center, an opportunity arose to perform an additional experiment at the test station built in the Big Karl area of the Institute of Nuclear Physics (IKP) in Forschungszentrum J{\"u}lich (FZJ), a newly build accelerator-driven neutron source~\cite{Zakalek:906983}. Here we investigated only the effect of putting advanced reflectors around the cold source (cryogenic methane).
%, with an accelerator-driven tantalum target as main source of neutrons. 
In this case, due to the short time between the invitation to participate and the allotted beamtime, no prior neutronic study was performed. However, the experience acquired by the engineers from the neutronic design for the experiment in Budapest was utilized in adapting this second experiment to the specific conditions related to the new facility. 

In general, the characteristics of the facility that are required for the experiment and were considered in the planning phase are:

\vspace{0.2cm}
\begin{itemize}[leftmargin=1cm]
    \item[-] The availability of a cold and stable moderator;
    \item[-] Sufficient room to install the prototype;
    \item[-] The availability of neutron detectors, with sufficient resolution to clearly quantify performance difference with a void prototype of same dimensions.
\end{itemize}
\vspace{0.2cm}

\noindent With these conditions in mind, in this section we describe both the experiment performed at the BNC reactor and the experiment performed at the JULIC accelerator in FZJ.

\subsection{Mock-up experiment at the Budapest Neutron Center}
\label{sec:design_experiment}
The proof-of-concept experiment at the CMTF facility at BNC was designed to test the reflective properties of both ND and \ce{MgH_2} in a representative beam-extraction configuration. In this section, the neutronic design of the experiment is discussed in some depth.

\subsubsection{Introduction to CMTF and the HighNESS experiment}
\label{subsec:intro_CMTF}
\begin{figure}[hb!]
\centering
\includegraphics[width=0.8\textwidth]{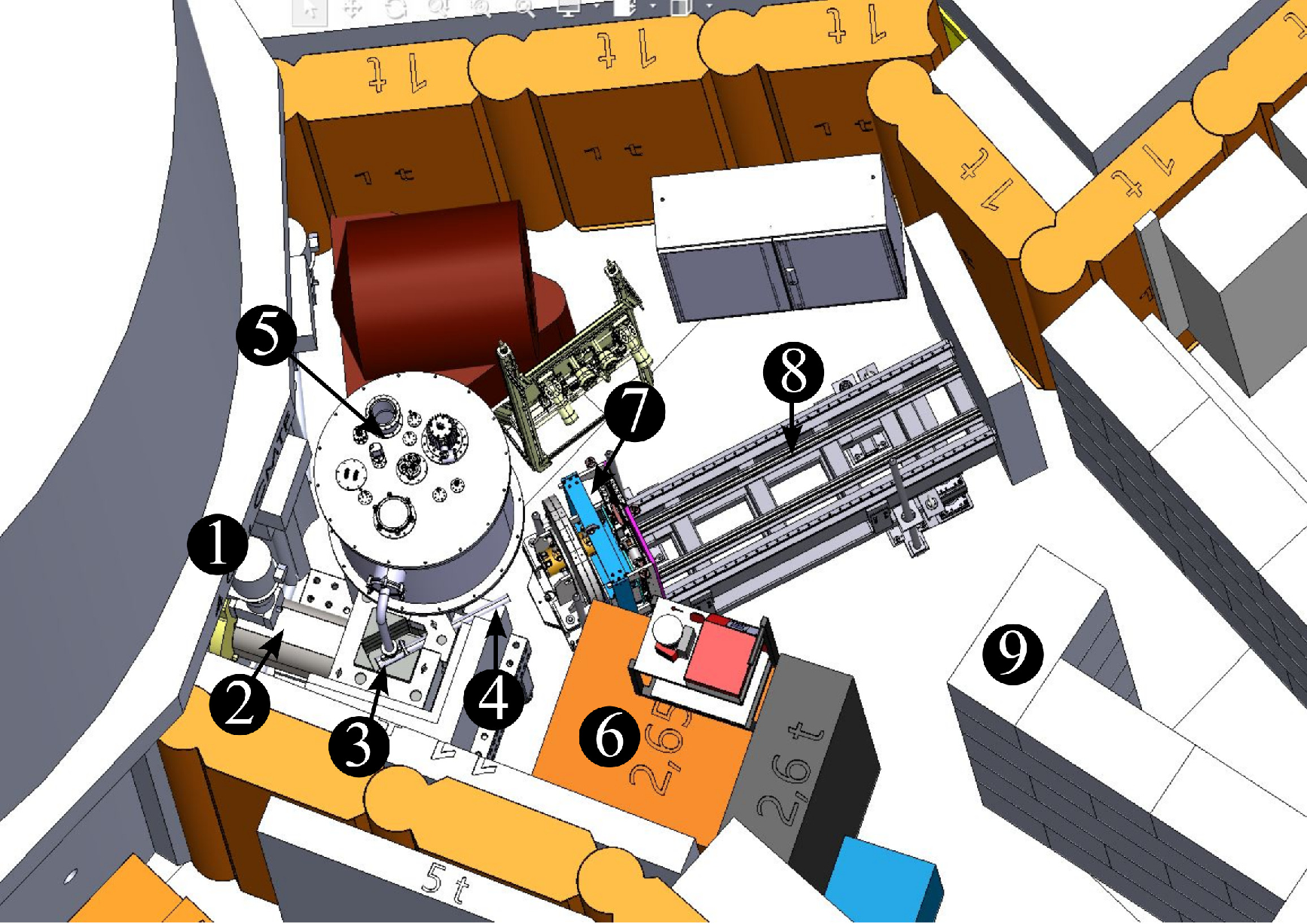}
\caption[Sketch of the beam channel \#4 beam line of the \SI{10}{MW} Budapest Research Reactor of BNC]{
Sketch of the  channel \#4 beam line of the \SI{10}{MW} Budapest Research Reactor of the Budapest Neutron Center (BNC). (1) Channel \#4 shutter drive, (2) primary carbon steel beam collimator, (3) lead reflector-moderator block, (4) extraction system, (5) cryo-cooler tank, (6) beam stop, (7) pin-hole assembly, (8) rail support system, (9) bunker's shielding walls.  
}\label{fig:bud_beamline}
\end{figure}

The BNC's objective is to provide users with a prototype of a low-dimensional liquid para-hydrogen cold neutron source with a significantly higher brightness than previously developed neutron moderators, in order to replace the existing moderator. The beamline, sketched in \cref{fig:bud_beamline}, has been built at channel 4 of the \SI{10}{MW} BNC reactor. The BNC reactor was commissioned in 1959 at 2.5\,MW, refurbished and then upgraded to 10\,MW in 1992. The maximum thermal flux is \SI{2.5e14}{n/cm^2s}.

The CMTF design starts with a steel collimator coupled with one of the reactor channel outlets and with a high-purity box-shaped lead reflector. A \SI{5}{cm} diameter beryllium disk is placed in the beam path to better diffuse the neutrons coming from the reactor. The moderator vessel consists of a 150-mm-long aluminum alloy tube with a diameter of \SI{25}{mm} filled with liquid parahydrogen at \SI{20}{K} placed at an angle of \SI{45}{\degree} with respect to the direct beam path. The beamline components of the moderator test facility include a collimator, a set of choppers, and a pin-hole imaging system to measure the brightness of the moderator cell based on the camera-obscura principle~\cite{FuziPinhole}.

\begin{figure}[tb!]      
    \begin{subfigure}[b]{0.5\textwidth}
        \centering
        \includegraphics[width=\textwidth]{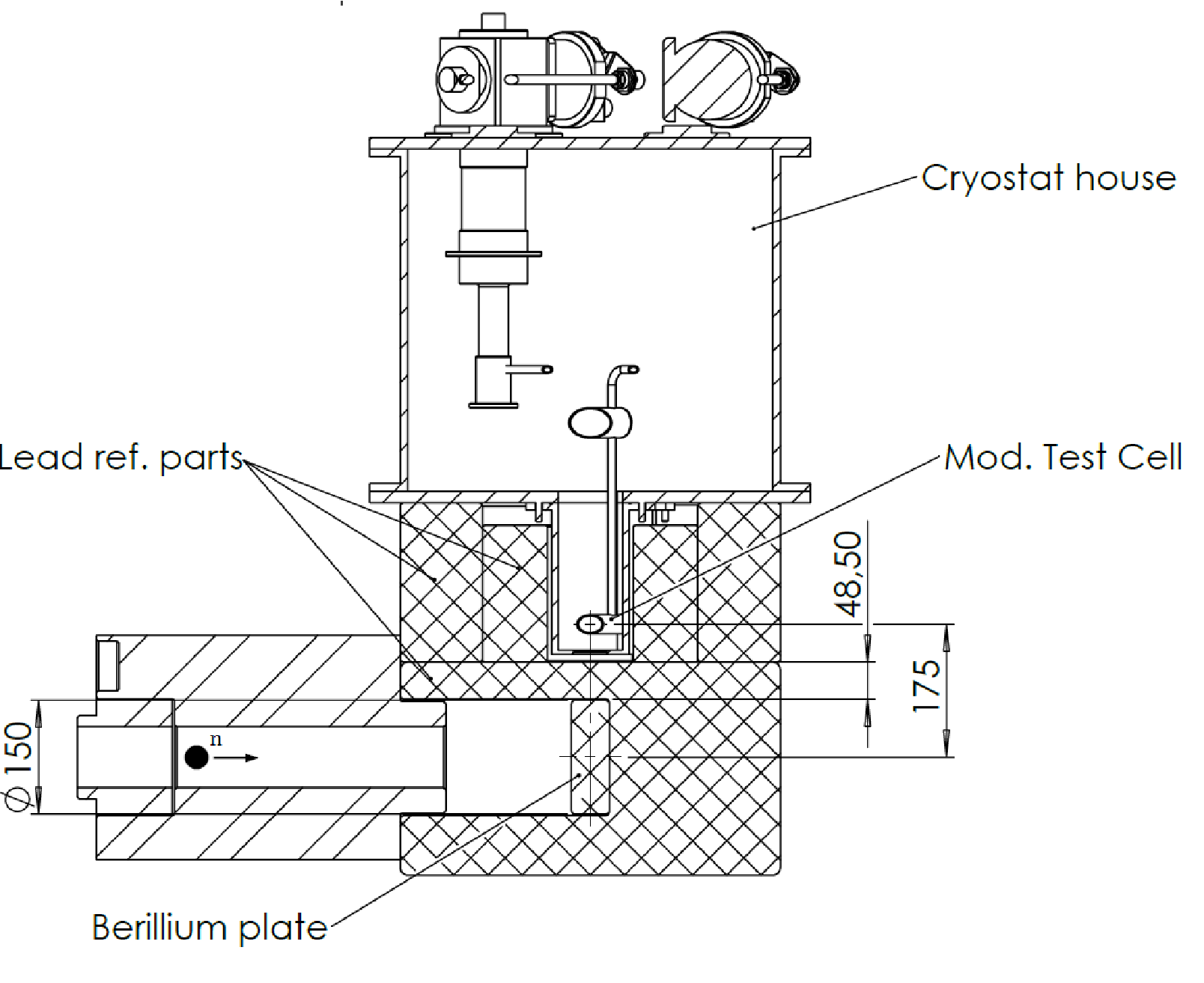}
        \subcaption{}
        \label{fig:budapest_CMTF_A}
    \end{subfigure}
    \hfill
    \begin{subfigure}[b]{0.5\textwidth}
        \centering        
        \includegraphics[width=\textwidth]{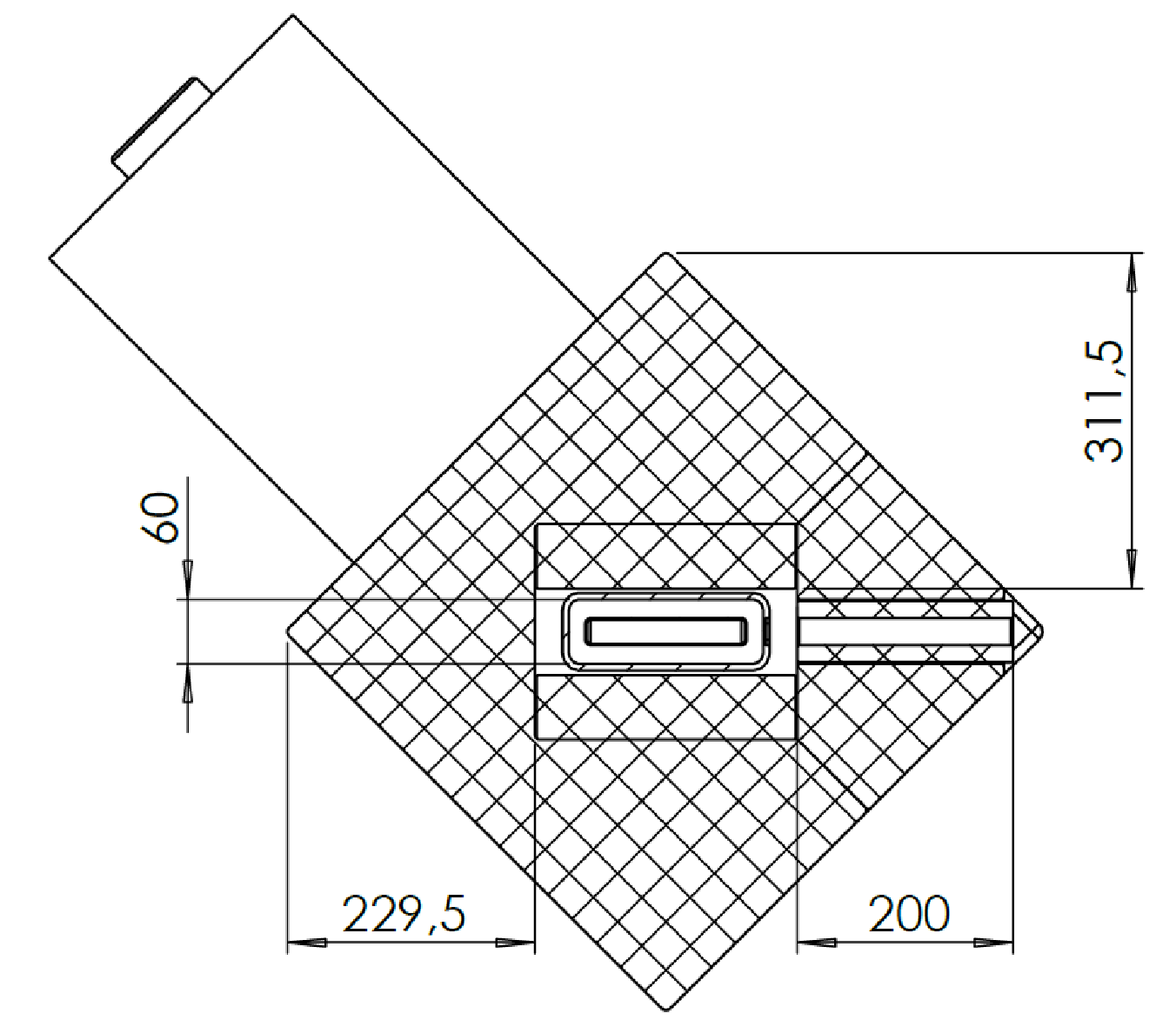}
        \subcaption{}
        \label{fig:budapest_CMTF_B}
    \end{subfigure}
\caption[CMTF section drawings]{Section drawings of the lead reflector-moderator block at the CMTF. (a) vertical section (b) horizontal section at the height of the moderator vessel.}
\label{fig:budapest_CMTF}
\end{figure}

Sectional views in~\cref{fig:budapest_CMTF} from the engineering drawings show important features of the reflector-moderator block. The lead reflector is composed of different blocks, which can be removed to accommodate more components, like advanced reflector jackets and an extraction system. The section just above the moderator is left uncovered and is meant to house the pipes of the cryogenic system. Finally, it is important to mention that this design is the result of a long process in the context of a facility that was still under construction. Many features have changed since it was first designed in 2021 and several details were re-defined during the construction (e.g. shielding).

The CMTF concept was the starting point for the design of the HighNESS experiment. The possibility for reusing some of the CMTF components and to adapt them to the HighNESS purposes saved on manufacturing costs, but also imposed some constraints on the neutronic design of the experiment. In particular, we decided to retain the collimator, the lead reflector, and the detection system, which meant that the best option for the source was to keep the same low-dimensional design based on parahydrogen and focus the work solely on the extraction system.

%The goal of the experiment was to test the reflective properties of \ce{MgH_2} and ND.
In the extraction channel, a thin layer of ND powder is expected to give a significant increase in the cold neutron flux~\cite{jamalipour_mostafa_2022}. This is mainly due to the quasi-specular reflectivity exhibited by ND in the cold energy range. Meanwhile, \ce{MgH_2} is expected to be much more effective than ND in reflecting the neutrons back into the source, where the environment is dominated by large-angle scattering, hence limiting the neutron leakage in the cold energy range~\cite{granada_studies_2020}. For these reasons, the proposed set-up for the HighNESS experiment makes the following additions to the CMTF design:

\vspace{0.2cm}
\begin{itemize}[leftmargin=1cm]
    \item[-]A reflector vessel that wraps the moderator tube on all sides, except the extraction side.
    \item[-]An extraction guide with a thin inner layer of ND powder, coupled with the moderator vessel.
\end{itemize}
\vspace{0.2cm}

The neutron beam from the reactor source enters the test station through an assembly of lead blocks and hits a plate made of beryllium (Be). This Be-plate (in red in \cref{fig:MCNP_CMTF}) is located directly below the test area for the moderator units and scatters the thermal neutrons, so that they are fed into the moderator and reflector assembly. After the neutrons are slowed down to cold neutron energies, they exit through the ND beam extraction system into the detector area.
In the next section, we present the neutronic calculations that lead us to the final dimensions for the prototype of such a system. 
\subsubsection{Neutronic modeling}
\label{subsec:neutronic_mod}
\begin{figure}[tb!]      
    \begin{subfigure}[b]{0.5\textwidth}
        \centering
        \includegraphics[width=\textwidth]{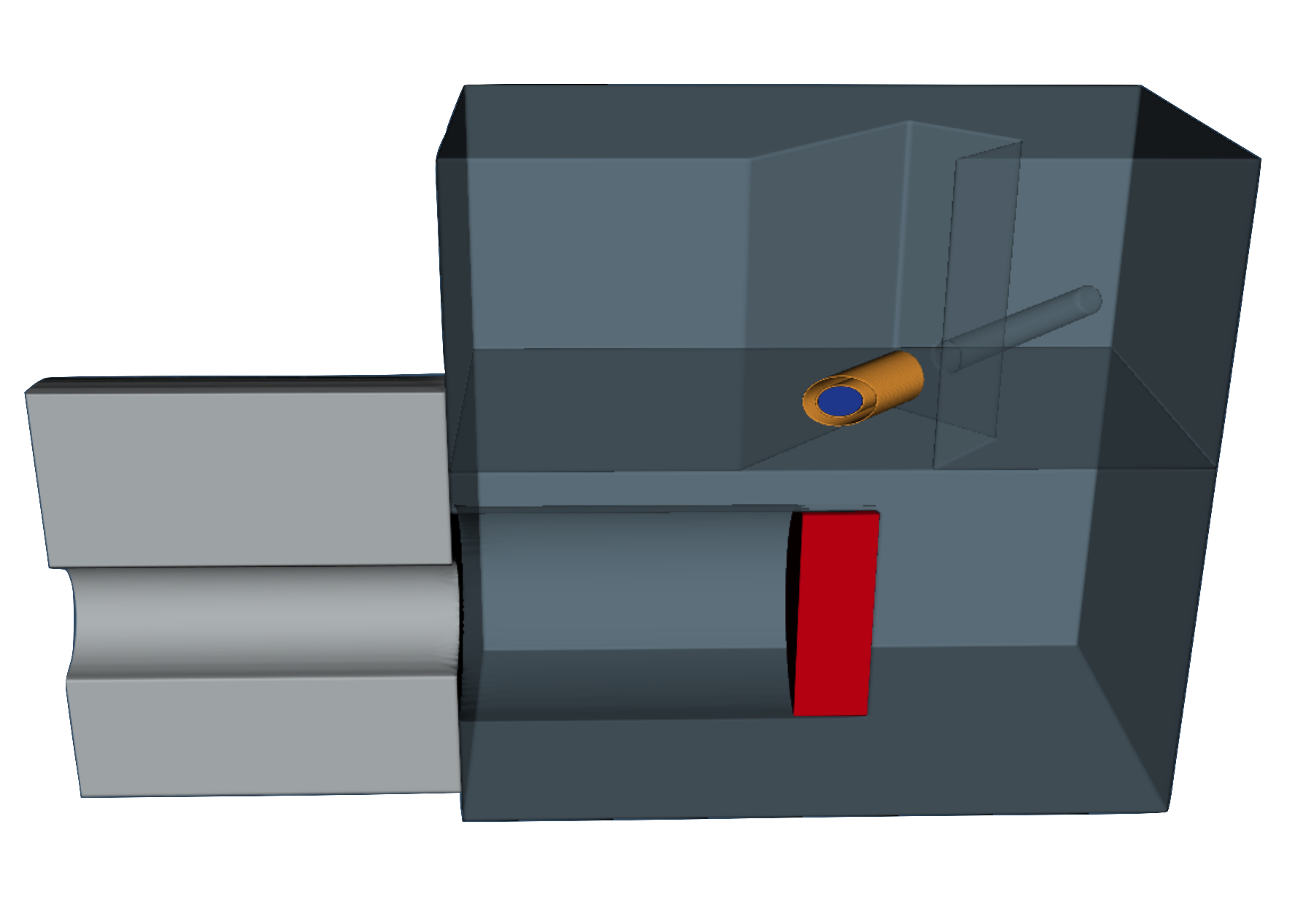}
        \subcaption{}
        \label{fig:geometry_CMTF_XZ}
    \end{subfigure}
    \hfill
    \begin{subfigure}[b]{0.5\textwidth}
        \centering        
        \includegraphics[width=\textwidth]{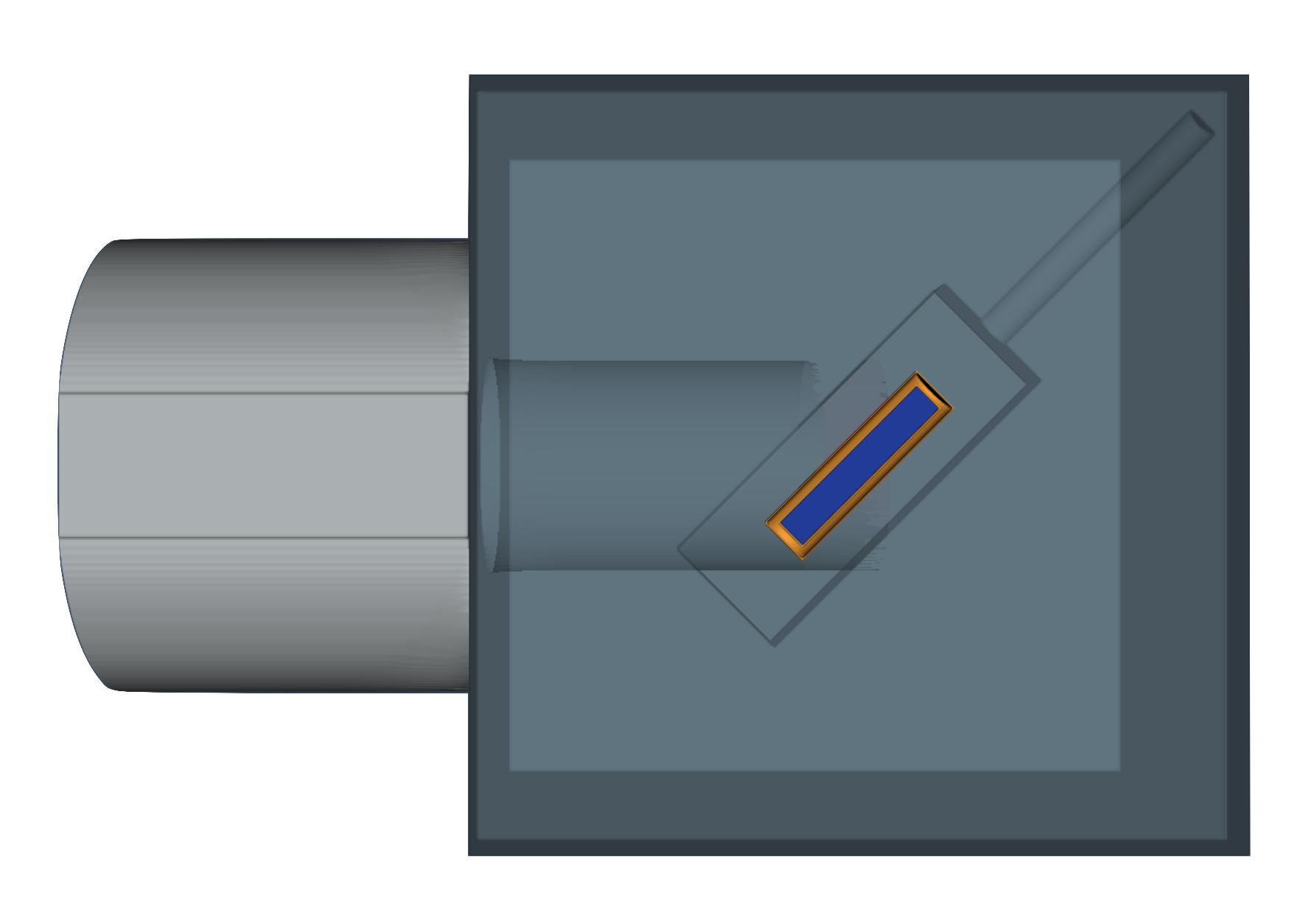}
        \subcaption{}
        \label{fig:geometry_CMTF_XY}
    \end{subfigure}
\caption[MCNP model of the CMTF and reflector-moderator block]{MCNP model of the lead reflector-moderator block at CMTF produced according to the engineering drawings in \cref{fig:budapest_CMTF}. (a) vertical section (b) horizontal section at the height of the moderator vessel.}
\label{fig:MCNP_CMTF}
\end{figure}

\begin{figure}[tb!]
\centering
\includegraphics[width=0.7\textwidth]{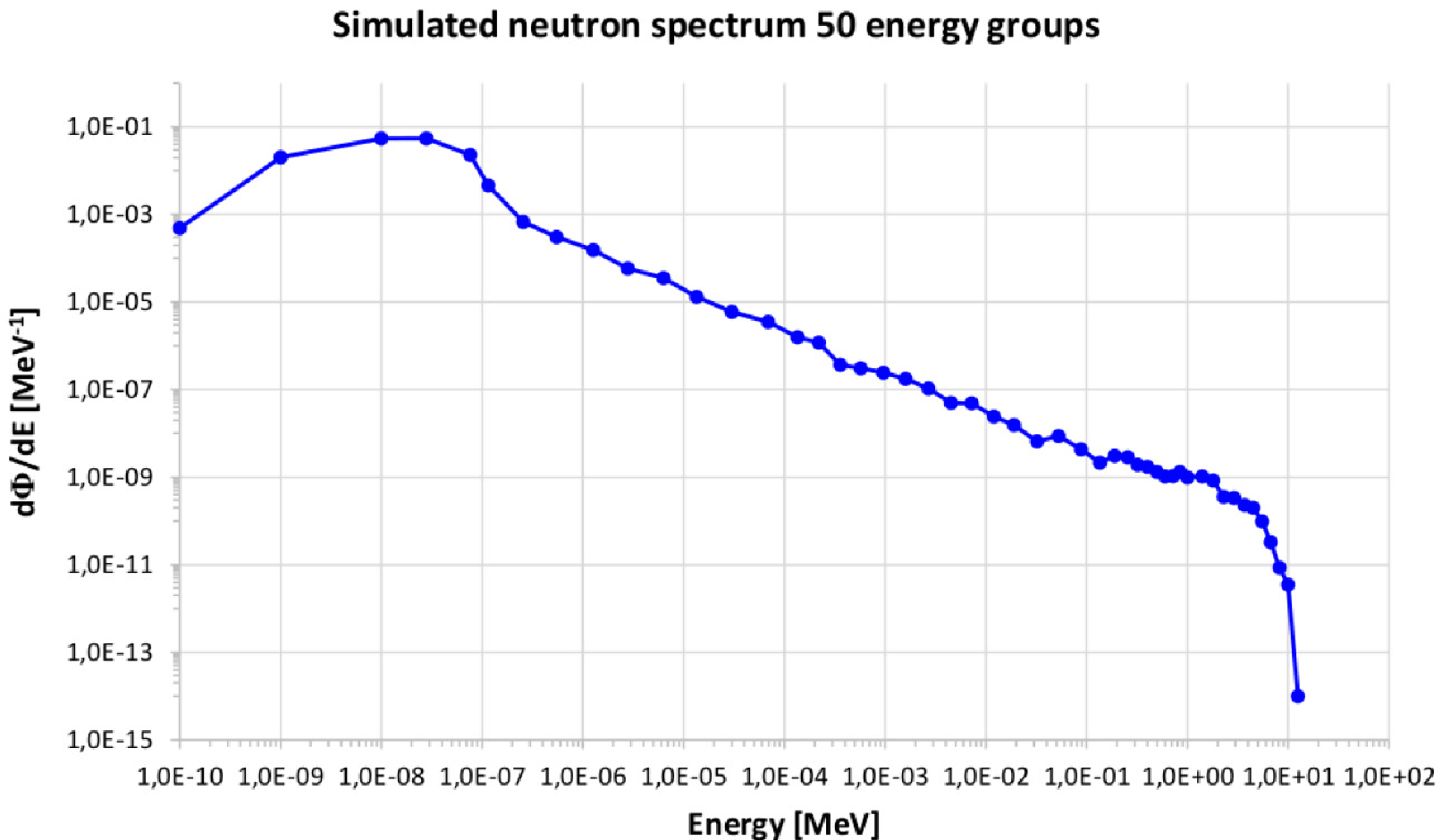}
\caption[Neutron source spectrum used in the calculations]{Neutron source spectrum used in the calculations and showed at ICNS 2022 \cite{ICNS2022_source}}\label{fig:ICNS2022_source}
\end{figure}

\begin{figure}[tb!]  
\centering
    \begin{subfigure}[b]{0.49\textwidth}
        \includegraphics[width=\textwidth]{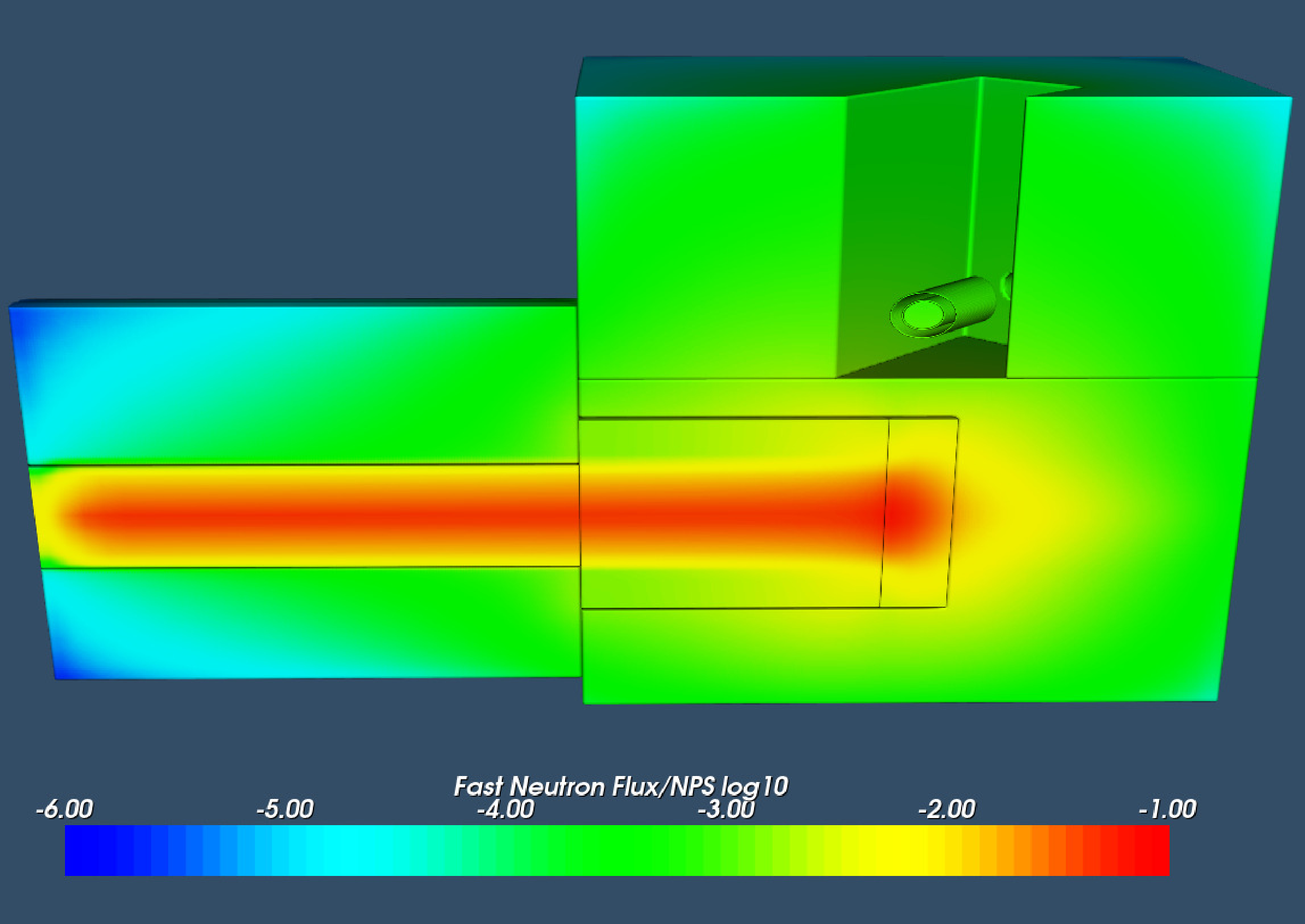}
        \subcaption{}
        \label{fig:fast_source}
    \end{subfigure}
    \hfill
    \begin{subfigure}[b]{0.49\textwidth}        
        \includegraphics[width=\textwidth]{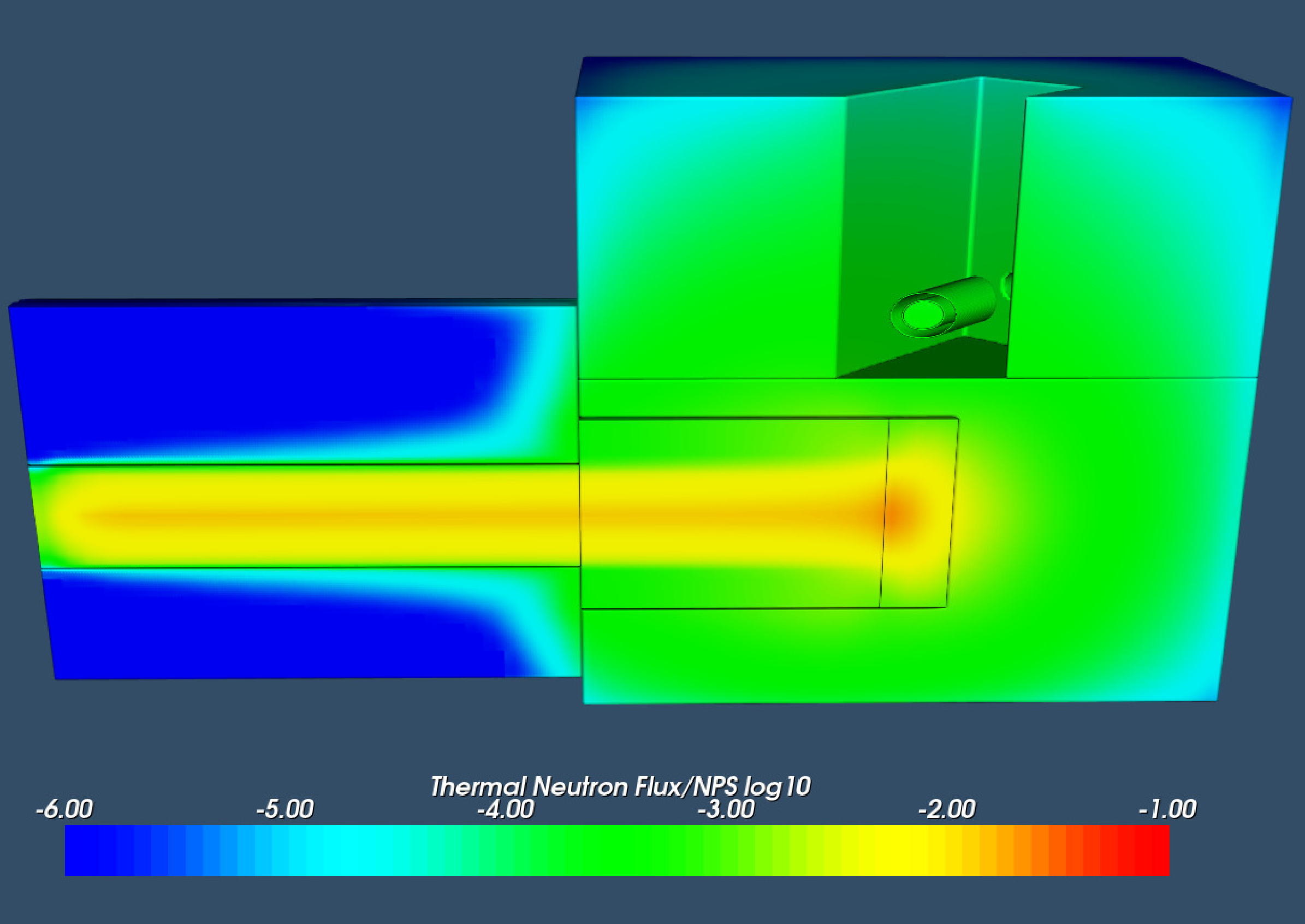}
        \subcaption{}
        \label{fig:thermal_source}
    \end{subfigure}

    \begin{subfigure}[b]{0.49\textwidth}        
        \includegraphics[width=\textwidth]{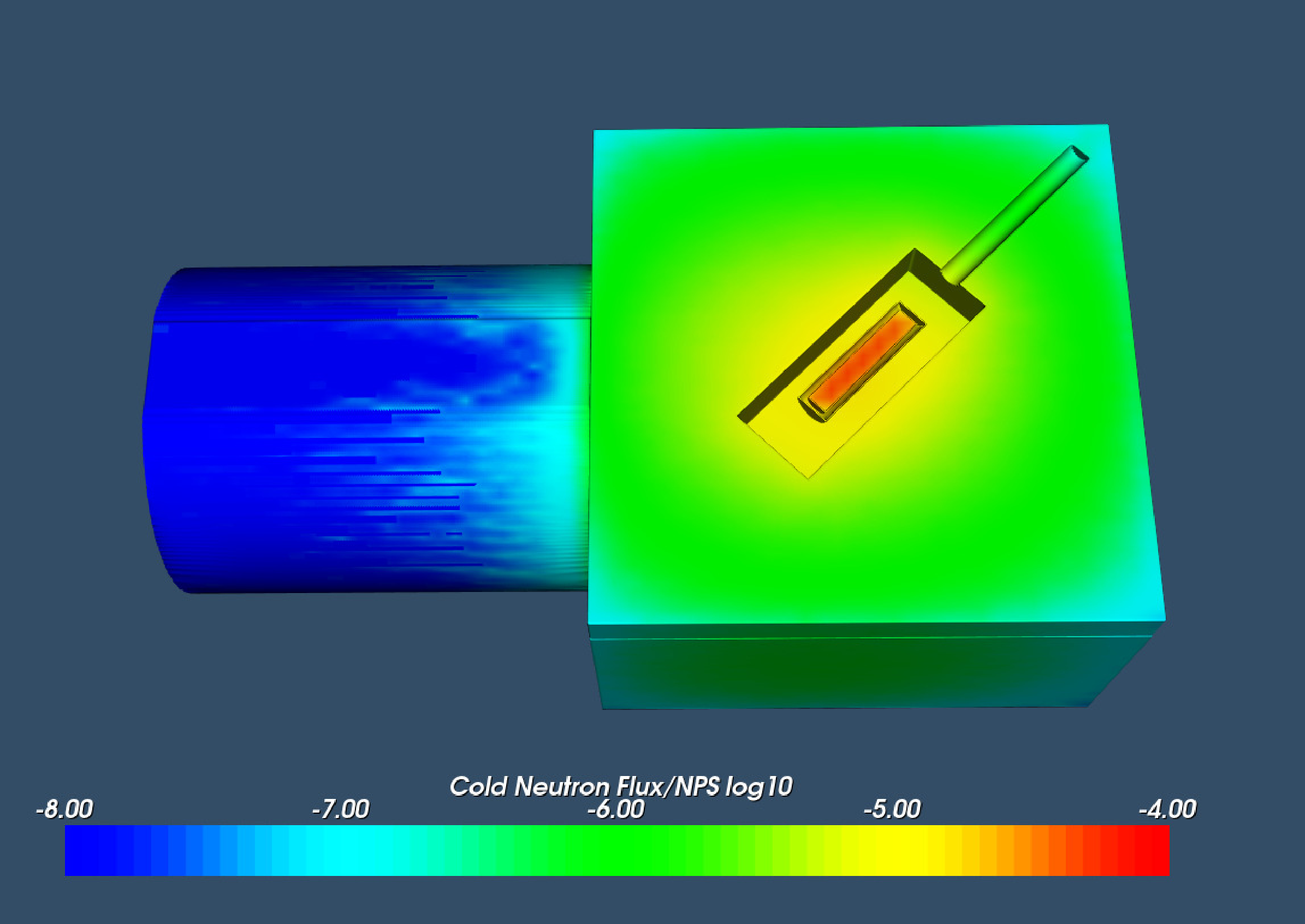}
        \subcaption{}
        \label{fig:cold_source}
    \end{subfigure}
\caption[Energy-integrated flux maps in \si{n/cm^2} per source neutron]{Energy-integrated flux maps in \si{n/cm^2} per source neutron. (a) Vertical section with energy integrated between \SI{81}{meV} and \SI{25}{MeV}. (b) Vertical section with energy integrated between \SI{13}{meV} and \SI{81}{meV} (c). Horizontal section at the moderator height with energy integrated below \SI{5.11}{meV} (\SI{4}{\angstrom}).}
\label{fig:fast_and_thermal_source}.
\end{figure}

The MCNP model of the CMTF moderator, produced based on the technical drawings provided by BNC (\cref{fig:budapest_CMTF}), is shown in \cref{fig:MCNP_CMTF}.
The source spectrum coming out of channel 4 (\cref{fig:ICNS2022_source}) used in the simulations was calculated from the full MCNPX reactor model and presented in \cite{ICNS2022_source}. The spectrum presents both a thermal and a fast peak around, respectively, \SI{70}{meV} and \SI{2.8}{MeV}, up to a maximum energy of \SI{12.5}{MeV}. A more accurate measurement of the source using gold foil activation is foreseen in the future.
Preliminary calculations on this starting model confirmed some of the neutronic features one would expect from such a concept:

\vspace{0.2cm}
\begin{itemize}[leftmargin=1cm]
    \item[-]Fast and thermal neutrons from the source are diffused by the beryllium disk  and the lead blocks. The efficient diffusion of fast and thermal neutrons, although it is optimal to increase the flux at the moderator, is also expected to give a significant contribution to the noise in the extraction channel (\cref{fig:fast_source} and \cref{fig:thermal_source});
    \item[-]In a similar way, an intense fast and thermal flux is coming out of the collimator and the reflector. However, the shielding around these components and, more generally, the issues regarding background and safety (e.g. activation after exposure) have been studied separately.
    \item[-]The flux map in the cold energy range, below \SI{5.11}{meV}, (\cref{fig:cold_source}) shows that the parahydrogen in the moderator vessel is indeed a cold source. The flux map also confirms that a cold neutron reflector capable of reducing the leakage around the moderator would considerably increase the performance of the system. 
\end{itemize}
\vspace{0.2cm}

\begin{figure}[!tb]      
    \begin{subfigure}[b]{0.5\textwidth}
        \centering
        \includegraphics[width=\textwidth]{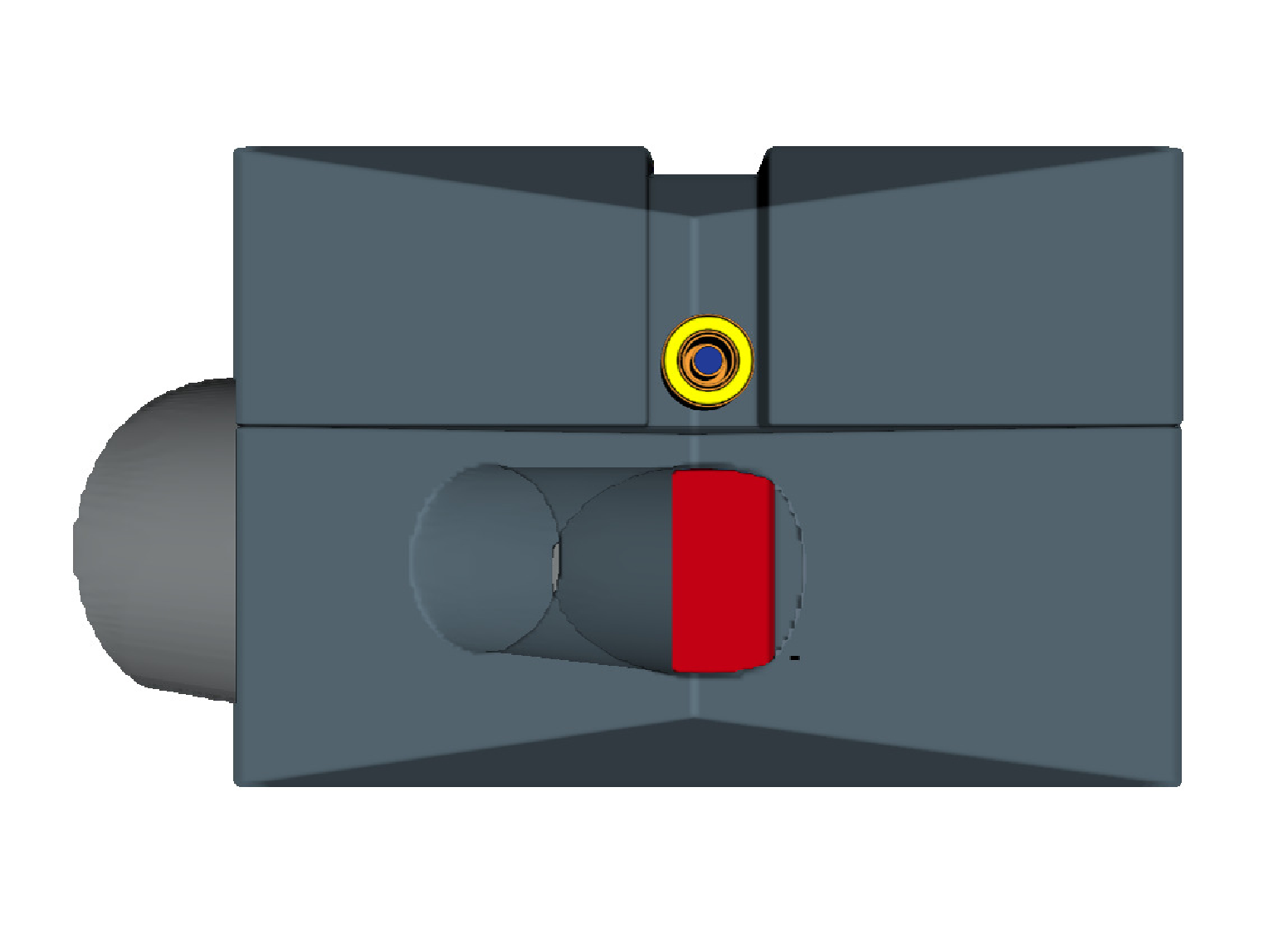}
        \subcaption{}
        \label{fig:MCNP_geo_highNESS_XY}
    \end{subfigure}
    \hfill
    \begin{subfigure}[b]{0.5\textwidth}
        \centering        
        \includegraphics[width=\textwidth]{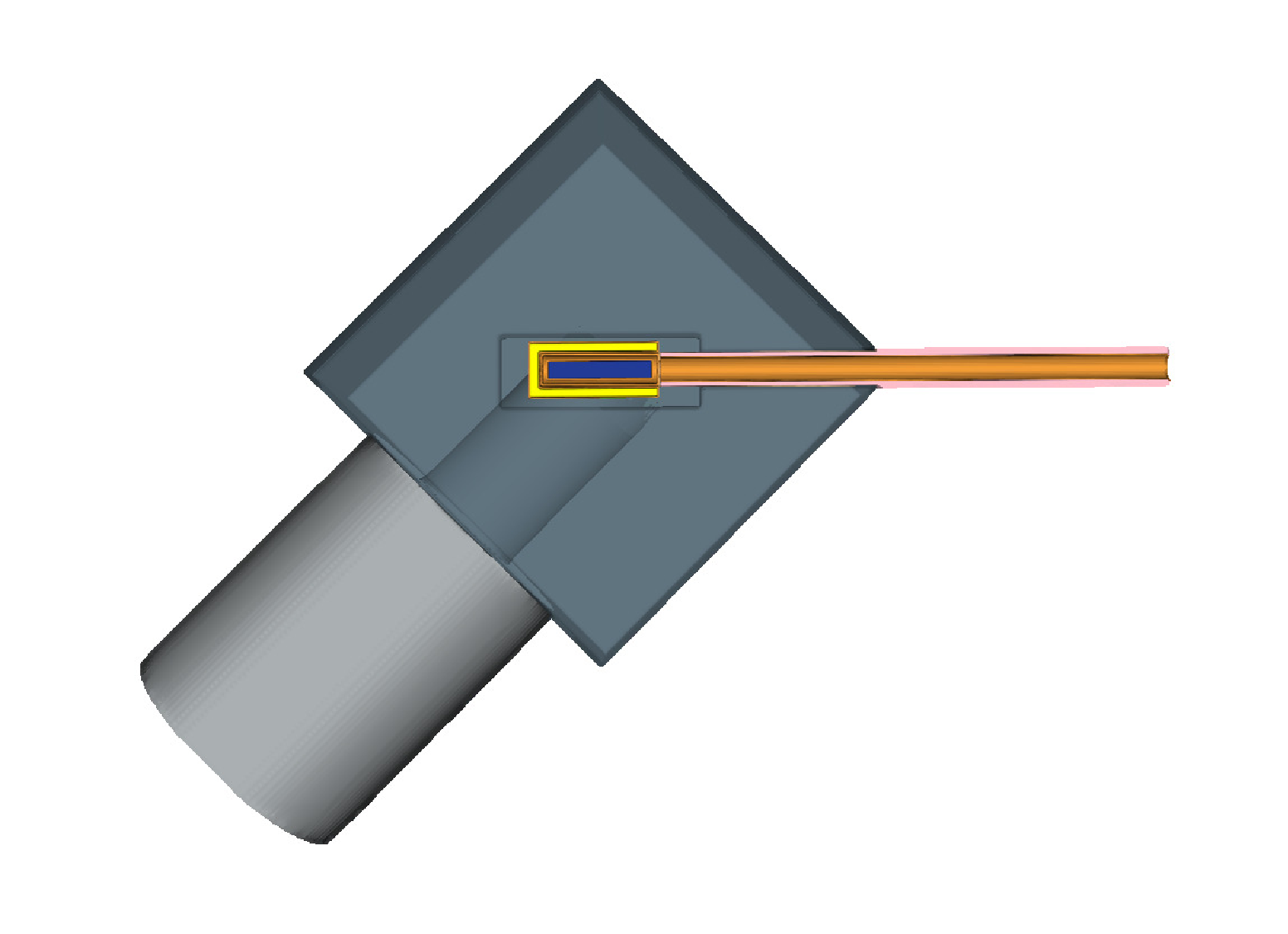}
        \subcaption{}
        \label{fig:MCNP_geo_highNESS_XZ}
    \end{subfigure}
\caption[MCNP model of the HighNESS moderator and extraction system ]{MCNP model of the lead reflector-moderator block for the HighNESS experiment. (a) vertical section with a view of the reflector (b) horizontal section at the height of the moderator vessel with a view of the extraction system.}
\label{fig:MCNP_geo_highNESS}
\end{figure}

\begin{figure}[!tb]      
    \begin{subfigure}[b]{0.5\textwidth}
        \centering
        \includegraphics[width=\textwidth]{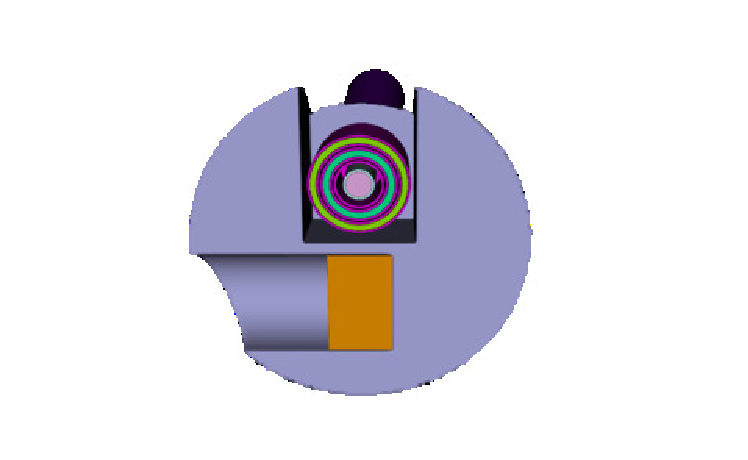}
        \subcaption{}
        \label{fig:toy_front}
    \end{subfigure}
    \hfill
    \begin{subfigure}[b]{0.5\textwidth}
        \centering        
        \includegraphics[width=\textwidth]{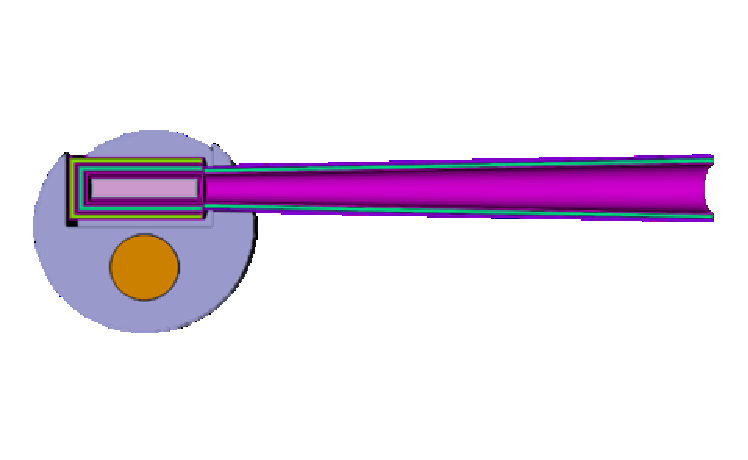}
        \subcaption{}
        \label{fig:toy_trumpet}
    \end{subfigure}
\caption[MCNP toy model of the moderator and extraction system]{MCNP toy model of the lead reflector-moderator block for the HighNESS experiment. (a) vertical section parallel to the neutron beam with a view of the reflector (b) section perpendicular to the neutron beam with a view of the extraction cone.}
\label{fig:MCNP_geo_toy}
\end{figure}

\indent The first test of the HighNESS concept was done with the model described in \cref{fig:MCNP_geo_highNESS}. The dimensions of the lead reflector were reduced in the inner section to make room for the reflector and the trumpet. In reality, the space for the installation of these external components came from the removal of lead blocks and does not precisely coincide with the simple assumption of the model. Also, details like cooling pipes and metal support structures were not taken into account at this stage of the modeling.\\
\indent The runs with this model were affected by a crippling low neutron transport efficiency from the source to the exit of the trumpet. As a result, the tallies far from the moderator exit could not statistically converge for any large amount of primary particles in the source. The initial explanation for the problem was a poor definition of geometry importance in the simulation, which would have made MCNP wasting time in tracking the diffused neutrons in the region of the large reflector far away from the extraction channel. 
%The attempt to define a suitable weight-window mesh to increase the transport efficiency, lead to an interesting study with ADVANTG reported in \cref{appendix:ADVANTG}. \\
The actual reason for this behavior was related to a wrong spatial definition of the source beam.
To simplify the problem, we decided to study the extraction with a simpler geometry that has all the essential characteristics of the full model. This corresponding {\it toy model} is shown in \cref{fig:MCNP_geo_toy}.\\
\indent The main difference between the toy model and the original one is the lead reflector. The original \qtyproduct{49x49x46}{cm} box-shaped reflector was substituted by a sphere with a radius of \SI{13}{cm} centered between the moderator and the beryllium disk. The moderator is also closer to the disk and forms a \SI{90}{\degree} angle with the source beam. The neutron source and the other relevant dimensions (e.g. moderator length and radius etc...) are identical to the ones in the original model. As we are studying relative gains due to the enhancing effect of advanced reflectors, the toy model appears to be a very convenient tool.
\subsubsection{Optimization of the source aperture}
\begin{figure}[tb]      
    \begin{subfigure}[b]{0.43\textwidth}
        \centering
        \includegraphics[width=\textwidth]{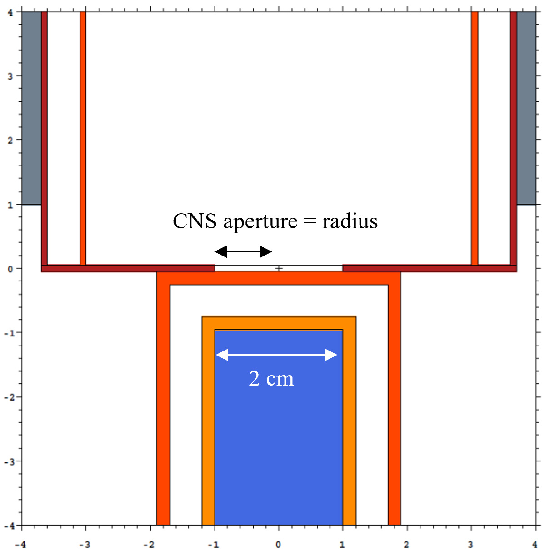}
        \subcaption{}
        \label{fig:CNS_aperture_1}
    \end{subfigure}
    \hfill
    \begin{subfigure}[b]{0.57\textwidth}
        \centering        
        \includegraphics[width=\textwidth]{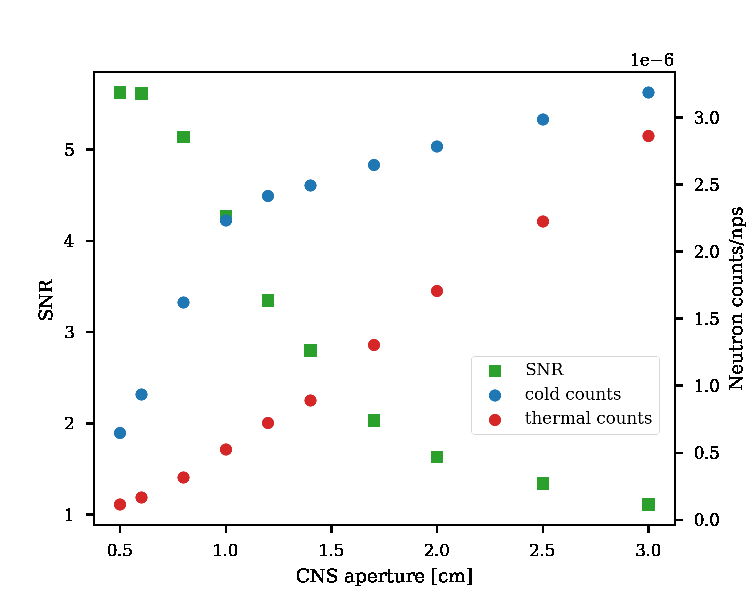}
        \subcaption{}
        \label{fig:SNR_CNS_aperture}
    \end{subfigure}
\caption[Optimization of the CNS aperture with the toy model]{Optimization of the CNS aperture with the toy model. (a) MCNP model without reflector and ND in the extraction. The cadmium plate is dark red. (b) SNR, CN, and thermal neutron count as a function of the CNS aperture. }
\label{fig:CNS_aperture}
\end{figure}

The moderator vessel is coupled with the extraction system through a cadmium plate with a circular opening. The reason to have a Cd window is to block the diffused thermal neutrons while allowing in the extraction tube mostly the cold neutrons coming from the moderator. To study separately the contributions from the components, we removed the advanced reflector around the source and in the extraction channel. A close-up of this interface in the MCNP model is shown in \cref{fig:CNS_aperture}. \\
\indent  The thickness of the cadmium layer between the moderator and the extraction system is \SI{1}{mm} in the simulation, but here we assumed perfect collimation, i.e. every neutron that enters the cell is absorbed. Similarly, we added a layer of cadmium around the extraction tube to prevent the thermal neutrons diffused in the lead reflector to reach the detector.
The radius of the empty tube at this step is \SI{3}{cm}, but the aperture in the cadmium is independent of the extraction and its radius can be optimized separately.
If we define cold neutrons as all the neutrons with $\lambda>\SI{1.8}{\angstrom}$ and thermal as the ones with $\SI{0.5}{\angstrom}<\lambda<\SI{1.8}{\angstrom}$, then the optimal aperture for the cold neutron source (CNS) should maximize the following signal-to-noise ratio (SNR):
\begin{equation}
    \mathrm{SNR} = \frac{C_c}{C_t}
\end{equation}
where $C_c$ and $C_t$ are the cold and thermal neutron counts respectively. It is clear that considering all the thermal neutrons as noise is a rough approximation since they are also part of the spectrum emitted by the cold moderator. However, it is reasonable to assume that the contribution from the diffused thermal field would quickly dominate as the aperture gets bigger. Hence, this simple definition still allows us to effectively optimize the aperture for the moderator's signal. The SNR can be calculated both at the aperture and at the end of the extraction tube. The plot in \cref{fig:SNR_CNS_aperture} shows the neutron counts and the SNR as a function of the CNS aperture when recording at \SI{60}{cm} from the emission surface (end of the extraction). The maximum value for the SNR is obtained for \SI{0.5}{cm}, which is simply the smallest radius chosen as input. However, this corresponds also to the minimum for the counts. A good compromise between SNR and count rate is found at the knee of the CN counts curve, which is precisely the radius of the moderator. In other words, it is not convenient to have the cadmium aperture larger than the moderator itself, since the expected increase in diffused thermal neutrons is not balanced anymore by the gain in the cold flux.
\subsubsection{Optimization of the source reflector}
\begin{figure}[tb!]       
    \begin{subfigure}[b]{0.43\textwidth}
        \centering
        \includegraphics[width=\textwidth]{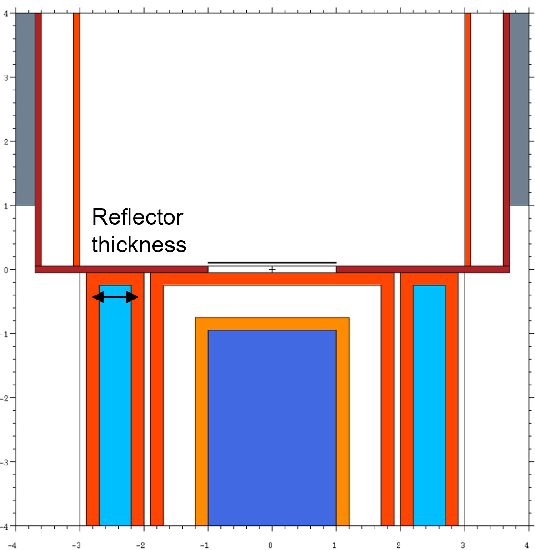}
        \subcaption{}
        \label{fig:CNS_reflector_geo}
    \end{subfigure}
    \hfill
    \begin{subfigure}[b]{0.57\textwidth}
        \centering        
        \includegraphics[width=\textwidth]{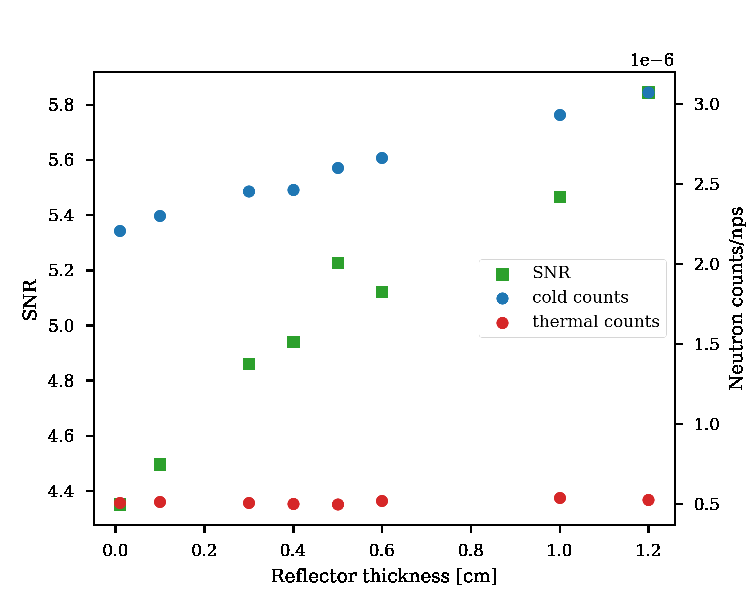}
        \subcaption{}
        \label{fig:SNR_CNR_MgH2}
    \end{subfigure}
\caption[Optimization of the reflector thickness with the toy model]{Optimization of the CNS reflector thickness with the toy model. (a) MCNP model without ND in the extraction. CNS aperture is \SI{1}{cm} (b) SNR, CN, and thermal neutron counts as a function of the reflector thickness.}
\label{fig:CNS_reflector}
\end{figure}

We fixed the CNS aperture to \SI{1}{cm} and study the optimal size for the advanced reflector around the CNS. The reflector jacket is made of aluminum with 2-mm-thick walls. The gap between the moderator vacuum jacket and the reflector is \SI{1}{mm}. The close-up of the geometry is shown in \cref{fig:CNS_reflector_geo}.\\
\indent We inserted room temperature \ce{MgH_2} in the simulation with increasing vessel thickness. The results on the cold and thermal counts and the SNR are shown in \cref{fig:SNR_CNR_MgH2}. It is clear that the increase in the cold neutron counts is correlated with the amount of \ce{MgH_2} in the system, and gains higher than 30\% are observed when putting \SI{1.2}{cm} of this material around the CNS. In this case, where the calculations show that more reflector material leads to better performance, the optimal value \SI{1}{cm} is the upper limit physically allowed inside the lead reflector. ND was also tested as reflector material, but no gain was observed in the cold range at any thickness value. However, an enhancing effect in the VCN range could not be observed due to poor tally convergence, but it is reasonable to expect a similar, if not higher, gain factor. For this reason, we prepared also a reflector jacket filled with ND.
\subsubsection{Optimization of the extraction system}
\begin{figure}[tb!]      
        \centering
    \includegraphics[width=.5\textwidth]{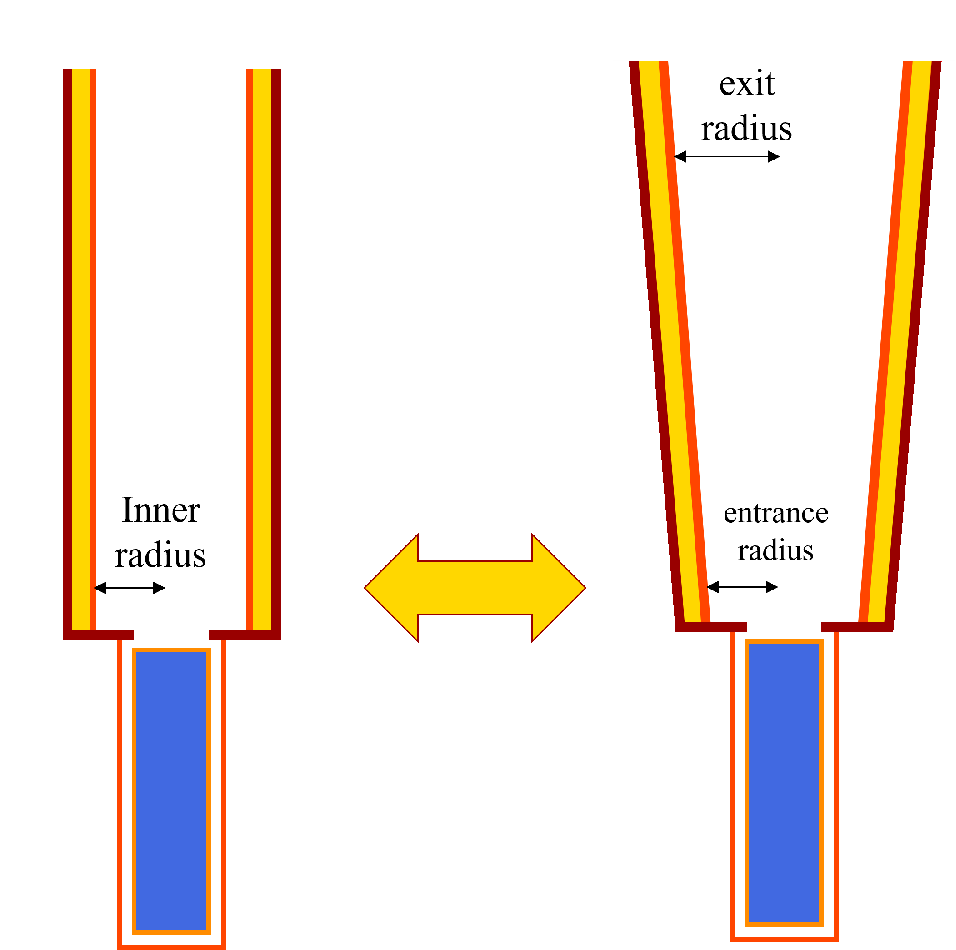}
\caption[Illustration of the two shapes for the extraction system]{Illustration of the two shapes for the extraction system, straight tube (left) and diverging cone (right) with their parameters. Not to scale. }
\label{fig:geom_extraction}
\end{figure}

\begin{figure}[tb!]
\centering
\includegraphics[width=\textwidth]{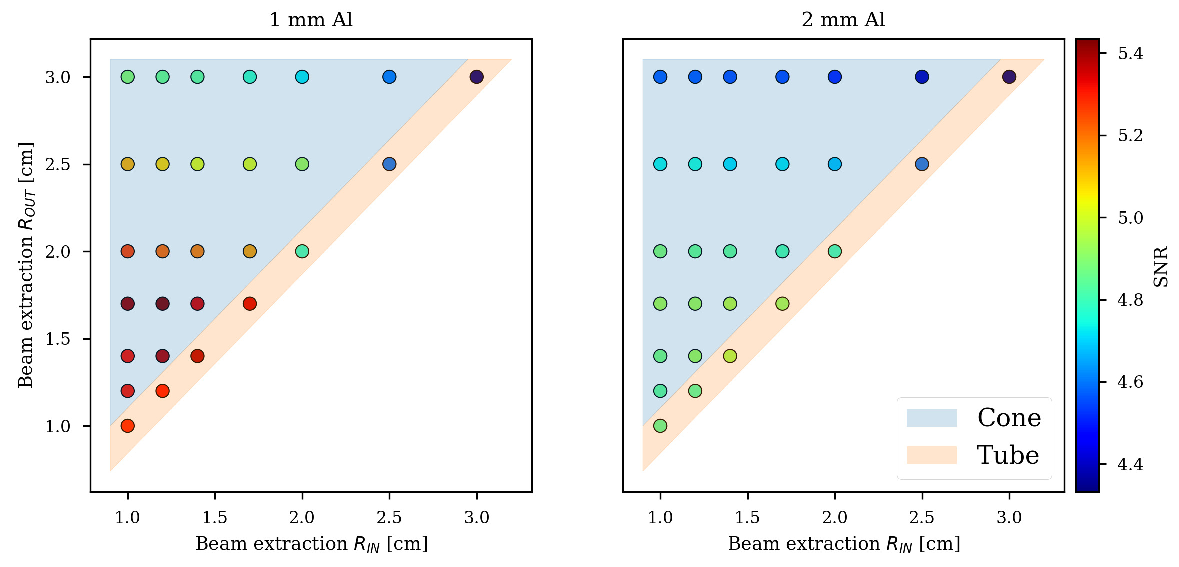}
\caption[SNR values for several configurations of the extraction system]{SNR values for several configurations of the extraction system. The points where $R_{IN}<R_{out}$ correspond to a divergent cone, while  $R_{IN}=R_{out}$ correspond to a straight tube. On the left, the inner Al walls separating the vacuum and the ND are \SI{1}{mm}-thick. On the right, the same inner walls are \SI{2}{mm}-thick. From an engineering point of view, a divergent cone with \SI{1}{mm} walls is not feasible.}\label{fig:SNR_extraction}
\end{figure}

The last component of the HighNESS experiment to study was the extraction system including the layer of ND powder. As a first approximation, we can fix the thickness of the ND layer to \SI{5}{mm} (and \SI{0.6}{g/cm^3} as density). The total length of the component was fixed to \SI{60}{cm} by the available space between the moderator-reflector assembly and the collimation of the pin-hole system. We can also ignore the CNS reflector to isolate the effect of the extraction. The CNS aperture is still \SI{1}{cm}. Two examples of extraction systems are sketched in \cref{fig:geom_extraction}.  \\
\indent From the neutronic point of view, the most convenient shape is a divergent cone. Similarly to what happens with the total reflection inside a mirror guide, quasi-specular reflections on divergent walls will progressively reduce the divergence of the beam at the expense of a higher emission surface. From the engineering point of view instead, thin aluminum walls are achievable only with cylindrical pipes, while cone-shaped ones can be manufactured with no less than \SI{2}{mm} walls. At the same time, the thickness of the internal wall can heavily influence the neutronic performance of the extraction system, since the path in the aluminum rapidly increases for grazing neutrons, hence increasing the chances of being absorbed. In \cref{fig:SNR_extraction} we report the SNR for several combinations of entrance and exit radii (size of the inner vacuum channel) for the extraction system and two Al wall thicknesses, of 1 and \SI{2}{mm}, respectively.\\
\indent When $R_{OUT}>R_{IN}$ the shape is a divergent cone, while for $R_{OUT}=R_{IN}$ it is a cylindrical tube. It is clear that the best neutronic performance is obtained with a small 1-mm-thick slightly divergent aluminum cone, but the difference with a tube with the same dimensions is relatively small while increasing the wall thickness to \SI{2}{mm} has a bigger impact. The best compromise between neutronic and engineering is then an extraction cylindrical tube with a radius of \SI{1.4}{cm} and wall thickness of \SI{1}{mm}. 
\subsubsection{Global optimization}
So far we have optimized the setup varying one or two parameters at a time while keeping the others constant. In this section, we wanted to study what would be the optimal solution in a more global sense. In other words, we tried to answer the question: what is the best setup when the CNS aperture, the entrance and the exit extraction radii, the reflector thickness, and the ND layer thickness are all free to vary?\\ 
\indent Once again we used Dakota to solve the problem in a fast and robust way. We first studied the parameters by optimizing one FOM at a time. Although they do not properly combine the needs of the experiment, these single-objective designs represent, within the statistical uncertainty, the best possible configurations for each FOM taken individually, so they can be used as maximum achievable values when making compromises in the multi-objective optimization. The results are summarized in \cref{tab:dakota_single_summary}. In light of the previous optimization effort, these values are not surprising: smallest CNS aperture and extraction radius for maximum SNR (opposite for CN counts) and tube over cone shape (\SI{1}{mm} over \SI{2}{mm} Al walls) for both FOMs. \\
\begin{table}[tb]
  \centering
%  \begin{threeparttable}
      \caption{Optimized designs for single-objective approach. The FOM in bold are the ones optimized for that run.} 
    \begin{tabular}{lcc}
    \toprule
    & \multicolumn{2}{c}{\textbf{Optimized FOM}} \\
    \cmidrule{2-3} 
    \textbf{Parameters [cm]} & \textbf{SNR} & \textbf{CN counts [n/nps]} \\
    \midrule
    CNS aperture & 1.0 &  3.0 \\
    Entrance extraction radius & 1.0 & 3.0 \\
    Exit extraction radius & 1.0 & 3.0 \\
    CNS reflector& 1.0 & 1.0 \\
    ND layer extraction & 1.0 & 1.0 \\
    \midrule
    \textbf{FOMs}$^{1}$ \\ %\tnote{1} \\
    \midrule
    SNR & \bfseries 7.33 & 0.98 \\
    CN counts [n/nps] & \num{4.61e-07} & \bfseries \num{7.32e-06} \\
    \bottomrule
    \end{tabular}\\%
%      \begin{tablenotes}
%    \item[1] In bold the values corresponding to the optimized FOM.
%  \end{tablenotes}
    $^{1}$In bold the values corresponding to the optimized FOM.
  \label{tab:dakota_single_summary}%
%  \end{threeparttable}

\end{table}% 
\indent The ultimate goal is to optimize the setup for both SNR and cold neutron counts, which means optimizing two objective functions at the same time. There are several approaches one can adopt, but the method we chose relies on the concept of the Pareto front. In a multi-objective problem, the optimal design is not a point. Rather, it is a set of points that satisfy the Pareto \textit{optimality criterion}, which is stated as follows in \cite{Dakota_6.18}: ``a feasible vector $X'$ is Pareto optimal if exists no other feasible vector $X$ which would improve some objective without causing a simultaneous worsening in at least one other objective''. A feasible point $X_0$ is said to be dominated if it can be improved on one or more objectives simultaneously. Points along the Pareto front are said to be non-dominated. \\
\indent The first step in the Pareto-set optimization method is reducing the multi-objective problem in a single-objective problem. Namely, Dakota optimizes a weighted sum of SNR and CN counts, each normalized by the maximum value achieved in the respective single-objective optimization. The normalization is essential since the SNR is orders of magnitude bigger than the counts per nps, hence, without the normalization, Dakota would be biased almost entirely toward the SNR. The weighted sum is the new objective function. Finally, the composition of the two FOMs was evaluated using a set of weights (SNR, CN counts). In \cref{tab:pareto_set_summary} the results of the simulations for a chosen weights set are summarized. As a sanity check, we made sure the results for the weights (1,0) and (0,1) were the same as the single-objective optimizations (within the statistical uncertainty). \\
    \begingroup
    \setlength{\tabcolsep}{10pt} 
    \begin{table}[tb]
    \centering
      \caption{Optimized designs from the Pareto-set approach.}
      \label{tab:pareto_set_summary}
    \begin{tabular}{lccccc}
    \toprule
    & \multicolumn{5}{c}{\textbf{Weight sets (SNR,CN counts)}}\\
    \cmidrule{2-6}
    \textbf{Parameters [cm]} & (0.75,0.25) & (0.65,0.35) & (0.5,0.5) & (0.35,0.65) & (0.25,0.75) \\
    \midrule
    CNS aperture & 1.0 & 1.0 & 1.0  & 2.8 & 3.0 \\
    Entrance extraction radius & 1.0 & 1.0 & 1.0 & 2.3 & 3.0 \\
    Exit extraction radius & 3.0 & 3.0 & 3.0 & 3.0 & 3.0 \\
    CNS reflector & 1.0 & 1.0 & 1.0 & 0.89 & 0.85 \\
    ND layer extraction & 0.67 & 0.73 & 2.3 & 2.0 & 2.5  \\
    \midrule
    \textbf{FOM ratio to max} \\
    \midrule
    SNR &         0.80 & 0.80 & 0.80 & 0.16 & 0.14 \\
    CN counts & 0.48 & 0.48 & 0.48 & 0.93 & 1.0 \\ 
    \bottomrule
    \end{tabular}%
\end{table}
\endgroup
\indent The analysis of the results of a multi-dimensional double-objective optimization is inevitably a complex task. We would like to point out some of the findings arising from this study:

\vspace{0.2cm}
\begin{itemize}[leftmargin=1cm]
\item[-]Despite the increasing importance of the CN counts weight, the CNS aperture is kept small to avoid a quick drop in the SNR. Only when the SNR has a small weight, the CNS aperture is ``allowed'' to increase. This shows how important it is to shield the extraction system from the intense diffused thermal field;
    \item[-]increasing the ND thickness in the extraction layer has a minimal impact on the CN counts;
    \item[-]almost half of the CN, compared to the single objective optimization, are obtained already with a small opening configuration. Hence,  there is no need to have a large extraction system to observe the enhancing effect.
\end{itemize} 
\vspace{0.2cm}

Taking into account all the previous optimization results and several iterations with the WP5 team, we decided to build a prototype with the following characteristics:
\begin{enumerate}
    \item[-]The CNS aperture in the cadmium disk is \SI{1}{cm}, matching the moderator size, to reduce the spurious thermal contribution to the detector;
    \item[-]the space around the cold source is enough to accommodate a reflector jacket \SI{1}{cm} thick; 
    \item[-]the extraction system is a tube with an inner radius of \SI{1.4}{cm} and wall thickness of \SI{1}{mm}.
\end{enumerate}
\subsubsection{Expected performance}
\label{subsec:expected_performance}
\begin{figure}
    \centering
    \includegraphics[width=0.9\textwidth]{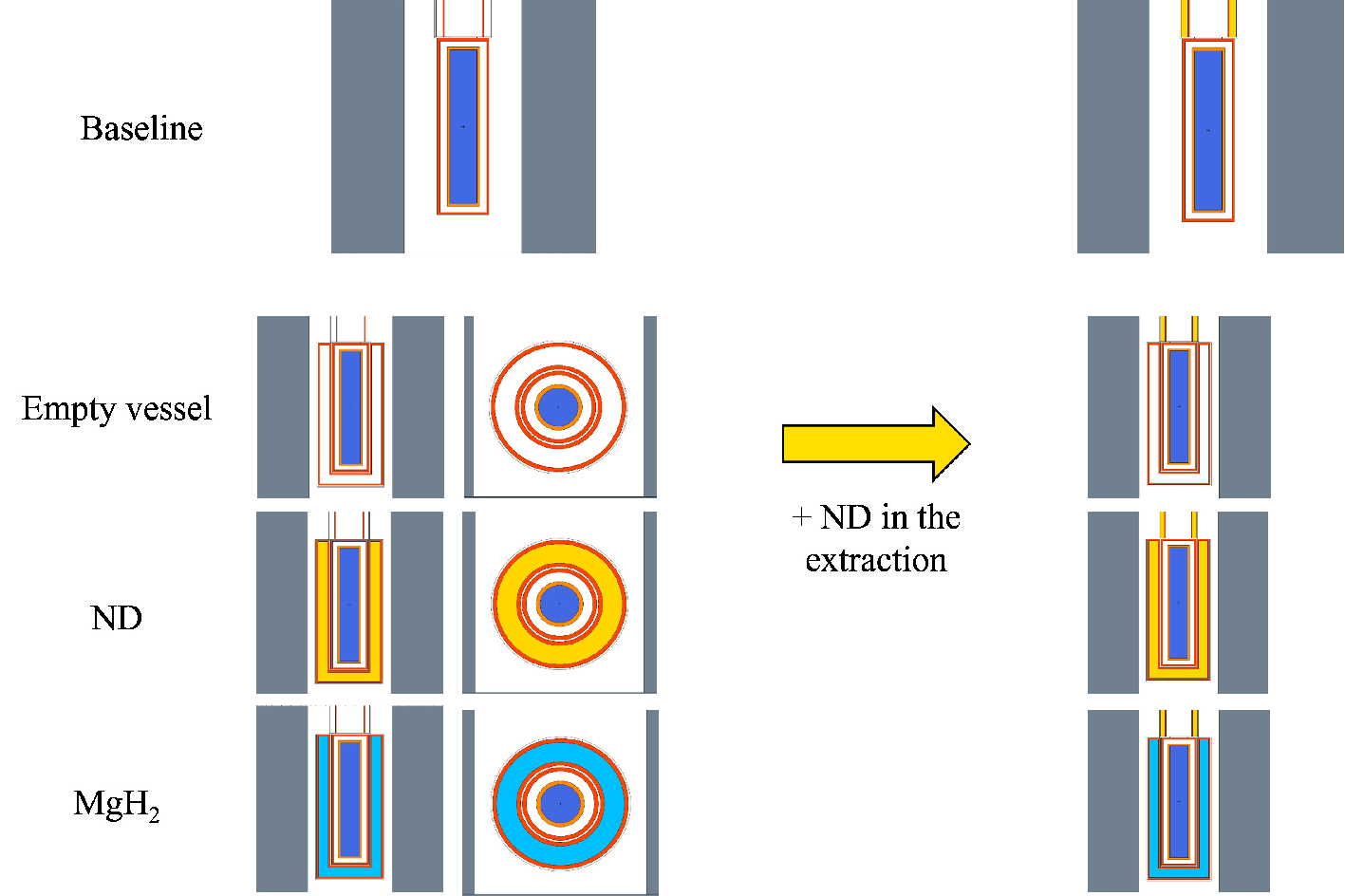}
    \caption[Scheme of measurements of the HighNESS experiment]{Graphical scheme of the measurements to perform at BNC facility in the context of the HighNESS experiment.}
    \label{fig:measure_scheme}
\end{figure}

\begin{figure}[tb!]       
    \begin{subfigure}[b]{\textwidth}
        \centering
        \includegraphics[width=\textwidth]{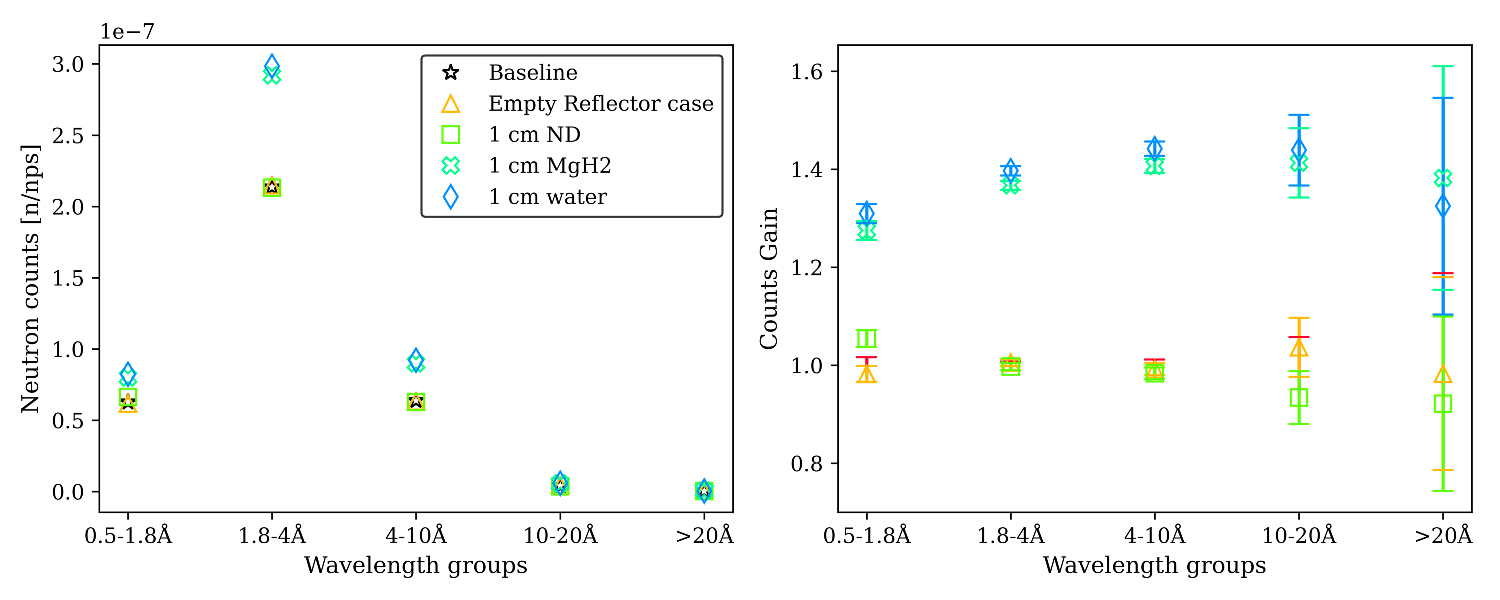}
        \subcaption{}
        \label{fig:gains_NoND}
    \end{subfigure}
    \begin{subfigure}[b]{\textwidth}
        \centering        
        \includegraphics[width=\textwidth]{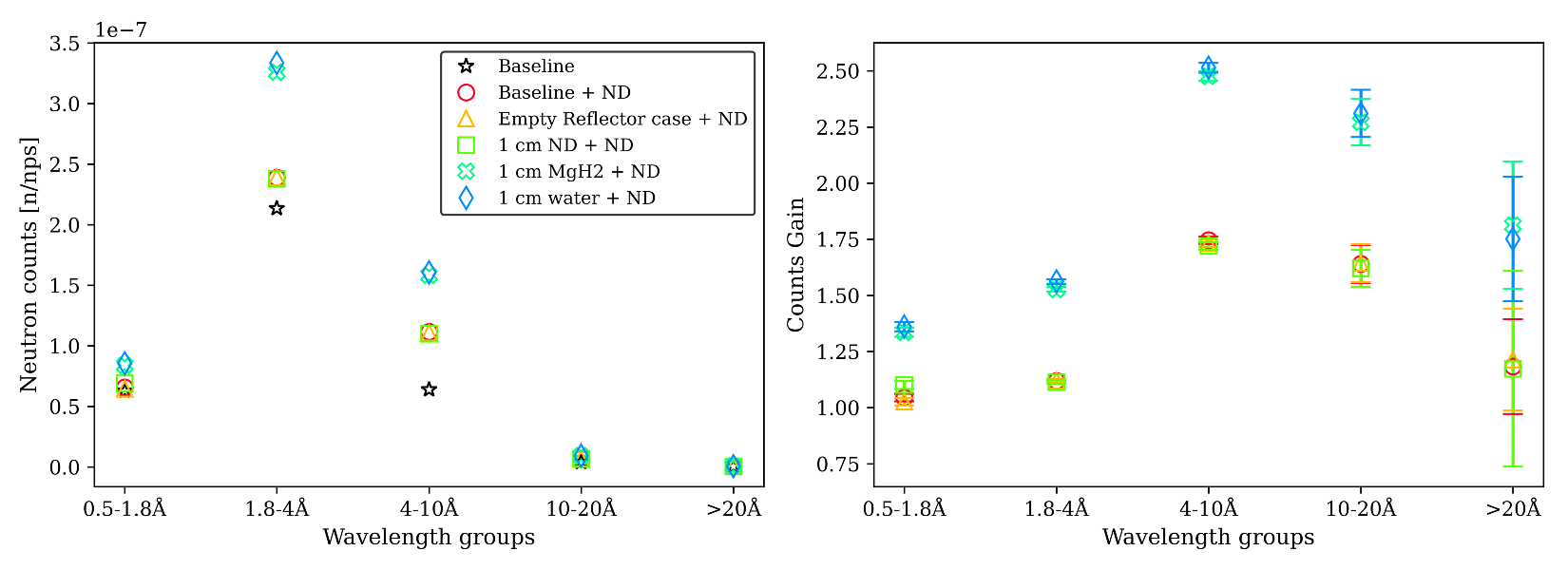}
        \subcaption{}
        \label{fig:gain_wND}
    \end{subfigure}
\caption[]{(left) Normalized neutron counts at the exit of the extraction tube for five wavelength groups for (a) all the cases without the thin ND layer in the extraction (b) with ND. (right) counts gains over the baseline}
\label{fig:gains_final}
\end{figure}

\begin{figure}[tb!]       
    \begin{subfigure}[b]{\textwidth}
        \centering
        \includegraphics[width=\textwidth]{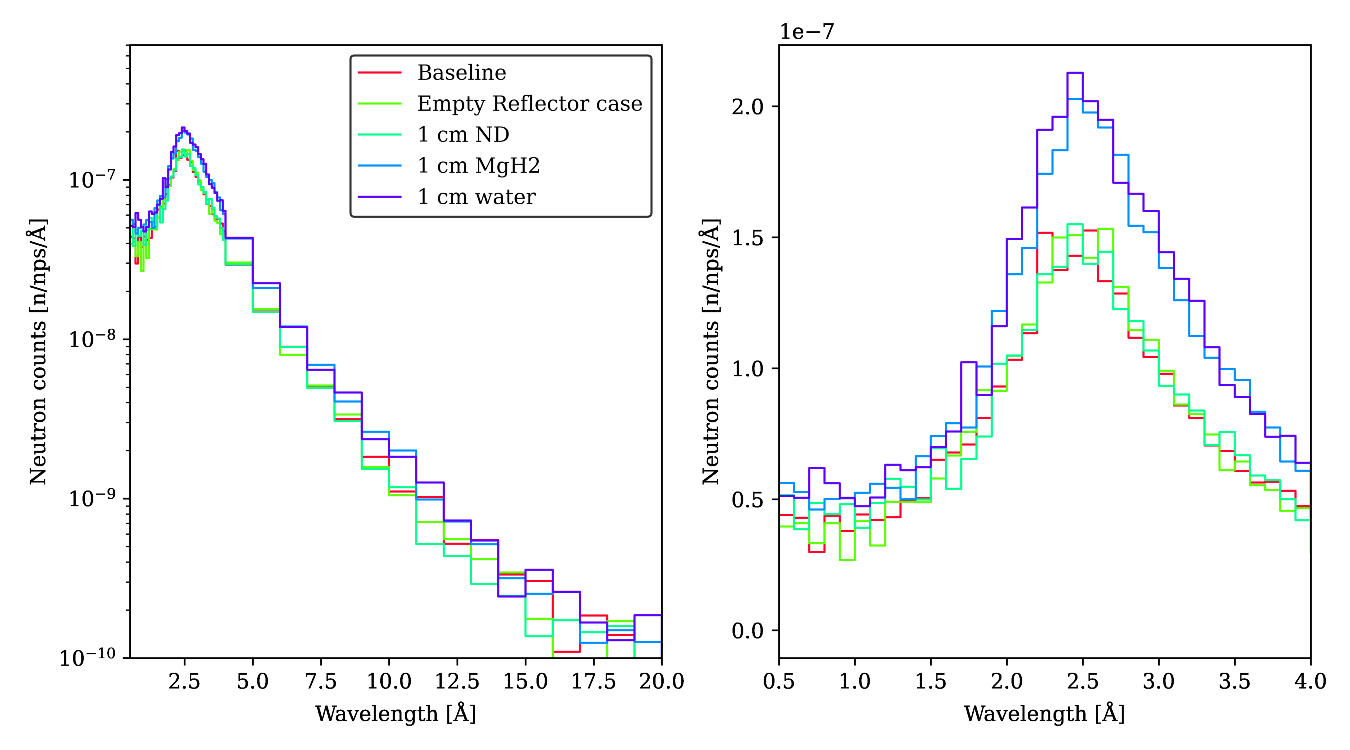}
        \subcaption{}
        \label{fig:spectra_noND}
    \end{subfigure}
    \begin{subfigure}[b]{\textwidth}
        \centering        
        \includegraphics[width=\textwidth]{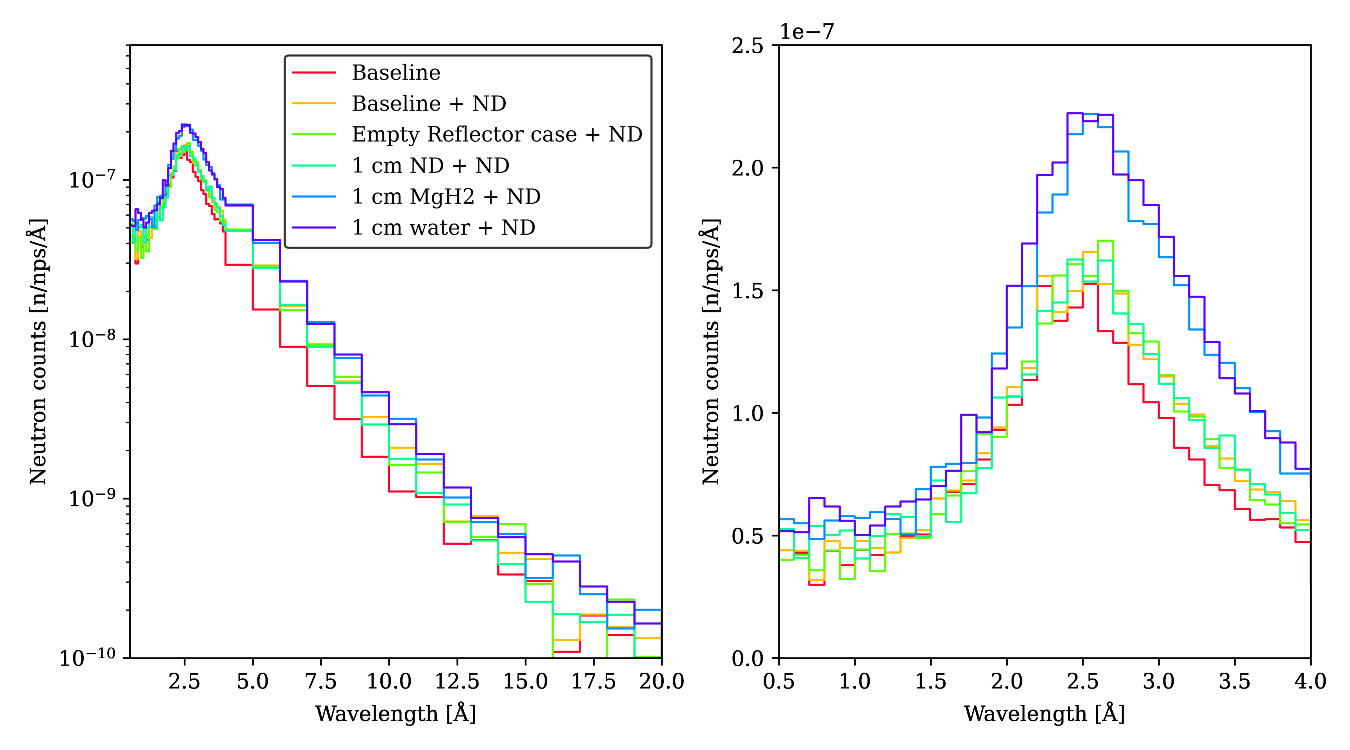}
        \subcaption{}
        \label{fig:spectra_ND}
    \end{subfigure}
\caption[Normalized neutron counts wavelength spectra at the exit of the extraction tube]{Normalized neutron counts wavelength spectra at the exit of the extraction tube for (a) all the cases without the thin ND layer in the extraction (b) with ND.}
\label{fig:spectra_final}
\end{figure}

The values found in the optimization phase are then inserted in the original full model (\cref{fig:MCNP_geo_highNESS}) to check the expected gains from the setup. We can divide the effect of the reflector from the effect of adding the extraction system filled with ND (sometimes indicated as +ND in the following summary). Hence, we define the common baseline as the model without both reflector and extraction tube and we calculate the gains after adding a 1-cm-thick jacket when it is empty, filled with ND, or with \ce{MgH_2}. All these cases, baseline included, are also calculated with ND in the extraction tube. A scheme summarizing the set of the eight measurements is shown in \cref{fig:measure_scheme}.\\
\indent We added one additcional case to this set to test the performance of a standard reflector, i.e. room temperature water. Hence we planned a total of ten measurements during the experiment. The gains we expect to observe are summarized in \cref{fig:gains_final}, while a more detailed overview of the simulation results can be found in \cref{tab:experiment_table}.\\
\indent In \cref{fig:gains_NoND}, we can see how more than 40\% gains over the baseline can be obtained with a \ce{MgH_2} reflector. However, similar performances are observed with standard room temperature water, while ND do not seem to have a measurable effect in the energy range studied. When adding ND to the extraction system, the gain factor goes well above 2 in the \SIrange{4}{10}{\angstrom} range, corresponding to the energy peak for quasi-specular reflection. The reason for similar results between water and \ce{MgH_2} is to be ascribed to the inelastic scattering on H. As shown in Figure 9 in \cite{granada_studies_2020}, for energies above \SI{0.1}{eV}, the inelastic cross section is dominated by down-scattering processes, hence in an intense thermal/fast field the dominating interaction with H is inelastic, while one could expect the reflector behavior to be dominant at low energies. Even though the effect is not purely due to the advanced cold reflector properties of \ce{MgH_2},  it is still interesting to study the material and its applicability in different environments.\\
\indent Finally, in \cref{fig:spectra_final}, we show the calculated spectra for all the cases, where an overview of the gains as a function of energy is given. 
\begin{sidewaystable}[!p]
  \centering
%\rotatebox{90}{%
%    \begin{minipage}{0.8\textheight}
      \def\arraystretch{1.2}
      \centering
      \setlength{\tabcolsep}{1.5mm}
      \small
      \centering
      \captionof{table}{Overview of the 10 simulations of measurements envisioned to be performed at the CMTF.}
\label{tab:experiment_table}
    \begin{tabular}{l c c c c c c c c c c}
    \toprule
	 \multirow{3}{*}{Counts/nps}& \multicolumn{10}{c}{Without ND in the extraction} \\			
     \cmidrule{2-11}
	& \multicolumn{2}{c}{$\lambda > \SI{20}{\angstrom}$} &       \multicolumn{2}{c}{$\SI{10}{\angstrom}<\lambda<\SI{20}{\angstrom}$} &	\multicolumn{2}{c}{$\SI{4}{\angstrom}<\lambda<\SI{10}{\angstrom}$} &	\multicolumn{2}{c}{$\SI{1.8}{\angstrom}<\lambda<\SI{4}{\angstrom}$}	&	\multicolumn{2}{c}{$\SI{1.8}{\angstrom}<\lambda<\SI{0.5}{\angstrom}$} \\	
\cmidrule{2-11}
 &	Value &	\% Rel. Err. &	Value &	\% Rel. Err..&	Value&	\% Rel. Err.&	Value&	\% Rel. Err.	& Value &	\% Rel. Err. \\
\midrule
Baseline &	\num{3.40E-10} &	9	 & \num{4.04E-09}	& 3 &	\num{6.39E-08} &	0.6 &	\num{2.14E-07} &	0.4	 & \num{6.28E-08}	& 0.8 \\
Empty vessel&	\num{3.34E-10} &	11 &	\num{4.18E-09} &	3 &	\num{6.34E-08} &	0.6 &	   \num{2.15E-07} &	0.4 &	\num{6.17E-08} &	0.8 \\
ND &	\num{3.13E-10} &	10 &\num{3.77E-09} &	3 &	\num{6.29E-08} &	0.6 &	   \num{2.13E-07} &	0.4 &	\num{6.62E-08} &	0.9 \\
\ce{MgH_2} &	\num{4.70E-10} &	7 &	   \num{5.70E-09} &	3 &	\num{8.99E-08} &	0.6 &   \num{2.92E-07} &	0.3 &	\num{8.01E-08} &	0.7 \\
Water &	\num{4.51E-10} &	7 &	   \num{5.81E-09} &	2 &	\num{9.22E-08} &	0.6 &   \num{2.98E-07} &	0.3 &	\num{8.22E-08} &	0.7 \\
\midrule

\multirow{3}{*}{Counts/nps}& \multicolumn{10}{c}{With ND in the extraction} \\			
     \cmidrule{2-11}
	& \multicolumn{2}{c}{$\lambda > \SI{20}{\angstrom}$} &       \multicolumn{2}{c}{$\SI{10}{\angstrom}<\lambda<\SI{20}{\angstrom}$} &	\multicolumn{2}{c}{$\SI{4}{\angstrom}<\lambda<\SI{10}{\angstrom}$} &	\multicolumn{2}{c}{$\SI{1.8}{\angstrom}<\lambda<\SI{4}{\angstrom}$}	&	\multicolumn{2}{c}{$\SI{1.8}{\angstrom}<\lambda<\SI{0.5}{\angstrom}$} \\	
\cmidrule{2-11}
 &	Value &	\% Rel. Err. &	Value &	\% Rel. Err..&	Value&	\% Rel. Err.&	Value&	\% Rel. Err.	& Value &	\% Rel. Err. \\
\midrule
Baseline	 & \num{4.02E-10} &	8 &	\num{6.62E-09}	& 2 & 	\num{1.12E-07} &	0.3 &	\num{2.39E-07}	& 0.4 &	\num{6.56E-08} &	0.8\\
Empty vessel & \num{4.13E-10} &	9 &	\num{6.63E-09}	& 2 & 	\num{1.11E-07} &	0.4 &	\num{2.39E-07}	& 0.4 &	\num{6.45E-08} &	0.8\\
ND 	         & \num{3.99E-10} &	28 &	\num{6.54E-09}	& 2 & 	\num{1.10E-07} &	0.4 &	\num{2.38E-07}	& 0.4 &	\num{6.91E-08} &	0.8\\
\ce{MgH_2} 	 & \num{6.17E-10} &	6 &	\num{9.17E-09}	& 2 & 	\num{1.58E-07} &	0.3 &	\num{3.26E-07}	& 0.3 &	\num{8.39E-08} &	0.7\\
Water 	     & \num{5.96E-10} &	6 &	\num{9.32E-09}	& 2 & 	\num{1.61E-07} &	0.3 &	\num{3.34E-07}	& 0.3 &	\num{8.55E-08} &	0.7 \\
\bottomrule
    \end{tabular}
%    \end{minipage}}
\end{sidewaystable}
\clearpage

\subsubsection{Dose and background calculations}
\label{subsec:dose_and_background}
\begin{figure}[hbt]
    \centering
    \includegraphics[width=\textwidth]{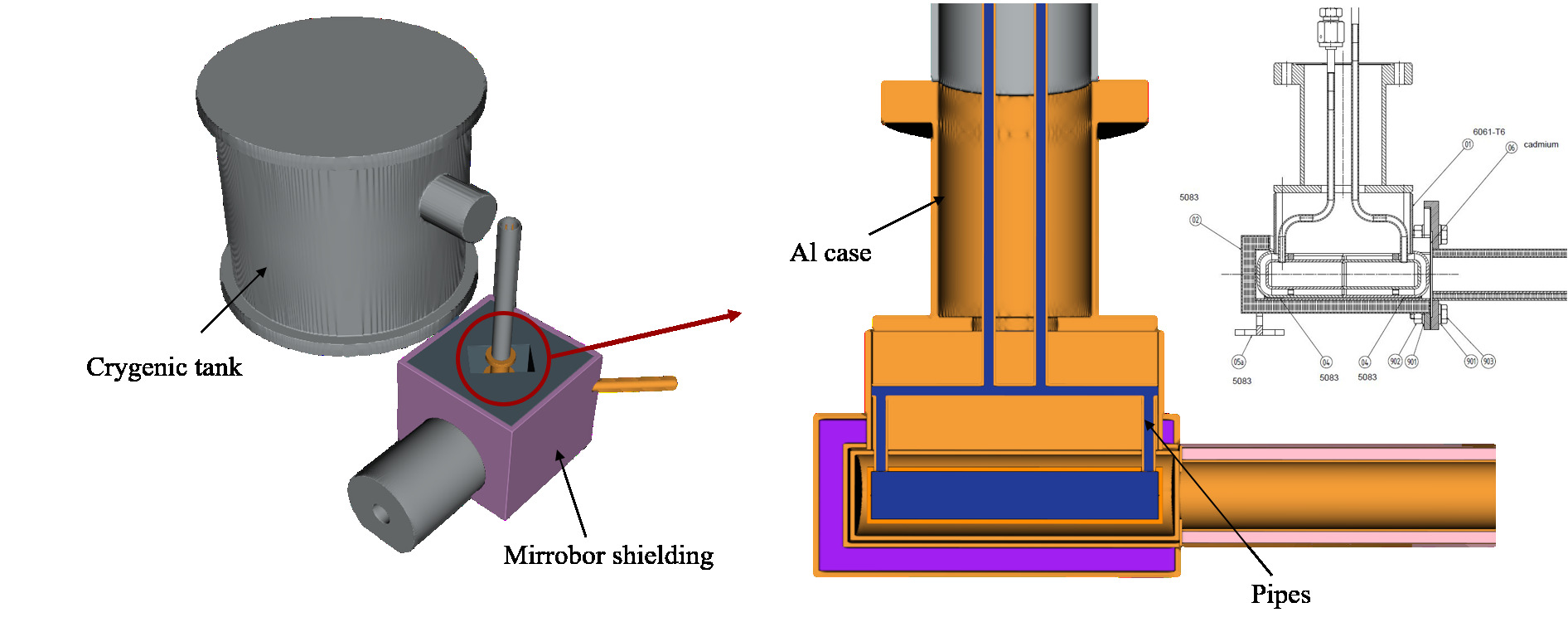}
    \caption[MCNP model of the cryogenic system for the HighNESS experiment]{MCNP model of the cryogenic system for the HighNESS experiment. The stainless steel cryogenic tank is shown on the left. On the right, an enlarged view of the moderator pipes and case according to the engineering CAD (top right).}
    \label{fig:cryo_model_MCNP}
\end{figure}

\begin{figure}[hbt]
    \centering
    \includegraphics[width=.7\textwidth]{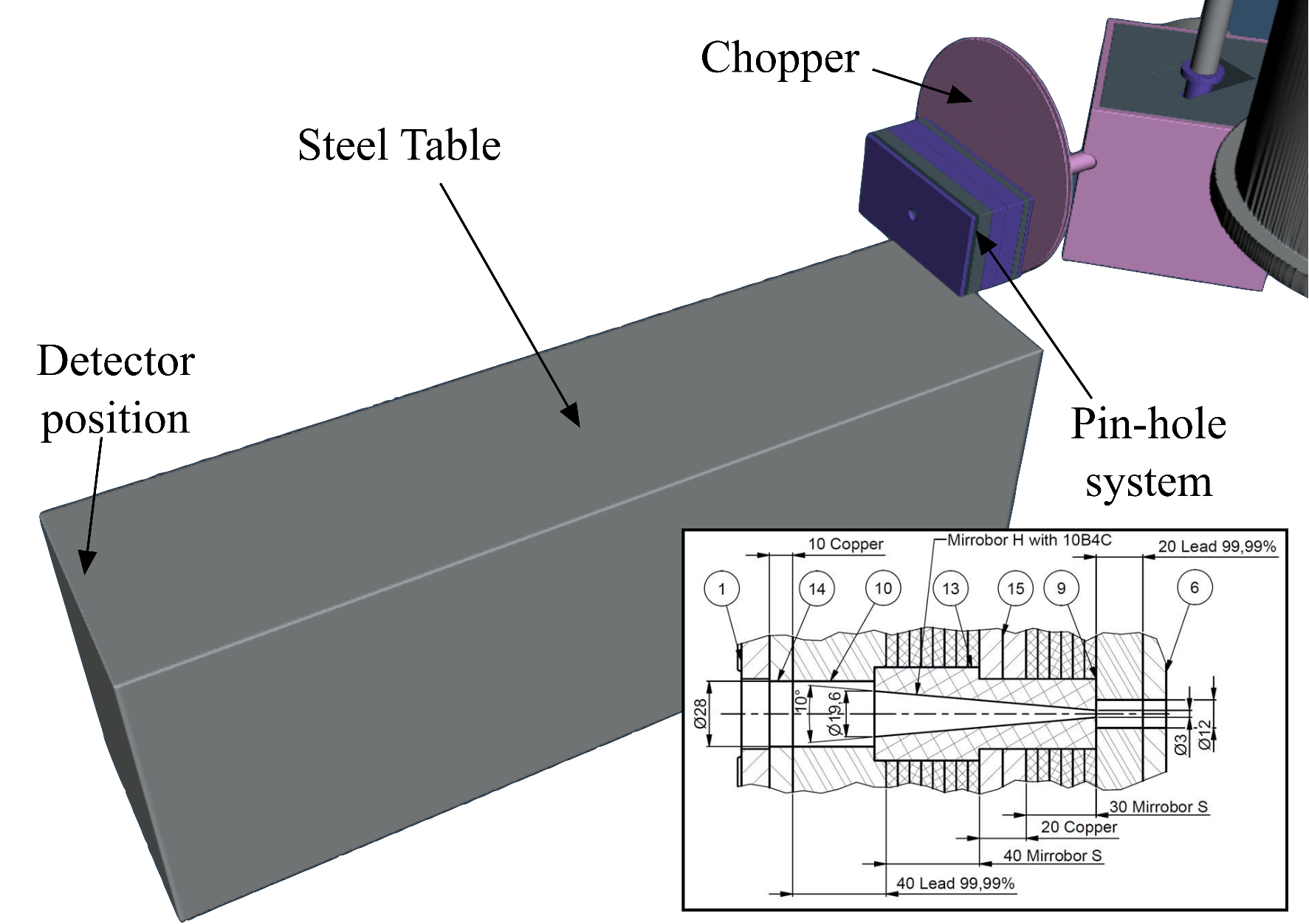}
    \caption[MCNP model of the detection system for the HighNESS experiment]{MCNP model of the detection system for the HighNESS experiment. The model has only the essential elements: the chopper, the pin-hole system, and the steel table with the support rails for the detector. The detailed drawings of the pin-hole system are reported in the bottom right corner.}
    \label{fig:detection_MCNP}
\end{figure}

\begin{figure}[tb!]
    \centering
    \includegraphics[width=.8\textwidth]{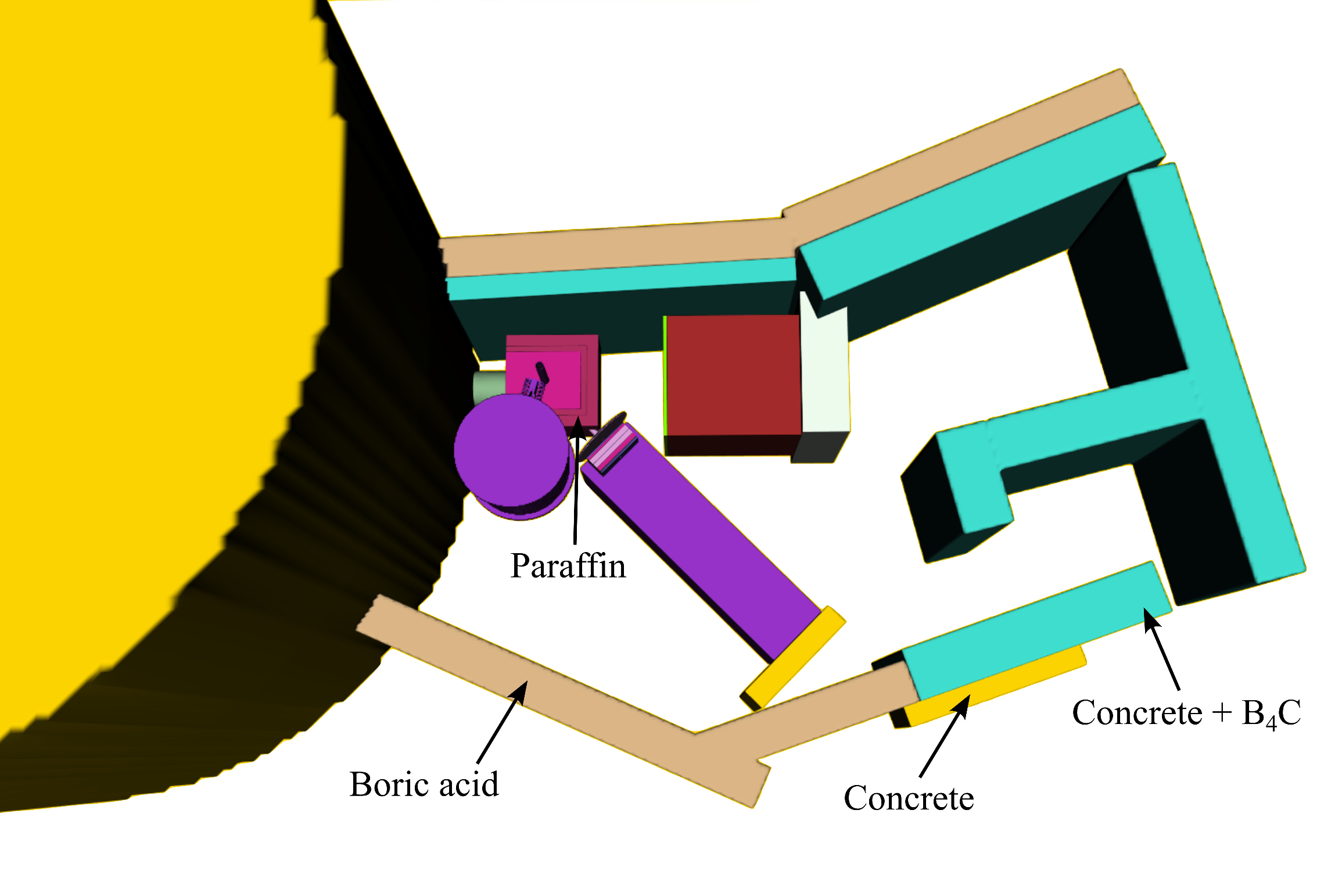}
    \caption[MCNP model of the experimental room for the HighNESS experiment.]{MCNP model of the experimental room for the HighNESS experiment. The different materials for the external and reflector shielding are highlighted. Mirrored along the horizontal direction compared to \cref{fig:bud_beamline}.}
    \label{fig:Bud_beamline_MCNP}
\end{figure}

The following step was to add increasing details to the model in order to simulate the whole experimental environment. In particular, we studied two quantities that are crucial for the measurements: the neutron and photon dose rates inside and outside the experimental hall and the background level at the detector compared to the signal from the moderator. \\
\indent The first step toward a more realistic model was made by adding the cryogenic system to the reflector-moderator assembly. In particular, we were interested in quantifying the effect on the neutronic of the piping case as well as the piping itself. These components are made of aluminum to reduce the impact on the performance, however, a strip of reflector must be removed to make space for them, which can be expected to further reduce the initial yield. The MCNP model, manually made from a simplified version of the CAD drawings, is shown in \cref{fig:cryo_model_MCNP}. The components further away from the moderator vessel, like the cryogenic tank and external transfer pipes, are made of stainless steel. The negative impact of the cryogenics at the exit of the extraction tube was estimated to be between 20\% to 30\% depending on the neutron energy (higher at lower wavelength) for the case with \ce{MgH_2} as CNS reflector and ND in the extraction. This effect is almost entirely due to the aluminum structure, hence should not affect the relative gains. \\
\indent To have a better estimate of the background at the detector position, we modeled the essential pieces of the detection system: the chopper, the pin-hole system, and the steel table according to the specifications provided (cfr. \cref{fig:bud_beamline}). In particular, the pin-hole system is made of several layers of metals, like lead and copper, and neutron shielding material like cadmium and boron plastic (Mirrobor\texttrademark). The result is shown in \cref{fig:detection_MCNP}.\\
The last piece necessary for the study is the model of the experimental bunker. A temporary model of the neighbor channel \#2 was used for preliminary calculations of the background at the detector position, which already highlighted how critical it was to carefully shield the reflector-moderator block and the detector from the intense diffused epithermal and thermal neutron flux. Later we received a close-to-final model of the experimental room designed by BNC, which included a concrete floor, a simple model for the heavy-concrete reactor, and external walls made of boric acid bath and Mirrotron boron concrete. The shielding around the reflector is made of mostly Mirrobor\texttrademark\,  sheets (80 wt\% \ce{B_4C} + \ce{CH_2O}) and paraffin wax. The complete MCNP model used for the exploratory analysis is shown in \cref{fig:Bud_beamline_MCNP}. In the figure, the geometry is mirrored along the horizontal direction compared to \cref{fig:bud_beamline}.\\
\indent At this point it should be noted that this model is still missing a design for the bunker ceiling and the shielding around the collimator at the beam port. At the time of performing the first simulations with this model, the CMTF construction phase was not fully completed and not including the design of the ceiling made of structural steel and several layers of borated HDPE. The space around the collimator, between the reactor's wall and the reflector, has been filled with paraffin wax blocks. However, these recent changes have not been implemented in the model yet.
\subsubsection{Background}
\begin{figure}[tb!]       
    \begin{subfigure}[b]{\textwidth}
        \centering
        \includegraphics[width=0.8\textwidth]{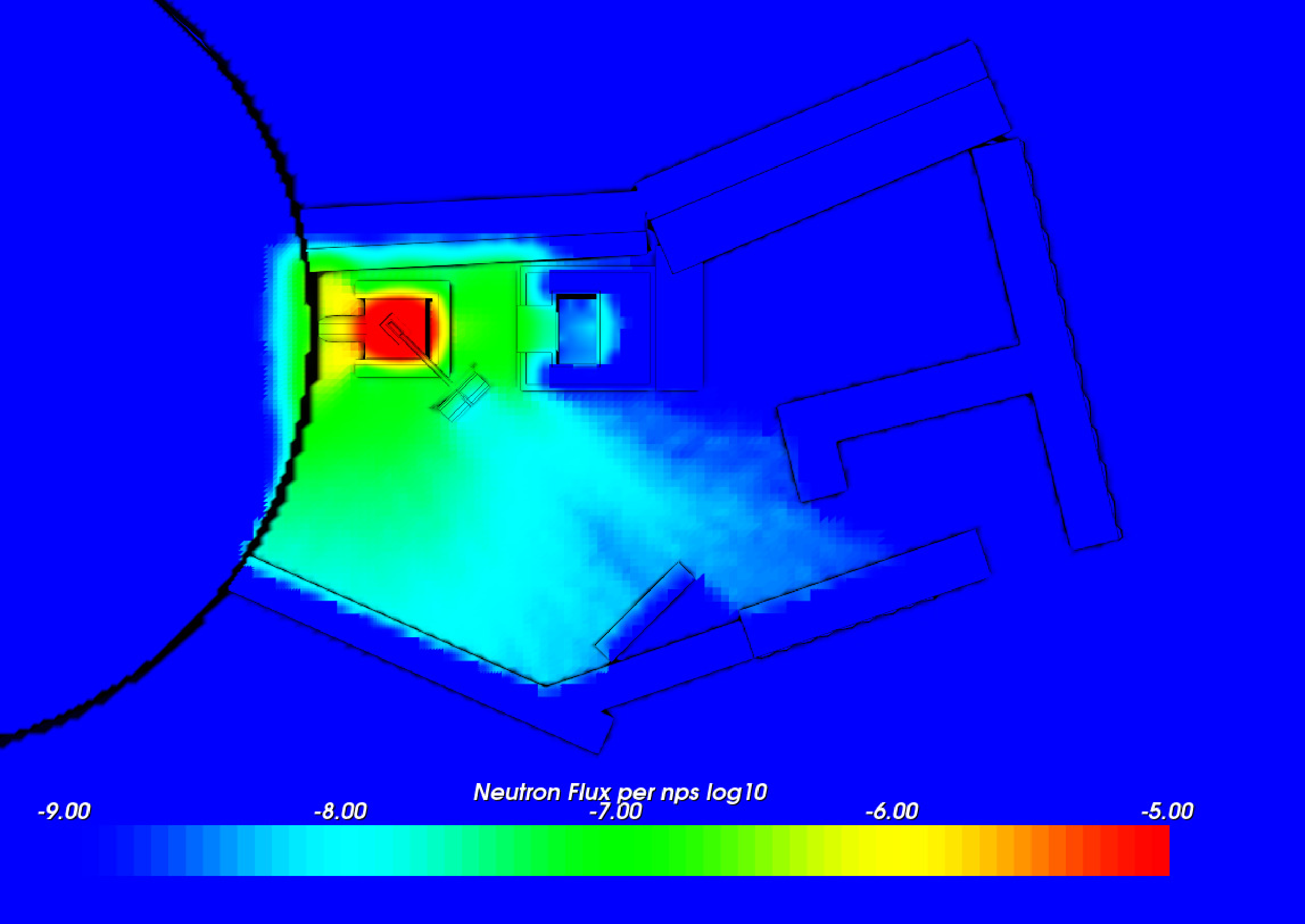}
        \subcaption{}
        \label{fig:fast}
    \end{subfigure}
    \begin{subfigure}[b]{\textwidth}
        \centering        
        \includegraphics[width=0.8\textwidth]{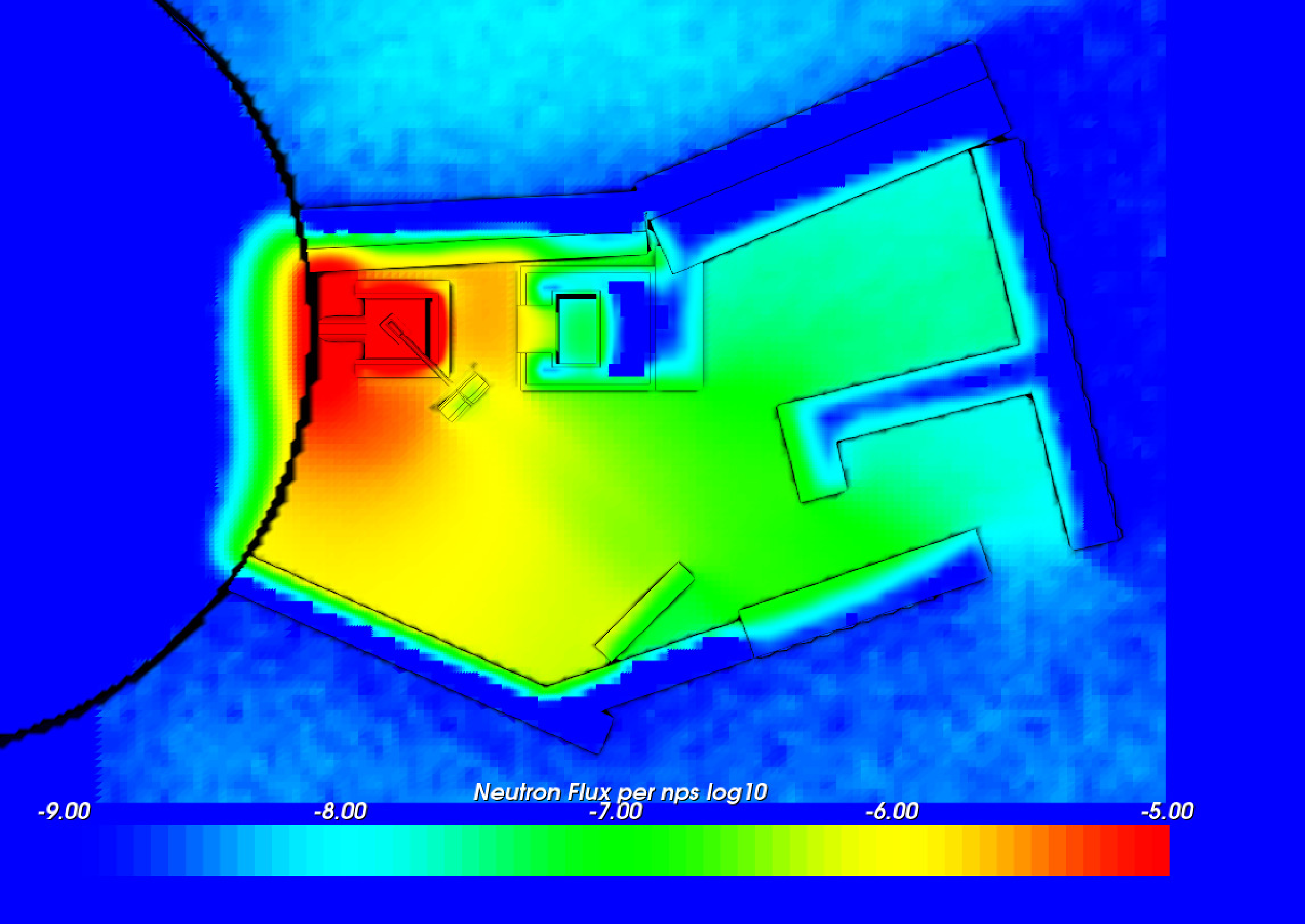}
        \subcaption{}
        \label{fig:epithermal}
    \end{subfigure}
\caption[Neutron flux map for fast and epithermal neutrons]{Neutron flux map at the moderator height with a spatial resolution of \qtyproduct{10x10x10}{cm} for (a) fast neutrons (\SIrange{1}{100}{MeV}) (b) epithermal neutrons (\SI{81.8}{meV} to \SI{1}{MeV}). Units are \si{n/cm^2/nps} in log scale.}
\label{fig:fast_and_epithermal_flux_map}
\end{figure}

\begin{figure}[tb!]       
    \begin{subfigure}[b]{\textwidth}
        \centering
        \includegraphics[width=0.8\textwidth]{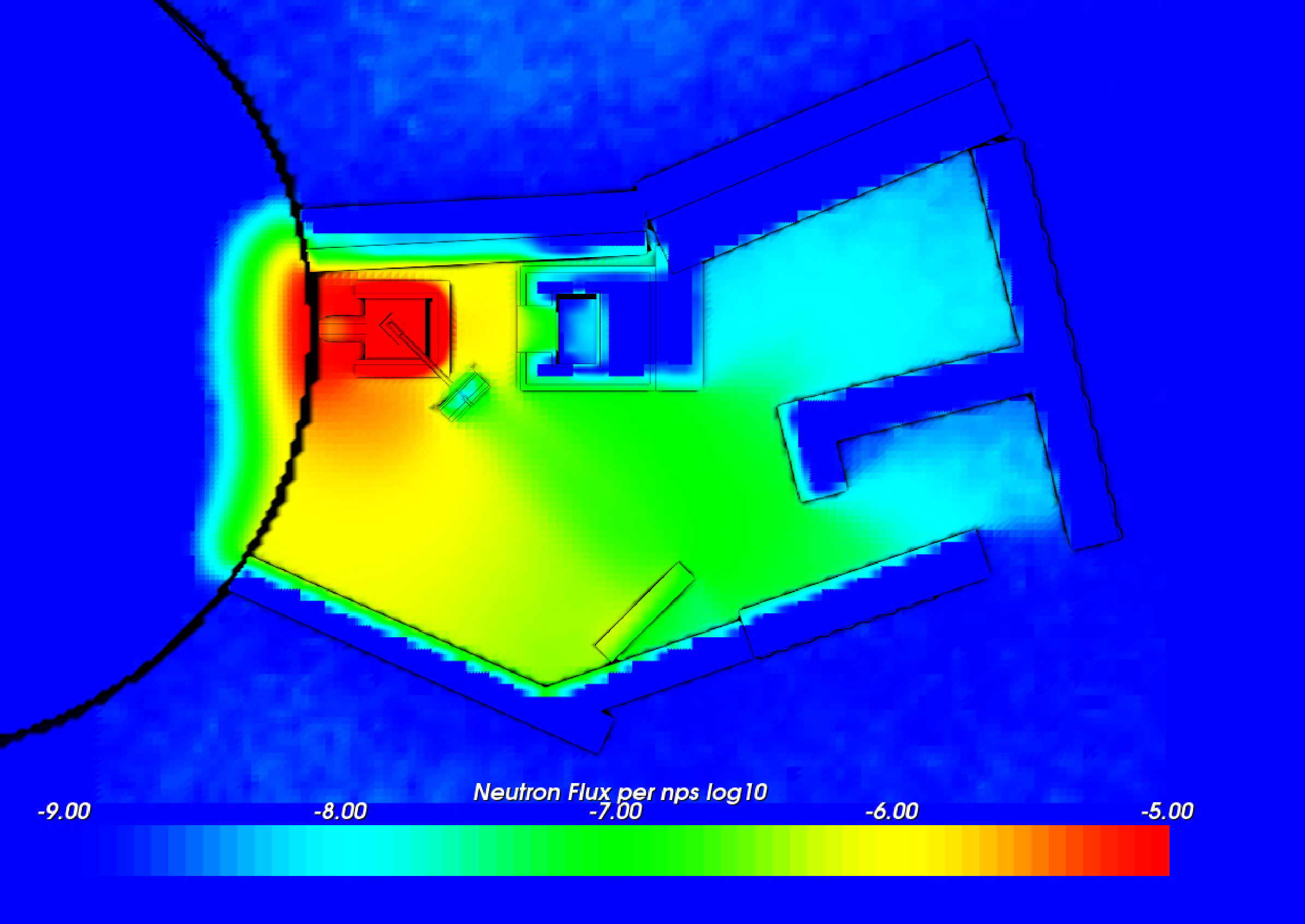}
        \subcaption{}
        \label{fig:thermal}
    \end{subfigure}
    \begin{subfigure}[b]{\textwidth}
        \centering        
        \includegraphics[width=0.8\textwidth]{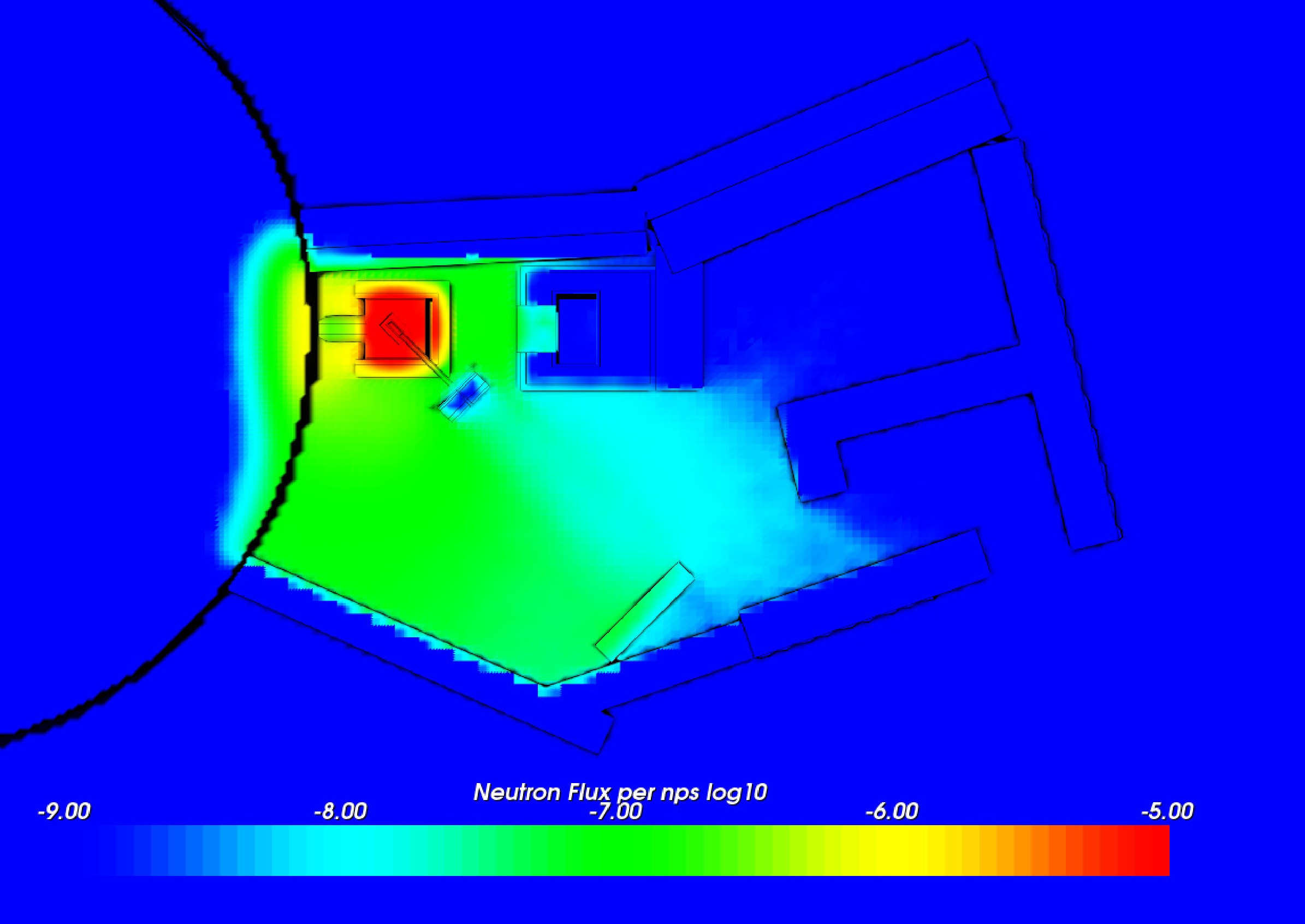}
        \subcaption{}
        \label{fig:cold}
    \end{subfigure}
\caption[Neutron flux map for thermal and cold neutrons]{Neutron flux map at the moderator height with a spatial resolution of \qtyproduct{10x10x10}{cm} for (a) thermal neutrons (\SI{13}{meV} to \SI{81.8}{meV}) (b) cold neutrons (\SI{5.11}{meV} to \SI{13}{meV}). Units are \si{n/cm^2/nps} in log scale.}
\label{fig:thermal_and_cold_flux_map}
\end{figure}

\begin{figure}[h!]
    \centering
    \includegraphics[width=.8\textwidth]{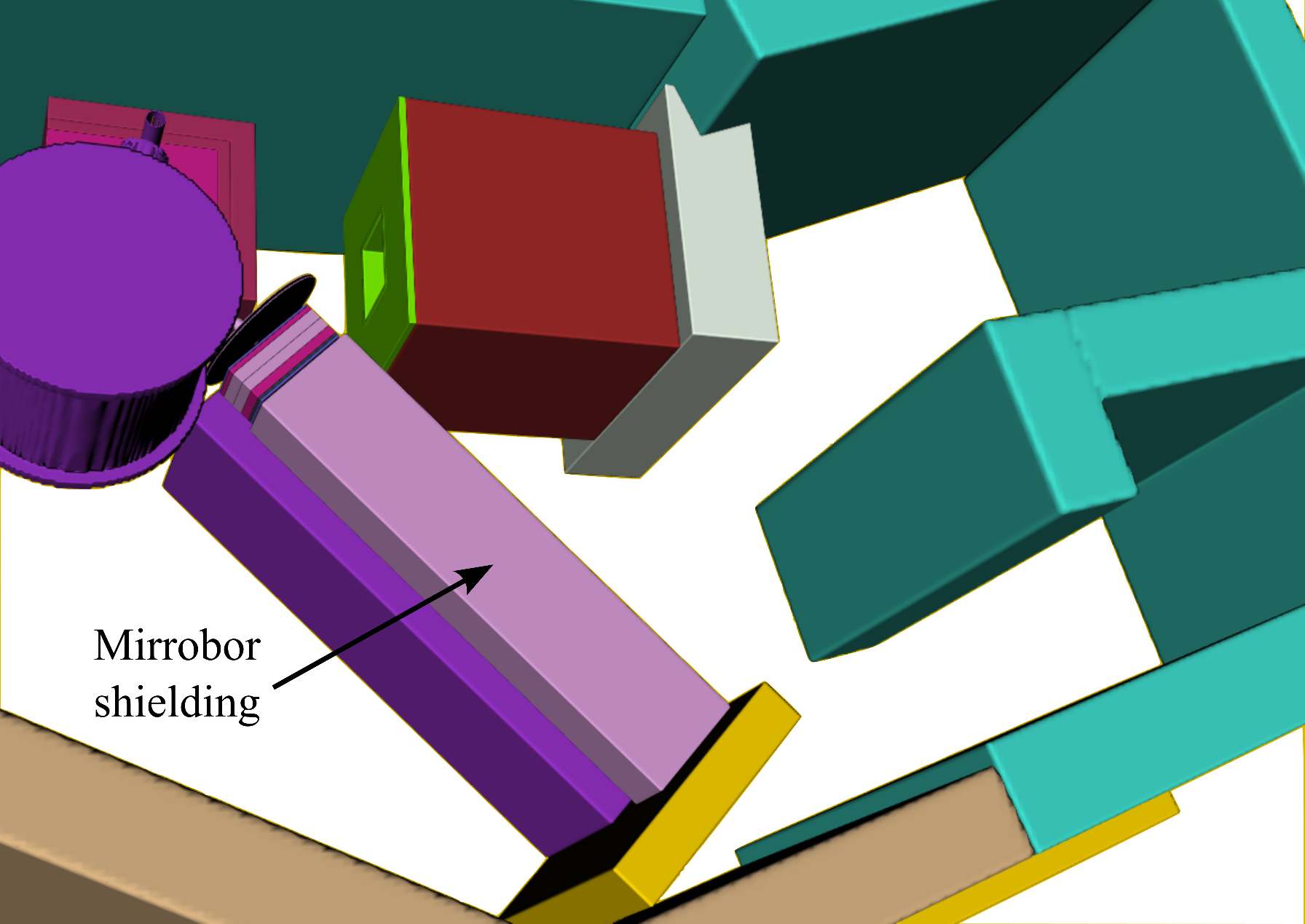}
    \caption[MCNP model of shielded detector for the HighNESS experiment.]{MCNP model of the shielded detector for the HighNESS experiment. The shielding is made of \SI{2}{cm} Mirrobor\texttrademark (80 wt\% \ce{B_4C} + \ce{CH_2O}), is \qtyproduct{20x20}{cm} and covers the gap from the pin-hole assembly to the detector position.}
    \label{fig:detector_shielding}
\end{figure}

A preliminary assessment of the background at the detector position with the full model shown in \cref{fig:Bud_beamline_MCNP} can be made by looking at the neutron flux map at different energy-groups. The neutron flux in \si{n/cm^2/nps} at the height of the moderator is shown in \cref{fig:fast_and_epithermal_flux_map} (fast and epithermal) and \cref{fig:thermal_and_cold_flux_map} (thermal and cold) with a spatial resolution of \qtyproduct{10x10x10}{cm}. The highly diffused neutron field in the epithermal (\SI{81.8}{meV} to \SI{1}{MeV}) and thermal (\SI{13}{meV} to \SI{81.8}{meV}) energy range appears to be 1 or 2 orders of magnitude higher than the cold flux in the detector position. While these flux values do not take the energy-dependent resolution of the detector into account, they still highlight the importance of reducing the diffusion out of both the collimator and the reflector. As an example, in the temporary model of the neighbor channel \#2 we observed that adding a layer of an ideal absorber (i.e. all neutrons get killed when entering the cell) around both the collimator and the lead block had the effect of reducing the high energy neutron flux by more than a factor 10. \\
\indent In the case of the latest model, we estimated the effect of shielding the 
 \qtyproduct{20x20}{cm} detector with \SI{2}{cm} Mirrobor\texttrademark\, sheets all the way from the pin-hole assembly. The model is shown in \cref{fig:detector_shielding}. The calculations are made with a point detector tally (next-event estimator) placed at the center of the detector position along the beam direction. Due to the incompatibility of the library with the next-event estimator, ND were removed from the extraction tube for this test. The results are reported in \cref{tab:shielding_det_test} with the same energy groups as the flux maps. From the reported values, it is clear that this solution is effective in shielding diffused cold and thermal neutrons, but it is not enough for epithermal neutrons.

 \begin{table}[!p]
     \centering
     \setlength{\tabcolsep}{10pt} 
     \begin{tabular}{c c c c c}
     \toprule
        Neutron flux [\si{n/cm^2/nps}] & Cold & Thermal & Epithermal & Fast  \\
     \midrule
      without shielding    & \num{2.43e-8} & \num{2.55e-7} & \num{3.75e-7} & \num{3.85e-9} \\
      with shielding    &    \num{2.79e-11} & \num{1.41e-10} & \num{4.79e-8} & \num{1.73e-9} \\
      \bottomrule
     \end{tabular}
     \caption[Neutron flux values measured at the detector position with next-event estimator]{Neutron flux values in \si{n/cm^2/nps} measured at the detector position with the next-event estimator (F5) tally with and without the Mirrobor\texttrademark shielding and for four different energy-groups: fast (\SIrange{1}{100}{MeV}), epithermal (\SI{81.8}{meV} to \SI{1}{MeV}), thermal (\SI{13}{meV} to \SI{81.8}{meV}) and cold neutrons (\SI{5.11}{meV} to \SI{13}{meV}). The relative error for all the values is 1\% . Due to the incompatibility of the library with the next-event estimator, ND were removed from the extraction tube for this test.}
     \label{tab:shielding_det_test}
 \end{table}
 \subsubsection{Dose rate}
\begin{figure}[h!]       
    \begin{subfigure}[b]{0.41\textwidth}
        \centering
        \includegraphics[width=\textwidth]{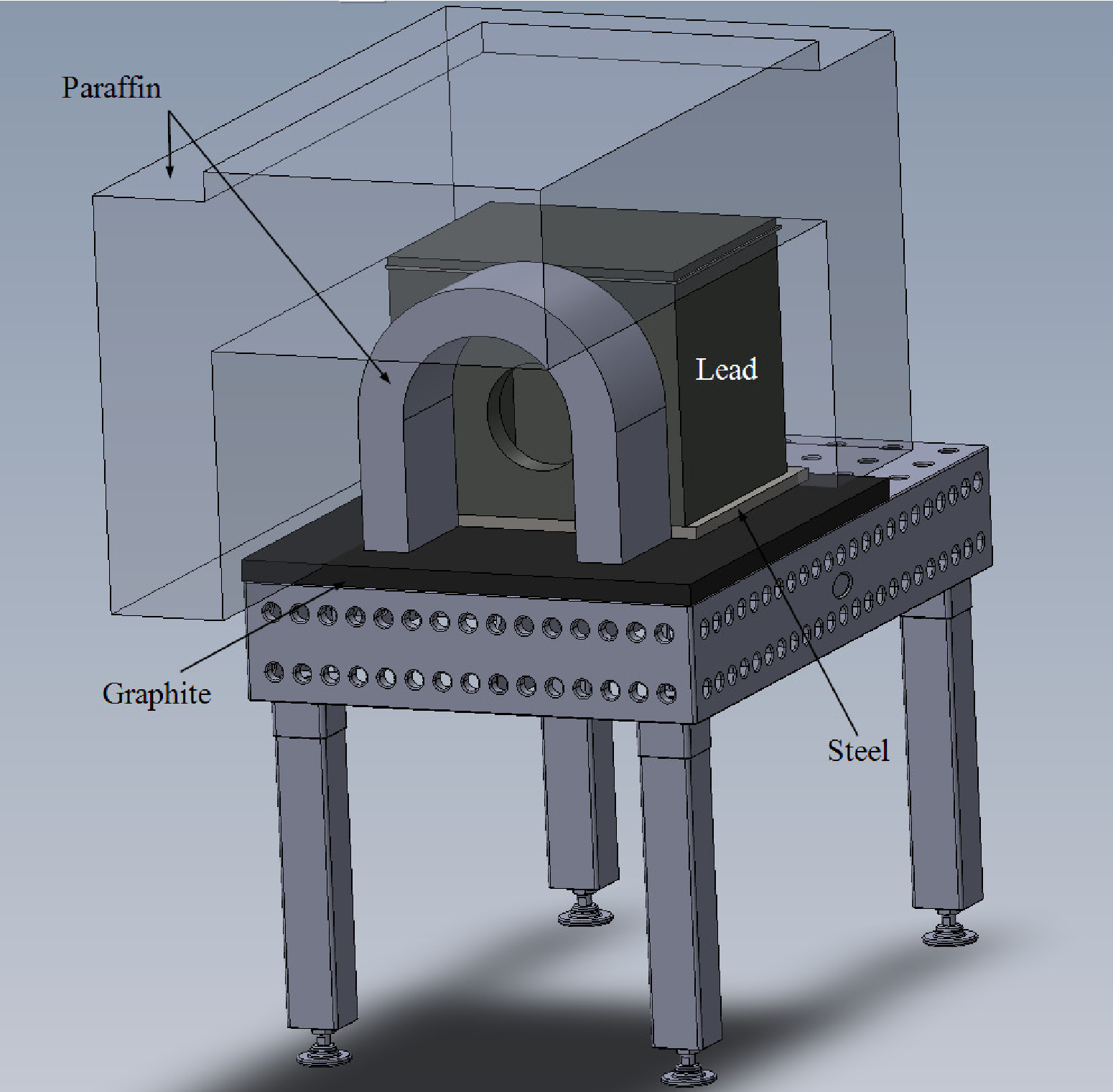}
        \subcaption{}
        \label{fig:dummy_lead}
    \end{subfigure}
    \begin{subfigure}[b]{0.562\textwidth}
        \centering        
        \includegraphics[width=\textwidth]{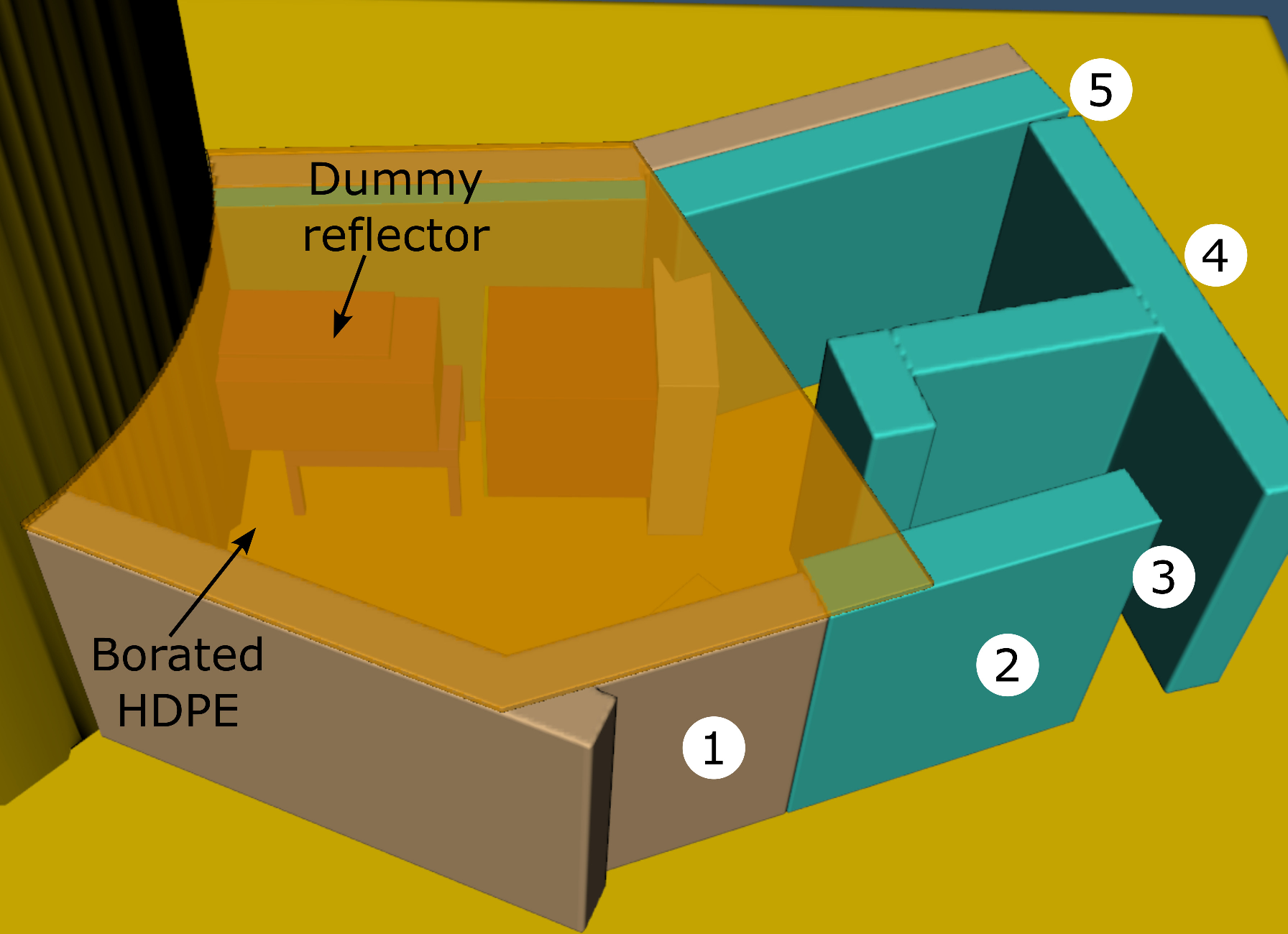}
        \subcaption{}
        \label{fig:dose_geo}
    \end{subfigure}
\caption[Dummy lead reflector and MCNP model of the experimental room for dose rates.]{(a) Dummy lead reflector with temporary paraffin wax shielding. (b) MCNP model of the experimental room adapted for the configuration of the dose rate measurements with the dummy reflector and the temporary roof. The five points where the dose rates were measured are also shown.}
\label{fig:dose_measurement}
\end{figure}

\begin{figure}[h!]
    \centering
    \includegraphics[width=.8\textwidth]{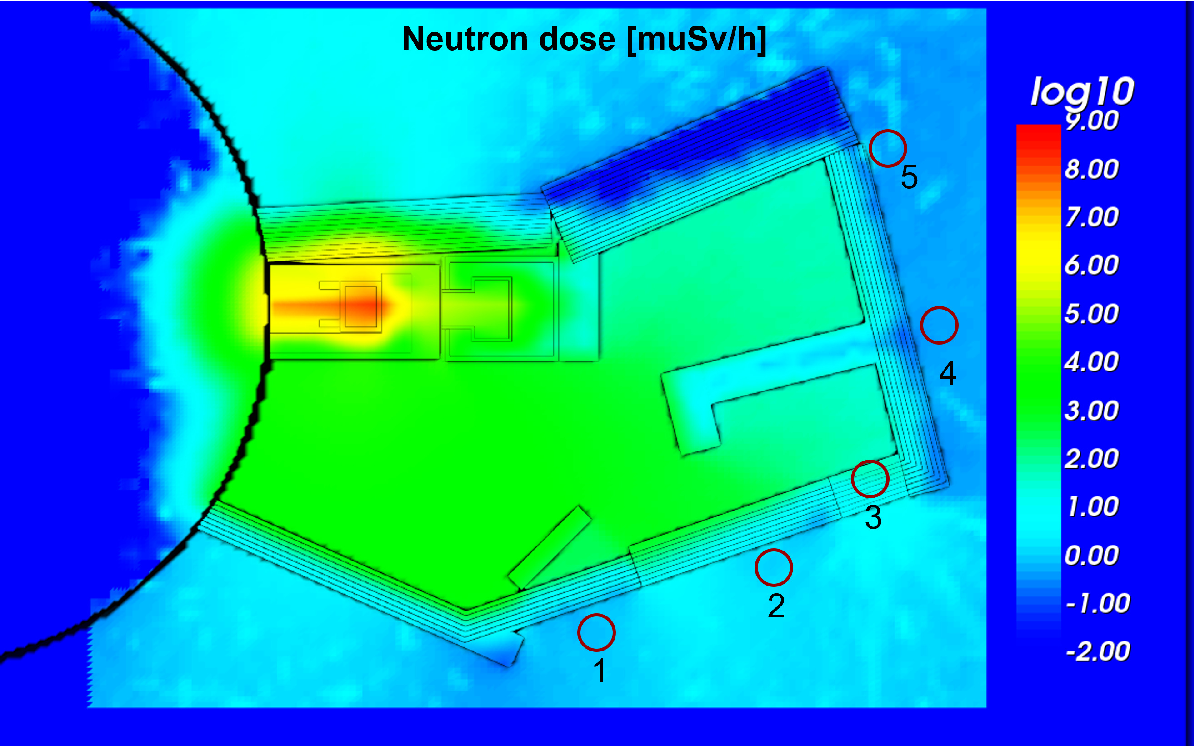}
    \caption[Neutron dose rate map]{Neutron dose rate map with a spatial resolution of \qtyproduct{10x10x10}{cm}. The dose rates are calculated using the Neutron Fluence-to-Dose Rate Conversion Factors based on ICRP-116 and assuming a neutron current of \SI{2.81e11}{n/s} through the beam port of Channel \#4. Units are \si{\micro Sv/h} in log scale. The five points where the dose rates were measured are also shown.}
    \label{fig:dose-no-roof}
\end{figure}

 \indent The measurement of the dose rates outside the bunker is an important task to guarantee the safety of the people working at the facility during the experiment. The estimation of the same dose rates with a model is needed not only to ensure the validity of the model itself but also to study what the worst case scenarios are and how to prevent them. A preliminary safety assessment with the channel \#4 beam shutter open and reactor power of both 1 and \SI{10}{MW} was performed by the radiation protection group in week 23 of 2023. A temporary lead reflector and paraffin wax shielding  (\cref{fig:dummy_lead}) was mounted in the room, and a temporary ceiling made of borated HDPE sheets was built. Neutron and gamma dose rates were measured at five points outside the boron concrete walls. In all the cases, the values were below the legal requirement of \SI{10}{\micro Sv/h} (see \cref{tab:dose_rates}).
 \begin{table}[!p]
     \centering
     \setlength{\tabcolsep}{10pt} 
     \begin{tabular}{c c c c c c }
     \toprule
       Dose rate [\si{\micro Sv/h}]   & P1 & P2 & P3 & P4 & P5  \\
     \midrule
      Neutron    & 3 & 3 & 8 & 6 & 1.6 \\
      Gamma     & 2 & 4 & 2.9 & 2.1 & 2.5 \\
      \bottomrule
     \end{tabular}
     \caption[Neutron and gamma dose rates]{Neutron and gamma dose rates measured by the reactor radiation protection group in week 23 of 2023. Reactor power was \SI{10}{MW}. Values obtained by private communication.}
     \label{tab:dose_rates}
 \end{table}
 We adapted the previous model by adding the dummy lead reflector and the temporary HDPE roof to compare the calculated dose rates with the measured one. The model and the five points measured are shown in \cref{fig:dose_geo}.\\
\indent The dose rates are calculated using the Neutron Fluence-to-Dose Rate Conversion Factors based on ICRP-116 and assuming a neutron current of \SI{2.81e11}{n/s} through the beam port of Channel \#4 (value calculated with the full reactor model in MCNPX). The resulting dose rate map is shown in \cref{fig:dose-no-roof}. Considering how preliminary these calculations are, we found a good agreement with the measured dose rate, at least in the order of magnitude. It should be noted that it was not possible to compare the gamma dose rate since no gamma source was included in the model. A pessimistic estimation of the gamma flux out of the beam port based on measurements in channel \#2 was proposed as a back-of-the-envelope solution to include gammas in the model. Moreover, it is clear that a dedicated study that measures precisely the flux in the five regions and improves the statistical convergence of the tallies is needed. 
\FloatBarrier
\subsubsection{Activation calculations}
Based on the neutronic model previously described, the activation of the experimental setup components was also studied, in particular the activation of the reflector vessels that have to be exchanged during the experiment to assess the effect of the different reflector materials and of the FZJ cryostat for shipping and future re-use purposes. The calculations were performed at the Oak Ridge National Laboratory.
The following hypotheses were considered:

\vspace{0.2cm}
\begin{itemize}[leftmargin=1cm]
   \item[-]40 hours of operation at \SI{2.81e11}{n/s}, which corresponds to 4-5 measurements per day of 1 to 2 hours
   \item[-]30 minutes, 1 h, 8 h, 1 day, 1 week and 1 month of decay time.
\end{itemize}
\vspace{0.2cm}

The dose rate limits to access the bunker are: 

\vspace{0.2cm}
\begin{itemize}[leftmargin=1cm]
   \item[-]In normal working conditions to access inside the bunker (reactor in operation, channel closed): below 30 $\mu$Sv/h,
   \item[-]Extreme conditions, e.g. for short immediate operations (reactor in operation, channel open): below 500 $\mu$Sv/h – max 1 hour cumulated dose for 2 months span.
\end{itemize}
\vspace{0.2cm}

\begin{figure}[tb!]    
        \centering
    \begin{subfigure}[b]{0.7\textwidth}
        \includegraphics[width=\textwidth]{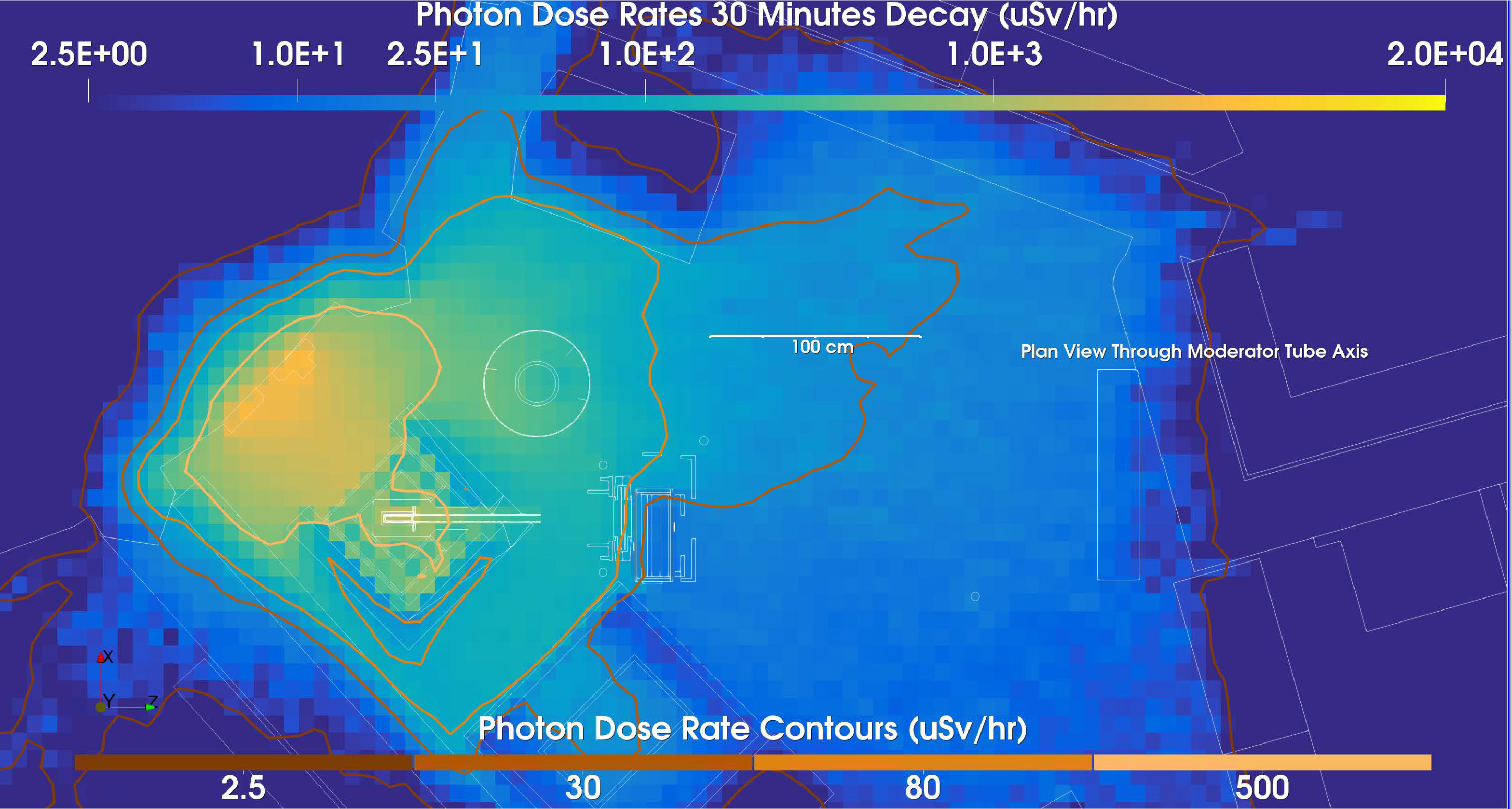}
        \subcaption{}
        \label{fig:Photon_dose_rate_30min}
    \end{subfigure}
    \begin{subfigure}[b]{0.7\textwidth}       
        \includegraphics[width=\textwidth]{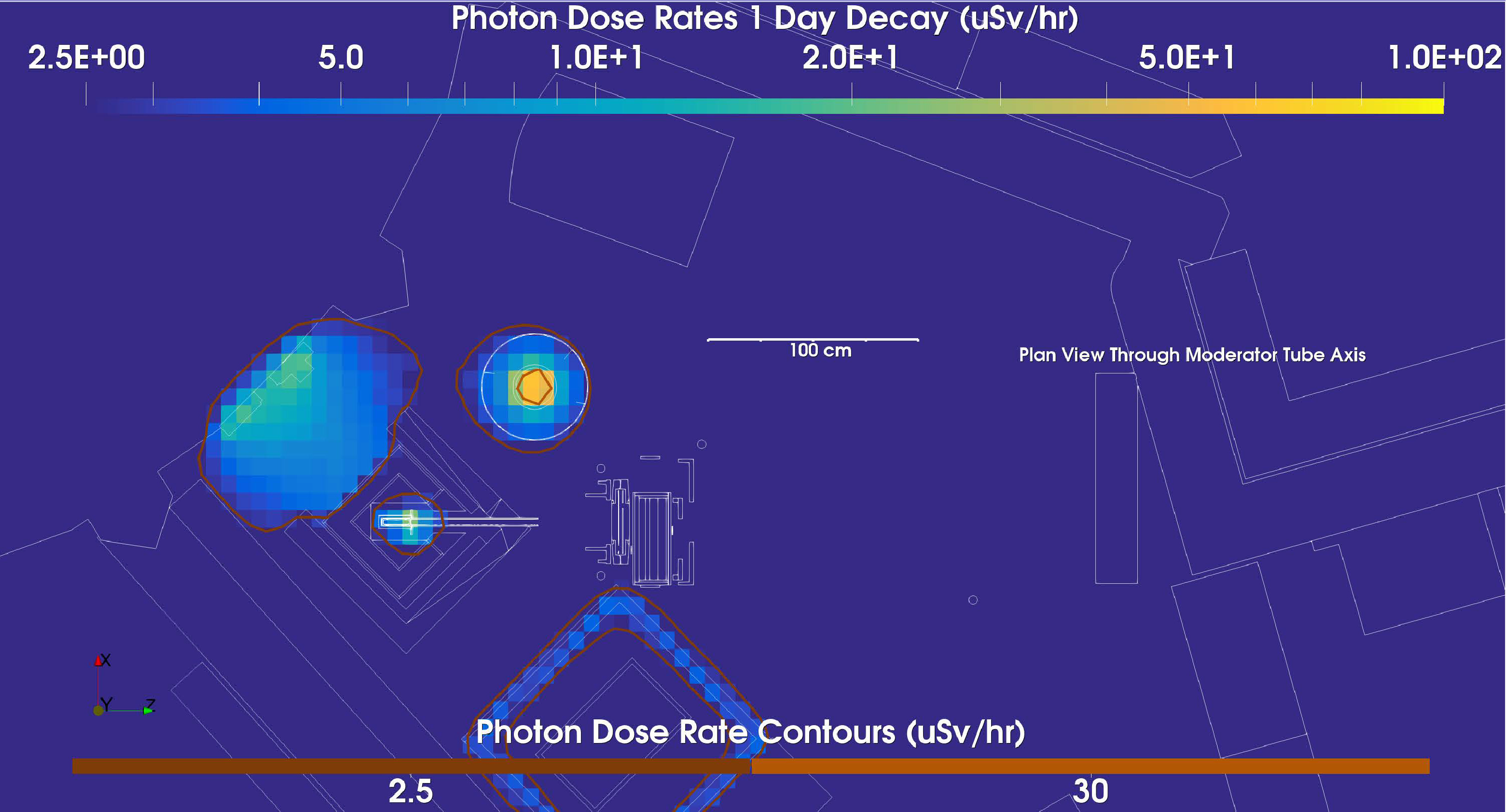}
        \subcaption{}
        \label{fig:Photon_dose_rate_1day}
    \end{subfigure}
    \caption{ Photon dose rates - 30 minutes of decay (a)  (b) 1 day of decay}
    \label{fig:Photon_dose_map}
\end{figure}

Due to issues in using the external supply line of \ce{H_2}, a smaller cryostat than the one initially described, self-supplied, was used for the experiment as described later on in \cref{fig:advrefl_1-9}. The cryostat was modelled according to the CAD description of this component.
The neutron fluxes were volume-averaged over whole parts. The constant 40 hours of operation were assuming that the chopper and the pin-hole were in closed positions. Regarding the materials definitions, a cobalt percentage of 0.05\% in carbon steels and of 0.2\% in stainless steel was considered.
\cref{fig:Photon_dose_map} presents photon dose rate maps through moderator center. After 30 minutes of decay, the limit of 500 $\mu$Sv/h corresponding to the extreme conditions is exceed in the area where the vessel filled with advanced material is located. It can be also seen that the dose rates in the area where the cryostat is located are exceeding the 30 $\mu$Sv/h limit for normal operating conditions. After one day of decay, this dose rate decreases to 2.5 $\mu$Sv/h for both areas of concern. After 8 hours of decay time, the vessels to be exchanged are still in an area where the calculated dose rates are more than 80 $\mu$Sv/h, which tends to indicate that a tool can be needed for remote handling of that component.
 
\begin{figure}[hbt!]
\begin{center}
\includegraphics[width=0.7\textwidth]{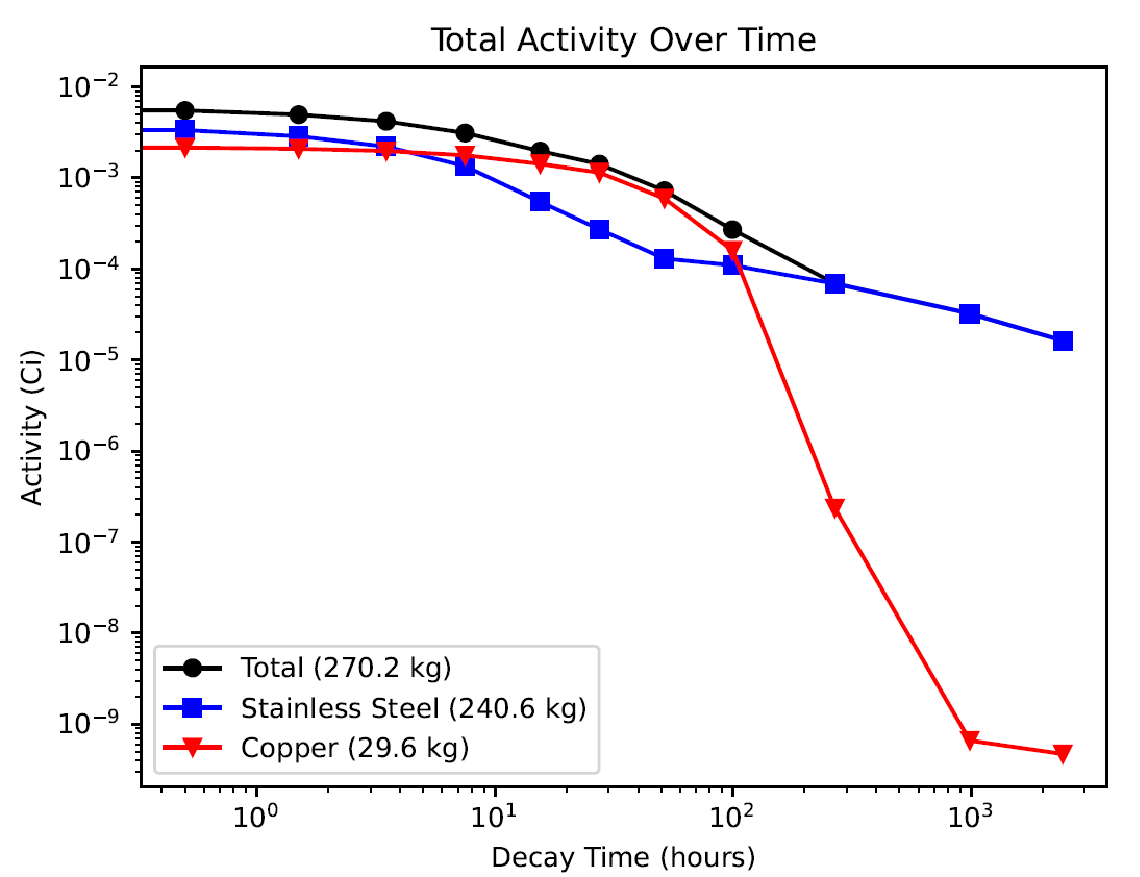}
\caption{Total activity over time of the FZJ cryostat}
\label{fig:Activation_cryostat_total}
\end{center}
\end{figure}

\begin{figure}[tb!]    
        \centering
    \begin{subfigure}[b]{0.5\textwidth}
        \includegraphics[width=\textwidth]{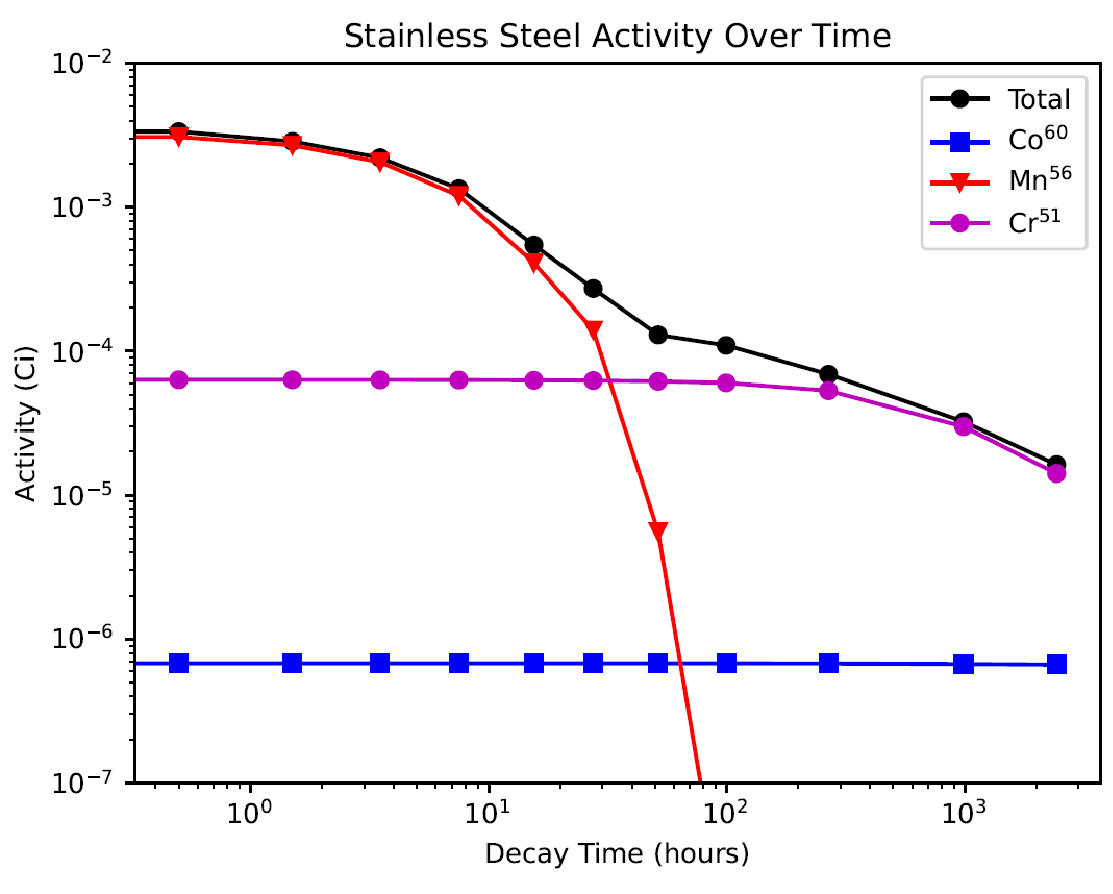}
        \subcaption{}
        \label{fig:Activation_SS}
    \end{subfigure}
    \begin{subfigure}[b]{0.5\textwidth}       
        \includegraphics[width=\textwidth]{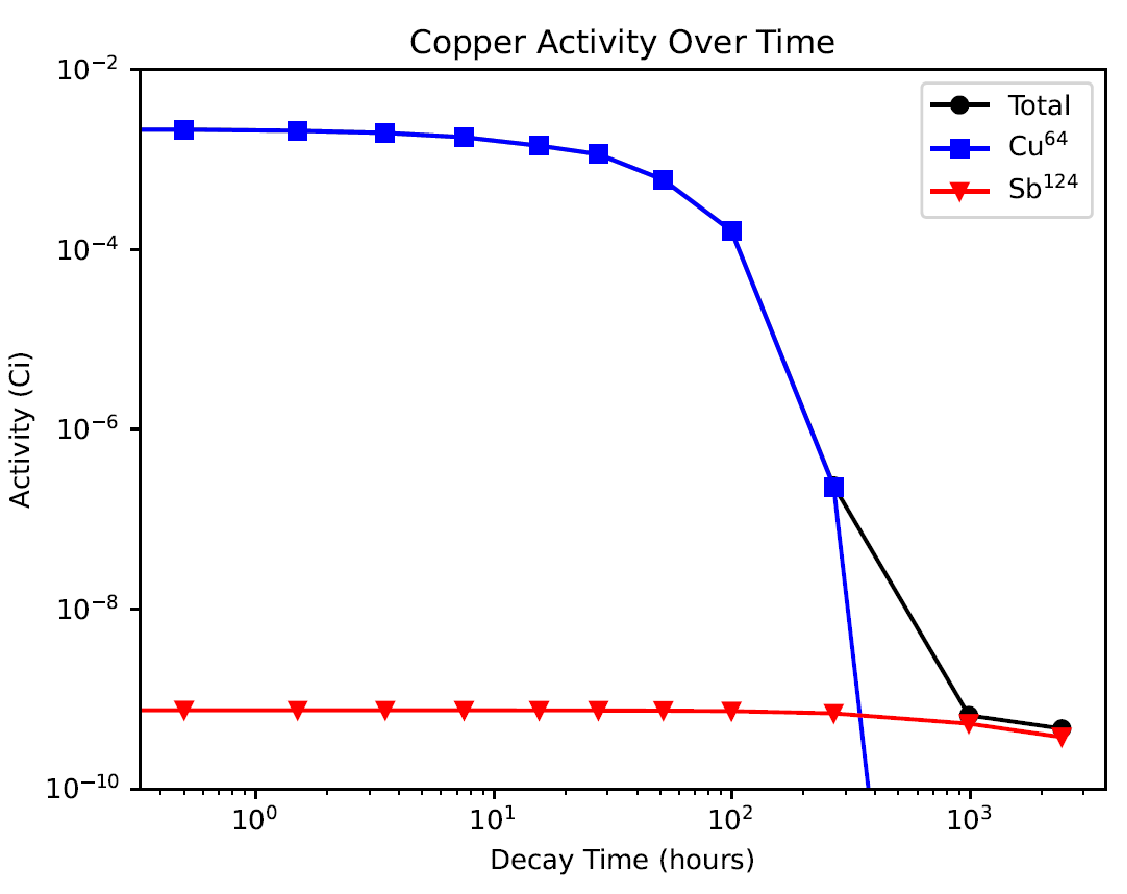}
        \subcaption{}
        \label{fig:Activation_copper}
    \end{subfigure}
    \caption{ (a) Activation of the FZJ cryostat - Contribution of Stainless steel (b) Contribution of copper}
    \label{fig:Activation_materials}
\end{figure}

After 40 hours of irradiation and 1 day of decay, the dose rate on contact at the surface of the cryostat is calculated as 2.5 $\mu$Sv/h. After a week of decay, the dose rate decreases by a factor of 10. 
The internal copper structure was taken into account. \cref{fig:Activation_cryostat_total} gives the total activity as a function of time of the cryostat. 
In stainless steel the activation is mainly dominated by \ce{^{56}Mn} as shown in \cref{fig:Activation_cryostat_total}. The two other main contributors are \ce{^{51}Mn} and \ce{^{60}Co}. In copper, the activation is mainly due to \ce{^{64}Cu} as reported in \cref{fig:Activation_materials}.
It should be noticed that this model is not taking into account any shielding around nor in the collimator.

\FloatBarrier
\subsection{Engineering of the experimental setup}
\begin{figure}[hbt!]
\begin{center}
\includegraphics[width=0.9\textwidth]{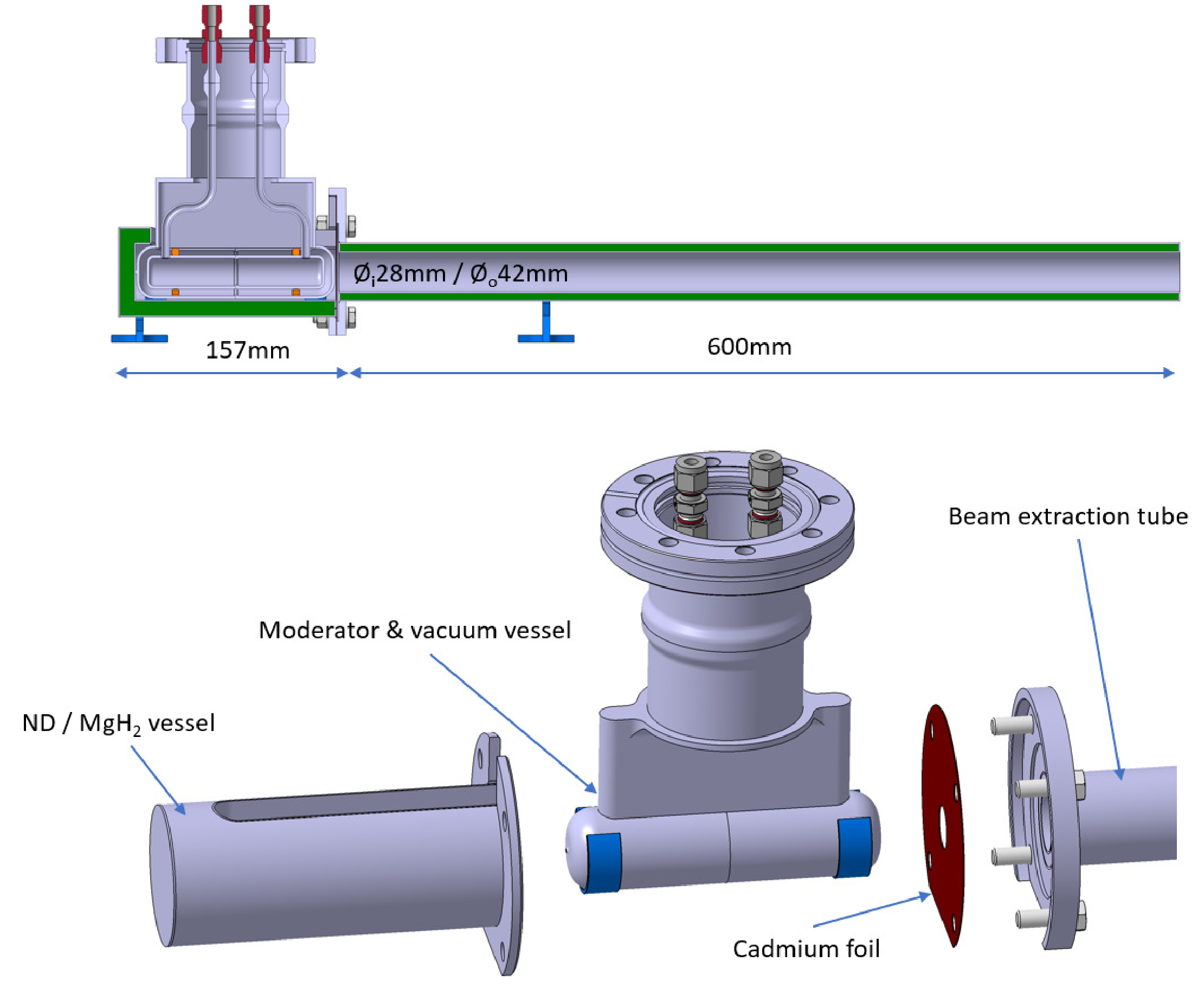}
\caption{Experimental moderator/reflector/guide setup with interchangeable configurations placed inside the neutron beam.}
\label{fig:advrefl_1-1}
\end{center}
\end{figure}

The engineering team (WP5) designed a moderator \& reflector unit, surrounded by advanced reflector materials as well as an associated beam extraction channel based on advanced reflector materials for the BNC experiment.
%In particular, the reflector and the extraction system was tested with the advanced reflector materials of nanodiamonds (ND) and magnesium hydride (MgH$_{2}$) at the moderator test facility of the Budapest Neutron Centre (BNC) in Hungary in late September 2023.%
To be able to quantify the effect of the advanced reflectors, a control measurement with an empty reflector vessel should be carried out first as reference. This initial measurement can then be followed by measurements performed with ND and MgH$_{2}$ vessel respectively. 
%The experimental facility and the planned test are described in all details in the Deliverable 6.2.%needs citation, see D4.1 or D4.2 in the BIB.
%"Neutronic design of the ESS beam extraction system and for experimental test".
Those interchangeable vessels filled with \SI{10}{mm} thick layers of ND or MgH$_{2}$ powder have to be placed around the cold para-hydrogen moderator, which is fed by neutrons of the Budapest research reactor. These reflectors can be exchanged to perform measurements with different configurations, shown in \cref{fig:advrefl_1-1}. The moderated neutrons of the moderator/reflector assembly are led into a \SI{600}{mm} long beam-tube, surrounded by a \SI{5}{mm} thick ND powder layer to extract the cold neutrons from the moderator. The spectra of the moderator/reflector/guide assembly are measured by a pin-hole (``camera obscura'') time-of-flight (TOF) setup. All structures which are placed in the neutron beam are manufactured of aluminum to reduce activation and allow fast exchanging of the reflector vessels. The H$_{2}$ supply lines are led into the cryostat system outside the bunker, where the LH$_{2}$ liquefaction takes place and a high para-H$_{2}$ concentration will be generated. %\cite{Santoro:2022tvi}

\subsubsection{Assembly and welding of the components}
\begin{figure}[hbt!]
\begin{center}
\includegraphics[width=0.9\textwidth]{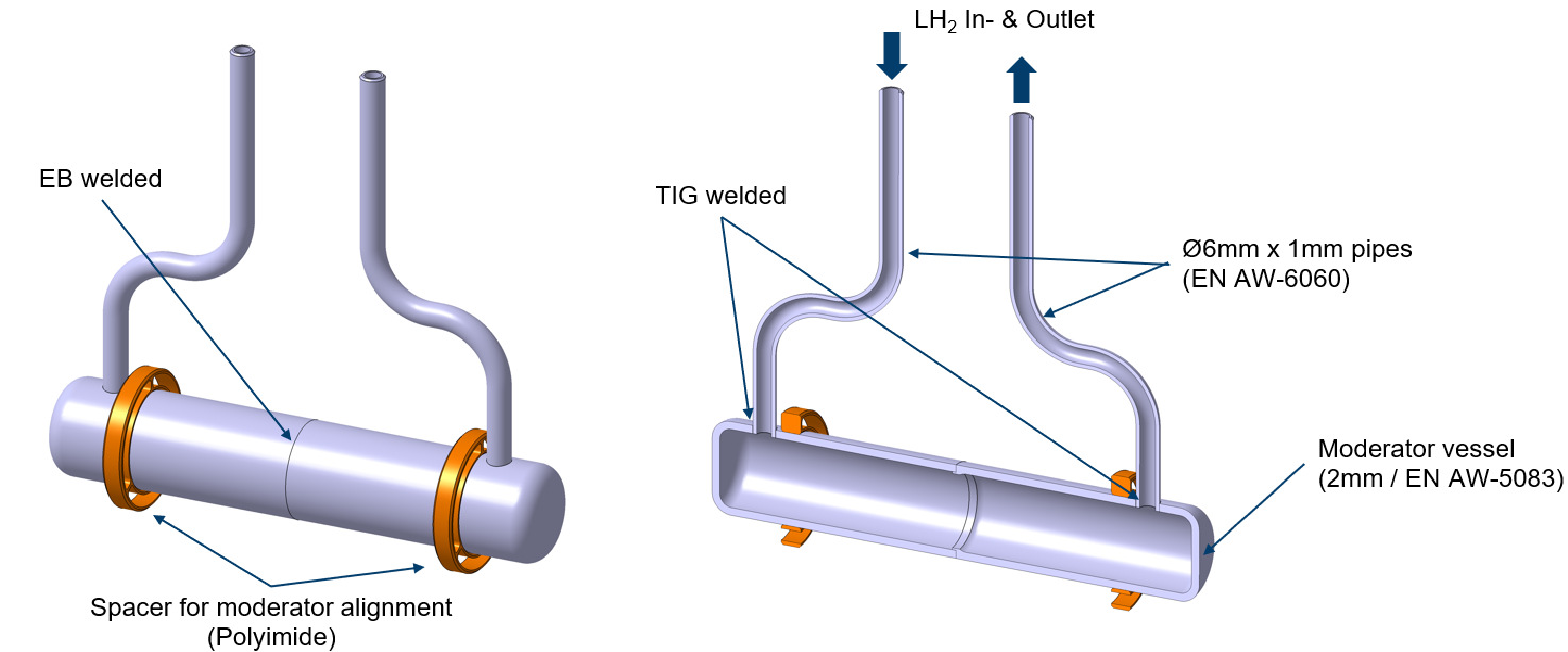}
\caption{Design of the LH$_{2}$ moderator.}
\label{fig:advrefl_1-2}
\end{center}
\end{figure}

The moderator itself consists of two half-tubes with back walls, which are milled from one piece of aluminum EN AW-5083 each. They are joined by EB welding to minimize the heat input during the process. This reduces bending and distortion issues during the welding process significantly. After the welding, spacer rings made of polyimide are put onto the vessel, to ensure proper alignment of the moderator in the surrounding vacuum vessel. Polyimide is chosen as material because of its low heat conductivity while being resistant to the low temperatures the moderator will reach, as it uses LH$_{2}$ as the moderating medium at a temperature of around \SI{20}{K}. This reduces the required cooling power of the cryostat, because it reduces thermal losses due to thermal insulation. In the next step, the supply pipes for the LH$_{2}$ are welded on by tungsten inert gas welding (TIG) (\cref{fig:advrefl_1-2} \& \cref{fig:advrefl_1-4} left).

\FloatBarrier

\begin{figure}[hbt!]
\begin{center}
\includegraphics[width=0.9\textwidth]{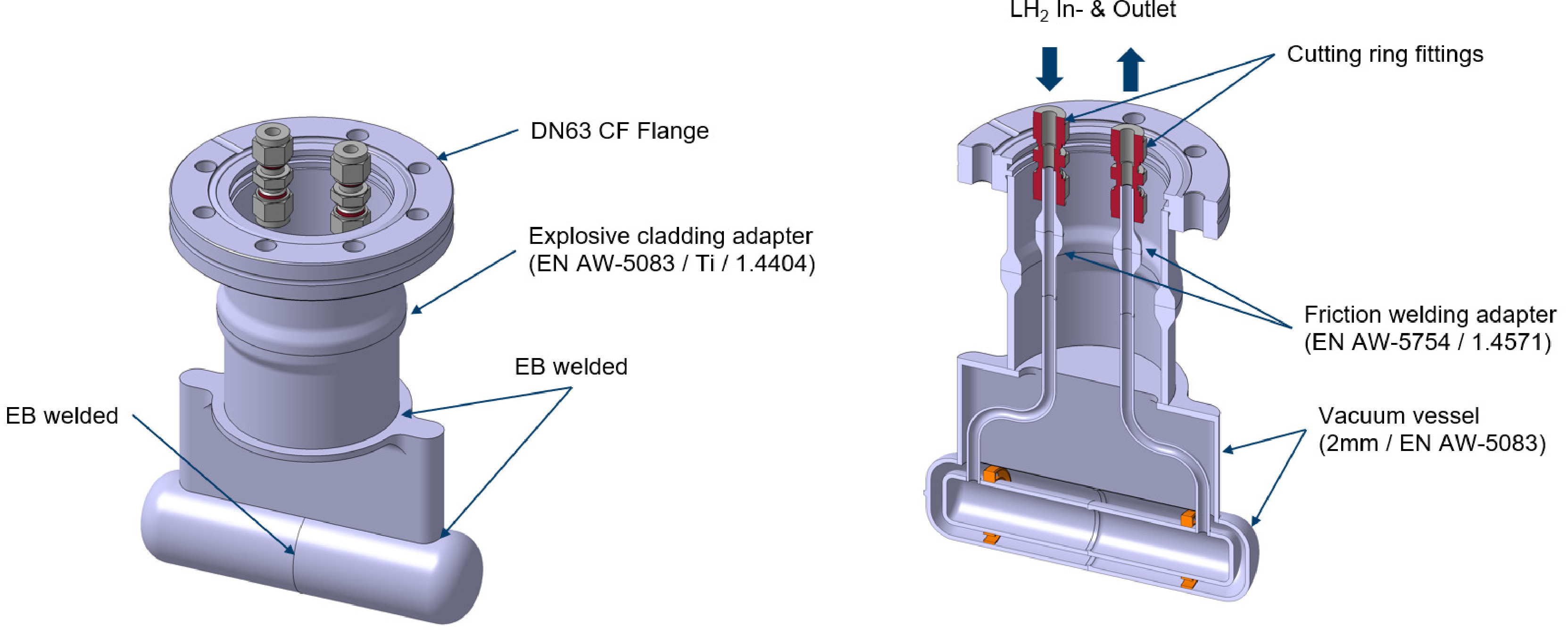}
\caption{Design of the vacuum vessel.}
\label{fig:advrefl_1-3}
\end{center}
\end{figure}

\begin{figure}[hbt!]
\begin{center}
\includegraphics[width=0.9\textwidth]{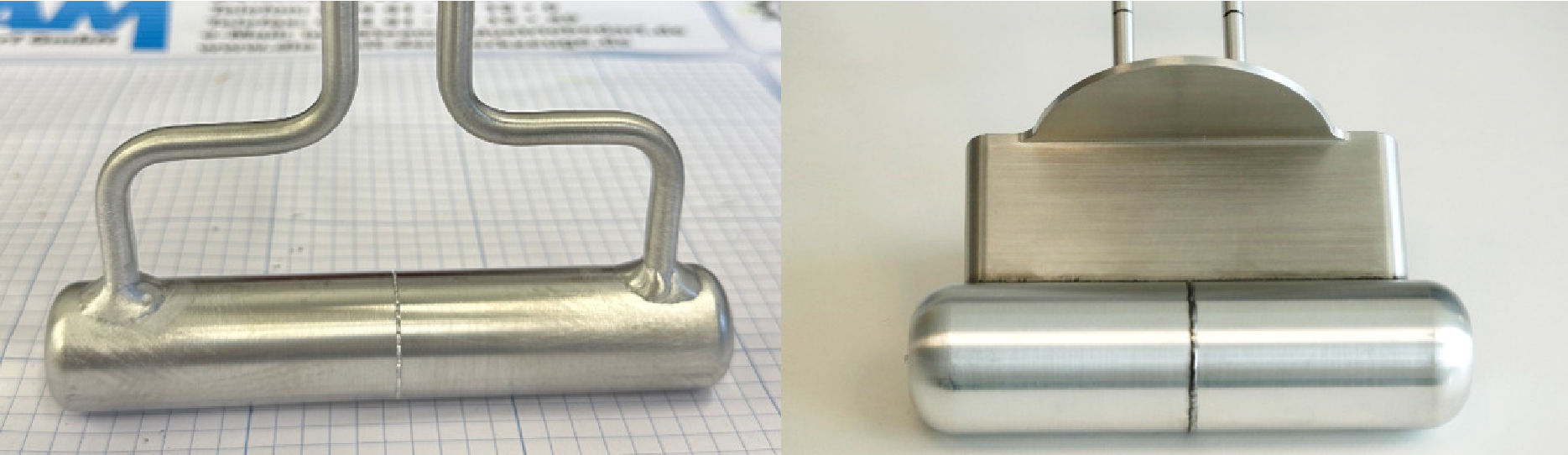}
\caption{Manufactured cold moderator (left: hydrogen vessel, right: vacuum jacket).}
\label{fig:advrefl_1-4}
\end{center}
\end{figure}

The vacuum vessel shown in \cref{fig:advrefl_1-3} and \cref{fig:advrefl_1-4} (right) ensures the insulation vacuum for the cryogenic parts of moderator system. This is necessary, because it affects the required cooling power that is needed for the low temperature of around \SI{20}{K} of the moderating medium LH$_{2}$. The jacket surrounds the moderator and the supply pipes. It is also joined by EB welding, for minimal heat input and less distortion when welding. There are two interfaces in this vacuum vessel. One is the connection of the LH$_{2}$ supply pipes and the other is the vacuum connection flange. On both interfaces, a transition from aluminum to stainless steel is needed, because aluminum is too soft to ensure a re-useable connection. Therefore, at the pipes, a friction welding adapter is used, so that the cutting ring fitting for the connection to the transfer-lines can be adapted to the stainless-steel end of the adapter. The CF-flange at the outer vacuum jacket is also made of stainless-steel. This transition from aluminum to stainless steel is solved with an explosive cladding adapter, which has a titanium (Ti) layer in between. This material mix is due to the cladding procedure via explosion itself and has no special reason for the usage of the adapter.

\FloatBarrier

\FloatBarrier

\begin{figure}[hbt!]
\begin{center}
\includegraphics[width=0.9\textwidth]{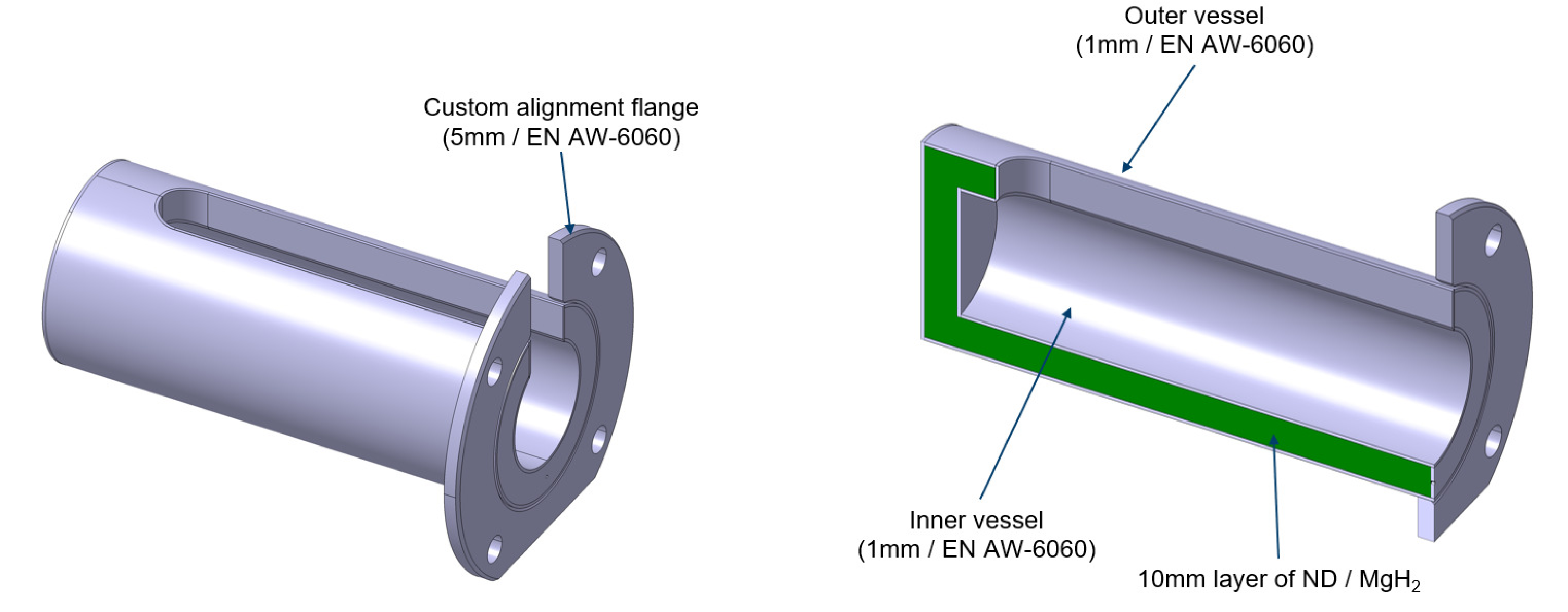}
\caption{Design of ND \& MgH$_{2}$ vessel.}
\label{fig:advrefl_1-6}
\end{center}
\end{figure}

The ND \& MgH$_{2}$ vessels are basically made of an inner and an outer part, as it is shown in \cref{fig:advrefl_1-6}. The inner and outer tubes are made of aluminum EN AW-6060 and are both closed with a disc also made of EN AW-6060, which is EB welded, due to the small wall-thicknesses of \SI{1}{mm}. They form a concentric vessel with a \SI{10}{mm} overall layer in between, that has been filled with either ND or MgH$_{2}$. The packing density of the ND vessels is \SI{0.25}{g/cm^3} and for MgH$_{2}$ \SI{0.60}{g/cm^3}.

%\begin{figure}[hbt!]
%\begin{center}
%\includegraphics[width=0.7\textwidth]{advrefl_WP5/1-5.eps}
%\caption{Filling of cap with ND.}
%\label{fig:advrefl_1-5}
%\label{fig:ND_cup}
%\end{center}
%\end{figure}

For a control measurement, one of the three vessel assemblies will stay empty. After the filling process, the vessels are closed with a ring-shaped disc by EB welding. To be able to connect the assembly to the beam extraction channel, a custom flange is tack welded onto the outer vessel part by laser beam welding.

\begin{figure}[hbt!]
\begin{center}
\includegraphics[width=0.7\textwidth]{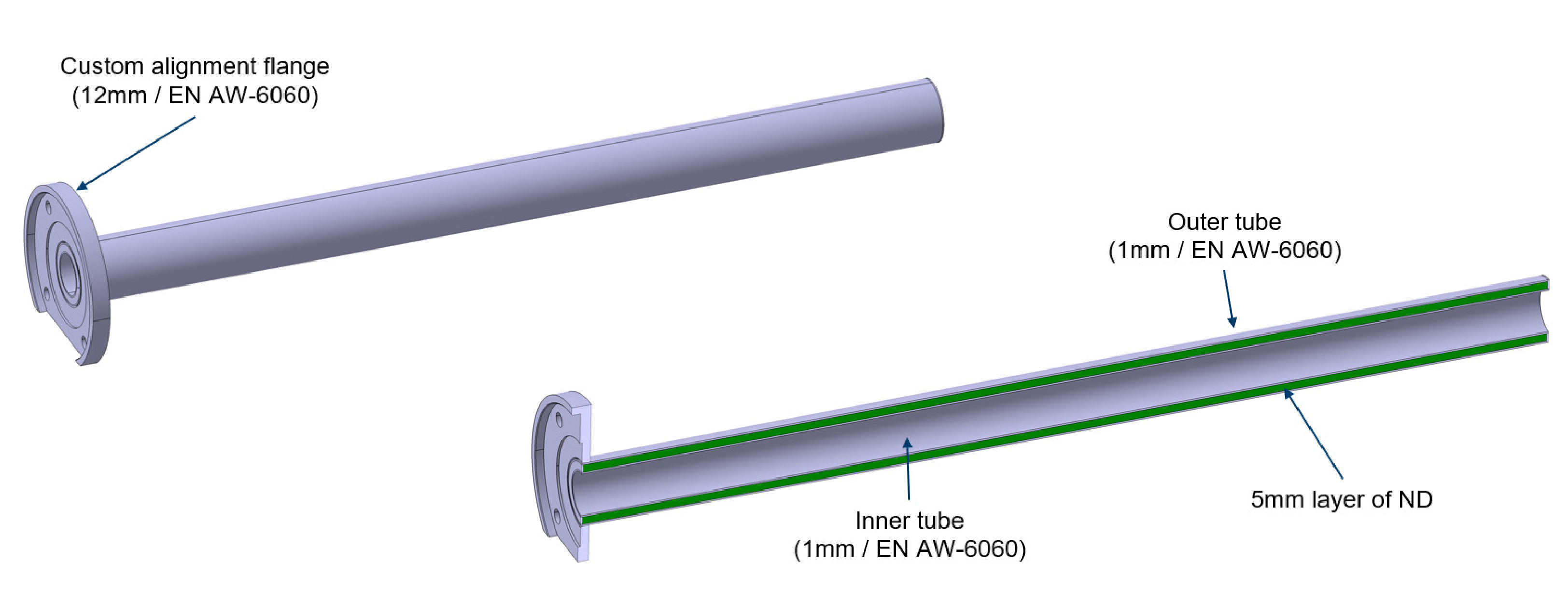}
\caption{Design of the ND beam extraction channel.}
\label{fig:advrefl_1-7}
\end{center}
\end{figure}

\begin{figure}[hbt!]
\begin{center}
\includegraphics[width=0.9\textwidth]{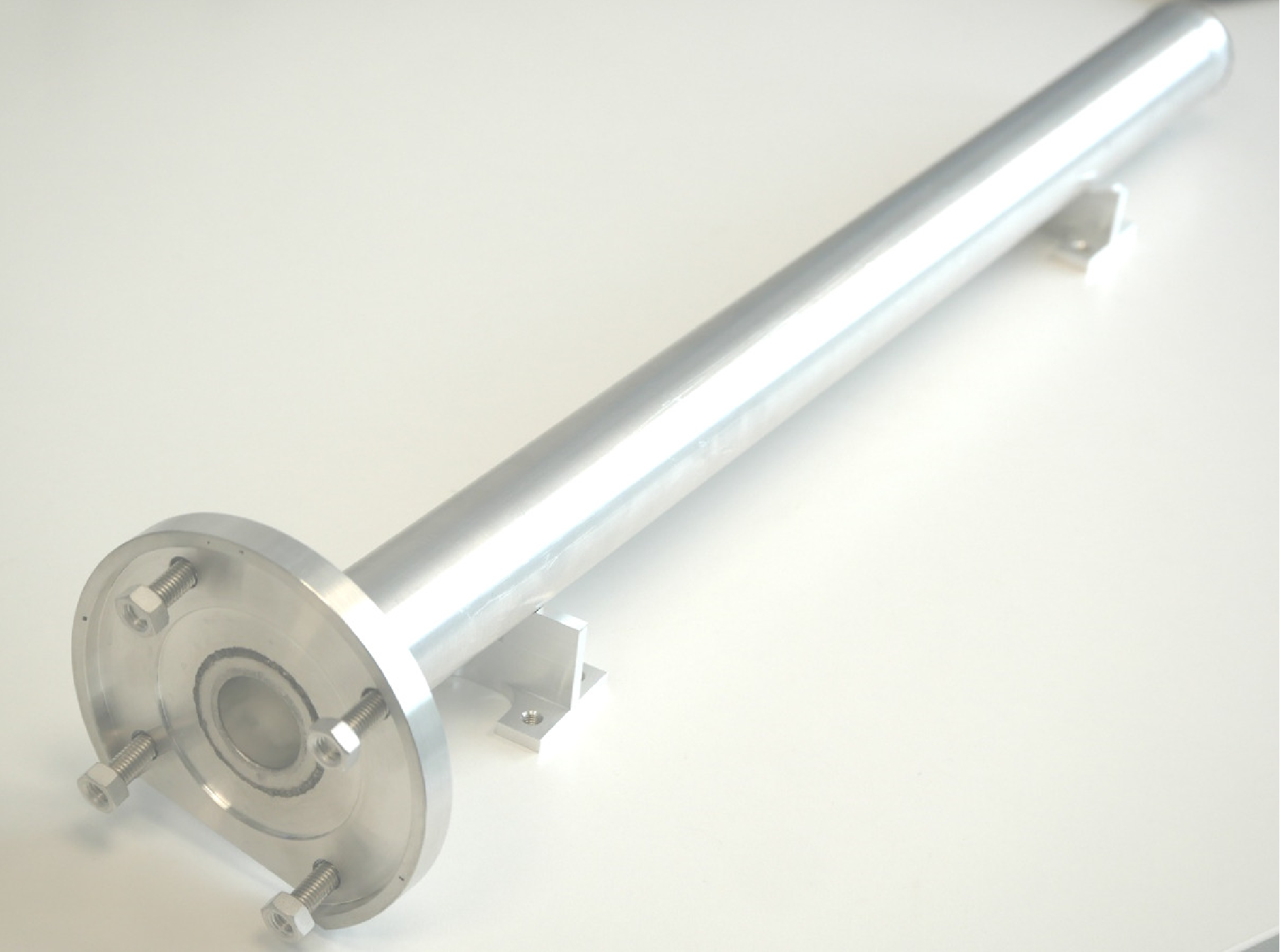}
\caption{Manufactured ND beam extraction channel.}
\label{fig:advrefl_1-8}
\end{center}
\end{figure}

The ND beam extraction channel in \cref{fig:advrefl_1-7} \& \cref{fig:advrefl_1-8} is generally of a very similar design, as the vessels described before in \cref{fig:advrefl_1-6}. It consists of an inner and an outer tube, which are in a first step closed from one side with a ring-shaped disc. They form a concentric vessel with a \SI{5}{mm} layer in between, which is filled with ND. The filling was carried out under a fume hood while the tube/vessel was placed on a shaker. After the filling process, the vessel is closed with a second ring-shaped disc. All parts are made of EN AW-6060 and for the joining process, EB-welding is used again, due to the small wall-thicknesses of \SI{1}{mm}. To be able to connect the assembly to the ND- \& MgH$_{2}$-vessels, a custom flange is tack welded onto the outer tube by laser beam welding.

\FloatBarrier

\FloatBarrier

\subsubsection{Assembly of the moderator \& beam extraction system into the test facility}

\begin{figure}[hbt!]
\begin{center}
\includegraphics[width=0.5\textwidth]{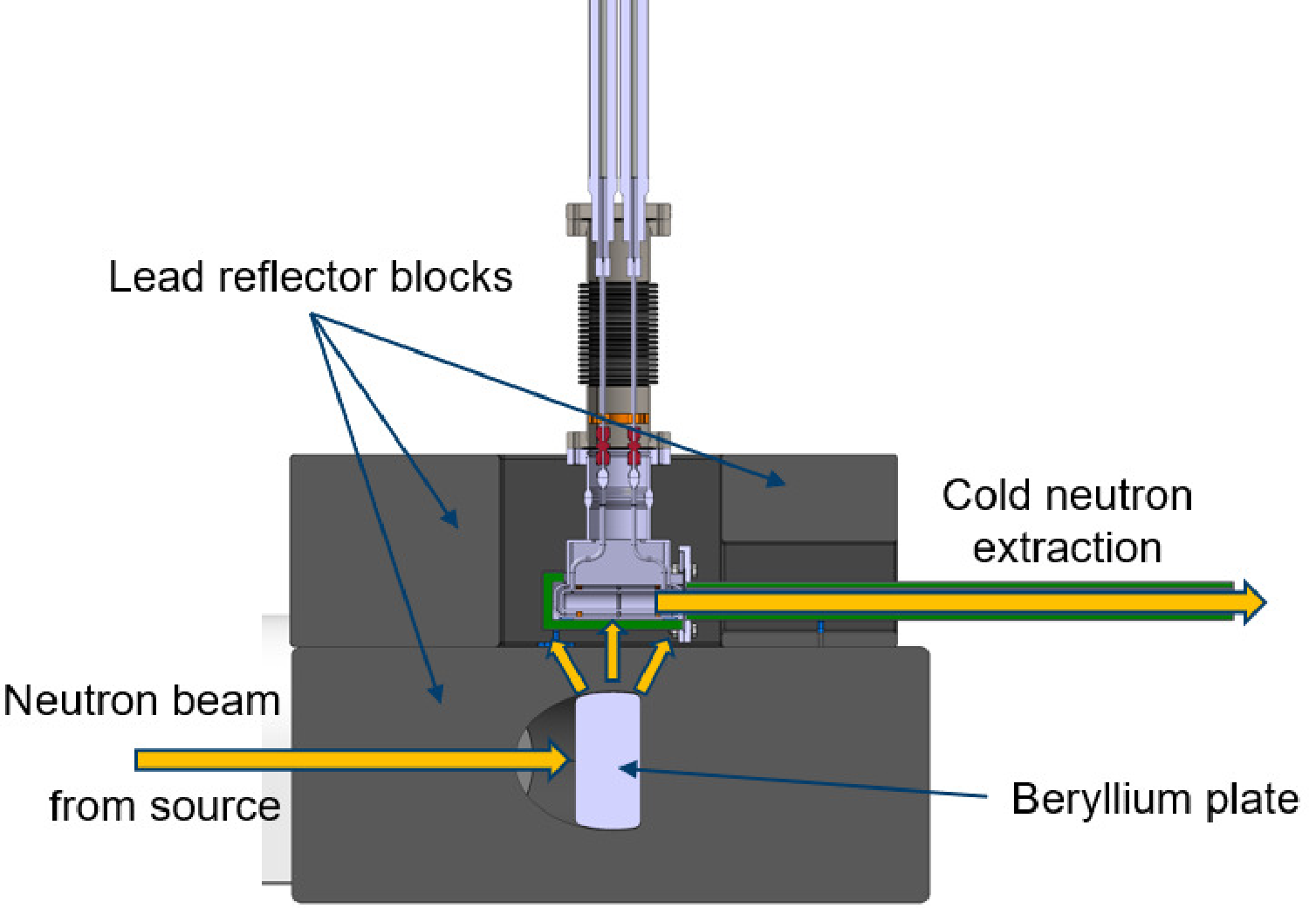}
\caption{Integration into moderator test facility}
\label{fig:advrefl_1-9}
\end{center}
\end{figure}

\cref{fig:advrefl_1-9} shows a side view cut-section of the moderator test facility at the Budapest Neutron Centre (BNC) with the integrated moderator and beam extraction systems. 

\FloatBarrier

\begin{figure}[hbt!]
\begin{center}
\includegraphics[width=0.8\textwidth]{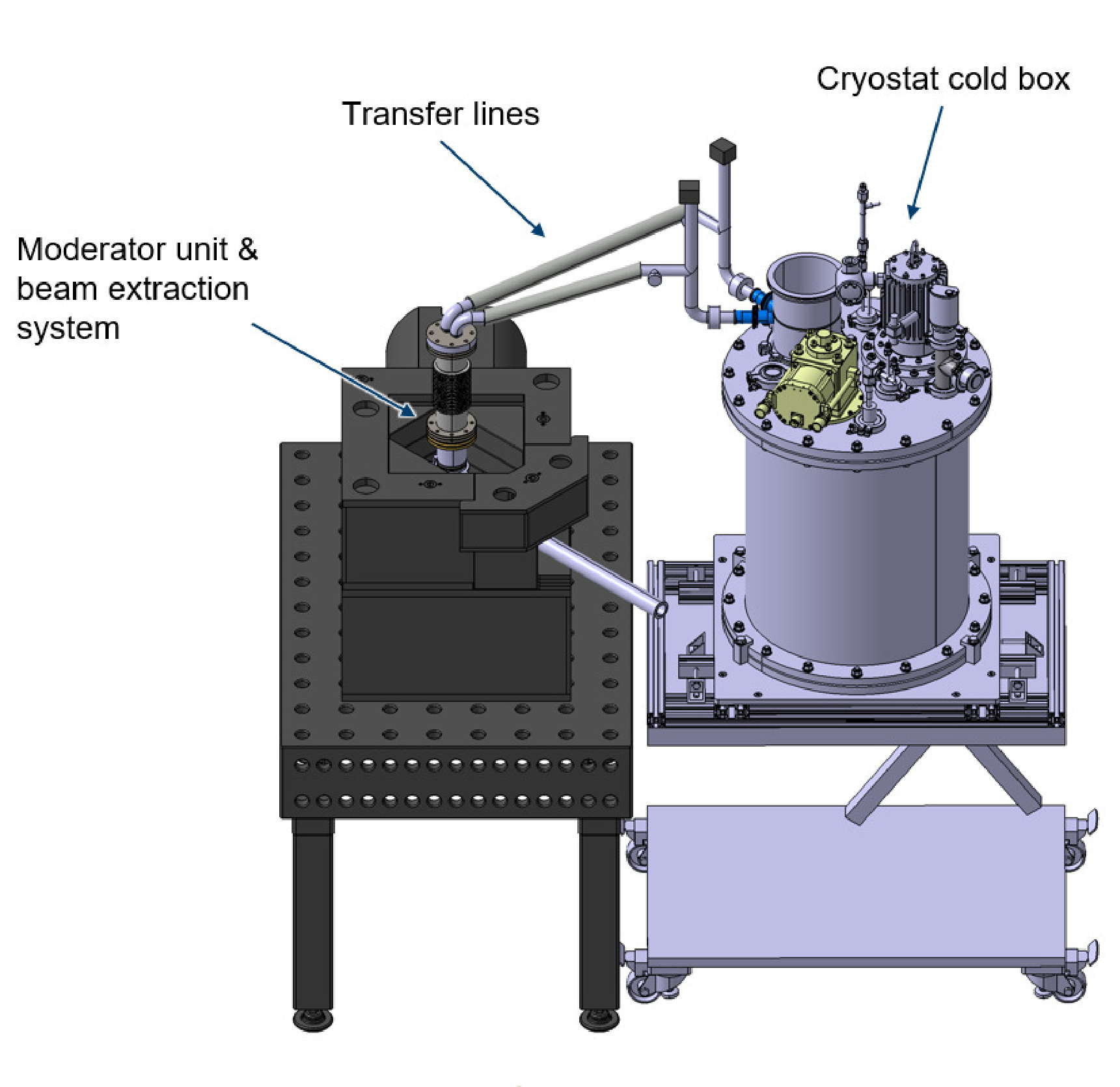}
\caption{Moderator system with cryostat cold box}
\label{fig:advrefl_1-10}
\end{center}
\end{figure}

\begin{figure}[hbt!]
\begin{center}
\includegraphics[width=0.8\textwidth]{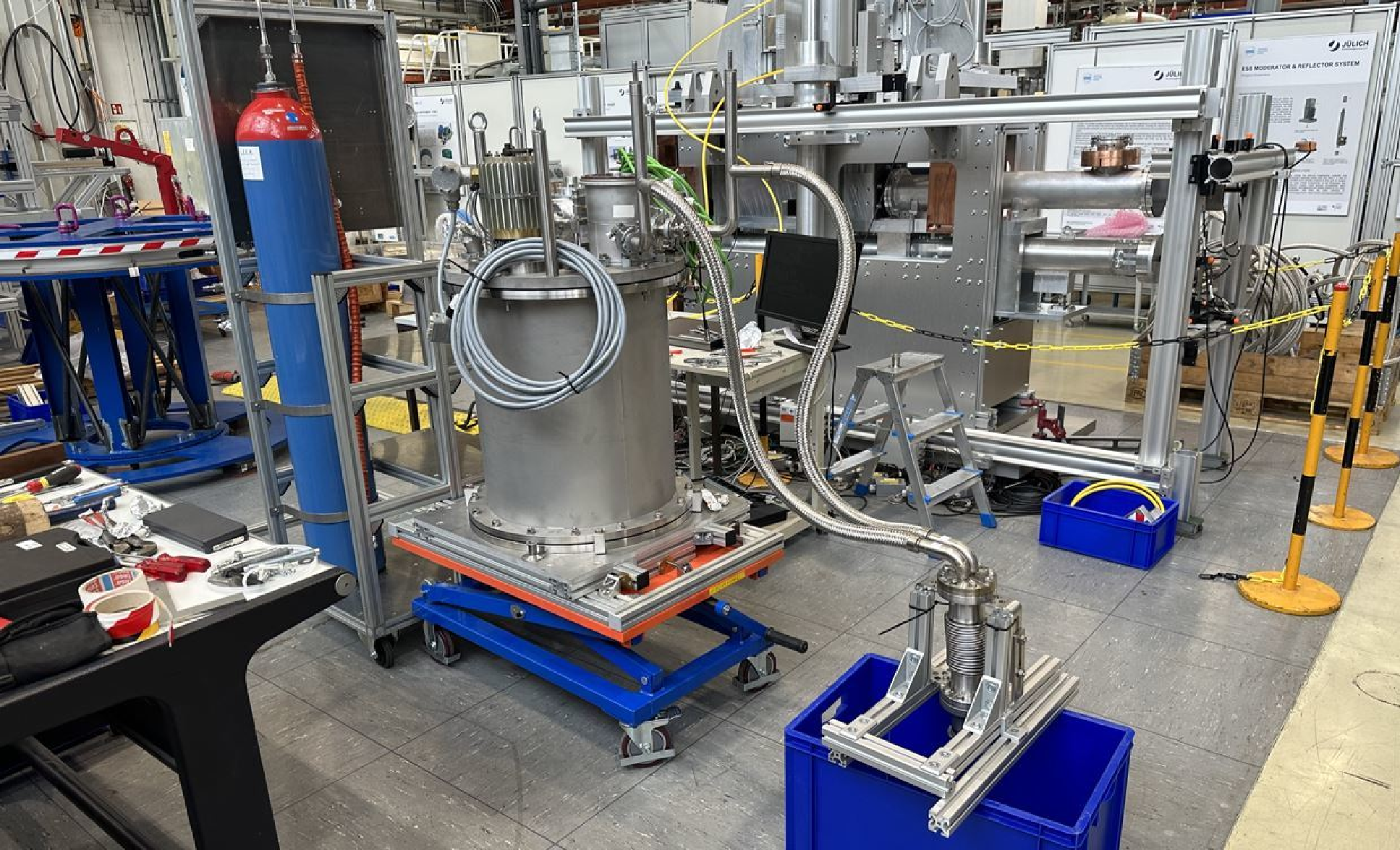}
\caption{Assembled cold moderator system}
\label{fig:advrefl_1-11}
\end{center}
\end{figure}

\cref{fig:advrefl_1-10} shows the moderator unit and the beam extraction system, located in the Budapest Neutron Centre (BNC) moderator test facility setup. \cref{fig:advrefl_1-11} is showing a view of the cryostat. The vacuum vessel surrounding the cold moderator and the LH$_{2}$ supply pipes are connected via vacuum insulated transfer lines with the cryostat cold box. This cryostat is also made and owned by Forschungszentrum J{\"u}lich and was used to perform the experiment. 
It is needed to supply the cold moderator with LH$_{2}$ in a closed loop using a circulation pump. A cold head with \SI{40}{W} @ \SI{20}{K} is used to liquefy and to cool down the hydrogen to \SI{20}{K} inside the cryostat via a heat exchanger. In addition, a catalyst ensures that the moderator is supplied with almost 100 \% para-hydrogen. The concentration ratio is monitored using Raman spectroscopy. It also supplies the insulation vacuum, which is necessary to maintain the cold temperatures inside the transfer lines and the vacuum jacket.

%\begin{figure}[hbt!]
%\begin{center}
%\includegraphics[width=0.9\textwidth]&{advrefl_WP5/1-12.eps}
%\caption{Bunker layout and component arrangement at the Budapest research reactor}
%\label{fig:advrefl_1-12}
%\end{center}
%\end{figure}

\FloatBarrier
\subsubsection{Results and discussion}
\begin{figure}[htb]
    \centering
    \includegraphics[width=0.4\textwidth]{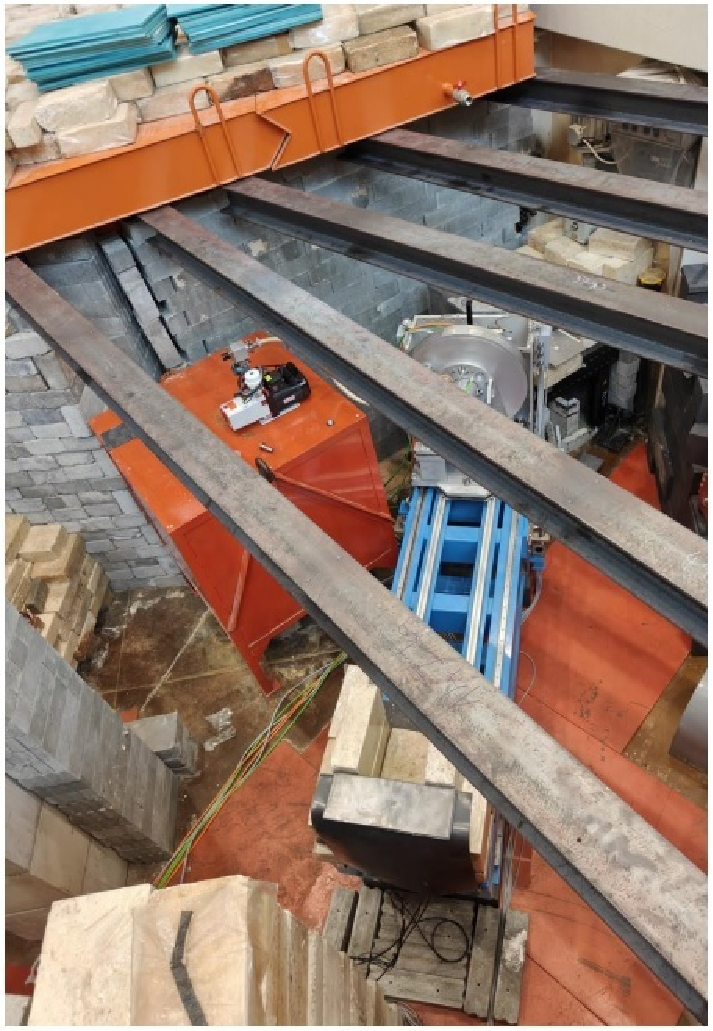}
    \caption{ CMTF beamline installed in bunker- June 2023, Courtesy of L. Rosta}
    \label{fig:Budapest1}
\end{figure}

\begin{figure}[tb!]    
        \centering
    \begin{subfigure}[b]{0.42\textwidth}
        \includegraphics[width=\textwidth]{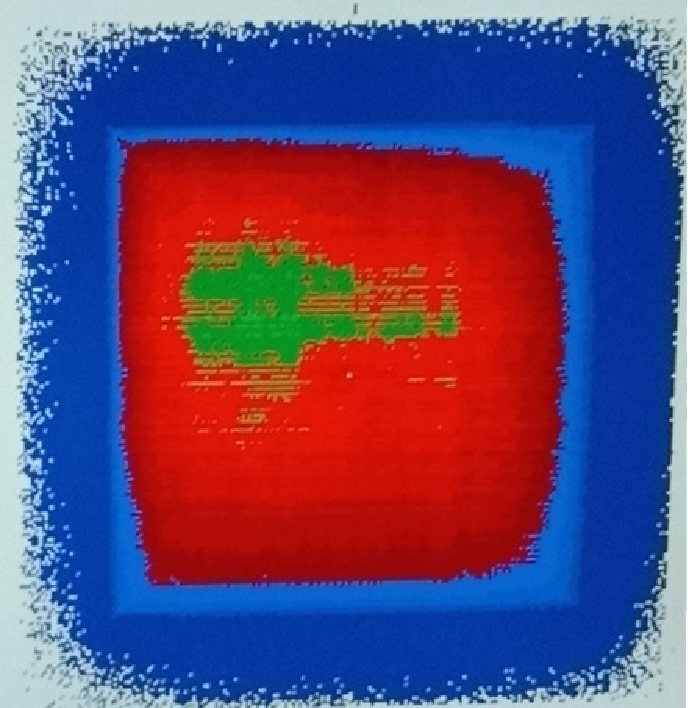}
        \subcaption{}
        \label{fig:2D_image}
    \end{subfigure}
    \begin{subfigure}[b]{0.42\textwidth}       
        \includegraphics[width=\textwidth]{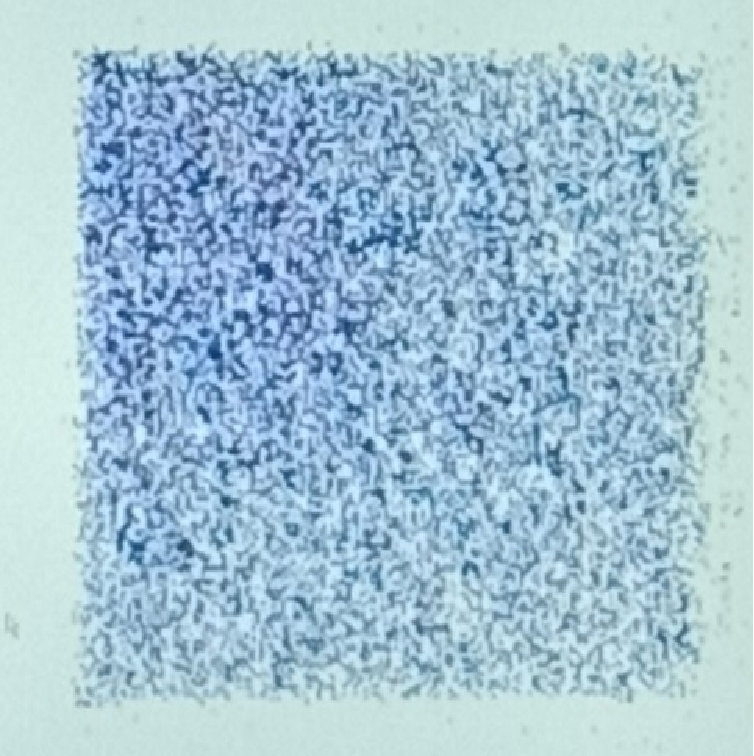}
        \subcaption{}
        \label{fig:Background_BNC}
    \end{subfigure}
    \caption{ (a) 2D image on the He-3 detector from the moderator at 10 MW (b) Background at closed beam.} %Courtesy of L. Rosta}
    \label{fig:Budapest2}
\end{figure}

\begin{figure}[tb!]    
        \centering
    \begin{subfigure}[b]{0.6\textwidth}
        \includegraphics[width=\textwidth]{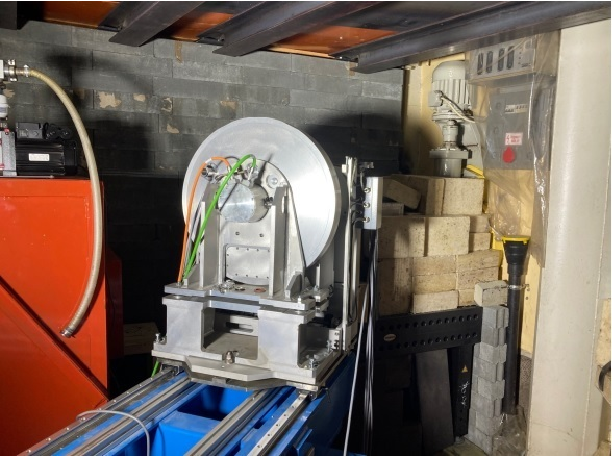}
        \subcaption{}
        \label{fig:shielding_week1_2_a}
    \end{subfigure}
    \begin{subfigure}[b]{0.6\textwidth}       
        \includegraphics[width=\textwidth]{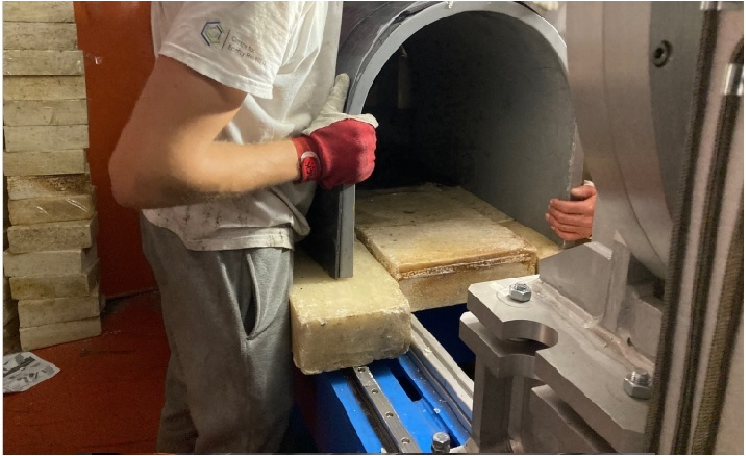}
        \subcaption{}
        \label{fig:shielding_week1_2_b}
    \end{subfigure}
    \caption{ (a) Paraffin blocks shielding (b) Shielding on the detector table.}
    \label{fig:Budapest_week1}
\end{figure}

In preparation for experiments to be performed using the channel \#4 beam line at the BNC reactor, the radiation protection environment was successfully tested on June 5$^\text{th}$ 2023 through neutron and gamma dose rates measurements performed at 1 MW reactor power at the channel outlet, near the reflector and outside the bunker area of the channel. The low level of the background allowed for a channel opening at 10 MW. The outside bunker measurements were found to be within the local regulation limits and gave information on how the shielding should be reinforced. In order to benefit from feedback for the future HighNESS measurements, WP4 and WP6 did participate in the setup of the  experimental campaign dedicated to the test of a low dimensional 
%water-filled 
moderator performed by the Mirrotron company from June 12$^\text{th}$ to June 23$^\text{rd}$ 2023. On June 23$^\text{rd}$ 2023, the CMTF installation was completed: channel-collimator, reflector, Be, outlet collimator, pin-hole system were fully installed and all components (pin-hole slit system, choppers and the detector) were working as shown in \cref{fig:Budapest1}. 
The roof on top of the experimental room was also completed. A 2D image from the moderator via the pin-hole system was recorded (see \cref{fig:Budapest2}). No TOF analysis was performed due to time limitation as the reactor had to stop but a fundamental milestone was reached at this point, showing a proof that the beamline can be operated safely with all the components installed, according to the design documents and the safety analysis report. First neutron measurements were performed at 10 MW reactor power with the completed assembly of the test facility. 
The HighNess measurement campaign took place from September 18$^\text{th}$ to September 29$^\text{th}$. The roof on top of the experiment room was completed and closed on and the beam was opened on September 21$^\text{st}$, allowing for first measurements. In particular, the dose rate measurements confirmed the results given by the activation calculations previously performed at the Oak Ridge National Laboratory. Considering the shutter closed, the dose rate recorded was 80 $\mu$Sv/h at the moderator position and 400 $\mu$Sv/h at the reflector surface, at shutdown after 1 hour of irradiation, decreasing to the level of 10 $\mu$Sv/h, which is below the safety limit, after few hours of cooling time. The measurement performed during the first week of the campaign clearly confirmed the need of reducing the background.  As illustrated in \cref{fig:Budapest_week1}, paraffin blocks were added to surround the collimator and the lead shielding encapsulating the experimental set-up. A tunnel made of Mirrobor\texttrademark\ plates (natural rubber and boron powder based flexible plates) was installed on the detector table, also combined with paraffin blocks.

Placing a layer of Mirrobor\texttrademark\ between the chopper and the detector led to a count rate decrease of 10\%, highlighting the fact that the background was mainly due to fast neutrons. 

\begin{figure}[tb!]    
        \centering
    \begin{subfigure}[b]{0.5\textwidth}
        \includegraphics[width=\textwidth]{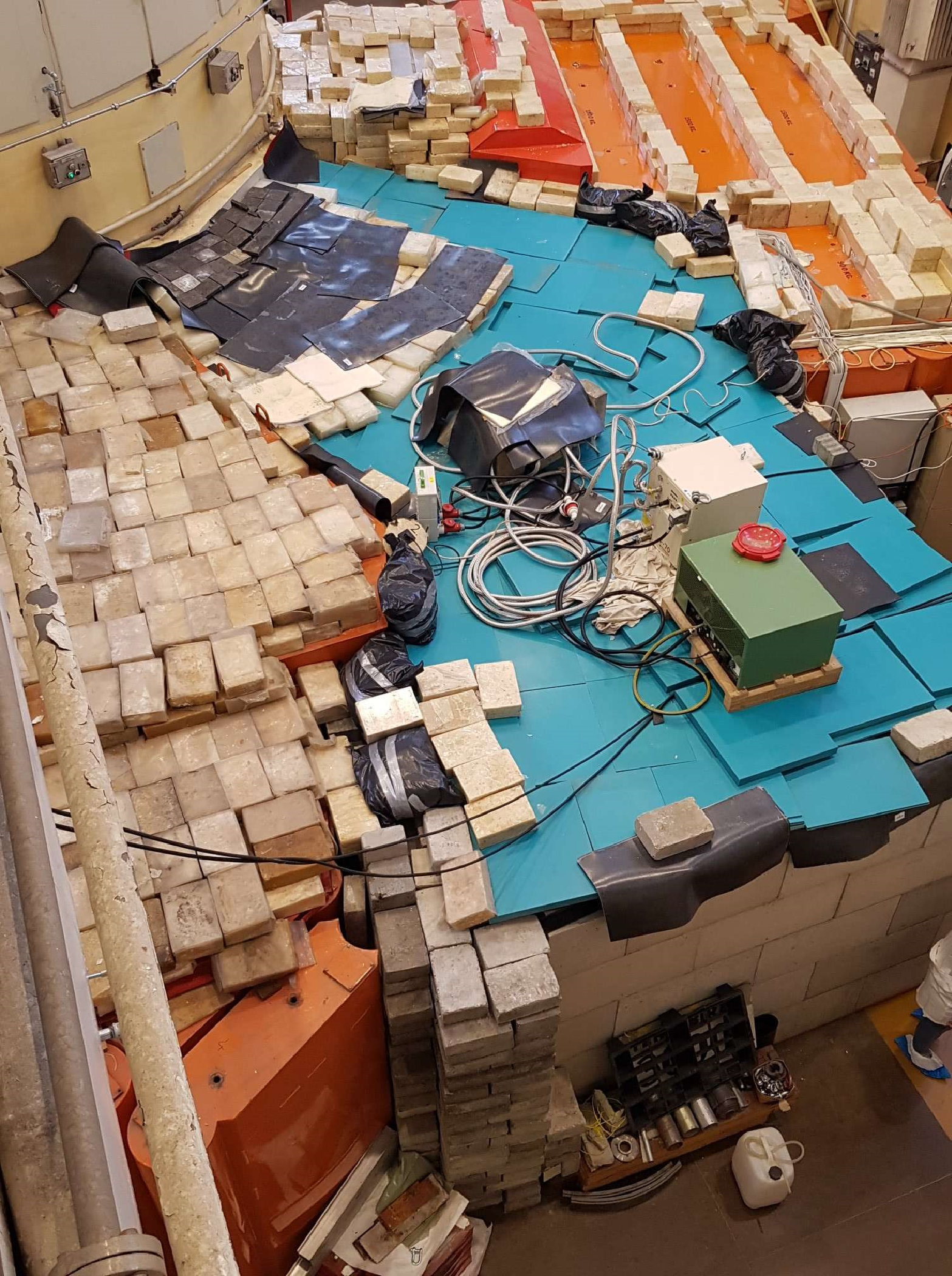}
        \subcaption{}
        \label{fig:Compressor}
    \end{subfigure}
    \begin{subfigure}[b]{0.5\textwidth}       
        \includegraphics[width=\textwidth]{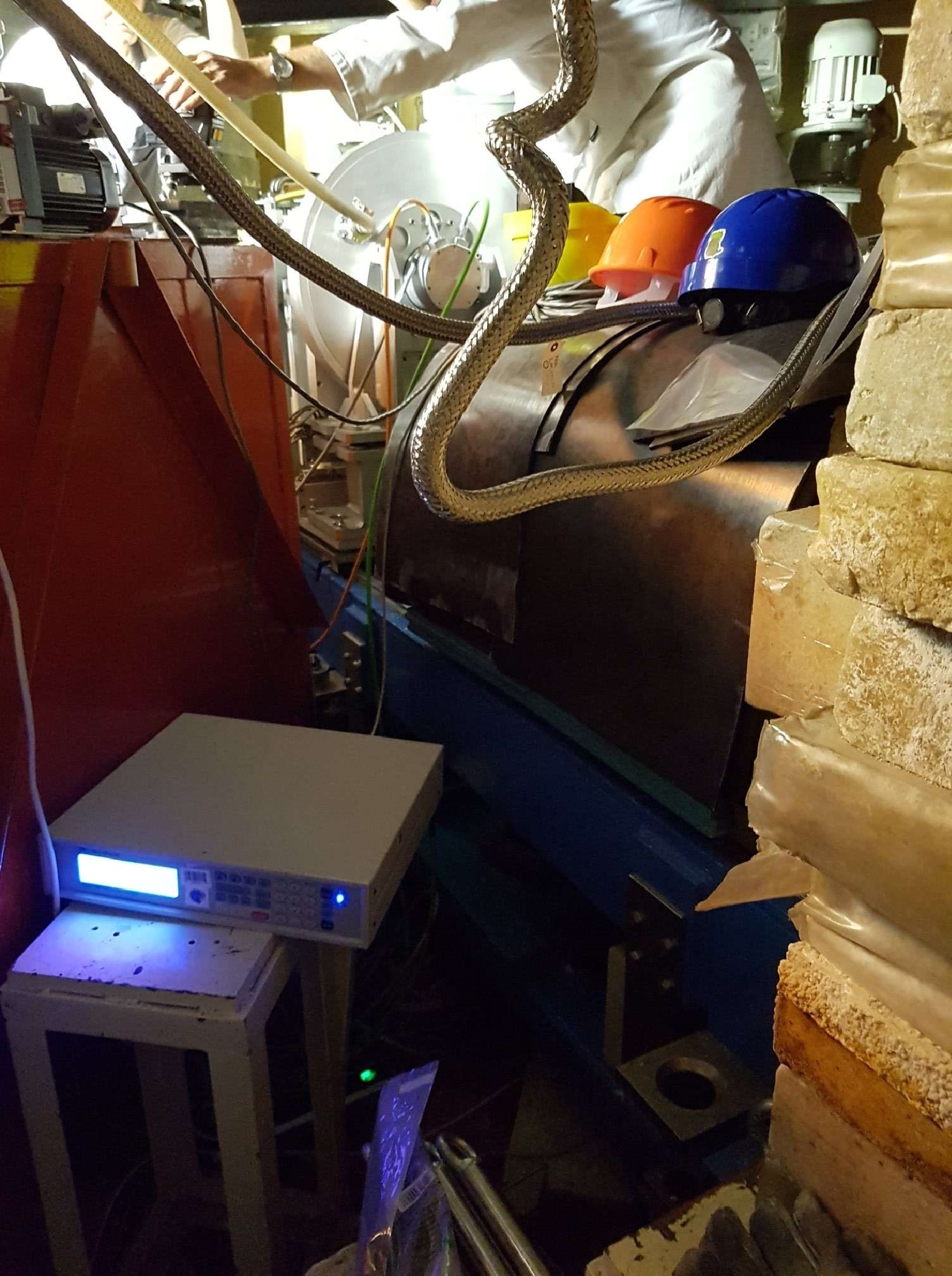}
        \subcaption{}
        \label{fig:Channel_4}
    \end{subfigure}
    \caption{ Status of BNC experiment on September 27$^\text{th}$ 2023 (a) View of the roof with compressor installed (b) Interior of the experimental room.}
    \label{fig:Budapest_week2}
\end{figure}

The reactor was stopped at 14:00 on September 22$^\text{nd}$ 2023. A last spectrum was recorded just before this shutdown. The efforts done to reinforce the shielding lead to a significant background reduction. Reinforcing the shielding around the detector and especially from its backside led to a reasonable background-to-noise ratio. In summary, the first week of the campaign was dedicated to the shielding improvement in order to reduce the background.
The cryogenic system was put in place during the second week. Cooling of hydrogen started on September 28$^\text{th}$. \cref{fig:Budapest_week2} illustrates the status of the facility at this time. 
Even if no additional measurements were possible in the time frame of the HighNESS project, the successful building of the beamline and manufacturing of the different components of the experimental setup and the first images of the moderator obtained are giving the proof that such measurements can be performed. This already an important milestone reached, and is promising for future campaigns.

%To be able to quantify the effect of the advanced reflectors, a control measurement with an empty reflector vessel was carried out first as reference. This initial measurement will be followed by a measurement with nanodiamonds reflector and finally a measurement with magnesium hydride reflector. 

\FloatBarrier
\subsection{Experiment at the JULIC Neutron Platform}
\begin{figure}[htb]
    \centering
    \includegraphics[width=0.9\textwidth]{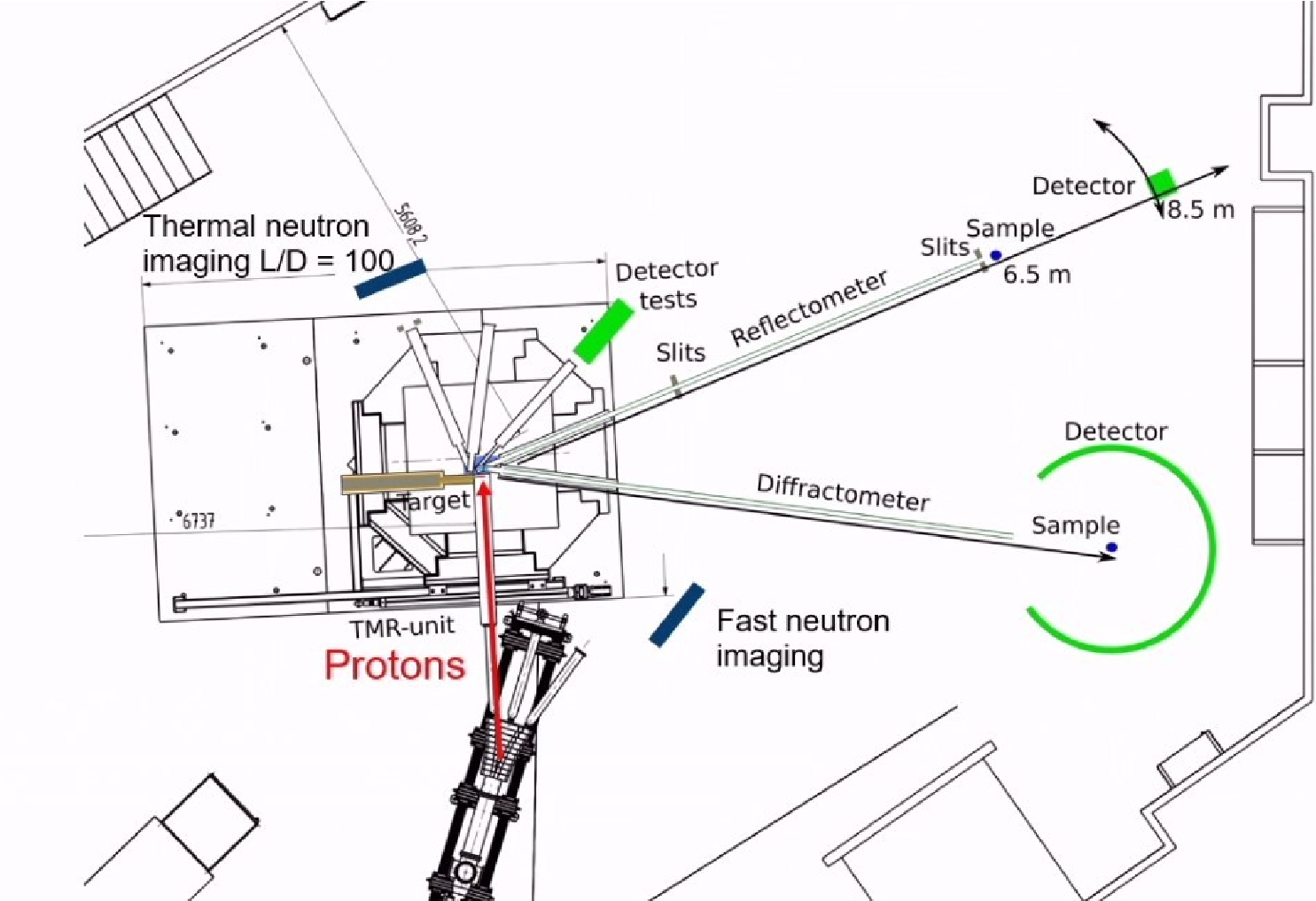}
    \caption{Schematic of the HiCANS test station at the Big Karl area of the IKP.}
    \label{fig:Big_Karl_plan}
\end{figure}

The construction of a test Target-Moderator-Reflector (TMR) unit is part of the High Brilliance Neutron Source (HBS) project \cite{zakalek2020high}, which aims to develop a High Current Accelerator-driven Neutron Source (HiCANS) which is competitive and cost-efficient. The test TMR station was developed over two years through a collaboration between JCNS and Central Institute of Engineering, Electronics and Analytics (ZEA-1) at FZJ and built in the ‘Big Karl’ area of the IKP. The neutrons are produced by \SI{45}{MeV} protons accelerated by the JUelich Light Ion Cyclotron (JULIC), delivered via a dedicated proton transfer beamline, and impinging on a newly developed high-power tantalum target with an internal microfluidic water-cooling loop. A schematic representation is shown in \cref{fig:Big_Karl_plan}. The subsequent fast neutrons emitted are moderated to thermal and cold range by, respectively, a high-density polyethylene (HDPE) premoderator and two cold sources, a low-dimensional hydrogen source and a methane moderator, both at cryogenic temperature. The TMR is surrounded by a 68-ton heavy concrete biological shielding with a unique modular design that can accommodate up to eight individual experiments. The first neutrons were produced on the target on December 12$^\text{th}$ 2022. Due to the low current of the proton beam (\SI{5}{\micro A} peak current), the neutron yield produced is relatively low ($\approx \SI{e10}{s^{-1}}$), so the facility is optimal for proof-of-principle experiments like the ones in the HighNESS project. The HighNESS experiment took place at the HiCANS test station from August 30$^\text{th}$ 2023 to September 3$^\text{rd}$ 2023.
\subsubsection{Experimental setup}
The protons hitting the target are extracted from the plasma of the ion source for the JULIC Cyclotron, the pre-accelerator of the COoler SYnchrotron (COSY), where they are accelerated from \SI{4}{keV} to \SI{45}{MeV}. During the period of the beamtime, the accelerator was operating in two modes:
\begin{enumerate}
    \item[-]\SI{25}{Hz} frequency and \SI{0.8}{ms} pulse length
    \item[-]\SI{10}{Hz} frequency and \SI{1} or \SI{3}{ms} pulse length 
\end{enumerate}
The first set of parameters was intended for most of the other experiments, with resolution-sensitive measurements where the cold and very cold tail of the moderator spectrum was not crucial. On the other hand, the low frequency and long pulse length mode was specifically tailored to the HighNESS needs, where fine resolution was not so important and the cold flux was essential. The accelerator operational schedule during the beamtime included the change of parameters early in the morning and in the afternoon after the HighNESS measurements were finished.  
During the preparation of the experiment, pulse lengths of \SI{1}{ms} and \SI{2}{ms} were also investigated in order to assess what would be the best trade-off between resolving the shorter wavelengths and having a high flux at long wavelengths. Since the factor 3 in neutron flux was deemed more important, only one short measurement at \SI{1}{ms} was performed before switching to \SI{3}{ms}. 
\begin{figure}[hbt]
    \centering
    \includegraphics[width=\textwidth]{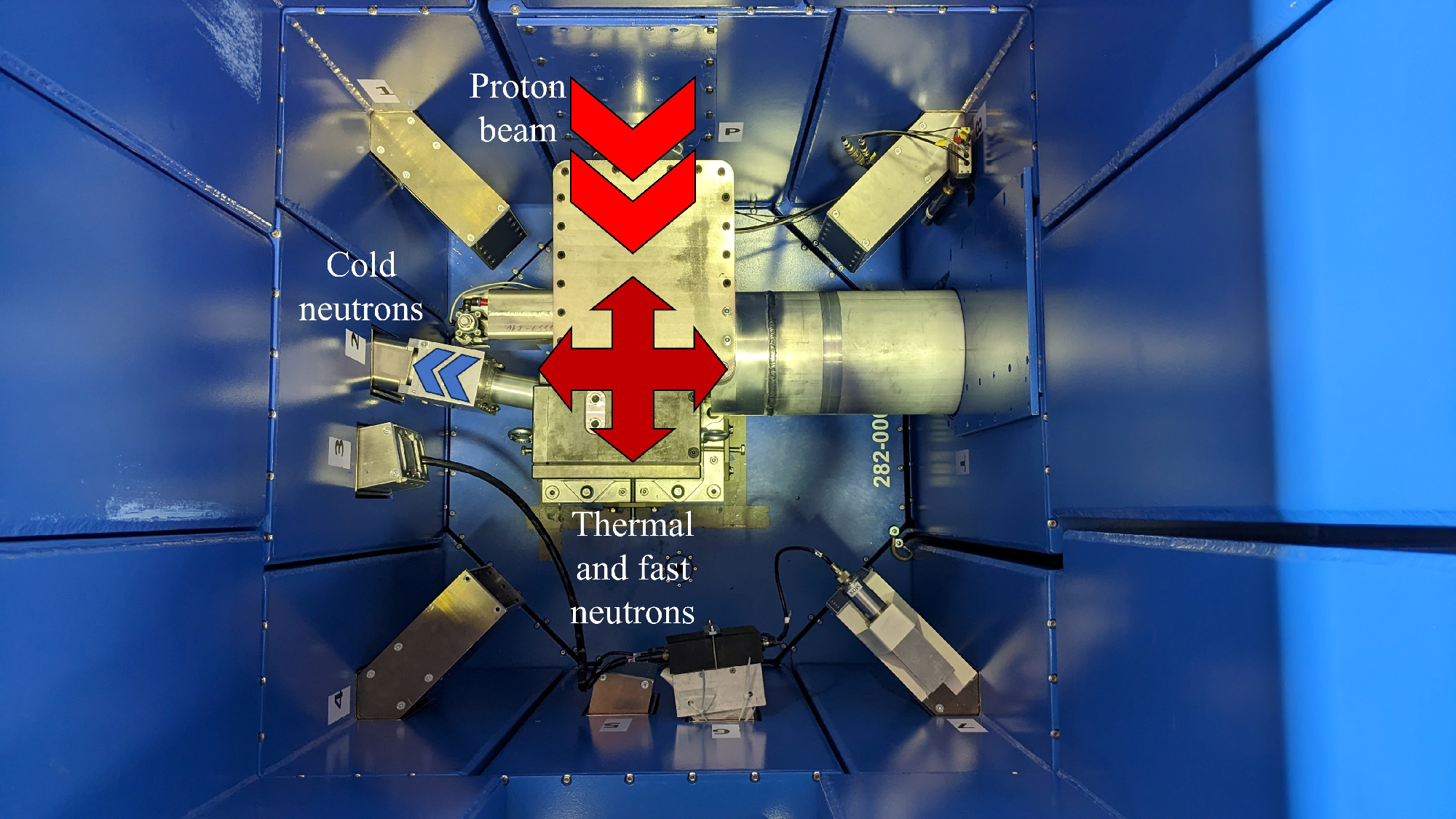}
    \caption{Target-Moderator-Reflector (TMR) area seen from above. Protons are coming from top. Cold neutrons moderated in the solid methane are extracted from beamline number 2.}
    \label{fig:inner_TMR_above}
\end{figure}
\begin{figure}[tb!]    
        \centering
    \begin{subfigure}[b]{0.42\textwidth}
        \includegraphics[width=\textwidth]{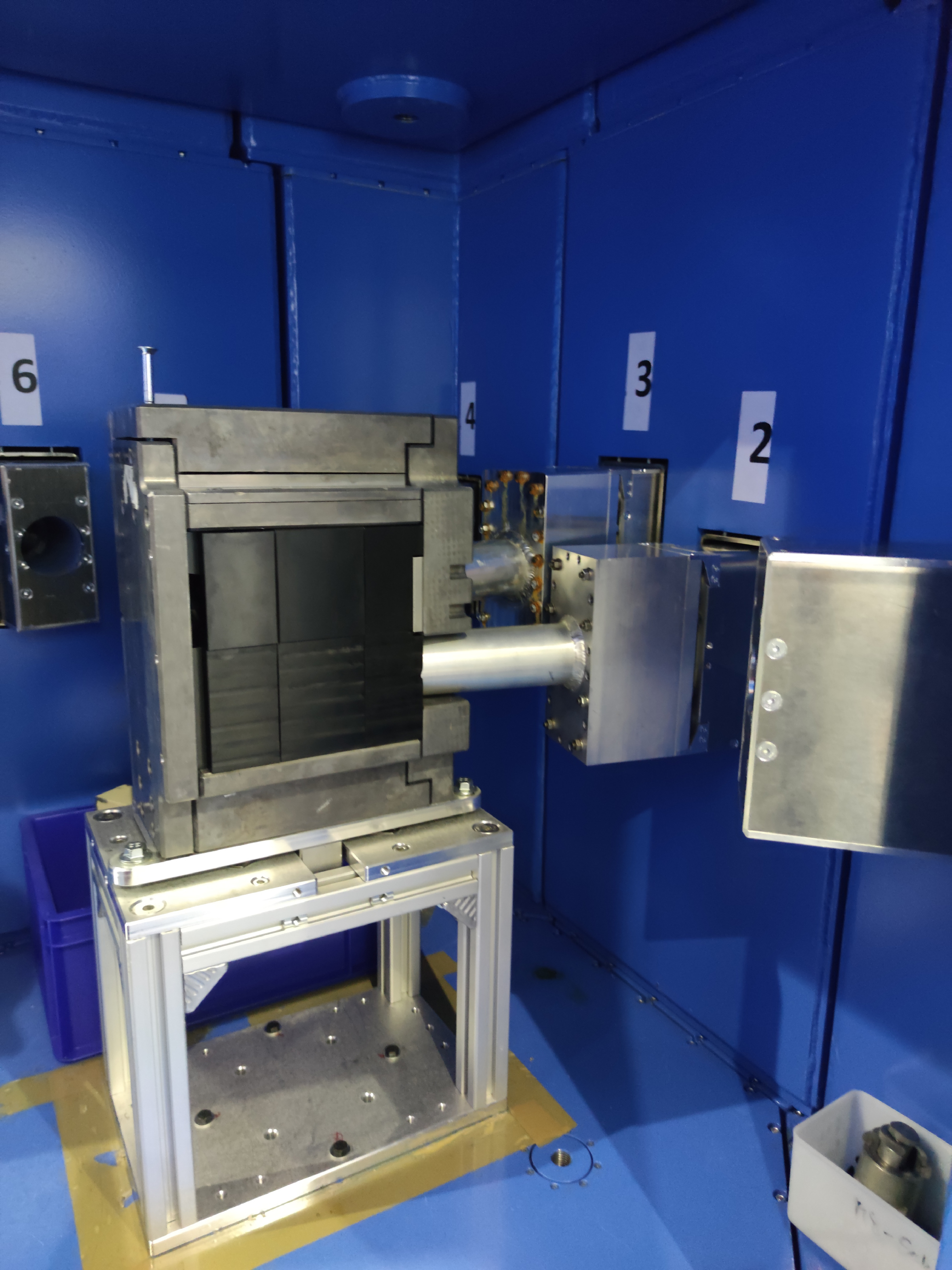}
        \subcaption{}
        \label{fig:TMR}
    \end{subfigure}
    \begin{subfigure}[b]{0.42\textwidth}       
        \includegraphics[width=\textwidth]{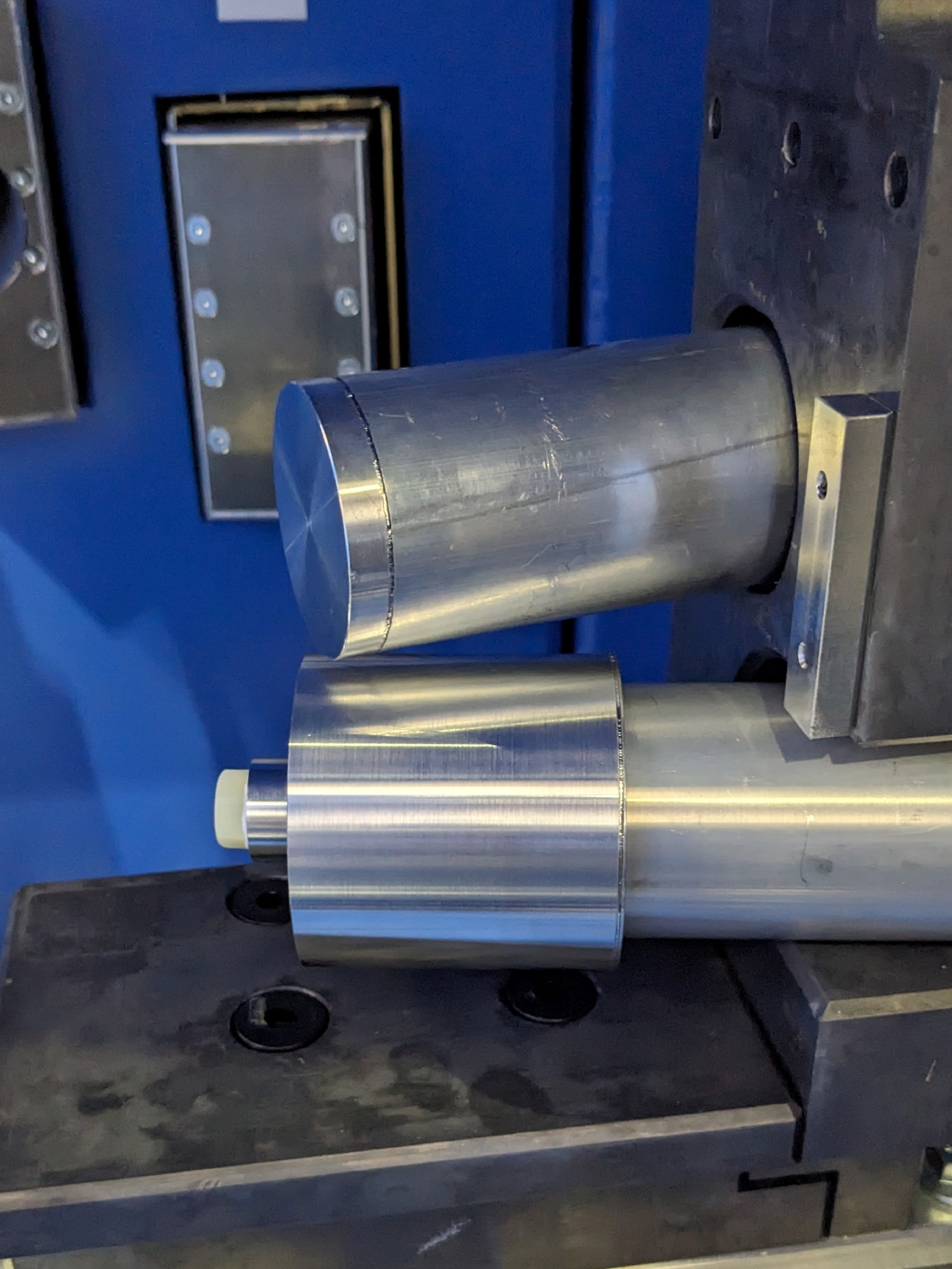}
        \subcaption{}
        \label{fig:cold_sources}
    \end{subfigure}
    \begin{subfigure}[b]{0.42\textwidth}
        \includegraphics[width=\textwidth]{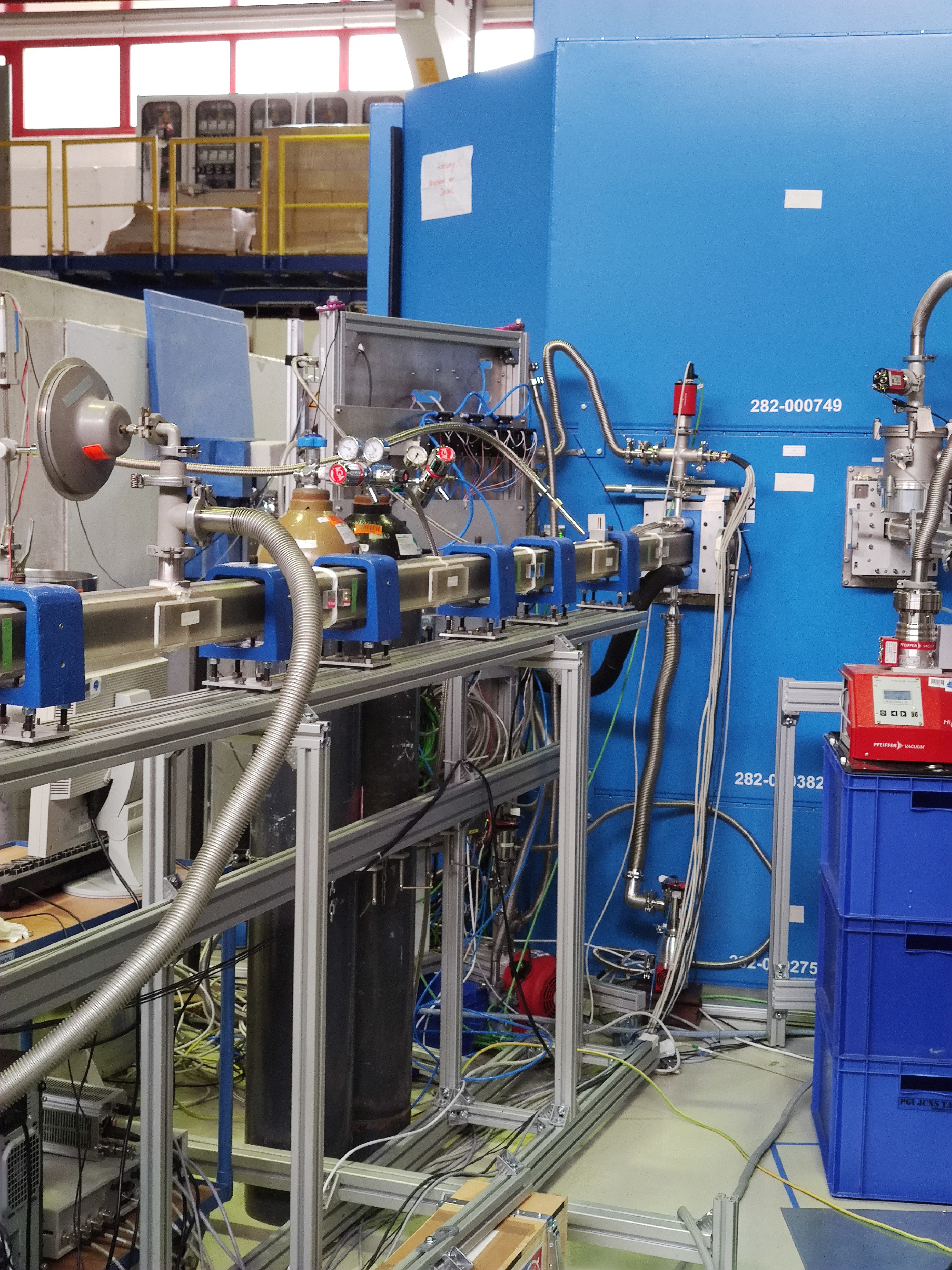}
        \subcaption{}
        \label{fig:guide}
    \end{subfigure}
    \begin{subfigure}[b]{0.42\textwidth}      
        \includegraphics[width=\textwidth]{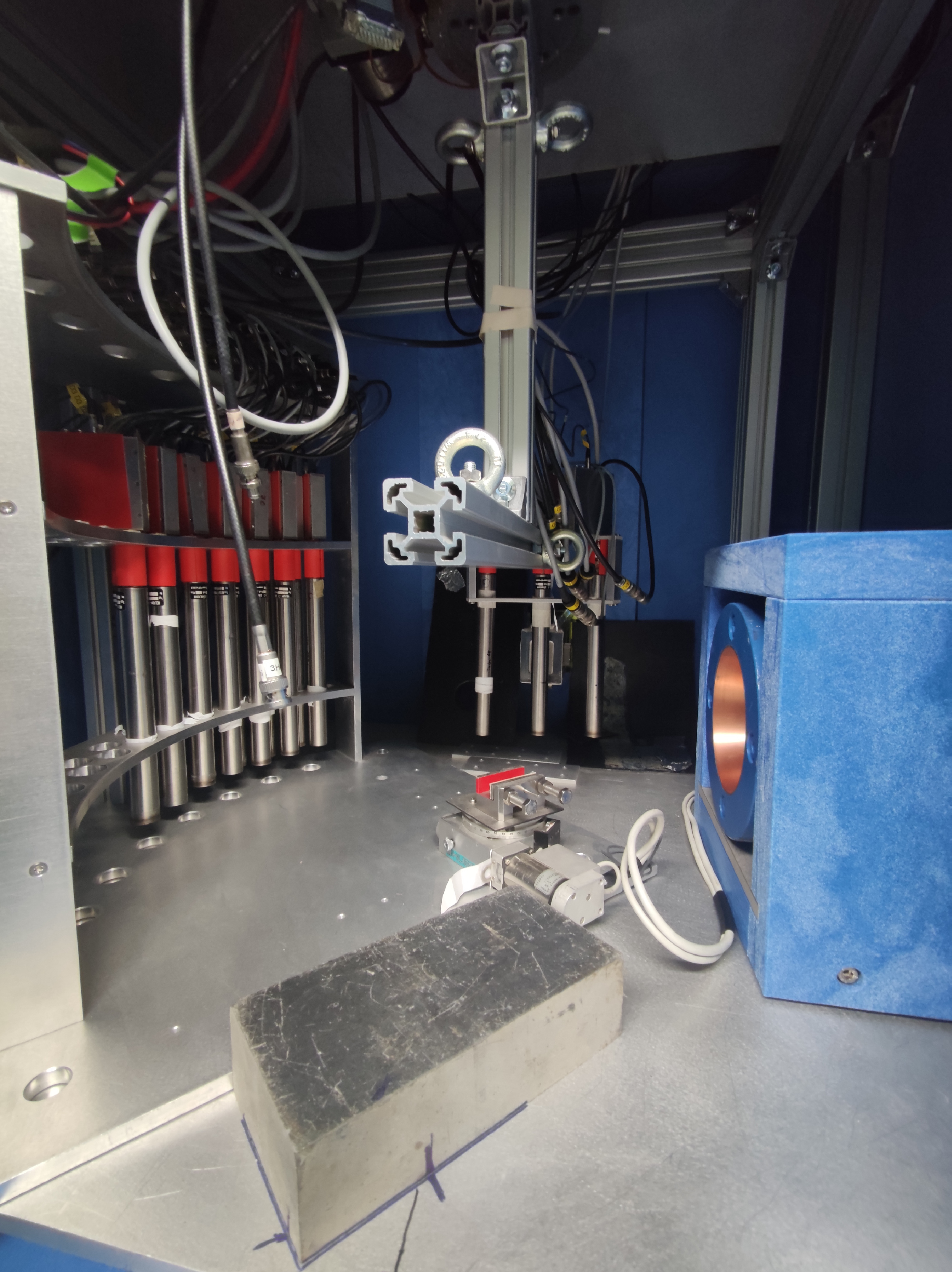}
        \subcaption{}
        \label{fig:detector}
    \end{subfigure}
    \caption{ Pictures of the (a) TMR with lead shielding around the HDPE premoderator and the extraction system for beamline 2 (b) Closer look at the two moderator outer vessels without the HDPE block. Above, the hydrogen moderator and below, the methane one with the HighNESS reflector cup surrounding it. (c) neutron optical guide (d) \ce{^3He} tube detectors at the sample area. The central tube in front of the guide exit is the detector used to record the signal. }
    \label{fig:guide+det}
\end{figure}

The TMR station presented in \cref{fig:TMR,fig:inner_TMR_above} is equipped with two cold sources: a liquid para-hydrogen moderator and a solid methane one (\cref{fig:cold_sources}). The two moderators are embedded in a HDPE block which acts as a premoderator. A lead case around the HDPE reflects the thermal neutrons and shields the sources from the gamma radiations.
The effect of the advanced reflectors was tested at the diffractometer of the facility (beamline number 2, see \cref{fig:Big_Karl_plan}), which looks into the solid methane cold source. The moderator vessel is an aluminum cylinder of \SI{36}{mm} in diameter and \SI{53}{mm} in length, with an inner cavity of \SI{22}{mm} in diameter and \SI{33}{mm} in length corresponding to the moderating volume. The source is kept at the cryogenics temperatures by liquid helium flowing in channels carved around the outer layer of the vessel. The poor thermal conductivity of solid methane is improved by inserting an aluminum foam which increases the thermal coupling with the vessel for a better heat removal. For the duration of the experiment, the source target temperature was set to \SI{22}{K}. The choice of the temperature was motivated by the need to keep the moderator conditions as stable as possible. Cooling the moderator at a lower temperature would have implied a higher cold flux, but also a much larger consumption of He to maintain that condition stable. This in turn would have lead to several He tank exchanges with a procedure requiring to warm up and re-solidify the methane, with no guarantee to go back to the same experimental conditions. During the measurements, the temperature stability was checked and noted in the logbook at the beginning and at the end of the run, with a resolution of \SI{1}{K}. Unfortunately, due to a software crash of the monitoring system, only the logs from the first day were available for the data analysis.
%\begin{figure}[tb!]    
 %       \centering
 %   \begin{subfigure}[b]{0.56\textwidth}
%        \centering
 %       \includegraphics[height=0.3\textheight]{advrefl_figs/bigkarl_moderator.eps}
 %       \subcaption{}
%        \label{fig:moderator}
%    \end{subfigure}
%    \begin{subfigure}[b]{0.42\textwidth} 
%        \centering
%        \includegraphics[height=0.3\textheight]{advrefl_figs/foam_methane.eps}
%        \subcaption{}
%        \label{fig:foam}
%    \end{subfigure}
%    \caption{Pictures of the (a) moderator vessel (b) aluminum foam used to increase heat conductivity of %solid methane. Courtesy of Monia El Barbari.}
%    \label{fig:moderator+foam}
%\end{figure}

The cold neutrons moderated inside the methane are transported out of the shielding to the sample area by a neutron guide with a \qtyproduct{30x44}{mm} section and coated by \ce{^58Ni} (m=1.18). The distance from the moderator surface to the guide exit is \SI{6.83}{m}. A \ce{^3He} tube placed in front of the window at the exit of the guide measures the intensity of the neutrons as a function of the time of flight. On both sides of the detector, two identical tubes measure the background in its proximity. Since the guide section is bigger than the dimensions of the moderator, also fast neutrons from the TMR are reaching the sample area. Here they can get thermalized or scattered by the borated polyethylene shielding, with a second chance to reach the detector as background. Thus, the tubes on the sides should give a good estimates of this small effect.

A set of useful monitors, synchronized with the same time-of-flight trigger as the detector for the signal, were scattered around the hall at the following locations: inside the TMR station (visible in the bottom right corner in \cref{fig:inner_TMR_above}), outside the biological shielding, and inside and outside the sample station shielding. In addition, the array of tubes of the banana detector of the diffractometer (visible in \cref{fig:detector}) was used. The time resolution was \SI{0.2}{ms} for a full-scale range of \SI{50}{ms}.
\subsubsection{Materials and Methods}
The experiment consisted in the insertion of several aluminum cups around the methane moderator, as shown in \cref{fig:cold_sources}, two of them had a cavity filled with, respectively, ND and \ce{MgH_2}, one had an empty cavity for comparison, acting as reference, and the last one was entirely made of polyethylene, the same material as the premoderator. The cups, manufactured by the WP5 engineering team in FZJ, are aluminum hollow cylinders with an outer diameter of \SI{74.5}{mm} and inner diameter of \SI{56.5}{mm}. In the circular crown, a \SI{5}{mm} gap hosts the reflector material in powder form, while the walls are \SI{2}{mm} thick. The powders were poured in the gap from a hole in the back of the cup, and sealed with a screw and a plastic ring.
%The powders were poured in the gap from a hole in the back of the cup, as shown in \cref{fig:ND_cup}, and sealed with a screw and a plastic ring.

\begin{figure}[htb!]    
        \centering
        \includegraphics[width=0.7\textwidth]{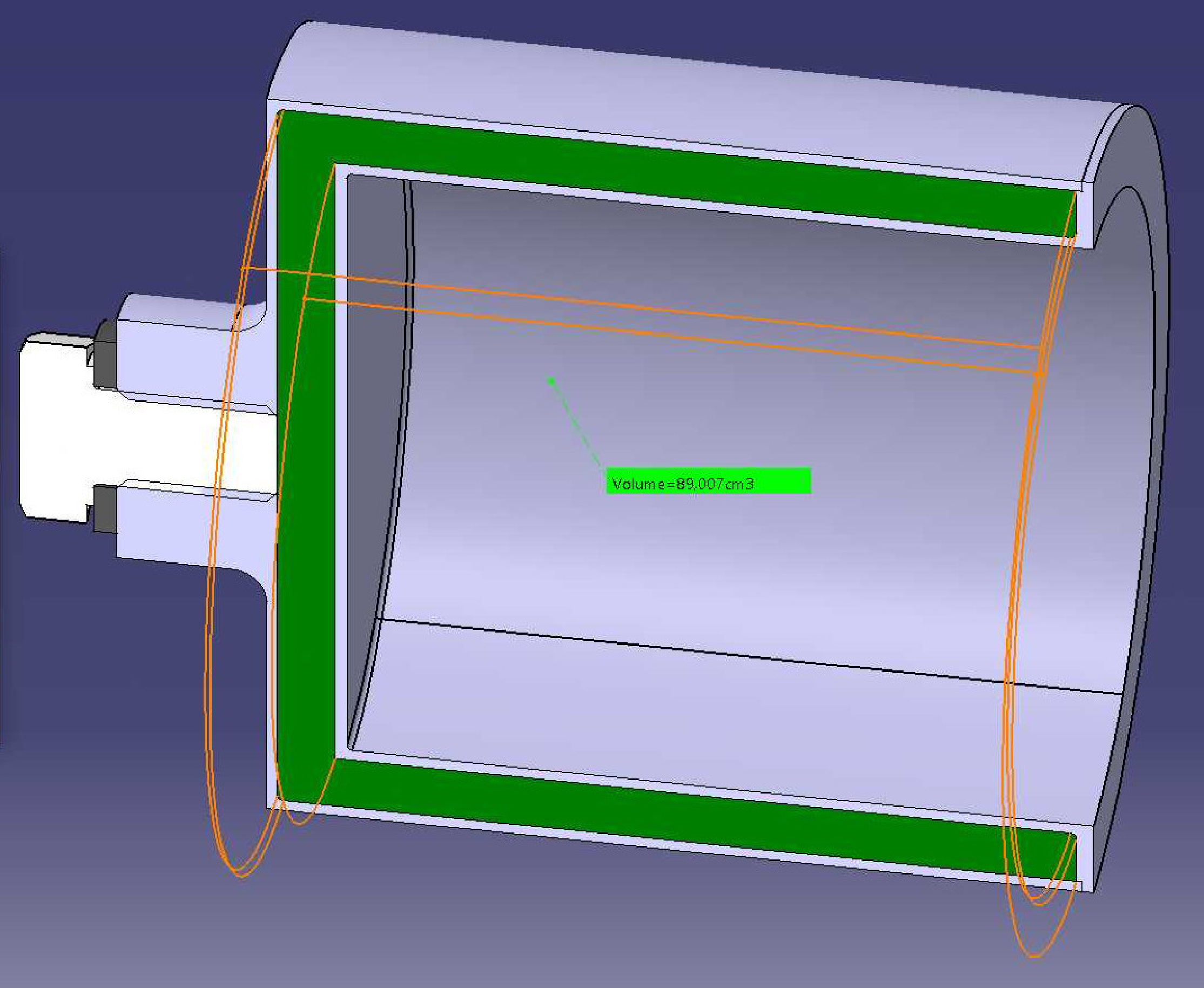}
    \caption{Diagram of the aluminum cap showing the calculated internal volume.}
    \label{fig:ND_cup_volume}
\end{figure}

%\begin{figure}[htb!]    
%        \centering
%        \includegraphics[width=0.7\textwidth]{advrefl_figs/cup_ND.eps}
%    \caption{Filling of the aluminum cup with ND powder done by Mathias Strothmann (WP5) at FZJ.}
%    \label{fig:ND_cup}
%\end{figure}

The nanodiamond powder used is commercial Detonation NanoDiamonds (DND) for nuclear applications \cite{ND_website}. The average particle size is \SI{5}{nm} and the purity declared by the manufacturer for the ash residues is $<\SI{0.1}{wt.\%}$. The size distribution of the same product was characterized in \cite{teshigawara2019measurement}, while an independent elemental analysis with inductively coupled plasma mass spectrometry (ICP-MS) technique performed in FZJ showed a total metallic impurity content of \SI{185}{mg/kg}, made up by traces of Ca (\SI{60}{mg/kg}), Mn (\SI{49}{mg/kg}), Na (\SI{27}{mg/kg}), Fe (\SI{23}{mg/kg}) and other elements in lower concentrations. After filling the cup, the mass of powder in the jacket was \SI{21.8}{g}. This means that, with a jacket volume of \SI{89}{cm^3} taken from the CAD (\cref{fig:ND_cup_volume}), the achieved fill density was \SI{0.245}{g/cm^3}.

Similarly, the \ce{MgH_2} was used in the form of a powder to fill the cavity of a second cup. The average size of the particles was \SI{15}{\micro m}, for a nominal density of \SI{1.45}{g/cm^3}. Also in this case, the manufacturer declared a purity of 99.9\% \cite{mgh2_website}. The powder density measured after filling the cup was \SI{0.66}{g/cm^3}.

\begin{figure}[!htb]
    \centering
    \includegraphics[width=0.6\textwidth]{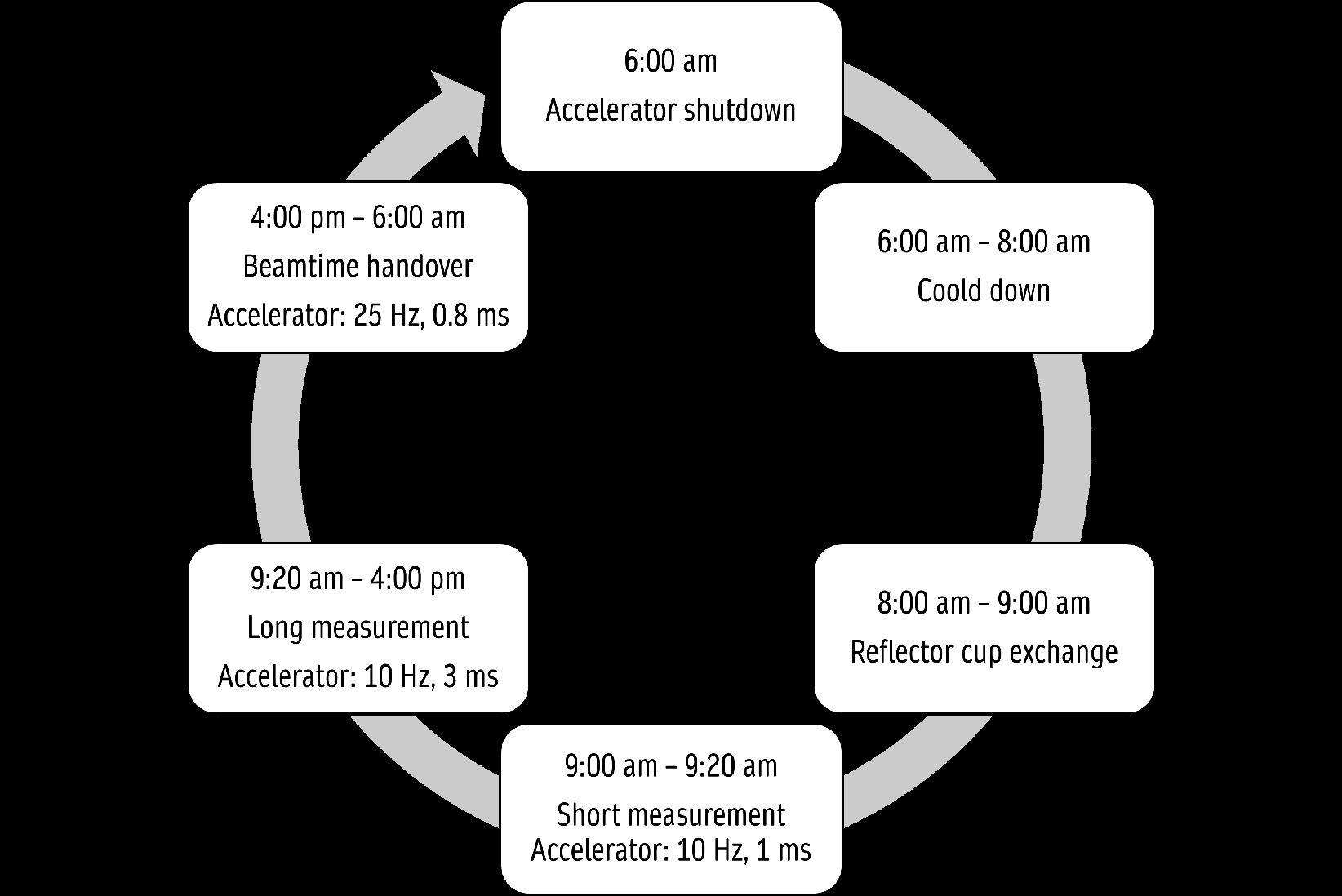}
    \bigskip
    \raisebox{.4\height}{\begin{tabular}[b]{l l}
    \toprule
      \multicolumn{2}{c}{Reflector Cup Schedule} \\ 
      \midrule
      Day 1 & Empty cup \\
      Day 2 & ND cup \\
      Day 3 & \ce{MgH_2} cup \\
      Day 4 & HDPE cup \\
      Day 5 & Empty/ND/Empty cup \\
      \bottomrule
    \end{tabular}}
    \captionlistentry[table]{A table beside a figure}
    \captionsetup{labelformat=andtable}
    \caption[]{(Left) Schematic of the experiment schedule for the first four days. (Right) Reflector cup used for each day of experiment.}
    \label{fig:schedule_accelerator}
  \end{figure}

\begin{figure}[hbt!]
    \centering
        \begin{subfigure}[b]{0.35\textwidth}
        \includegraphics[width=\textwidth]{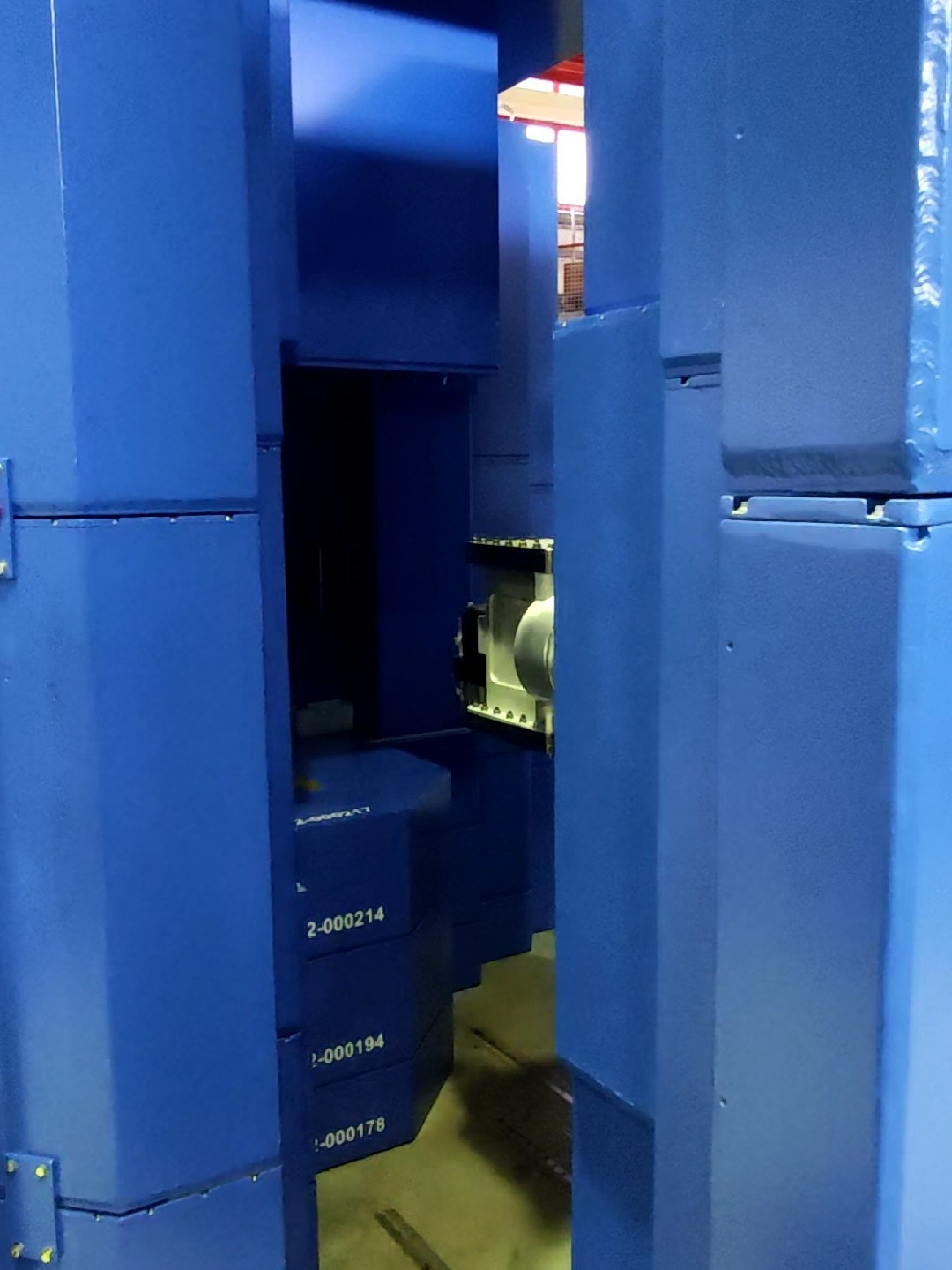}
        \subcaption{}
        \label{fig:open_shielding}
    \end{subfigure}
    \begin{subfigure}[b]{0.35\textwidth}      
        \includegraphics[width=\textwidth]{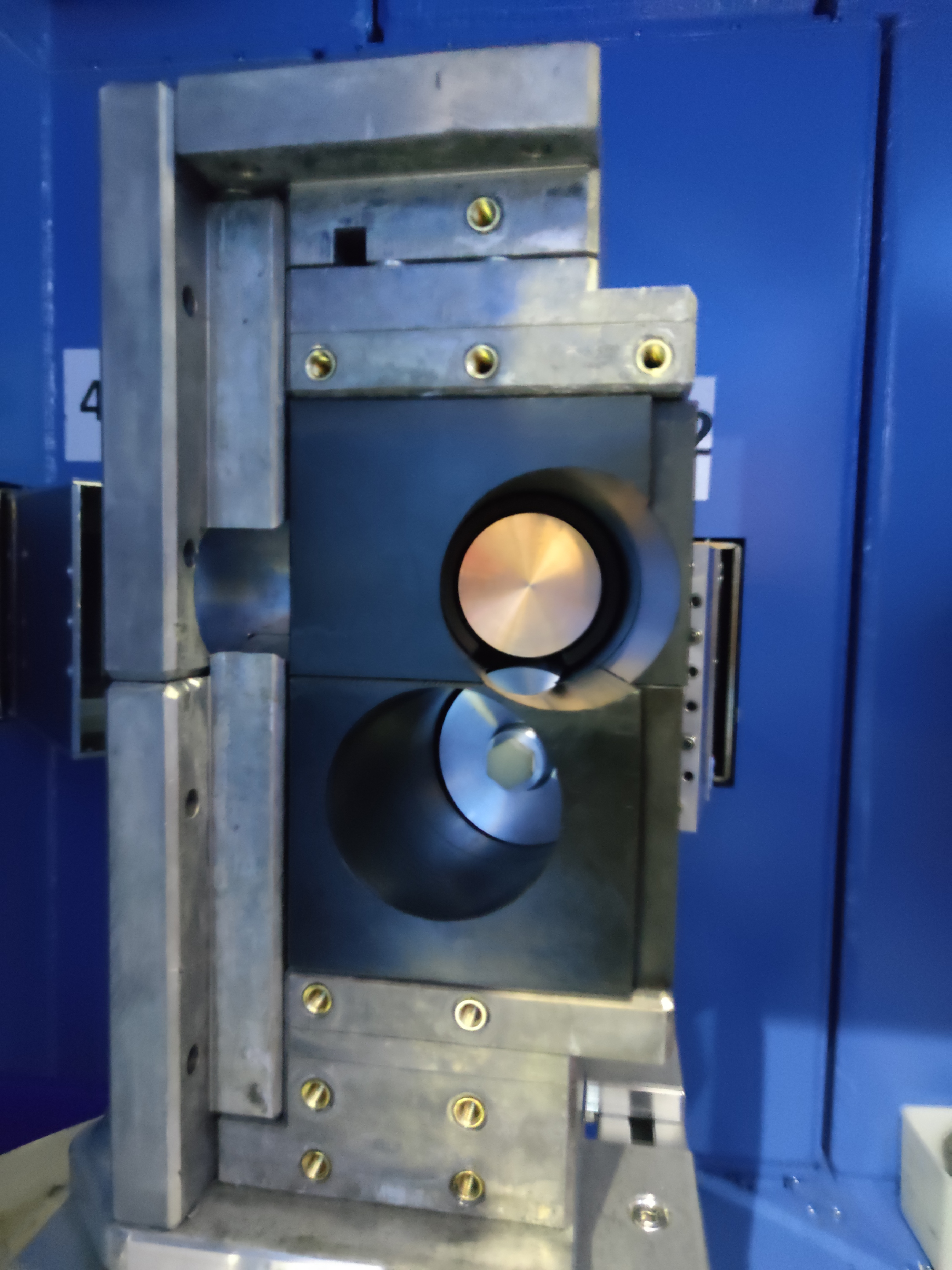}
        \subcaption{}
        \label{fig:cup_exchange_back_1}
    \end{subfigure}
    \caption{(a) Opened modular biological shielding with target (b) Example of the view of the TMR when the back lead panel of the reflector is removed for the cup exchange.}
    \label{fig:cup_exchange_back}
\end{figure}

During the five days of the beamtime, the schedule of the experiment was meant to limit the change of the accelerator parameters as much as possible, while giving the same share of time to all the users of the facility. The schedule for the first four days is summarized in \cref{fig:schedule_accelerator}, while for the last day the accelerator parameters were fixed to a \SI{10}{Hz} and \SI{3}{ms} with two cup exchanges. The procedure for the cup exchange consisted in shutting down the accelerator, letting the target decay for at least two hours, perform a radiation survey of the area, entering into the modular biological shielding, removing the back lead panel of the reflector, extracting the old cup from behind and inserting the new one (\cref{fig:cup_exchange_back}). Manipulation of activated components was performed by radiologically qualified personnel from FZJ and under supervision of radiation protection officers.

\subsubsection{Data Analysis}
\begin{figure}[bt!]
    \centering
    \includegraphics[width=0.8\textwidth]{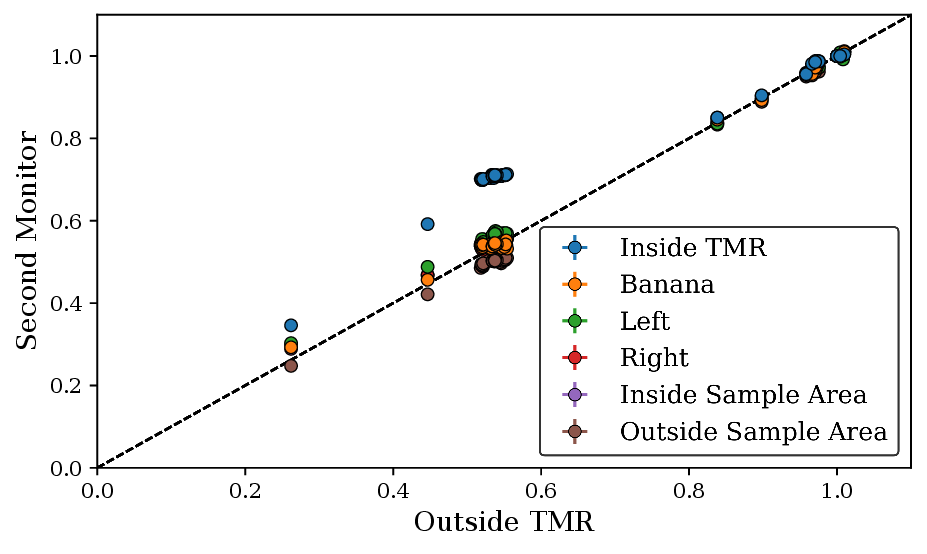}
    \caption{Plot showing the correlation between the integrated counts of the monitor outside the TMR shielding and all the other monitors. The counts are normalized by both the acquisition time and the counts for the first run (T0).}
    \label{fig:correlation_with_Outside_TMR}
\end{figure}

\begin{figure}[bt!]
    \centering
    \includegraphics[width=\textwidth]{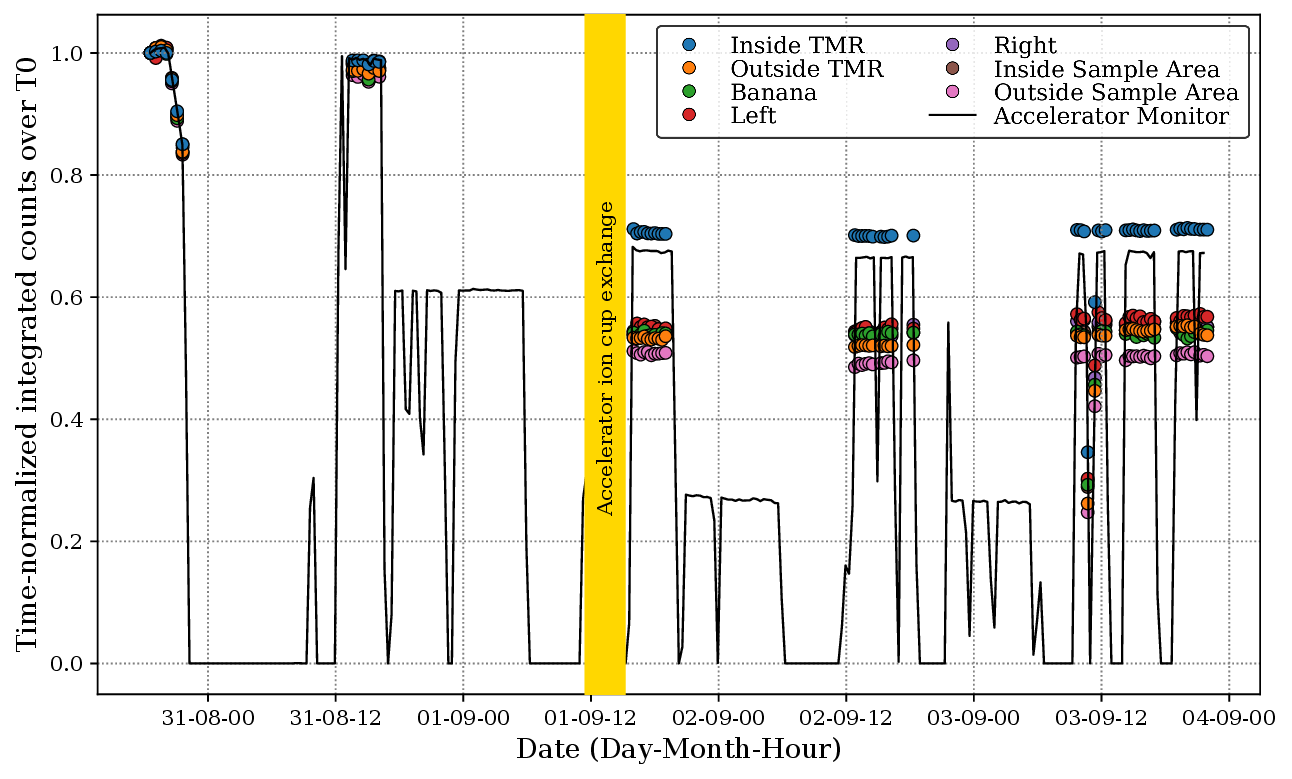}
    \caption{Plot showing the correlation between the accelerator monitor counts and all the other monitors. The counts are normalized by both the running time and the counts for the first run at T0. The time window for the exchange of the accelerator cup is shown. The lower peaks in-between the HighNESS experiment correspond to the shorter pulse length regime.}
    \label{fig:correlation_with_accelerator}
\end{figure}

The measurement of the time-of-flight spectrum was subdivided into multiple acquisitions of 20-\SI{30}{min} each. In this way the stability of the source during the data acquisition could be studied. After subtracting the background contribution, which is almost negligible for cold neutrons with this setup, the spectra are normalized with the respective integrated monitor counts. Some of the runs were affected by statistically meaningful deviations from the average measurement, which are related to malfunctioning parts of the setup (more details will be given on this topic in the discussion part). Once identified the cause of the systematic deviation, those runs were discarded from the analysis. The resulting averaged and corrected spectrum was calculated as:
\begin{equation}
\label{eq:average}
    \overline{S^{*}}(t)=\sum_i^{N}\frac{1}{N}\cfrac{S^{i}(t)-B^i(t)}{\sum_t M^{i}(t)}
\end{equation}
where $S^i(t)$ is the signal from the main detector for the $i$-th run, $B^i(t)$ are the average background counts from the two \ce{^3He} tubes on the left and on the right of the main detector, and $M^i(t)$ are the monitor counts used for normalization.
The conversion from time-of-flight bin in ms to wavelength bin in Å is obtained as:
\begin{equation}
    \lambda = \frac{\SI{3956}{Åm\per\second}}{\SI{6.83}{m}}*t\left[\si{ms}\right]*0.001 = \SI{0.5792}{\angstrom\per\second}*t\left[\si{ms}\right]
\end{equation}
In order to study the effect of the advanced reflector around the moderator on the cold spectrum, the ratio between the cases with the reflector and the empty cup is evaluated:
\begin{equation}
\label{ratio}
    r(\lambda) = \frac{\overline{S^{*}_R}(\lambda)}{\overline{S^{*}}_{empty}(\lambda)} 
\end{equation}
It is clear that the stability of both the signal and the monitor is critical in evaluating $r(\lambda)$, since uncorrelated systematic drifts in the two terms could cause ratios greater or smaller than 1 even when there is no real effect. In \cref{fig:correlation_with_Outside_TMR}, the sum of the counts for the monitor situated outside the TMR for all runs, normalized by both the acquisition time and the counts of the first run (T0), is plotted against the same quantity for all the other monitors. 
Variations in absolute counts can be caused by fluctuations of the proton current, but the correlation between the monitors is expected to stay constant, which is not the case for the monitor located inside the TMR. The two largest clusters of runs are separated by an intensity gap due to a decrease in the accelerator current that occurred after the second day and was caused by the exchange of the accelerator ion cup, after which the current went from \SI{8}{\micro A} to \SI{6.3}{\micro A}. This effect is more visible in \cref{fig:correlation_with_accelerator}, where the time-normalized integrated counts for all monitors, further divided by the counts at T0, are plotted as a function of the time and compared with the accelerator monitor, also normalized in the same way.
After the exchange of the ion cup, the 20\% decrease in current results in 30\% less counts for the monitor inside the TMR area and almost 50\% for the other monitors. The accelerator monitor was not considered reliable for normalization by the instrument scientist since, after a change in the ADC threshold to allegedly avoid overflow, no extensive testing was conducted on the relation between the proton current and the monitor. In this respect, the analysis of the correlations among the network of monitors highlights that the monitor located inside the TMR area is not reliable for normalization either. A possible explanation for this effect could hence be that both monitors are underestimating the count drop after the ion cup exchange due to overflow. Thus, the monitor outside the TMR shielding is used for normalization in the data analysis.

\begin{figure}[bt!]
    \centering
        \begin{subfigure}[b]{0.49\textwidth}
        \includegraphics[width=\textwidth]{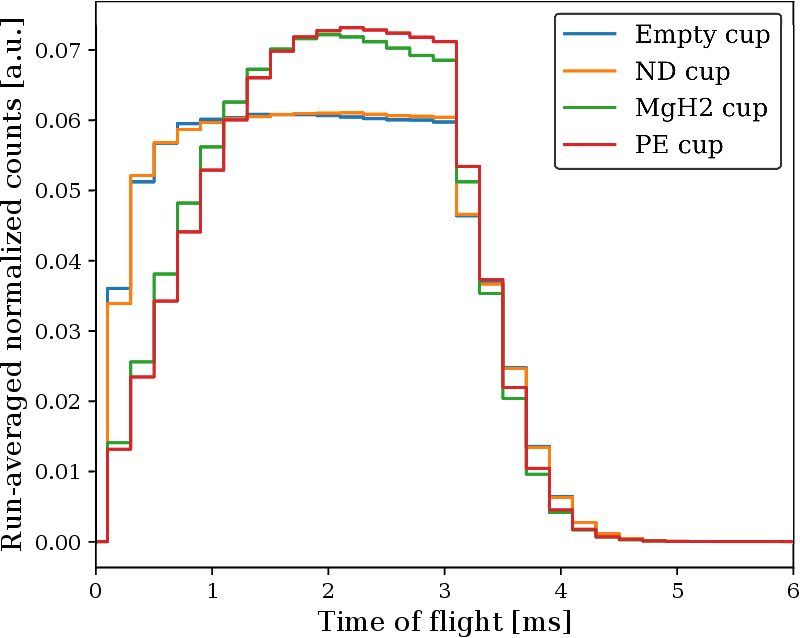}
        \subcaption{}
        \label{fig:monitor_inside_inc}
    \end{subfigure}
    \begin{subfigure}[b]{0.5\textwidth}      
        \includegraphics[width=\textwidth]{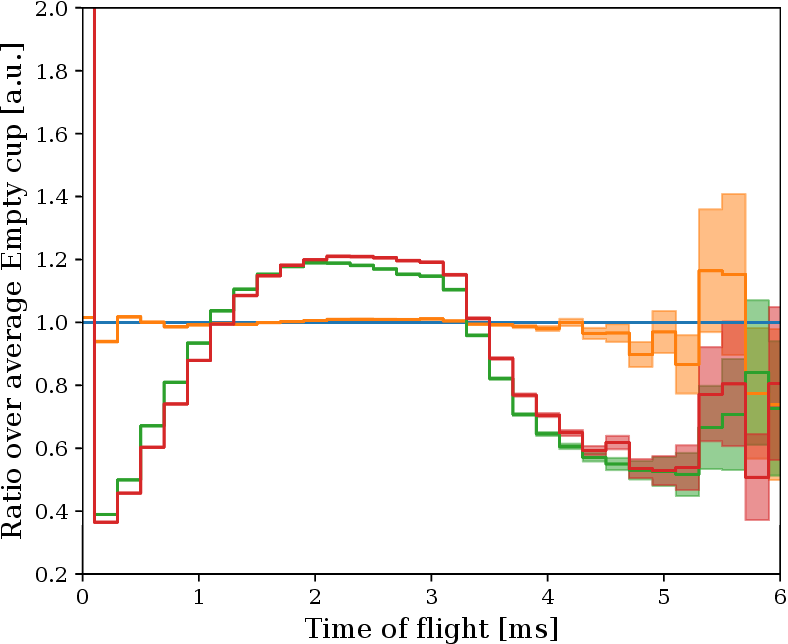}
        \subcaption{}
        \label{fig:monitor_inside_inc_ratio}
    \end{subfigure}
    \caption{(a) Monitor counts inside the TMR area as a function of time, divided by the total counts and averaged over the stable runs for measurements from Day 1-4. (b) ratio of the same curves over the empty cup. The shaded area indicates the standard deviation. }
    \label{fig:monitor_inside}
\end{figure}

\begin{figure}[hbt!]
    \centering
        \begin{subfigure}[b]{0.49\textwidth}
        \includegraphics[width=\textwidth]{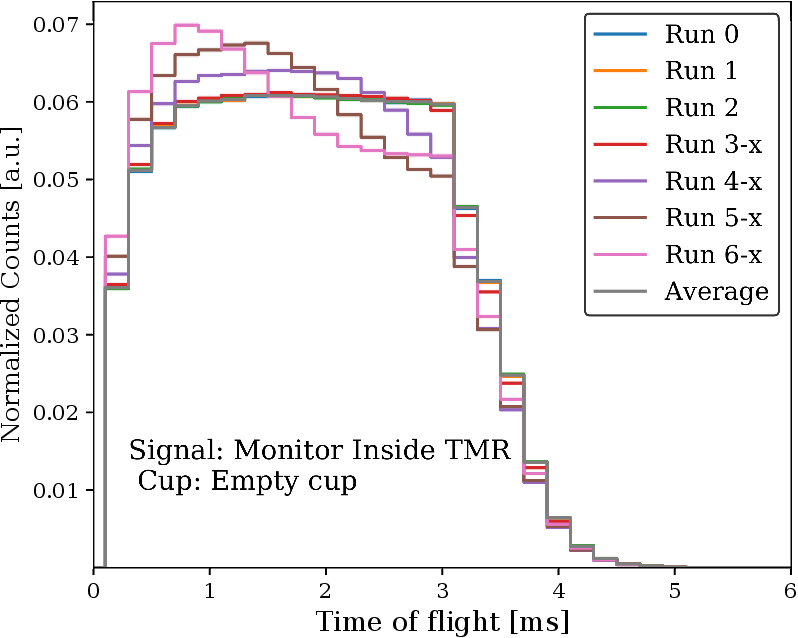}
        \subcaption{}
        \label{fig:monitor_inside_empty_inc}
    \end{subfigure}
    \begin{subfigure}[b]{0.49\textwidth}      
        \includegraphics[width=\textwidth]{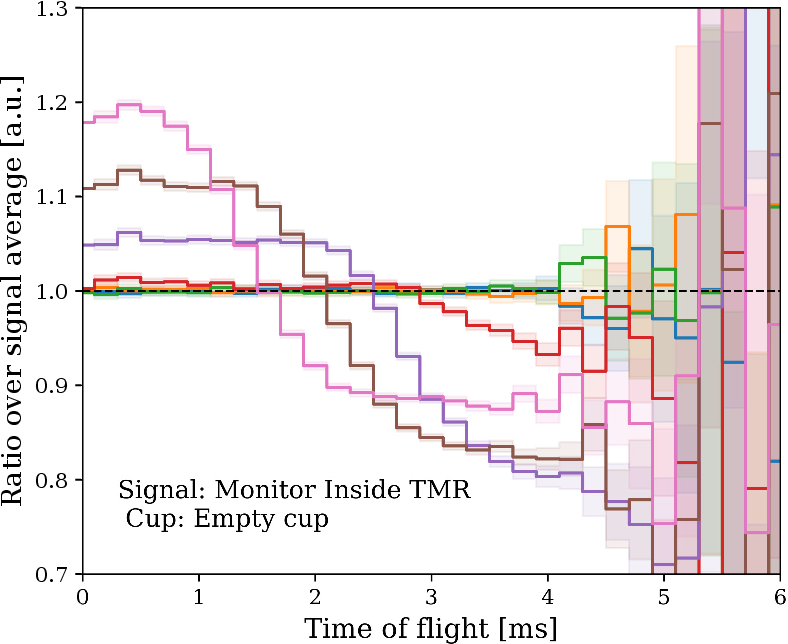}
        \subcaption{}
        \label{fig:monitor_inside_empty_inc_ratio}
    \end{subfigure}
    \begin{subfigure}[b]{0.49\textwidth}
        \includegraphics[width=\textwidth]{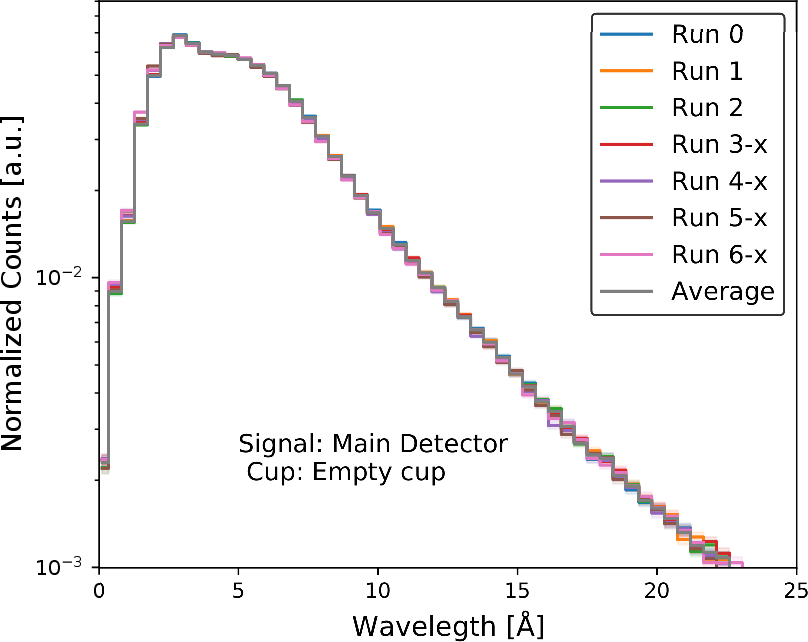}
        \subcaption{}
        \label{fig:main_empty_inc}
    \end{subfigure}
    \begin{subfigure}[b]{0.49\textwidth}      
        \includegraphics[width=\textwidth]{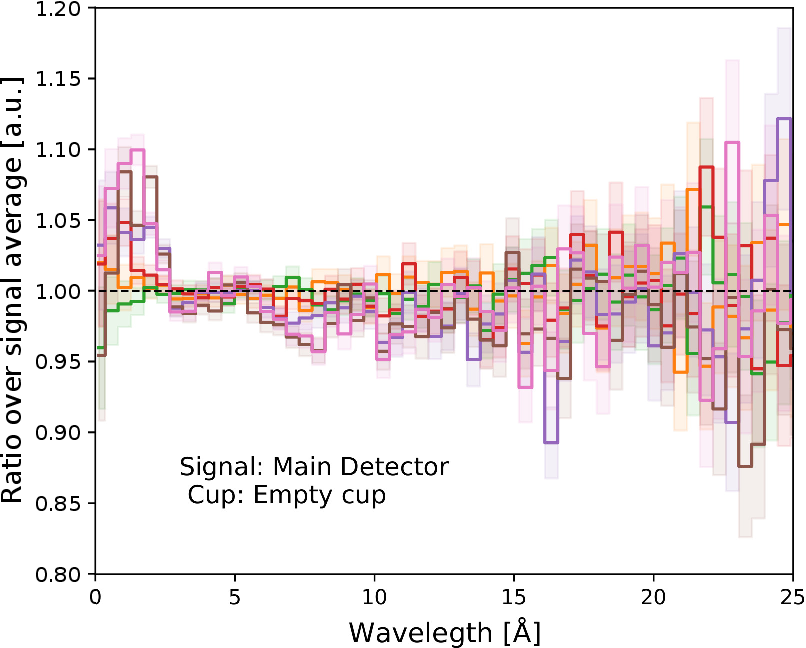}
        \subcaption{}
        \label{fig:main_empty_inc_ratio}
    \end{subfigure}
    \caption{Normalized counts for each run and ratio over average for the empty cup measurement for: (a) and (b) Monitor inside the TMR area, (c) and (d) main detector. The signal from the main detector is re-binned by a factor 4 (\SI{0.46}{\angstrom}). All the runs not included in the stable average are marked by an x.  }
    \label{fig:inc_empty}
\end{figure}

The integrated counts are not the only measure of the system stability. After the exchange of the accelerator ion cup, a different pulse shape was observed. In \cref{fig:monitor_inside_inc} the counts of the monitor inside the TMR area, divided by the total counts and averaged over the stable runs, are shown as a function of the time, while in \cref{fig:monitor_inside_inc_ratio} the ratio of the same curves over the empty cup was calculated. 
As an example, the monitor inside the TMR area is shown because the pulse shape is easily recognizable, although the same behavior is observed in all the monitors. Also, for the sake of clarity, only the runs from the first four days are shown. The ion cup was exchanged between the ND and \ce{MgH_2} cup measurements. The question that arises is if this could produce a systematic impact on both the spectra and, most importantly, the ratio with the empty cup. Since a similar effect was also observed the first day during the measurements with the empty cup (\cref{fig:inc_empty}), we can study that case to isolate as much variables as possible and try to answer the initial question. 

\begin{figure}[htb!]
    \centering
    \includegraphics[width=\textwidth]{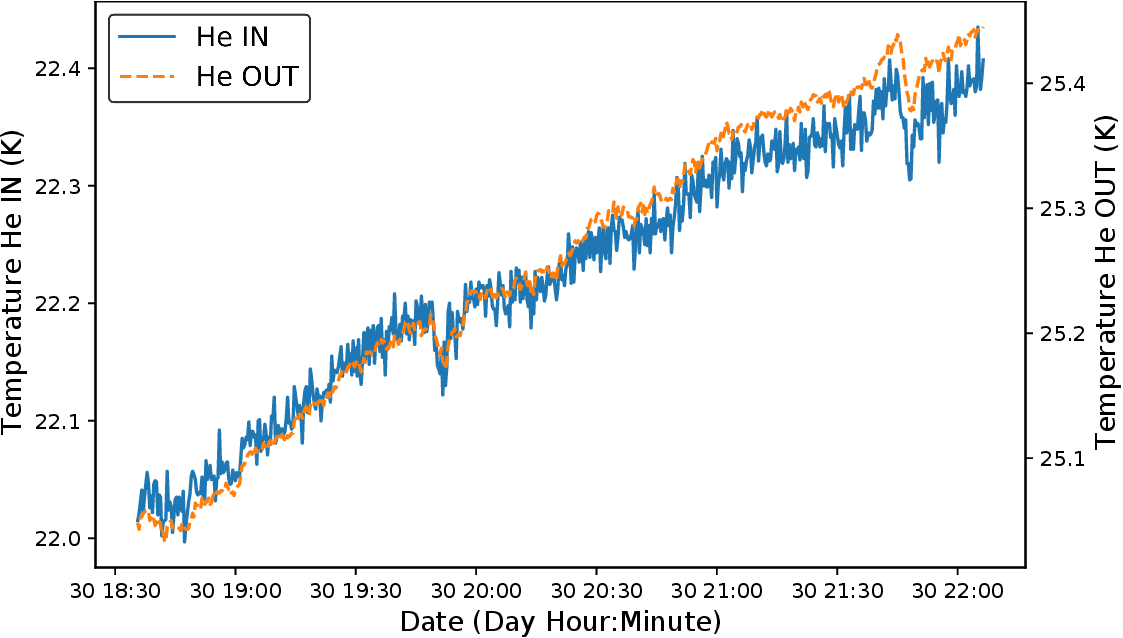}
    \caption{Variation of the He temperature flowing in and out of the moderator during the empty cup measurement on Day 1.}
    \label{fig:temp_he_empty}
\end{figure}

The change of the proton pulse shape on the first day was due to two decoupled clocks signals that were slowly drifting out of synchronization and cutting the proton pulse. Tis drift was only discovered the day after, and the problem was quickly fixed before starting the new measurements. In \cref{fig:monitor_inside_empty_inc}, this drift is clearly visible in the runs 4 to 6, althoug it probably already started in run 3. These runs are hence discarded from the calculation of the average and marked with an x in the plots. The normalization hides the differences in absolute counts (lower for runs 3 to 6), but highlights the different shapes and allows to calculate the ratios in  \cref{fig:monitor_inside_empty_inc_ratio}. The study of the signal from the main detector, calculated with \cref{eq:average}, in \cref{fig:main_empty_inc,fig:main_empty_inc_ratio} shows that a drift of 10\%-20\% in the pulse shape is positively correlated to a statistically significant variation of 5\%-10\% in the counts of the cold signal up to $\approx\SI{20}{\angstrom}$. A potential confounding factor worth to study in this case is the temperature of the moderator. The only log data available is  the temperature as a function of time of the liquid He flowing in and out of the moderator vessel during the first measurements. This set of data is reported in \cref{fig:temp_he_empty}. According to the responsible of the source, the nominal temperature of the moderator is the one of the inflow. While a warming trend is clearly visible, an absolute variation of \SI{0.4}{K} would be hardly appreciable in the final spectrum. This conclusion comes from the experiences with temperature drifts during the last day of beamtime, which suggest that differences of 2-\SI{3}{K} are necessary to start obtaining a measurable effect on the spectra. 
In conclusion, since it cannot be ruled out that the pulse shape has an effect on the signal, the results from day 1 and day 2 cannot be compared with the measurements taken after the ion cup exchange. The first set of results represents the first attempt to measure the effect of the ND cup over the empty cup of Day 1. The second set, including \ce{MgH_2}, polyethylene, and a second ND attempt, uses the first measurement of the empty cup from Day 5.

\subsubsection{Results}
\begin{figure}[hbt!]
    \centering
        \begin{subfigure}[b]{0.49\textwidth}
        \includegraphics[width=\textwidth]{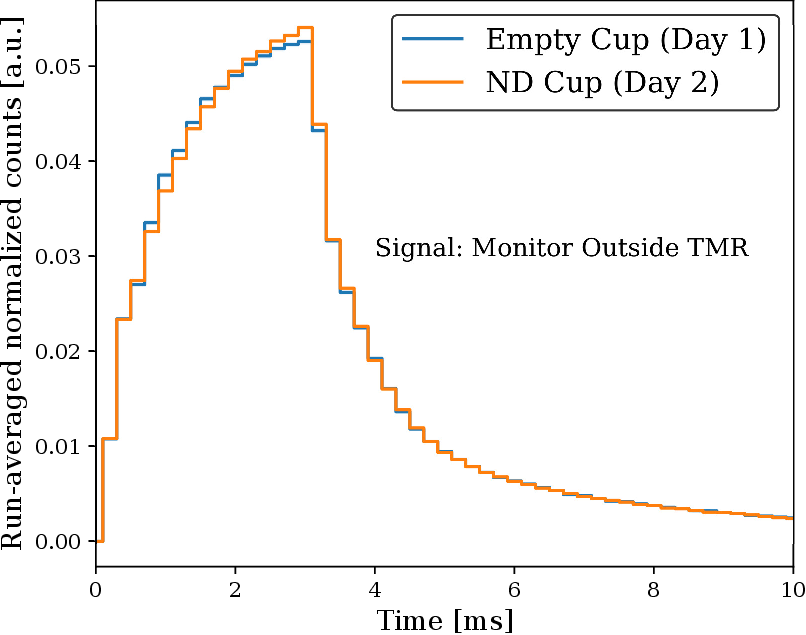}
        \subcaption{}
        \label{fig:Monitor_ND_Empty_1}
    \end{subfigure}
    \begin{subfigure}[b]{0.49\textwidth}      
        \includegraphics[width=\textwidth]{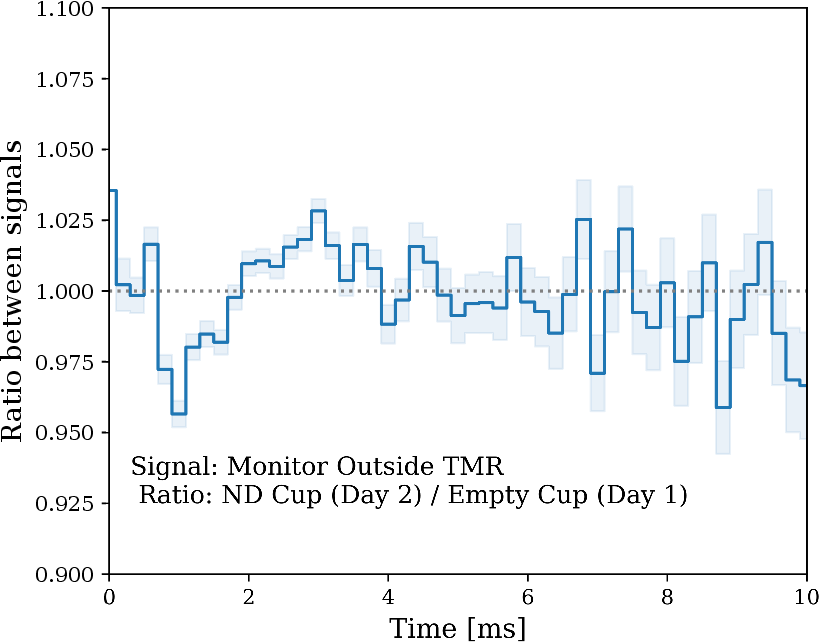}
        \subcaption{}
        \label{fig:Monitor_ND_Empty_ratio_1}
    \end{subfigure}
    \begin{subfigure}[b]{0.49\textwidth}
        \includegraphics[width=\textwidth]{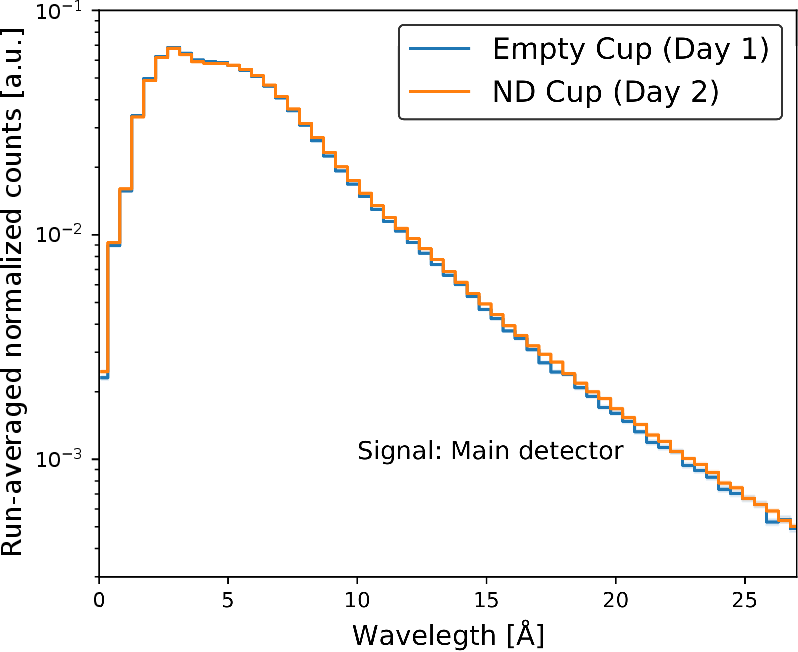}
        \subcaption{}
        \label{fig:Signal_ND_Empty_1}
    \end{subfigure}
    \begin{subfigure}[b]{0.49\textwidth}      
        \includegraphics[width=\textwidth]{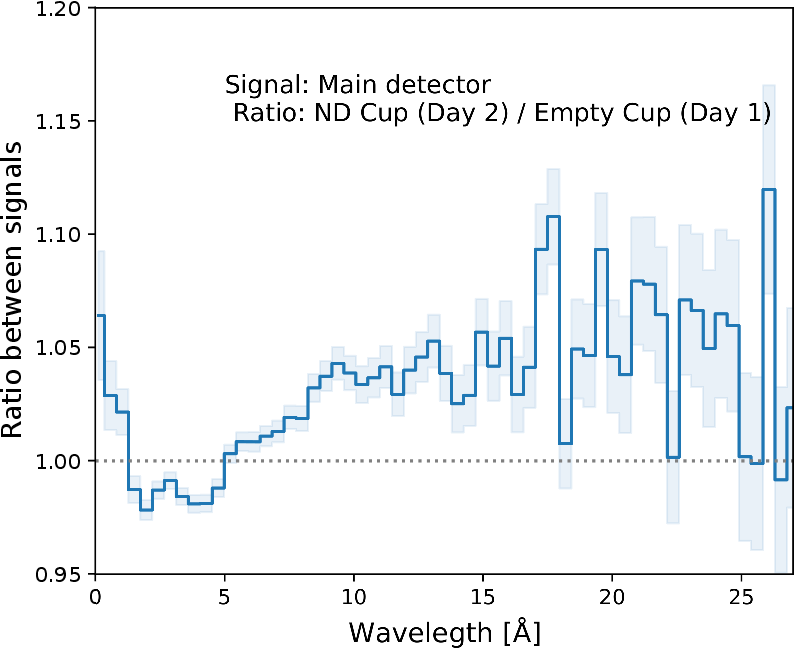}
        \subcaption{}
        \label{fig:Signal_ND_Empty_ratio_1}
    \end{subfigure}
    \caption{Run-averaged normalized counts and ratio between signals: (a) and (b) Monitor inside the TMR area, (c) and (d) main detector for the ND cup and Empty cup cases. The signal from the main detector is re-binned by a factor 4 (\SI{0.46}{\angstrom}).}
    \label{fig:ND_EMPTY_1}
\end{figure}

\begin{figure}[hbt!]
    \centering
        \begin{subfigure}[b]{0.49\textwidth}
        \includegraphics[width=\textwidth]{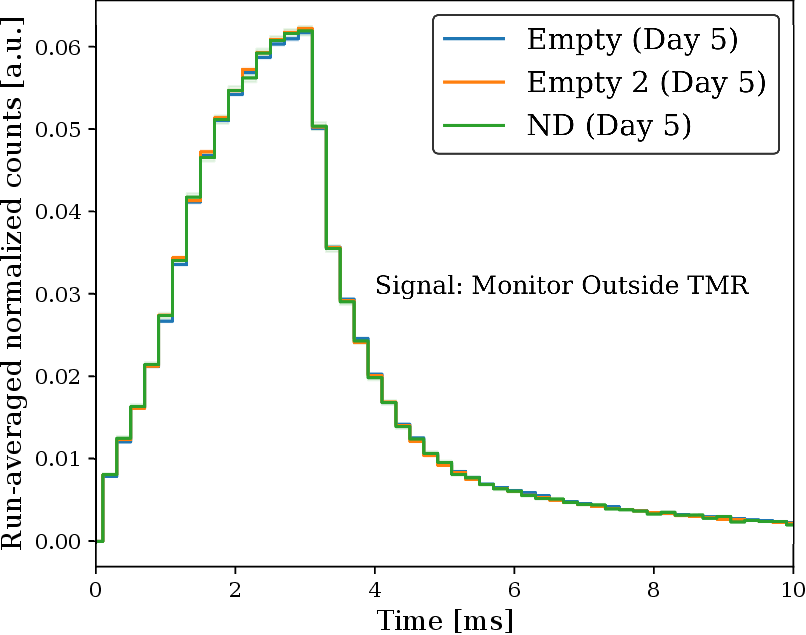}
        \subcaption{}
        \label{fig:Monitor_ND_Empty_2}
    \end{subfigure}
    \begin{subfigure}[b]{0.49\textwidth}      
        \includegraphics[width=\textwidth]{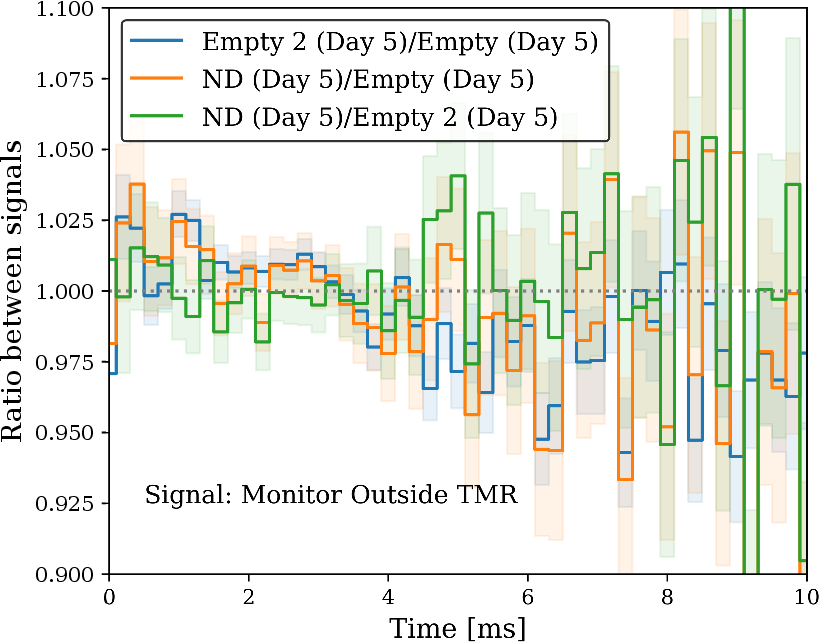}
        \subcaption{}
        \label{fig:Monitor_ND_Empty_ratio_2}
    \end{subfigure}
    \begin{subfigure}[b]{0.49\textwidth}
        \includegraphics[width=\textwidth]{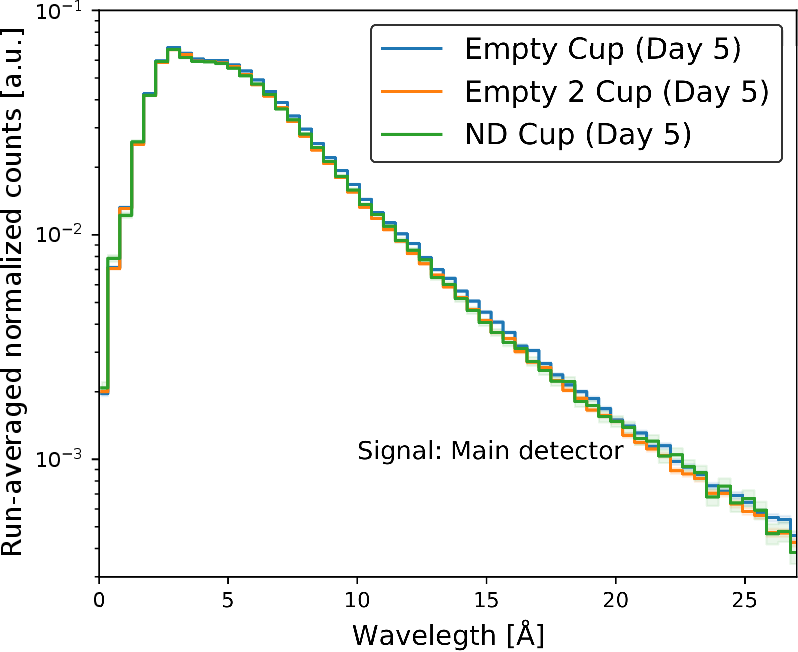}
        \subcaption{}
        \label{fig:Signal_ND_Empty_2}
    \end{subfigure}
    \begin{subfigure}[b]{0.49\textwidth}      
        \includegraphics[width=\textwidth]{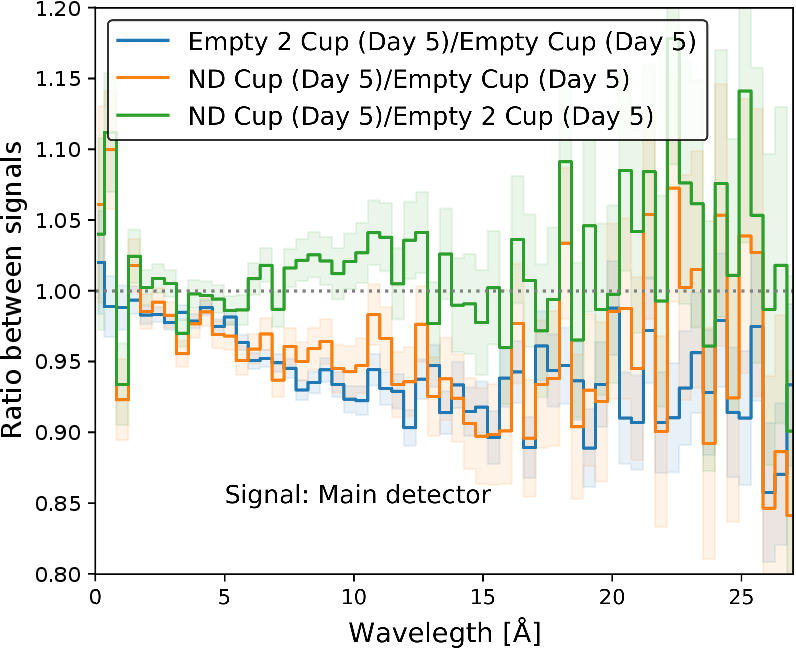}
        \subcaption{}
        \label{fig:Signal_ND_Empty_ratio_2}
    \end{subfigure}
    \caption{Run-averaged normalized counts and ratio between signals: (a) and (b) Monitor inside the TMR area, (c) and (d) main detector for the ND cup and Empty cups cases of Day 5. The signal from the main detector is re-binned by a factor 4 (\SI{0.46}{\angstrom}).}
    \label{fig:ND_EMPTY_2}
\end{figure}

\begin{figure}[hbt!]
    \centering
        \begin{subfigure}[b]{0.49\textwidth}
        \includegraphics[width=\textwidth]{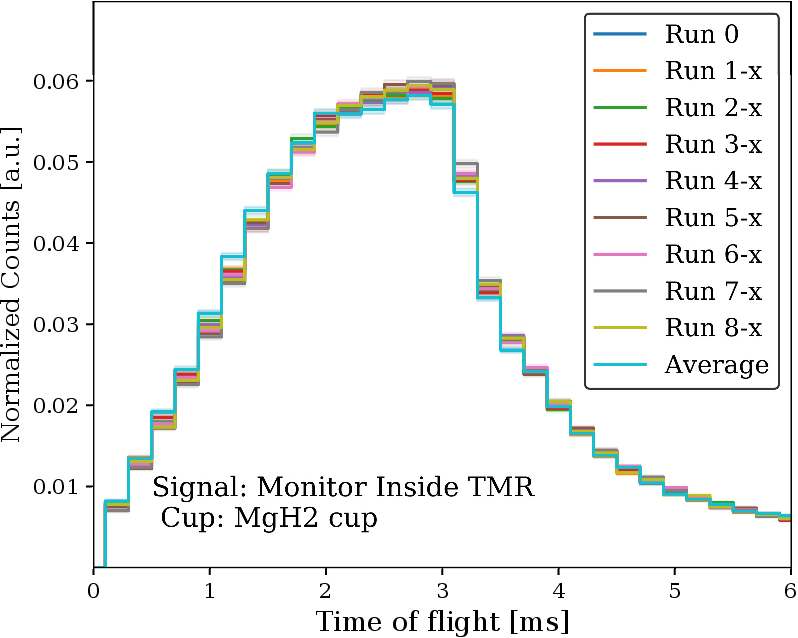}
        \subcaption{}
        \label{fig:monitor_inside_mgh2_inc}
    \end{subfigure}
    \begin{subfigure}[b]{0.49\textwidth}      
        \includegraphics[width=\textwidth]{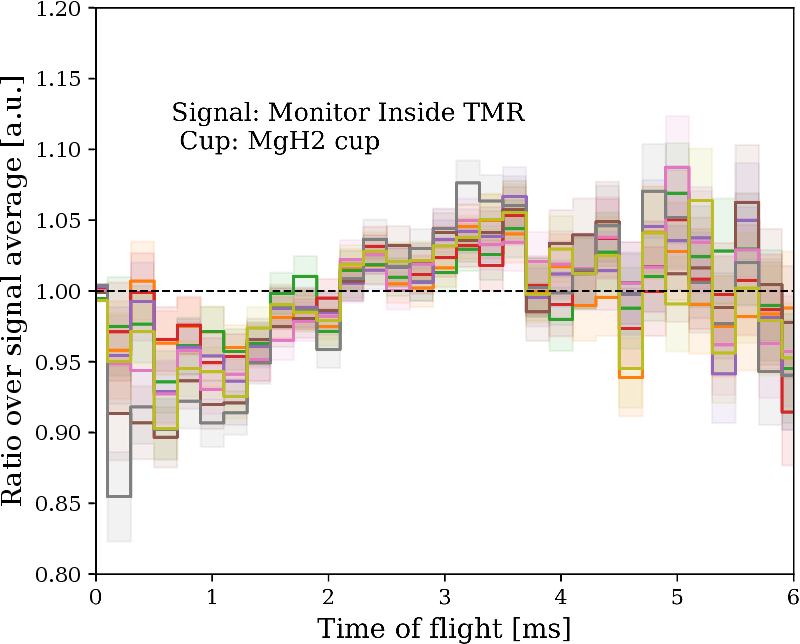}
        \subcaption{}
        \label{fig:monitor_inside_mgh2_inc_ratio}
    \end{subfigure}
    \begin{subfigure}[b]{0.49\textwidth}
        \includegraphics[width=\textwidth]{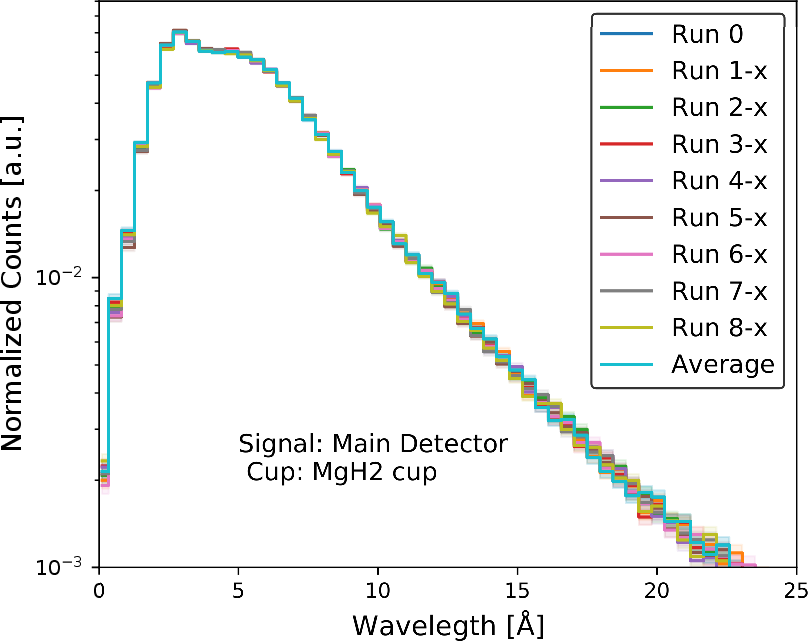}
        \subcaption{}
        \label{fig:Signal_mgh2_inc}
    \end{subfigure}
    \begin{subfigure}[b]{0.49\textwidth}      
        \includegraphics[width=\textwidth]{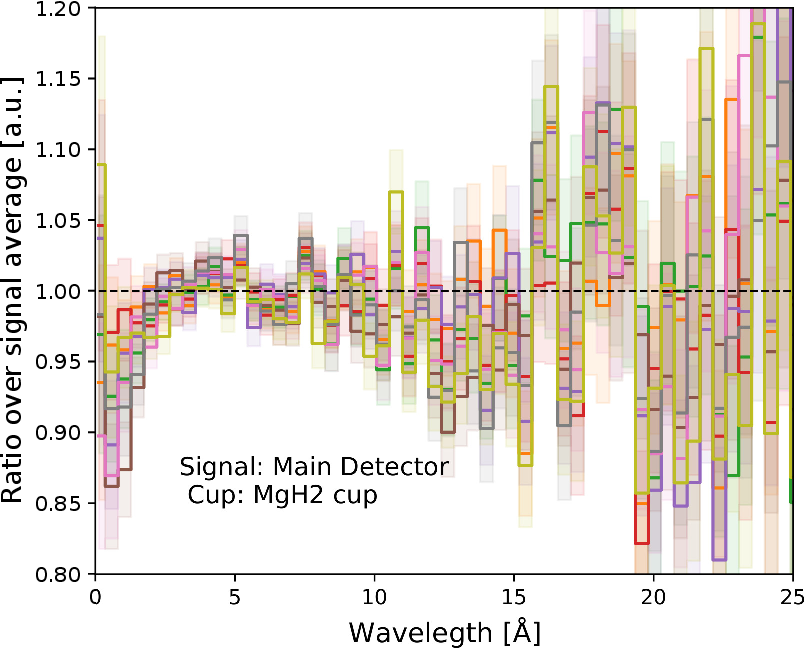}
        \subcaption{}
        \label{fig:Signal_mgh2_inc_ratio}
    \end{subfigure}
    \caption{Normalized counts for each run and ratio over average for the \ce{MgH_2} cup measurement for: (a) and (b) Monitor inside the TMR area, (c) and (d) main detector. The signal from the main detector is re-binned by a factor 4 (\SI{0.46}{\angstrom}). Only the first run is used for the average, so all the other runs are marked by an x.}
    \label{fig:MgH2_inc}
\end{figure}

\begin{figure}[hbt!]
    \centering
        \begin{subfigure}[b]{0.49\textwidth}
        \includegraphics[width=\textwidth]{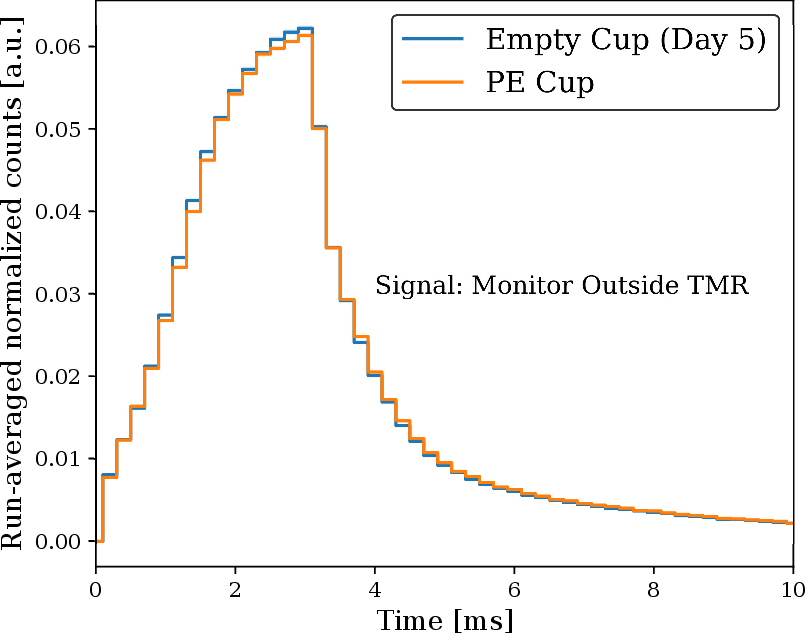}
        \subcaption{}
        \label{fig:Monitor_PE_Empty_1}
    \end{subfigure}
    \begin{subfigure}[b]{0.49\textwidth}      
        \includegraphics[width=\textwidth]{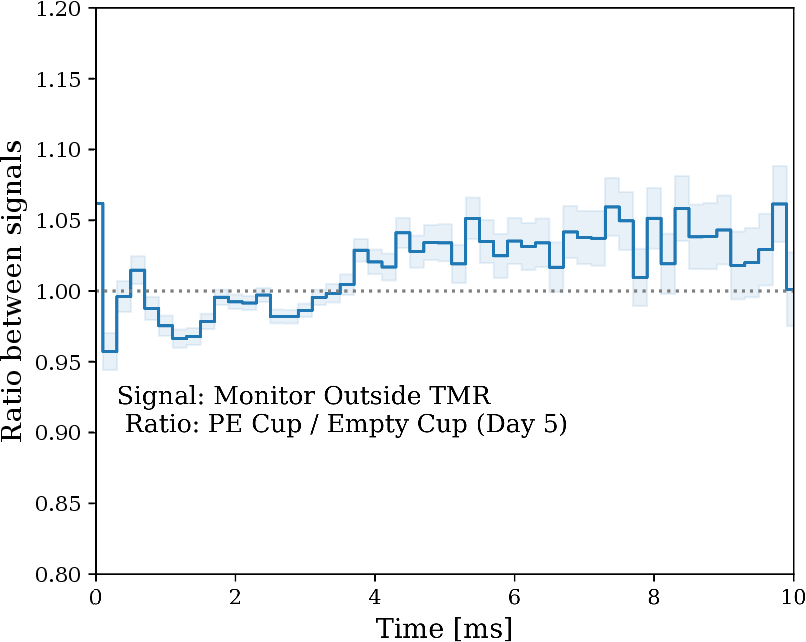}
        \subcaption{}
        \label{fig:Monitor_PE_Empty_ratio_1}
    \end{subfigure}
    \begin{subfigure}[b]{0.49\textwidth}
        \includegraphics[width=\textwidth]{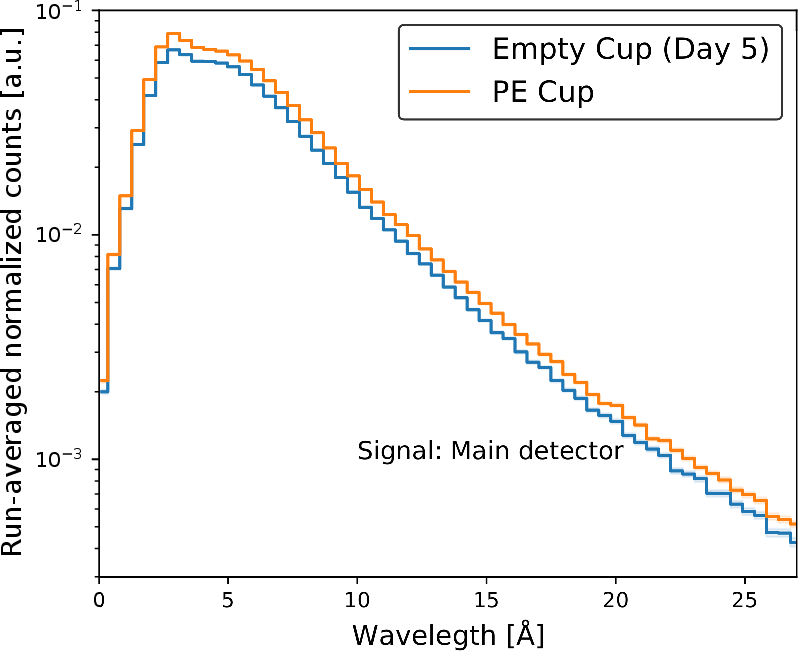}
        \subcaption{}
        \label{fig:Signal_PE_Empty_1}
    \end{subfigure}
    \begin{subfigure}[b]{0.49\textwidth}      
        \includegraphics[width=\textwidth]{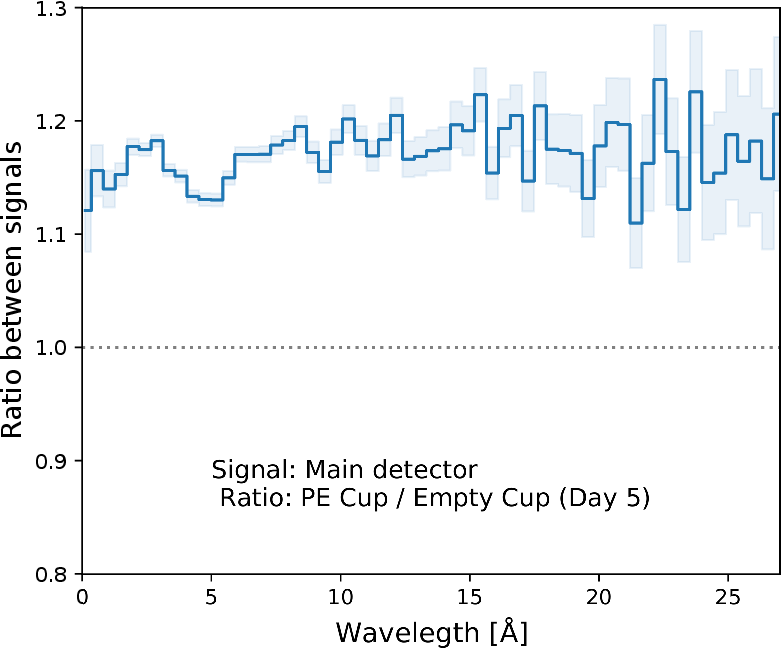}
        \subcaption{}
        \label{fig:Signal_PE_Empty_ratio_1}
    \end{subfigure}
    \caption{Run-averaged normalized counts and ratio between signals: (a) and (b) Monitor inside the TMR area, (c) and (d) main detector for the PE cup and first empty cup case of Day 5. The signal from the main detector is re-binned by a factor 4 (\SI{0.46}{\angstrom}).}
    \label{fig:PE_EMPTY_1}
\end{figure}

\paragraph{First attempt with ND cup}
The first two days we measured the empty cup and the ND cup for the first time. As already mentioned, at the end of the measurement with the empty cup, the accelerator pulse was unstable, but it was promptly fixed by the end of the day. The data collected during the accelerator drifts were not included in the average. For the ND cup, the first thing noticed after the cup exchange was an increase in the moderator temperature. The monitoring system was reporting a liquid He in-flow temperature of \SI{23.8}{K}. Before starting the runs, the temperature was restored to \SI{22}{K} by increasing the flow of the coolant. In \cref{fig:ND_EMPTY_1} the results are presented for both the monitor outside the TMR and the signal from the main detector. 
The ratio in \cref{fig:Signal_ND_Empty_ratio_1} is calculated with \cref{ratio} and shows a small, but statistically significant, increase in counts with the ND cup over the empty cup above \SI{5}{\angstrom}. The increase reaches up to 5\% at longer wavelength. Even if there is a difference in pulse shape in \cref{fig:Monitor_ND_Empty_ratio_1}, it is probably too small to produce the observed effect in the signal.
In any case, the increment at such a short wavelengths needs to be investigated further, since the reflectivity of the ND is expected to strongly increase above $\approx\SI{20}{\angstrom}$.
\paragraph{Second attempt with ND cup}
During the last day, three measurements with two cup exchanges were performed: empty, ND and empty cup a second time. The reason to plan the first two in the same day was to prevent as much as possible any source of instability, e.g. changes in the accelerator parameters. Unfortunately, even these measurements were not flawless in terms of stability.  During the first shutdown to exchange the cup, the moderator cooling control system crashed and the moderator warmed up to \SI{70}{K}. The methane mass flow was still 0, so most likely it was still frozen by the time it was brought back to \SI{22}{K} for ND cup runs. After 9 runs with the ND cup, a rise in the moderator temperature to \SI{24}{K} was noticed because the He flow was getting smaller. This caused a systematic drift in the different ND runs that produced, by the time of the last run, 4\% less counts integrated above \SI{5}{\angstrom}. For this reason, only the first \SI{20}{min} run, the closest to the nominal temperature of \SI{22}{K}, were used. For the second cup exchange, the empty cup was put back to rule out any possible spurious contribution from the first re-cooling of the day. The results for both the empty cups and the ND cup are presented in \cref{fig:ND_EMPTY_2}. The ratio between the empty cups in \cref{fig:Signal_ND_Empty_ratio_2} suggests that the two measurements are not equivalent and the unexpected warm up had, indeed, a negative effect on the initial conditions. In particular, the same trend is also visible when the ND cup signal is compared with the first empty cup. In this regard, the ratio between the signals from the first ND run and the second empty cup is probably the most reliable. The lower statistics, resulting from keeping only the first ND run, does not allow to measure any statistically significant effect above unity. Thus, we could not confirm the result from the first days (\cref{fig:Signal_ND_Empty_ratio_1}).
\paragraph{Magnesium hydride cup}
The cup with the \ce{MgH_2} was put in right before the accelerator ion cup broke. After the exchange and the restart of the accelerator, many tests were conducted on the accelerator parameters. The inconsistencies we report in \cref{fig:MgH2_inc}, where we assume that the average is given by the first run only, are most likely due to an unsettled accelerator pulse that lasted for the whole measurement. Unfortunately, we were not able to catch it on time to find a solution. Thus, we conclude that the ratio with the first empty cup signal taken on Day 5 would not be meaningful.  
\paragraph{Polyethylene cup}
The PE cup inserted on Day 4 benefited from stable accelerator conditions that remained until the next day when the first empty cup measurement on Day 5 was done. However, toward the end of the day the cooling control system crashed, and the moderator melted. After the restart of the system and the slow cooling of the source back to \SI{22}{K}, one more 20-min-long run was taken. The spectrum from this run was still distinctively a warm-up spectrum, so it is not considered for the average value. Early in the morning the next day, a quick run before the shutdown with the accelerator at \SI{10}{Hz} and \SI{1}{ms} was acquired and compared with the equivalent short runs of day before, showing good agreement. This suggests that the cooling process was most likely not over by the time of the last measurement, but it eventually reached the starting point over the course of the evening. Hence, the most reliable data set is the one taken before the melting of the moderator. The results shown in \cref{fig:PE_EMPTY_1}, and in particular \cref{fig:Signal_PE_Empty_1,fig:Signal_PE_Empty_ratio_1}, suggest that the polyethylene cup uniformly increases the thermal and cold production of the source by 15\%. Due to its hydrogen content, the removal of pre-moderation material in an intense high-energy neutron field to accommodate an advanced reflector is a key factor to take into account. 

\subsubsection{Discussion}
The experimental campaign at the HiCANS test station was part of a series of efforts by HighNESS to characterize the advanced reflectors. The models developed and used in the simulations so far confirmed their promising application to cold, very cold and ultracold neutron sources. The experiment, first attempt to verify these effects, was not originally part of the HighNESS plan and it was arranged at the eleventh hour when the opportunity arose. The insights gained from the design of the experiment at BNC were transposed to the setup for the test station, but the impromptu decision made it hard to perform an in-depth study beforehand. This fact may have contributed to the inconclusive results from the beamtime. Nevertheless, this beamtime helped in pinning down critical technical challenges involved in performing this kind of experiment and possible improvements:

\vspace{0.2cm}
\begin{itemize}[leftmargin=1cm]
    \item[-] The stability of the source should be the first concern, in terms of both the accelerator and moderator. A careful examination of the options has to take into account how to minimize the risk of disruptions and malfunctioning. In order to quantify the stability sought, the setup can be studied beforehand with simulations that include the relevant features;
    \item[-]the monitoring of stability of the systems can be improved with a calibrated measurement of the accelerator pulse and a reliable temperature control system;
    \item[-] The procedure of exchanging the cup required long shutdown/restart time of the facility. An out-of-pile source may possibly be advantageous in this regard, but background could become a problem. Minimizing the exchanges and maximizing the time window for each measurement would be undoubtedly beneficial; 
    \item[-] Studying the possibility of compressing the powder inside the cups, or inserting bigger cups, would increase the relative effect, hence relaxing the requirements on the stability. This contribution should also be investigated with appropriate simulations;
    \item[-] In order to better understand the limits of application, it would be interesting to test the cup concept in an in-beam setup with a pure cold source. This setup can possibly better highlight the properties of the advanced reflector over common hydrogen-rich materials, e.g. polyethylene, which in almost all cases can provide higher hydrogen density than advanced reflector powders. While this is undoubtedly advantageous for an in-pile source in a high-energy neutron field, a pure cold beam could emphasize the reflector properties without the premoderator component. 
\end{itemize}
\vspace{0.2cm}

%% file: conclusions.tex
\section{Conclusions}
\label{sec:conclusions}

%NOTE: open the chat in the upper-right hand corner for live commenting
%we may want to either shorten or delete the "which is place above..." clause in this first sentence or have an explicit discussion of both the uppar and lower locations.
The primary goal of the HighNESS project was the design of a second neutron source to complement the high-brightness bi-spectral source which is placed above the spallation target and serves all the neutron scattering instruments of the initial suite planned for ESS. To achieve this objective, the HighNESS project outlined a scientific case in its proposal, encompassing various neutron scattering techniques and including the NNBAR experiment, which aims to detect neutron to antineutron oscillation.

At the heart of the project is the design of a high-intensity cold source, which will directly serve instruments and experiments, but also secondary VCN and UCN sources. Without a source delivering at least several times the intensity of the upper moderator, it would not be possible to achieve the ambitious scientific goals outlined in the proposal. The choice of liquid deuterium, used for decades at reactors and continuous spallation sources, was confirmed to be the only viable option for such a source. 
The final design is able to deliver an intensity above 4\,\AA~of a factor of 10 greater than the upper moderator, thus exceeding the proposal expectations.  This remarkable outcome is crucial for enabling the NNBAR experiment to reach its anticipated sensitivity levels, as discussed further in HighNESS Conceptual Design Report Volume II. Additionally, the cold source features an extended emission surface compared to the upper moderator. This property has been duly exploited in the design of novel SANS and imaging instruments that will complement and outperform the ones under construction at ESS .

The design of the cold source was an iterative process between the neutronic and engineering teams. A detailed engineering study of the cold moderator was performed including stress analysis and fluid dynamics simulations, to operate and cool the liquid deuterium, the aluminum structure and the cold beryllium filter that increases the cold flux for NNBAR. The design was reviewed by an international team of experts and a few further investigations are recommended, which are detailed at the end of  
\cref{ch:1}.

% WP7 section, this seemed like the best spot in the conclusion. Review. Approved by Markus
To guide the optimization of the cold source, the HighNESS project included a design of a small suite of conceptual instruments that takes advantage of the increased source intensity. The suite included two SANS instruments -- one using a conventional design and one based on focusing with Wolter optics -- as well as an imaging instrument. 
The performance of these instruments were evaluated with different proposed models for the source, and the results were considered when deciding which source models to pursue further.
With the current design of the cold source it was found that the Wolter optics based SANS instrument in particular would have world-leading capabilities with impressive flux, bandwidth, minimum $Q$, and resolution simultaneously. The larger source also allowed for a simple pinhole imaging beamline that would have a large homogeneous field of view, still offering modest wavelength resolution for advanced imaging techniques, and thus complementing the ESS instrument ODIN, which is under construction.

In addition to the cold source developments, the HighNESS project aimed to ambitiously extend the range of available neutrons for ESS users into the VCN and UCN ranges. The project achieved significant results in both of these areas. 

The desire for an intense source of VCNs has been a longstanding aspiration within the user community, spanning at least two decades. The realization of such a source has faced challenges primarily due to the lack of knowledge regarding the properties of low-temperature materials, which were considered the most promising candidates for such a source. To enable reliable Monte Carlo simulations and, consequently, the design of a neutron source, the availability of thermal scattering cross-section libraries is essential. In the HighNESS project, these libraries were developed for candidate VCN moderator and reflector materials, including deuterated clathrate hydrates and nanodiamonds. Additionally, new libraries for solid deuterium were developed through collaboration with another EU program called EURIZON.
The development of these libraries stands as a significant achievement within the HighNESS project. It can be deemed a major breakthrough in the field of source design. HighNESS has not only created these thermal scattering libraries but also made them readily accessible to the community, adhering to the open-source policy outlined in the project proposal. This step was taken to ensure the wider dissemination of the project's outcomes. This contribution unfolded during a period of notably vibrant development, partly due to recent advancements in compact neutron source design.

Thanks to these developments, it was possible to tackle the task of designing a VCN source. The results were very compelling. As in the case of the cold source development, another {\it factor of 10} manifested: this time, it is the increase in the brightness for neutrons above 40 \AA, compared to the brightness of a conventional cold source placed in the same location below the spallation target. Such a breakthrough is due to the use of solid deuterium at 5\,K, in combination with nanodiamonds, and to its reduced upscattering of VCNs, compared to liquid deuterium at 22\,K which is used for the intense cold source.  If duly exploited, this type of dedicated VCN source could be a game changer for several neutron scattering techniques such as SANS and spin-echo. 
% WP7 sentence on VCN optics % Review
However, both SANS and spin-echo beamlines using a VCN source would need optics that counteract the effects of gravity. Such systems were already investigated as a part of the project and highly encouraging results were obtained with focusing optics corrected by prisms. 

The scientific case, along with general design concepts for VCN and UCN sources, underwent extensive discussions during two dedicated workshops {\footnote{\url{https://indico.esss.lu.se/event/2810/}} {\footnote{\url{https://indico.esss.lu.se/event/3195/}} organized at ESS by HighNESS in collaboration with the League of European Neutron Sources (LENS). These workshops saw active participation from over 100 members of the community. One of the most encouraging outcomes of these workshops was the multitude of ideas generated for UCN sources at ESS. This not only underscores the potential that ESS offers but also validates the feasibility of the upgrade paths explored by HighNESS. Following the workshop's directions, the HighNESS team conducted neutronic design studies for most of the proposed UCN source options. Both in-pile and in-beam designs were considered, utilizing two commonly used materials for UCN production: solid deuterium and superfluid helium. The results have been highly promising, positioning ESS for a world-leading UCN source.

The choice of the optimal UCN source for ESS depends on various factors, including technical feasibility (e.g., cooling of the source and UCN extraction) and the integration of the three source categories developed in HighNESS. The project explored at least eight potential integration options for CN, VCN, and UCN sources, as discussed in detail in \cref{sourceintegretion} . These options should be subject to further study and development in future projects, and they represent a valuable legacy of the HighNESS project.

%The HighNESS project has thoroughly and carefully exploited a feature unique only to ESS: the large beamport (LBP), which offers a solid angle of view of the source, orders of magnitude larger than a conventional neutron guide. The LBP was originally intended for NNBAR. However, it proves to be an exceptional location for UCN sources as well; in fact, two designs of UCN sources, both based on He-II, have been performed at the LBP. This would mean, however, that a UCN source placed in the LBP would have to start installation and operation after the NNBAR experiment is completed, which could be a problem since the UCN community would have to wait several years before starting its scientific program. A careful coordination effort between the different communities would be required, but this is certainly a possibility.

%Another example concerns the VCN source. As mentioned above, we have found that the best design consists of a solid deuterium moderator, which would completely replace the high-intensity liquid deuterium moderator. In that scenario, one would envision a program based on the use of a high-intensity cold source, which would eventually be replaced with the VCN source, which is still quite competitive as cold moderator, but with a brightness for VCNs an order of magnitude higher. Such an operations program would flow logically in parallel to the required development of the VCN source. %i understand this last sentence, but it might work better without repeating "development"

To achieve its goals, the HighNESS project also conducted an extensive experimental program. This program involved the measurement of cross sections for various materials relevant to the project, with a particular focus on clathrate hydrates and graphite compounds. These measurements were conducted at facilities such as ILL (using the IN5 and PANTHER beamlines) and PSI (utilizing the BOA beamline).

Another aspect of the experimental program involved the measurement of prototype advanced reflectors designed to enhance the VCN flux, as outlined in the HighNESS proposal. In this case, a significant challenge arose due to the limited availability of operational facilities where prototype measurements could be conducted. At the time the proposal was written, there was only one such facility, which was the LENS facility in Indiana. Unfortunately, this facility shut down at the beginning of the HighNESS project. Another facility under construction was located at the Budapest Neutron Center, and the HighNESS project provided strong support for its development. As part of this effort, the project organized a collaborative measurement involving a cold moderator with advanced reflectors made of nanodiamonds and MgH$_2$. However, it is important to note that this moderator test facility was still in the developmental phase throughout the duration of the HighNESS project. During this time, only preliminary measurements involving dose rates and background were conducted using an ambient water moderator. The project team remains hopeful that the planned tests can be carried out in the near future once the facility reaches full operational capacity. Another facility that came online recently (December 2022) is the JULIC Neutron Platform in Forschungszentrum J\"ulich. The HighNESS team had the opportunity to perform a first pilot measurement of a simplified setup of the Budapest measurement. 
%Such exploratory measurements were benefici the first steps towards a strong experimental program for the development of the proposed moderators.

%we use "fully" again in the next sentence at the beginning of "Outlook", so i am deleting it here even though it works well -Ben

Finally, in addition to delivering outstanding results as promised in the proposal, the HighNESS consortium has devoted considerable time and effort to training young scientists, including both Ph.D. students and postdoctoral researchers. This training involved not only supervisory roles but also dedicated activities such as organizing the first International School on Thermal Neutron Scattering Kernel Generation \footnote{\url{https://indico.esss.lu.se/event/3096/}}, which took place at ESS from May 22 to May 26, 2023 and attracted 40 participants world wide.

%\begin{itemize}

%\item Consolidation of the scientific case

%\item{Development of thermal scattering libraries}

%\item{Measurement of cross sections}

%\item{design of intense cold source}

%\item{Breakthrough design of VCN source}

%\item{design of several UCN sources}

%\item{development scenarios of HighNESS sources and integration with main source}

%\item{enngineering design of CN source}

%\item{Engineering design of UCN source}

%\item{Neutron scattering instruments design}

%\item{NNBAR}

%\end{itemize}

\section{Outlook}

There are a number of developments recommended to fully capitalize on the success of the HighNESS project, and some of them have already been hinted in the previous section.

\textbf{Source design.} One aspect concerns the technical challenges related to operating the sources at the high power of ESS. ESS will operate for many years at 2\,MW, with an eventual upgrade to 5\,MW. The liquid deuterium moderator will most likely operate at 2\,MW 
but currently, there is no design that can guarantee its functionality at 5\,MW. Toward the conclusion of the HighNESS project, the design was significantly simplified, including a simpler shape for the reentrant hole and fewer flow channels inside the vessel. However, further work is required in this area to develop a high-performing design that can operate reliably at 5\,MW. 

For a solid deuterium moderator the situation is even more uncertain, since such a moderator has never been operated at the MW power level. We have identified and started to investigate some possible engineering solutions however the development of a VCN source capable of operating in a high radiation environment, would require a major R\&D program.

\textbf{Instrument design.} A successful design of a VCN source should be accompanied by a dedicated design of experiments using VCN beams. In fact, most of the instrument design in HighNESS concerned the high intensity cold neutron beams delivered by the liquid deuterium moderator. The study of an optics system to transport VCNs and counteract the effect of gravity mentioned above showed promising results, and could be the basis to begin designing dedicated VCN instruments for neutron scattering. The exploration of the NNBAR with VCN design should be accompanied by the development of a dedicated VCN fundamental physics program.

\textbf{Experimental program.} HighNESS has provided major input toward the development of new facilities to test moderators, in particular the moderator test facility at the Budapest Research Center and the Big Karl facility in J\"ulich. 
A key necessary step in the continuation of the research initiated by HighNESS would be the realization of prototypes of VCN sources based on solid deuterium and nanodiamonds, as well as deuterated clathrate hydrates. Prior to the implementation of such prototypes in the in-pile positions of the aforementioned moderator test facilities, characterization with cold and very cold neutron beams are envisaged. Suitable sites for such investigations are the cold neutron beam facility PF1B and the very cold neutron beam of PF2 at the ILL.
%not abbreviating cold neutron and very cold neutron in the sentences above makes sense, because it's in another context, but we might want to keep it all abbreviated, for clarity. -ben
%Additionally, further prototyping and tests of neutron optical transport by nested mirror systems for an in-beam UCN source and an engineering study of implementation of such a source at the ESS large beam-port would be needed.
%, and research would  be required even for an in-beam solution, where the cooling of the source is much easier but there might be other challenges.

\textbf{Measurement and calculation of cross sections.} HighNESS has established a process to calculate thermal scattering libraries and perform experimental validations. More developments are certainly recommended to deepen the experimental and theoretical knowledge of such materials, as well as to support the inclusion of these materials in Monte Carlo simulations. Notable examples are the measurement of the cross sections of solid deuterium at 5\,K for VCNs, and superfluid helium below 1\,K for UCNs. Furthermore, measurements on binary clathrate hydrates, hosting both fully deuterated THF and molecular oxygen are needed to benchmark the developed scattering kernels for these compounds. 
%This can be done using the instrumentation described above, namely neutron time of flight spectroscopy, accompanied by neutron diffraction. 

\section{Acknowledgements }
%WP2
We would like to acknowledge the Paul Scherrer Institute for access to beam time at the facility under proposal numbers 20212829, 20221100 and 20222657. We acknowledge MAX IV Laboratory for time on Beamline COSAXS under Proposal 20221553. Research conducted at MAX IV, a Swedish national user facility, is supported by the Swedish Research council under contract 2018-07152, the Swedish Governmental Agency for Innovation Systems under contract 2018-04969, and Formas under contract 2019-02496. We would also like to acknowledge the support by ILL services in performing the experiments carried out there.
We would also like to thank the ESS Manufacturing Workshop and the ESS sample environment group for machining the triaxial and cylindrical sample holders.
HighNESS is funded by the European Union Framework Programme for Research and Innovation Horizon 2020, under grant agreemeent 951782